\documentclass{ldbc}

\usepackage{multirow}
\usepackage{float}
\usepackage{amsfonts}
\usepackage{rotating}
\usepackage{pifont}
\usepackage{rotating}
\usepackage{subcaption}
\usepackage{amsmath}
\usepackage{longtable}
\usepackage{tabularx}
\usepackage{varwidth}
\usepackage{hyperref}
\usepackage{xspace}
\usepackage[table]{xcolor}
\usepackage{listings}
\usepackage[export]{adjustbox}
\usepackage{array}
\usepackage{enumitem}
\usepackage{etoolbox}

\usepackage{todonotes}
\presetkeys{todonotes}{inline}{}

\usepackage[normalem]{ulem}
\usepackage{cellspace}
\usepackage{rotate}
\usepackage{MnSymbol}
\usepackage{xifthen}
\usepackage{numprint}
\usepackage[scale=0.78,ttdefault=true]{AnonymousPro}

\usepackage{bm}

\usepackage{caption}
\usepackage{tikz}
\usetikzlibrary{arrows}
\usetikzlibrary{arrows.meta}
\usetikzlibrary{patterns}

\usepackage[eprint=false]{biblatex}
\renewbibmacro*{doi+eprint+url}{%
	\printfield{doi}%
	\newunit\newblock%
	\iftoggle{bbx:eprint}{%
		\usebibmacro{eprint}%
	}{}%
	\newunit\newblock%
	\iffieldundef{doi}{%
		\usebibmacro{url+urldate}}%
	{}%
}

\usepackage{ccicons}
\usepackage{circledsteps}

\providecommand{\tightlist}{%
    \setlength{\itemsep}{0pt}\setlength{\parskip}{0pt}}

\setlist[]{noitemsep, topsep=5pt}

\newcommand{\asc}{\uparrow}
\newcommand{\desc}{\downarrow}

\newcommand{\yedscale}{0.52}
\newcommand{\patternscale}{0.43}

\newcommand{\datagen}{Datagen\xspace}
\newcommand{\ldbcsnb}{LDBC SNB\xspace}

\newcommand{\snbbi}{SNB BI\xspace}

\newcommand{\yes}{$\bigotimes$\xspace}

\newcommand{\no}{$\bigcircle$\xspace}

\newcommand{\interval}[2]{\ensuremath{\textcolor{green}{\big[ #1}, \ \textcolor{red}{#2 \big)} }}

\newcommand{\type}[1]{\textsf{#1}}
\newcommand{\constant}[1]{\textbf{#1}}
\newcommand{\variable}[1]{\mathit{#1}}
\newcommand{\instance}[1]{\mathsf{#1}}

\newcommand{\created}{\ast}
\newcommand{\deleted}{\dagger}

\newcommand{\varbound}[2]{{#2}\variable{#1}}
\newcommand{\varc}[1]{\varbound{#1}{\created}}
\newcommand{\vard}[1]{\varbound{#1}{\deleted}}

\newcommand{\varn}[1]{\variable{n}_\mathrm{#1}}
\newcommand{\varcn}[1]{\varbound{\varn{#1}}{\created}}
\newcommand{\vardn}[1]{\varbound{\varn{#1}}{\deleted}}

\newcommand{\instbound}[3]{{#3}\instance{#1}_\instance{#2}}
\newcommand{\instc}[2]{\instbound{#1}{#2}{\created}}
\newcommand{\instd}[2]{\instbound{#1}{#2}{\deleted}}

\newcommand{\xSS}{\constant{SS}\xspace}
\newcommand{\xSE}{\constant{SE}\xspace}
\newcommand{\xNC}{\constant{NC}\xspace}
\newcommand{\xBL}{\constant{BL}\xspace}

\newcommand{\eComment}{\variable{comm}\xspace}
\newcommand{\eFlashmobEvent}{\variable{fme}\xspace}
\newcommand{\eForum}{\variable{forum}\xspace}
\newcommand{\eHasMember}{\variable{hm}\xspace}
\newcommand{\eHasModerator}{\variable{hmd}\xspace}
\newcommand{\eKnows}{\variable{knows}\xspace}
\newcommand{\eLikes}{\variable{likes}\xspace}
\newcommand{\eMessage}{\variable{m}\xspace}
\newcommand{\ePerson}{\variable{p}\xspace}
\newcommand{\ePost}{\variable{post}\xspace}

\newcommand{\tCity}{\type{City}\xspace}
\newcommand{\tCities}{\type{Cities}\xspace}
\newcommand{\tCountry}{\type{Country}\xspace}
\newcommand{\tCountries}{\type{Countries}\xspace}
\newcommand{\tComment}{\type{Comment}\xspace}

\newcommand{\tComments}{\type{Comments}\xspace}

\newcommand{\tForum}{\type{Forum}\xspace}
\newcommand{\tForums}{\type{Forums}\xspace}

\newcommand{\tHasMember}{\type{hasMember}\xspace}
\newcommand{\tHasModerator}{\type{hasModerator}\xspace}
\newcommand{\tHasTag}{\type{hasTag}\xspace}
\newcommand{\tKnows}{\type{knows}\xspace}
\newcommand{\tLikes}{\type{likes}\xspace}
\newcommand{\tMessage}{\type{Message}\xspace}
\newcommand{\tMessages}{\type{Messages}\xspace}
\newcommand{\tPerson}{\type{Person}\xspace}
\newcommand{\tPersons}{\type{Persons}\xspace}
\newcommand{\tPost}{\type{Post}\xspace}
\newcommand{\tPosts}{\type{Posts}\xspace}
\newcommand{\tTag}{\type{Tag}\xspace}
\newcommand{\tTags}{\type{Tags}\xspace}

\newcommand{\tPhotos}{\type{Photos}\xspace}
\newcommand{\tReplyOf}{\type{replyOf}\xspace}

\def\shadedBox(#1,#2,#3){

  \fill[pattern=north west lines,pattern color=grey] (#1,#2) --  (#1,#2 - #3) -- (#1 + 0.3,#2 - #3) --  (#1 + 0.3,#2);
   \draw [grey,thin,dashed] (#1,#2)  -- (#1,#2 - #3);
   \draw [grey,thin,dashed] (#1 + 0.3,#2) -- (#1 + 0.3,#2 - #3);
   \draw [grey,line width=0.6mm] (#1,#2 - #3) -- node[midway,below,grey] {$\Delta$} (#1 + 0.3,#2 - #3);

}

\definecolor{Person}{HTML}{fdb462}
\definecolor{Message}{HTML}{bebada}
\definecolor{Forum}{HTML}{b3de69}
\definecolor{Comment}{HTML}{80b1d3}
\definecolor{Post}{HTML}{fb8072}
\definecolor{Company}{HTML}{ccebc5}
\definecolor{University}{HTML}{ffed6f}
\definecolor{City}{HTML}{8dd3c7}
\definecolor{Tag}{HTML}{fccde5}
\definecolor{Country}{HTML}{ffffb3}

\definecolor{ldbcpale}{HTML}{86cd7d}
\definecolor{ldbc}{HTML}{439539}
\definecolor{grey}{rgb}{0.52, 0.52, 0.51}
\definecolor{red}{rgb}{0.7, 0.11, 0.11}
\definecolor{blue}{rgb}{0.0, 0.0, 0.55}
\definecolor{green}{rgb}{0.0, 0.42, 0.24}

\definecolor{mydarkyellow}{HTML}{ffc329}
\definecolor{mylightyellow}{HTML}{fee090}

\definecolor{mydarkblue}{HTML}{80d6ed}
\definecolor{mylightblue}{HTML}{e0f3f8}

\newcommand{\variant}[1]{\textit{#1}\xspace}
\newcommand{\variantA}{\variant{a}}
\newcommand{\variantB}{\variant{b}}

\newcommand{\tpcH}{\mbox{TPC-H}\xspace}
\newcommand{\tpcDS}{\mbox{TPC-DS}\xspace}

\newtoggle{StandaloneWorkloadSpecification}
\togglefalse{StandaloneWorkloadSpecification}

\newcommand{\param}[1]{\texttt{\$#1}{}}

\newcommand{\interactivevone}{Interactive~v1\xspace}
\newcommand{\interactivevtwo}{Interactive~v2\xspace}

\newcommand{\snbinteractivevtwo}{SNB Interactive~v2\xspace}

\newcommand{\snbOperation}[1]{{\ttfamily \fontseries{m}\selectfont #1}\xspace}
\newcommand{\INS}[1][]{\snbOperation{INS#1}}
\newcommand{\DEL}[1][]{\snbOperation{DEL#1}}
\newcommand{\CR}[1][]{\snbOperation{IC #1}}
\newcommand{\SR}[1][]{\snbOperation{IS #1}}

\newcommand{\Messages}{\type{Messages}\xspace}
\newcommand{\Person}{\type{Person}\xspace}

\newcommand{\Persons}{\type{Persons}\xspace}

\newcommand{\Countries}{\type{Countries}\xspace}

\newcommand{\hasCreator}{\type{hasCreator}\xspace}

\newcommand{\knows}{\type{knows}\xspace}

\newcommand{\snbperson}{\texttt{person}\xspace}
\newcommand{\snbfriend}{\texttt{friend}\xspace}
\newcommand{\snbcomment}{\texttt{comment}\xspace}

\definecolor{parameter}{HTML}{e41a1c}
\definecolor{result}{HTML}{377eb8}
\definecolor{sort}{HTML}{4daf4a}

\definecolor{IC}{HTML}{ffffcc}
\definecolor{IS}{HTML}{fed9a6}
\definecolor{INS}{HTML}{e5d8bd}
\definecolor{BI}{HTML}{decbe4}
\definecolor{DEL}{HTML}{33ccff}

\newboolean{standalone}

\reversemarginpar
\newcommand{\currentQueryCard}{n/a}
\newcommand{\queryRefCard}[3]{
	\ifthenelse{
		\equal{\currentQueryCard}{#1}
	}{%
		\colorbox{white}{\tt #2 #3}%
	}{%
		\ifthenelse{
			\boolean{standalone}
		}{%
			\href{https://ldbcouncil.org/ldbc_snb_docs/#1.pdf}{\colorbox{#2}{\tt #2 #3}}%
		}{%
			\hyperref[sec:#1]{\colorbox{#2}{\tt #2 #3}}%
		}
	}%
}

\newcommand{\attributeNumberWidth}{0.33cm}
\newcommand{\attributeColumnWidth}{2.5cm}
\newcommand{\typeColumnWidth}{2.7cm}
\newcommand{\descriptionColumnWidth}{10.3cm}
\newcommand{\largeDescriptionColumnWidth}{13cm}

\newcommand{\tableHeaderFirst}[1]{\multicolumn{1}{|c|}{\bf #1}}
\newcommand{\tableHeader}[1]{\multicolumn{1}{c|}{\bf #1}}

\newcommand{\queryCardWidth}{17cm}
\newcommand{\queryPropertyCell}{\small \sf \centering}
\newcommand{\queryPropertyCellWidth}{1.48cm}

\newcommand{\attributeCardWidth}{14.66cm}
\newcommand{\typeWidth}{2.04cm}

\newcommand{\paramNumberCell}{\cellcolor{parameter}\color{white}\footnotesize}
\newcommand{\resultNumberCell}{\cellcolor{result}\color{white}\footnotesize}
\newcommand{\sortNumberCell}{\cellcolor{sort}\color{white}\footnotesize}

\newcommand{\directionCell}{\cellcolor{gray!20}}
\newcommand{\resultOriginCell}{\tt}
\newcommand{\edgeDirectionCell}{\tt}

\newcommand{\varNameText}{\tt}
\newcommand{\varNameCell}{\varNameText}

\newcommand{\typeText}{\footnotesize\sf}
\newcommand{\typeCellBase}{\cellcolor{gray!20}\typeText}
\newcommand{\typeCell}{\typeCellBase\raggedright}

\newcommand{\chokePoint}[1]{\hyperref[choke_point_#1]{#1}}

\newcommand{\innerCardVSpace}{\vspace{1.1ex}}
\newcommand{\queryCardVSpace}{\vspace{2ex}}

\newcolumntype{Y}{>{\raggedright\arraybackslash}X}

\newcolumntype{C}[1]{>{\centering\let\newline\\\arraybackslash\hspace{0pt}}m{#1}}

\newcolumntype{M}{>{\begin{varwidth}{3.8cm}}Sl<{\end{varwidth}}}

\newcommand{\attributeTable}[3]{
	\vspace{1ex}
	\begin{tabularx}{\linewidth}{|l|Y|}
	\hline
	\bf Attribute   & \varNameCell  #1 \\ \hline
	\bf Type        & \typeCellBase #2 \\ \hline 
	\bf Description &               #3 \\
	\hline
	\end{tabularx}}

\definecolor{lightgray}{RGB}{242,242,242}
\definecolor{keywordcolor}{RGB}{0,0,160}
\definecolor{commentcolor}{RGB}{0,128,64}
\definecolor{instruction}{HTML}{107762}

\lstdefinelanguage{cypher}
{
	morekeywords={
		MATCH, OPTIONAL, WHERE, NOT, AND, OR, XOR, RETURN, DISTINCT, ORDER, BY, ASC, ASCENDING, DESC, DESCENDING, UNWIND, AS, UNION, WITH, ALL, CREATE, DELETE, DETACH, REMOVE, SET, MERGE, SET, SKIP, LIMIT, IN, CALL, CASE, WHEN,
		INDEX, DROP, UNIQUE, CONSTRAINT, EXPLAIN, PROFILE, START, FOREACH, 
		GROUP, HAVING,
	},
	sensitive=true,
	morecomment=[l]{//},
	morecomment=[s]{/*}{*/},
	morestring=[b]{"},
	literate=*
	  {<<}{\color{instruction}\guillemotleft{}}{1}
	  {>>}{\textcolor{instruction}{\guillemotright{}}\color{black}}{1}
}

\lstdefinelanguage{sparql}{
	morekeywords={SELECT, DISTINCT, WHERE, OPTIONAL, FILTER, NOT, EXISTS, MINUS, sameTerm, bound},
}

\WPTitle{Social Network Benchmark Task Force}

\renewcommand{\delIDText}{}

\dissPU 

\natR 

\author{LDBC Social Network Benchmark Task Force}


\bibliography{ms}

\setboolean{standalone}{false}

\abstract{
LDBC's Social Network Benchmark (\ldbcsnb) is an effort intended to test
various functionalities of systems used for graph-like data management. For this,
\ldbcsnb uses the recognizable scenario of operating a social network, characterized by
its graph-shaped data.

\ldbcsnb consists of two workloads that focus on different
functionalities: the Interactive workload (interactive transactional queries)
and the Business Intelligence workload (analytical queries). 

This document contains the definition of both workloads. This includes a detailed
explanation of the data used in the \ldbcsnb benchmark, a detailed description
for all queries, and instructions on how to generate the data and run the
benchmark with the provided software.
}

\execSummary{

The new data economy era, based on complexly structured, distributed and large
datasets, has brought on new demands on data management and analytics.  As a
consequence, new industry actors have appeared, offering technologies specially
built for the management of graph-like data. Also, traditional database
technologies, such as relational databases, are being adapted to the new
demands to remain competitive.

LDBC's Social Network Benchmark (\ldbcsnb) is an industrial and academic
initiative, formed by principal actors in the field of graph-like data
management. Its goal is to define a framework where different graph based
technologies can be fairly tested and compared, that can drive the
identification of systems' bottlenecks and required functionalities, and can
help researchers to open new research frontiers.

The philosophy around which \ldbcsnb is designed is to be easy to
understand, flexible and cheap to adopt. For all these reasons,
\ldbcsnb will propose different workloads representing all the usage scenarios
of graph-like database technologies, hence, targeting systems of different
nature and characteristics.  In order increase its adoption by industry and
research institutions, \ldbcsnb provides all necessary software, which are
designed to be easy to use and deploy at a small cost.

This document contains:
\begin{itemize}
\item A detailed specification of the data used in the whole \ldbcsnb benchmark.
\item A detailed specification of the workloads.
\item A detailed specification of the execution rules of the benchmark.
\item A detailed specification of the auditing rules and the full disclosure
  report's required contents.
\end{itemize}
}

\begin{document}

\maketitle

\chapter*{Acknowledgments}
Special thanks to all the people that have contributed to the development of this benchmark suite:
\begin{itemize}
  \item Renzo Angles (Universidad de Talca)
  \item J\'anos Benjamin Antal (Budapest University of Technology and Economics)
  \item Alex Averbuch (Neo4j)
  \item Altan Birler (TUM)
  \item Peter Boncz (Vrije Universiteit Amsterdam, CWI)
  \item M\'arton B\'ur (McGill University)
  \item Orri Erling (OpenLink Software)
  \item Andrey Gubichev (Technische Universit\"at M\"unchen)
  \item Vlad Haprian (Oracle Labs)
  \item Moritz Kaufmann (Technische Universit\"at M\"unchen)
  \item Josep Llu\'is Larriba Pey (Universitat Polit\`ecnica de Catalunya)
  \item Norbert Mart\'inez (Huawei Technologies)
  \item J\'ozsef Marton (Budapest University of Technology and Economics)
  \item Marcus Paradies (SAP, DLR)
  \item Minh-Duc Pham (Altran)
  \item Arnau Prat-P\'erez (DAMA UPC, Sparsity Technologies)
  \item David P\"uroja (CWI)
  \item Mirko Spasi\'c (OpenLink Software)
  \item Benjamin A. Steer (Queen Mary University of London, Pometry)
  \item D\'avid Szak\'allas
  \item G\'abor Sz\'arnyas (MTA-BME Lend\"ulet Research Group on Cyber-Physical Systems, Budapest University of Technology and Economics, CWI)
  \item Jack Waudby (Newcastle University)
  \item Mingxi Wu (TigerGraph)
  \item Yuchen Zhang (TigerGraph)
\end{itemize}

\chapter*{Definitions}

This section defines fundamental concepts used in the LDBC benchmark terminology. Part of the definitions below are repeated from the LDBC benchmark specification document.

\begin{description}
    \item[\ldbcsnb] The Linked Data Benchmark Council's Social Network Benchmark suite which currently consists of the Interactive workload and a preliminary version of the Business Intelligence workload.
    
    \item[System Under Test (SUT)] This is the totality of the hardware and software that participates in a benchmark run, excluding parts that are exclusively used for driving the workload. If the parts driving the workload are collocated on the same operating system instance as the SUT, then this is also considered a part of the SUT. In client-server configurations where the test driver is not on a machine hosting any DBMS function the SUT is not considered to encompass the hardware or software which exclusively serves to drive the test workload.
    
    \item[\datagen] This module is provided by LDBC SNB and produces the standard benchmark datasets to be loaded into the SUT for the benchmark. The data generation phase is not part of running the benchmark.
 
    \item[Test Driver (Benchmark Driver, Driver)] The test driver refers to the parts of the benchmark run that coordinate query execution and, if prescribed by a given benchmark, data loading.
    
    \item[Workload (Benchmark)] This is the totality of the tasks a particular benchmark performs against an SUT. This includes data loading as well as the query/update workload. This does not include preparatory stages such as generating benchmark data with a data generator or transferring the data to the platform constituting the SUT. 
    The terms workload and benchmark are synonyms in this context. 

    \item[Time Compression Ratio (TCR)]
    This parameter of the Interactive workload compresses (or stretches) durations between operation start times to increase (or decrease) operation rate, thereby allowing systems to reach their maximum throughput for a given workload. The smaller this number is, the higher compression ratio it represents (\eg 2.0 means run benchmark $2\times$ slower, while 0.1 = run benchmark $10\times$ faster). Systems are expected to compete on achieving the \emph{lowest possible} TCR (\ie the highest $\text{TCR}^{-1}=\frac{1}{\text{TCR}}$).
    
    \item[Query mix] The ratio of read and update queries of a workload, and the frequency at which they are issued.
    
    \item[Scale Factor (SF)] The \ldbcsnb is designed to target systems of different size and scale. The scale factor determines the size of the data used to run the benchmark. The scale factor refers to the measured size of the data in Gigabytes when serialised in CsvSingularProjectedFK.

    \item[Validation Step] The benchmark specifies a scale factor for which ACID test cases are executed and the query results are compared to a reference result set (\ie expected output). This step is required to use the very same set of queries and data structures (this includes both PDS, IADS and EADS -- defined below) that are used in the actual benchmark runs. 
    
    \item[Schema (Database Schema)] A schema is the totality of the non-built-in declarations which are fed into the SUT prior to running a workload. For a relational system, the schema consists of tables, indices, views, materialised views and declarative constraints (\eg foreign key and not null constraints). An ontology for an RDF system counts as a schema if it is loaded on the SUT. An RDF SUT may have no schema at all and still run the workload. However, any declaration or setting (\eg indices) that is not on by default in the SUT, but is used in at least one case of the benchmark run counts as part of the schema.
    The schema does not include stored procedures, triggers, or other imperative (procedural) application specific code that may reside on the SUT and could impact the benchmark results. The schema is required to be the same across all benchmark runs using the same scale factor for a given workload.

    \item[Primary Data Structure (PDS)] 
    This is anything that may influence the result of a database query or may be changed by an update of the database. These may be resident in RAM or durable media or both. Examples of data structures are database base tables and adjacency lists.

    \item[Implicit Auxiliary Data Structure (IADS)] This is a data structure for providing more efficient access to all or parts of the primary data structure. IADS are created by the DBMS automatically and the system may allow them to be turned off. 
    \begin{quote}
        Some systems, such as many RDF stores have multiple covering indices on the primary data structure. The definition in this case is that the primary data structure consists of all the differently ordered full copies of the base table; a table of subject predicate object graph (SPOG) in the RDF case. In this same instance, Auxiliary data structures comprise any data structure which materialise a subset of the SPOG.    
    \end{quote}
    
    \item[Explicit Auxiliary Data Structure (EADS)] These are any application or workload profile specific structures that are declared in addition to the PDSs and IADSs managed by the SUT. These duplicate the data and are created with explicit statements. Secondary indices, materialised views, with or without aggregates, are all examples of this in a relational context. 
    The decision about the used EADS is always part of the schema declaration.
    \begin{quote}
        In the case of relational systems, an ADS may be an index from primary key values to a heap table, if the system in question has such concepts. 
        A secondary index of a relational table, in its memory based and durable media based manifestations is an example for EADS. Such a secondary index is not considered an ADS since it must be declared, which makes its creation explicit. An ADS must be implicit and not created by any specific DDL statement or directive. In the case of RDF systems, if the implementation supports user definable index schemes, as long as these are defined once and apply to all triples/quads, such structures are designated as ADS. If an RDF system selectively makes data structures which apply to some quads but not to others, then such structures are designated as EADS.
    \end{quote}

    \item[SUT-Resident Logic] This is any application specific code that is resident on the SUT, whether by static linking, dynamic loading, JIT, interpretation or any other means of embedding application specific logic into a generic DBMS. Examples of this are stored procedures, hosting Java, CLR or other run times in the SUT process (or processes), loading application specific libraries to extend native functions or data structures etc. A special case is that of a database exclusively accessed via an in-process API. In these cases, any code that is not the test driver or a workload implementation expressed against a generally supported API of the DBMS is deemed SUT resident logic in addition to any other code which may fit the above definitions.

    \item[Test Sponsor] The party which initiates an audit of a benchmark implementation over an SUT. This is typically the vendor of a key component of the SUT, \eg DBMS or hardware.

    \item[Full Disclosure Report (FDR)] This is a document which allows reproduction of any audited benchmark result by a third party. It contains complete description of the circumstances of the benchmark run, including version and configuration of SUT, dataset and test driver.
    
\end{description}

\chapter{Introduction}
\label{sec:introduction}


\section{Motivation for the Benchmark}

The new era of data economy, based on large, distributed, and complexly
structured datasets, has brought on new and complex challenges in the field of
data management and analytics. These datasets, usually modeled as large
graphs, have attracted both industry and academia, due to new
opportunities in research and innovation they offer. This situation has also
opened the door for new companies to emerge, offering new non-relational and
graph-like technologies that are called to play a significant role in upcoming
years.

The change in the data paradigm calls for new benchmarks to test these new
emerging technologies, as they set a framework where different systems can
compete and be compared in a fair way, they let technology providers identify
the bottlenecks and gaps of their systems and, in general, drive the research
and development of new information technology solutions. Without them, the
uptake of these technologies is at risk by not providing the industry with
clear, user-driven targets for performance and functionality.

The Linked Data Benchmark Council's~\cite{DBLP:journals/corr/abs-2307-04350} Social Network Benchmark (\ldbcsnb) aims at being a comprehensive
benchmark by setting the rules for the evaluation of graph-like data management
technologies.  \ldbcsnb is designed to be a plausible look-alike of all the
aspects of operating a social network site, as one of the most representative
and relevant use cases of modern graph-like applications.

\ldbcsnb includes the Interactive
workload~\cite{DBLP:conf/sigmod/ErlingALCGPPB15}, which consists of user-centric
transactional-like interactive queries, and the Business Intelligence workload,
which includes analytic queries to respond to business-critical questions.
Initially, a graph analytics workload was also included in the roadmap of
\ldbcsnb, but this was finally delegated to the Graphalytics benchmark
project~\cite{DBLP:journals/pvldb/IosupHNHPMCCSAT16,DBLP:journals/corr/abs-2011-15028}, which was adopted as an official LDBC graph
analytics benchmark. \ldbcsnb and Graphalytics combined target a broad range of
systems with different nature and characteristics.  \ldbcsnb and Graphalytics
aim at capturing the essential features of these scenarios while
abstracting away details of specific business deployments.

This document contains the definition of the Interactive workload and the first
draft of the Business Intelligence workload. This includes a detailed
explanation of the data used in the \ldbcsnb benchmark, a detailed description
for all queries, and instructions on how to generate the data and run the
benchmark with the provided software.


\section{Relevance to the Industry}

\ldbcsnb is intended to provide the following value to different stakeholders:

\begin{itemize}
 \item For \textbf{end users} facing graph processing tasks, \ldbcsnb provides
     a recognizable scenario against which it is possible to compare merits of
     different products and technologies.  By covering a wide variety of scales
     and price points, \ldbcsnb can serve as an aid to technology selection.
 \item For \textbf{vendors} of graph database technology, \ldbcsnb provides a
     checklist of features and performance characteristics that helps in
     product positioning and can serve to guide new development.
 \item For \textbf{researchers}, both industrial and academic, the \ldbcsnb
     dataset and workload provide interesting challenges in multiple
     choke point areas, such as query optimization, (distributed) graph
     analysis, transactional throughput, and provides a way to objectively
     compare the effectiveness and efficiency of new and existing technology in
     these areas.
\end{itemize}

The technological scope of \ldbcsnb comprises all systems that one might
conceivably use to perform social network data management tasks:

\begin{itemize}
 \item \textbf{Graph database management systems} (\eg Neo4j, TigerGraph, AWS Neptune)
     are novel technologies aimed at storing property graphs,
     \ie graphs with labels and properties (attributes) on nodes and edges.
     They support graph traverals, typically by means of APIs, though
     some of them also support dedicated graph-oriented query languages (\eg
     Neo4j's Cypher and TigerGraph's GSQL, as well as the GQL and SQL/PGQ standards).
     These systems' internal structures are typically designed
     to store dynamic graphs that change over time.  They offer support for
     transactional queries with some degree of consistency, and value-based
     indices to quickly locate nodes and edges. Finally, their architecture is
     typically single-machine (non-cluster). These systems can
     potentially implement all three workloads, though Interactive and Business Intelligence
     workloads are where they will presumably be more competitive.
 \item \textbf{Graph processing frameworks} (\eg Giraph, Signal/Collect,
     GraphLab, Green Marl) are designed to perform global graph
     computations, executed in parallel or in a lockstep fashion. These computations are typically
     long latency, involving many nodes and edges and often consist of approximation
     answers to NP-complete problems. These systems expose an API, sometimes following
     a vertex-centric paradigm, and their architecture targets both single-machine and
     cluster systems. These systems will likely implement the Graph Analytics workload.
 \item \textbf{RDF database systems} (\eg OWLIM, Virtuoso, Stardog, AWS Neptune)
      are systems that implement the SPARQL~1.1 query
     language, similar in complexity to \mbox{SQL-92}, which allows for structured
     queries, and simple traversals. RDF database systems often come with
     additional support for simple reasoning (sameAs, subClass), text search, and
     geospatial predicates.  RDF database systems generally support
     transactions, but not always with full concurrency and serializability and
     their supposed strength is integrating multiple data sources (\eg
     DBpedia). Their architecture is both single-machine and clustered, and
     they will likely target Interactive and Business Intelligence workloads.
\item \textbf{Relational database systems} (\eg PostgreSQL, MySQL, Oracle, IBM Db2,
     Microsoft SQL Server, Virtuoso, MonetDB, Vectorwise, Vertica, DuckDB but also Hive and
     Impala) treat graph data relationally, and queries are formulated in SQL and/or
     PL/SQL. Both single-machine and cluster systems exist. They do not
     normally support recursion or stateful recursive algorithms, which makes     them not at home in the Graph Analytics workloads.
\end{itemize}


\section{General Benchmark Overview}

\ldbcsnb aims at being a complete benchmark, designed with the following goals in mind:

\begin{itemize}
 \item \textbf{Rich coverage.} \ldbcsnb is intended to cover most demands
     encountered in the management of complexly structured data.
 \item \textbf{Modularity.} \ldbcsnb is broken into parts that can be
     individually addressed. In this manner \ldbcsnb
     stimulates innovation without imposing an overly high threshold for
     participation.
 \item \textbf{Reasonable implementation cost.} For a product offering relevant
     functionality, the effort for obtaining initial results with SNB should be
     small, in the order of days.
 \item \textbf{Relevant selection of challenges.} Benchmarks are known to
     direct product development in certain directions. \ldbcsnb is informed by
     the state-of-the-art in database research so as to offer optimization
     challenges for years to come while not having a prohibitively high
     threshold for entry.
 \item \textbf{Reproducibility and documentation of results.} \ldbcsnb
     will specify the rules for full disclosure of benchmark execution and for
     auditing of benchmark runs in accordance with the LDBC Byelaws~\cite{ldbc_byelaws}.
     The workloads may be run on any equipment
     but the exact configuration and price of the hardware and software must be
     disclosed.
\end{itemize}

\ldbcsnb benchmark is modeled around the operation of a real social network
site. A social network site represents a relevant use case for the following
reasons:

\begin{itemize}
    \item It is simple to understand for a large audience, as it is
        arguably present in our every-day life in different shapes and forms.
    \item It allows testing a complete range of interesting
        challenges, by means of different workloads targeting systems of
        different nature and characteristics.
    \item A social network can be scaled, allowing the design of a
        scalable benchmark targeting systems of different sizes and budgets.
\end{itemize}

\autoref{sec:benchmark-specification} summarizes LDBC's benchmark design philosophy.

In \autoref{sec:data}, we define the schema of the data used in
the benchmark. The schema represents a realistic social network, including
people and their activities in the social network during a period of time.
Personal information of each person, such as name, birthday, interests
or places where people work or study, is included. A person's activity is
represented in the form of friendship relationships and content sharing (\ie
messages and pictures). \ldbcsnb provides a scalable synthetic data generator
based on the MapReduce paradigm, which produces networks with the
described schema with distributions and correlations similar to those expected
in a real social network. Furthermore, the data generator is designed to be
user-friendly. The proposed data schema is shared by all the different proposed
workloads, those we currently have, and those that will be proposed in the future.

In \autoref{sec:workloads}, an overview of the workloads is provided.
All SNB workloads are designed to mimic
the different usage scenarios found in operating a real social network site, and
each of them targets one or more types of systems. Each workload defines a set
of queries and query mixes, designed to stress the SUTs in different choke point
areas, while being credible and realistic. The Interactive workload reproduces the
interaction between the users of the social network by including lookups and
transactions, which update small portions of the database. These queries are
designed to be interactive and target systems capable of responding to such queries
with low latency for multiple concurrent users. The Business Intelligence workload
represents analytic queries a social network company would
like to perform in the social network, to take advantage of the data and to
discover new business opportunities. This workload explores moderate to large
portions of the graph from different entities, and performs more resource-intensive
operations.

All workloads provide an execution test driver, which is responsible for executing
the workloads and gathering the results. The driver is designed with simplicity
and portability in mind to ease the implementation on systems with different
nature and characteristics at a low implementation cost. Furthermore, it
automatically handles the validation of the queries by means of a validation
dataset provided by LDBC.  The overall philosophy of \ldbcsnb is to provide
the necessary software tools to run the benchmark, and therefore to reduce the
benchmark's entry point as much as possible.

\autoref{sec:update-operations} defines the update operations used in the SNB workloads. \autoref{sec:interactive-v1}, \autoref{sec:interactive-v2}, and \autoref{sec:bi} define the SNB \interactivevone, \interactivevtwo, and BI workloads, respectively.
\autoref{sec:auditing} contains the SNB auditing policies.
\autoref{sec:acid-test-suite} defines the ACID test suite.
\autoref{sec:related-work} summarized the related work on graph processing benchmarks.


\section{Related Projects}

Along the Social Network Benchmark, LDBC~\cite{DBLP:journals/sigmod/AnglesBLF0ENMKT14} provides other benchmarks as well:

\begin{itemize}
	\item The Semantic Publishing Benchmark (SPB)~\cite{DBLP:conf/semweb/SpasicJP16} measures the performance of \emph{semantic databases} operating on RDF datasets.
	\item The Graphalytics benchmark~\cite{DBLP:journals/pvldb/IosupHNHPMCCSAT16} measures the performance of \emph{graph analysis} operations (\eg PageRank, local clustering coefficient).
\end{itemize}


\section{Participation of Industry and Academia}

The list of institutions that take part in the definition and development
of \ldbcsnb is formed by relevant actors from both the industry and academia in
the field of linked data management. All the participants have contributed with
their experience and expertise in the field, making a credible and relevant
benchmark, which meets all the desired needs.


\section{Technical Report}

This technical report is available on arXiv~\cite{DBLP:journals/corr/abs-2001-02299} and is updated upon new releases of the SNB.

\chapter{Benchmark Specification}
\label{sec:benchmark-specification}

\begin{figure}[htbp]
    \centering
    \includegraphics[scale=\yedscale]{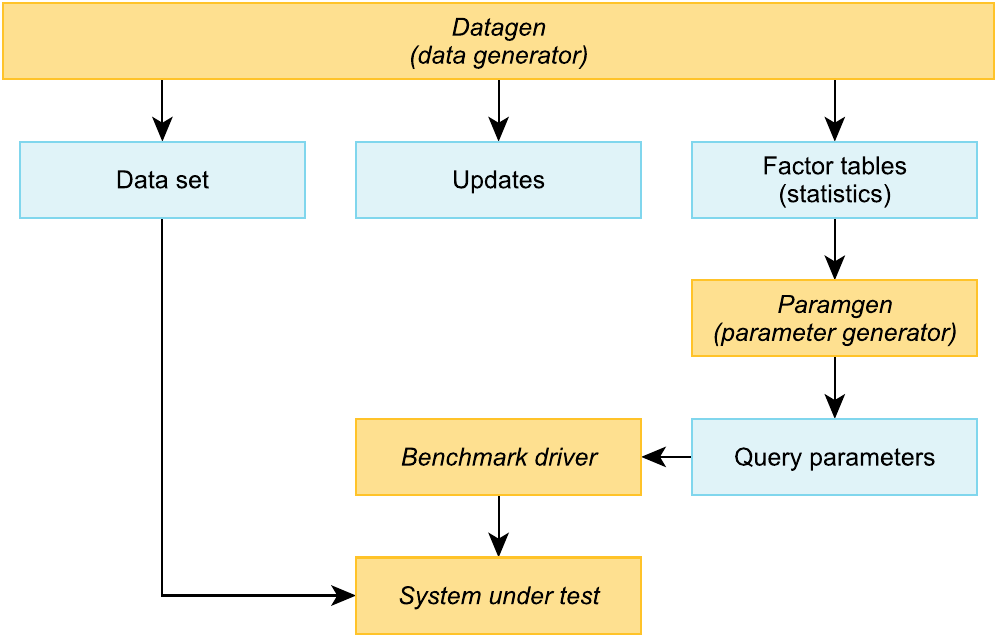}
    \caption{
        High-level overview of the frameworks implementing each LDBC Social Network Benchmark workload.
        Legend:
        \fcolorbox{mydarkyellow}{mylightyellow}{{\scriptsize \textit{\textsf{Software component}}}}
        \fcolorbox{mydarkblue}{mylightblue}{{\scriptsize \textsf{Data artifact\vphantom{p}}}}
    }
    \label{fig:snb-overview}
\end{figure}

\paragraph{Overview}
The LDBC Social Network Benchmark workloads require several components to generate data and updates, produce query substitution parameters, run the benchmark on the system under test, \etc
\autoref{fig:snb-overview} shows a blueprint for the frameworks implementing the \ldbcsnb benchmark workloads.

\paragraph*{Portability}
\ldbcsnb is designed to be flexible and to have an affordable entry point. \ldbcsnb aims to accommodate systems from single node in-memory systems to large distributed multi-node clusters. Therefore, the requirements to fulfill for executing \ldbcsnb are limited to pure software requirements to be able to run the tools.
While the benchmark specification aims to be portable, the software provided by \ldbcsnb have been primarily tested on Linux-based operating systems (\eg Ubuntu LTS). The driver and clients for the reference implementations were implemented in Java. The generator has two versions: the Hadoop-based one was written in Java, while the Spark-based one is written in a mix of Java and Scala. 

\paragraph*{Auditable systems}
\ldbcsnb does not impose the usage of any specific type of system, as it targets systems of different nature and characteristics, from graph databases, graph processing frameworks and RDF systems, to traditional relational database management systems. Therefore, data can be stored in the most convenient manner the test sponsor may decide, as long as it conforms with the execution rules.

\chapter{Data sets and data generation}
\label{sec:data}

This chapter introduces the data used by \ldbcsnb. This includes the different
data types, the data schema, how it is generated and the different scale
factors.

\textbf{Warning.} This chapter describes the latest variant of the data set.
If you are looking for information on the Interactive workload, please also check 
\autoref{sec:legacy-data-sets}.

\section{Data Types}
\autoref{table:types} describes the different data types used in the benchmark.

\begin{table}[h]
    \centering
    \begin{tabular}{|>{\typeCell}p{\attributeColumnWidth}|p{\largeDescriptionColumnWidth}|}
        \hline
        \tableHeaderFirst{Type} & \tableHeader{Description} \\
        \hline
        ID &  integer type with 64-bit precision. All IDs within a single entity type (\eg Person, Message) are unique, but different entity types (\eg a Forum and a Tag) might have the same ID.\\
        \hline
        32-bit Integer &  integer type with 32-bit precision\\
        \hline
        64-bit Integer &  integer type with 64-bit precision\\
        \hline
        32-bit Float &  integer type with 32-bit precision\\
        \hline
        64-bit Float &  integer type with 64-bit precision\\
        \hline
        String & variable length text of size 80 Unicode characters\\
        \hline
        Long String & variable length text of size 256 Unicode characters\\
        \hline
        Text &  variable length text of size 2000 Unicode characters\\
        \hline
        Date &  date with a precision of a day, encoded as a string with the following format: \texttt{yyyy-mm-dd}, where \texttt{yyyy} is a four-digit integer representing the year,
        the year, \texttt{mm} is a two-digit integer representing the month and \texttt{dd} is a two-digit integer representing the day. \\
        \hline
        DateTime &  date with a precision of milliseconds, encoded as a string with the following format: \texttt{yyyy-mm-ddTHH:MM:ss.sss+00:00}, where \texttt{yyyy} is a four-digit integer representing the year,
        the year, \texttt{mm} is a two-digit integer representing the month and \texttt{dd} is a two-digit integer representing the day, \texttt{HH} is a two-digit integer representing the hour, \texttt{MM} is a two
        digit integer representing the minute and \texttt{ss.sss} is a five digit fixed point real number representing the seconds up to millisecond precision. Finally, the \texttt{+00:00} of the end represents the
        timezone, which should always be GMT (both for inputs and outputs).\\
        \hline
        Boolean &  logical type, taking the value of either \texttt{True} of \texttt{False}\\
        \hline
    \end{tabular}
    \caption{Description of the data types. Some types such as 32-bit  Float and 64-bit Integer are currently not used in the benchmark.}
    \label{table:types}
\end{table}

\section{Data Schema}

\begin{figure}[htbp]
    \centering
    \includegraphics[width=\textwidth]{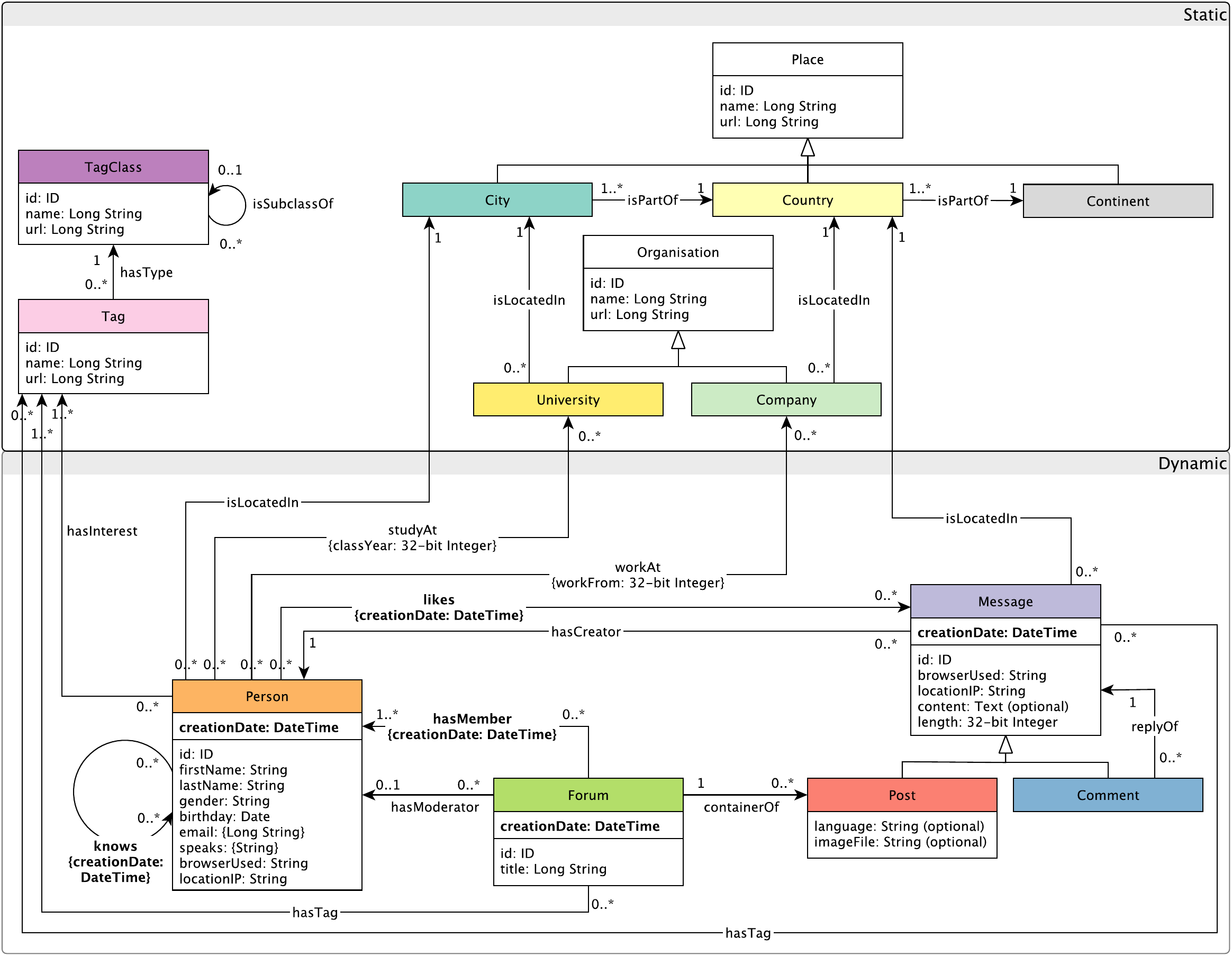}
    \caption{UML class diagram-style depiction of the \ldbcsnb graph schema. Note that the \textsf{knows} edges should be treated as undirected (but are serialized only in a single direction). The cardinality of the \textsf{hasModerator} edge has changed between version~1 (where it was exactly 1) and version~2 (where it is 0..1).}
    \label{fig:schema}
\end{figure}

\autoref{fig:schema} shows the data schema in UML. The schema defines the
structure of the data used in the benchmark in terms of entities and their
relations. Data represents a snapshot of the activity of a social network
during a period of time. Data includes entities such as Persons, Organisations,
and Places. The schema also models the way persons interact, by means of the
friendship relations established with other persons, and the sharing of content
such as Messages (both textual and images), replies to Messages and likes to
Messages.  People form groups to talk about specific topics, which are
represented as Tags\footnote{Tags are basically equivalent to \emph{hashtags}
on contemporary social media sites. In this document, we occasionally use the term
\emph{topic} to refer to tags}. An example graph conforming the SNB schema is shown in \autoref{sec:example-graph}.

\ldbcsnb has been designed to be flexible and to target systems of different
nature and characteristics. As such, it does not force any particular internal
representation of the schema. The \datagen component
supports multiple output data formats to
fit the needs of different types of systems, including RDF, relational DBMS and
graph DBMS.

The schema specifies different entities, their attributes and their relations.
All of them are described in the following sections.

\subsubsection*{Textual Restrictions and Notes}
\begin{itemize}
    \item Messages always have a non-empty \texttt{content} attribute.
    \item Posts have either a \texttt{content} or an \texttt{imageFile} attribute (\ie they always have exactly one of them.) The one they do not have is represented with an empty string or with \texttt{NULL}.
    \item Posts in a forum can be created by a non-member person if and only if that person is the moderator of the Forum.
\end{itemize}

\subsection{Entities (Nodes)}

{\flushleft \textbf{City:}} a sub-class of a Place, and represents a
city of the real world. City entities are used to specify where persons live,
as well as where universities operate.

{\flushleft \textbf{Comment:}} a sub-class of a Message, and represents a
comment made by a person to an existing message (either a Post or a Comment).

{\flushleft \textbf{Company:}} a sub-class of an Organisation, and represents a company where persons work.

{\flushleft \textbf{Continent:}} a sub-class of a Place, and represents a continent of the real world.

{\flushleft \textbf{Country:}} a sub-class of a Place, and represents a country of the real world.
Countries are selected as the place of operation for Companies as well as the location of Messages.

{\flushleft \textbf{Forum:}} a meeting point where people
post messages. Forums are characterized by the topics (represented as tags)
people in the forum are talking about. Although from the schema's perspective
it is not evident, there exist three different types of
forums.  They are distinguished by their titles:

\begin{itemize}
    \item personal walls: ``Wall of \ldots''
    \item image albums: ``Album $k$ of \ldots''
    \item groups: ``Group for \ldots''
\end{itemize}

\autoref{table:forum} shows the attributes of the Forum entity.

\begin{table}[H]
    \begin{tabular}{|>{\varNameCell}p{\attributeColumnWidth}|>{\typeCell}p{\typeColumnWidth}|p{\descriptionColumnWidth}|}
        \hline
        \tableHeaderFirst{Attribute} & \tableHeader{Type} & \tableHeader{Description} \\
        \hline
        id & ID  & The identifier of the forum.\\
        \hline
        title & Long String  & The title of the forum.\\
        \hline
        creationDate & DateTime  & The date the forum was created.\\
        \hline
    \end{tabular}
    \caption{Attributes of the Forum entity.}
    \label{table:forum}
\end{table}

{\flushleft \textbf{Message:}} an abstract entity that represents a message
created by a person. \autoref{table:message} shows the attributes of the Message
abstract entity.

\begin{table}[H]
    \begin{tabular}{|>{\varNameCell}p{\attributeColumnWidth}|>{\typeCell}p{\typeColumnWidth}|p{\descriptionColumnWidth}|}
        \hline
        \tableHeaderFirst{Attribute} & \tableHeader{Type} & \tableHeader{Description} \\
        \hline
        id & ID  & The identifier of the message.\\
        \hline
        browserUsed & String  & The browser used by the Person to create the message.\\
        \hline
        creationDate & DateTime  & The date the message was created.\\
        \hline
        locationIP & String  & The IP of the location from which the message was created.\\
        \hline
        content & Text (optional)  & The content of the message.\\
        \hline
        length & 32-bit Integer  & The length of the content.\\
        \hline
    \end{tabular}
    \caption{Attributes of the Message interface.}
    \label{table:message}
\end{table}

{\flushleft \textbf{Organisation:}} an institution of the real
world. \autoref{table:organisation} shows the attributes of the Organisation
entity.

\begin{table}[H]
    \begin{tabular}{|>{\varNameCell}p{\attributeColumnWidth}|>{\typeCell}p{\typeColumnWidth}|p{\descriptionColumnWidth}|}
        \hline
        \tableHeaderFirst{Attribute} & \tableHeader{Type} & \tableHeader{Description} \\
        \hline
        id & ID  & The identifier of the organisation.\\
        \hline
        name & Long String  & The name of the organisation.\\
        \hline
        url & Long String  & The URL of the organisation.\\
        \hline
    \end{tabular}
    \caption{Attributes of the Organisation entity.}
    \label{table:organisation}
\end{table}

{\flushleft \textbf{Person:}} the avatar a real world person creates
when he/she joins the network, and contains various information about the
person as well as network related information. \autoref{table:person} shows
the attributes of the Person entity.

\begin{table}[H]
    \begin{tabular}{|>{\varNameCell}p{\attributeColumnWidth}|>{\typeCell}p{\typeColumnWidth}|p{\descriptionColumnWidth}|}
        \hline
        \tableHeaderFirst{Attribute} & \tableHeader{Type} & \tableHeader{Description} \\
        \hline
        id & ID  & The identifier of the person.\\
        \hline
        firstName & String  & The first name of the person.\\
        \hline
        lastName & String  & The last name of the person.\\
        \hline
        gender & String  & The gender of the person.\\
        \hline
        birthday & Date  & The birthday of the person.\\
        \hline
        email & \{Long String\}  & The set of emails the person has (cardinality: at least one).\\
        \hline
        speaks & \{String\}  & The set of languages the person speaks (cardinality: at least one).\\
        \hline
        browserUsed & String  & The browser used by the person when he/she registered to the social network.\\
        \hline
        locationIP & String  & The IP of the location from which the person was registered to the social network.\\
        \hline
        creationDate & DateTime  & The date the person joined the social network.\\
        \hline
    \end{tabular}
    \caption{Attributes of the Person entity.}
    \label{table:person}
\end{table}

{\flushleft \textbf{Place:}} a place in the world.
\autoref{table:place} shows the attributes of the Place entity. Note, each Place has additional parameters: longitude and latitude, which are not exposed. These are used internally for sorting places.

\begin{table}[H]
    \begin{tabular}{|>{\varNameCell}p{\attributeColumnWidth}|>{\typeCell}p{\typeColumnWidth}|p{\descriptionColumnWidth}|}
        \hline
        \tableHeaderFirst{Attribute} & \tableHeader{Type} & \tableHeader{Description} \\
        \hline
        id & ID  & The identifier of the place.\\
        \hline
        name & Long String  & The name of the place.\\
        \hline
        url & Long String  & The URL of the place.\\
        \hline
    \end{tabular}
    \caption{Attributes of the Place entity.}
    \label{table:place}
\end{table}

{\flushleft \textbf{Post:}} a sub-class of Message, that is posted in a
forum. Posts are created by persons into the forums where they belong.
Posts contain either content or imageFile, always one of them but never both.
The one they do not have is an empty string.
\autoref{table:post} shows the attributes of the Post entity.

\begin{table}[H]
    \begin{tabular}{|>{\varNameCell}p{\attributeColumnWidth}|>{\typeCell}p{\typeColumnWidth}|p{\descriptionColumnWidth}|}
        \hline
        \tableHeaderFirst{Attribute} & \tableHeader{Type} & \tableHeader{Description} \\
        \hline
        language & String (optional) & The language of the post. Mutually exclusive with \texttt{imageFile}. \\
        \hline
        imageFile & String (optional) & The image file of the post. Mutually exclusive with \texttt{language}.\\
        \hline
    \end{tabular}
    \caption{Attributes of the Post entity.}
    \label{table:post}
\end{table}

{\flushleft \textbf{Tag:}} a topic or a concept. Tags are used to
specify the topics of forums and posts, as well as the topics a person is
interested in. \autoref{table:tag} shows the atltributes of the Tag entity.

\begin{table}[H]
    \begin{tabular}{|>{\varNameCell}p{\attributeColumnWidth}|>{\typeCell}p{\typeColumnWidth}|p{\descriptionColumnWidth}|}
        \hline
        \tableHeaderFirst{Attribute} & \tableHeader{Type} & \tableHeader{Description} \\
        \hline
        id & ID  & The identifier of the tag.\\
        \hline
        name & Long String  &  The name of the tag.\\
        \hline
        url & Long String  &  The URL of the tag.\\
        \hline
    \end{tabular}
    \caption{Attributes of the Tag entity.}
    \label{table:tag}
\end{table}

{\flushleft \textbf{TagClass:}} a class used to build a hierarchy of tags. \autoref{table:tagclass} shows the attributes of the TagClass entity.

\begin{table}[H]
    \begin{tabular}{|>{\varNameCell}p{\attributeColumnWidth}|>{\typeCell}p{\typeColumnWidth}|p{\descriptionColumnWidth}|}
        \hline
        \tableHeaderFirst{Attribute} & \tableHeader{Type} & \tableHeader{Description} \\
        \hline
        id & ID  & The identifier of the tagclass.\\
        \hline
        name & Long String  &  The name of the tagclass.\\
        \hline
        url & Long String  &  The URL of the tagclass.\\
        \hline
    \end{tabular}
    \caption{Attributes of the TagClass entity.}
    \label{table:tagclass}
\end{table}

{\flushleft \textbf{University:}} a sub-class of Organisation,
and represents an institution where persons study.

\subsection{Relations (Edges)}

Relations (edges) connect entities of different types. The endpoint entities are defined by their ``id'' attribute.
Edge instances starting from or ending in a given node are treated as a set, \ie no ordering is defined on the edges.
Multiple edges (\ie edges of the same type between two entity instances) are not allowed in SNB graphs.

\begin{longtable}{|>{\varNameCell}p{2.5cm}|>{\typeCell}p{2.5cm}|>{\typeCell}p{2.5cm}|>{\edgeDirectionCell}c|p{6.5cm}|}
    \hline
     \tableHeaderFirst{Name} & \tableHeader{Source} & \tableHeader{Destination} & \tableHeader{Type} & \tableHeader{Description} \\
     \hline
     containerOf & Forum[1] & Post[0..*] & D & A Forum and a Post contained in it\\
     \hline
     hasCreator & Message[0..*] & Person[1] & D & A Message and its creator (Person)\\
     \hline
     hasInterest & Person[0..*] & Tag[1..*] & D & A Person and a Tag representing a topic the person is interested in\\
     \hline
     hasMember & Forum[0..*] &  Person[1..*] & D & A Forum and its member (Person)

     In version 1:

     \attributeTable{joinDate}{DateTime}{The Date the person joined the Forum}

     In version 2:

     \attributeTable{creationDate}{DateTime}{The Date the person joined the Forum}

     \\
     \hline
     hasModerator & Forum[0..*] &
     \textrm{In version 1:}
     
     Person[1]
     
     \textrm{In version 2:}

     Person[0..1]
     & D & A Forum and its moderator (Person) \\
     \hline
     hasTag & Message[0..*] & Tag[0..*] & D & A Message and a Tag representing the message's topic \\
     \hline
     hasTag & Forum[0..*] & Tag[1..*] & D & A Forum and a Tag representing the forum's topic \\
     \hline
     hasType & Tag[0..*] & TagClass[1] & D & A Tag and a TagClass the tag belongs to \\
     \hline
     isLocatedIn & Company[0..*] & Country[1] & D & A Company and its home Country \\
     \hline
     isLocatedIn & Message[0..*] & Country[1] & D & A Message and the Country from which it was issued \\
     \hline
     isLocatedIn & Person[0..*] & City[1] & D & A Person and their home City \\
     \hline
     isLocatedIn & University[0..*] & City[1] & D &  A University and the City where the university is \\
     \hline
     isPartOf & City[1..*] & Country[1] & D & A City and the Country it is part of \\
     \hline
     isPartOf & Country[1..*] & Continent[1] & D & A Country and the Continent it is part of \\
     \hline
     isSubclassOf & TagClass[0..*] & TagClass[0..1] & D & A TagClass and its parent TagClass \\
     \hline
     knows & Person[0..*] & Person[0..*] & U & Two Persons that know each other.
     Note that
     (1)~the knows edges are undirected (all other edge types are directed and
     (2)~to avoid duplications, these edges are only serialized to one direction and it is the responsibility of the loader/implementation component to treat them as undirected.

     In this specification document, we use the terms ``knows'' and ``friends (with/of/etc.)'' interchangeably.

     \attributeTable{creationDate}{DateTime}{The date the knows relation was established}

     \\
     \hline
     likes & Person[0..*] & Message[0..*] & D & A Person that likes a Message

     \attributeTable{creationDate}{DateTime}{The date the like was issued}

     \\
     \hline
     replyOf & Comment[0..*] & Message[1] & D & A Comment and the Message it replies \\
     \hline
     studyAt & Person[0..*] & University[0..*] & D & A Person and a University it has studied

     \attributeTable{classYear}{32-bit Integer}{The year the person graduated}

     \\
     \hline
     workAt & Person[0..*] & Company[0..*] & D & A Person and a Company it works

     \attributeTable{workFrom}{32-bit Integer}{The year the person started to work at that Company}

     \\
     \hline
 \caption{Description of the data relations. Type -- \texttt{D}: directed edge, \texttt{U}: undirected edge.}
 \label{table:relations}
\end{longtable}

\subsection{Domain Concepts}

A \emph{thread} consists of Messages, starting with a single Post and the Comments that -- either directly or transitively -- reply to that Post.

\section{Data Generation}
\label{sec:data_generation}

\ldbcsnb provides \datagen (Data Generator), which produces synthetic
datasets following the schema described above. Data
produced mimics a social network's activity during a period of time. Three
parameters determine the generated data: the number of persons, the number of
years simulated, and the starting year of simulation. \datagen is defined by the
following characteristics:

\begin{itemize}
    \item \textbf{Realism.} Data generated by \datagen mimics the
        characteristics of those found in a real social network. In \datagen,
        output attributes, cardinalities, correlations and distributions have
        been finely tuned to reproduce a real social network in each of its
        aspects. On the one hand, it is aware of the  data and link distributions
        found in a real social network such as Facebook. On the other hand, it
        uses real data from DBpedia, such as property dictionaries, which are
        used to ensure that attribute values are realistic and correlated.
    \item \textbf{Scalability.} Since \ldbcsnb targets systems of different
        scales and budgets, \datagen is capable of generating datasets of
        different sizes, from a few Gigabytes to Terabytes. \datagen is
        implemented following the MapReduce parallel paradigm, allowing the
        generation of small datasets in single node machines, as well as large
        datasets on commodity clusters.
    \item \textbf{Determinism.} \datagen is deterministic regardless of the number
        of cores/machines used to produce the data. This important feature
        guarantees that all Test Sponsors will face the same dataset,
        thus, making the comparisons between different systems fair and the
        benchmarks' results reproducible.
    \item \textbf{Usability.} \ldbcsnb is designed to have an affordable entry
        point. As such, \datagen's design is  severely influenced by this
        philosophy, and therefore it is designed to be as easy to use as
        possible.
\end{itemize}

\subsection{Resource Files}

\datagen uses a set of resource files with data
extracted from DBpedia. Conceptually, \datagen generates attribute's
values following a property dictionary model that is defined by

\begin{itemize}
    \item a dictionary $D$
    \item a ranking function $R$
    \item a probability function $F$
\end{itemize}

Dictionary $D$ is a fixed set of values. The ranking function $R$ is a bijection
that assigns to each value in a dictionary a unique rank between 1 and $|D|$.
The probability density function $F$ specifies how the data generator chooses
values from dictionary $D$ using the rank for each term in the dictionary. The
idea to have a separate ranking and probability function is motivated by the
need of generating correlated values: in particular, the ranking function is
typically parameterized by some parameters: different parameter values result
in different rankings. For example, in the case of a dictionary of property
firstName, the popularity of first names might depend on the gender, country
and birthday properties. Thus, the fact that the popularity of first names in
different countries and times is different, is reflected by the different ranks
produced by function $R$ over the full dictionary of names.  \datagen uses a
dictionary for each literal property, as well as ranking functions for all
literal properties. These are materialized in a set of resource files, which
are described in \autoref{table:property_dictionaries}.

\begin{table}[htb]
\begin{tabular}{|p{4cm}|p{12cm}|}
    \hline
    \tableHeaderFirst{Resource Name} & \tableHeader{Description} \\
    \hline
    Browsers & Contains a list of web browsers and their probability to be used. It is used to set the browsers used by the users.\\
    \hline
    Cities by Country & Contains a list of cites and the country they belong. It is used to assign cities to users and universities.\\
    \hline
    Companies by Country & Contains the set of companies per country. It is used to set the countries where companies operate.\\
    \hline
    Countries & Contains a list of countries and their populations. It is used to obtain the amount of people generated for each country.\\
    \hline
    Emails & Contains the set of email providers. It is used to generate the email accounts of persons.\\
    \hline
    IP Zones & Contains the set of IP ranges assigned to each country. It is used to assign the IP addresses to users.\\
    \hline
    Languages by Country & Contains the set of languages spoken in each country. It is used to set the languages spoken by each user.\\
    \hline
    Name by Country & Contains the set of names and the probability to appear in each country. It is used to assign names to persons, correlated with their countries.\\
    \hline
    Popular places by Country & Contains the set of popular places per country. These are used to set where images attached to posts are taken from.\\
    \hline
    Surnames' by Country & Contains the set of surnames and the probability to appear in each country. It is used to assign surnames to persons, correlated with their countries.\\
    \hline
    Tags by Country & Contains a set of tags and their probability to appear in each country. It is used to assign the interests to persons and forums.\\
    \hline
    Tag Classes & Contains, for each tag, the classes it belongs to.\\
    \hline
    Tag Hierarchies & Contains, for each tagClass, their parent tagClass.\\
    \hline
    Tag Matrix & Contains, for each tag, the correlation probability with the other tags. It is used enrich the tags associated to messages.\\
    \hline
    Tag Text & Contains, for each tag, a text. This is used to generate the text for messages.\\
    \hline
    Universities by City & Contains the set of universities per city. It is used to set the cities where universities operate.\\
    \hline
\end{tabular}
\caption{Resource files.}
\label{table:property_dictionaries}
\end{table}

\subsection{Graph Generation}

\begin{figure}[htb]
    \centering
    \includegraphics[scale=\yedscale]{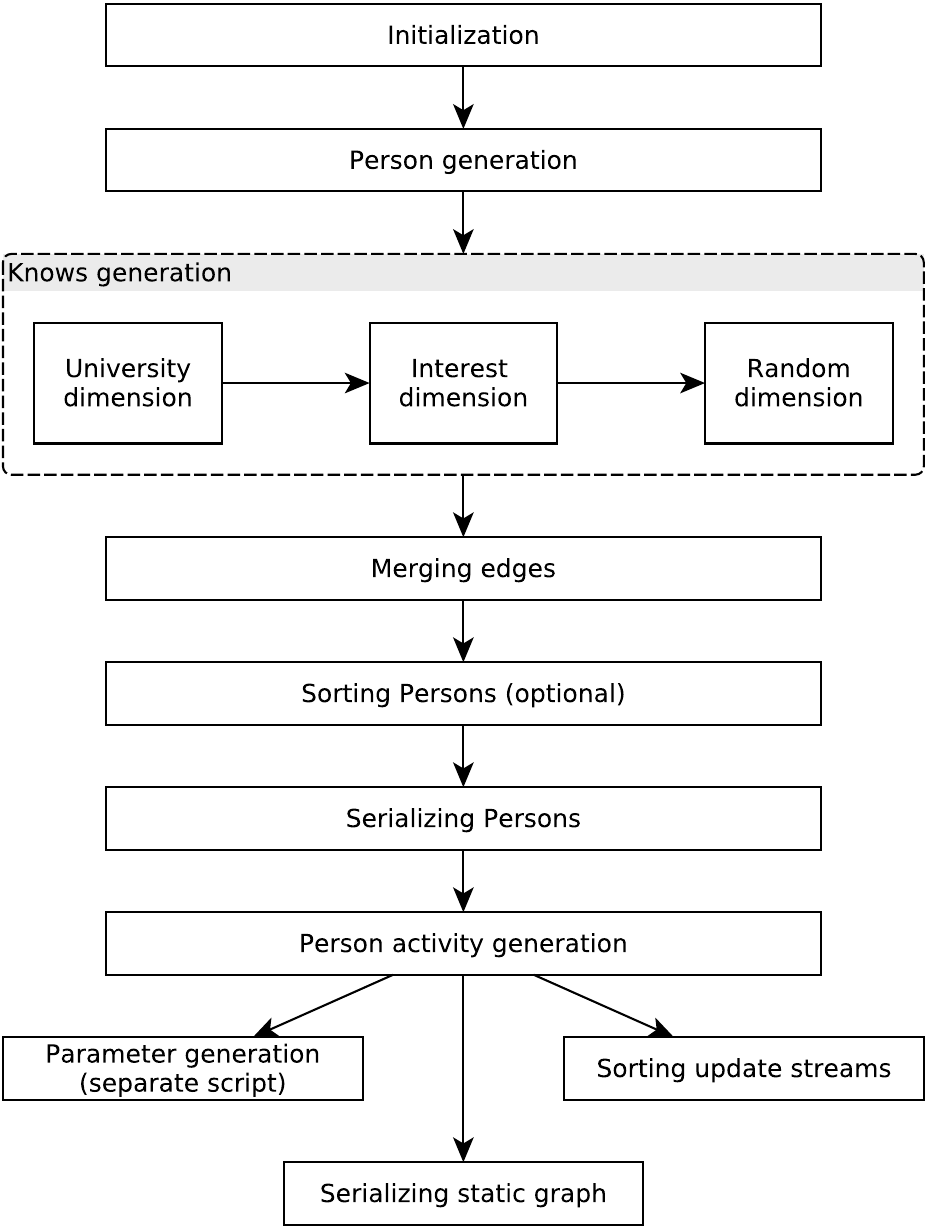}
    \caption{The \datagen generation process.}
    \label{fig:generation_process}
\end{figure}

\autoref{fig:generation_process} conceptually depicts the full data
generation process. The first step loads all the dictionaries and resource
files, and initializes the \datagen parameters.  Second, it generates all the
Persons in the graph, and the minimum necessary information to operate. Part of
this information are the interests of the persons, and the number of knows
relationships of every person, which is guided by a degree distribution
function similar to that found in Facebook~\cite{facebook_anatomy}.

The next three steps are devoted to the creation of knows relationships.  An
important aspect of real social networks, is the fact that similar persons
(with similar interests and behaviors) tend to be connected. This is known as
the Homophily principle~\cite{mcpherson2001birds,DBLP:journals/socnet/BaroneC18}, and implies the presence of
a larger amount of triangles than that expected in a random network. In order
to reproduce this characteristic, \datagen generates the edges by means of
correlation dimensions.  Given a person, the probability to be connected to
another person is typically skewed with respect to some similarity between the
persons. That is, for a person $p$ and for a small set of persons that are
somehow similar to it, there is a high connectivity probability, whereas for
most other persons, this probability is quite low. This knowledge is
exploited by \datagen to reproduce correlations.

Given a similarity function $M(p) : p \rightarrow [0, \infty)$ that gives a score to a person,
with the characteristic that two similar persons will have similar scores, we
can sort all the persons by function $M$ and compare a person $p$ against only the
$K$ neighbouring persons in the sorted array. The consequence of this approach is
that similar persons are grouped together, and the larger the
distance between two persons indicates a monotonic increase in their similarity
difference. In order to choose the persons to connect, \datagen uses a geometric
probability distribution that provides a probability for picking persons to
connect, that are between 1 and $K$ positions apart in the similarity
ranking.

Similarity functions and probability distribution functions over ranked
distance drive what kind of persons will be connected with an edge, not how
many. As stated above, the number of friends of a person is determined by a
Facebook-like distribution. The edges that will be connected to a person $p$,
are selected by randomly picking the required number of edges according to the
correlated probability distributions as discussed before. In the case that
multiple correlations exist, another probability function is used to divide the
intended number of edges between the various correlation dimensions. In \datagen,
three correlated dimensions are chosen: the first one depends on where the
person studied and when, and the second correlation dimension depends on the
interests of the person, and the third one is random (to reproduce the random
noise present in real data). Thus, \datagen has a Facebook-like distributed node
degree, and a predictable (but not fixed) average split between the reasons for
creating edges.

In the next step, a person's activity, in the form of forums, posts and comments
is created. \datagen reproduces the fact that people with a larger number of
friends have a higher activity, and hence post more photos and comments to a
larger number of posts. Another important characteristic of real persons'
activity in social network, are time correlations.  Usually, person' posts
creation in a social network is driven by real world events.  For
instance, one may think about an important event such as the elections in a
country, or a natural disaster. Around the time these events occur, network
activity about these events' topics sees an increase in volume. \datagen
reproduces these characteristics with the simulation of what we name as
flashmob events.  Several events are generated randomly at the beginning of the
generation process, which are assigned a random tag, and are given a time and
an intensity which represents the repercussion of the event in the real world.
When persons' posts are created, some of them are classified as flashmob posts,
and their topics and dates are assigned based on the generated flashmob events.
The volume of activity around this events is modeled following a model similar
to that described in~\cite{DBLP:conf/kdd/LeskovecBKT08}. Furthermore, in order to reproduce the
more uniform every day person activity, \datagen also generates posts uniformly
distributed along all the simulated time.

Finally, in the last step the data is serialized into the output files.

\subsection{Distributions, Parameters, and Quirks}
\label{sec:distr-param}

\begin{figure}[htb]
  \centering
  \includegraphics[scale=\yedscale]{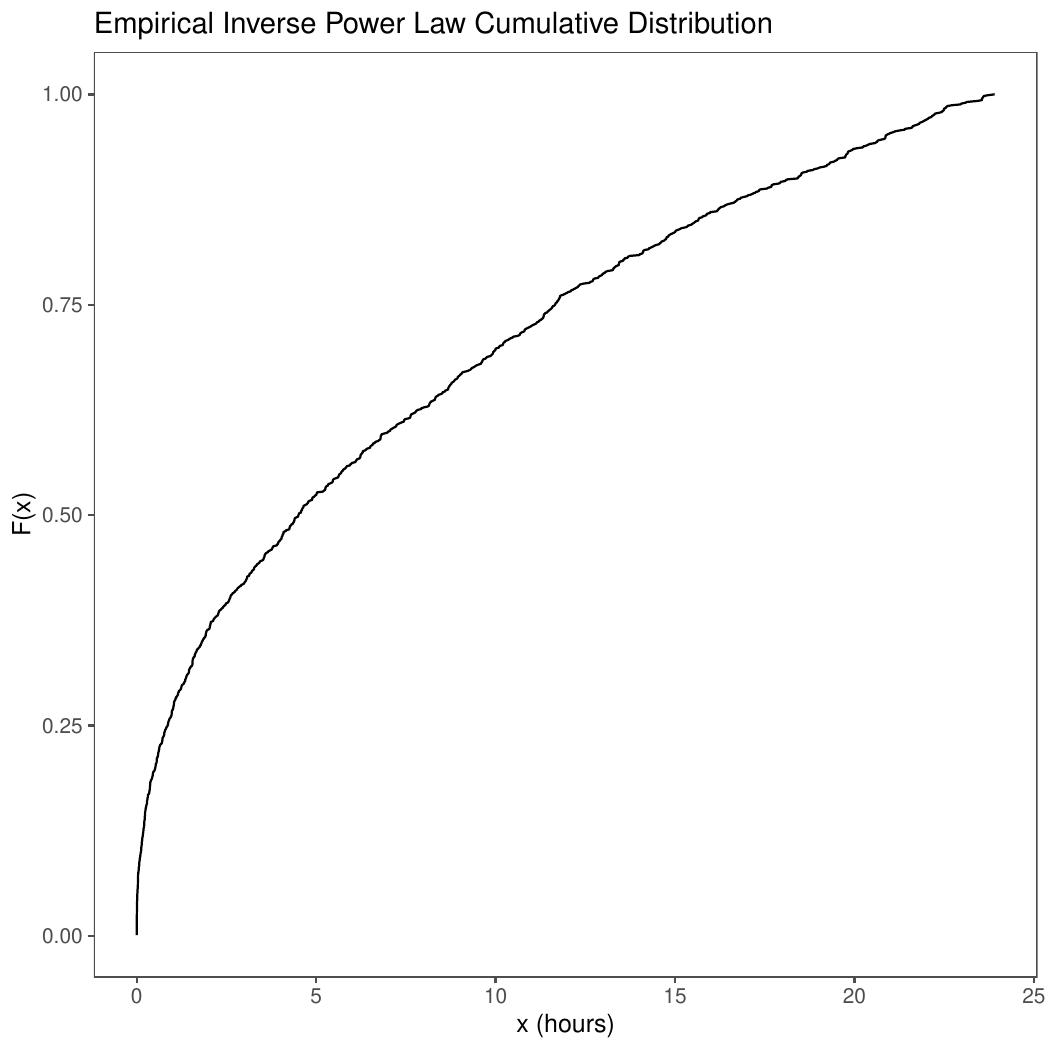}
  \caption{The power-law used to generate comments.}
  \label{fig:comments_dist}
\end{figure}

\begin{figure}[htb]
  \centering
  \includegraphics[scale=\yedscale]{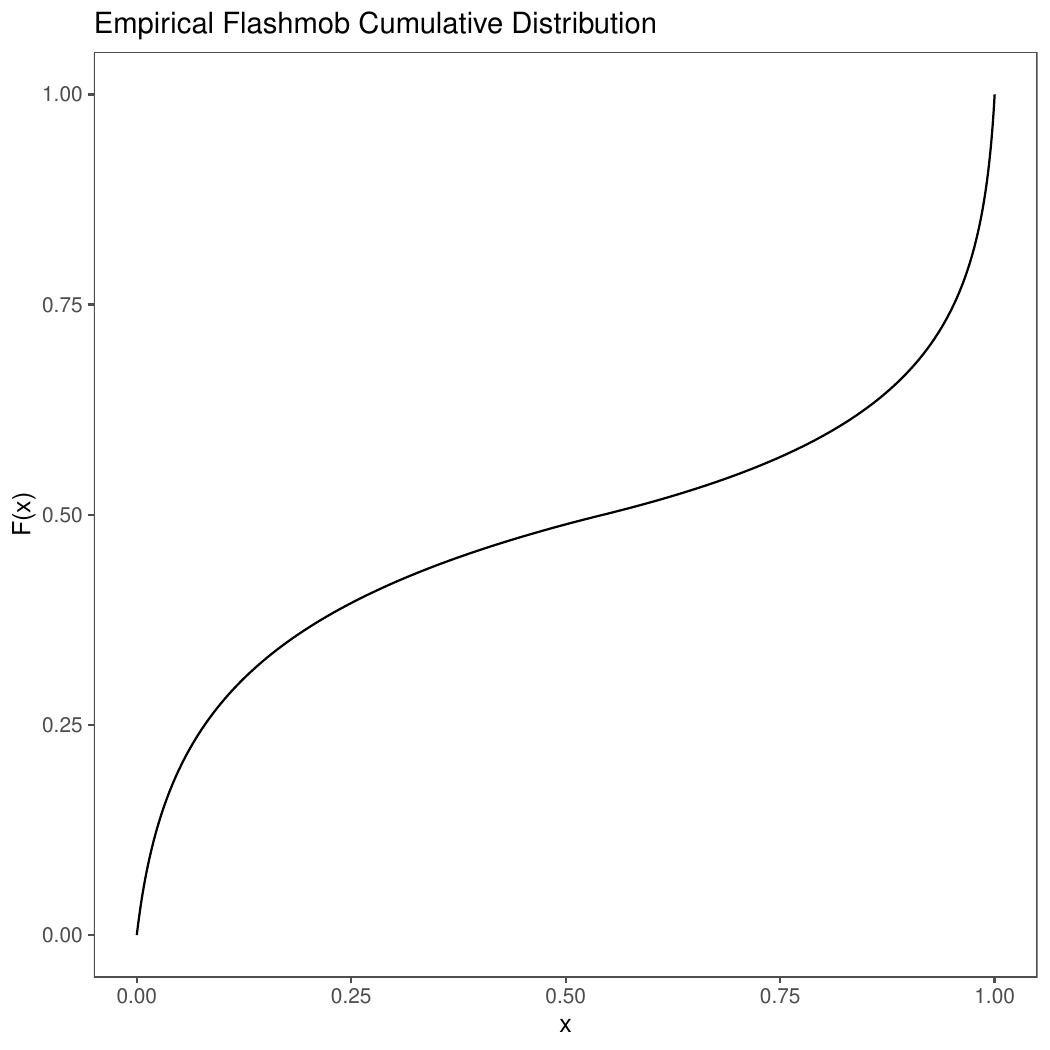}
  \caption{The distribution used to generate posts during flashmob events.}
  \label{fig:flashmob_dist}
\end{figure}

Interesting quirks:
\begin{itemize}
\item A Person is not a member of their Wall but they are its moderator, they do not have a hasMember edge.
\item Each Album generated for Person will have approximately 70\% of their friends as members.
\item A given Person has a 5\% chance of being a moderator of a set of groups.
\item Group membership is composed of 30\% from the moderator's friends and the remainder is chosen randomly (from the block the person is in).
\item Comments are only made in Walls and Groups.
\item Messages can only receive likes during a 7-day window after their creation.
\item Comments always occur within one day of Message they are replying to. The creation date is generated following the power-law distribution in Figure \ref{fig:comments_dist}. The mean delay between Comments and their parent Message is 6.85 hours.
\item Flashmob events span a 72-hour time window with a specific event time in the middle of the window; there are 36 hours each side of the specific event time. Following the distribution in Figure \ref{fig:flashmob_dist} posts are generated either side of flashmob event time, posts are clustered around the specific event time.
\end{itemize}

\subsection{Implementation Details}

\datagen is implemented using the MapReduce parallel paradigm. In MapReduce, a
Map function runs on different parts of the input data, in parallel and on many
node clusters. This function processes the input data and produces for each
result a key. Reduce functions then obtain this data and Reducers run in
parallel on many cluster nodes. The produced key simply determines the Reducer
to which the results are sent. The use of the MapReduce paradigm allows the
generator to scale considerably, allowing the generation of huge datasets by
using clusters of machines.

In the case of \datagen, the overall process is divided into three MapReduce jobs.
In the first job, each mapper generates a subset of the persons of the graph. A
key is assigned to each person using one of the similarity functions described
above. Then, reducers receive the key-value pairs sorted by the key,
generate the knows relations following the described windowing process, and
assign to each person a new key based on another similarity function, for the
next MapReduce pass.  This process can be successively repeated for additional
correlation dimension.  Finally, the last reducer generates the remaining
information such as forums, posts and comments.

\section{Output Data}

For each scale factor, \datagen produces three different artefacts:
\begin{itemize}
  \item \textbf{Dataset:} The dataset to be bulk loaded by the SUT. In the Interactive workload, it corresponds to roughly the 90\% of the total generated network.
  \item \textbf{Update Streams:} A set of update streams containing update
    queries, which are used by the driver to generate the update queries of the
    workloads. This update
    streams correspond to the remaining 10\% of the generated dataset.
  \item \textbf{Substitution Parameters:} A set of files containing the
    different parameter bindings that will be used by the driver to generate the
    read queries of the workloads.
\end{itemize}

\subsection{Scale Factors}
\label{sec:scale-factors}

\ldbcsnb defines a set of scale factors (SFs), targeting systems of different sizes and budgets.
SFs are computed based on the ASCII size \emph{in Gibibytes} of the generated output files using the \textsf{csv-singular-merged-fk} serializer (see \autoref{sec:serializers}).\footnote{This way of characterizing the size of the data set is identical to the scaling of TPC-H and TPC-DS.}
For both workloads, the SF1 data set is 1~GiB, SF100 is 100~GiB, and SF\numprint{10000} is \numprint{10000} GiB (not 10 TiB).

However, \textbf{note that the two SNB workloads have different data sets} due with different folder structures.

The data sets sizes are established as follows:
For both workloads, we use the default settings for the splitting the data into an intial (bulk-loaded) data set and the later update operations (``streams'').
For Interactive, both the 90\% initial data and the 10\% update streams count towards the total size and the \textsf{csv-singular-merged-fk} serializer is used.
For BI, the sum of the initial snapshot (97\%) and the update operations (daily inserts and deletes) are measured and the default CSV serializers (\textsf{composite-merged-fk}) is used.

It is important to note that for a given workload and scale factor, data sets generated using different serializers contain exactly the same data, the only difference is in how they are represented.%

The currently available SFs are the following: 1, 3, 10, 30, 100, 300, \numprint{1000}, \numprint{3000}, \numprint{10000}, \numprint{30000}.
Additionally, three small SFs, 0.003, 0.1, and 0.3 are provided to help initial testing and validation efforts.

The Test Sponsor may select the SF that better fits their needs, by properly configuring the \datagen, as described in \autoref{sec:data_generation}.
The size of the resulting dataset is mainly affected by the following configuration parameters: the number of persons and the number of years simulated.
By default, all SFs are defined over a period of three years, starting from 2010, and SFs are computed by scaling the number of Persons in the network.
\autoref{tab:number-of-entities-bi-raw} shows the number of entities for SFs 1, \ldots, \numprint{30000} data sets.


\begin{table}[htb]
    \scriptsize
    \setlength{\tabcolsep}{.3em}
    \tiny
    \begin{tabular}{|>{\tt}l||r|r|r|r|r|r|r|r|r|r|r|r|}
        \hline
        \tableHeaderFirst{File}             & \tableHeader{SF1}  & \tableHeader{SF3}  & \tableHeader{SF10}  & \tableHeader{SF30}   & \tableHeader{SF100}  & \tableHeader{SF300}  & \tableHeader{SF\numprint{1000}} & \tableHeader{SF\numprint{3000}} & \tableHeader{SF\numprint{10000}} & \tableHeader{SF\numprint{30000}} \\ \hline
        \hline\hline
        static/Organisation                 & \numprint{7955}    & \numprint{7955}    & \numprint{7955}     & \numprint{7955}      & \numprint{7955}      & \numprint{7955}      & \numprint{7955}                 & \numprint{7955}                 & \numprint{7955}                  & \numprint{7955}                  \\\hline
        static/Place                        & \numprint{1460}    & \numprint{1460}    & \numprint{1460}     & \numprint{1460}      & \numprint{1460}      & \numprint{1460}      & \numprint{1460}                 & \numprint{1460}                 & \numprint{1460}                  & \numprint{1460}                  \\\hline
        static/Tag                          & \numprint{16080}   & \numprint{16080}   & \numprint{16080}    & \numprint{16080}     & \numprint{16080}     & \numprint{16080}     & \numprint{16080}                & \numprint{16080}                & \numprint{16080}                 & \numprint{16080}                 \\\hline
        static/TagClass                     & \numprint{71}      & \numprint{71}      & \numprint{71}       & \numprint{71}        & \numprint{71}        & \numprint{71}        & \numprint{71}                   & \numprint{71}                   & \numprint{71}                    & \numprint{71}                    \\\hline
        dynamic/Comment                     & \numprint{2391707} & \numprint{7275929} & \numprint{24318240} & \numprint{71971437}  & \numprint{238859896} & \numprint{698717507} & \numprint{2305141269}           & \numprint{6788314573}           & \numprint{22203530429}           & \numprint{68078584186}           \\
        dynamic/Comment\_hasTag\_Tag        & \numprint{2903970} & \numprint{8957968} & \numprint{30193298} & \numprint{90186505}  & \numprint{300936421} & \numprint{885843849} & \numprint{2934823389}           & \numprint{8669809939}           & \numprint{28414179030}           & \numprint{87250551072}           \\\hline
        dynamic/Forum                       & \numprint{106594}  & \numprint{259629}  & \numprint{705629}   & \numprint{1754332}   & \numprint{4876750}   & \numprint{12314071}  & \numprint{35084033}             & \numprint{92411437}             & \numprint{272234669}             & \numprint{770847855}             \\
        dynamic/Forum\_hasMember\_Person    & \numprint{3260692} & \numprint{9831062} & \numprint{33637572} & \numprint{100176831} & \numprint{336799532} & \numprint{992219233} & \numprint{3299845513}           & \numprint{9734943439}           & \numprint{31952684743}           & \numprint{98131214167}           \\
        dynamic/Forum\_hasTag\_Tag          & \numprint{342040}  & \numprint{841153}  & \numprint{2294050}  & \numprint{5682315}   & \numprint{15787515}  & \numprint{39868135}  & \numprint{113622479}            & \numprint{299293084}            & \numprint{881501639}             & \numprint{2495628126}            \\\hline
        dynamic/Person                      & \numprint{10620}   & \numprint{25870}   & \numprint{70800}    & \numprint{175950}    & \numprint{487700}    & \numprint{1230500}   & \numprint{3505000}              & \numprint{9232000}              & \numprint{27200000}              & \numprint{77000000}              \\
        dynamic/Person\_hasInterest\_Tag    & \numprint{246066}  & \numprint{607394}  & \numprint{1659221}  & \numprint{4103933}   & \numprint{11398465}  & \numprint{28784564}  & \numprint{82043446}             & \numprint{216113647}            & \numprint{636466970}             & \numprint{1801780271}            \\
        dynamic/Person\_knows\_Person       & \numprint{219450}  & \numprint{668431}  & \numprint{2304951}  & \numprint{6880584}   & \numprint{23116805}  & \numprint{68313982}  & \numprint{227125539}            & \numprint{670962543}            & \numprint{2201852957}            & \numprint{6763316230}            \\
        dynamic/Person\_likes\_Comment      & \numprint{1616891} & \numprint{5469630} & \numprint{20401119} & \numprint{66391084}  & \numprint{243335846} & \numprint{776234551} & \numprint{2796244391}           & \numprint{8801761184}           & \numprint{30518383179}           & \numprint{97396567634}           \\
        dynamic/Person\_likes\_Post         & \numprint{844544}  & \numprint{2659885} & \numprint{9328362}  & \numprint{29137595}  & \numprint{105650858} & \numprint{335953318} & \numprint{1210202589}           & \numprint{3822741245}           & \numprint{13258168236}           & \numprint{42113297722}           \\
        dynamic/Person\_studyAt\_University & \numprint{8562}    & \numprint{20755}   & \numprint{56777}    & \numprint{140829}    & \numprint{390266}    & \numprint{984945}    & \numprint{2804285}              & \numprint{7386305}              & \numprint{21760681}              & \numprint{61607278}              \\
        dynamic/Person\_workAt\_Company     & \numprint{22766}   & \numprint{55826}   & \numprint{154122}   & \numprint{383107}    & \numprint{1061627}   & \numprint{2678190}   & \numprint{7627121}              & \numprint{20093569}             & \numprint{59188556}              & \numprint{167544307}             \\\hline
        dynamic/Post                        & \numprint{1192942} & \numprint{3056157} & \numprint{8781335}  & \numprint{22948816}  & \numprint{67764850}  & \numprint{181024990} & \numprint{548192276}            & \numprint{1516905453}           & \numprint{4693293319}            & \numprint{13820145527}           \\
        dynamic/Post\_hasTag\_Tag           & \numprint{778511}  & \numprint{2384596} & \numprint{8112750}  & \numprint{24116550}  & \numprint{80572324}  & \numprint{237819624} & \numprint{789063560}            & \numprint{2330311354}           & \numprint{7634983368}            & \numprint{23442869026}           \\\hline
    \end{tabular}
    \centering
    \caption{Properties of data sets for each scale factor for the \emph{raw data sets} produced the Spark-based generator, used as a basis of the data sets of SNB Interactive v2 and SNB BI.}
    \label{tab:number-of-entities-bi-raw}
\end{table}


\begin{table}
    \setlength{\tabcolsep}{.3em}
    \footnotesize
    \centering
    \begin{tabular}{|l|l|c|C{1.2cm}C{1.2cm}|C{1.2cm}C{1.2cm}|}
        \hline
        \multicolumn{1}{|c|}{\multirow{2}{*}{\bf Serializer name (v2.x)}} & \multicolumn{1}{c|}{\multirow{2}{*}{\bf Legacy serializer name (v0.x and v1.x)}} & \multirow{2}{*}{\bf Nodes} & \multicolumn{2}{c|}{\bf Attributes} & \multicolumn{2}{c|}{\bf Edges}                                        \\
                                                                           &                                                                         &                            & \bf single-valued                   & \bf multi-valued               & \bf one- to-many & \bf many- to-many \\ \hline
        \textsf{csv-singular-projected-fk}                                 & CsvBasic                                                                & \yes                       & \no                                 & \yes                           & \yes             & \yes              \\
        \textsf{csv-composite-projected-fk}                                & CsvComposite                                                            & \yes                       & \no                                 & \no                            & \yes             & \yes              \\
        \textsf{csv-singular-merged-fk}                                    & CsvMergeForeign                                                         & \yes                       & \no                                 & \yes                           & \no              & \yes              \\
        \textsf{csv-composite-merged-fk}                                   & CsvCompositeMergeForeign                                                & \yes                       & \no                                 & \no                            & \no              & \yes              \\ \hline
    \end{tabular}
    \caption{Attributes and edges serialized to separate files the different CSV serializers.}
    \label{tab:csv-serializers}
\end{table}

\autoref{tab:csv-serializers} shows how each CSV serializer handles attributes/edges of different cardinalities, demonstrating why \textsf{csv-singular-projected-fk} has the most files and \textsf{csv-composite-merged-fk} has the least number of files.

\begin{table}[htb]
    \scriptsize
    \setlength{\tabcolsep}{.5em}
    \centering
    \begin{tabularx}{\linewidth}{|>{\sffamily}c|>{\tt}l|>{\tt}X|}
        \hline
        \tableHeaderFirst{C} & \tableHeader{File}                      & \tableHeader{Content}                                                                                      \\
        \hline\hline
        N                    & static/Organisation/part-*.csv                      & id | type | name | url \\
        E                    & static/Organisation\_isLocatedIn\_Place/part-*.csv  & OrganisationId | PlaceId \\
        \hline
        N                    & static/Place/part-*.csv                             & id | name | url | type \\
        E                    & static/Place\_isPartOf\_Place/part-*.csv            & Place1Id | Place2Id \\
        \hline
        N                    & static/Tag/part-*.csv                               & id | name | url \\
        E                    & static/Tag\_hasType\_TagClass/part-*.csv            & TagId | TagClassId \\
        \hline
        N                    & static/TagClass/part-*.csv                          & id | name | url \\
        E                    & static/TagClass\_isSubclassOf\_TagClass/part-*.csv  & TagClass1Id | TagClass2Id \\
        \hline\hline
        N                    & dynamic/Comment/part-*.csv                          & creationDate | id | locationIP | browserUsed | content | length \\
        E                    & dynamic/Comment\_hasCreator\_Person/part-*.csv      & creationDate | CommentId | PersonId \\
        E                    & dynamic/Comment\_hasTag\_Tag/part-*.csv             & creationDate | CommentId | TagId \\
        E                    & dynamic/Comment\_isLocatedIn\_Country/part-*.csv    & creationDate | CommentId | CountryId \\
        E                    & dynamic/Comment\_replyOf\_Comment/part-*.csv        & creationDate | Comment1Id | Comment2Id \\
        E                    & dynamic/Comment\_replyOf\_Post/part-*.csv           & creationDate | CommentId | PostId \\
        \hline
        N                    & dynamic/Forum/part-*.csv                            & creationDate | id | title \\
        E                    & dynamic/Forum\_containerOf\_Post/part-*.csv         & creationDate | ForumId | PostId \\
        E                    & dynamic/Forum\_hasMember\_Person/part-*.csv         & creationDate | ForumId | PersonId \\
        E                    & dynamic/Forum\_hasModerator\_Person/part-*.csv      & creationDate | ForumId | PersonId \\
        E                    & dynamic/Forum\_hasTag\_Tag/part-*.csv               & creationDate | ForumId | TagId \\
        \hline
        N                    & dynamic/Person/part-*.csv                           & creationDate | id | firstName | lastName | gender | birthday | locationIP | browserUsed | language | email \\
        E                    & dynamic/Person\_hasInterest\_Tag/part-*.csv         & creationDate | personId | interestId \\
        E                    & dynamic/Person\_isLocatedIn\_City/part-*.csv        & creationDate | PersonId | CityId \\
        E                    & dynamic/Person\_knows\_Person/part-*.csv            & creationDate | Person1Id | Person2Id \\
        E                    & dynamic/Person\_likes\_Comment/part-*.csv           & creationDate | PersonId | CommentId \\
        E                    & dynamic/Person\_likes\_Post/part-*.csv              & creationDate | PersonId | PostId \\
        E                    & dynamic/Person\_studyAt\_University/part-*.csv      & creationDate | PersonId | UniversityId | classYear \\
        E                    & dynamic/Person\_workAt\_Company/part-*.csv          & creationDate | PersonId | CompanyId | workFrom \\
        \hline
        N                    & dynamic/Post/part-*.csv                             & creationDate | id | imageFile | locationIP | browserUsed | language | content | length \\
        E                    & dynamic/Post\_hasCreator\_Person/part-*.csv         & creationDate | PostId | PersonId \\
        E                    & dynamic/Post\_hasTag\_Tag/part-*.csv                & creationDate | PostId | TagId \\
        E                    & dynamic/Post\_isLocatedIn\_Country.csv              & creationDate | PostId | CountryId \\
        \hline
    \end{tabularx}
    \caption{Files output by the \texttt{csv-composite-projected-fk} serializer (31 in total). The first part of the table contains the static entites, the second part contains the dynamic ones.
        Notation -- \textsf{C}: entity category, \textsf{N}: node, \textsf{E}: edge.}
    \label{table:csv-composite-projected-fk}
\end{table}

\begin{table}[htb]
    \scriptsize
    \centering
    \begin{tabularx}{\linewidth}{|>{\sffamily}c|>{\tt}l|>{\tt}X|}
        \hline
        \tableHeaderFirst{C} & \tableHeader{File}                   & \tableHeader{Content}                                                                                               \\
        \hline\hline
        N                    & static/Organisation/part-*.csv                   & id | type | name | url | LocationPlaceId \\
        \hline
        N                    & static/Place/part-*.csv                          & id | name | url | type | PartOfPlaceId \\
        \hline
        N                    & static/Tag/part-*.csv                            & id | name | url | TypeTagClassId \\
        \hline
        N                    & static/TagClass/part-*.csv                       & id | name | url | SubclassOfTagClassId \\
        \hline\hline
        N                    & dynamic/Comment/part-*.csv                       & creationDate | id | locationIP | browserUsed | content | length | CreatorPersonId | LocationCountryId | ParentPostId | ParentCommentId \\
        E                    & dynamic/Comment\_hasTag\_Tag/part-*.csv          & creationDate | CommentId | TagId \\
        \hline
        N                    & dynamic/Forum/part-*.csv                         & creationDate | id | title | ModeratorPersonId \\
        E                    & dynamic/Forum\_hasMember\_Person/part-*.csv      & creationDate | ForumId | PersonId \\
        E                    & dynamic/Forum\_hasTag\_Tag/part-*.csv            & creationDate | ForumId | TagId \\
        \hline
        N                    & dynamic/Person/part-*.csv                        & creationDate | id | firstName | lastName | gender | birthday | locationIP | browserUsed | LocationCityId | language | email \\
        E                    & dynamic/Person\_hasInterest\_Tag/part-*.csv      & creationDate | personId | interestId \\
        E                    & dynamic/Person\_knows\_Person/part-*.csv         & creationDate | Person1Id | Person2Id \\
        E                    & dynamic/Person\_likes\_Comment/part-*.csv        & creationDate | PersonId | CommentId \\
        E                    & dynamic/Person\_likes\_Post/part-*.csv           & creationDate | PersonId | PostId \\
        E                    & dynamic/Person\_studyAt\_University/part-*.csv   & creationDate | PersonId | UniversityId | classYear \\
        E                    & dynamic/Person\_workAt\_Company/part-*.csv       & creationDate | PersonId | CompanyId | workFrom \\
        \hline
        N                    & dynamic/Post/part-*.csv                          & creationDate | id | imageFile | locationIP | browserUsed | language | content | length | CreatorPersonId | ContainerForumId | LocationCountryId \\
        E                    & dynamic/Post\_hasTag\_Tag/part-*.csv             & creationDate | PostId | TagId \\
        \hline
    \end{tabularx}
    \caption{Files output by the \texttt{csv-composite-merged-fk} serializer (18 in total). The first part of the table contains the static entites, the second part contains the dynamic ones.
        Notation -- \textsf{C}: entity category, \textsf{N}: node, \textsf{E}: edge.}
    \label{table:csv-composite-merged-fk}
\end{table}

\begin{table}[htb]
    \scriptsize
    \centering
    \begin{tabularx}{\linewidth}{|>{\sffamily}c|>{\tt}l|>{\tt}X|}
        \hline
        \tableHeaderFirst{C} & \tableHeader{File}                   & \tableHeader{Content}                                                                                                                                                                    \\
        \hline\hline
        N                    & static/Organisation/part-*.csv                   & id | type | name | url | LocationPlaceId \\
        \hline
        N                    & static/Place/part-*.csv                          & id | name | url | type | PartOfPlaceId \\
        \hline
        N                    & static/Tag/part-*.csv                            & id | name | url | TypeTagClassId \\
        \hline
        N                    & static/TagClass/part-*.csv                       & id | name | url | SubclassOfTagClassId \\
        \hline\hline
        N                    & dynamic/Comment/part-*.csv                       & creationDate | deletionDate | explicitlyDeleted | id | locationIP | browserUsed | content | length | CreatorPersonId | LocationCountryId | ParentPostId | ParentCommentId \\
        E                    & dynamic/Comment\_hasTag\_Tag/part-*.csv          & creationDate | deletionDate | CommentId | TagId \\
        \hline
        N                    & dynamic/Forum/part-*.csv                         & creationDate | deletionDate | explicitlyDeleted | id | title | ModeratorPersonId \\
        E                    & dynamic/Forum\_hasMember\_Person/part-*.csv      & creationDate | deletionDate | explicitlyDeleted | ForumId | PersonId \\
        E                    & dynamic/Forum\_hasTag\_Tag/part-*.csv            & creationDate | deletionDate | ForumId | TagId \\
        \hline
        N                    & dynamic/Person/part-*.csv                        & creationDate | deletionDate | explicitlyDeleted | id | firstName | lastName | gender | birthday | locationIP | browserUsed | LocationCityId | language | email \\
        E                    & dynamic/Person\_hasInterest\_Tag/part-*.csv      & creationDate | deletionDate | personId | interestId \\
        E                    & dynamic/Person\_knows\_Person/part-*.csv         & creationDate | deletionDate | explicitlyDeleted | Person1Id | Person2Id \\
        E                    & dynamic/Person\_likes\_Comment/part-*.csv        & creationDate | deletionDate | explicitlyDeleted | PersonId | CommentId \\
        E                    & dynamic/Person\_likes\_Post/part-*.csv           & creationDate | deletionDate | explicitlyDeleted | PersonId | PostId \\
        E                    & dynamic/Person\_studyAt\_University/part-*.csv   & creationDate | deletionDate | PersonId | UniversityId | classYear \\
        E                    & dynamic/Person\_workAt\_Company/part-*.csv       & creationDate | deletionDate | PersonId | CompanyId | workFrom \\
        \hline
        N                    & dynamic/Post/part-*.csv                          & creationDate | deletionDate | explicitlyDeleted | id | imageFile | locationIP | browserUsed | language | content | length | CreatorPersonId | ContainerForumId | LocationCountryId \\
        E                    & dynamic/Post\_hasTag\_Tag/part-*.csv             & creationDate | deletionDate | PostId | TagId \\
        \hline
    \end{tabularx}
    \caption{Directories created by the raw serializer (18 in total). The first part of the table contains the static entites, the second part contains the dynamic ones.
        Notation -- \textsf{C}: entity category, \textsf{N}: node, \textsf{E}: edge.
        The entities with the \texttt{explicitlyDeleted} attribute -- Comment, Forum, Post nodes, and hasMember, knows, likes (Comment/Post) edges -- denote whether the entity is deleted as part of an explicit delete operation or implicitly through a cascading delete operation.}
    \label{table:csv-raw}
\end{table}

\subsection{Serializers}
\label{sec:serializers}

The datasets are generated in the \texttt{social\_network/} directory, split into static and dynamic parts (\autoref{fig:schema}).
The filenames (without the extension) end in \texttt{\_i\_j} where \texttt{i} is the block id and \texttt{j} is the partition id (set by \texttt{numThreads}).
The SUT has to take care only of the generated Dataset to be bulk loaded. Using \texttt{NULL} values for storing optional values is allowed.

\datagen currently only supports CSV-based serializers.
These produce CSV-like text files which use the pipe character ``\texttt{|}'' as the primary field separator and the semicolon character ``\texttt{;}'' as a separator for multi-valued attributes (for the composite serializers).
The CSV files are stored in two subdirectories: \texttt{static/} and \texttt{dynamic/}.
Depending on the number of threads used for generating the dataset, the number of files varies, since there is a file generated per thread. We indicate this with ``\texttt{part-*}'' in the specification.

The following CSV variants are supported:
    \begin{itemize}
      \item \textsf{csv-composite-projected-fk:}
      Each relation with a cardinality larger than one are output in a separate file.
      Generated files and their schemas as shown in \autoref{table:csv-composite-projected-fk}.

      \item \textsf{csv-composite-merged-fk:}
      This serializer is similar to \textsf{csv-composite-projected-fk}, but relations that have a cardinality of 1-to-N are merged in the entity files as a foreign keys.
      There are 13~such relations in total:
      \begin{itemize}
        \item Comment\_hasCreator\_Person, Comment\_isLocatedIn\_Country, Comment\_replyOf\_Comment, Comment\_replyOf\_Post (merged to Comment);
        \item Forum\_hasModerator\_Person (merged to Forum);
        \item Organisation\_isLocatedIn\_Place (merged to Organisation);
        \item Person\_isLocatedIn\_City (merged to Person);
        \item Place\_isPartOf\_Place (merged to Place);
        \item Post\_hasCreator\_Person, Post\_isLocatedIn\_Country, Forum\_containerOf\_Post (merged to Post);
        \item Tag\_hasType\_TagClass (merged to Tag);
        \item TagClass\_isSubclassOf\_TagClass (merged to TagClass)
      \end{itemize}
      Generated files and their schemas as shown in \autoref{table:csv-composite-merged-fk}.
    
      \item \textsf{csv-singular-merged-fk}:
      Similar to the \textsf{csv-composite-merged-fk} but multi-valued attributes (\texttt{Person.email} and \texttt{Person.speaks}) are stored as separate directories (\texttt{Person\_email\_EmailAddress} and \texttt{Person\_speaks\_Language}, resp.).
    
      \item \textsf{csv-singular-projected-fk}:
      Similar to the \textsf{csv-composite-projected-fk} but multi-valued attributes (\texttt{Person.email} and \texttt{Person.speaks}) are stored as separate directories (\texttt{Person\_email\_EmailAddress} and \texttt{Person\_speaks\_Language}, resp.).

      \item \texttt{raw} mode:
      The file names are the same as in \texttt{composite-merged-fk} but there are two important differences:
      (1)~additional attributes, \eg \texttt{deletionDate}, \texttt{explicitlyDeleted}, and \texttt{weight} (used for weighted graphs in Graphalytics~\cite{DBLP:journals/corr/abs-2011-15028}), are included,
      (2)~all data is included, \ie if a Forum is created and deleted before sampling the initial data set, it is included in this data set.
      Generated files and their schemas as shown in \autoref{table:csv-raw}.
    \end{itemize}

\paragraph{Inheritance}
    
The inheritance hierarchies in the schema have two important characteristics
(1)~all subclasses use the same id space, \eg there cannot be a Comment and a Post with id 1 at the same time,
(2)~they are serialized to CSVs using either the \emph{map hierarchy to single table} or \emph{map each concrete class to its own table} strategies\footnote{\url{http://www.agiledata.org/essays/mappingObjects.html}}:

\begin{description}
    \item[Message = Comment | Post]
    \emph{Map each concrete class to its own table} is used \ie separate CSV files are used for the Comment and the Post classes.

    \item[Place = City | Country | Continent]
    \emph{Map hierarchy to single table} is used, \ie all Place node are serialized in a single file. A discriminator attribute ``type'' is used with the value denoting the concrete class, \eg ``Country''.

    \item[Organisation = Company | University]
    \emph{Map hierarchy to single table} is used, \ie all Organisation nodes are serialized in a single fiel. A discriminator attribute ``type'' is used with the value denoting the concrete class, \eg ``Company''.
\end{description}

\subsection{Interactive Update Streams (Inserts)}

The generic schema for the Interactive update streams is given in \autoref{table:update_stream_generic_schema}, while the concrete schemas of each insert operations is given in \autoref{table:update_stream_schemas}.
The update stream files are generated in the \texttt{social\_network/} directory and are grouped as follows:

\begin{itemize}
    \item \texttt{updateStream\_*\_person.csv} files contain update operation 1: \queryRefCard{insert-01}{INS}{1}
    \item \texttt{updateStream\_*\_forum.csv} files contain update operations 2--8: %
    \queryRefCard{insert-02}{INS}{2}
    \queryRefCard{insert-03}{INS}{3}
    \queryRefCard{insert-04}{INS}{4}
    \queryRefCard{insert-05}{INS}{5}
    \queryRefCard{insert-06}{INS}{6}
    \queryRefCard{insert-07}{INS}{7}
    \queryRefCard{insert-08}{INS}{8}
\end{itemize}

Remark: update streams in version 1 only contain inserts, while in version 2, they contain both inserts and deletes.

\subsection{Substitution Parameters}

The substitution parameters are generated in the \texttt{substitution\_parameters/} directory.
Each parameter file is named \texttt{\{interactive|bi\}\_<id>\_param.txt}, corresponding to an operation of
Interactive complex reads (\queryRefCard{interactive-complex-read-01}{IC}{1}--\queryRefCard{interactive-complex-read-14-v2}{IC}{14v2}) and
BI reads (\queryRefCard{bi-read-01}{BI}{1}--\queryRefCard{bi-read-20}{BI}{20}).
The schemas of these files are defined by the operator, \eg the schema of \queryRefCard{interactive-complex-read-01}{IC}{1} is ``\texttt{personId|firstName}''.


\section{Introducing Delete Operations}

\paragraph{Challenge for systems}
To support deletion operations graph processing systems need to solve numerous technical challenges:
\begin{enumerate}
\item Users should be able to \emph{express deletion operations} using the database API, preferably using a high-level declarative query language with clear semantics~\cite{Green2019}.
\item Deletion operations \emph{limit the algorithms and data structures} that can be used by a system. Certain dynamic graph algorithms are significantly more expensive to recompute in the presence of deletes~\cite{DBLP:conf/soda/Roditty13} or only support either insert or deletions but not both~\cite{DBLP:conf/esa/RodittyZ04}. A number of updatable matrix storage formats only support efficient insertions but not deletions~\cite{DBLP:conf/hpec/BusatoGBB18}. Meanwhile some graph databases might be able to exploit indices to speed up deletions~\cite[Sec.~4.4.2]{Besta2019}
\item \emph{Distributed graph databases} need to employ specialized protocols to enforce consistency of deletions~\cite{Waudby2020}.
\end{enumerate}

\paragraph{Challenge for benchmarks}
Due to their importance and challenging nature, we found it necessary to incorporate delete operations into LDBC benchmarks.
However, doing so is a non-trivial task as it impacts on each component in the benchmark workflow:
workload specifications, data generation, parameter curation, and the workload driver.
This section focuses primarily on data generation.

The initial step in generating delete operations is to define the semantics of the desired operations.
To understand common behaviour of deletes we informally surveyed several social networks, the findings of which motivated the design
of 8~delete operations described in~\autoref{sec:bi-delete-operations}.

The next step was to generate \emph{delete events} within LDBC's synthetic data generator and ensure that they follow a logic order in
the social network.
For example, a delete \tKnows edge event can only occur after both \tPersons join the network and become friends,
but before either \tPerson leaves the network.
To achieve this Datagen was extended to support \emph{dynamic entities}.
Dynamic entities have a \emph{creation date} and a \emph{deletion date}, which together comprise an entity's \emph{lifespan}.
Once generated this allows for the extraction of deletion operations, which can be utilized by LDBC workloads.
Deriving valid lifespans for dynamic entities was the subject of a short paper published at the GRADES-NDA~2020
workshop~\cite{DBLP:conf/sigmod/WaudbySPS20} and is presented in~\autoref{sec:lifespan-management}.

Next it was important to distinguish between \emph{implicit} and \emph{explicit} delete events.
Continuing with the \tKnows edge example, once created the connection exists until either \tPerson leaves the network,
at which point the connection is implicitly deleted, as per the semantics of delete \tPerson~(\autoref{sec:delete-01}).
Alternatively, at any time up until this event, the friendship can be explicitly deleted,
\ie two people have a disagreement and ``unfriend'' each other, but both continue using the social network.
Distinguishing between these types of events is important as only explicit delete events should become delete operations
in the workload.

To achieve this each dynamic entity is assigned a probability of being explicitly deleted, if selected the entity is marked as such;
this is used to ensure the correct serialization of delete events into delete operations.
For entities selected for explicit deletion the next step is to determine a realistic time at which the event occurs.
For example, a post has a higher probability of being deleted soon after it was posted compared to after 5 days.
To achieve this each dynamic entity is assigned a realistic distribution to select delete event timestamps from,
which respects the bounds imposed by the valid lifespans.
The probability distributions used to determine if a dynamic entities is explicitly deleted and then when that event occurs is discussed
in~\autoref{sec:ensuring-realism}.

Once generated dynamic entities must be correctly serialized.
Depending on its creation date, deletion date, and if the entity is explicitly deleted it can,
(i) spawn an insert and delete operation,
(ii) be included in the bulk load component and spawn a delete operation,
(iii) just be included in the bulk load component,
(iv) spawn only an insert operation, and
(v) not be serialized at all!
The approach for doing this is presented in~\autoref{sec:conv-delete-events}.

We summarize the numerous challenges supporting the generation of dynamic entities and thus delete operations poses below:
\begin{enumerate}
\item \textbf{Validity.} The generator should produce \emph{valid lifespans},
where each generated dynamic entity guarantees that
(a)~events in the graph follow a logical order: \eg in a social network, two people can become friends only after both persons joined the network and before either person leaves the network,
(b)~the graph never violates the cardinality constraints prescribed by its schema, and
(c)~the graph continuously satisfies the semantic constraints required by the application domain (\eg no isolated comments in a social network).
\item \textbf{Realism.} The generator should create a graph with a realistic correlations and distribution of entities over time.
For example, in a social network the distribution of activity is non-uniform over time, real-world events such as elections or controversial posts 
can drive spikes of posts and unfollowings respectively~\cite{DBLP:conf/www/MyersL14}.
In addition, deletions can be correlated with certain attributes: \eg the likelihood a person leaves the network may be correlated with their number of friends~\cite{Lorincz2019}.
Also, there are often temporal correlations between entity creation and deletion: \eg posts have an increased chance of deletion immediately following creation compared to after a 3~month period.
\item \textbf{Serialization.} Care must be taken to distinguish between implicit and explicit delete events when creating the bulk load component, insert operations, and delete operations.
\item \textbf{Scalability.} A graph with dynamic entities should be generated at scale (up to billions of edges).
\end{enumerate}

\section{Lifespan Management}
\label{sec:lifespan-management}

This section is based on the short paper published at the GRADES-NDA~2020 workshop~\cite{DBLP:conf/sigmod/WaudbySPS20} authored by the task force members.


In this section, we define the constraints for generating dynamic entities in a social network. Each dynamic entity gets a \emph{lifespan}, represented by two \emph{lifespan attributes}, a \emph{creation date} and a \emph{deletion date}.
We first briefly review the data generator, introduce our notation and define the parameters of the generation process. Then, we define the semantic constraints which regulate the participation in certain relationships along with the constraints for selecting intervals. We illustrate an application of these with two examples, shown in \autoref{fig:example-graph} and \autoref{fig:example-graph-complex}.

\paragraph{Graph schema}

The LDBC Datagen component~\cite{Pham2012,Datagen} is responsible for generating the graph used in the benchmarks. It produces a synthetic dataset modelling a social network's activity. Its graph schema has 11~concrete node types connected by 20~edge types, and its entities (nodes/edges) are classified as either dynamic or static (\autoref{fig:schema}).
The dynamic part of the graph comprises of a fully connected \tPerson graph and a number of \tMessage trees under \tForums.

\paragraph{Notation}
To describe lifespans and related constraints, we use the following notation.
Constants are uppercase bold, \eg $\constant{NC}$.
Entity types are monospaced, \eg \tPerson, \tHasMember.
Variables are lowercase italic, \eg $\variable{pers}, \variable{hm}$.
Entities are sans-serif, \eg $\instance{P_1}, \instance{HM}$.
For an entity $x$, $\varc{x}{}$ denotes its creation date, while $\vard{x}{}$ denotes its deletion date.
In most cases, both the creation and the deletion date are selected from an interval, \eg $\varc{x}{} \in \interval{d_1}{d_2}$ means that entity $\variable{x}$ should be created between dates $d_1$ (inclusive) and $d_2$ (exclusive).
The selected creation and deletion dates together form an interval that represents the lifespan of its entity.
If any of the intervals for selecting the lifespan attributes of an entity are empty, \ie $d_2 \leq d_1$, the entity should be discarded.
As illustrated later, this interval is often used to determine the intervals where the creation and deletion dates of dependant entities are selected.

\paragraph{Parameters} 
We parameterize the generator as follows.
The network is created in 2010 and exists for 10~years at which point the network collapses ($\xNC = 2020$).
Data is simulated for a 3-year period, between the simulation start, $\xSS = 2010$ and the simulation end, $\xSE = 2013$.
In order to allow \emph{windowed execution} by the LDBC SNB driver (used for multi-threaded and distributed operation), we define a sufficiently large amount of time that needs to pass between consecutive operations on an entity as $\Delta = 10\text{s}$.


\subsection{General Rules}

In this section, we define general rules that must be satisfied by all entities in the graph. In subsequent sections, we refine these with domain-specific constraints.
For a node $\varn{1}$, we always require that:
\begin{itemize}
    \item $\varcn{1} \in \interval{\xSS}{\xSE}$, the node must be created between the simulation start and the simulation end.
    \item $\vardn{1} \in \interval{\varcn{1} + \Delta}{\xNC}$, the node must exist for at least $\Delta$ time and must be deleted before the network collapse.
\end{itemize}

To enforce referential integrity constraints (\ie prevent dangling edges), the lifespan of edge $\variable{e}$ between nodes $\varn{1}$ and $\varn{2}$ must always satisfy the following criteria:

\begin{itemize}
    \item $ \varc{e} \in \interval
        {\max(\varcn{1}, \varcn{2})}
        {\min(\vardn{1}, \vardn{2}, \xSE)} $,
        in other terms,
        the edge must be created no sooner than both of its endpoints
        but before
        any of its endpoints are deleted.
    \item $ \vard{e} \in \interval
        {\varc{e} + \Delta}
        {\min(\vardn{1}, \vardn{2})} $,
        \ie the edge must exist for at least $\Delta$ time and
        deleted no later than
        any of its endpoints.
\end{itemize}

To further refine the constraints for edges, we distinguish between two main cases.

(1)~The endpoints of edge $\variable{e}$ are existing node $\varn{1}$ and node $\varn{2}$ which is inserted at the same time as the edge:
\begin{itemize}
    \item $ \varc{e} = \varcn{2} $ 
    \item $ \vard{e} = \min(\vardn{1}, \vardn{2}) $.
    In case of edges with \emph{containment semantics} (node $\varn{1}$ contains $\varn{2}$ through edge $e$),
    node $\varn{2}$ must always be deleted at the same time as edge $\variable{e}$,
    therefore
    $\vard{e} = \vardn{2}$ and $\vardn{2} \leq \vardn{1}$.
    %
\end{itemize}

(2)~In other cases, the edge must be created when both of its endpoints already exist and must be deleted no later than them:

\begin{itemize}
    \item $ \varc{e} \in \interval
        {\max(\varcn{1}, \varcn{2}) + \Delta}
        {\min(\vardn{1}, \vardn{2}, \xSE)} $
    \item $ \vard{e} \in \interval
        {\varc{e} + \Delta}
        {\min(\vardn{1}, \vardn{2})} $
\end{itemize}

These constraints capture the ``minimum'' (\ie most relaxed) set of constraints that must be enforced in all domains.
Next, we introduce additional constraints specific to our social network schema.

\subsection{Person}

A \tPerson $\ePerson$ is the avatar a real-world person creates when they join the network. A \tPerson joins the network, $\varc{\ePerson}$, during the simulation period and they leave the network, $\vard{\ePerson}$, during the network lifetime:

\begin{itemize}
    \item $\varc{\ePerson} \in \interval{\xSS}{\xSE}$
    \item $\vard{\ePerson} \in \interval{\varc{\ePerson} + \Delta}{\xNC}$
\end{itemize}

For the edges of \tPerson nodes pointing to a static node
($\type{isLocatedIn}$,
$\type{studyAt}$,
$\type{workAt}$, and
$\type{hasInterest}$),
we assign the creation and deletion date from $\varc{\ePerson}$ and $\vard{\ePerson}$, resp.


\begin{figure}[ht]
  \centering
  \begin{subfigure}{\linewidth}
    \tikzstyle{vertex} = [circle,minimum width=45pt,draw]
    \centering
    \begin{tikzpicture}[node distance=2cm]
      \node (v1) [vertex,xshift=0cm,yshift=0cm,fill=Person] {\scriptsize{$\instance{P_1}:\type{Person}$}};
      \node (v2) [vertex,xshift=5cm,yshift=0cm,fill=Person] {\scriptsize{$\instance{P_2}:\type{Person}$}};
      \node [below of=v1,yshift=1cm] {\tiny{\texttt{$\instc{P}{1}$: Feb 22 2010}}};
      \node [below of=v1,yshift=0.7cm] {\tiny{\texttt{$\instd{P}{1}$: Jul 26 2014}}};
      \node [below of=v2,yshift=1cm] {\tiny{\texttt{$\instc{P}{2}$: Mar 07 2010}}};
      \node [below of=v2,yshift=0.7cm] {\tiny{\texttt{$\instd{P}{2}$: Oct 17 2012}}};
      \draw[thick,->,>=stealth] (v1) --
      node [midway,above] {\scriptsize{$\instance{knows_{1,2}} : \type{Knows}$}}
      node [midway,below,align=right,text width=2.5cm] {\tiny{\texttt{$\instc{knows}{1,2}$: Dec 01 2011}}}
      node [midway,below,yshift=-0.3cm,align=right,text width=2.5cm] {\tiny{\texttt{$\instd{knows}{1,2}$: Jun 05 2012}}} (v2);
    \end{tikzpicture}
    \caption{An instance of a \tKnows edge connecting two \tPerson nodes. \emph{Creation} and \emph{deletion} dates are shown for each entity.}
    \label{fig:person-knows-graph}
    \end{subfigure}
    \begin{subfigure}{\linewidth}
    \footnotesize
    \centering
    \begin{tikzpicture}[node distance=2cm,thick,every node/.style={transform shape}]

      \draw [thick,->,>=stealth] node [above,black] {} (-1,0/1.5) -- (6,0/1.5); 
      \draw [grey,thin] (-0.5,0/1.5) node [above,black] {$\xSS$} -- (-0.5,-6/1.5);
      \draw [grey,thin]  (5.5,0/1.5) node [above,black] {$\xNC$} -- (5.5,-6/1.5);
      \draw [grey,thin]  (4.5,0/1.5) node [above,black] {$\xSE$} -- (4.5,-6/1.5);

      \draw[mark=*,mark size=2pt,mark options={color=green}] plot coordinates {(0.5,-1/1.5)} node [left] {$\instance{P_1}$}
      -- plot[mark=*,mark size=2pt,mark options={color=red}] coordinates {(5.0,-1/1.5)};

      \draw[mark=*,mark size=2pt,mark options={color=green}] plot coordinates {(1,-2/1.5)}  node [left] {$\instance{P_2}$}
      -- plot[mark=*,mark size=2pt,mark options={color=red}] coordinates {(4.0,-2/1.5)};

      \draw[thin,grey,<->] plot coordinates {(1.3,-3/1.5)} node [left,black,xshift=-0.2cm] {\textcolor{grey}{$\instc{knows}{1,2}$}} node [align=right,xshift=-3.2cm,text width=2.55cm] {{\textcolor{green}{$\max(\instc{P}{1}, \instc{P}{2}) + \Delta$}}}
      -- plot[mark=*,mark size=2pt,mark options={color=blue}] coordinates {(2.5,-3/1.5)}
      -- plot coordinates {(4.0,-3/1.5)} node [xshift=3cm,text width=2.5cm] {{\textcolor{red}{$\min(\instd{P}{1}, \instd{P}{2}, \xSE)$}}};

     \draw [grey,thin,dashed] (4.0,-2/1.5) -- (4.0,-3/1.5);
     \shadedBox(1,-2/1.5,1/1.5);

     \draw[thin,grey,<->] plot coordinates {(2.8,-4/1.5)} node [left,black,xshift=-0.2cm] {\textcolor{grey}{$\instd{knows}{1,2}$}} node [align=right,xshift=-4.7cm,text width=2.55cm] {{\textcolor{green}{$\instc{knows}{1,2} + \Delta$}}}
     -- plot[mark=*,mark options={color=blue}] coordinates {(3.5,-4/1.5)}
     -- plot coordinates {(4.0,-4/1.5)} node [xshift=3cm,text width=2.5cm] {{\textcolor{red}{$\min (\instd{P}{1},  \instd{P}{2})$}}}; 

     \draw [grey,thin,dashed] (4.0,-3/1.5) -- (4.0,-4/1.5);
     \shadedBox(2.5,-3/1.5,1/1.5);

      \draw[mark=*,mark size=2pt,mark options={color=green}] plot coordinates {(2.5,-5/1.5)}  node [left] {$\instance{knows_{1,2}}$}
      -- plot[mark=*,mark size=2pt,mark options={color=red}] coordinates {(3.5,-5/1.5)};
     \draw [thin,dashed,grey] (2.5,-4/1.5) -- (2.5,-5/1.5);
     \draw [thin,dashed,grey] (3.5,-4/1.5) -- (3.5,-5/1.5);

    \end{tikzpicture}
    \caption{
      Illustration of the intervals in which the \emph{creation dates} \textcolor{green}{$\bullet$} and the \emph{deletion dates} \textcolor{red}{$\bullet$} can be selected.
      Thick black lines represent entity lifespans and thin grey lines represent valid intervals that dates can be selected in; \textcolor{blue}{$\bullet$} indicates the selected times (spanning the lifespan interval of the given entity).
      On the thin grey lines, thicker sections represent the minimal amount of time that must pass before selecting a value.
      In case of creation dates, this is used to ensure that the dependant entity exists for at least $\Delta$ time.
      In case of deletion dates, it is used to ensure that the entity exists for at least $\Delta$ time.
    }
    \label{fig:person-knows}
  \end{subfigure}

  \caption{Example graph and its intervals.}
  \label{fig:example-graph}
\end{figure}

\subsubsection{Knows}

The \tKnows edge connects two \tPersons $p_i$ and $p_j$ that know each other in the network.
The intervals where the creation and deletion dates can be generated in are illustrated in \autoref{fig:person-knows} and defined below:
\begin{itemize}
    \item $\varc{\eKnows_{i,j}} \in \interval{\max(\varc{\ePerson}_{i},\varc{\ePerson}_{j}) + \Delta}{\min(\vard{\ePerson}_{i},\vard{\ePerson}_{j}, \xSE)}$
    \item $\vard{\eKnows_{i,j}} \in \interval{\varc{\eKnows_{i,j}} + \Delta}{\min (\vard{\ePerson}_{i}, \vard{\ePerson}_{j})}$
\end{itemize}

\subsection{Forum and Message}

The rules for \tForum and \tMessage nodes along with their edges are given in \autoref{sec:forum} and \autoref{sec:message}, respectively, and illustrated in \autoref{fig:example-graph-complex}.


\subsection{Forum}
\label{sec:forum}
\label{sec:hasModerator}

A \tForum is a meeting point where people post \tMessages.
There exists three categories of \tForums:
Wall ($\eForum_\textsf{w}$),
Album ($\eForum_\textsf{a}$),
and Group ($\eForum_\textsf{g}$).
Each \tForum has a set of \tPersons connected via \tHasMember edges, a set of \tTags connected via \tHasTag edges, a single moderator connected by a \tHasModerator edge and a set of \tMessages (discussed in Section \ref{sec:message}).
For all \tForums the outgoing \tHasTag edges get their creation date and deletion date from $\varc{\eForum}$ and $\vard{\eForum}$, respectively.

\subsubsection{Groups}
Groups are public places for people that share interests, any \tPerson can create a Group $\eForum_\textsf{g}$ during their lifespan. A Group can be deleted anytime after it was created.
\begin{itemize}
    \item $\varc{\eForum}_{\mathsf{g}} \in \interval{\varc{\ePerson} + \Delta}{\min (\vard{\ePerson}, \xSE)}$
    \item $\vard{\eForum}_{\mathsf{g}} \in \interval{\varc{\eForum}_{\mathsf{g}} + \Delta}{\xNC}$
\end{itemize}

\paragraph{Group Moderator}
The initial \tHasModerator $\eHasModerator_{\mathsf{g}}$ is the Group creator. If the moderator leaves the Group, the Group will have no moderator (this is allowed in the schema of version 0.4.0+, see \autoref{fig:schema}).
\begin{itemize}
\item $\varc{\eHasModerator}_{\mathsf{g}} \in \interval{\varc{\eForum}_{\mathsf{g}} + \Delta}{\min (\vard{\eForum}_{\mathsf{g}}, \vard{\ePerson}, \xSE)}$
\item $\vard{\eHasModerator}_{\mathsf{g}} \in \interval{ \varc{\eHasModerator}_{\mathsf{g}} + \Delta}{\min (\vard{\eForum}_{\mathsf{g}}, \vard{\ePerson})}$
\end{itemize}

\paragraph{Group Membership}
Any \tPerson $\ePerson$ can become a member of a given group. The \tHasMember $\eHasMember_{\mathsf{g}}$ creation is generated from the interval in which the \tPerson and \tForum lifespans overlap. The deletion date is generated from the interval between the membership creation date (incremented by $\Delta$) and the minimum of the \tPerson and \tForum deletion dates.
\begin{itemize}
    \item $\varc{\eHasMember}_{\mathsf{g}} \in \interval{\max ( \varc{\eForum}_{\mathsf{g}}, \varc{\ePerson}) + \Delta}{\min (\vard{\eForum}, \vard{\ePerson}, \xSE)} $
    \item $\vard{\eHasMember}_{\mathsf{g}} \in \interval{\varc{\eHasMember}_{\mathsf{g}} + \Delta}{\min (\vard{\eForum}_{\mathsf{g}}, \vard{\ePerson})}$
\end{itemize}

\subsubsection{Walls}
Every \tPerson $\ePerson$, has a Wall $\eForum_\textsf{w}$ which is created when the \tPerson joins the social network. The wall is deleted when the \tPerson is deleted.
\begin{itemize}
    \item $\varc{\eForum}_{\mathsf{w}} = \varc{\ePerson} + \Delta$
    \item $\vard{\eForum}_{\mathsf{w}} = \vard{\ePerson}$
\end{itemize}

\paragraph{Wall Moderator}
Each \tPerson has a \tHasModerator $\eHasModerator_{\mathsf{w}}$ edge to their wall, which gets the creation date (incremented by $\Delta$) and deletion date from $\eForum_\textsf{w}$.
Note, only the moderator can create \tPost nodes on the wall and the connecting \tTag nodes are set based on the interest of the moderator.
\begin{itemize}
    \item $\varc{\eHasModerator}_{\mathsf{w}} = \varc{\eForum}_{\mathsf{w}} + \Delta$
    \item $\vard{\eHasModerator}_{\mathsf{w}} = \vard{\eForum}_{\mathsf{w}}$
\end{itemize}

\paragraph{Wall Membership}
For a \tPerson $p_i$, all their friends $p_j$ (\tPerson nodes connected via a \tKnows edge) become members of $\eForum_\textsf{w}$ at the time the \tKnows edge is created. Hence, a \tHasMember $\eHasMember_{\mathsf{w}}$ edge gets the creation date of \tKnows incremented by $\Delta$. The deletion date is derived from the minimum of the \tForum deletion date and \tKnows deletion date.
\begin{itemize}
    \item $\varc{\eHasMember}_{\mathsf{w}} = \varc{\eKnows_{i,j}} + \Delta$
    \item $\vard{\eHasMember}_{\mathsf{w}} = \min(\vard{\eForum}_{\mathsf{w}}, \vard{\eKnows_{i,j}})$
\end{itemize}

\subsubsection{Albums}
A \tPerson can create multiple Albums ($\eForum_\textsf{a}$) containing a set of \tPhotos{}. Albums can be created and then deleted at any point during the lifespan of the \tPerson.
\begin{itemize}
    \item $\varc{\eForum}_\textsf{a} \in \interval{\varc{\ePerson} + \Delta}{\min (\vard{\ePerson}, \xSE)}$
    \item $\vard{\eForum}_\textsf{a} \in \interval{ \varc{\eForum}_\textsf{a} + \Delta}{\vard{\ePerson}}$
\end{itemize}

\paragraph{Album Moderator}
The \tPerson is the moderator for any Album they create. Album ownership cannot change hence \tHasModerator $\eHasModerator_{\mathsf{a}}$ gets the creation date (incremented by $\Delta$) and deletion date from $\varc{\eForum}_\textsf{a}$ and $\vard{\eForum}_\textsf{a}$ respectively.
\begin{itemize}
\item $\varc{\eHasModerator}_{\mathsf{a}} = \varc{\eForum}_{\mathsf{a}} + \Delta$
\item $\vard{\eHasModerator}_{\mathsf{a}} = \vard{\eForum}_{\mathsf{a}}$
\end{itemize}

\paragraph{Album Membership}
Only friends $\ePerson_i$ of a \tPerson $\ePerson_j$ can become members of Albums created by $\ePerson_j$. The \tHasMember $\eHasMember_{\mathsf{a}}$ edge creation date is derived from the Album and $\type{knows}$ creation dates. The deletion is derived from the $\type{Forum}$ and $\type{knows}$ deletion dates.
\begin{itemize}
    \item $\varc{\eHasMember}_\textsf{a} = \max ( \varc{\eForum}_\textsf{a}, \varc{\eKnows}_{i,j} ) + \Delta $
    \item $\vard{\eHasMember}_{\mathsf{w}} = \min ( \vard{\eForum}_\textsf{a}, \vard{\eKnows}_{i,j} ) $
\end{itemize}

\subsection{Message}
\label{sec:message}

A \tMessage is an abstract entity that represents a message created by a \tPerson.
There are two \tMessage subtypes: \tPost and \tComment.
A \tPost is created in a \tForum and a \tComment represents a comment made by a \tPerson to an existing \tMessage (either a \tPost or a \tComment).
In a \tForum the set of \tMessage nodes form a \emph{tree} with a \tPost node at the root and \tComment nodes for the rest.

\subsubsection{Post}

A \tPost can be created by a \tPerson in a \tForum.
Only the moderator (\ie owner) can post on a Wall or in an Album (\tHasModerator),
whereas all members including the moderator (\tHasMember/\tHasModerator) can post in a Group.
These relationships are captured with the $\eHasMember$ variable in the formulas.
\tPosts are divided in three categories, \emph{regular posts}, \emph{photos}, and \emph{flashmob posts}.

\paragraph{Regular Posts and Photos}

Regular posts capture the standard daily activity in a Group or on a Wall.
Photos are created in Albums. (Interaction with Photos is limited to \tLikes, see details in \autoref{sec:likes}). The creation date for these is determined as follows:
$$\varc{\ePost} \in \interval{\varc{\eHasMember} + \Delta}{\min (\vard{\eHasMember}, \xSE) }$$

\paragraph{Flashmob Posts}

Flashmob posts are generated around events that attract significant interest
(such as elections) that result in a spike in activity.
These events span a $2\phi$-hour time window centered around a specific event time, flashmob event $\eFlashmobEvent$, in the middle of the window; there are $\phi$ hours each side of the specific event time.
$$
\varc{\ePost} \in \interval{\max(\varc{\eHasMember} + \Delta,\eFlashmobEvent - \phi\textrm{\,h})}{\min (\vard{\eHasMember},\eFlashmobEvent + \phi\textrm{\,h},\xSE)}
$$

The deletion dates for all categories of \tPosts are determined as:
$$\vard{\ePost} \in \interval{\varc{\ePost} + \Delta}{\vard{\eHasMember}}$$

\paragraph{containerOf edge}

Each \tPost node has an incoming $\type{containerOf}$ edge which gets the same lifespan attributes as the \tPost.

\subsubsection{Comment}

A \tComment $\eComment$ is created by \tPerson $\ePerson$ as a reply to \tMessage $\eMessage$. \tComments are only made in Walls and Groups. \tComment always occur within $\gamma$~days of their parent message following a power-law distribution with mean 6.85 hours.

\begin{itemize}
    \item $\varc{\eComment} \in \interval{\max (\varc{\eMessage}, \varc{\eHasMember}) + \Delta}{\min (\vard{\eMessage}, \vard{\eHasMember}, \varc{\eMessage} + \gamma\textrm{\,d}, \xSE)}$
    \item $\vard{\eComment} \in \interval{\varc{\eComment} + \Delta}{\min (\vard{\eMessage}, \vard{\eHasMember})}$
\end{itemize}

\paragraph{replyOf edge}

\tComments always have an outgoing \tReplyOf edge with containment semantics, \ie the target \tMessage contains the \tComment. These edges get the same lifespan as their source \tComment.

\subsubsection{likes}
\label{sec:likes}

A \tLikes edge $\eLikes$ can exist between \tPerson $\ePerson$ and \tMessage $\eMessage$. Messages can only receive likes during a $\mu$-day window after their creation at which point no more activity is generated.

\begin{itemize}
    \item $\varc{\eLikes} \in \interval{\max(\varc{\ePerson}, \varc{\eMessage}) + \Delta}{\min (\vard{\ePerson}, \vard{\eMessage}, \varc{\eMessage} + \mu\textrm{\,d}, \xSE)}$
    \item $\vard{\eLikes} \in \interval{\varc{\eLikes} + \Delta}{\min (\vard{\ePerson}, \vard{\eMessage})}$
\end{itemize}

\subsection{Complex Example}
\label{sec:complex-example}

In \autoref{fig:example-graph-complex}, a complex example graph is shown with the corresponding intervals.
Both \emph{the intervals for selecting the creation and deletion date} attributes and the selected \emph{lifespan intervals} are shown.

\begin{figure*}[htp]
  \centering
  \begin{subfigure}{\linewidth}
    \tikzstyle{vertex} = [circle, minimum width=40pt,draw,inner sep=0pt]
\tikzstyle{edge} = [thick,->,>=stealth,out=180,in=0]
\usetikzlibrary{positioning}

\centering
\begin{tikzpicture}[node distance=4cm,scale=0.85,every node/.style={transform shape}]
  \node (forum) [vertex,xshift=0cm,yshift=0cm,fill=Forum] {\scriptsize{$\instance{F_{1}}:\type{Forum}$}};
  \node (post) [vertex,xshift=0cm,yshift=0cm,fill=Post,right of=forum] {\scriptsize{$\instance{Post_{1}}:\type{Post}$}};
  \node (comment1) [vertex,xshift=0cm,yshift=0cm,fill=Comment,right of=post] {\scriptsize{$\instance{C_{1}}:\type{Comm}$}};
  \node (comment2) [vertex,xshift=0cm,yshift=0cm,fill=Comment,right of=comment1] {\scriptsize{$\instance{C_{2}}:\type{Comm}$}};

  \node (person1) [vertex,xshift=0cm,yshift=2cm,fill=Person,left of=forum] {\scriptsize{$\instance{P_{1}}:\type{Person}$}};
  \node (person2) [vertex,xshift=0cm,yshift=0cm,fill=Person,left of=forum] {\scriptsize{$\instance{P_{2}}:\type{Person}$}};
  \node (person3) [vertex,xshift=0cm,yshift=0.5cm,fill=Person,below of=forum] {\scriptsize{$\instance{P_{3}}:\type{Person}$}};

  \draw [edge] (forum) to [out=145,in=-15] node [midway,above,sloped] {\scriptsize{$\instance{hmod} : \type{hasModerator}$}}
  node [midway,above,xshift=1.2cm,yshift=0.6cm,align=center,text width=2.8cm] {\scriptsize{$\instc{hmod}{}$: \texttt{Apr 01 2010}}}
  node [midway,above,xshift=1.2cm,yshift=0.2cm,align=center,text width=2.8cm] {\scriptsize{$\instd{hmod}{}$: \texttt{Oct 02 2012}}}
  (person1);
  \draw [edge] (forum) to [out=175,in=-45] node [midway,below,sloped] {\scriptsize{$\instance{hmem_{1}} : \type{hasMember}$}} (person1);
  \draw [edge] (forum) to [out=-155,in=-25]
  node [midway,below] {\scriptsize{$\instance{hmem_{2}} : \type{hasMember}$}}
  node [midway,below,yshift=-0.5cm,align=center,text width=3.5cm] {\scriptsize{$\instc{hmem}{2}$: \texttt{Jun 15 2010}}}
  node [midway,below,yshift=-0.9cm,align=center,text width=3.5cm] {\scriptsize{$\instd{hmem}{2}$: \texttt{Jul 26 2012}}}
  (person2);

  \draw [edge] (forum) --
  node [midway,above,xshift=1.5cm,yshift=0.5cm] {\scriptsize{$\instance{hmem_{3}} : \type{hasMember}$}}
  node [midway,above,xshift=1.5cm,yshift=0.1cm,align=right,text width=2.8cm] {\scriptsize{$\instc{hmem}{3}$: \texttt{Dec 08 2010}}}
  node [midway,above,xshift=1.5cm,yshift=-0.3cm,align=right,text width=2.8cm] {\scriptsize{$\instd{hmem}{3}$: \texttt{Feb 29 2012}}} (person3);

  \draw [edge] (forum) -- node [midway,above,grey] {\scriptsize{$\instance{cO} : \type{containerOf}$}} (post);

  \draw [edge] (post) -- node [midway,above,grey] {\scriptsize{$\instance{rO_{1}} : \type{replyOf}$}} (comment1);
  \draw [edge] (comment1) to [out=0,in=180] node [midway,above,grey] {\scriptsize{$\instance{rO_{2}} : \type{replyOf}$}} (comment2);
  \draw [edge] (comment1) to [out=145,in=12]  node [near start,above,grey,sloped] {\scriptsize{$\instance{hC_{1}} : \type{hasCreator}$}} (person1);
  \draw [edge] (comment2) to [out=-135,in=-90] node [near start,above,grey,sloped] {\scriptsize{$\instance{hC_{2}} : \type{hasCreator}$}} (person2);
  \draw [edge] (post) to [out=-90,in=45] node [midway,below,grey,sloped,xshift=0.2cm] {\scriptsize{$\instance{hC_{3}} : \type{hasCreator}$}}  (person3);

  \node [below of=comment1,yshift=3cm,text width=2.5cm,align=center] {\scriptsize{$\instc{C}{1}$: \texttt{Dec 17 2010}}};
  \node [below of=comment1,yshift=2.6cm,text width=2.5cm,align=center] {\scriptsize{$\instd{C}{1}$: \texttt{Dec 18 2010}}};

  \node [above of=comment2,yshift=-2.6cm,text width=2.5cm,align=center] {\scriptsize{$\instc{C}{2}$: \texttt{Dec 18 2010}}};
  \node [above of=comment2,yshift=-3cm,text width=2.5cm,align=center] {\scriptsize{$\instd{C}{2}$: \texttt{Dec 18 2010}}};

  \node [above of=post,yshift=-2.6cm,text width=2.5cm,align=center] {\scriptsize{$\instc{Post}{1}$: \texttt{Dec 16 2010}}};
  \node [above of=post,yshift=-3cm,text width=2.5cm,align=center] {\scriptsize{$\instd{Post}{1}$: \texttt{Dec 12 2011}}};

  \node [above right=0.6 cm and -0.5cm of forum,text width=2.5cm,align=center] {\scriptsize{$\instc{F}{1}$: \texttt{Apr 01 2010}}};
  \node [above right=0.2 cm and -0.5cm of forum,text width=2.5cm,align=center] {\scriptsize{$\instd{F}{1}$: \texttt{Oct 02 2012}}};

  \node [left of=person2,xshift=2cm,yshift=0cm,text width=2.5cm,align=center] {\scriptsize{$\instc{P}{2}$: \texttt{Jan 29 2010}}};
  \node [left of=person2,xshift=2cm,yshift=-0.4cm,text width=2.5cm,align=center] {\scriptsize{$\instd{P}{2}$: \texttt{Nov 15 2012}}};

  \node [left of=person1,xshift=2cm,yshift=0cm,text width=2.5cm,align=center] {\scriptsize{$\instc{P}{1}$: \texttt{Feb 08 2010}}};
  \node [left of=person1,xshift=2cm,yshift=-0.4cm,text width=2.5cm,align=center] {\scriptsize{$\instd{P}{1}$: \texttt{Dec 23 2016}}};

  \node [right of=person3,xshift=-2cm,yshift=0cm,text width=2.5cm,align=center] {\scriptsize{$\instc{P}{3}$: \texttt{Jul 21 2010}}};
  \node [right of=person3,xshift=-2cm,yshift=-0.4cm,text width=2.5cm,align=center] {\scriptsize{$\instd{P}{3}$: \texttt{Apr 17 2012}}};

\end{tikzpicture}
    \caption{Example graph with an instance of a \tForum containing a \tMessage tree of depth 3 and its \tPerson members. Lifespan attributes (\emph{creation} and \emph{deletion dates}) are shown for each dynamic entity. Edges in grey get their lifespan attributes as per \autoref{fig:schema} and \autoref{sec:hasModerator}.}
    \label{fig:comments-graph}
  \end{subfigure}
  \begin{subfigure}{\linewidth}
    \centering
    \footnotesize
    \begin{tikzpicture}[node distance=2cm,thick,scale=0.78,every node/.style={transform shape}]
      \draw [thick,->,>=stealth] node [above,black] {} (-1,0/1.8) -- (13,0/1.8);
      \draw [grey,thin] (-0.5,0/1.8) node [above,black] {$\xSS$} -- (-0.5,-22/1.8);
      \draw [grey,thin] (12.5,0/1.8) node [above,black] {$\xNC$} -- (12.5,-22/1.8);
      \draw [grey,thin] (11.8,0/1.8) node [above,black] {$\xSE$} -- (11.8,-22/1.8);

      \draw[mark=*,mark size=2pt,mark options={color=green}] plot coordinates {(0.5,-1/1.8)} node [left] {$\instance{P_1}$}
      -- plot[mark=*,mark size=2pt,mark options={color=red}] coordinates {(12,-1/1.8)};
      \shadedBox(0.5,-1/1.8,1/1.8);

      \draw[thin,grey,<->] plot coordinates {(0.8,-2/1.8)} node [left,black,xshift=-0.2cm] {\textcolor{grey}{$\instc{F}{1}$}} node [align=right,xshift=-3.5cm,text width=4cm] {{\textcolor{green}{$\instc{P}{1} + \Delta$}}}
      -- plot[mark=*,mark size=2pt,mark options={color=blue}] coordinates {(1,-2/1.8)}
      -- plot coordinates {(11.8,-2/1.8)} node [xshift=3.0cm,text width=4cm] {{\textcolor{red}{$\min(\instd{P}{1},\xSE)$}}};
      \shadedBox(1,-2/1.8,1/1.8);

      \draw[thin,grey,<->] plot coordinates {(1.3,-3/1.8)} node [left,black,xshift=-0.2cm] {\textcolor{grey}{$\instd{F}{1}$}} node [align=right,xshift=-4cm,text width=4cm] {{\textcolor{green}{$\instc{F}{1} + \Delta$}}}
      -- plot[mark=*,mark size=2pt,mark options={color=blue}] coordinates {(11,-3/1.8)}
      -- plot coordinates {(12.5,-3/1.8)} node [xshift=2.3cm,text width=4cm] {{\textcolor{red}{$\xNC$}}};
      \draw [grey,thin,dashed] (1,-3/1.8) -- (1,-4/1.8);
      \draw [grey,thin,dashed] (11,-3/1.8) -- (11,-4/1.8);

      \draw[mark=*,mark size=2pt,mark options={color=green}] plot coordinates {(1,-4/1.8)}  node [left] {$\instance{F_{1}}$}
      -- plot[mark=*,mark size=2pt,mark options={color=red}] coordinates {(11,-4/1.8)};
      \draw[mark=*,mark size=2pt,mark options={color=green}] plot coordinates {(1,-4.5/1.8)}  node [left] {$\instance{hmem_{1}}$}
      -- plot[mark=*,mark size=2pt,mark options={color=red}] coordinates {(11,-4.5/1.8)};
      \draw [thin,dashed,grey]   (1.0,-4/1.8) -- (1,-5/1.8);

      \draw [thin,dashed,gray] (2.5,-4/1.8) -- (2.5,-9/1.8);
      \draw [thin,dashed,gray] (10.5,-4/1.8) -- (10.5,-11/1.8);
      \draw [thin,dashed,gray]   (11.0,-4/1.8) -- (11,-7/1.8);

      \draw[mark=*,mark size=2pt,mark options={color=green}] plot coordinates {(0.2,-5/1.8)}  node [left] {$\instance{P_2}$}
      -- plot[mark=*,mark size=2pt,mark options={color=red}] coordinates {(11.25,-5/1.8)};
      \shadedBox(1,-5/1.8,1/1.8)

      \draw[thin,grey,<->] plot coordinates {(1.3,-6/1.8)} node [left,black,xshift=-0.2cm] {\textcolor{grey}{$\instc{hmem}{2}$}} node [align=right,xshift=-4cm,text width=4cm] {{\textcolor{green}{{$\max ( \instc{F}{1}, \instc{P}{2}) + \Delta$}}}}
      -- plot[mark=*,mark size=2pt,mark options={color=blue}] coordinates {(1.7,-6/1.8)}
      -- plot coordinates {(11,-6/1.8)} node [xshift=3.8cm,text width=4cm] {{\textcolor{red}{$\min(\instd{F}{1},\instd{P}{2},\xSE)$}}};
      \shadedBox(1.7,-6/1.8,1/1.8);

      \draw[thin,grey,<->] plot coordinates {(2,-7/1.8)} node [left,black,xshift=-0.2cm] {\textcolor{grey}{$\instd{hmem}{2}$}} node [align=right,xshift=-4.7cm,text width=4cm] {{\textcolor{green}{$\instc{hmem}{2} + \Delta$}}}
      -- plot[mark=*,mark size=2pt,mark options={color=blue}] coordinates {(10.7,-7/1.8)}
      -- plot coordinates {(11,-7/1.8)} node [xshift=3.8cm,text width=4cm] {{\textcolor{red}{$\min(\instd{F}{1}, \instd{P}{2})$}}};
      \draw [grey,thin,dashed] (1.7,-7/1.8) -- (1.7,-8/1.8);
      \draw [grey,thin,dashed] (10.7,-7/1.8) -- (10.7,-8/1.8);

      \draw[mark=*,mark size=2pt,mark options={color=green}] plot coordinates {(1.7,-8/1.8)}  node [left] {$\instance{hmem_{2}}$}
      -- plot[mark=*,mark size=2pt,mark options={color=red}] coordinates {(10.7,-8/1.8)};

      \draw[mark=*,mark size=2pt,mark options={color=green}] plot coordinates {(2.5,-9/1.8)}  node [left] {$\instance{P_3}$}
      -- plot[mark=*,mark size=2pt,mark options={color=red}] coordinates {(10.5,-9/1.8)};
      \shadedBox(2.5,-9/1.8,1/1.8);

      \draw[thin,grey,<->] plot coordinates {(2.8,-10/1.8)} node [left,black,xshift=-0.2cm] {\textcolor{grey}{$\instc{hmem}{3}$}}
      -- plot[mark=*,mark size=2pt,mark options={color=blue}] coordinates {(3.5,-10/1.8)} node [align=right,xshift=-6.2cm,text width=4cm] {{\textcolor{green}{{$\max ( \instc{F}{1}, \instc{P}{3}) + \Delta$}}}}
      -- plot coordinates {(10.5,-10/1.8)} node [xshift=4.3cm,text width=4cm] {{\textcolor{red}{$\min(\instd{F}{1}, \instd{P}{3}, \xSE)$}}};
      \shadedBox(3.5,-10/1.8,1/1.8);

      \draw[thin,grey,<->] plot coordinates {(3.8,-11/1.8)} node [left,xshift=-0.2cm,black] {\textcolor{grey}{$\instd{hmem}{3}$}} node [align=right,xshift=-6.5cm,text width=4cm] {{\textcolor{green}{$\instc{hmem}{3} + \Delta$}}}
      -- plot[mark=*,mark size=2pt,mark options={color=blue}] coordinates {(10,-11/1.8)}
      -- plot coordinates {(10.5,-11/1.8)} node [xshift=4.3cm,text width=4cm] {{\textcolor{red}{$\min(\instd{F}{1}, \instd{P}{3})$}}};
      \draw [grey,thin,dashed] (3.5,-11/1.8) -- (3.5,-12/1.8);
      \draw [grey,thin,dashed] (10,-11/1.8) -- (10,-12/1.8);


      \draw[mark=*,mark size=2pt,mark options={color=green}] plot coordinates {(3.5,-12/1.8)}  node [left] {$\instance{hmem_{3}}$}
      -- plot[mark=*,mark size=2pt,mark options={color=red}] coordinates {(10,-12/1.8)};
      \shadedBox(3.5,-12/1.8,1/1.8);
      \draw [thin,dashed,grey] (10,-12/1.8)  -- (10,-14/1.8);

      \draw[thin,grey,<->] plot coordinates {(3.8,-13/1.8)} node [left,black,xshift=-0.2cm] {\textcolor{grey}{$\instc{Post}{1}$}} node [align=right,xshift=-6.5cm,text width=4cm] {{\textcolor{green}{{$\instc{hmem}{3} + \Delta$}}}}
      -- plot[mark=*,mark size=2pt,mark options={color=blue}] coordinates {(4,-13/1.8)}
      -- plot coordinates {(10,-13/1.8)} node [xshift=4.8cm,text width=4cm] {{\textcolor{red}{$\min (\instd{hmem}{3}, \xSE)$}}};
      \shadedBox(4,-13/1.8,1/1.8);

      \draw[thin,grey,<->] plot coordinates {(4.3,-14/1.8)} node [left,black,xshift=-0.2cm] {\textcolor{grey}{$\instd{Post}{1}$}} node [align=right,xshift=-7cm,text width=4cm] {{\textcolor{green}{$\instc{Post}{1} + \Delta$}}}
      -- plot[mark=*,mark size=2pt,mark options={color=blue}] coordinates {(9.5,-14/1.8)}
      -- plot coordinates {(10,-14/1.8)} node [xshift=4.8cm,text width=4cm] {{\textcolor{red}{$\instd{hmem}{3}$}}}; 
      \draw [grey,thin,dashed] (4,-14/1.8) -- (4,-15/1.8);
      \draw [grey,thin,dashed] (9.5,-14/1.8) -- (9.5,-15/1.8);

      \draw[mark=*,mark size=2pt,mark options={color=green}] plot coordinates {(4,-15/1.8)}  node [left] {$\instance{Post_1}$}
      -- plot[mark=*,mark size=2pt,mark options={color=red}] coordinates {(9.5,-15/1.8)};
      \draw [thin,dashed,grey] (5.3,-15/1.8)  -- (5.3,-16/1.8);

      \draw[thin,grey,<->] plot coordinates {(4.3,-16/1.8)} node [left,black,xshift=-0.2cm] {\textcolor{grey}{$\instc{C}{1}$}} node [align=right,xshift=-7cm,text width=4cm] {{\textcolor{green}{{$\max (\instc{Post}{1}, \instc{hmem}{1}) + \Delta$}}}}
      -- plot[mark=*,mark size=2pt,mark options={color=blue}] coordinates {(4.8,-16/1.8)}
      -- plot coordinates {(5.3,-16/1.8)} node [xshift=9.5cm,text width=4cm] {{\textcolor{red}{$\min (\instc{Post}{1} + \gamma\,\mathrm{d}, \instd{hmem}{1}, \xSE)$}}};
      \shadedBox(4,-15/1.8,1/1.8);

      \draw[thin,grey,<->] plot coordinates {(5.1,-17/1.8)} node [left,black,xshift=-0.2cm] {\textcolor{grey}{$\instd{C}{1}$}} node [align=right,xshift=-7.8cm,text width=4cm] {{\textcolor{green}{{$\instc{Post}{1} + \Delta$}}}}
      -- plot[mark=*,mark size=2pt,mark options={color=blue}] coordinates {(7.8,-17/1.8)}
      -- plot coordinates {(9.5,-17/1.8)} node [xshift=5.3cm,text width=4cm] {{\textcolor{red}{$\min (\instd{Post}{1}, \instd{hmem}{1})$}}}; 
      \draw [grey,thin,dashed] (4.8,-17/1.8) -- (4.8,-18/1.8);
      \draw [grey,thin,dashed] (7.8,-17/1.8) -- (7.8,-18/1.8);

      \shadedBox(4.8,-16/1.8,1/1.8);
      \draw [thin,dashed,grey] (9.5,-15/1.8)  -- (9.5,-17/1.8);

      \draw[mark=*,mark size=2pt,mark options={color=green}] plot coordinates {(4.8,-18/1.8)}  node [left] {$\instance{C_1}$}
      -- plot[mark=*,mark size=2pt,mark options={color=red}] coordinates {(7.8,-18/1.8)};

      \draw[thin,grey,<->] plot coordinates {(5.1,-19/1.8)} node [left,black,xshift=-0.2cm] {\textcolor{grey}{$\instc{C}{2}$}} node[align=right,xshift=-7.8cm,text width=4cm] {{\textcolor{green}{{$\max (\instc{C}{1}, \instc{hmem}{2}) + \Delta$}}}}
      -- plot[mark=*,mark size=2pt,mark options={color=blue}] coordinates {(5.5,-19/1.8)}
      -- plot coordinates {(6.1,-19/1.8)}  node [xshift=8.7cm,text width=4cm] {{\textcolor{red}{$\min (\instc{C}{1} + \gamma\,\mathrm{d}, \instd{hmem}{2}, \xSE)$}}};

      \shadedBox(4.8,-18/1.8,1/1.8);
      \draw [thin,dashed,grey] (6.1,-18/1.8)  -- (6.1,-19/1.8);
      \draw [thin,dashed,grey] (7.8,-18/1.8)  -- (7.8,-20/1.8);

      \draw[thin,grey,<->] plot coordinates {(5.8,-20/1.8)} node [left,black,xshift=-0.2cm] {\textcolor{grey}{$\instd{C}{2}$}} node [align=right,xshift=-8.5cm,text width=4cm] {{\textcolor{green}{{$\instc{C}{1} + \Delta$}}}}
      -- plot[mark=*,mark size=2pt,mark options={color=blue}] coordinates {(6.5,-20/1.8)}
      -- plot coordinates {(7.8,-20/1.8)} node [xshift=7cm,text width=4cm] {{\textcolor{red}{$\min (\instd{C}{1}, \instd{hmem}{2})$}}}; 
      \shadedBox(5.5,-19/1.8,1/1.8);
      \draw [grey,thin,dashed] (5.5,-20/1.8) -- (5.5,-21/1.8);
      \draw [grey,thin,dashed] (6.5,-20/1.8) -- (6.5,-21/1.8);

      \draw[mark=*,mark size=2pt,mark options={color=green}] plot coordinates {(5.5,-21/1.8)}  node [left] {$\instance{C_2}$}
      -- plot[mark=*,mark size=2pt,mark options={color=red}] coordinates {(6.5,-21/1.8)};

    \end{tikzpicture}
    \caption{Illustration of the intervals in which the \emph{creation dates} \textcolor{green}{$\bullet$} and the \emph{deletion dates} \textcolor{red}{$\bullet$} of entities can be selected. Thick black lines represent entity lifespans and thin grey lines represent valid intervals that dates can be selected in; \textcolor{blue}{$\bullet$} indicates the selected times (spanning the lifespan interval of the given entity). On the thin grey lines, thicker sections represent the minimal amount of time that must pass before selecting a value. In case of creation dates, this is used to ensure that the dependant entity exists for at least $\Delta$ time. In case of deletion dates, it is used to ensure that the entity exists for at least $\Delta$ time.}
    \label{fig:comment-interval}
  \end{subfigure}
  \caption{Example graph and time intervals for selecting lifespan attributes, \emph{creation} and \emph{deletion dates}.}
  \label{fig:example-graph-complex}
\end{figure*}


\section{Ensuring Realism}
\label{sec:ensuring-realism}

Capturing realistic deletion behaviour was broken down into two dimensions.
Firstly, each dynamic entity is assigned a probability of being explicitly deleted.
Second, if selected for explicit deletion, a deletion event date is selected using a distribution bound by the valid lifespan of that entity.
To make informed choices of deletion probabilities and deletion date distributions, where possible, real-world data was used.

\paragraph{Delete Person}

\begin{figure}[htb]
  \centering
  \includegraphics[scale=\yedscale]{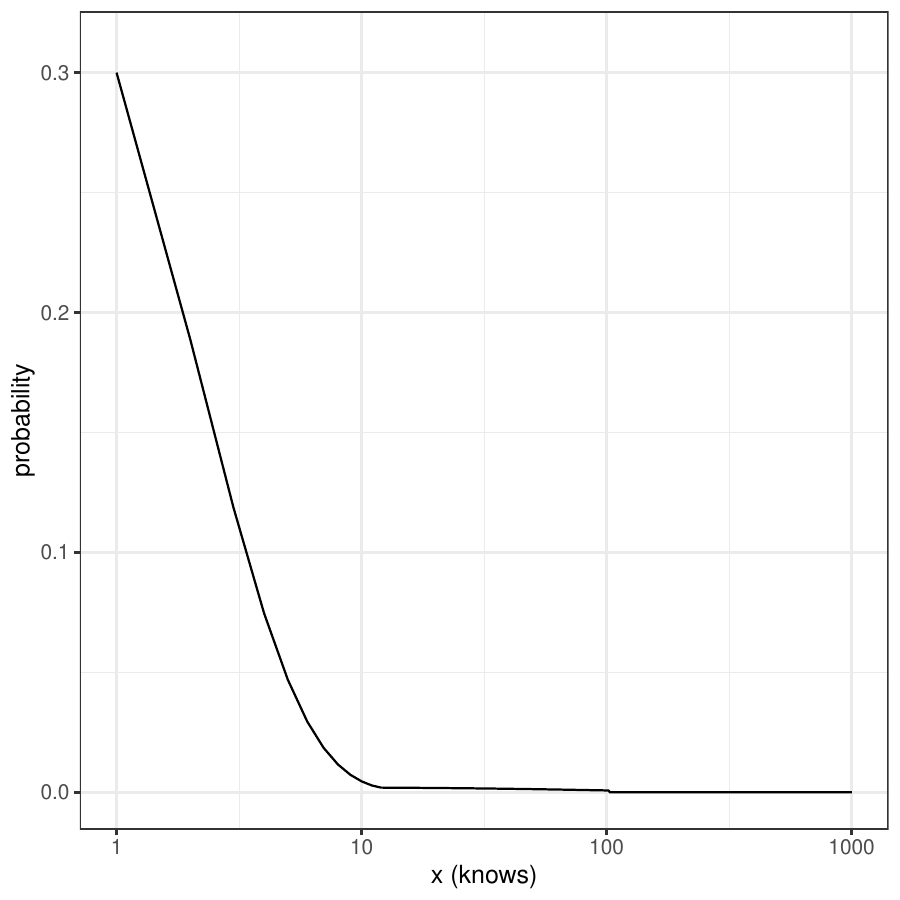}
  \caption{Distribution for determining the probability a \tPerson is deleted given their number of connections.}
  \label{fig:if-person}
\end{figure}

Lorincz \etal~\cite{Lorincz2019} have analyzed iWiW, a now-defunct Hungarian social network, observing that people with more connections are less likely to leave a social network.
When a \tPerson is generated they are assigned a \emph{maxKnows} value which indicates the amount of \tKnows connections they
will make across the lifetime of the network.
This information is then utilized to determine the probability a person is explicitly deleted using the distribution provided
in~\cite{Lorincz2019}, reproduced in~\autoref{fig:if-person}.
A deletion event date is then selected uniformly from the person's valid lifespan.
On average 3.5\% of \tPersons are deleted across the simulation period.

\paragraph{Delete Knows}

Myers and Leskovec~\cite{DBLP:conf/www/MyersL14} analysed 1.2 billion tweets from 13.1 million Twitter users.
These users made 112.3 million new connections, and deleted 39.2 million connections; a 3:1 follow:unfollow ratio.
As Datagen models a generic social media platform we have chosen a different ratio of 20:1
(on average 5\% of \tKnows edges are deleted), rather than overcapture behavior that may be unique to a single site.
\cite{DBLP:conf/www/MyersL14} also finds a constant background flux of follows and unfollows interleaved with bursts in such activity.
Currently, Datagen has no follow bursts, thus, we have decided not to incorporate unfollow bursts.
They also find less similar friends have a high probability of being unfollowed; modelling this relationship is work in progress.
If a \tKnows edge is selected for explicit deletion then a deletion date is then selected uniformly from the edge's valid lifespan.

\paragraph{Delete Post/Comment and Delete Post/Comment Like}

Posts in groups and walls are produced via a uniform generator and a flashmob generator, capturing background events and bursts in
events respectively.
A comment generator is then used to produce a tree of comments on each post.
Posts in albums are referred to as photos, they are produced by a different generator and do not have flashmob events nor do they have comment trees.
Additionally, all posts and comments have a number of likes generated for it.

Almuhimedi \etal~\cite{DBLP:conf/cscw/AlmuhimediWLSA13} tracked 292K Twitter users for 1 week.
They found 2.4\% of 67.2M tweets were deleted across 4 categories: status posts, retweets, replies, and mentions of other users that
were not replies.
In order to apply these findings to Datagen and obtain the average percentage of \tMessages and \tLikes deleted across the simulation
period, tweet categories were mapped to Datagen \tMessage types.
\autoref{table:almuhimedi-mapping} gives the mapping and the percentage deleted across the simulation period within each category.

\begin{table}[htb]
  \centering
  \begin{tabular}{ |l|l|r| }
    \hline
    \multicolumn{1}{|c|}{\textbf{Message type~\cite{DBLP:conf/cscw/AlmuhimediWLSA13}}} &
    \multicolumn{1}{c|}{\textbf{Datagen}} &
    \multicolumn{1}{c|}{\textbf{\% Deleted}} \\
    \hline\hline
    Status updates & Post/Photo & 2.7 \\
    \hline
    Non-reply mentions &  Post/Photo & 2.7 \\
    \hline
    Replies & Comment & 1.8 \\
    \hline
    Retweets & Post/Photo/Comment Likes & 2.4 \\
    \hline
  \end{tabular}
  \centering
  \caption{Mapping of~\cite{DBLP:conf/cscw/AlmuhimediWLSA13} message types to LDBC's schema.}
  \label{table:almuhimedi-mapping}
\end{table}

\begin{figure}[htb]
  \centering
  \includegraphics[scale=\yedscale]{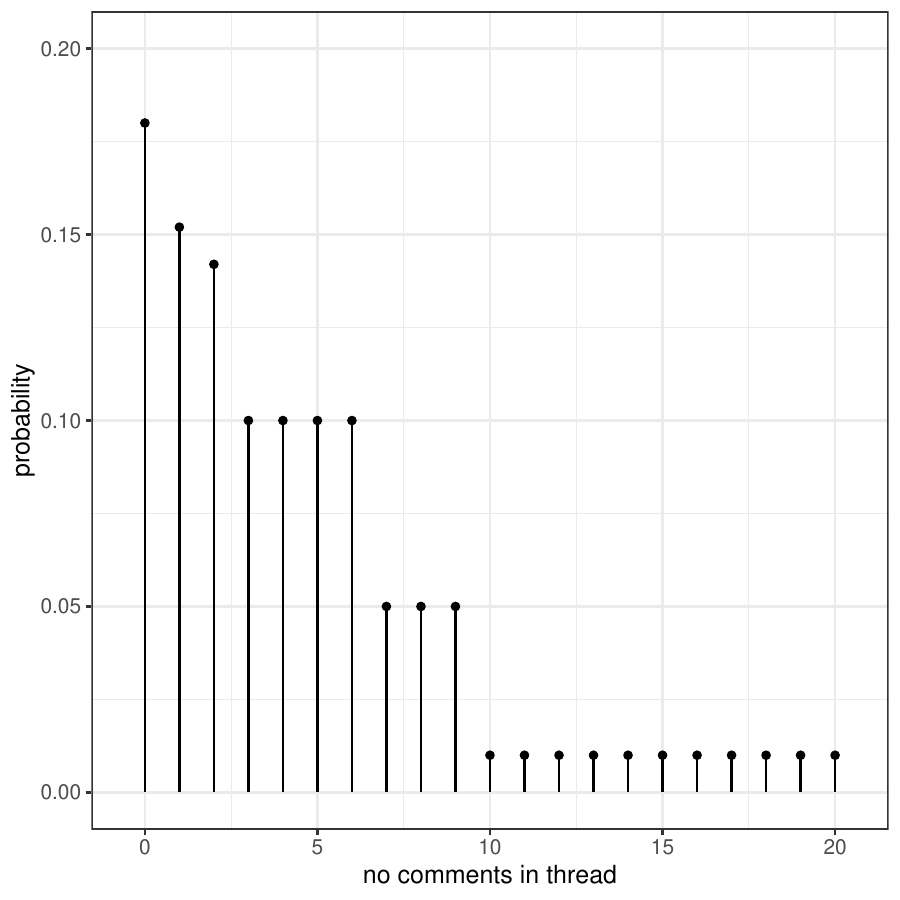}
  \caption{Probability a post is deleted given the number of comments in its thread.}
  \label{fig:if-post}
\end{figure}

Additionally, \cite{DBLP:conf/cscw/AlmuhimediWLSA13} identified not all users delete messages, with around 50\% of users doing so.
Thus, each \tPerson in the network has a 50\% chance of being marked a \emph{messageDeleter}, who subsequently, may or may not, delete post, comments, or likes.
\cite{DBLP:conf/cscw/AlmuhimediWLSA13} also identify a relationship between the depth of replies to a tweet and the chance the tweet
is deleted -- a tweet with less replies is more likely to be deleted.
We apply this relationship to the number of  \tComments in a \tPosts thread using the distribution in \autoref{fig:if-post}.
Note, this distribution has an average of 2.7\% aligning with \autoref{table:almuhimedi-mapping}.

\begin{figure}[htb]
  \centering
  \includegraphics[scale=\yedscale]{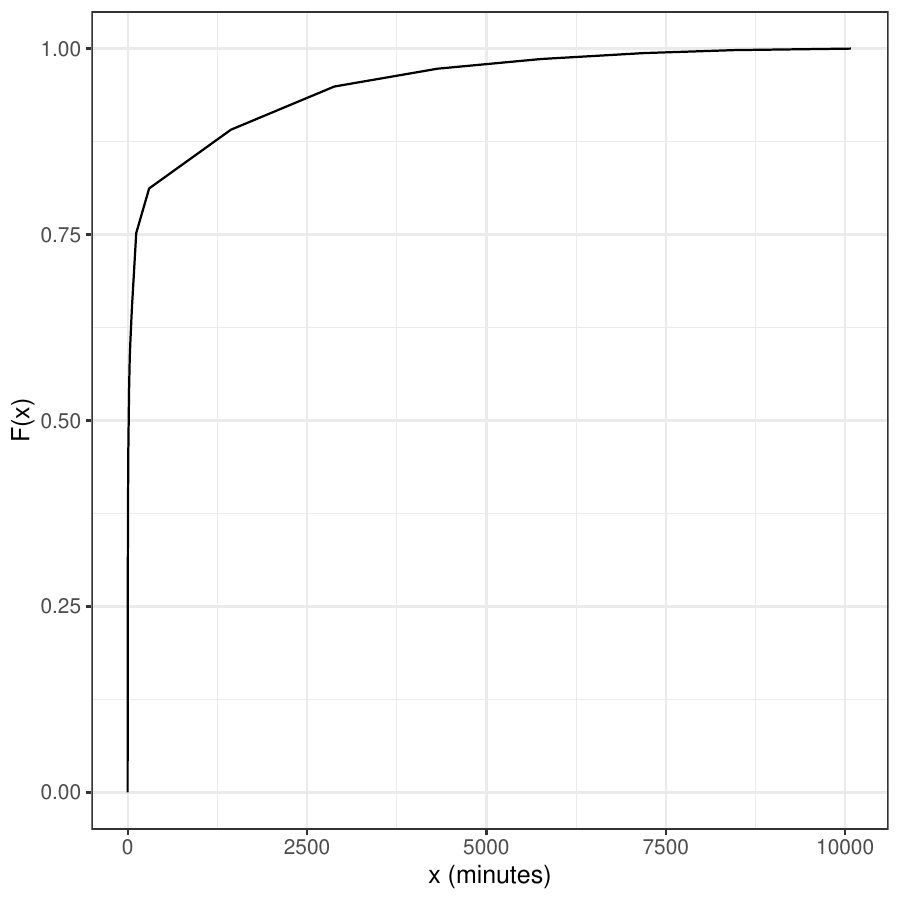}
  \caption{Cumulative probability density function of when a post, comment, or like is deleted after it is created (x = 0).}
  \label{fig:when-activity}
\end{figure}

Almuhimedi \etal also observe a temporal relationship for when a tweet is deleted -- a tweet has a higher chance of being deleted soon
after it was created.
They found 50\% of all deleted tweets where removed within 8~minutes of creation.
We have recreated the temporal distribution in~\cite{DBLP:conf/cscw/AlmuhimediWLSA13} and use it to generate deletion dates from
the valid lifespan intervals for posts, comments, and likes that are selected for explicit deletion~\autoref{fig:when-activity}.

\paragraph{Delete Forum and Delete Forum Membership}

We currently do not have empirical evidence to motivate realistic behaviour of \tForum deletion.
Forums have 3 types: walls, groups, and albums.
Groups and albums can be explicitly deleted, walls cannot.
The target proportion of groups and albums that are deleted across the simulation period is 1\%.

Additionally, we currently do not have empirical evidence to motivate realistic behaviour of \tHasMember edge deletion.
Only membership of groups can be explicitly deleted.
The target proportion of group memberships that are deleted across the simulation period is 5\%.

\section{Converting Delete Events into Delete Operations}
\label{sec:conv-delete-events}
Datagen supports 3 modes, each having different output:
\begin{itemize}
\item \textbf{Interactive}. Produces the data necessary for the Interactive workload. Includes a set of bulk load csv files and a number of update streams, which contain only insert operations.
\item \textbf{BI}. Produces the data necessary for the Business Intelligence workload. Includes a set of bulk load csv files and a number of update batches, which contain insert and delete operations.
\item \textbf{Raw}. Produces a fully dynamic graph without insert or delete operations. Includes a set of bulk load csv files (covering whole simulation period), with each dynamic entity having creation and deletion date attributes serialized. This mode is not intended for use with any LDBC workload.
\end{itemize}

When run in Interactive mode Datagen produces a graph that monotonically increases in size over the simulation period with insert-only operations, \eg once \tPerson joins the network they never leave, not delete a post nor unlike a picture.
This is mode is supported for backward compatibility with the Interactive workload.

The modes BI and raw use the dynamic graph containing creation events and deletion events.
Raw mode effectively serializes the graph to a bulk component and has a slightly different schema, with each entity having creation date and deletion date fields.
This mode was developed for testing, yet may be useful to users that require a dynamic graph data set for purposes other than benchmarking.

For the BI mode the generated data must be converted into a bulk load component and a series of update batches (containing insert and delete operations).
\autoref{fig:serialization-conds} displays the possible creation and deletion dates a dynamic entity can have with respect to the bulk load cut off, simulation end, and network collapse, which determines the target file the entity should be serialized to.
For example, if a \tPost is created after the bulk load and deleted before the simulation end this should result in a insert and a delete operation in the update batch data set.
If an entity is marked for explicit deletion then, if the conditions in \autoref{fig:serialization-conds} are satisfied then a deletion operation is serialized into the update batches.

\begin{figure}[ht]
  \centering
  \begin{subfigure}{\linewidth}
    \centering
    \begin{tikzpicture}[node distance=2cm,thick,every node/.style={transform shape}]
      \draw [thick,->,>=stealth] node [above,black] {} (-1,0) -- (6,0);
      \draw [thin] (-0.5,0.3) node [above,black] {$\xSS$} -- (-0.5,-0.3);
      \draw[mark=*,mark size=2pt,mark options={color=green}] plot coordinates {(0.5,0)};
      \draw[mark=*,mark size=2pt,mark options={color=red}] plot coordinates {(2.5,0)};
      \draw [thin]  (3.5,0.3) node [above,black] {$\xBL$} -- (3.5,-0.3);
      \draw [thin]  (4.5,0.3) node [above,black] {$\xSE$} -- (4.5,-0.3);
      \draw [thin]  (5.5,0.3) node [above,black] {$\xNC$} -- (5.5,-0.3);
    \end{tikzpicture}
    \caption{Dynamic entity has creation and deletion dates before the bulk load cut off. This entity is not serialized.}
    \label{fig:cond-1}
    \end{subfigure}
    \begin{subfigure}{\linewidth}
    \centering
    \begin{tikzpicture}[node distance=2cm,thick,every node/.style={transform shape}]
      \draw [thick,->,>=stealth] node [above,black] {} (-1,0) -- (6,0); 
      \draw [thin] (-0.5,0.3) node [above,black] {$\xSS$} -- (-0.5,-0.3);
      \draw[mark=*,mark size=2pt,mark options={color=green}] plot coordinates {(0.5,0)};
      \draw [thin]  (3.5,0.3) node [above,black] {$\xBL$} -- (3.5,-0.3);
      \draw[mark=*,mark size=2pt,mark options={color=red}] plot coordinates {(4,0)};
      \draw [thin]  (4.5,0.3) node [above,black] {$\xSE$} -- (4.5,-0.3);
      \draw [thin]  (5.5,0.3) node [above,black] {$\xNC$} -- (5.5,-0.3);
    \end{tikzpicture}
        \caption{Dynamic entity has creation date before the bulk load cut off and a deletion date after the bulk load cut off, but before the simulation end. Such an entity is serialized into the bulk load component and spawns a delete operation.}
    \label{fig:cond-2}
  \end{subfigure}
    \begin{subfigure}{\linewidth}
    \centering
    \begin{tikzpicture}[node distance=2cm,thick,every node/.style={transform shape}]
      \draw [thick,->,>=stealth] node [above,black] {} (-1,0) -- (6,0); 
      \draw [thin] (-0.5,0.3) node [above,black] {$\xSS$} -- (-0.5,-0.3);
      \draw[mark=*,mark size=2pt,mark options={color=green}] plot coordinates {(0.5,0)};
      \draw [thin]  (3.5,0.3) node [above,black] {$\xBL$} -- (3.5,-0.3);
      \draw [thin]  (4.5,0.3) node [above,black] {$\xSE$} -- (4.5,-0.3);
      \draw[mark=*,mark size=2pt,mark options={color=red}] plot coordinates {(5,0)};
      \draw [thin]  (5.5,0.3) node [above,black] {$\xNC$} -- (5.5,-0.3);
    \end{tikzpicture}
        \caption{Dynamic entity has creation date before the bulk load cut off and a deletion date after the simulation end. Such an entity is in serialized only into the bulk load component.}
    \label{fig:cond-3}
  \end{subfigure}
    \begin{subfigure}{\linewidth}
    \centering
    \begin{tikzpicture}[node distance=2cm,thick,every node/.style={transform shape}]
      \draw [thick,->,>=stealth] node [above,black] {} (-1,0) -- (6,0); 
      \draw [thin] (-0.5,0.3) node [above,black] {$\xSS$} -- (-0.5,-0.3);
      \draw [thin]  (3.5,0.3) node [above,black] {$\xBL$} -- (3.5,-0.3);
      \draw[mark=*,mark size=2pt,mark options={color=green}] plot coordinates {(3.75,0)};
      \draw[mark=*,mark size=2pt,mark options={color=red}] plot coordinates {(4.35,0)};
      \draw [thin]  (4.5,0.3) node [above,black] {$\xSE$} -- (4.5,-0.3);
      \draw [thin]  (5.5,0.3) node [above,black] {$\xNC$} -- (5.5,-0.3);
    \end{tikzpicture}
    \caption{Dynamic entity has creation date after the bulk load cut off and a deletion date before the simulation end. Such an entity produces an insert operation and a delete operation.}
    \label{fig:cond-4}
  \end{subfigure}
    \begin{subfigure}{\linewidth}
    \centering
    \begin{tikzpicture}[node distance=2cm,thick,every node/.style={transform shape}]
      \draw [thick,->,>=stealth] node [above,black] {} (-1,0) -- (6,0); 
      \draw [thin] (-0.5,0.3) node [above,black] {$\xSS$} -- (-0.5,-0.3);
      \draw [thin]  (3.5,0.3) node [above,black] {$\xBL$} -- (3.5,-0.3);
      \draw[mark=*,mark size=2pt,mark options={color=green}] plot coordinates {(4,0)};
      \draw [thin]  (4.5,0.3) node [above,black] {$\xSE$} -- (4.5,-0.3);
      \draw[mark=*,mark size=2pt,mark options={color=red}] plot coordinates {(5,0)};
      \draw [thin]  (5.5,0.3) node [above,black] {$\xNC$} -- (5.5,-0.3);
    \end{tikzpicture}
        \caption{Dynamic entity has creation date after the bulk load cut off, but before the simulation end, and a deletion date after the simulation end. Such an entity produces only an insert operation.}
    \label{fig:cond-5}
  \end{subfigure}
  \caption{Possible dynamic entity \emph{creation} \textcolor{green}{$\bullet$} and \emph{deletion} \textcolor{red}{$\bullet$} dates with respect to simulation start, bulk load cut off, simulation end, and network collapse.}
  \label{fig:serialization-conds}
\end{figure}

\chapter{Workloads}
\label{sec:workloads}


\section{Query Description Format}
\label{sub:queries_structure}

Queries are described in natural language using a well-defined structure that consists of three sections:
\textit{description}, a concise textual description of the query,
\textit{parameters}, a list of input parameters and their types;
\textit{results}, a list of expected results and their types.
Additionally, queries returning multiple results specify \emph{sorting criteria} and a \emph{limit} (to return top-$k$ results).
For strings, the sorting criteria should be interpreted as a binary comparison of the strings.%
\footnote{\texttt{C} or \texttt{POSIX} collation in PostgreSQL, see \url{https://www.postgresql.org/docs/13/locale.html}}%
\footnote{\texttt{BINARY} collation in DuckDB, see \url{https://duckdb.org/docs/sql/expressions/collations}}

We use the following notation:

\begin{itemize}
	\item \textbf{Node type}: node type in the dataset.
		One word, possibly constructed by appending multiple words together, starting with an uppercase character and following the camel case notation,
        \eg \textsf{TagClass} represents an entity of type ``TagClass''.
    \item \textbf{Edge type}: edge type in the dataset.
        One word, possibly constructed by appending multiple words together, starting with a lowercase character and following the camel case notation
        \eg \mbox{\textsf{workAt}} represents an edge of type ``workAt''.
    \item \textbf{Attribute}: attribute of a node or an edge in the dataset.
        One word, possibly constructed by appending multiple words together, starting with a lowercase character and following the camel case notation,
        and prefixed by a ``.'' to dereference the node/edge,
        \eg \textsf{person.firstName} refers to ``firstName'' attribute on the ``person'' entity,
        and \mbox{\textsf{studyAt.classYear}} refers to ``classYear'' attribute on the ``studyAt'' edge.
    \item \textbf{Unordered Set}: an unordered collection of distinct elements.
        Surrounded by \{ and \} braces, with the element type between them,
        \eg \textsf{\{String\}} refers to a set of strings.
    \item \textbf{Ordered List}: an ordered collection where duplicate elements are allowed.
        Surrounded by [ and ] braces, with the element type between them,
        \eg \textsf{[String]} refers to a list of strings.
    \item \textbf{Ordered Tuple}: a fixed-length, fixed-order list of elements, where elements at each position of the tuple have predefined, possibly different, types.
        Surrounded by < and > braces, with the element types between them in a specific order
        \eg \textsf{<String, Boolean>} refers to a 2-tuple containing a string value in the first element and a boolean value in the second,
        and \textsf{[<String, Boolean>]} is an ordered list of those 2-tuples.
\end{itemize}

\paragraph{Categorization of results.}

Results are categorized according to their source of origin:

\begin{itemize}
	\item \textbf{Raw} (\texttt{R}), if the result attribute is returned with an unmodified value and type.
	\item \textbf{Calculated} (\texttt{C}), if the result is calculated from attributes using arithmetic operators, functions, boolean conditions, etc.
	\item \textbf{Aggregated} (\texttt{A}), if the result is an aggregated value, \eg a count or a sum of another value. If a result is both calculated and aggregated (\eg \lstinline{count(x) + count(y)} or \lstinline{avg(x + y)}), it is considered an aggregated result.
	\item \textbf{Meta} (\texttt{M}), if the result is based on type information, \eg the type of a node.
\end{itemize}


\section{Conventions for Query Definitions}

\paragraph{Interval notations.}

Closed interval boundaries are denoted with 
\texttt{[} 
and \texttt{]}, while open interval boundaries are denoted with \texttt{(} and 
\texttt{)}. For example, \texttt{[0, 1)} denotes an interval between 0 and 1, 
closed on the left and open on the right.

\paragraph{Comparing Date and DateTime values.}

Some query specifications (\eg \queryRefCard{bi-read-01}{BI}{1}) require implementations to compare a
\textsf{DateTime} value with a \textsf{Date} value. In these cases, the 
\textsf{Date} value should be implicitly converted \textsf{DateTime} value 
with a time of 00:00:00.000+00:00 (\ie with the timezone of GMT).

\paragraph{Matching semantics.}

Unless noted otherwise, the specification uses \emph{homomorphic} matching 
semantics~\cite{DBLP:journals/csur/AnglesABHRV17}, \ie both nodes and edges can 
occur multiple times in a match. Note that for variable-length path, duplicate 
edges are not allowed.

\paragraph{Aggregation semantics.}

The \lstinline{count} aggregation always requires the query to determine the number of \emph{distinct} elements (nodes or edges). For example, this can be achieved in the Cypher, SPARQL and SQL query languages with the \lstinline[language=sql]{count(DISTINCT ...)} construct.

\paragraph{Graph patterns.}

To illustrate queries, we use graph patterns such as \autoref{fig:example-graph-pattern} with the following notation:

\begin{figure}[ht]
	\begin{center}
		\includegraphics[scale=\yedscale,margin=0cm .2cm]{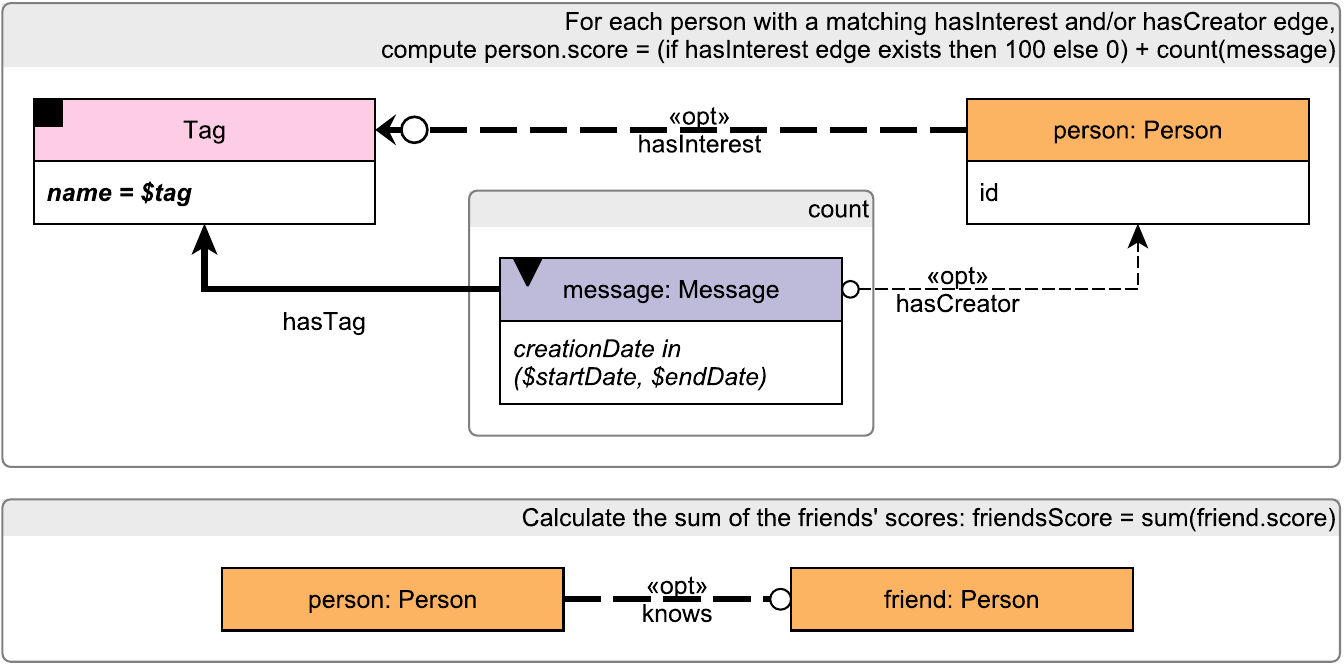}
		\caption{Example graph pattern.}
		\label{fig:example-graph-pattern}
	\end{center}
\end{figure}

\begin{itemize}
	\item Nodes in the pattern are shown with rectangular boxes with their type name stated at the top and emphasized with colour coding.
	\item A black square $\blacksquare$ in the node's top left corner and a \emph{\textbf{bold italic}} condition denote that the node is uniquely specified by the query parameters (\eg by using an identifier or a unique attribute such as \textsf{URL}).
	\item Attributes of nodes and edges can be subject to range constraints (\eg date within a given range, birthday larger than a given date, \etc). These are denoted with the \raisebox{0.3ex}{$\blacktriangledown$} symbol.
	\item Nodes in the pattern are captioned with \textsf{entityName: EntityType} (camel case 
	notation for both, starting with a lowercase character for the first and an 
	uppercase character for the second). If the \textsf{entityName} is neither returned in the query results (in raw, aggregated, or calculated form), nor referenced in the query specification, the \textsf{entityName} can be omitted.
	\item Edges in the graph pattern use the following notation:
	\begin{itemize}
		\item Regular edges, \ie edges that must be present in the subgraph, are denoted with \uline{solid black lines}.
		\item Negative edges, \ie edges that must not be present in the subgraph, are denoted with \textcolor{red}{\dashuline{dashed red lines}} and the \guillemotleft neg\guillemotright\ keyword.
		\item Optional edges, \ie edges that may or may not be in the subgraph, are denoted with \dashuline{dashed black lines}, the \guillemotleft opt\guillemotright\ keyword, and a circle symbol $\circ$ at the optional end of the edge.
		\item Edges without direction have no arrows. Their semantics is that there must be an edge in \emph{the least one of the (incoming, outgoing) directions}.
	\end{itemize}
	\begin{center}
		\tikzstyle{vertex} = [rectangle,minimum height=3mm,minimum width=6mm]
		\begin{tikzpicture}[node distance=15mm]
		\node[draw] (v1) [vertex] {};
		\node[draw] (v2) [vertex,right of=v1] {};
		\draw[solid,->,>=stealth]
		(v1) -- node [midway,above] {} (v2);
		\end{tikzpicture}
		\qquad
		\tikzstyle{vertex} = [rectangle,minimum height=3mm,minimum width=6mm]
		\begin{tikzpicture}[node distance=15mm]
		\node[draw] (v1) [vertex] {};
		\node[draw] (v2) [vertex,right of=v1] {};
		\draw[densely dashed,->,>=stealth,color=red]
		(v1) -- node [midway,above] {\scriptsize{\guillemotleft neg\guillemotright}} (v2);
		\end{tikzpicture}
		\qquad
		\tikzstyle{vertex} = [rectangle,minimum height=3mm,minimum width=6mm]
		\begin{tikzpicture}[node distance=15mm]
		\node[draw] (v1) [vertex] {};
		\node[draw] (v2) [vertex,right of=v1] {};
		\draw[densely dashed,o->,>=stealth]
		(v1) -- node [midway,above] {\scriptsize{\guillemotleft opt\guillemotright}} (v2);
		\end{tikzpicture}
		\qquad
		\tikzstyle{vertex} = [rectangle,minimum height=3mm,minimum width=6mm]
		\begin{tikzpicture}[node distance=15mm]
		\node[draw] (v1) [vertex] {};
		\node[draw] (v2) [vertex,right of=v1] {};
		\draw[-]
		(v1) -- node [midway,above] {} (v2);
		\end{tikzpicture}
	\end{center}
	\item Edges with many-to-many cardinalities are denoted with thicker lines, emphasizing that they may contribute more results in the result set.
	\item Filtering conditions are typeset in \textit{italic}, \eg $\textit{id = 
	\textdollar tag}$.
	\item Attributes that should be returned are denoted in sans-serif font, \eg \textsf{name}.
	\item Variable length paths, \ie edges that can be traversed multiple times 
	are denoted with \textsf{*min\ldots{}max}, \eg \textsf{replyOf*} or 
	\textsf{knows*1\ldots{}2}. By default, the value of \textsf{min} is 1, 
	and the value of \textsf{max} is unlimited.
	\item Aggregations are shown in boxes with a grey strip on their top describing the type of aggregation (\lstinline{count}, \lstinline{sum}, \lstinline{average}, etc.).
\end{itemize}

\paragraph{Keywords.} The pattern notation uses a small set of keywords:

\begin{itemize}
	\item Aggregation operations:
	\lstinline{avg},
	\lstinline{count}, 
	\lstinline{sum}.
	\item Functions:
	\begin{itemize}
		\item \lstinline{floor(x)}: returns $\lfloor x \rfloor$,
		\item \lstinline{year(date)}: extracts the year from a given date,
		\item \lstinline{month(date)}: extracts the month from a given date.
		\item \lstinline{day(date)}: extracts the day (of the month) from a given date.
	\end{itemize}
\end{itemize}

\paragraph{Deletions.}
Deletions of a single element are denoted with a red cross~\includegraphics[scale=0.25]{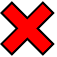},
while recursive deletions are denoted with a purple cross~\includegraphics[scale=0.25]{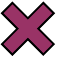}.

\paragraph{Resolving ambiguity.}

Note that if the textual description and the graph pattern are different for a particular query (either due to an error or the lack of sophistication in the graphical syntax), \emph{the textual description takes precedence}.


\section{Substitution Parameters}
\label{sec:substitution-parameters}

Together with the dataset, \datagen produces a set of parameters per
query type. Parameter generation is designed in such a way that for each query
type, all of the generated parameters yield similar runtime behaviour of that
query.

Specifically, the selection of parameters for a query template guarantees the following properties of the resulting queries:
\begin{enumerate}
\item[P1:] the query runtime has a bounded variance: the average runtime corresponds to the behavior of the majority of the queries
\item[P2:] the runtime distribution is stable: different samples of (\eg 10) parameter bindings used in different query streams result in an identical runtime distribution across streams
\item[P3:] the optimal logical plan (optimal operator order) of the queries is the same: this ensures that a specific query template tests the system's behavior under the well-chosen technical difficulty (\eg handling voluminous joins or proper cardinality estimation for subqueries, \etc)
\end{enumerate}

As a result, the amount of data that the query touches is roughly the
same for every parameter binding, assuming that the query optimizer figures out a
reasonable execution plan for the query. This is done to avoid bindings that
cause unexpectedly long or short runtimes of queries, or even result in a
completely different optimal execution plan. Such effects could arise due to
the data skew and correlations between values in the generated dataset.

In order to get the parameter bindings for each of the queries, we have designed a \textit{Parameter Curation} procedure that works in two stages:

\begin{enumerate}
\item for each query template for all possible parameter bindings, we determine the size of intermediate results in the {\em intended} query plan. Intermediate result size heavily influences the runtime of a query, so two queries with the same operator tree and similar intermediate result sizes at every level of this operator tree are expected to have similar runtimes. This analysis is effectively a side effect of data generation, that is we keep all the necessary counts (number of friends per user, number of posts of friends \etc) as we create the dataset.
\item then, a greedy algorithm selects (``curates'') those parameters with similar intermediate result counts from the domain of all the parameters.
\end{enumerate}

Parameter bindings are stored in the \texttt{substitution\_parameters} folder
inside the data generator directory. Each query gets its bindings in a separate
file. Every line of a parameter file is a JSON-formatted collection of
key-value pairs (name of the parameter and its value). For example, the Query 1
parameter bindings are stored in file \texttt{query\_1\_param.txt}, and one of
its lines may look like this:

\begin{lstlisting}
{"PersonID": 1, "Name": "Lei", "PersonURI": "http://www.ldbc.eu/ldbc_socialnet/1.0/data/pers1"}
\end{lstlisting}

Depending on implementation, the SUT may refer to persons either by IDs
(relational and graph databases) or URIs (RDF systems), so we provide both
values for the Person parameter.  Finally, parameters for short reads are taken
from those in complex reads and inserts.


\section{Return Values}
\label{sec:return-values}

Return values are subject to the following rules:

\begin{itemize}
	\item \textsf{DateTime} and \textsf{Date} values should use GMT timezone (or they should be converted by the client to GMT).
\end{itemize}

\chapter{Update operations}
\label{sec:update-operations}

This chapter contains the specifications of the Insert and Delete operations in the SNB suite. Inserts are used in the BI workload as well as the Interactive v1 and v2 workloads. Deletes are only used in the \interactivevtwo and BI workloads.

\section{Insert Operations}
\label{sec:insert-operations}

Each insert operations creates

\begin{enumerate}
    \item either a single node of a certain type, along with its edges to other existing nodes
    \item or a single edge of a certain type between two existing nodes.
\end{enumerate}
In \interactivevone, these operations were called ``updates''.
In \interactivevtwo, they are called ``inserts''.

\renewcommand*{\arraystretch}{1.1}

\subsection*{Updates / insert / 1}
\label{sec:insert-01}

\let\oldemph\emph
\renewcommand{\emph}[1]{{\footnotesize \sf #1}}
\let\oldtextbf\textbf
\renewcommand{\textbf}[1]{{\it #1}}

\renewcommand{\currentQueryCard}{insert-01}
\marginpar{
	\raggedleft
	\vspace{0.22ex}

	\queryRefCard{insert-01}{INS}{1}\\
	\queryRefCard{insert-02}{INS}{2}\\
	\queryRefCard{insert-03}{INS}{3}\\
	\queryRefCard{insert-04}{INS}{4}\\
	\queryRefCard{insert-05}{INS}{5}\\
	\queryRefCard{insert-06}{INS}{6}\\
	\queryRefCard{insert-07}{INS}{7}\\
	\queryRefCard{insert-08}{INS}{8}\\
}

\noindent\begin{tabularx}{\queryCardWidth}{|>{\queryPropertyCell}p{\queryPropertyCellWidth}|X|}
	\hline
	query & Updates / insert / 1 \\ \hline
	title & Add person \\ \hline
	pattern & \centering \includegraphics[scale=\patternscale,margin=0cm .2cm]{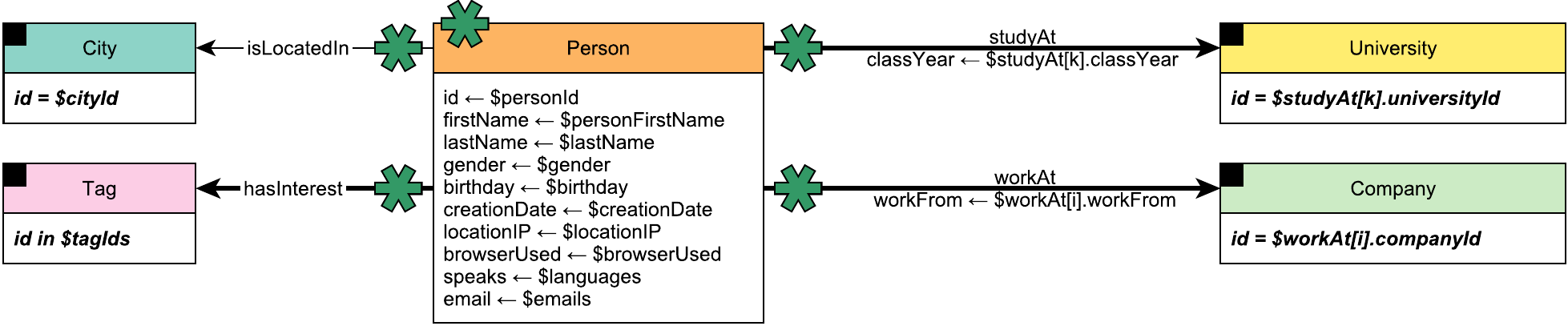} \tabularnewline \hline
	description & Add a \emph{Person} \textbf{node}, connected to the network by 4
possible \textbf{edge} types.
 \\ \hline

		params &
		\innerCardVSpace{\begin{tabularx}{\attributeCardWidth}{|>{\paramNumberCell}C{\attributeNumberWidth}|>{\varNameCell}M|>{\typeCell}m{\typeWidth}|Y|} \hline
		$\mathsf{1}$ & \$personId
 & ID
 &  \\ \hline
		$\mathsf{2}$ & \$personFirstName
 & String
 &  \\ \hline
		$\mathsf{3}$ & \$personLastName
 & String
 &  \\ \hline
		$\mathsf{4}$ & \$gender
 & String
 &  \\ \hline
		$\mathsf{5}$ & \$birthday
 & Date
 &  \\ \hline
		$\mathsf{6}$ & \$creationDate
 & DateTime
 &  \\ \hline
		$\mathsf{7}$ & \$locationIP
 & String
 &  \\ \hline
		$\mathsf{8}$ & \$browserUsed
 & String
 &  \\ \hline
		$\mathsf{9}$ & \$cityId
 & ID
 &  \\ \hline
		$\mathsf{10}$ & \$languages
 & \{String\}
 &  \\ \hline
		$\mathsf{11}$ & \$emails
 & \{Long String\}
 &  \\ \hline
		$\mathsf{12}$ & \$tagIds
 & \{ID\}
 &  \\ \hline
		$\mathsf{13}$ & \$studyAt
 & \{\textless ID, 32-bit Integer\textgreater\}
 & \texttt{\{\textless{}universityId,\ classYear\textgreater{}\}}
 \\ \hline
		$\mathsf{14}$ & \$workAt
 & \{\textless ID, 32-bit Integer\textgreater\}
 & \texttt{\{\textless{}companyId,\ workFrom\textgreater{}\}}
 \\ \hline
		\end{tabularx}}\innerCardVSpace \\ \hline

	CPs &
	\multicolumn{1}{>{\raggedright}l|}{
		\chokePoint{9.1}, 
		\chokePoint{9.2}
		} \\ \hline
\end{tabularx}
\queryCardVSpace

\let\emph\oldemph
\let\textbf\oldtextbf

\renewcommand{\currentQueryCard}{0}
\renewcommand*{\arraystretch}{1.1}

\subsection*{Updates / insert / 2}
\label{sec:insert-02}

\let\oldemph\emph
\renewcommand{\emph}[1]{{\footnotesize \sf #1}}
\let\oldtextbf\textbf
\renewcommand{\textbf}[1]{{\it #1}}

\renewcommand{\currentQueryCard}{insert-02}
\marginpar{
	\raggedleft
	\vspace{0.22ex}

	\queryRefCard{insert-01}{INS}{1}\\
	\queryRefCard{insert-02}{INS}{2}\\
	\queryRefCard{insert-03}{INS}{3}\\
	\queryRefCard{insert-04}{INS}{4}\\
	\queryRefCard{insert-05}{INS}{5}\\
	\queryRefCard{insert-06}{INS}{6}\\
	\queryRefCard{insert-07}{INS}{7}\\
	\queryRefCard{insert-08}{INS}{8}\\
}

\noindent\begin{tabularx}{\queryCardWidth}{|>{\queryPropertyCell}p{\queryPropertyCellWidth}|X|}
	\hline
	query & Updates / insert / 2 \\ \hline
	title & Add like to post \\ \hline
	pattern & \centering \includegraphics[scale=\patternscale,margin=0cm .2cm]{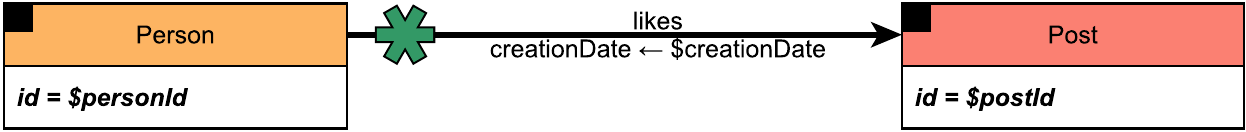} \tabularnewline \hline
	description & Add a \emph{likes} \textbf{edge} to a \emph{Post}.
 \\ \hline

		params &
		\innerCardVSpace{\begin{tabularx}{\attributeCardWidth}{|>{\paramNumberCell}C{\attributeNumberWidth}|>{\varNameCell}M|>{\typeCell}m{\typeWidth}|Y|} \hline
		$\mathsf{1}$ & \$personId
 & ID
 &  \\ \hline
		$\mathsf{2}$ & \$postId
 & ID
 &  \\ \hline
		$\mathsf{3}$ & \$creationDate
 & DateTime
 &  \\ \hline
		\end{tabularx}}\innerCardVSpace \\ \hline

	CPs &
	\multicolumn{1}{>{\raggedright}l|}{
		\chokePoint{9.2}
		} \\ \hline
\end{tabularx}
\queryCardVSpace

\let\emph\oldemph
\let\textbf\oldtextbf

\renewcommand{\currentQueryCard}{0}
\renewcommand*{\arraystretch}{1.1}

\subsection*{Updates / insert / 3}
\label{sec:insert-03}

\let\oldemph\emph
\renewcommand{\emph}[1]{{\footnotesize \sf #1}}
\let\oldtextbf\textbf
\renewcommand{\textbf}[1]{{\it #1}}

\renewcommand{\currentQueryCard}{insert-03}
\marginpar{
	\raggedleft
	\vspace{0.22ex}

	\queryRefCard{insert-01}{INS}{1}\\
	\queryRefCard{insert-02}{INS}{2}\\
	\queryRefCard{insert-03}{INS}{3}\\
	\queryRefCard{insert-04}{INS}{4}\\
	\queryRefCard{insert-05}{INS}{5}\\
	\queryRefCard{insert-06}{INS}{6}\\
	\queryRefCard{insert-07}{INS}{7}\\
	\queryRefCard{insert-08}{INS}{8}\\
}

\noindent\begin{tabularx}{\queryCardWidth}{|>{\queryPropertyCell}p{\queryPropertyCellWidth}|X|}
	\hline
	query & Updates / insert / 3 \\ \hline
	title & Add like to comment \\ \hline
	pattern & \centering \includegraphics[scale=\patternscale,margin=0cm .2cm]{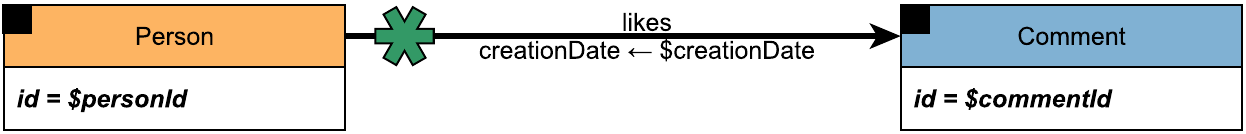} \tabularnewline \hline
	description & Add a \emph{likes} \textbf{edge} to a \emph{Comment}.
 \\ \hline

		params &
		\innerCardVSpace{\begin{tabularx}{\attributeCardWidth}{|>{\paramNumberCell}C{\attributeNumberWidth}|>{\varNameCell}M|>{\typeCell}m{\typeWidth}|Y|} \hline
		$\mathsf{1}$ & \$personId
 & ID
 &  \\ \hline
		$\mathsf{2}$ & \$commentId
 & ID
 &  \\ \hline
		$\mathsf{3}$ & \$creationDate
 & DateTime
 &  \\ \hline
		\end{tabularx}}\innerCardVSpace \\ \hline

	CPs &
	\multicolumn{1}{>{\raggedright}l|}{
		\chokePoint{9.2}
		} \\ \hline
\end{tabularx}
\queryCardVSpace

\let\emph\oldemph
\let\textbf\oldtextbf

\renewcommand{\currentQueryCard}{0}
\renewcommand*{\arraystretch}{1.1}

\subsection*{Updates / insert / 4}
\label{sec:insert-04}

\let\oldemph\emph
\renewcommand{\emph}[1]{{\footnotesize \sf #1}}
\let\oldtextbf\textbf
\renewcommand{\textbf}[1]{{\it #1}}

\renewcommand{\currentQueryCard}{insert-04}
\marginpar{
	\raggedleft
	\vspace{0.22ex}

	\queryRefCard{insert-01}{INS}{1}\\
	\queryRefCard{insert-02}{INS}{2}\\
	\queryRefCard{insert-03}{INS}{3}\\
	\queryRefCard{insert-04}{INS}{4}\\
	\queryRefCard{insert-05}{INS}{5}\\
	\queryRefCard{insert-06}{INS}{6}\\
	\queryRefCard{insert-07}{INS}{7}\\
	\queryRefCard{insert-08}{INS}{8}\\
}

\noindent\begin{tabularx}{\queryCardWidth}{|>{\queryPropertyCell}p{\queryPropertyCellWidth}|X|}
	\hline
	query & Updates / insert / 4 \\ \hline
	title & Add forum \\ \hline
	pattern & \centering \includegraphics[scale=\patternscale,margin=0cm .2cm]{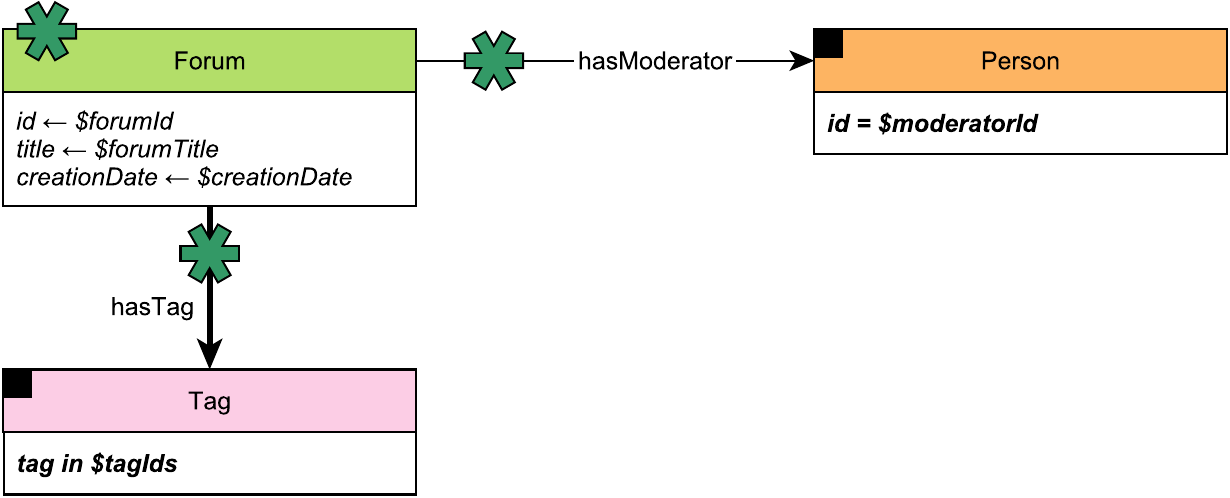} \tabularnewline \hline
	description & Add a \emph{Forum} \textbf{node}, connected to the network by 2 possible
\textbf{edge} types.
 \\ \hline

		params &
		\innerCardVSpace{\begin{tabularx}{\attributeCardWidth}{|>{\paramNumberCell}C{\attributeNumberWidth}|>{\varNameCell}M|>{\typeCell}m{\typeWidth}|Y|} \hline
		$\mathsf{1}$ & \$forumId
 & ID
 &  \\ \hline
		$\mathsf{2}$ & \$forumTitle
 & Long String
 &  \\ \hline
		$\mathsf{3}$ & \$creationDate
 & DateTime
 &  \\ \hline
		$\mathsf{4}$ & \$moderatorId
 & ID
 &  \\ \hline
		$\mathsf{5}$ & \$tagIds
 & \{ID\}
 &  \\ \hline
		\end{tabularx}}\innerCardVSpace \\ \hline

	CPs &
	\multicolumn{1}{>{\raggedright}l|}{
		\chokePoint{9.1}, 
		\chokePoint{9.2}
		} \\ \hline
\end{tabularx}
\queryCardVSpace

\let\emph\oldemph
\let\textbf\oldtextbf

\renewcommand{\currentQueryCard}{0}
\renewcommand*{\arraystretch}{1.1}

\subsection*{Updates / insert / 5}
\label{sec:insert-05}

\let\oldemph\emph
\renewcommand{\emph}[1]{{\footnotesize \sf #1}}
\let\oldtextbf\textbf
\renewcommand{\textbf}[1]{{\it #1}}

\renewcommand{\currentQueryCard}{insert-05}
\marginpar{
	\raggedleft
	\vspace{0.22ex}

	\queryRefCard{insert-01}{INS}{1}\\
	\queryRefCard{insert-02}{INS}{2}\\
	\queryRefCard{insert-03}{INS}{3}\\
	\queryRefCard{insert-04}{INS}{4}\\
	\queryRefCard{insert-05}{INS}{5}\\
	\queryRefCard{insert-06}{INS}{6}\\
	\queryRefCard{insert-07}{INS}{7}\\
	\queryRefCard{insert-08}{INS}{8}\\
}

\noindent\begin{tabularx}{\queryCardWidth}{|>{\queryPropertyCell}p{\queryPropertyCellWidth}|X|}
	\hline
	query & Updates / insert / 5 \\ \hline
	title & Add forum membership \\ \hline
	pattern & \centering \includegraphics[scale=\patternscale,margin=0cm .2cm]{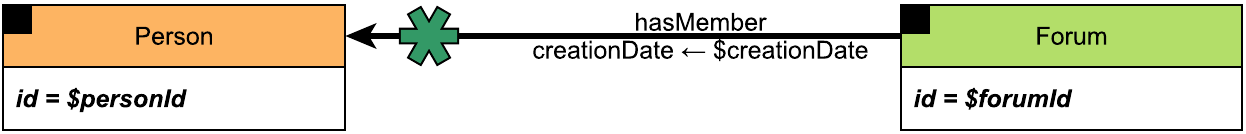} \tabularnewline \hline
	description & Add a \emph{Forum} membership \textbf{edge} (\emph{hasMember}) to a
\emph{Person}.
 \\ \hline

		params &
		\innerCardVSpace{\begin{tabularx}{\attributeCardWidth}{|>{\paramNumberCell}C{\attributeNumberWidth}|>{\varNameCell}M|>{\typeCell}m{\typeWidth}|Y|} \hline
		$\mathsf{1}$ & \$personId
 & ID
 &  \\ \hline
		$\mathsf{2}$ & \$forumId
 & ID
 &  \\ \hline
		$\mathsf{3}$ & \$creationDate
 & DateTime
 &  \\ \hline
		\end{tabularx}}\innerCardVSpace \\ \hline

	CPs &
	\multicolumn{1}{>{\raggedright}l|}{
		\chokePoint{9.1}, 
		\chokePoint{9.2}
		} \\ \hline
\end{tabularx}
\queryCardVSpace

\let\emph\oldemph
\let\textbf\oldtextbf

\renewcommand{\currentQueryCard}{0}
\renewcommand*{\arraystretch}{1.1}

\subsection*{Updates / insert / 6}
\label{sec:insert-06}

\let\oldemph\emph
\renewcommand{\emph}[1]{{\footnotesize \sf #1}}
\let\oldtextbf\textbf
\renewcommand{\textbf}[1]{{\it #1}}

\renewcommand{\currentQueryCard}{insert-06}
\marginpar{
	\raggedleft
	\vspace{0.22ex}

	\queryRefCard{insert-01}{INS}{1}\\
	\queryRefCard{insert-02}{INS}{2}\\
	\queryRefCard{insert-03}{INS}{3}\\
	\queryRefCard{insert-04}{INS}{4}\\
	\queryRefCard{insert-05}{INS}{5}\\
	\queryRefCard{insert-06}{INS}{6}\\
	\queryRefCard{insert-07}{INS}{7}\\
	\queryRefCard{insert-08}{INS}{8}\\
}

\noindent\begin{tabularx}{\queryCardWidth}{|>{\queryPropertyCell}p{\queryPropertyCellWidth}|X|}
	\hline
	query & Updates / insert / 6 \\ \hline
	title & Add post \\ \hline
	pattern & \centering \includegraphics[scale=\patternscale,margin=0cm .2cm]{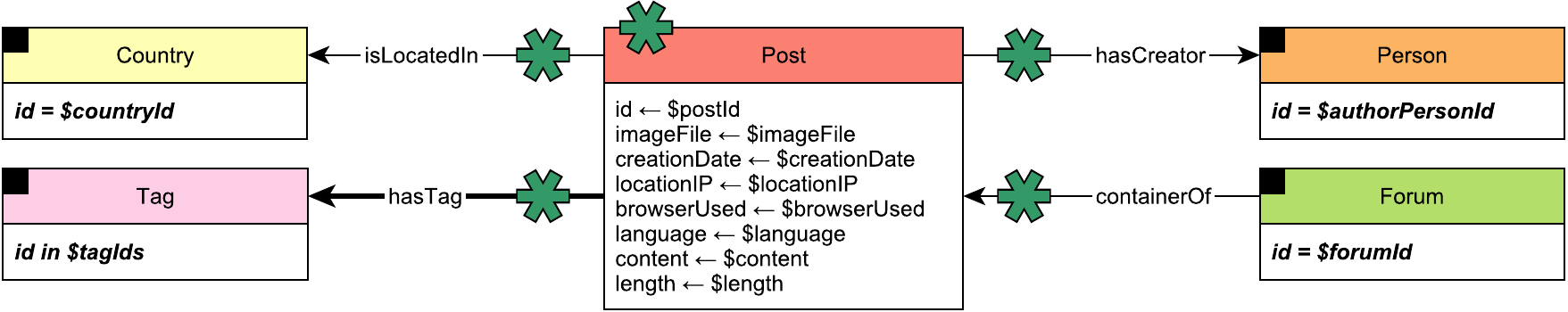} \tabularnewline \hline
	description & Add a \emph{Post} \textbf{node} connected to the network by 4 possible
\textbf{edge} types (\emph{hasCreator}, \emph{containerOf},
\emph{isLocatedIn}, \emph{hasTag}).
 \\ \hline

		params &
		\innerCardVSpace{\begin{tabularx}{\attributeCardWidth}{|>{\paramNumberCell}C{\attributeNumberWidth}|>{\varNameCell}M|>{\typeCell}m{\typeWidth}|Y|} \hline
		$\mathsf{1}$ & \$postId
 & ID
 &  \\ \hline
		$\mathsf{2}$ & \$imageFile
 & String
 &  \\ \hline
		$\mathsf{3}$ & \$creationDate
 & DateTime
 &  \\ \hline
		$\mathsf{4}$ & \$locationIP
 & String
 &  \\ \hline
		$\mathsf{5}$ & \$browserUsed
 & String
 &  \\ \hline
		$\mathsf{6}$ & \$language
 & String
 &  \\ \hline
		$\mathsf{7}$ & \$content
 & Text
 &  \\ \hline
		$\mathsf{8}$ & \$length
 & 32-bit Integer
 &  \\ \hline
		$\mathsf{9}$ & \$authorPersonId
 & ID
 &  \\ \hline
		$\mathsf{10}$ & \$forumId
 & ID
 &  \\ \hline
		$\mathsf{11}$ & \$countryId
 & ID
 &  \\ \hline
		$\mathsf{12}$ & \$tagIds
 & \{ID\}
 &  \\ \hline
		\end{tabularx}}\innerCardVSpace \\ \hline

	CPs &
	\multicolumn{1}{>{\raggedright}l|}{
		\chokePoint{9.1}, 
		\chokePoint{9.2}
		} \\ \hline
\end{tabularx}
\queryCardVSpace

\let\emph\oldemph
\let\textbf\oldtextbf

\renewcommand{\currentQueryCard}{0}
\renewcommand*{\arraystretch}{1.1}

\subsection*{Updates / insert / 7}
\label{sec:insert-07}

\let\oldemph\emph
\renewcommand{\emph}[1]{{\footnotesize \sf #1}}
\let\oldtextbf\textbf
\renewcommand{\textbf}[1]{{\it #1}}

\renewcommand{\currentQueryCard}{insert-07}
\marginpar{
	\raggedleft
	\vspace{0.22ex}

	\queryRefCard{insert-01}{INS}{1}\\
	\queryRefCard{insert-02}{INS}{2}\\
	\queryRefCard{insert-03}{INS}{3}\\
	\queryRefCard{insert-04}{INS}{4}\\
	\queryRefCard{insert-05}{INS}{5}\\
	\queryRefCard{insert-06}{INS}{6}\\
	\queryRefCard{insert-07}{INS}{7}\\
	\queryRefCard{insert-08}{INS}{8}\\
}

\noindent\begin{tabularx}{\queryCardWidth}{|>{\queryPropertyCell}p{\queryPropertyCellWidth}|X|}
	\hline
	query & Updates / insert / 7 \\ \hline
	title & Add comment \\ \hline
	pattern & \centering \includegraphics[scale=\patternscale,margin=0cm .2cm]{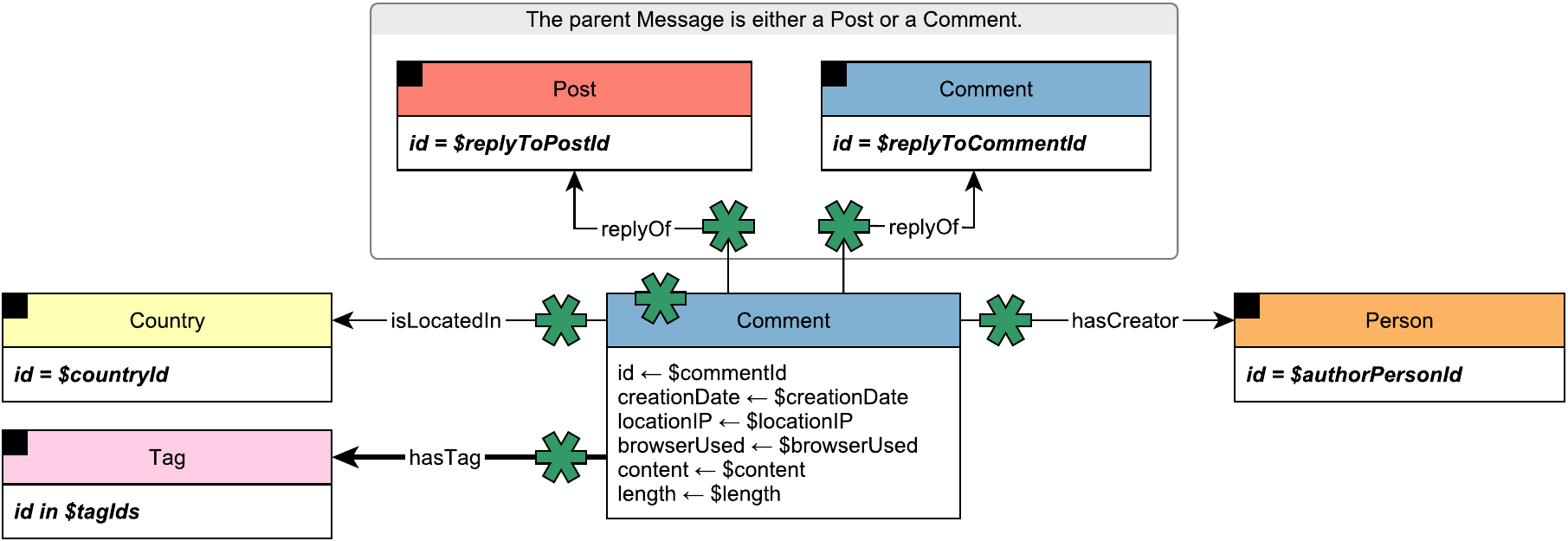} \tabularnewline \hline
	description & Add a \emph{Comment} \textbf{node} replying to a
\emph{Post}/\emph{Comment}, connected to the network by 4 possible
\textbf{edge} types (\emph{replyOf}, \emph{hasCreator},
\emph{isLocatedIn}, \emph{hasTag}).
 \\ \hline

		params &
		\innerCardVSpace{\begin{tabularx}{\attributeCardWidth}{|>{\paramNumberCell}C{\attributeNumberWidth}|>{\varNameCell}M|>{\typeCell}m{\typeWidth}|Y|} \hline
		$\mathsf{1}$ & \$commentId
 & ID
 &  \\ \hline
		$\mathsf{2}$ & \$creationDate
 & DateTime
 &  \\ \hline
		$\mathsf{3}$ & \$locationIP
 & String
 &  \\ \hline
		$\mathsf{4}$ & \$browserUsed
 & String
 &  \\ \hline
		$\mathsf{5}$ & \$content
 & Text
 &  \\ \hline
		$\mathsf{6}$ & \$length
 & 32-bit Integer
 &  \\ \hline
		$\mathsf{7}$ & \$authorPersonId
 & ID
 &  \\ \hline
		$\mathsf{8}$ & \$countryId
 & ID
 &  \\ \hline
		$\mathsf{9}$ & \$replyToPostId
 & ID
 & \textbf{old version:} \(-1\) if the \emph{Comment} is a reply of a
\emph{Comment}; \textbf{new version:} null if the \emph{Comment} is a
reply of a \emph{Post}
 \\ \hline
		$\mathsf{10}$ & \$replyToCommentId
 & ID
 & \textbf{old version:} \(-1\) if the \emph{Comment} is a reply of a
\emph{Post}; \textbf{new version:} null if the \emph{Comment} is a reply
of a \emph{Post}
 \\ \hline
		$\mathsf{11}$ & \$tagIds
 & \{ID\}
 &  \\ \hline
		\end{tabularx}}\innerCardVSpace \\ \hline

	CPs &
	\multicolumn{1}{>{\raggedright}l|}{
		\chokePoint{9.1}, 
		\chokePoint{9.2}
		} \\ \hline
\end{tabularx}
\queryCardVSpace

\let\emph\oldemph
\let\textbf\oldtextbf

\renewcommand{\currentQueryCard}{0}
\renewcommand*{\arraystretch}{1.1}

\subsection*{Updates / insert / 8}
\label{sec:insert-08}

\let\oldemph\emph
\renewcommand{\emph}[1]{{\footnotesize \sf #1}}
\let\oldtextbf\textbf
\renewcommand{\textbf}[1]{{\it #1}}

\renewcommand{\currentQueryCard}{insert-08}
\marginpar{
	\raggedleft
	\vspace{0.22ex}

	\queryRefCard{insert-01}{INS}{1}\\
	\queryRefCard{insert-02}{INS}{2}\\
	\queryRefCard{insert-03}{INS}{3}\\
	\queryRefCard{insert-04}{INS}{4}\\
	\queryRefCard{insert-05}{INS}{5}\\
	\queryRefCard{insert-06}{INS}{6}\\
	\queryRefCard{insert-07}{INS}{7}\\
	\queryRefCard{insert-08}{INS}{8}\\
}

\noindent\begin{tabularx}{\queryCardWidth}{|>{\queryPropertyCell}p{\queryPropertyCellWidth}|X|}
	\hline
	query & Updates / insert / 8 \\ \hline
	title & Add friendship \\ \hline
	pattern & \centering \includegraphics[scale=\patternscale,margin=0cm .2cm]{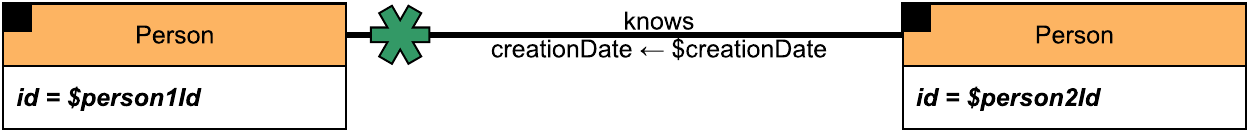} \tabularnewline \hline
	description & Add a friendship \textbf{edge} (\emph{knows}) between two
\emph{Persons}.
 \\ \hline

		params &
		\innerCardVSpace{\begin{tabularx}{\attributeCardWidth}{|>{\paramNumberCell}C{\attributeNumberWidth}|>{\varNameCell}M|>{\typeCell}m{\typeWidth}|Y|} \hline
		$\mathsf{1}$ & \$person1Id
 & ID
 &  \\ \hline
		$\mathsf{2}$ & \$person2Id
 & ID
 &  \\ \hline
		$\mathsf{3}$ & \$creationDate
 & DateTime
 &  \\ \hline
		\end{tabularx}}\innerCardVSpace \\ \hline

	CPs &
	\multicolumn{1}{>{\raggedright}l|}{
		\chokePoint{9.2}
		} \\ \hline
\end{tabularx}
\queryCardVSpace

\let\emph\oldemph
\let\textbf\oldtextbf

\renewcommand{\currentQueryCard}{0}

\section{Delete Operations}
\label{sec:delete-operations}

Each delete operation removes

\begin{enumerate}
    \item a single edge between two existing nodes
    \item or a node, all its edges and, in certain cases, nodes and edges that are transitively reachable on a certain path (thus performing a cascading delete).
\end{enumerate}

\renewcommand*{\arraystretch}{1.1}

\subsection*{Updates / delete / 1}
\label{sec:delete-01}

\let\oldemph\emph
\renewcommand{\emph}[1]{{\footnotesize \sf #1}}
\let\oldtextbf\textbf
\renewcommand{\textbf}[1]{{\it #1}}

\renewcommand{\currentQueryCard}{delete-01}
\marginpar{
	\raggedleft
	\vspace{0.22ex}

	\queryRefCard{delete-01}{DEL}{1}\\
	\queryRefCard{delete-02}{DEL}{2}\\
	\queryRefCard{delete-03}{DEL}{3}\\
	\queryRefCard{delete-04}{DEL}{4}\\
	\queryRefCard{delete-05}{DEL}{5}\\
	\queryRefCard{delete-06}{DEL}{6}\\
	\queryRefCard{delete-07}{DEL}{7}\\
	\queryRefCard{delete-08}{DEL}{8}\\
}

\noindent\begin{tabularx}{\queryCardWidth}{|>{\queryPropertyCell}p{\queryPropertyCellWidth}|X|}
	\hline
	query & Updates / delete / 1 \\ \hline
	title & Remove person and its personal forums and message (sub)threads \\ \hline
	pattern & \centering \includegraphics[scale=\patternscale,margin=0cm .2cm]{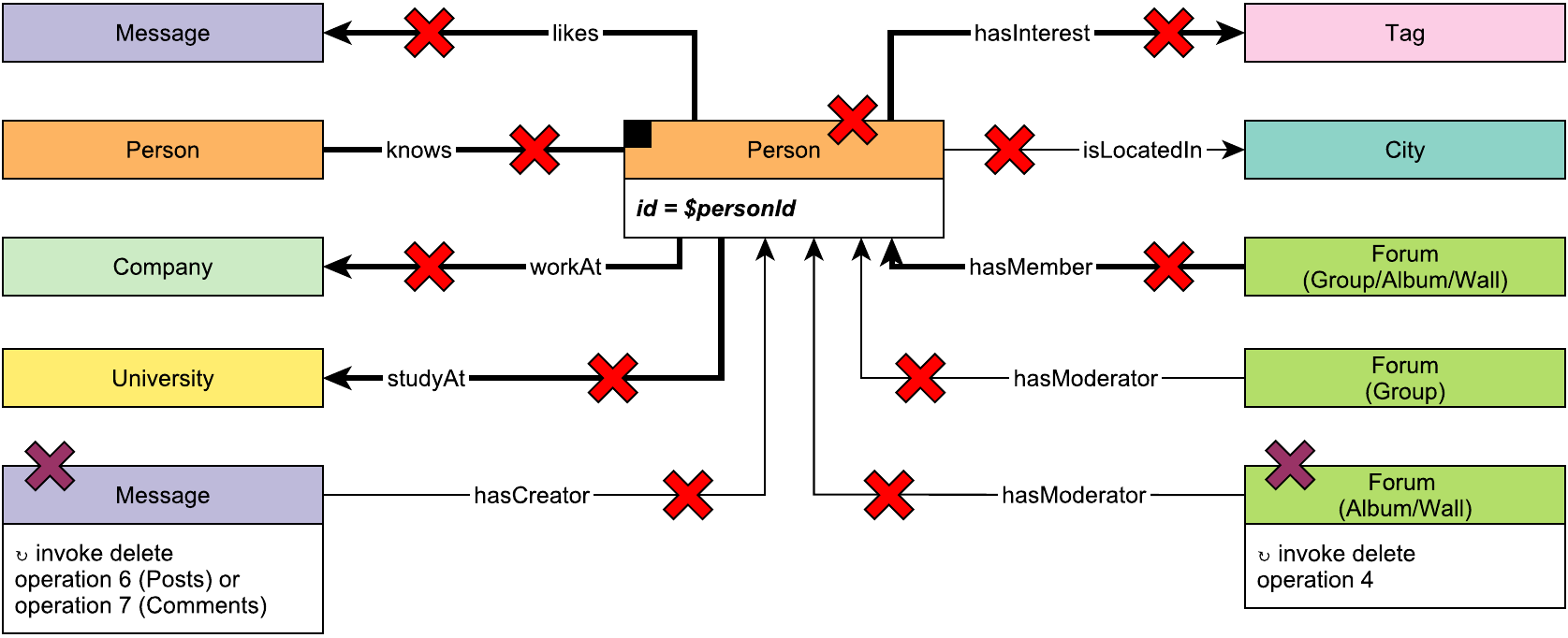} \tabularnewline \hline
	description & Remove a \emph{Person} with ID \texttt{\$personId} and its edges
(\emph{isLocatedIn}, \emph{studyAt}, \emph{workAt}, \emph{hasInterest},
\emph{likes}, \emph{knows}, \emph{hasMember}, \emph{hasModerator},
\emph{hasCreator}). Additionally, remove the Album and Wall
\emph{Forums} whose moderator is the \emph{Person} and remove all
\emph{Messages} the \emph{Person} has created in the rest of the
\emph{Forums} (Groups).
 \\ \hline

		params &
		\innerCardVSpace{\begin{tabularx}{\attributeCardWidth}{|>{\paramNumberCell}C{\attributeNumberWidth}|>{\varNameCell}M|>{\typeCell}m{\typeWidth}|Y|} \hline
		$\mathsf{1}$ & \$personId
 & ID
 &  \\ \hline
		\end{tabularx}}\innerCardVSpace \\ \hline

	CPs &
	\multicolumn{1}{>{\raggedright}l|}{
		\chokePoint{9.3}, 
		\chokePoint{9.4}, 
		\chokePoint{9.5}
		} \\ \hline
	relevance &
		\footnotesize \begin{itemize}
\tightlist
\item
  Removal of a \emph{Person} removes \emph{Forums} of type ``Walls'' and
  ``Albums'' but not ``Groups'', which can continue if even the founder
  has left the network. For Groups, the \emph{hasModerator} edge is
  deleted. We have discussed various approaches to appoint a new
  moderator, e.g.

  \begin{enumerate}
  \def\labelenumi{\arabic{enumi}.}
  \tightlist
  \item
    choose member at random from the set of existing group members or
  \item
    the member with the oldest group join date becomes the moderator.
    However, to keep the generator and the workload simple, currently no
    moderator is selected, leaving the group without a moderator.
  \end{enumerate}
\item
  Removal of a \emph{Person} removes all \emph{Posts}/\emph{Comments}
  they are creator of this could result in the removal of a
  \emph{Comment} in the middle of a thread.
\end{itemize}
 \\ \hline%
\end{tabularx}
\queryCardVSpace

\let\emph\oldemph
\let\textbf\oldtextbf

\renewcommand{\currentQueryCard}{0}
\renewcommand*{\arraystretch}{1.1}

\subsection*{Updates / delete / 2}
\label{sec:delete-02}

\let\oldemph\emph
\renewcommand{\emph}[1]{{\footnotesize \sf #1}}
\let\oldtextbf\textbf
\renewcommand{\textbf}[1]{{\it #1}}

\renewcommand{\currentQueryCard}{delete-02}
\marginpar{
	\raggedleft
	\vspace{0.22ex}

	\queryRefCard{delete-01}{DEL}{1}\\
	\queryRefCard{delete-02}{DEL}{2}\\
	\queryRefCard{delete-03}{DEL}{3}\\
	\queryRefCard{delete-04}{DEL}{4}\\
	\queryRefCard{delete-05}{DEL}{5}\\
	\queryRefCard{delete-06}{DEL}{6}\\
	\queryRefCard{delete-07}{DEL}{7}\\
	\queryRefCard{delete-08}{DEL}{8}\\
}

\noindent\begin{tabularx}{\queryCardWidth}{|>{\queryPropertyCell}p{\queryPropertyCellWidth}|X|}
	\hline
	query & Updates / delete / 2 \\ \hline
	title & Remove post like \\ \hline
	pattern & \centering \includegraphics[scale=\patternscale,margin=0cm .2cm]{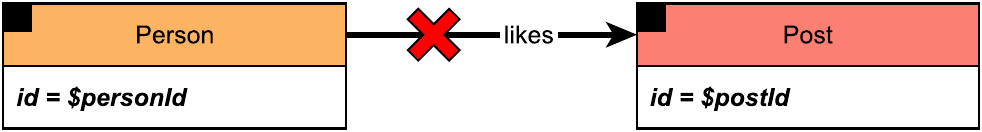} \tabularnewline \hline
	description & Given a \emph{Person} with ID \texttt{\$personId} and a \emph{Post} with
ID \texttt{\$postId}, remove the \emph{likes} edge between them.
 \\ \hline

		params &
		\innerCardVSpace{\begin{tabularx}{\attributeCardWidth}{|>{\paramNumberCell}C{\attributeNumberWidth}|>{\varNameCell}M|>{\typeCell}m{\typeWidth}|Y|} \hline
		$\mathsf{1}$ & \$personId
 & ID
 &  \\ \hline
		$\mathsf{2}$ & \$postId
 & ID
 &  \\ \hline
		\end{tabularx}}\innerCardVSpace \\ \hline

	CPs &
	\multicolumn{1}{>{\raggedright}l|}{
		\chokePoint{9.4}
		} \\ \hline
	relevance &
		\footnotesize Removal of a \emph{likes} edge is a rare event, e.g.~people accidently
liking a \emph{Post}, this can be reflected by the relative frequency of
the operation.
 \\ \hline%
\end{tabularx}
\queryCardVSpace

\let\emph\oldemph
\let\textbf\oldtextbf

\renewcommand{\currentQueryCard}{0}
\renewcommand*{\arraystretch}{1.1}

\subsection*{Updates / delete / 3}
\label{sec:delete-03}

\let\oldemph\emph
\renewcommand{\emph}[1]{{\footnotesize \sf #1}}
\let\oldtextbf\textbf
\renewcommand{\textbf}[1]{{\it #1}}

\renewcommand{\currentQueryCard}{delete-03}
\marginpar{
	\raggedleft
	\vspace{0.22ex}

	\queryRefCard{delete-01}{DEL}{1}\\
	\queryRefCard{delete-02}{DEL}{2}\\
	\queryRefCard{delete-03}{DEL}{3}\\
	\queryRefCard{delete-04}{DEL}{4}\\
	\queryRefCard{delete-05}{DEL}{5}\\
	\queryRefCard{delete-06}{DEL}{6}\\
	\queryRefCard{delete-07}{DEL}{7}\\
	\queryRefCard{delete-08}{DEL}{8}\\
}

\noindent\begin{tabularx}{\queryCardWidth}{|>{\queryPropertyCell}p{\queryPropertyCellWidth}|X|}
	\hline
	query & Updates / delete / 3 \\ \hline
	title & Remove comment like \\ \hline
	pattern & \centering \includegraphics[scale=\patternscale,margin=0cm .2cm]{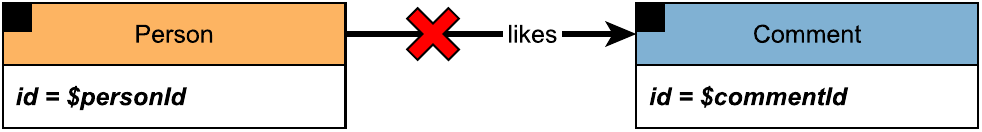} \tabularnewline \hline
	description & Given a \emph{Person} with ID \texttt{\$personId} and a \emph{Comment}
with ID \texttt{\$commentId}, remove the \emph{likes} edge between them.
 \\ \hline

		params &
		\innerCardVSpace{\begin{tabularx}{\attributeCardWidth}{|>{\paramNumberCell}C{\attributeNumberWidth}|>{\varNameCell}M|>{\typeCell}m{\typeWidth}|Y|} \hline
		$\mathsf{1}$ & \$personId
 & ID
 &  \\ \hline
		$\mathsf{2}$ & \$commentId
 & ID
 &  \\ \hline
		\end{tabularx}}\innerCardVSpace \\ \hline

	CPs &
	\multicolumn{1}{>{\raggedright}l|}{
		\chokePoint{9.4}
		} \\ \hline
	relevance &
		\footnotesize Removal of a \emph{likes} edge is a rare event, e.g.~people accidently
liking a \emph{Comment}, this can be reflected by the relative frequency
of the operation.
 \\ \hline%
\end{tabularx}
\queryCardVSpace

\let\emph\oldemph
\let\textbf\oldtextbf

\renewcommand{\currentQueryCard}{0}
\renewcommand*{\arraystretch}{1.1}

\subsection*{Updates / delete / 4}
\label{sec:delete-04}

\let\oldemph\emph
\renewcommand{\emph}[1]{{\footnotesize \sf #1}}
\let\oldtextbf\textbf
\renewcommand{\textbf}[1]{{\it #1}}

\renewcommand{\currentQueryCard}{delete-04}
\marginpar{
	\raggedleft
	\vspace{0.22ex}

	\queryRefCard{delete-01}{DEL}{1}\\
	\queryRefCard{delete-02}{DEL}{2}\\
	\queryRefCard{delete-03}{DEL}{3}\\
	\queryRefCard{delete-04}{DEL}{4}\\
	\queryRefCard{delete-05}{DEL}{5}\\
	\queryRefCard{delete-06}{DEL}{6}\\
	\queryRefCard{delete-07}{DEL}{7}\\
	\queryRefCard{delete-08}{DEL}{8}\\
}

\noindent\begin{tabularx}{\queryCardWidth}{|>{\queryPropertyCell}p{\queryPropertyCellWidth}|X|}
	\hline
	query & Updates / delete / 4 \\ \hline
	title & Remove forum and its content \\ \hline
	pattern & \centering \includegraphics[scale=\patternscale,margin=0cm .2cm]{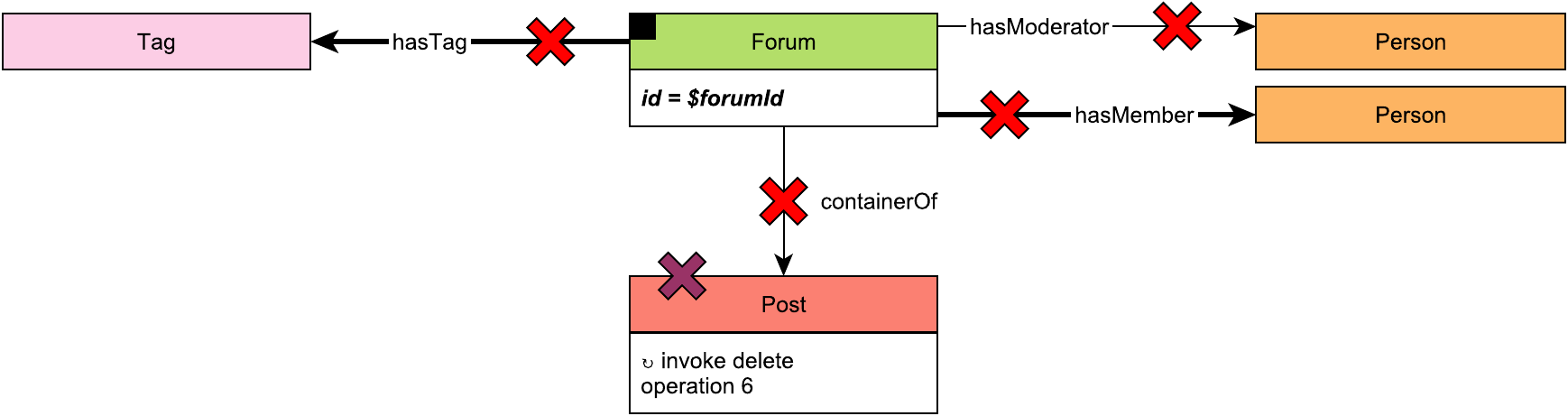} \tabularnewline \hline
	description & Remove a \emph{Forum} with ID \texttt{\$forumId} and its edges
(\emph{hasModerator}, \emph{hasMember}, \emph{hasTag}) and all
\emph{Posts} in the \emph{Forum} (connected by \emph{containerOf} edges)
and their direct and transitive \emph{Comments}.
 \\ \hline

		params &
		\innerCardVSpace{\begin{tabularx}{\attributeCardWidth}{|>{\paramNumberCell}C{\attributeNumberWidth}|>{\varNameCell}M|>{\typeCell}m{\typeWidth}|Y|} \hline
		$\mathsf{1}$ & \$forumId
 & ID
 &  \\ \hline
		\end{tabularx}}\innerCardVSpace \\ \hline

	CPs &
	\multicolumn{1}{>{\raggedright}l|}{
		\chokePoint{9.3}, 
		\chokePoint{9.4}, 
		\chokePoint{9.5}
		} \\ \hline
	relevance &
		\footnotesize n/a
 \\ \hline%
\end{tabularx}
\queryCardVSpace

\let\emph\oldemph
\let\textbf\oldtextbf

\renewcommand{\currentQueryCard}{0}
\renewcommand*{\arraystretch}{1.1}

\subsection*{Updates / delete / 5}
\label{sec:delete-05}

\let\oldemph\emph
\renewcommand{\emph}[1]{{\footnotesize \sf #1}}
\let\oldtextbf\textbf
\renewcommand{\textbf}[1]{{\it #1}}

\renewcommand{\currentQueryCard}{delete-05}
\marginpar{
	\raggedleft
	\vspace{0.22ex}

	\queryRefCard{delete-01}{DEL}{1}\\
	\queryRefCard{delete-02}{DEL}{2}\\
	\queryRefCard{delete-03}{DEL}{3}\\
	\queryRefCard{delete-04}{DEL}{4}\\
	\queryRefCard{delete-05}{DEL}{5}\\
	\queryRefCard{delete-06}{DEL}{6}\\
	\queryRefCard{delete-07}{DEL}{7}\\
	\queryRefCard{delete-08}{DEL}{8}\\
}

\noindent\begin{tabularx}{\queryCardWidth}{|>{\queryPropertyCell}p{\queryPropertyCellWidth}|X|}
	\hline
	query & Updates / delete / 5 \\ \hline
	title & Remove forum membership \\ \hline
	pattern & \centering \includegraphics[scale=\patternscale,margin=0cm .2cm]{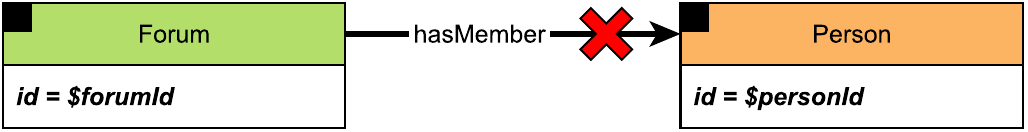} \tabularnewline \hline
	description & Given a \emph{Forum} with ID \texttt{\$forumId} and a \emph{Person} with
ID \texttt{\$personId}, remove the \emph{hasMember} edge between them.
 \\ \hline

		params &
		\innerCardVSpace{\begin{tabularx}{\attributeCardWidth}{|>{\paramNumberCell}C{\attributeNumberWidth}|>{\varNameCell}M|>{\typeCell}m{\typeWidth}|Y|} \hline
		$\mathsf{1}$ & \$forumId
 & ID
 &  \\ \hline
		$\mathsf{2}$ & \$personId
 & ID
 &  \\ \hline
		\end{tabularx}}\innerCardVSpace \\ \hline

	CPs &
	\multicolumn{1}{>{\raggedright}l|}{
		\chokePoint{9.4}
		} \\ \hline
	relevance &
		\footnotesize n/a
 \\ \hline%
\end{tabularx}
\queryCardVSpace

\let\emph\oldemph
\let\textbf\oldtextbf

\renewcommand{\currentQueryCard}{0}
\renewcommand*{\arraystretch}{1.1}

\subsection*{Updates / delete / 6}
\label{sec:delete-06}

\let\oldemph\emph
\renewcommand{\emph}[1]{{\footnotesize \sf #1}}
\let\oldtextbf\textbf
\renewcommand{\textbf}[1]{{\it #1}}

\renewcommand{\currentQueryCard}{delete-06}
\marginpar{
	\raggedleft
	\vspace{0.22ex}

	\queryRefCard{delete-01}{DEL}{1}\\
	\queryRefCard{delete-02}{DEL}{2}\\
	\queryRefCard{delete-03}{DEL}{3}\\
	\queryRefCard{delete-04}{DEL}{4}\\
	\queryRefCard{delete-05}{DEL}{5}\\
	\queryRefCard{delete-06}{DEL}{6}\\
	\queryRefCard{delete-07}{DEL}{7}\\
	\queryRefCard{delete-08}{DEL}{8}\\
}

\noindent\begin{tabularx}{\queryCardWidth}{|>{\queryPropertyCell}p{\queryPropertyCellWidth}|X|}
	\hline
	query & Updates / delete / 6 \\ \hline
	title & Remove post thread \\ \hline
	pattern & \centering \includegraphics[scale=\patternscale,margin=0cm .2cm]{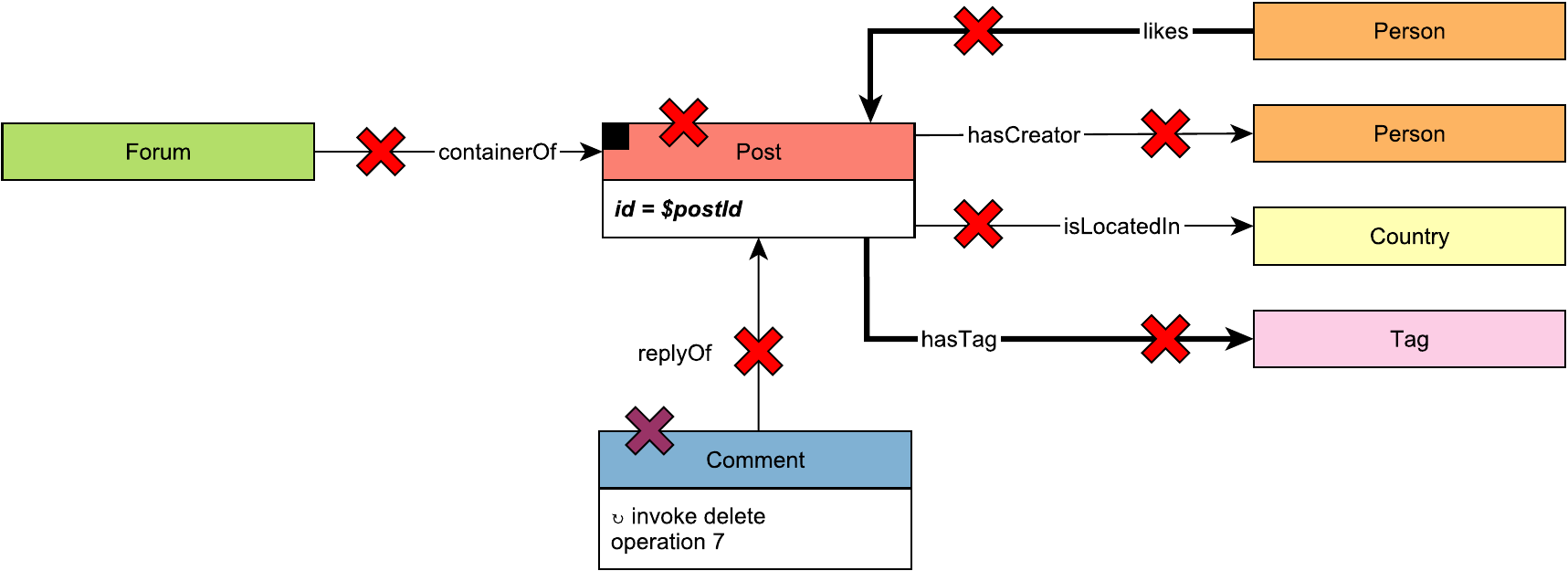} \tabularnewline \hline
	description & Remove a \emph{Post} node with ID \texttt{\$postId} and its edges
(\emph{isLocatedIn}, \emph{likes}, \emph{hasCreator}, \emph{hasTag},
\emph{containerOf}). Remove all replies to the \emph{Post} and the
connecting \emph{replyOf} edges. In addition, remove all transitive
reply \emph{Comments} to the \emph{Post} and their edges.
 \\ \hline

		params &
		\innerCardVSpace{\begin{tabularx}{\attributeCardWidth}{|>{\paramNumberCell}C{\attributeNumberWidth}|>{\varNameCell}M|>{\typeCell}m{\typeWidth}|Y|} \hline
		$\mathsf{1}$ & \$postId
 & ID
 &  \\ \hline
		\end{tabularx}}\innerCardVSpace \\ \hline

	CPs &
	\multicolumn{1}{>{\raggedright}l|}{
		\chokePoint{9.3}, 
		\chokePoint{9.4}, 
		\chokePoint{9.5}
		} \\ \hline
	relevance &
		\footnotesize n/a
 \\ \hline%
\end{tabularx}
\queryCardVSpace

\let\emph\oldemph
\let\textbf\oldtextbf

\renewcommand{\currentQueryCard}{0}
\renewcommand*{\arraystretch}{1.1}

\subsection*{Updates / delete / 7}
\label{sec:delete-07}

\let\oldemph\emph
\renewcommand{\emph}[1]{{\footnotesize \sf #1}}
\let\oldtextbf\textbf
\renewcommand{\textbf}[1]{{\it #1}}

\renewcommand{\currentQueryCard}{delete-07}
\marginpar{
	\raggedleft
	\vspace{0.22ex}

	\queryRefCard{delete-01}{DEL}{1}\\
	\queryRefCard{delete-02}{DEL}{2}\\
	\queryRefCard{delete-03}{DEL}{3}\\
	\queryRefCard{delete-04}{DEL}{4}\\
	\queryRefCard{delete-05}{DEL}{5}\\
	\queryRefCard{delete-06}{DEL}{6}\\
	\queryRefCard{delete-07}{DEL}{7}\\
	\queryRefCard{delete-08}{DEL}{8}\\
}

\noindent\begin{tabularx}{\queryCardWidth}{|>{\queryPropertyCell}p{\queryPropertyCellWidth}|X|}
	\hline
	query & Updates / delete / 7 \\ \hline
	title & Remove comment subthread \\ \hline
	pattern & \centering \includegraphics[scale=\patternscale,margin=0cm .2cm]{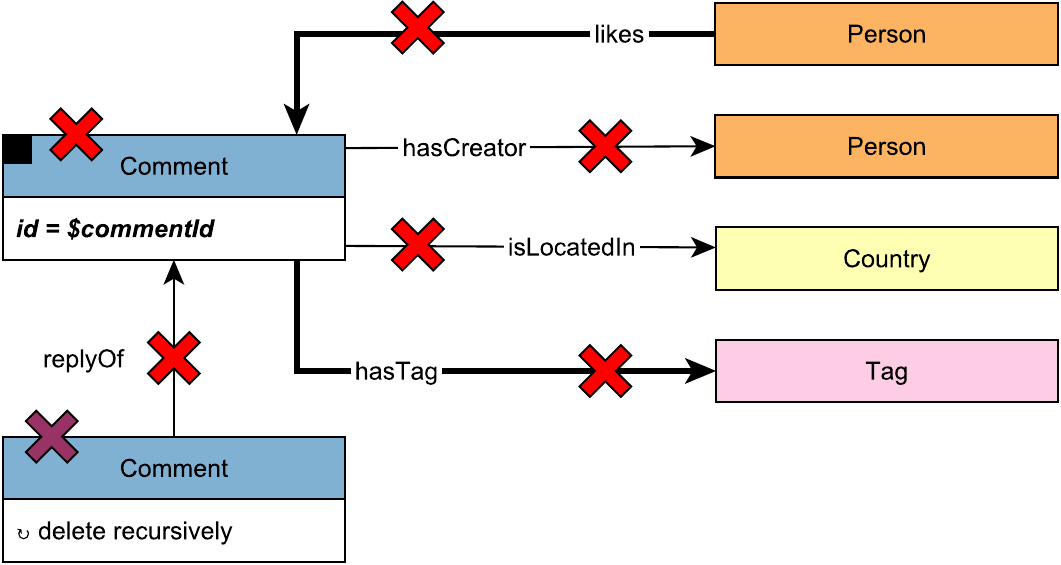} \tabularnewline \hline
	description & Remove a \emph{Comment} node with ID \texttt{\$commentId} and its
\textbf{edges} (\emph{isLocatedIn}, \emph{likes}, \emph{hasCreator},
\emph{hasTag}). In addition, remove all replies to the \emph{Comment}
connected by \emph{replyOf} and their \textbf{edges}.
 \\ \hline

		params &
		\innerCardVSpace{\begin{tabularx}{\attributeCardWidth}{|>{\paramNumberCell}C{\attributeNumberWidth}|>{\varNameCell}M|>{\typeCell}m{\typeWidth}|Y|} \hline
		$\mathsf{1}$ & \$commentId
 & ID
 &  \\ \hline
		\end{tabularx}}\innerCardVSpace \\ \hline

	CPs &
	\multicolumn{1}{>{\raggedright}l|}{
		\chokePoint{9.3}, 
		\chokePoint{9.4}, 
		\chokePoint{9.5}
		} \\ \hline
	relevance &
		\footnotesize n/a
 \\ \hline%
\end{tabularx}
\queryCardVSpace

\let\emph\oldemph
\let\textbf\oldtextbf

\renewcommand{\currentQueryCard}{0}
\renewcommand*{\arraystretch}{1.1}

\subsection*{Updates / delete / 8}
\label{sec:delete-08}

\let\oldemph\emph
\renewcommand{\emph}[1]{{\footnotesize \sf #1}}
\let\oldtextbf\textbf
\renewcommand{\textbf}[1]{{\it #1}}

\renewcommand{\currentQueryCard}{delete-08}
\marginpar{
	\raggedleft
	\vspace{0.22ex}

	\queryRefCard{delete-01}{DEL}{1}\\
	\queryRefCard{delete-02}{DEL}{2}\\
	\queryRefCard{delete-03}{DEL}{3}\\
	\queryRefCard{delete-04}{DEL}{4}\\
	\queryRefCard{delete-05}{DEL}{5}\\
	\queryRefCard{delete-06}{DEL}{6}\\
	\queryRefCard{delete-07}{DEL}{7}\\
	\queryRefCard{delete-08}{DEL}{8}\\
}

\noindent\begin{tabularx}{\queryCardWidth}{|>{\queryPropertyCell}p{\queryPropertyCellWidth}|X|}
	\hline
	query & Updates / delete / 8 \\ \hline
	title & Remove friendship \\ \hline
	pattern & \centering \includegraphics[scale=\patternscale,margin=0cm .2cm]{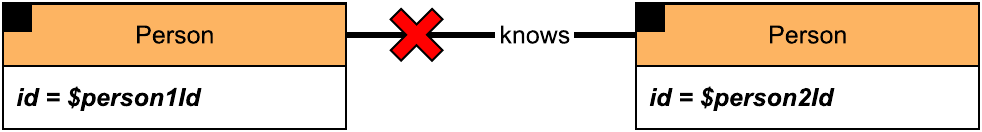} \tabularnewline \hline
	description & Given two \emph{Person} nodes with IDs \texttt{\$person1Id} and
\texttt{\$person2Id}, remove the \emph{knows} edge between them.
 \\ \hline

		params &
		\innerCardVSpace{\begin{tabularx}{\attributeCardWidth}{|>{\paramNumberCell}C{\attributeNumberWidth}|>{\varNameCell}M|>{\typeCell}m{\typeWidth}|Y|} \hline
		$\mathsf{1}$ & \$person1Id
 & ID
 &  \\ \hline
		$\mathsf{2}$ & \$person2Id
 & ID
 &  \\ \hline
		\end{tabularx}}\innerCardVSpace \\ \hline

	CPs &
	\multicolumn{1}{>{\raggedright}l|}{
		\chokePoint{9.4}
		} \\ \hline
	relevance &
		\footnotesize n/a
 \\ \hline%
\end{tabularx}
\queryCardVSpace

\let\emph\oldemph
\let\textbf\oldtextbf

\renewcommand{\currentQueryCard}{0}

\chapter{Interactive v1 Workload}
\label{sec:interactive-v1}

The Interactive v1 workload consists of a set of relatively complex read-only queries, that touch a significant
amount of data -- often the two-step friendship neighbourhood and associated messages --, but typically in close proximity to a single node. Hence, the query complexity is sublinear to the dataset size.

The LDBC SNB Interactive workload consists of three query classes:

\begin{itemize}
\item \textbf{Complex read-only queries.} See \autoref{sec:interactive-v1-complex-reads}.
\item \textbf{Short read-only queries.} See \autoref{sec:interactive-v1-short-reads}.
\item \textbf{Insert operations.} See \autoref{sec:insert-operations}.
\end{itemize}

\subsection*{Related Publications}

A detailed description of the workload (covering reads and inserts) is available in the paper published at \mbox{SIGMOD} 2015~\cite{DBLP:conf/sigmod/ErlingALCGPPB15}. The ACID Test Suite was first published at TPCTC 2020~\cite{DBLP:conf/tpctc/WaudbySKMBS20}.
\iftoggle{StandaloneWorkloadSpecification}{}{
    It is part of this specification in \autoref{sec:acid-test-suite}.
}

\subsection*{Related Software Components}

\begin{itemize}
    \item Datagen (Hadoop-based): \url{https://github.com/ldbc/ldbc_snb_datagen_hadoop}
    \item Driver: \url{https://github.com/ldbc/ldbc_snb_interactive_v1_driver}
    \item Reference implementations: \url{https://github.com/ldbc/ldbc_snb_interactive_v1_impls}
\end{itemize}


\section{Complex Reads}
\label{sec:interactive-v1-complex-reads}

\renewcommand*{\arraystretch}{1.1}

\subsection*{Interactive / complex / 1}
\label{sec:interactive-complex-read-01}

\let\oldemph\emph
\renewcommand{\emph}[1]{{\footnotesize \sf #1}}
\let\oldtextbf\textbf
\renewcommand{\textbf}[1]{{\it #1}}\renewcommand{\currentQueryCard}{interactive-complex-read-01}
\marginpar{
	\raggedleft
	\vspace{0.22ex}

	\queryRefCard{interactive-complex-read-01}{IC}{1}\\
	\queryRefCard{interactive-complex-read-02}{IC}{2}\\
	\queryRefCard{interactive-complex-read-03}{IC}{3}\\
	\queryRefCard{interactive-complex-read-04}{IC}{4}\\
	\queryRefCard{interactive-complex-read-05}{IC}{5}\\
	\queryRefCard{interactive-complex-read-06}{IC}{6}\\
	\queryRefCard{interactive-complex-read-07}{IC}{7}\\
	\queryRefCard{interactive-complex-read-08}{IC}{8}\\
	\queryRefCard{interactive-complex-read-09}{IC}{9}\\
	\queryRefCard{interactive-complex-read-10}{IC}{10}\\
	\queryRefCard{interactive-complex-read-11}{IC}{11}\\
	\queryRefCard{interactive-complex-read-12}{IC}{12}\\
	\queryRefCard{interactive-complex-read-13}{IC}{13}\\
	\queryRefCard{interactive-complex-read-14-v1}{IC}{14v1}\\
	\queryRefCard{interactive-complex-read-14-v2}{IC}{14v2}\\
}

\noindent\begin{tabularx}{\queryCardWidth}{|>{\queryPropertyCell}p{\queryPropertyCellWidth}|X|}
	\hline
	query & Interactive / complex / 1 \\ \hline
	title & Transitive friends with a certain name \\ \hline
	pattern & \centering \includegraphics[scale=\patternscale,margin=0cm .2cm]{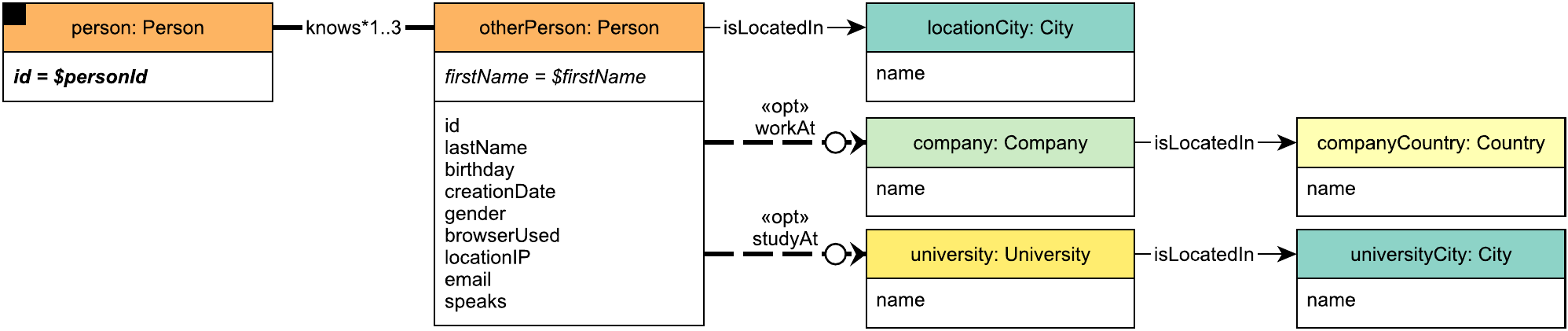} \tabularnewline \hline
	description & Given a start \emph{Person} with ID \texttt{\$personId}, find
\emph{Persons} with a given first name (\texttt{\$firstName}) that the
start \emph{Person} is connected to (excluding start \emph{Person}) by
at most 3 steps via the \emph{knows} relationships. Return
\emph{Persons}, including the distance (1..3), summaries of the
\emph{Persons} workplaces and places of study.
 \\ \hline

		params &
		\innerCardVSpace{\begin{tabularx}{\attributeCardWidth}{|>{\paramNumberCell}C{\attributeNumberWidth}|>{\varNameCell}M|>{\typeCell}m{\typeWidth}|Y|} \hline
		$\mathsf{1}$ & \$personId
 & ID
 &  \\ \hline
		$\mathsf{2}$ & \$firstName
 & String
 &  \\ \hline
		\end{tabularx}}\innerCardVSpace \\ \hline

		result &
		\innerCardVSpace{\begin{tabularx}{\attributeCardWidth}{|>{\resultNumberCell}C{\attributeNumberWidth}|>{\varNameCell}M|>{\typeCell}m{\typeWidth}|>{\resultOriginCell}c|Y|} \hline
		$\mathsf{1}$ & otherPerson.id & ID & R &
				 \\ \hline
		$\mathsf{2}$ & otherPerson.lastName & String & R &
				 \\ \hline
		$\mathsf{3}$ & distanceFromPerson & 32-bit Integer & C &
				 \\ \hline
		$\mathsf{4}$ & otherPerson.birthday & Date & R &
				 \\ \hline
		$\mathsf{5}$ & otherPerson.creationDate & DateTime & R &
				 \\ \hline
		$\mathsf{6}$ & otherPerson.gender & String & R &
				 \\ \hline
		$\mathsf{7}$ & otherPerson.browserUsed & String & R &
				 \\ \hline
		$\mathsf{8}$ & otherPerson.locationIP & String & R &
				 \\ \hline
		$\mathsf{9}$ & otherPerson.email & \{Long String\} & R &
				 \\ \hline
		$\mathsf{10}$ & otherPerson.speaks & \{String\} & R &
				 \\ \hline
		$\mathsf{11}$ & locationCity.name & String & R &
				 \\ \hline
		$\mathsf{12}$ & universities & \{\textless String, 32-bit Integer, String\textgreater\} & A &
				\texttt{\{\textless{}university.name,\ studyAt.classYear,\ universityCity.name\textgreater{}\}}
 \\ \hline
		$\mathsf{13}$ & companies & \{\textless String, 32-bit Integer, String\textgreater\} & A &
				\texttt{\{\textless{}company.name,\ workAt.workFrom,\ companyCountry.name\textgreater{}\}}
 \\ \hline
		\end{tabularx}}\innerCardVSpace \\ \hline

		sort		&
		\innerCardVSpace{\begin{tabularx}{\attributeCardWidth}{|>{\sortNumberCell}C{\attributeNumberWidth}|>{\varNameCell}M|>{\directionCell}c|Y|} \hline
		$\mathsf{1}$ & distanceFromPerson
 & $\asc
$ &  \\ \hline
		$\mathsf{2}$ & otherPerson.lastName
 & $\asc
$ &  \\ \hline
		$\mathsf{3}$ & otherPerson.id
 & $\asc
$ &  \\ \hline
		\end{tabularx}}\innerCardVSpace \\ \hline
	limit & 20 \\ \hline
	CPs &
	\multicolumn{1}{>{\raggedright}l|}{
		\chokePoint{2.1}, 
		\chokePoint{5.3}, 
		\chokePoint{8.2}
		} \\ \hline
	relevance &
		\footnotesize This query is a representative of a simple navigational query. It is
interesting for several aspects. (1) It requires for a complex
aggregation for returning the concatenation of universities, companies,
languages and email information of the \emph{Person}. (2) It tests the
ability of the optimizer to move the evaluation of sub-queries
functionally dependant on the \emph{Person}, after the evaluation of the
top-k. (3) Its performance is highly sensitive to properly estimating
the cardinalities in each transitive path, and paying attention not to
explore already visited \emph{Persons}.
 \\ \hline%
\end{tabularx}
\queryCardVSpace

\let\emph\oldemph
\let\textbf\oldtextbf

\renewcommand{\currentQueryCard}{0}
\renewcommand*{\arraystretch}{1.1}

\subsection*{Interactive / complex / 2}
\label{sec:interactive-complex-read-02}

\let\oldemph\emph
\renewcommand{\emph}[1]{{\footnotesize \sf #1}}
\let\oldtextbf\textbf
\renewcommand{\textbf}[1]{{\it #1}}\renewcommand{\currentQueryCard}{interactive-complex-read-02}
\marginpar{
	\raggedleft
	\vspace{0.22ex}

	\queryRefCard{interactive-complex-read-01}{IC}{1}\\
	\queryRefCard{interactive-complex-read-02}{IC}{2}\\
	\queryRefCard{interactive-complex-read-03}{IC}{3}\\
	\queryRefCard{interactive-complex-read-04}{IC}{4}\\
	\queryRefCard{interactive-complex-read-05}{IC}{5}\\
	\queryRefCard{interactive-complex-read-06}{IC}{6}\\
	\queryRefCard{interactive-complex-read-07}{IC}{7}\\
	\queryRefCard{interactive-complex-read-08}{IC}{8}\\
	\queryRefCard{interactive-complex-read-09}{IC}{9}\\
	\queryRefCard{interactive-complex-read-10}{IC}{10}\\
	\queryRefCard{interactive-complex-read-11}{IC}{11}\\
	\queryRefCard{interactive-complex-read-12}{IC}{12}\\
	\queryRefCard{interactive-complex-read-13}{IC}{13}\\
	\queryRefCard{interactive-complex-read-14-v1}{IC}{14v1}\\
	\queryRefCard{interactive-complex-read-14-v2}{IC}{14v2}\\
}

\noindent\begin{tabularx}{\queryCardWidth}{|>{\queryPropertyCell}p{\queryPropertyCellWidth}|X|}
	\hline
	query & Interactive / complex / 2 \\ \hline
	title & Recent messages by your friends \\ \hline
	pattern & \centering \includegraphics[scale=\patternscale,margin=0cm .2cm]{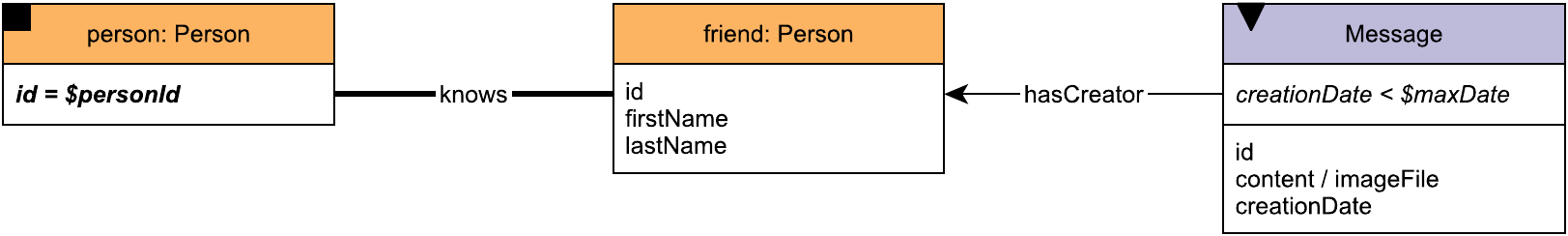} \tabularnewline \hline
	description & Given a start \emph{Person} with ID \texttt{\$personId}, find the most
recent \emph{Messages} from all of that \emph{Person}'s friends
(\texttt{friend} nodes). Only consider \emph{Messages} created before
the given \texttt{\$maxDate} (excluding that day).
 \\ \hline

		params &
		\innerCardVSpace{\begin{tabularx}{\attributeCardWidth}{|>{\paramNumberCell}C{\attributeNumberWidth}|>{\varNameCell}M|>{\typeCell}m{\typeWidth}|Y|} \hline
		$\mathsf{1}$ & \$personId
 & ID
 &  \\ \hline
		$\mathsf{2}$ & \$maxDate
 & Date
 &  \\ \hline
		\end{tabularx}}\innerCardVSpace \\ \hline

		result &
		\innerCardVSpace{\begin{tabularx}{\attributeCardWidth}{|>{\resultNumberCell}C{\attributeNumberWidth}|>{\varNameCell}M|>{\typeCell}m{\typeWidth}|>{\resultOriginCell}c|Y|} \hline
		$\mathsf{1}$ & friend.id & ID & R &
				 \\ \hline
		$\mathsf{2}$ & friend.firstName & String & R &
				 \\ \hline
		$\mathsf{3}$ & friend.lastName & String & R &
				 \\ \hline
		$\mathsf{4}$ & message.id & ID & R &
				 \\ \hline
		$\mathsf{5}$ & message.content or message.imageFile (for photos) & Text & R &
				 \\ \hline
		$\mathsf{6}$ & message.creationDate & DateTime & R &
				 \\ \hline
		\end{tabularx}}\innerCardVSpace \\ \hline

		sort		&
		\innerCardVSpace{\begin{tabularx}{\attributeCardWidth}{|>{\sortNumberCell}C{\attributeNumberWidth}|>{\varNameCell}M|>{\directionCell}c|Y|} \hline
		$\mathsf{1}$ & message.creationDate
 & $\desc
$ &  \\ \hline
		$\mathsf{2}$ & message.id
 & $\asc
$ &  \\ \hline
		\end{tabularx}}\innerCardVSpace \\ \hline
	limit & 20 \\ \hline
	CPs &
	\multicolumn{1}{>{\raggedright}l|}{
		\chokePoint{1.1}, 
		\chokePoint{2.2}, 
		\chokePoint{2.3}, 
		\chokePoint{3.2}, 
		\chokePoint{8.5}
		} \\ \hline
	relevance &
		\footnotesize This is a navigational query looking for paths of length two, starting
from a given \emph{Person}, going to their friends and from them, moving
to their published \emph{Posts} and \emph{Comments}. This query
exercices both the optimizer and how data is stored. It tests the
ability to create execution plans taking advantage of the orderings
induced by some operators to avoid performing expensive sorts. This
query requires selecting \emph{Posts} and \emph{Comments} based on their
creation date, which might be correlated with their identifier and
therefore, having intermediate results with interesting orders. Also,
messages could be stored in an order correlated with their creation date
to improve data access locality. Finally, as many of the attributes
required in the projection are not needed for the execution of the
query, it is expected that the query optimizer will move the projection
to the end.
 \\ \hline%
\end{tabularx}
\queryCardVSpace

\let\emph\oldemph
\let\textbf\oldtextbf

\renewcommand{\currentQueryCard}{0}
\renewcommand*{\arraystretch}{1.1}

\subsection*{Interactive / complex / 3}
\label{sec:interactive-complex-read-03}

\let\oldemph\emph
\renewcommand{\emph}[1]{{\footnotesize \sf #1}}
\let\oldtextbf\textbf
\renewcommand{\textbf}[1]{{\it #1}}\renewcommand{\currentQueryCard}{interactive-complex-read-03}
\marginpar{
	\raggedleft
	\vspace{0.22ex}

	\queryRefCard{interactive-complex-read-01}{IC}{1}\\
	\queryRefCard{interactive-complex-read-02}{IC}{2}\\
	\queryRefCard{interactive-complex-read-03}{IC}{3}\\
	\queryRefCard{interactive-complex-read-04}{IC}{4}\\
	\queryRefCard{interactive-complex-read-05}{IC}{5}\\
	\queryRefCard{interactive-complex-read-06}{IC}{6}\\
	\queryRefCard{interactive-complex-read-07}{IC}{7}\\
	\queryRefCard{interactive-complex-read-08}{IC}{8}\\
	\queryRefCard{interactive-complex-read-09}{IC}{9}\\
	\queryRefCard{interactive-complex-read-10}{IC}{10}\\
	\queryRefCard{interactive-complex-read-11}{IC}{11}\\
	\queryRefCard{interactive-complex-read-12}{IC}{12}\\
	\queryRefCard{interactive-complex-read-13}{IC}{13}\\
	\queryRefCard{interactive-complex-read-14-v1}{IC}{14v1}\\
	\queryRefCard{interactive-complex-read-14-v2}{IC}{14v2}\\
}

\noindent\begin{tabularx}{\queryCardWidth}{|>{\queryPropertyCell}p{\queryPropertyCellWidth}|X|}
	\hline
	query & Interactive / complex / 3 \\ \hline
	title & Friends and friends of friends that have been to given countries \\ \hline
	pattern & \centering \includegraphics[scale=\patternscale,margin=0cm .2cm]{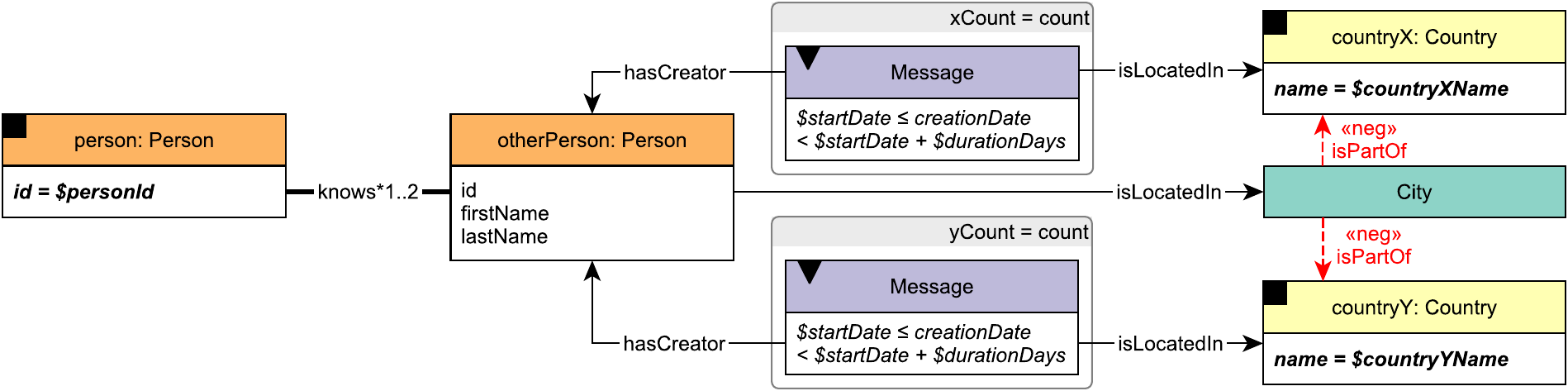} \tabularnewline \hline
	description & Given a start \emph{Person} with ID \texttt{\$personId}, find
\emph{Persons} that are their friends and friends of friends (excluding
the start \emph{Person}) that have made \emph{Posts} / \emph{Comments}
in both of the given \emph{Countries} (named \texttt{\$countryXName} and
\texttt{\$countryYName}), within
\texttt{{[}\$startDate,\ \$startDate\ +\ \$durationDays)} (closed-open
interval). Only \emph{Persons} that are foreign to these
\emph{Countries} are considered, that is \emph{Persons} whose location
\emph{Country} is neither named \texttt{\$countryXName} nor
\texttt{\$countryYName}.
 \\ \hline

		params &
		\innerCardVSpace{\begin{tabularx}{\attributeCardWidth}{|>{\paramNumberCell}C{\attributeNumberWidth}|>{\varNameCell}M|>{\typeCell}m{\typeWidth}|Y|} \hline
		$\mathsf{1}$ & \$personId
 & ID
 &  \\ \hline
		$\mathsf{2}$ & \$countryXName
 & String
 & In SNB Interactive v2, this query has two variants:

\texttt{(a)} Correlated \emph{Countries}

\texttt{(b)} Anti-correlated \emph{Countries}
 \\ \hline
		$\mathsf{3}$ & \$countryYName
 & String
 &  \\ \hline
		$\mathsf{4}$ & \$startDate
 & Date
 & Beginning of requested period
 \\ \hline
		$\mathsf{5}$ & \$durationDays
 & 32-bit Integer
 & Duration of requested period, in days. The interval
\texttt{{[}\$startDate,\ \$startDate\ +\ \$durationDays)} is closed-open
 \\ \hline
		\end{tabularx}}\innerCardVSpace \\ \hline

		result &
		\innerCardVSpace{\begin{tabularx}{\attributeCardWidth}{|>{\resultNumberCell}C{\attributeNumberWidth}|>{\varNameCell}M|>{\typeCell}m{\typeWidth}|>{\resultOriginCell}c|Y|} \hline
		$\mathsf{1}$ & otherPerson.id & ID & R &
				 \\ \hline
		$\mathsf{2}$ & otherPerson.firstName & String & R &
				 \\ \hline
		$\mathsf{3}$ & otherPerson.lastName & String & R &
				 \\ \hline
		$\mathsf{4}$ & xCount & 32-bit Integer & A &
				Number of \emph{Messages} from \emph{Country} named
\texttt{\$countryXName} created by the \emph{Person} within the given
time
 \\ \hline
		$\mathsf{5}$ & yCount & 32-bit Integer & A &
				Number of \emph{Messages} from \emph{Country} named
\texttt{\$countryYName} created by the \emph{Person} within the given
time
 \\ \hline
		$\mathsf{6}$ & count & 32-bit Integer & A &
				\texttt{count} = \texttt{xCount} + \texttt{yCount}
 \\ \hline
		\end{tabularx}}\innerCardVSpace \\ \hline

		sort		&
		\innerCardVSpace{\begin{tabularx}{\attributeCardWidth}{|>{\sortNumberCell}C{\attributeNumberWidth}|>{\varNameCell}M|>{\directionCell}c|Y|} \hline
		$\mathsf{1}$ & count
 & $\desc
$ &  \\ \hline
		$\mathsf{2}$ & otherPerson.id
 & $\asc
$ &  \\ \hline
		\end{tabularx}}\innerCardVSpace \\ \hline
	limit & 20 \\ \hline
	CPs &
	\multicolumn{1}{>{\raggedright}l|}{
		\chokePoint{2.1}, 
		\chokePoint{3.1}, 
		\chokePoint{5.1}, 
		\chokePoint{8.2}, 
		\chokePoint{8.5}
		} \\ \hline
	relevance &
		\footnotesize This query looks for paths of length two and three, starting from a
\emph{Person}, going to friends or friends of friends, and then moving
to \emph{Messages}. This query tests the ability of the query optimizer
to select the most efficient join ordering, which will depend on the
cardinalities of the intermediate results. Many friends of friends can
be duplicate, then it is expected to eliminate duplicates and those
people prior to access the \emph{Post} and \emph{Comments}, as well as
eliminate those friends from \emph{Countries} named
\texttt{\$countryXName} and \texttt{\$countryYName}, as the size of the
intermediate results can be severely affected. A possible structural
optimization could be to materialize the number of \emph{Posts} and
\emph{Comments} created by a \emph{Person}, and progressively filter
those people that could not even fall in the top 20 even having all
their posts in the \emph{Countries} named \texttt{\$countryXName} and
\texttt{\$countryYName}.
 \\ \hline%
\end{tabularx}
\queryCardVSpace

\let\emph\oldemph
\let\textbf\oldtextbf

\renewcommand{\currentQueryCard}{0}
\renewcommand*{\arraystretch}{1.1}

\subsection*{Interactive / complex / 4}
\label{sec:interactive-complex-read-04}

\let\oldemph\emph
\renewcommand{\emph}[1]{{\footnotesize \sf #1}}
\let\oldtextbf\textbf
\renewcommand{\textbf}[1]{{\it #1}}\renewcommand{\currentQueryCard}{interactive-complex-read-04}
\marginpar{
	\raggedleft
	\vspace{0.22ex}

	\queryRefCard{interactive-complex-read-01}{IC}{1}\\
	\queryRefCard{interactive-complex-read-02}{IC}{2}\\
	\queryRefCard{interactive-complex-read-03}{IC}{3}\\
	\queryRefCard{interactive-complex-read-04}{IC}{4}\\
	\queryRefCard{interactive-complex-read-05}{IC}{5}\\
	\queryRefCard{interactive-complex-read-06}{IC}{6}\\
	\queryRefCard{interactive-complex-read-07}{IC}{7}\\
	\queryRefCard{interactive-complex-read-08}{IC}{8}\\
	\queryRefCard{interactive-complex-read-09}{IC}{9}\\
	\queryRefCard{interactive-complex-read-10}{IC}{10}\\
	\queryRefCard{interactive-complex-read-11}{IC}{11}\\
	\queryRefCard{interactive-complex-read-12}{IC}{12}\\
	\queryRefCard{interactive-complex-read-13}{IC}{13}\\
	\queryRefCard{interactive-complex-read-14-v1}{IC}{14v1}\\
	\queryRefCard{interactive-complex-read-14-v2}{IC}{14v2}\\
}

\noindent\begin{tabularx}{\queryCardWidth}{|>{\queryPropertyCell}p{\queryPropertyCellWidth}|X|}
	\hline
	query & Interactive / complex / 4 \\ \hline
	title & New topics \\ \hline
	pattern & \centering \includegraphics[scale=\patternscale,margin=0cm .2cm]{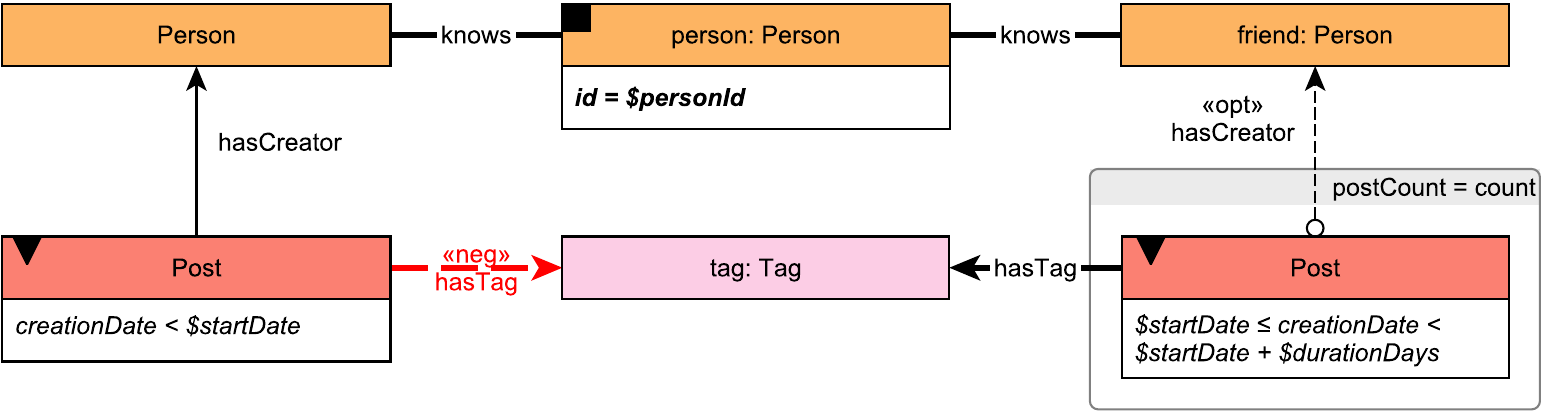} \tabularnewline \hline
	description & Given a start \emph{Person} with ID \texttt{\$personId}, find
\emph{Tags} that are attached to \emph{Posts} that were created by that
\emph{Person}'s friends. Only include \emph{Tags} that were attached to
friends' \emph{Posts} created within a given time interval
\texttt{{[}\$startDate,\ \$startDate\ +\ \$durationDays)} (closed-open)
and that were never attached to friends' \emph{Posts} created before
this interval.
 \\ \hline

		params &
		\innerCardVSpace{\begin{tabularx}{\attributeCardWidth}{|>{\paramNumberCell}C{\attributeNumberWidth}|>{\varNameCell}M|>{\typeCell}m{\typeWidth}|Y|} \hline
		$\mathsf{1}$ & \$personId
 & ID
 &  \\ \hline
		$\mathsf{2}$ & \$startDate
 & Date
 &  \\ \hline
		$\mathsf{3}$ & \$durationDays
 & 32-bit Integer
 & Duration of requested period, in days. The interval
\texttt{{[}\$startDate,\ \$startDate\ +\ \$durationDays)} is closed-open
 \\ \hline
		\end{tabularx}}\innerCardVSpace \\ \hline

		result &
		\innerCardVSpace{\begin{tabularx}{\attributeCardWidth}{|>{\resultNumberCell}C{\attributeNumberWidth}|>{\varNameCell}M|>{\typeCell}m{\typeWidth}|>{\resultOriginCell}c|Y|} \hline
		$\mathsf{1}$ & tag.name & Long String & R &
				 \\ \hline
		$\mathsf{2}$ & postCount & 32-bit Integer & A &
				Number of \emph{Posts} made within the given time interval that have
\texttt{tag}
 \\ \hline
		\end{tabularx}}\innerCardVSpace \\ \hline

		sort		&
		\innerCardVSpace{\begin{tabularx}{\attributeCardWidth}{|>{\sortNumberCell}C{\attributeNumberWidth}|>{\varNameCell}M|>{\directionCell}c|Y|} \hline
		$\mathsf{1}$ & postCount
 & $\desc
$ &  \\ \hline
		$\mathsf{2}$ & tag.name
 & $\asc
$ &  \\ \hline
		\end{tabularx}}\innerCardVSpace \\ \hline
	limit & 10 \\ \hline
	CPs &
	\multicolumn{1}{>{\raggedright}l|}{
		\chokePoint{2.3}, 
		\chokePoint{8.2}, 
		\chokePoint{8.5}
		} \\ \hline
	relevance &
		\footnotesize This query looks for paths of length two, starting from a given
\emph{Person}, moving to \emph{Posts} and then to \emph{Tags}. It tests
the ability of the query optimizer to properly select the usage of hash
joins or index based joins, depending on the cardinality of the
intermediate results. These cardinalities are clearly affected by the
input \emph{Person}, the number of friends, the variety of \emph{Tags},
the time interval and the number of \emph{Posts}.
 \\ \hline%
\end{tabularx}
\queryCardVSpace

\let\emph\oldemph
\let\textbf\oldtextbf

\renewcommand{\currentQueryCard}{0}
\renewcommand*{\arraystretch}{1.1}

\subsection*{Interactive / complex / 5}
\label{sec:interactive-complex-read-05}

\let\oldemph\emph
\renewcommand{\emph}[1]{{\footnotesize \sf #1}}
\let\oldtextbf\textbf
\renewcommand{\textbf}[1]{{\it #1}}\renewcommand{\currentQueryCard}{interactive-complex-read-05}
\marginpar{
	\raggedleft
	\vspace{0.22ex}

	\queryRefCard{interactive-complex-read-01}{IC}{1}\\
	\queryRefCard{interactive-complex-read-02}{IC}{2}\\
	\queryRefCard{interactive-complex-read-03}{IC}{3}\\
	\queryRefCard{interactive-complex-read-04}{IC}{4}\\
	\queryRefCard{interactive-complex-read-05}{IC}{5}\\
	\queryRefCard{interactive-complex-read-06}{IC}{6}\\
	\queryRefCard{interactive-complex-read-07}{IC}{7}\\
	\queryRefCard{interactive-complex-read-08}{IC}{8}\\
	\queryRefCard{interactive-complex-read-09}{IC}{9}\\
	\queryRefCard{interactive-complex-read-10}{IC}{10}\\
	\queryRefCard{interactive-complex-read-11}{IC}{11}\\
	\queryRefCard{interactive-complex-read-12}{IC}{12}\\
	\queryRefCard{interactive-complex-read-13}{IC}{13}\\
	\queryRefCard{interactive-complex-read-14-v1}{IC}{14v1}\\
	\queryRefCard{interactive-complex-read-14-v2}{IC}{14v2}\\
}

\noindent\begin{tabularx}{\queryCardWidth}{|>{\queryPropertyCell}p{\queryPropertyCellWidth}|X|}
	\hline
	query & Interactive / complex / 5 \\ \hline
	title & New groups \\ \hline
	pattern & \centering \includegraphics[scale=\patternscale,margin=0cm .2cm]{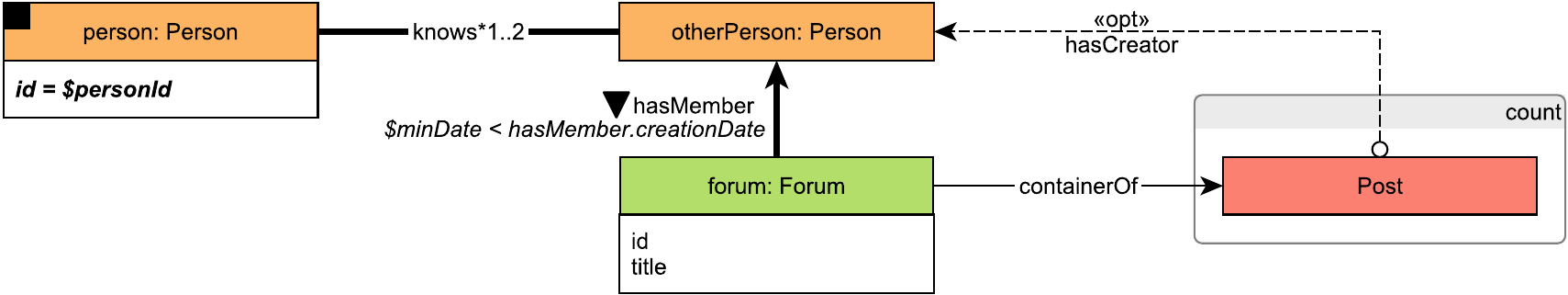} \tabularnewline \hline
	description & Given a start \emph{Person} with ID \texttt{\$personId}, denote their
friends and friends of friends (excluding the start \emph{Person}) as
\texttt{otherPerson}.

Find \emph{Forums} that any \emph{Person} \texttt{otherPerson} became a
\emph{member} of after a given date (\texttt{\$minDate}). For each of
those \emph{Forums}, count the number of \emph{Posts} that were created
by the \emph{Person} \texttt{otherPerson}.
 \\ \hline

		params &
		\innerCardVSpace{\begin{tabularx}{\attributeCardWidth}{|>{\paramNumberCell}C{\attributeNumberWidth}|>{\varNameCell}M|>{\typeCell}m{\typeWidth}|Y|} \hline
		$\mathsf{1}$ & \$personId
 & ID
 &  \\ \hline
		$\mathsf{2}$ & \$minDate
 & Date
 &  \\ \hline
		\end{tabularx}}\innerCardVSpace \\ \hline

		result &
		\innerCardVSpace{\begin{tabularx}{\attributeCardWidth}{|>{\resultNumberCell}C{\attributeNumberWidth}|>{\varNameCell}M|>{\typeCell}m{\typeWidth}|>{\resultOriginCell}c|Y|} \hline
		$\mathsf{1}$ & forum.title & Long String & R &
				 \\ \hline
		$\mathsf{2}$ & postCount & 32-bit Integer & A &
				Number of \emph{Posts} made in \texttt{forum} that were created by the
\emph{Person} \texttt{otherPerson}
 \\ \hline
		\end{tabularx}}\innerCardVSpace \\ \hline

		sort		&
		\innerCardVSpace{\begin{tabularx}{\attributeCardWidth}{|>{\sortNumberCell}C{\attributeNumberWidth}|>{\varNameCell}M|>{\directionCell}c|Y|} \hline
		$\mathsf{1}$ & postCount
 & $\desc
$ &  \\ \hline
		$\mathsf{2}$ & forum.id
 & $\asc
$ &  \\ \hline
		\end{tabularx}}\innerCardVSpace \\ \hline
	limit & 20 \\ \hline
	CPs &
	\multicolumn{1}{>{\raggedright}l|}{
		\chokePoint{2.3}, 
		\chokePoint{3.3}, 
		\chokePoint{8.2}, 
		\chokePoint{8.5}
		} \\ \hline
	relevance &
		\footnotesize This query looks for paths of length two and three, starting from a
given \emph{Person}, moving to friends and friends of friends, and then
getting the \emph{Forums} they are members of. Besides testing the
ability of the query optimizer to select the proper join operator, it
rewards the usage of indices, but their accesses will be presumably
scattered due to the two/three-hop search space of the query, leading to
unpredictable and scattered index accesses. Having efficient
implementations of such indices will be highly beneficial.
 \\ \hline%
\end{tabularx}
\queryCardVSpace

\let\emph\oldemph
\let\textbf\oldtextbf

\renewcommand{\currentQueryCard}{0}
\renewcommand*{\arraystretch}{1.1}

\subsection*{Interactive / complex / 6}
\label{sec:interactive-complex-read-06}

\let\oldemph\emph
\renewcommand{\emph}[1]{{\footnotesize \sf #1}}
\let\oldtextbf\textbf
\renewcommand{\textbf}[1]{{\it #1}}\renewcommand{\currentQueryCard}{interactive-complex-read-06}
\marginpar{
	\raggedleft
	\vspace{0.22ex}

	\queryRefCard{interactive-complex-read-01}{IC}{1}\\
	\queryRefCard{interactive-complex-read-02}{IC}{2}\\
	\queryRefCard{interactive-complex-read-03}{IC}{3}\\
	\queryRefCard{interactive-complex-read-04}{IC}{4}\\
	\queryRefCard{interactive-complex-read-05}{IC}{5}\\
	\queryRefCard{interactive-complex-read-06}{IC}{6}\\
	\queryRefCard{interactive-complex-read-07}{IC}{7}\\
	\queryRefCard{interactive-complex-read-08}{IC}{8}\\
	\queryRefCard{interactive-complex-read-09}{IC}{9}\\
	\queryRefCard{interactive-complex-read-10}{IC}{10}\\
	\queryRefCard{interactive-complex-read-11}{IC}{11}\\
	\queryRefCard{interactive-complex-read-12}{IC}{12}\\
	\queryRefCard{interactive-complex-read-13}{IC}{13}\\
	\queryRefCard{interactive-complex-read-14-v1}{IC}{14v1}\\
	\queryRefCard{interactive-complex-read-14-v2}{IC}{14v2}\\
}

\noindent\begin{tabularx}{\queryCardWidth}{|>{\queryPropertyCell}p{\queryPropertyCellWidth}|X|}
	\hline
	query & Interactive / complex / 6 \\ \hline
	title & Tag co-occurrence \\ \hline
	pattern & \centering \includegraphics[scale=\patternscale,margin=0cm .2cm]{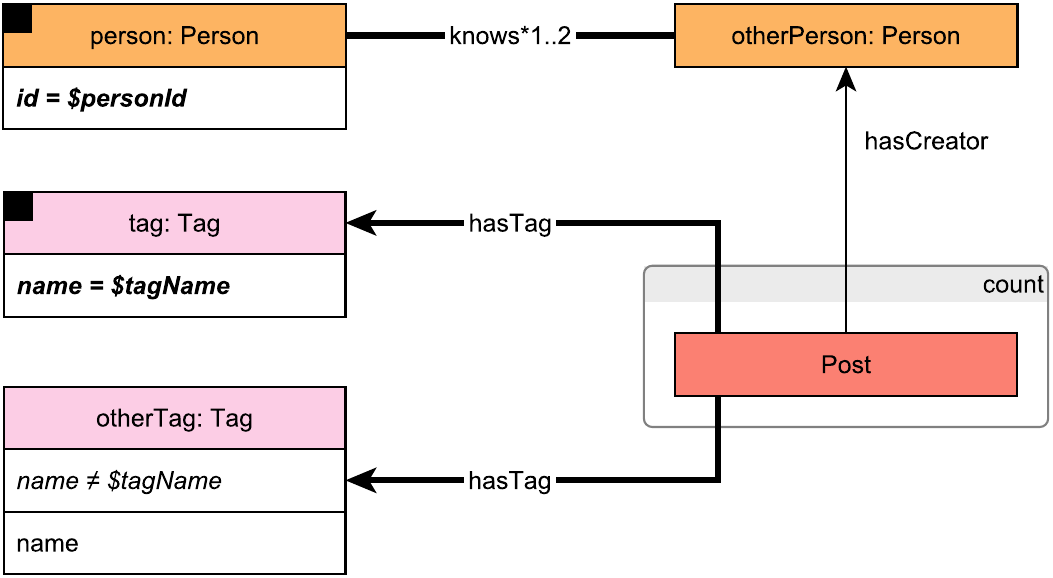} \tabularnewline \hline
	description & Given a start \emph{Person} with ID \texttt{\$personId} and a \emph{Tag}
with name \texttt{\$tagName}, find the other \emph{Tags} that occur
together with this \emph{Tag} on \emph{Posts} that were created by start
\emph{Person}'s friends and friends of friends (excluding start
\emph{Person}). Return top 10 \emph{Tags}, and the count of \emph{Posts}
that were created by these \emph{Persons}, which contain both this
\emph{Tag} and the given \emph{Tag}.
 \\ \hline

		params &
		\innerCardVSpace{\begin{tabularx}{\attributeCardWidth}{|>{\paramNumberCell}C{\attributeNumberWidth}|>{\varNameCell}M|>{\typeCell}m{\typeWidth}|Y|} \hline
		$\mathsf{1}$ & \$personId
 & ID
 &  \\ \hline
		$\mathsf{2}$ & \$tagName
 & Long String
 &  \\ \hline
		\end{tabularx}}\innerCardVSpace \\ \hline

		result &
		\innerCardVSpace{\begin{tabularx}{\attributeCardWidth}{|>{\resultNumberCell}C{\attributeNumberWidth}|>{\varNameCell}M|>{\typeCell}m{\typeWidth}|>{\resultOriginCell}c|Y|} \hline
		$\mathsf{1}$ & otherTag.name & Long String & R &
				 \\ \hline
		$\mathsf{2}$ & postCount & 32-bit Integer & A &
				Number of \emph{Posts} that were created by friends and friends of
friends, which have the \emph{Tag} \texttt{otherTag}
 \\ \hline
		\end{tabularx}}\innerCardVSpace \\ \hline

		sort		&
		\innerCardVSpace{\begin{tabularx}{\attributeCardWidth}{|>{\sortNumberCell}C{\attributeNumberWidth}|>{\varNameCell}M|>{\directionCell}c|Y|} \hline
		$\mathsf{1}$ & postCount
 & $\desc
$ &  \\ \hline
		$\mathsf{2}$ & otherTag.name
 & $\asc
$ &  \\ \hline
		\end{tabularx}}\innerCardVSpace \\ \hline
	limit & 10 \\ \hline
	CPs &
	\multicolumn{1}{>{\raggedright}l|}{
		\chokePoint{5.1}, 
		\chokePoint{8.2}
		} \\ \hline
	relevance &
		\footnotesize This query looks for paths of lengths three or four, starting from a
given \emph{Person}, moving to friends or friends of friends, then to
\emph{Posts} and finally ending at a given \emph{Tag}.
 \\ \hline%
\end{tabularx}
\queryCardVSpace

\let\emph\oldemph
\let\textbf\oldtextbf

\renewcommand{\currentQueryCard}{0}
\renewcommand*{\arraystretch}{1.1}

\subsection*{Interactive / complex / 7}
\label{sec:interactive-complex-read-07}

\let\oldemph\emph
\renewcommand{\emph}[1]{{\footnotesize \sf #1}}
\let\oldtextbf\textbf
\renewcommand{\textbf}[1]{{\it #1}}\renewcommand{\currentQueryCard}{interactive-complex-read-07}
\marginpar{
	\raggedleft
	\vspace{0.22ex}

	\queryRefCard{interactive-complex-read-01}{IC}{1}\\
	\queryRefCard{interactive-complex-read-02}{IC}{2}\\
	\queryRefCard{interactive-complex-read-03}{IC}{3}\\
	\queryRefCard{interactive-complex-read-04}{IC}{4}\\
	\queryRefCard{interactive-complex-read-05}{IC}{5}\\
	\queryRefCard{interactive-complex-read-06}{IC}{6}\\
	\queryRefCard{interactive-complex-read-07}{IC}{7}\\
	\queryRefCard{interactive-complex-read-08}{IC}{8}\\
	\queryRefCard{interactive-complex-read-09}{IC}{9}\\
	\queryRefCard{interactive-complex-read-10}{IC}{10}\\
	\queryRefCard{interactive-complex-read-11}{IC}{11}\\
	\queryRefCard{interactive-complex-read-12}{IC}{12}\\
	\queryRefCard{interactive-complex-read-13}{IC}{13}\\
	\queryRefCard{interactive-complex-read-14-v1}{IC}{14v1}\\
	\queryRefCard{interactive-complex-read-14-v2}{IC}{14v2}\\
}

\noindent\begin{tabularx}{\queryCardWidth}{|>{\queryPropertyCell}p{\queryPropertyCellWidth}|X|}
	\hline
	query & Interactive / complex / 7 \\ \hline
	title & Recent likers \\ \hline
	pattern & \centering \includegraphics[scale=\patternscale,margin=0cm .2cm]{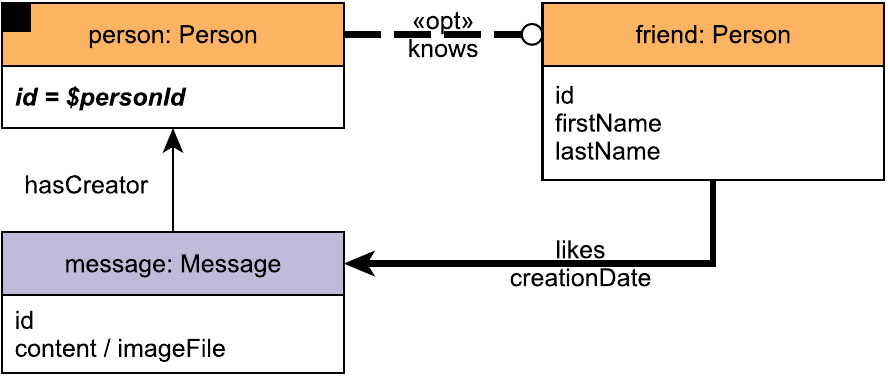} \tabularnewline \hline
	description & Given a start \emph{Person} with ID \texttt{\$personId}, find the most
recent \texttt{likes} on any of start \emph{Person}'s \emph{Messages}.
Find \emph{Persons} that liked (\texttt{likes} edge) any of start
\emph{Person}'s \emph{Messages}, the \emph{Messages} they liked most
recently, the creation date of that like, and the latency in minutes
(\texttt{minutesLatency}) between creation of \emph{Messages} and like.
Additionally, for each \emph{Person} found return a flag indicating
(\texttt{isNew}) whether the liker is a friend of start \emph{Person}.
In case that a \emph{Person} liked multiple \emph{Messages} at the same
time, return the \emph{Message} with lowest identifier.

\textbf{Validation rule:} Depending on whether the system-under-test
supports leap seconds or uses UTC-SLS (UTC with Smoothed Leap Seconds),
a difference of 1 minute can occur between the \texttt{minutesLatency}
results of two correct implementations when the time interval includes
June 30, 2012, when there was a leap second. Therefore, the
\texttt{minutesLatency} value is validated using a tolerance of 1
minute.
 \\ \hline

		params &
		\innerCardVSpace{\begin{tabularx}{\attributeCardWidth}{|>{\paramNumberCell}C{\attributeNumberWidth}|>{\varNameCell}M|>{\typeCell}m{\typeWidth}|Y|} \hline
		$\mathsf{1}$ & \$personId
 & ID
 &  \\ \hline
		\end{tabularx}}\innerCardVSpace \\ \hline

		result &
		\innerCardVSpace{\begin{tabularx}{\attributeCardWidth}{|>{\resultNumberCell}C{\attributeNumberWidth}|>{\varNameCell}M|>{\typeCell}m{\typeWidth}|>{\resultOriginCell}c|Y|} \hline
		$\mathsf{1}$ & friend.id & ID & R &
				friend.id = personId is allowed
 \\ \hline
		$\mathsf{2}$ & friend.firstName & String & R &
				 \\ \hline
		$\mathsf{3}$ & friend.lastName & String & R &
				 \\ \hline
		$\mathsf{4}$ & likes.creationDate & DateTime & R &
				 \\ \hline
		$\mathsf{5}$ & message.id & ID & R &
				 \\ \hline
		$\mathsf{6}$ & message.content or message.imageFile (for photos) & Text & R &
				 \\ \hline
		$\mathsf{7}$ & minutesLatency & 32-bit Integer & C &
				Duration between the creation of the \emph{Message} and the creation of
the like, in minutes.
 \\ \hline
		$\mathsf{8}$ & isNew & Boolean & C &
				\texttt{False} if \texttt{person} and \texttt{friend} know each other,
\texttt{True} otherwise
 \\ \hline
		\end{tabularx}}\innerCardVSpace \\ \hline

		sort		&
		\innerCardVSpace{\begin{tabularx}{\attributeCardWidth}{|>{\sortNumberCell}C{\attributeNumberWidth}|>{\varNameCell}M|>{\directionCell}c|Y|} \hline
		$\mathsf{1}$ & likes.creationDate
 & $\desc
$ &  \\ \hline
		$\mathsf{2}$ & friend.id
 & $\asc
$ &  \\ \hline
		\end{tabularx}}\innerCardVSpace \\ \hline
	limit & 20 \\ \hline
	CPs &
	\multicolumn{1}{>{\raggedright}l|}{
		\chokePoint{2.2}, 
		\chokePoint{2.3}, 
		\chokePoint{3.3}, 
		\chokePoint{5.1}, 
		\chokePoint{8.1}, 
		\chokePoint{8.3}
		} \\ \hline
	relevance &
		\footnotesize This query looks for paths of length two, starting from a given
\emph{Person}, moving to its published messages and then to
\emph{Persons} who liked them. It tests several aspects related to join
optimization, both at query optimization plan level and execution engine
level. On the one hand, many of the columns needed for the projection
are only needed in the last stages of the query, so the optimizer is
expected to delay the projection until the end. This query implies
accessing two-hop data, and as a consequence, index accesses are
expected to be scattered. We expect to observe variate cardinalities,
depending on the characteristics of the input parameter, so properly
selecting the join operators will be crucial. This query has a lot of
correlated sub-queries, so it is testing the ability to flatten the
query execution plans.
 \\ \hline%
\end{tabularx}
\queryCardVSpace

\let\emph\oldemph
\let\textbf\oldtextbf

\renewcommand{\currentQueryCard}{0}
\renewcommand*{\arraystretch}{1.1}

\subsection*{Interactive / complex / 8}
\label{sec:interactive-complex-read-08}

\let\oldemph\emph
\renewcommand{\emph}[1]{{\footnotesize \sf #1}}
\let\oldtextbf\textbf
\renewcommand{\textbf}[1]{{\it #1}}\renewcommand{\currentQueryCard}{interactive-complex-read-08}
\marginpar{
	\raggedleft
	\vspace{0.22ex}

	\queryRefCard{interactive-complex-read-01}{IC}{1}\\
	\queryRefCard{interactive-complex-read-02}{IC}{2}\\
	\queryRefCard{interactive-complex-read-03}{IC}{3}\\
	\queryRefCard{interactive-complex-read-04}{IC}{4}\\
	\queryRefCard{interactive-complex-read-05}{IC}{5}\\
	\queryRefCard{interactive-complex-read-06}{IC}{6}\\
	\queryRefCard{interactive-complex-read-07}{IC}{7}\\
	\queryRefCard{interactive-complex-read-08}{IC}{8}\\
	\queryRefCard{interactive-complex-read-09}{IC}{9}\\
	\queryRefCard{interactive-complex-read-10}{IC}{10}\\
	\queryRefCard{interactive-complex-read-11}{IC}{11}\\
	\queryRefCard{interactive-complex-read-12}{IC}{12}\\
	\queryRefCard{interactive-complex-read-13}{IC}{13}\\
	\queryRefCard{interactive-complex-read-14-v1}{IC}{14v1}\\
	\queryRefCard{interactive-complex-read-14-v2}{IC}{14v2}\\
}

\noindent\begin{tabularx}{\queryCardWidth}{|>{\queryPropertyCell}p{\queryPropertyCellWidth}|X|}
	\hline
	query & Interactive / complex / 8 \\ \hline
	title & Recent replies \\ \hline
	pattern & \centering \includegraphics[scale=\patternscale,margin=0cm .2cm]{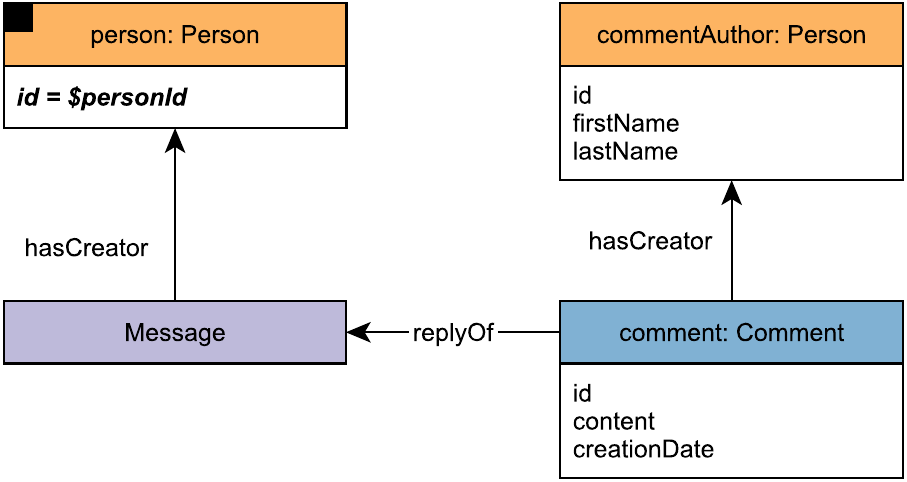} \tabularnewline \hline
	description & Given a start \emph{Person} with ID \texttt{\$personId}, find the most
recent \emph{Comments} that are replies to \emph{Messages} of the start
\emph{Person}. Only consider direct (single-hop) replies, not the
transitive (multi-hop) ones. Return the reply \emph{Comments}, and the
\emph{Person} that created each reply \emph{Comment}.
 \\ \hline

		params &
		\innerCardVSpace{\begin{tabularx}{\attributeCardWidth}{|>{\paramNumberCell}C{\attributeNumberWidth}|>{\varNameCell}M|>{\typeCell}m{\typeWidth}|Y|} \hline
		$\mathsf{1}$ & \$personId
 & ID
 &  \\ \hline
		\end{tabularx}}\innerCardVSpace \\ \hline

		result &
		\innerCardVSpace{\begin{tabularx}{\attributeCardWidth}{|>{\resultNumberCell}C{\attributeNumberWidth}|>{\varNameCell}M|>{\typeCell}m{\typeWidth}|>{\resultOriginCell}c|Y|} \hline
		$\mathsf{1}$ & commentAuthor.id & ID & R &
				 \\ \hline
		$\mathsf{2}$ & commentAuthor.firstName & String & R &
				 \\ \hline
		$\mathsf{3}$ & commentAuthor.lastName & String & R &
				 \\ \hline
		$\mathsf{4}$ & comment.creationDate & DateTime & R &
				 \\ \hline
		$\mathsf{5}$ & comment.id & ID & R &
				 \\ \hline
		$\mathsf{6}$ & comment.content & Text & R &
				 \\ \hline
		\end{tabularx}}\innerCardVSpace \\ \hline

		sort		&
		\innerCardVSpace{\begin{tabularx}{\attributeCardWidth}{|>{\sortNumberCell}C{\attributeNumberWidth}|>{\varNameCell}M|>{\directionCell}c|Y|} \hline
		$\mathsf{1}$ & comment.creationDate
 & $\desc
$ &  \\ \hline
		$\mathsf{2}$ & comment.id
 & $\asc
$ &  \\ \hline
		\end{tabularx}}\innerCardVSpace \\ \hline
	limit & 20 \\ \hline
	CPs &
	\multicolumn{1}{>{\raggedright}l|}{
		\chokePoint{2.4}, 
		\chokePoint{3.3}, 
		\chokePoint{5.3}
		} \\ \hline
	relevance &
		\footnotesize This query looks for paths of length two, starting from a given
\emph{Person}, going through its created \emph{Messages} and finishing
at their replies. In this query there is temporal locality between the
replies being accessed. Thus the top-k order by this can interact with
the selection, i.e.~do not consider older \emph{Posts} than the 20th
oldest seen so far.
 \\ \hline%
\end{tabularx}
\queryCardVSpace

\let\emph\oldemph
\let\textbf\oldtextbf

\renewcommand{\currentQueryCard}{0}
\renewcommand*{\arraystretch}{1.1}

\subsection*{Interactive / complex / 9}
\label{sec:interactive-complex-read-09}

\let\oldemph\emph
\renewcommand{\emph}[1]{{\footnotesize \sf #1}}
\let\oldtextbf\textbf
\renewcommand{\textbf}[1]{{\it #1}}\renewcommand{\currentQueryCard}{interactive-complex-read-09}
\marginpar{
	\raggedleft
	\vspace{0.22ex}

	\queryRefCard{interactive-complex-read-01}{IC}{1}\\
	\queryRefCard{interactive-complex-read-02}{IC}{2}\\
	\queryRefCard{interactive-complex-read-03}{IC}{3}\\
	\queryRefCard{interactive-complex-read-04}{IC}{4}\\
	\queryRefCard{interactive-complex-read-05}{IC}{5}\\
	\queryRefCard{interactive-complex-read-06}{IC}{6}\\
	\queryRefCard{interactive-complex-read-07}{IC}{7}\\
	\queryRefCard{interactive-complex-read-08}{IC}{8}\\
	\queryRefCard{interactive-complex-read-09}{IC}{9}\\
	\queryRefCard{interactive-complex-read-10}{IC}{10}\\
	\queryRefCard{interactive-complex-read-11}{IC}{11}\\
	\queryRefCard{interactive-complex-read-12}{IC}{12}\\
	\queryRefCard{interactive-complex-read-13}{IC}{13}\\
	\queryRefCard{interactive-complex-read-14-v1}{IC}{14v1}\\
	\queryRefCard{interactive-complex-read-14-v2}{IC}{14v2}\\
}

\noindent\begin{tabularx}{\queryCardWidth}{|>{\queryPropertyCell}p{\queryPropertyCellWidth}|X|}
	\hline
	query & Interactive / complex / 9 \\ \hline
	title & Recent messages by friends or friends of friends \\ \hline
	pattern & \centering \includegraphics[scale=\patternscale,margin=0cm .2cm]{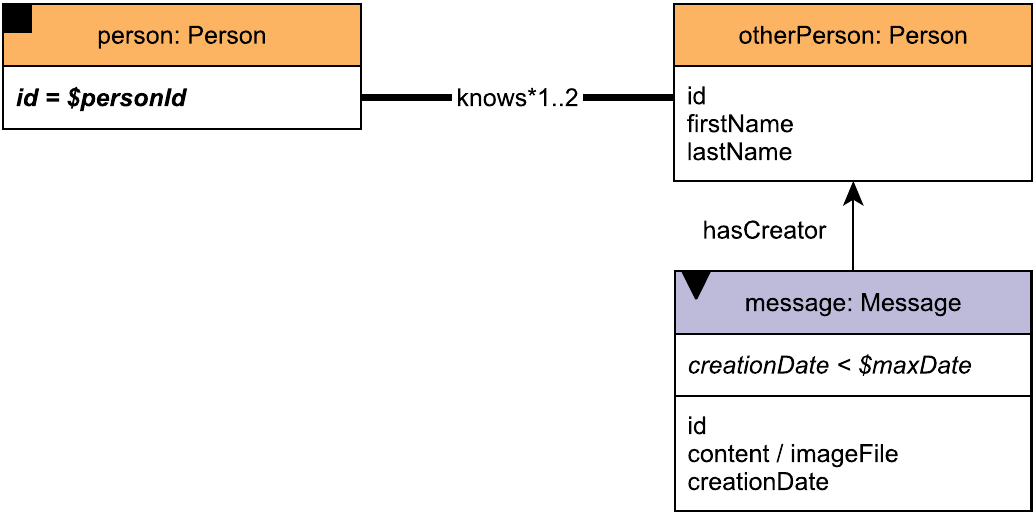} \tabularnewline \hline
	description & Given a start \emph{Person} with ID \texttt{\$personId}, find the most
recent \emph{Messages} created by that \emph{Person}'s friends or
friends of friends (excluding the start \emph{Person}). Only consider
\emph{Messages} created before the given \texttt{\$maxDate} (excluding
that day).
 \\ \hline

		params &
		\innerCardVSpace{\begin{tabularx}{\attributeCardWidth}{|>{\paramNumberCell}C{\attributeNumberWidth}|>{\varNameCell}M|>{\typeCell}m{\typeWidth}|Y|} \hline
		$\mathsf{1}$ & \$personId
 & ID
 &  \\ \hline
		$\mathsf{2}$ & \$maxDate
 & Date
 &  \\ \hline
		\end{tabularx}}\innerCardVSpace \\ \hline

		result &
		\innerCardVSpace{\begin{tabularx}{\attributeCardWidth}{|>{\resultNumberCell}C{\attributeNumberWidth}|>{\varNameCell}M|>{\typeCell}m{\typeWidth}|>{\resultOriginCell}c|Y|} \hline
		$\mathsf{1}$ & otherPerson.id & ID & R &
				 \\ \hline
		$\mathsf{2}$ & otherPerson.firstName & String & R &
				 \\ \hline
		$\mathsf{3}$ & otherPerson.lastName & String & R &
				 \\ \hline
		$\mathsf{4}$ & message.id & ID & R &
				 \\ \hline
		$\mathsf{5}$ & message.content or message.imageFile (for photos) & Text & R &
				 \\ \hline
		$\mathsf{6}$ & message.creationDate & DateTime & R &
				 \\ \hline
		\end{tabularx}}\innerCardVSpace \\ \hline

		sort		&
		\innerCardVSpace{\begin{tabularx}{\attributeCardWidth}{|>{\sortNumberCell}C{\attributeNumberWidth}|>{\varNameCell}M|>{\directionCell}c|Y|} \hline
		$\mathsf{1}$ & message.creationDate
 & $\desc
$ &  \\ \hline
		$\mathsf{2}$ & message.id
 & $\asc
$ &  \\ \hline
		\end{tabularx}}\innerCardVSpace \\ \hline
	limit & 20 \\ \hline
	CPs &
	\multicolumn{1}{>{\raggedright}l|}{
		\chokePoint{1.1}, 
		\chokePoint{1.2}, 
		\chokePoint{2.2}, 
		\chokePoint{2.3}, 
		\chokePoint{3.2}, 
		\chokePoint{3.3}, 
		\chokePoint{8.5}
		} \\ \hline
	relevance &
		\footnotesize This query looks for paths of length two or three, starting from a given
\emph{Person}, moving to its friends and friends of friends, and ending
at their created \emph{Messages}. This is one of the most complex
queries, as the list of choke points indicates. This query is expected
to touch variable amounts of data with entities of different
characteristics, and therefore, properly estimating cardinalities and
selecting the proper operators will be crucial.
 \\ \hline%
\end{tabularx}
\queryCardVSpace

\let\emph\oldemph
\let\textbf\oldtextbf

\renewcommand{\currentQueryCard}{0}
\renewcommand*{\arraystretch}{1.1}

\subsection*{Interactive / complex / 10}
\label{sec:interactive-complex-read-10}

\let\oldemph\emph
\renewcommand{\emph}[1]{{\footnotesize \sf #1}}
\let\oldtextbf\textbf
\renewcommand{\textbf}[1]{{\it #1}}\renewcommand{\currentQueryCard}{interactive-complex-read-10}
\marginpar{
	\raggedleft
	\vspace{0.22ex}

	\queryRefCard{interactive-complex-read-01}{IC}{1}\\
	\queryRefCard{interactive-complex-read-02}{IC}{2}\\
	\queryRefCard{interactive-complex-read-03}{IC}{3}\\
	\queryRefCard{interactive-complex-read-04}{IC}{4}\\
	\queryRefCard{interactive-complex-read-05}{IC}{5}\\
	\queryRefCard{interactive-complex-read-06}{IC}{6}\\
	\queryRefCard{interactive-complex-read-07}{IC}{7}\\
	\queryRefCard{interactive-complex-read-08}{IC}{8}\\
	\queryRefCard{interactive-complex-read-09}{IC}{9}\\
	\queryRefCard{interactive-complex-read-10}{IC}{10}\\
	\queryRefCard{interactive-complex-read-11}{IC}{11}\\
	\queryRefCard{interactive-complex-read-12}{IC}{12}\\
	\queryRefCard{interactive-complex-read-13}{IC}{13}\\
	\queryRefCard{interactive-complex-read-14-v1}{IC}{14v1}\\
	\queryRefCard{interactive-complex-read-14-v2}{IC}{14v2}\\
}

\noindent\begin{tabularx}{\queryCardWidth}{|>{\queryPropertyCell}p{\queryPropertyCellWidth}|X|}
	\hline
	query & Interactive / complex / 10 \\ \hline
	title & Friend recommendation \\ \hline
	pattern & \centering \includegraphics[scale=\patternscale,margin=0cm .2cm]{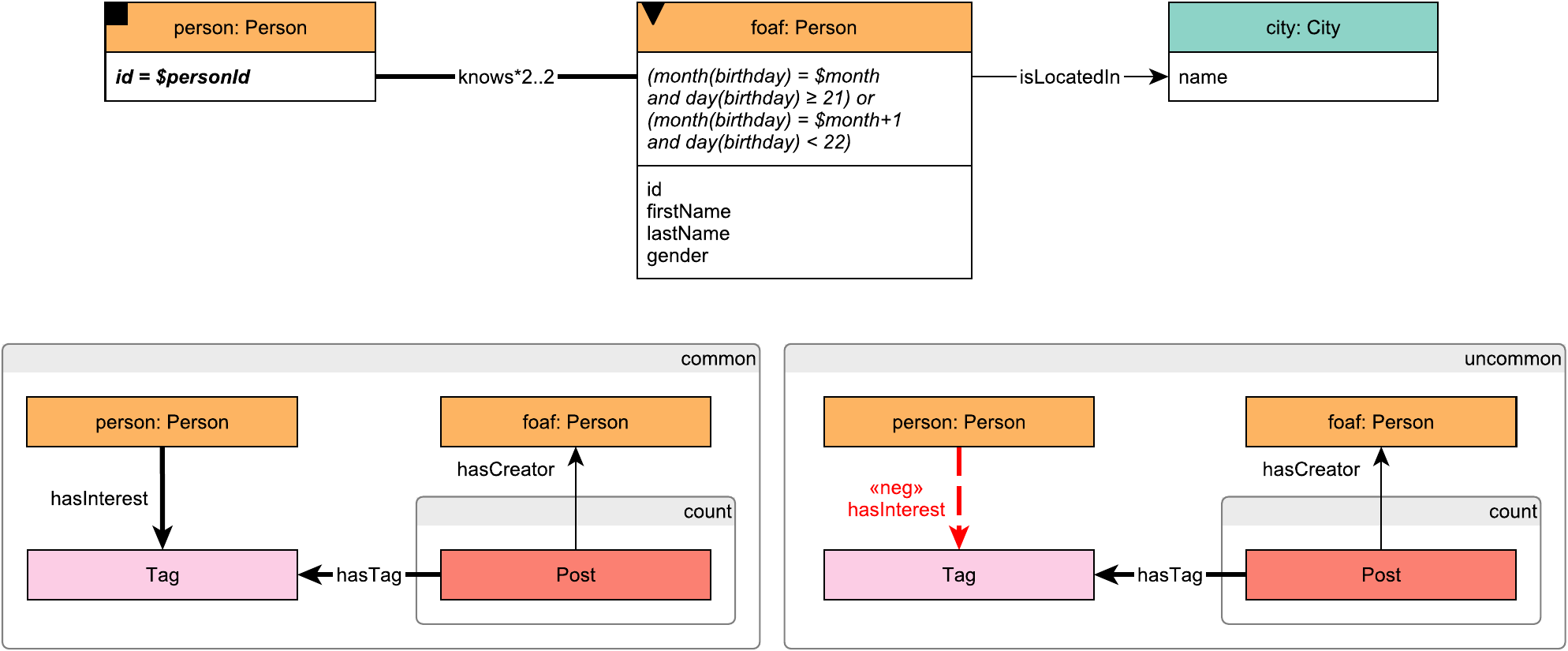} \tabularnewline \hline
	description & Given a start \emph{Person} with ID \texttt{\$personId}, find that
\emph{Person}'s friends of friends (\texttt{foaf}) -- excluding the
start \emph{Person} and his/her immediate friends --, who were born on
or after the 21st of a given \texttt{\$month} (in any year) and before
the 22nd of the following month. Calculate the similarity between each
\texttt{friend} and the start \texttt{person}, where
\texttt{commonInterestScore} is defined as follows:

\begin{itemize}
\tightlist
\item
  \texttt{common} = number of \emph{Posts} created by \texttt{friend},
  such that the \emph{Post} has a \emph{Tag} that the start
  \texttt{person} is interested in
\item
  \texttt{uncommon} = number of \emph{Posts} created by \texttt{friend},
  such that the \emph{Post} has no \emph{Tag} that the start
  \texttt{person} is interested in
\item
  \texttt{commonInterestScore} = \texttt{common} - \texttt{uncommon}
\end{itemize}
 \\ \hline

		params &
		\innerCardVSpace{\begin{tabularx}{\attributeCardWidth}{|>{\paramNumberCell}C{\attributeNumberWidth}|>{\varNameCell}M|>{\typeCell}m{\typeWidth}|Y|} \hline
		$\mathsf{1}$ & \$personId
 & ID
 &  \\ \hline
		$\mathsf{2}$ & \$month
 & 32-bit Integer
 & Between 1 and 12. Implementations may also pass the next month as an
additional \texttt{\$nextMonth} parameter
 \\ \hline
		\end{tabularx}}\innerCardVSpace \\ \hline

		result &
		\innerCardVSpace{\begin{tabularx}{\attributeCardWidth}{|>{\resultNumberCell}C{\attributeNumberWidth}|>{\varNameCell}M|>{\typeCell}m{\typeWidth}|>{\resultOriginCell}c|Y|} \hline
		$\mathsf{1}$ & foaf.id & ID & R &
				 \\ \hline
		$\mathsf{2}$ & foaf.firstName & String & R &
				 \\ \hline
		$\mathsf{3}$ & foaf.lastName & String & R &
				 \\ \hline
		$\mathsf{4}$ & commonInterestScore & 32-bit Integer & A &
				 \\ \hline
		$\mathsf{5}$ & foaf.gender & String & R &
				 \\ \hline
		$\mathsf{6}$ & city.name & String & R &
				 \\ \hline
		\end{tabularx}}\innerCardVSpace \\ \hline

		sort		&
		\innerCardVSpace{\begin{tabularx}{\attributeCardWidth}{|>{\sortNumberCell}C{\attributeNumberWidth}|>{\varNameCell}M|>{\directionCell}c|Y|} \hline
		$\mathsf{1}$ & commonInterestScore
 & $\desc
$ &  \\ \hline
		$\mathsf{2}$ & foaf.id
 & $\asc
$ &  \\ \hline
		\end{tabularx}}\innerCardVSpace \\ \hline
	limit & 10 \\ \hline
	CPs &
	\multicolumn{1}{>{\raggedright}l|}{
		\chokePoint{2.3}, 
		\chokePoint{3.3}, 
		\chokePoint{4.1}, 
		\chokePoint{4.2}, 
		\chokePoint{5.1}, 
		\chokePoint{5.2}, 
		\chokePoint{6.1}, 
		\chokePoint{7.1}, 
		\chokePoint{8.6}
		} \\ \hline
	relevance &
		\footnotesize This query looks for paths of length two, starting from a \emph{Person}
and ending at the friends of their friends. It does widely scattered
graph traversal, and one expects no locality of in friends of friends,
as these have been acquired over a long time and have widely scattered
identifiers. The join order is simple but one must see that the
anti-join for ``not in my friends'' is better with hash. Also the last
pattern in the scalar sub-queries joining or anti-joining the
\emph{Tags} of the candidate's \emph{Posts} to interests of self should
be by hash.
 \\ \hline%
\end{tabularx}
\queryCardVSpace

\let\emph\oldemph
\let\textbf\oldtextbf

\renewcommand{\currentQueryCard}{0}
\renewcommand*{\arraystretch}{1.1}

\subsection*{Interactive / complex / 11}
\label{sec:interactive-complex-read-11}

\let\oldemph\emph
\renewcommand{\emph}[1]{{\footnotesize \sf #1}}
\let\oldtextbf\textbf
\renewcommand{\textbf}[1]{{\it #1}}\renewcommand{\currentQueryCard}{interactive-complex-read-11}
\marginpar{
	\raggedleft
	\vspace{0.22ex}

	\queryRefCard{interactive-complex-read-01}{IC}{1}\\
	\queryRefCard{interactive-complex-read-02}{IC}{2}\\
	\queryRefCard{interactive-complex-read-03}{IC}{3}\\
	\queryRefCard{interactive-complex-read-04}{IC}{4}\\
	\queryRefCard{interactive-complex-read-05}{IC}{5}\\
	\queryRefCard{interactive-complex-read-06}{IC}{6}\\
	\queryRefCard{interactive-complex-read-07}{IC}{7}\\
	\queryRefCard{interactive-complex-read-08}{IC}{8}\\
	\queryRefCard{interactive-complex-read-09}{IC}{9}\\
	\queryRefCard{interactive-complex-read-10}{IC}{10}\\
	\queryRefCard{interactive-complex-read-11}{IC}{11}\\
	\queryRefCard{interactive-complex-read-12}{IC}{12}\\
	\queryRefCard{interactive-complex-read-13}{IC}{13}\\
	\queryRefCard{interactive-complex-read-14-v1}{IC}{14v1}\\
	\queryRefCard{interactive-complex-read-14-v2}{IC}{14v2}\\
}

\noindent\begin{tabularx}{\queryCardWidth}{|>{\queryPropertyCell}p{\queryPropertyCellWidth}|X|}
	\hline
	query & Interactive / complex / 11 \\ \hline
	title & Job referral \\ \hline
	pattern & \centering \includegraphics[scale=\patternscale,margin=0cm .2cm]{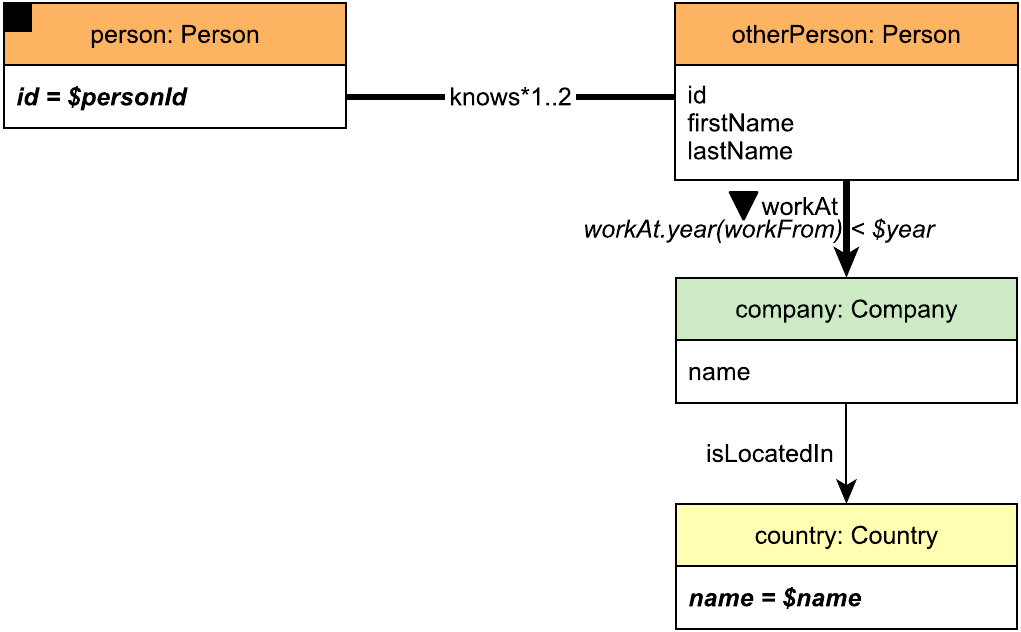} \tabularnewline \hline
	description & Given a start \emph{Person} with ID \texttt{\$personId}, find that
\emph{Person}'s friends and friends of friends (excluding start
\emph{Person}) who started working in some \emph{Company} in a given
\emph{Country} with name \texttt{\$countryName}, before a given date
(\texttt{\$workFromYear}).
 \\ \hline

		params &
		\innerCardVSpace{\begin{tabularx}{\attributeCardWidth}{|>{\paramNumberCell}C{\attributeNumberWidth}|>{\varNameCell}M|>{\typeCell}m{\typeWidth}|Y|} \hline
		$\mathsf{1}$ & \$personId
 & ID
 &  \\ \hline
		$\mathsf{2}$ & \$countryName
 & String
 &  \\ \hline
		$\mathsf{3}$ & \$workFromYear
 & 32-bit Integer
 &  \\ \hline
		\end{tabularx}}\innerCardVSpace \\ \hline

		result &
		\innerCardVSpace{\begin{tabularx}{\attributeCardWidth}{|>{\resultNumberCell}C{\attributeNumberWidth}|>{\varNameCell}M|>{\typeCell}m{\typeWidth}|>{\resultOriginCell}c|Y|} \hline
		$\mathsf{1}$ & otherPerson.id & ID & R &
				 \\ \hline
		$\mathsf{2}$ & otherPerson.firstName & String & R &
				 \\ \hline
		$\mathsf{3}$ & otherPerson.lastName & String & R &
				 \\ \hline
		$\mathsf{4}$ & company.name & String & R &
				 \\ \hline
		$\mathsf{5}$ & workAt.workFrom & 32-bit Integer & R &
				 \\ \hline
		\end{tabularx}}\innerCardVSpace \\ \hline

		sort		&
		\innerCardVSpace{\begin{tabularx}{\attributeCardWidth}{|>{\sortNumberCell}C{\attributeNumberWidth}|>{\varNameCell}M|>{\directionCell}c|Y|} \hline
		$\mathsf{1}$ & workAt.workFrom
 & $\asc
$ &  \\ \hline
		$\mathsf{2}$ & otherPerson.id
 & $\asc
$ &  \\ \hline
		$\mathsf{3}$ & company.name
 & $\desc
$ &  \\ \hline
		\end{tabularx}}\innerCardVSpace \\ \hline
	limit & 10 \\ \hline
	CPs &
	\multicolumn{1}{>{\raggedright}l|}{
		\chokePoint{1.3}, 
		\chokePoint{2.3}, 
		\chokePoint{2.4}, 
		\chokePoint{3.3}, 
		\chokePoint{4.2}
		} \\ \hline
	relevance &
		\footnotesize This query looks for paths of length two or three, starting from a
\emph{Person}, moving to friends or friends of friends, and ending at a
\emph{Company}. In this query, there are selective joins and a top-k
order by that can be exploited for optimizations.
 \\ \hline%
\end{tabularx}
\queryCardVSpace

\let\emph\oldemph
\let\textbf\oldtextbf

\renewcommand{\currentQueryCard}{0}
\renewcommand*{\arraystretch}{1.1}

\subsection*{Interactive / complex / 12}
\label{sec:interactive-complex-read-12}

\let\oldemph\emph
\renewcommand{\emph}[1]{{\footnotesize \sf #1}}
\let\oldtextbf\textbf
\renewcommand{\textbf}[1]{{\it #1}}\renewcommand{\currentQueryCard}{interactive-complex-read-12}
\marginpar{
	\raggedleft
	\vspace{0.22ex}

	\queryRefCard{interactive-complex-read-01}{IC}{1}\\
	\queryRefCard{interactive-complex-read-02}{IC}{2}\\
	\queryRefCard{interactive-complex-read-03}{IC}{3}\\
	\queryRefCard{interactive-complex-read-04}{IC}{4}\\
	\queryRefCard{interactive-complex-read-05}{IC}{5}\\
	\queryRefCard{interactive-complex-read-06}{IC}{6}\\
	\queryRefCard{interactive-complex-read-07}{IC}{7}\\
	\queryRefCard{interactive-complex-read-08}{IC}{8}\\
	\queryRefCard{interactive-complex-read-09}{IC}{9}\\
	\queryRefCard{interactive-complex-read-10}{IC}{10}\\
	\queryRefCard{interactive-complex-read-11}{IC}{11}\\
	\queryRefCard{interactive-complex-read-12}{IC}{12}\\
	\queryRefCard{interactive-complex-read-13}{IC}{13}\\
	\queryRefCard{interactive-complex-read-14-v1}{IC}{14v1}\\
	\queryRefCard{interactive-complex-read-14-v2}{IC}{14v2}\\
}

\noindent\begin{tabularx}{\queryCardWidth}{|>{\queryPropertyCell}p{\queryPropertyCellWidth}|X|}
	\hline
	query & Interactive / complex / 12 \\ \hline
	title & Expert search \\ \hline
	pattern & \centering \includegraphics[scale=\patternscale,margin=0cm .2cm]{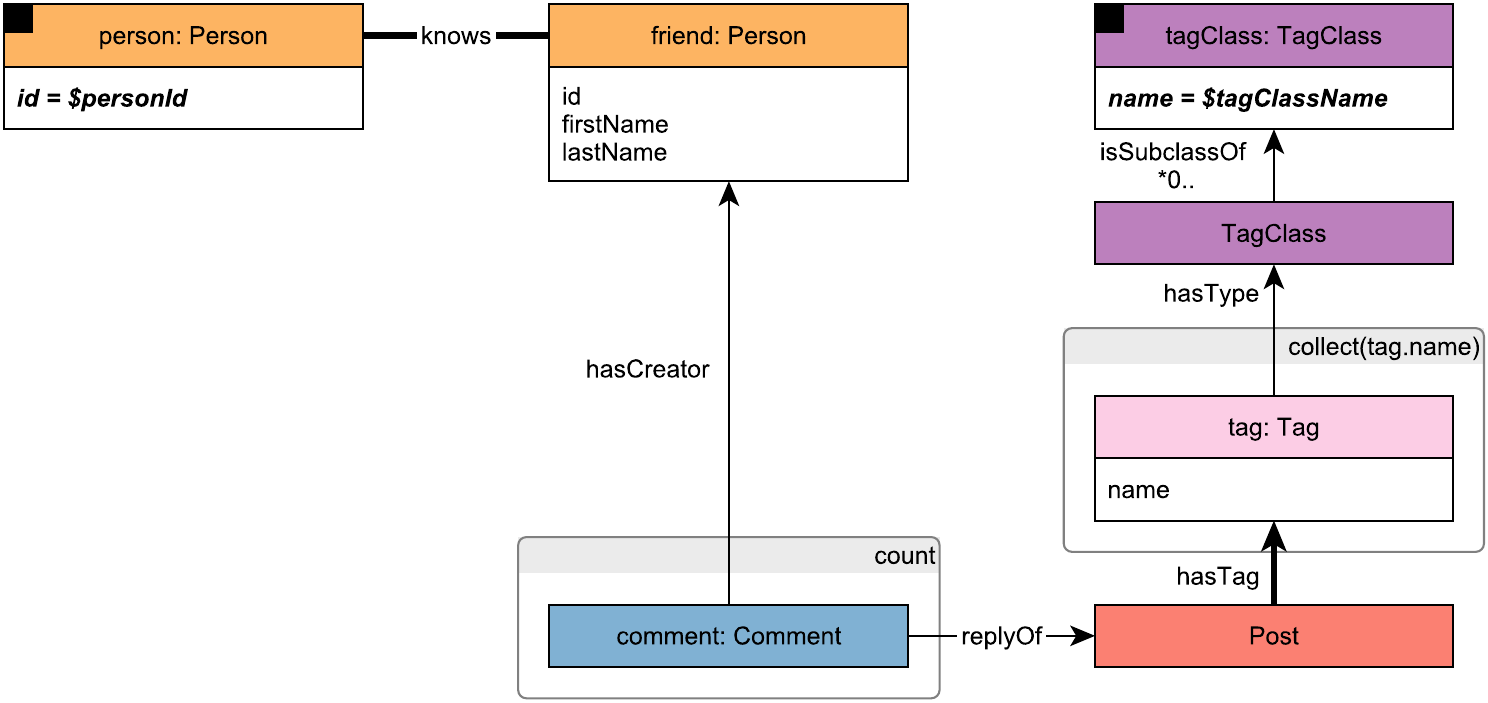} \tabularnewline \hline
	description & Given a start \emph{Person} with ID \texttt{\$personId}, find the
\emph{Comments} that this \emph{Person}'s friends made in reply to
\emph{Posts}, considering only those \emph{Comments} that are direct
(single-hop) replies to \emph{Posts}, not the transitive (multi-hop)
ones. Only consider \emph{Posts} with a \emph{Tag} in a given
\emph{TagClass} with name \texttt{\$tagClassName} or in a descendent of
that \emph{TagClass}. Count the number of these reply \emph{Comments},
and collect the \emph{Tags} that were attached to the \emph{Posts} they
replied to, but only collect \emph{Tags} with the given \emph{TagClass}
or with a descendant of that \emph{TagClass}. Return \emph{Persons} with
at least one reply, the reply count, and the collection of \emph{Tags}.
 \\ \hline

		params &
		\innerCardVSpace{\begin{tabularx}{\attributeCardWidth}{|>{\paramNumberCell}C{\attributeNumberWidth}|>{\varNameCell}M|>{\typeCell}m{\typeWidth}|Y|} \hline
		$\mathsf{1}$ & \$personId
 & ID
 &  \\ \hline
		$\mathsf{2}$ & \$tagClassName
 & Long String
 &  \\ \hline
		\end{tabularx}}\innerCardVSpace \\ \hline

		result &
		\innerCardVSpace{\begin{tabularx}{\attributeCardWidth}{|>{\resultNumberCell}C{\attributeNumberWidth}|>{\varNameCell}M|>{\typeCell}m{\typeWidth}|>{\resultOriginCell}c|Y|} \hline
		$\mathsf{1}$ & friend.id & ID & R &
				 \\ \hline
		$\mathsf{2}$ & friend.firstName & String & R &
				 \\ \hline
		$\mathsf{3}$ & friend.lastName & String & R &
				 \\ \hline
		$\mathsf{4}$ & tagNames & \{Long String\} & A &
				 \\ \hline
		$\mathsf{5}$ & replyCount & 32-bit Integer & A &
				 \\ \hline
		\end{tabularx}}\innerCardVSpace \\ \hline

		sort		&
		\innerCardVSpace{\begin{tabularx}{\attributeCardWidth}{|>{\sortNumberCell}C{\attributeNumberWidth}|>{\varNameCell}M|>{\directionCell}c|Y|} \hline
		$\mathsf{1}$ & replyCount
 & $\desc
$ &  \\ \hline
		$\mathsf{2}$ & friend.id
 & $\asc
$ &  \\ \hline
		\end{tabularx}}\innerCardVSpace \\ \hline
	limit & 20 \\ \hline
	CPs &
	\multicolumn{1}{>{\raggedright}l|}{
		\chokePoint{3.3}, 
		\chokePoint{7.2}, 
		\chokePoint{7.3}, 
		\chokePoint{8.2}
		} \\ \hline
	relevance &
		\footnotesize This query starts at a \emph{Person}, moves to its friends, and the to
their \emph{Comments} and their root \emph{Posts}. Then, it gets the
\emph{Tag} of each \emph{Post} and checks whether it (directly or
transitively) belongs to the specified \emph{TagClass}. This can be
thought of a bidirectional search between the \emph{Person} and the
\emph{TagClass}. The difficulty of this query is determining the optimal
direction of this traversal.
 \\ \hline%
\end{tabularx}
\queryCardVSpace

\let\emph\oldemph
\let\textbf\oldtextbf

\renewcommand{\currentQueryCard}{0}
\renewcommand*{\arraystretch}{1.1}

\subsection*{Interactive / complex / 13}
\label{sec:interactive-complex-read-13}

\let\oldemph\emph
\renewcommand{\emph}[1]{{\footnotesize \sf #1}}
\let\oldtextbf\textbf
\renewcommand{\textbf}[1]{{\it #1}}\renewcommand{\currentQueryCard}{interactive-complex-read-13}
\marginpar{
	\raggedleft
	\vspace{0.22ex}

	\queryRefCard{interactive-complex-read-01}{IC}{1}\\
	\queryRefCard{interactive-complex-read-02}{IC}{2}\\
	\queryRefCard{interactive-complex-read-03}{IC}{3}\\
	\queryRefCard{interactive-complex-read-04}{IC}{4}\\
	\queryRefCard{interactive-complex-read-05}{IC}{5}\\
	\queryRefCard{interactive-complex-read-06}{IC}{6}\\
	\queryRefCard{interactive-complex-read-07}{IC}{7}\\
	\queryRefCard{interactive-complex-read-08}{IC}{8}\\
	\queryRefCard{interactive-complex-read-09}{IC}{9}\\
	\queryRefCard{interactive-complex-read-10}{IC}{10}\\
	\queryRefCard{interactive-complex-read-11}{IC}{11}\\
	\queryRefCard{interactive-complex-read-12}{IC}{12}\\
	\queryRefCard{interactive-complex-read-13}{IC}{13}\\
	\queryRefCard{interactive-complex-read-14-v1}{IC}{14v1}\\
	\queryRefCard{interactive-complex-read-14-v2}{IC}{14v2}\\
}

\noindent\begin{tabularx}{\queryCardWidth}{|>{\queryPropertyCell}p{\queryPropertyCellWidth}|X|}
	\hline
	query & Interactive / complex / 13 \\ \hline
	title & Single shortest path \\ \hline
	pattern & \centering \includegraphics[scale=\patternscale,margin=0cm .2cm]{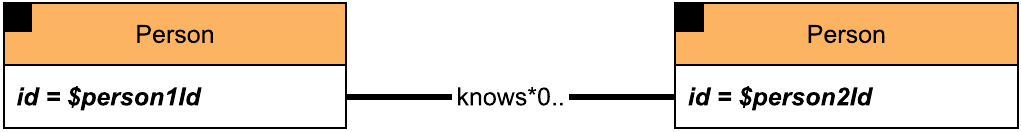} \tabularnewline \hline
	description & Given two \emph{Persons} with IDs \texttt{\$person1Id} and
\texttt{\$person2Id}, find the shortest path between these two
\emph{Persons} in the subgraph induced by the \emph{knows} edges. Return
the length of this path:

\begin{itemize}
\tightlist
\item
  \(-1\): no path found
\item
  \(0\): start person = end person
\item
  \(> 0\): path found (start person \(\neq\) end person)
\end{itemize}
 \\ \hline

		params &
		\innerCardVSpace{\begin{tabularx}{\attributeCardWidth}{|>{\paramNumberCell}C{\attributeNumberWidth}|>{\varNameCell}M|>{\typeCell}m{\typeWidth}|Y|} \hline
		$\mathsf{1}$ & \$person1Id
 & ID
 & In SNB Interactive v2, this query has two variants:

\texttt{(b)} Guaranteed that there is no path between the two
\emph{Persons}

\texttt{(b)} Guaranteed that there is a 4-hop path between the two
\emph{Persons}
 \\ \hline
		$\mathsf{2}$ & \$person2Id
 & ID
 &  \\ \hline
		\end{tabularx}}\innerCardVSpace \\ \hline

		result &
		\innerCardVSpace{\begin{tabularx}{\attributeCardWidth}{|>{\resultNumberCell}C{\attributeNumberWidth}|>{\varNameCell}M|>{\typeCell}m{\typeWidth}|>{\resultOriginCell}c|Y|} \hline
		$\mathsf{1}$ & shortestPathLength & 32-bit Integer & C &
				 \\ \hline
		\end{tabularx}}\innerCardVSpace \\ \hline

	CPs &
	\multicolumn{1}{>{\raggedright}l|}{
		\chokePoint{3.3}, 
		\chokePoint{7.2}, 
		\chokePoint{7.3}, 
		\chokePoint{7.5}, 
		\chokePoint{7.8}, 
		\chokePoint{8.1}, 
		\chokePoint{8.6}
		} \\ \hline
	relevance &
		\footnotesize This query looks for a variable length path, starting at a given
\emph{Person} and finishing at an another given \emph{Person}. Proper
cardinality estimation and search space pruning, will be crucial. This
query also allows for possible parallel implementations.
 \\ \hline%
\end{tabularx}
\queryCardVSpace

\let\emph\oldemph
\let\textbf\oldtextbf

\renewcommand{\currentQueryCard}{0}
\renewcommand*{\arraystretch}{1.1}

\subsection*{Interactive / complex / 14v1}
\label{sec:interactive-complex-read-14-v1}

\let\oldemph\emph
\renewcommand{\emph}[1]{{\footnotesize \sf #1}}
\let\oldtextbf\textbf
\renewcommand{\textbf}[1]{{\it #1}}\renewcommand{\currentQueryCard}{interactive-complex-read-14-v1}
\marginpar{
	\raggedleft
	\vspace{0.22ex}

	\queryRefCard{interactive-complex-read-01}{IC}{1}\\
	\queryRefCard{interactive-complex-read-02}{IC}{2}\\
	\queryRefCard{interactive-complex-read-03}{IC}{3}\\
	\queryRefCard{interactive-complex-read-04}{IC}{4}\\
	\queryRefCard{interactive-complex-read-05}{IC}{5}\\
	\queryRefCard{interactive-complex-read-06}{IC}{6}\\
	\queryRefCard{interactive-complex-read-07}{IC}{7}\\
	\queryRefCard{interactive-complex-read-08}{IC}{8}\\
	\queryRefCard{interactive-complex-read-09}{IC}{9}\\
	\queryRefCard{interactive-complex-read-10}{IC}{10}\\
	\queryRefCard{interactive-complex-read-11}{IC}{11}\\
	\queryRefCard{interactive-complex-read-12}{IC}{12}\\
	\queryRefCard{interactive-complex-read-13}{IC}{13}\\
	\queryRefCard{interactive-complex-read-14-v1}{IC}{14v1}\\
	\queryRefCard{interactive-complex-read-14-v2}{IC}{14v2}\\
}

\noindent\begin{tabularx}{\queryCardWidth}{|>{\queryPropertyCell}p{\queryPropertyCellWidth}|X|}
	\hline
	query & Interactive / complex / 14v1 \\ \hline
	title & Trusted connection paths (v1) \\ \hline
	pattern & \centering \includegraphics[scale=\patternscale,margin=0cm .2cm]{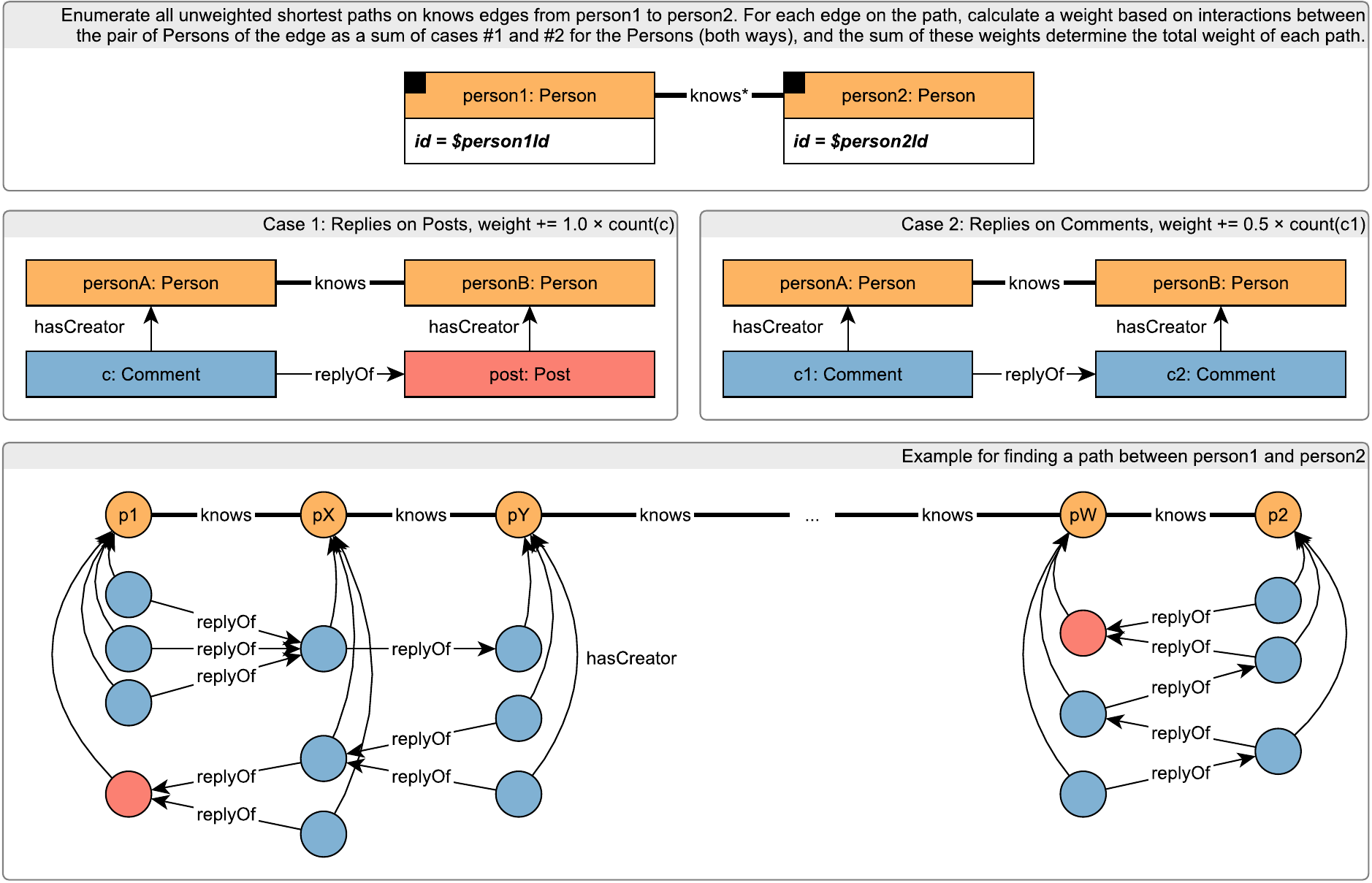} \tabularnewline \hline
	description & \textbf{This query is used in SNB Interactive v1.}

Given two \emph{Persons} with IDs \texttt{\$person1Id} and
\texttt{\$person2Id}, find all (unweighted) shortest paths between these
two \emph{Persons}, in the subgraph induced by the \emph{knows}
relationship.

Then, for each path calculate a weight. The nodes in the path are
\emph{Persons}, and the weight of a path is the sum of weights between
every pair of consecutive \emph{Person} nodes in the path.

The weight for a pair of \emph{Persons} is calculated based on their
interactions:

\begin{itemize}
\tightlist
\item
  Every direct reply (by one of the \emph{Persons}) to a \emph{Post} (by
  the other \emph{Person}) is 1.0.
\item
  Every direct reply (by one of the \emph{Persons}) to a \emph{Comment}
  (by the other \emph{Person}) is 0.5.
\end{itemize}

Note that interactions are counted both ways (e.g.~if Alice writes 2
\emph{Post} replies and 1 \emph{Comment} reply to Bob, while Bob writes
3 \emph{Post} replies and 4 \emph{Comment} replies to Alice, their
interaction score is
\(2 \times 1.0 + 1 \times 0.5 + 3 \times 1.0 + 4 \times 0.5 = 7.5\)).

Return all the paths with shortest length and their weights. Do not
return any rows if there is no path between the two \emph{Persons}.
 \\ \hline

		params &
		\innerCardVSpace{\begin{tabularx}{\attributeCardWidth}{|>{\paramNumberCell}C{\attributeNumberWidth}|>{\varNameCell}M|>{\typeCell}m{\typeWidth}|Y|} \hline
		$\mathsf{1}$ & \$person1Id
 & ID
 &  \\ \hline
		$\mathsf{2}$ & \$person2Id
 & ID
 &  \\ \hline
		\end{tabularx}}\innerCardVSpace \\ \hline

		result &
		\innerCardVSpace{\begin{tabularx}{\attributeCardWidth}{|>{\resultNumberCell}C{\attributeNumberWidth}|>{\varNameCell}M|>{\typeCell}m{\typeWidth}|>{\resultOriginCell}c|Y|} \hline
		$\mathsf{1}$ & personIdsInPath & {[}ID{]} & C &
				Identifiers representing an ordered sequence of the \emph{Persons} in
the path
 \\ \hline
		$\mathsf{2}$ & pathWeight & 64-bit Float & C &
				 \\ \hline
		\end{tabularx}}\innerCardVSpace \\ \hline

		sort		&
		\innerCardVSpace{\begin{tabularx}{\attributeCardWidth}{|>{\sortNumberCell}C{\attributeNumberWidth}|>{\varNameCell}M|>{\directionCell}c|Y|} \hline
		$\mathsf{1}$ & pathWeight
 & $\desc
$ & The order of paths with the same weight is unspecified
 \\ \hline
		\end{tabularx}}\innerCardVSpace \\ \hline
	CPs &
	\multicolumn{1}{>{\raggedright}l|}{
		\chokePoint{3.3}, 
		\chokePoint{5.3}, 
		\chokePoint{7.2}, 
		\chokePoint{7.3}, 
		\chokePoint{7.5}, 
		\chokePoint{7.7}, 
		\chokePoint{8.1}, 
		\chokePoint{8.2}, 
		\chokePoint{8.3}, 
		\chokePoint{8.6}
		} \\ \hline
	relevance &
		\footnotesize This query looks for a variable length path, starting at a given
\emph{Person} and finishing at an another given \emph{Person}. This is a
more complex query as it not only requires computing the path length,
but returning it and computing a weight. To compute this weight one must
look for smaller sub-queries with paths of length three, formed by the
two \emph{Persons} at each step, a \emph{Post} and a \emph{Comment}.
 \\ \hline%
\end{tabularx}
\queryCardVSpace

\let\emph\oldemph
\let\textbf\oldtextbf

\renewcommand{\currentQueryCard}{0}


\section{Short Reads}
\label{sec:interactive-v1-short-reads}

\renewcommand*{\arraystretch}{1.1}

\subsection*{Interactive / short / 1}
\label{sec:interactive-short-read-01}

\let\oldemph\emph
\renewcommand{\emph}[1]{{\footnotesize \sf #1}}
\let\oldtextbf\textbf
\renewcommand{\textbf}[1]{{\it #1}}\renewcommand{\currentQueryCard}{interactive-short-read-01}
\marginpar{
	\raggedleft
	\vspace{0.22ex}

	\queryRefCard{interactive-short-read-01}{IS}{1}\\
	\queryRefCard{interactive-short-read-02}{IS}{2}\\
	\queryRefCard{interactive-short-read-03}{IS}{3}\\
	\queryRefCard{interactive-short-read-04}{IS}{4}\\
	\queryRefCard{interactive-short-read-05}{IS}{5}\\
	\queryRefCard{interactive-short-read-06}{IS}{6}\\
	\queryRefCard{interactive-short-read-07}{IS}{7}\\
}

\noindent\begin{tabularx}{\queryCardWidth}{|>{\queryPropertyCell}p{\queryPropertyCellWidth}|X|}
	\hline
	query & Interactive / short / 1 \\ \hline
	title & Profile of a person \\ \hline
	pattern & \centering \includegraphics[scale=\patternscale,margin=0cm .2cm]{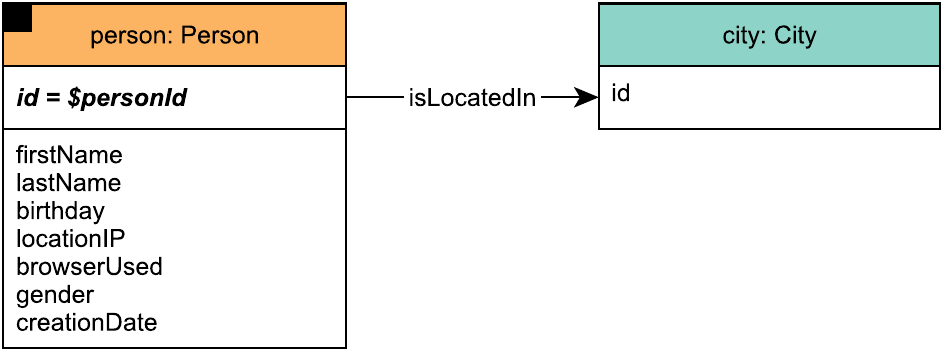} \tabularnewline \hline
	description & Given a start \emph{Person} with ID \texttt{\$personId}, retrieve their
first name, last name, birthday, IP address, browser, and city of
residence.
 \\ \hline

		params &
		\innerCardVSpace{\begin{tabularx}{\attributeCardWidth}{|>{\paramNumberCell}C{\attributeNumberWidth}|>{\varNameCell}M|>{\typeCell}m{\typeWidth}|Y|} \hline
		$\mathsf{1}$ & \$personId
 & ID
 &  \\ \hline
		\end{tabularx}}\innerCardVSpace \\ \hline

		result &
		\innerCardVSpace{\begin{tabularx}{\attributeCardWidth}{|>{\resultNumberCell}C{\attributeNumberWidth}|>{\varNameCell}M|>{\typeCell}m{\typeWidth}|>{\resultOriginCell}c|Y|} \hline
		$\mathsf{1}$ & person.firstName & String & R &
				 \\ \hline
		$\mathsf{2}$ & person.lastName & String & R &
				 \\ \hline
		$\mathsf{3}$ & person.birthday & Date & R &
				 \\ \hline
		$\mathsf{4}$ & person.locationIP & String & R &
				 \\ \hline
		$\mathsf{5}$ & person.browserUsed & String & R &
				 \\ \hline
		$\mathsf{6}$ & city.id & ID & R &
				 \\ \hline
		$\mathsf{7}$ & person.gender & String & R &
				 \\ \hline
		$\mathsf{8}$ & person.creationDate & DateTime & R &
				 \\ \hline
		\end{tabularx}}\innerCardVSpace \\ \hline

\end{tabularx}
\queryCardVSpace

\let\emph\oldemph
\let\textbf\oldtextbf

\renewcommand{\currentQueryCard}{0}
\renewcommand*{\arraystretch}{1.1}

\subsection*{Interactive / short / 2}
\label{sec:interactive-short-read-02}

\let\oldemph\emph
\renewcommand{\emph}[1]{{\footnotesize \sf #1}}
\let\oldtextbf\textbf
\renewcommand{\textbf}[1]{{\it #1}}\renewcommand{\currentQueryCard}{interactive-short-read-02}
\marginpar{
	\raggedleft
	\vspace{0.22ex}

	\queryRefCard{interactive-short-read-01}{IS}{1}\\
	\queryRefCard{interactive-short-read-02}{IS}{2}\\
	\queryRefCard{interactive-short-read-03}{IS}{3}\\
	\queryRefCard{interactive-short-read-04}{IS}{4}\\
	\queryRefCard{interactive-short-read-05}{IS}{5}\\
	\queryRefCard{interactive-short-read-06}{IS}{6}\\
	\queryRefCard{interactive-short-read-07}{IS}{7}\\
}

\noindent\begin{tabularx}{\queryCardWidth}{|>{\queryPropertyCell}p{\queryPropertyCellWidth}|X|}
	\hline
	query & Interactive / short / 2 \\ \hline
	title & Recent messages of a person \\ \hline
	pattern & \centering \includegraphics[scale=\patternscale,margin=0cm .2cm]{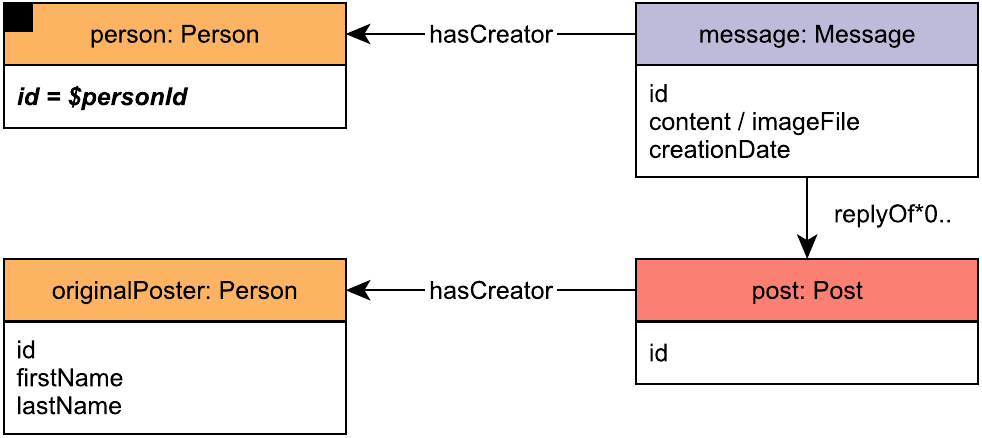} \tabularnewline \hline
	description & Given a start \emph{Person} with ID \texttt{\$personId}, retrieve the
last 10 \emph{Messages} created by that user. For each \emph{Message},
return that \emph{Message}, the original \emph{Post} in its conversation
(\texttt{post}), and the author of that \emph{Post}
(\texttt{originalPoster}). If any of the \emph{Messages} is a
\emph{Post}, then the original \emph{Post} (\texttt{post}) will be the
same \emph{Message}, i.e.~that \emph{Message} will appear twice in that
result.
 \\ \hline

		params &
		\innerCardVSpace{\begin{tabularx}{\attributeCardWidth}{|>{\paramNumberCell}C{\attributeNumberWidth}|>{\varNameCell}M|>{\typeCell}m{\typeWidth}|Y|} \hline
		$\mathsf{1}$ & \$personId
 & ID
 &  \\ \hline
		\end{tabularx}}\innerCardVSpace \\ \hline

		result &
		\innerCardVSpace{\begin{tabularx}{\attributeCardWidth}{|>{\resultNumberCell}C{\attributeNumberWidth}|>{\varNameCell}M|>{\typeCell}m{\typeWidth}|>{\resultOriginCell}c|Y|} \hline
		$\mathsf{1}$ & message.id & ID & R &
				 \\ \hline
		$\mathsf{2}$ & message.content or message.imageFile (for photos) & Text & R &
				 \\ \hline
		$\mathsf{3}$ & message.creationDate & DateTime & R &
				 \\ \hline
		$\mathsf{4}$ & post.id & ID & R &
				 \\ \hline
		$\mathsf{5}$ & originalPoster.id & ID & R &
				 \\ \hline
		$\mathsf{6}$ & originalPoster.firstName & String & R &
				 \\ \hline
		$\mathsf{7}$ & originalPoster.lastName & String & R &
				 \\ \hline
		\end{tabularx}}\innerCardVSpace \\ \hline

		sort		&
		\innerCardVSpace{\begin{tabularx}{\attributeCardWidth}{|>{\sortNumberCell}C{\attributeNumberWidth}|>{\varNameCell}M|>{\directionCell}c|Y|} \hline
		$\mathsf{1}$ & message.creationDate
 & $\desc
$ &  \\ \hline
		$\mathsf{2}$ & message.id
 & $\desc
$ &  \\ \hline
		\end{tabularx}}\innerCardVSpace \\ \hline
	limit & 10 \\ \hline
\end{tabularx}
\queryCardVSpace

\let\emph\oldemph
\let\textbf\oldtextbf

\renewcommand{\currentQueryCard}{0}
\renewcommand*{\arraystretch}{1.1}

\subsection*{Interactive / short / 3}
\label{sec:interactive-short-read-03}

\let\oldemph\emph
\renewcommand{\emph}[1]{{\footnotesize \sf #1}}
\let\oldtextbf\textbf
\renewcommand{\textbf}[1]{{\it #1}}\renewcommand{\currentQueryCard}{interactive-short-read-03}
\marginpar{
	\raggedleft
	\vspace{0.22ex}

	\queryRefCard{interactive-short-read-01}{IS}{1}\\
	\queryRefCard{interactive-short-read-02}{IS}{2}\\
	\queryRefCard{interactive-short-read-03}{IS}{3}\\
	\queryRefCard{interactive-short-read-04}{IS}{4}\\
	\queryRefCard{interactive-short-read-05}{IS}{5}\\
	\queryRefCard{interactive-short-read-06}{IS}{6}\\
	\queryRefCard{interactive-short-read-07}{IS}{7}\\
}

\noindent\begin{tabularx}{\queryCardWidth}{|>{\queryPropertyCell}p{\queryPropertyCellWidth}|X|}
	\hline
	query & Interactive / short / 3 \\ \hline
	title & Friends of a person \\ \hline
	pattern & \centering \includegraphics[scale=\patternscale,margin=0cm .2cm]{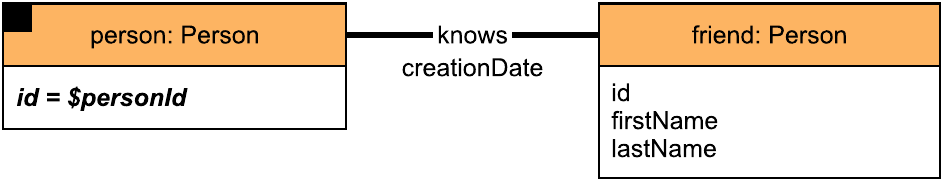} \tabularnewline \hline
	description & Given a start \emph{Person} with ID \texttt{\$personId}, retrieve all of
their friends, and the date at which they became friends.
 \\ \hline

		params &
		\innerCardVSpace{\begin{tabularx}{\attributeCardWidth}{|>{\paramNumberCell}C{\attributeNumberWidth}|>{\varNameCell}M|>{\typeCell}m{\typeWidth}|Y|} \hline
		$\mathsf{1}$ & \$personId
 & ID
 &  \\ \hline
		\end{tabularx}}\innerCardVSpace \\ \hline

		result &
		\innerCardVSpace{\begin{tabularx}{\attributeCardWidth}{|>{\resultNumberCell}C{\attributeNumberWidth}|>{\varNameCell}M|>{\typeCell}m{\typeWidth}|>{\resultOriginCell}c|Y|} \hline
		$\mathsf{1}$ & friend.id & ID & R &
				 \\ \hline
		$\mathsf{2}$ & friend.firstName & String & R &
				 \\ \hline
		$\mathsf{3}$ & friend.lastName & String & R &
				 \\ \hline
		$\mathsf{4}$ & knows.creationDate & DateTime & R &
				 \\ \hline
		\end{tabularx}}\innerCardVSpace \\ \hline

		sort		&
		\innerCardVSpace{\begin{tabularx}{\attributeCardWidth}{|>{\sortNumberCell}C{\attributeNumberWidth}|>{\varNameCell}M|>{\directionCell}c|Y|} \hline
		$\mathsf{1}$ & knows.creationDate
 & $\desc
$ &  \\ \hline
		$\mathsf{2}$ & friend.id
 & $\asc
$ &  \\ \hline
		\end{tabularx}}\innerCardVSpace \\ \hline
\end{tabularx}
\queryCardVSpace

\let\emph\oldemph
\let\textbf\oldtextbf

\renewcommand{\currentQueryCard}{0}
\renewcommand*{\arraystretch}{1.1}

\subsection*{Interactive / short / 4}
\label{sec:interactive-short-read-04}

\let\oldemph\emph
\renewcommand{\emph}[1]{{\footnotesize \sf #1}}
\let\oldtextbf\textbf
\renewcommand{\textbf}[1]{{\it #1}}\renewcommand{\currentQueryCard}{interactive-short-read-04}
\marginpar{
	\raggedleft
	\vspace{0.22ex}

	\queryRefCard{interactive-short-read-01}{IS}{1}\\
	\queryRefCard{interactive-short-read-02}{IS}{2}\\
	\queryRefCard{interactive-short-read-03}{IS}{3}\\
	\queryRefCard{interactive-short-read-04}{IS}{4}\\
	\queryRefCard{interactive-short-read-05}{IS}{5}\\
	\queryRefCard{interactive-short-read-06}{IS}{6}\\
	\queryRefCard{interactive-short-read-07}{IS}{7}\\
}

\noindent\begin{tabularx}{\queryCardWidth}{|>{\queryPropertyCell}p{\queryPropertyCellWidth}|X|}
	\hline
	query & Interactive / short / 4 \\ \hline
	title & Content of a message \\ \hline
	pattern & \centering \includegraphics[scale=\patternscale,margin=0cm .2cm]{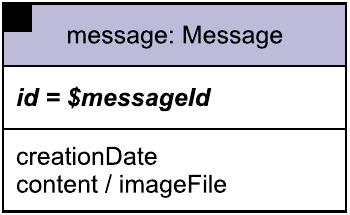} \tabularnewline \hline
	description & Given a \emph{Message} with ID \texttt{\$messageId}, retrieve its
content and creation date.
 \\ \hline

		params &
		\innerCardVSpace{\begin{tabularx}{\attributeCardWidth}{|>{\paramNumberCell}C{\attributeNumberWidth}|>{\varNameCell}M|>{\typeCell}m{\typeWidth}|Y|} \hline
		$\mathsf{1}$ & \$messageId
 & ID
 &  \\ \hline
		\end{tabularx}}\innerCardVSpace \\ \hline

		result &
		\innerCardVSpace{\begin{tabularx}{\attributeCardWidth}{|>{\resultNumberCell}C{\attributeNumberWidth}|>{\varNameCell}M|>{\typeCell}m{\typeWidth}|>{\resultOriginCell}c|Y|} \hline
		$\mathsf{1}$ & message.creationDate & DateTime & R &
				\texttt{messageCreationDate}
 \\ \hline
		$\mathsf{2}$ & message.content or message.imageFile (for photos) & Text & R &
				\texttt{messageContent}
 \\ \hline
		\end{tabularx}}\innerCardVSpace \\ \hline

\end{tabularx}
\queryCardVSpace

\let\emph\oldemph
\let\textbf\oldtextbf

\renewcommand{\currentQueryCard}{0}
\renewcommand*{\arraystretch}{1.1}

\subsection*{Interactive / short / 5}
\label{sec:interactive-short-read-05}

\let\oldemph\emph
\renewcommand{\emph}[1]{{\footnotesize \sf #1}}
\let\oldtextbf\textbf
\renewcommand{\textbf}[1]{{\it #1}}\renewcommand{\currentQueryCard}{interactive-short-read-05}
\marginpar{
	\raggedleft
	\vspace{0.22ex}

	\queryRefCard{interactive-short-read-01}{IS}{1}\\
	\queryRefCard{interactive-short-read-02}{IS}{2}\\
	\queryRefCard{interactive-short-read-03}{IS}{3}\\
	\queryRefCard{interactive-short-read-04}{IS}{4}\\
	\queryRefCard{interactive-short-read-05}{IS}{5}\\
	\queryRefCard{interactive-short-read-06}{IS}{6}\\
	\queryRefCard{interactive-short-read-07}{IS}{7}\\
}

\noindent\begin{tabularx}{\queryCardWidth}{|>{\queryPropertyCell}p{\queryPropertyCellWidth}|X|}
	\hline
	query & Interactive / short / 5 \\ \hline
	title & Creator of a message \\ \hline
	pattern & \centering \includegraphics[scale=\patternscale,margin=0cm .2cm]{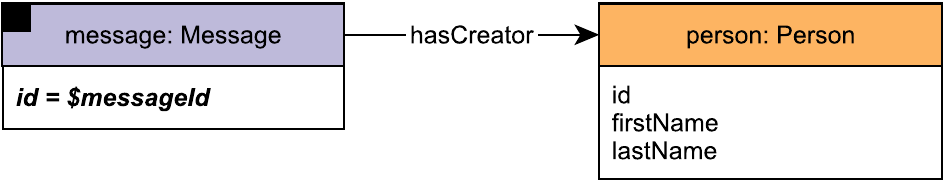} \tabularnewline \hline
	description & Given a \emph{Message} with ID \texttt{\$messageId}, retrieve its
author.
 \\ \hline

		params &
		\innerCardVSpace{\begin{tabularx}{\attributeCardWidth}{|>{\paramNumberCell}C{\attributeNumberWidth}|>{\varNameCell}M|>{\typeCell}m{\typeWidth}|Y|} \hline
		$\mathsf{1}$ & \$messageId
 & ID
 &  \\ \hline
		\end{tabularx}}\innerCardVSpace \\ \hline

		result &
		\innerCardVSpace{\begin{tabularx}{\attributeCardWidth}{|>{\resultNumberCell}C{\attributeNumberWidth}|>{\varNameCell}M|>{\typeCell}m{\typeWidth}|>{\resultOriginCell}c|Y|} \hline
		$\mathsf{1}$ & person.id & ID & R &
				 \\ \hline
		$\mathsf{2}$ & person.firstName & String & R &
				 \\ \hline
		$\mathsf{3}$ & person.lastName & String & R &
				 \\ \hline
		\end{tabularx}}\innerCardVSpace \\ \hline

\end{tabularx}
\queryCardVSpace

\let\emph\oldemph
\let\textbf\oldtextbf

\renewcommand{\currentQueryCard}{0}
\renewcommand*{\arraystretch}{1.1}

\subsection*{Interactive / short / 6}
\label{sec:interactive-short-read-06}

\let\oldemph\emph
\renewcommand{\emph}[1]{{\footnotesize \sf #1}}
\let\oldtextbf\textbf
\renewcommand{\textbf}[1]{{\it #1}}\renewcommand{\currentQueryCard}{interactive-short-read-06}
\marginpar{
	\raggedleft
	\vspace{0.22ex}

	\queryRefCard{interactive-short-read-01}{IS}{1}\\
	\queryRefCard{interactive-short-read-02}{IS}{2}\\
	\queryRefCard{interactive-short-read-03}{IS}{3}\\
	\queryRefCard{interactive-short-read-04}{IS}{4}\\
	\queryRefCard{interactive-short-read-05}{IS}{5}\\
	\queryRefCard{interactive-short-read-06}{IS}{6}\\
	\queryRefCard{interactive-short-read-07}{IS}{7}\\
}

\noindent\begin{tabularx}{\queryCardWidth}{|>{\queryPropertyCell}p{\queryPropertyCellWidth}|X|}
	\hline
	query & Interactive / short / 6 \\ \hline
	title & Forum of a message \\ \hline
	pattern & \centering \includegraphics[scale=\patternscale,margin=0cm .2cm]{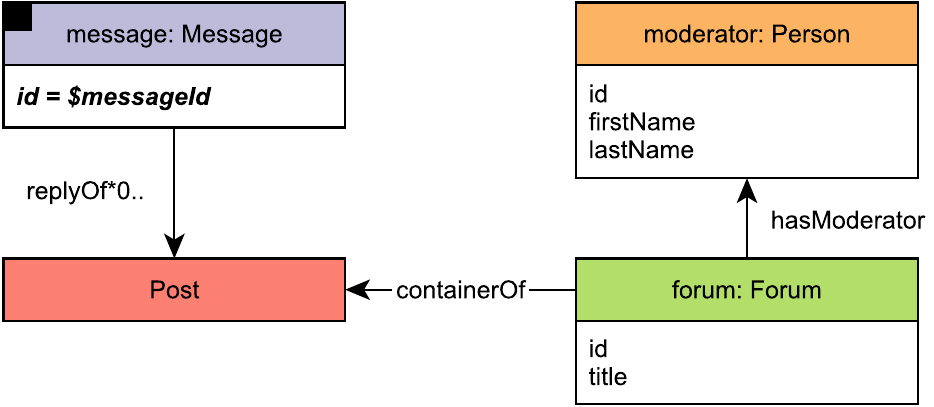} \tabularnewline \hline
	description & Given a \emph{Message} with ID \texttt{\$messageId}, retrieve the
\emph{Forum} that contains it and the \emph{Person} that moderates that
\emph{Forum}. Since \emph{Comments} are not directly contained in
\emph{Forums}, for \emph{Comments}, return the \emph{Forum} containing
the original \emph{Post} in the thread which the \emph{Comment} is
replying to.
 \\ \hline

		params &
		\innerCardVSpace{\begin{tabularx}{\attributeCardWidth}{|>{\paramNumberCell}C{\attributeNumberWidth}|>{\varNameCell}M|>{\typeCell}m{\typeWidth}|Y|} \hline
		$\mathsf{1}$ & \$messageId
 & ID
 &  \\ \hline
		\end{tabularx}}\innerCardVSpace \\ \hline

		result &
		\innerCardVSpace{\begin{tabularx}{\attributeCardWidth}{|>{\resultNumberCell}C{\attributeNumberWidth}|>{\varNameCell}M|>{\typeCell}m{\typeWidth}|>{\resultOriginCell}c|Y|} \hline
		$\mathsf{1}$ & forum.id & ID & R &
				 \\ \hline
		$\mathsf{2}$ & forum.title & Long String & R &
				 \\ \hline
		$\mathsf{3}$ & moderator.id & ID & R &
				 \\ \hline
		$\mathsf{4}$ & moderator.firstName & String & R &
				 \\ \hline
		$\mathsf{5}$ & moderator.lastName & String & R &
				 \\ \hline
		\end{tabularx}}\innerCardVSpace \\ \hline

\end{tabularx}
\queryCardVSpace

\let\emph\oldemph
\let\textbf\oldtextbf

\renewcommand{\currentQueryCard}{0}
\renewcommand*{\arraystretch}{1.1}

\subsection*{Interactive / short / 7}
\label{sec:interactive-short-read-07}

\let\oldemph\emph
\renewcommand{\emph}[1]{{\footnotesize \sf #1}}
\let\oldtextbf\textbf
\renewcommand{\textbf}[1]{{\it #1}}\renewcommand{\currentQueryCard}{interactive-short-read-07}
\marginpar{
	\raggedleft
	\vspace{0.22ex}

	\queryRefCard{interactive-short-read-01}{IS}{1}\\
	\queryRefCard{interactive-short-read-02}{IS}{2}\\
	\queryRefCard{interactive-short-read-03}{IS}{3}\\
	\queryRefCard{interactive-short-read-04}{IS}{4}\\
	\queryRefCard{interactive-short-read-05}{IS}{5}\\
	\queryRefCard{interactive-short-read-06}{IS}{6}\\
	\queryRefCard{interactive-short-read-07}{IS}{7}\\
}

\noindent\begin{tabularx}{\queryCardWidth}{|>{\queryPropertyCell}p{\queryPropertyCellWidth}|X|}
	\hline
	query & Interactive / short / 7 \\ \hline
	title & Replies of a message \\ \hline
	pattern & \centering \includegraphics[scale=\patternscale,margin=0cm .2cm]{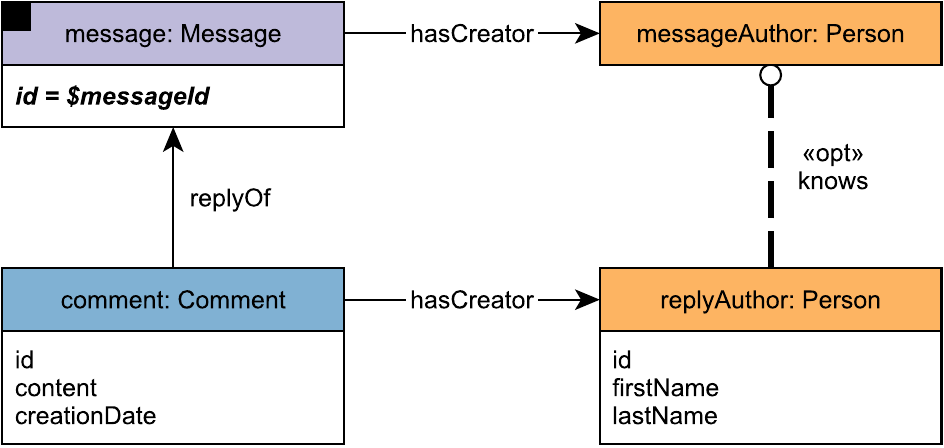} \tabularnewline \hline
	description & Given a \emph{Message} with ID \texttt{\$messageId}, retrieve the
(1-hop) \emph{Comments} that reply to it.

In addition, return a boolean flag \texttt{knows} indicating if the
author of the reply (\texttt{replyAuthor}) knows the author of the
original message (\texttt{messageAuthor}). If author is same as original
author, return \texttt{False} for \texttt{knows} flag.
 \\ \hline

		params &
		\innerCardVSpace{\begin{tabularx}{\attributeCardWidth}{|>{\paramNumberCell}C{\attributeNumberWidth}|>{\varNameCell}M|>{\typeCell}m{\typeWidth}|Y|} \hline
		$\mathsf{1}$ & \$messageId
 & ID
 &  \\ \hline
		\end{tabularx}}\innerCardVSpace \\ \hline

		result &
		\innerCardVSpace{\begin{tabularx}{\attributeCardWidth}{|>{\resultNumberCell}C{\attributeNumberWidth}|>{\varNameCell}M|>{\typeCell}m{\typeWidth}|>{\resultOriginCell}c|Y|} \hline
		$\mathsf{1}$ & comment.id & ID & R &
				 \\ \hline
		$\mathsf{2}$ & comment.content & Text & R &
				 \\ \hline
		$\mathsf{3}$ & comment.creationDate & DateTime & R &
				 \\ \hline
		$\mathsf{4}$ & replyAuthor.id & ID & R &
				 \\ \hline
		$\mathsf{5}$ & replyAuthor.firstName & String & R &
				 \\ \hline
		$\mathsf{6}$ & replyAuthor.lastName & String & R &
				 \\ \hline
		$\mathsf{7}$ & knows & Boolean & C &
				\texttt{True} if the \emph{knows} edge exists between the
\texttt{replyAuthor} and the \texttt{messageAuthor} nodes,
\texttt{False} otherwise (including the case when the two nodes are the
same)
 \\ \hline
		\end{tabularx}}\innerCardVSpace \\ \hline

		sort		&
		\innerCardVSpace{\begin{tabularx}{\attributeCardWidth}{|>{\sortNumberCell}C{\attributeNumberWidth}|>{\varNameCell}M|>{\directionCell}c|Y|} \hline
		$\mathsf{1}$ & comment.creationDate
 & $\desc
$ &  \\ \hline
		$\mathsf{2}$ & replyAuthor.id
 & $\asc
$ &  \\ \hline
		\end{tabularx}}\innerCardVSpace \\ \hline
\end{tabularx}
\queryCardVSpace

\let\emph\oldemph
\let\textbf\oldtextbf

\renewcommand{\currentQueryCard}{0}


\iftoggle{StandaloneWorkloadSpecification}{
    \section{Insert Operations}
    \label{sec:insert-operations}
    
}


\section{Workload Definition}
\label{sec:interactive-workload-definition}

The \emph{Test Driver} is in charge of the execution of the Interactive Workload.
At the beginning of the execution, the Test Driver creates a query mix by
assigning to each query instance, a query issue time and a set of parameters
taken from the generated substitution parameter set described above.  

Query issue times have to be carefully assigned. Although substitution
parameters are chosen in such a way that queries of the same type take similar
time, not all query types have the same complexity and touch the same amount of
data, which causes them to scale differently for the different scale factors.
Therefore, if all query instances, regardless of their type, are issued
at the same rate, those more complex queries will dominate the execution's
result, making faster query types purposeless. To avoid this situation, each
query type is executed at a different rate. The way the execution rate is decided,
also depends on the nature of the query: complex read, short read or update.

Update queries' issue times are taken from the update streams generated by the
data generator. These are the times where the actual event happened during the
simulation of the social network. Complex reads' times are expressed in terms
of update operations. For each complex read query type, a frequency value is
assigned which specifies the relation between the number of updates performed
per complex read. \autoref{table:freqs} shows the frequencies for each complex query and SF used in the \interactivevone workload (\autoref{sec:interactive-v1}).

\begin{table}[htb]
    \centering
    \begin{tabular}{|r|r|r|r|r|r|r|r|}
        \hline
        \textbf{Query} & \textbf{SF1} & \textbf{SF3} & \textbf{SF10} & \textbf{SF30} & \textbf{SF100} & \textbf{SF300} & \textbf{SF\numprint{1000}} \\
        \hline
        \hline
        1              & 26           & 26           & 26            & 26            & 26             & 26             & 26                         \\
        \hline
        2              & 37           & 37           & 37            & 37            & 37             & 37             & 37                         \\
        \hline
        3              & 69           & 79           & 92            & 106           & 123            & 142            & 165                        \\
        \hline
        4              & 36           & 36           & 36            & 36            & 36             & 36             & 36                         \\
        \hline
        5              & 57           & 61           & 66            & 72            & 78             & 84             & 91                         \\
        \hline
        6              & 129          & 172          & 236           & 316           & 434            & 580            & 796                        \\
        \hline
        7              & 87           & 72           & 54            & 48            & 38             & 32             & 25                         \\
        \hline
        8              & 45           & 27           & 15            & 9             & 5              & 3              & 1                          \\
        \hline
        9              & 157          & 209          & 287           & 384           & 527            & 705            & 967                        \\
        \hline
        10             & 30           & 32           & 35            & 37            & 40             & 44             & 47                         \\
        \hline
        11             & 16           & 17           & 19            & 20            & 22             & 24             & 26                         \\
        \hline
        12             & 44           & 44           & 44            & 44            & 44             & 44             & 44                         \\
        \hline
        13             & 19           & 19           & 19            & 19            & 19             & 19             & 19                         \\
        \hline
        14             & 49           & 49           & 49            & 49            & 49             & 49             & 49                         \\
        \hline
    \end{tabular}
    \caption{Frequencies for each Interactive complex query and SF.}
    \label{table:freqs}
\end{table}

Finally, short reads are inserted in order to balance the ratio between reads
and writes, and to simulate the behavior of a real user of the social network.
For each complex read instance, a sequence of short reads is planned. There are two
types of short read sequences: Person centric and Message centric. Depending on
the type of the complex read, one of them is chosen. Each sequence consists of
a set of short reads which are issued in a row. The issue time assigned to each
short read in the sequence is determined at run time, and is based on the
completion time of the complex read it depends on. 
The substitution parameters for short reads are taken from the results of previously
executed queries, including both complex and short reads:

\begin{itemize}
\item Complex reads:
    \queryRefCard{interactive-complex-read-01}{IC}{1}
    \queryRefCard{interactive-complex-read-02}{IC}{2}
    \queryRefCard{interactive-complex-read-03}{IC}{3}
    \queryRefCard{interactive-complex-read-07}{IC}{7}
    \queryRefCard{interactive-complex-read-08}{IC}{8}
    \queryRefCard{interactive-complex-read-09}{IC}{9}
    \queryRefCard{interactive-complex-read-10}{IC}{10}
    \queryRefCard{interactive-complex-read-11}{IC}{11}
    \queryRefCard{interactive-complex-read-12}{IC}{12}
    \queryRefCard{interactive-complex-read-14-v1}{IC}{14v1}
    \queryRefCard{interactive-complex-read-14-v2}{IC}{14v2}
\item Short reads:
    \queryRefCard{interactive-short-read-02}{IS}{2}
    \queryRefCard{interactive-short-read-03}{IS}{3}
    \queryRefCard{interactive-short-read-05}{IS}{5}
    \queryRefCard{interactive-short-read-06}{IS}{6}
    \queryRefCard{interactive-short-read-07}{IS}{7}
\end{itemize}

To see which short and complex queries can potentially trigger additional short query queries, see \autoref{tab:short-read-queries-triggered}.

Once a short read sequence is issued (and provided that sufficient substitution parameters 
exist), there is a probability that another short read sequence is issued. This probability decreases for each new sequence issued.%
\footnote{The probability can be adjusted using the \texttt{ldbc.snb.interactive.short\_read\_dissipation} configuration option.}
Since the same random number generator seed is used across
executions, the workload is deterministic.

\begin{table}[htbp]
    \centering
    \begin{tabular}{|l|c|c|c|c|c|c|c|}
        \hline
        \bf                                                        &
        \bf \queryRefCard{interactive-short-read-01}{IS}{1}        &                                                %
        \bf \queryRefCard{interactive-short-read-02}{IS}{2}        &                                                %
        \bf \queryRefCard{interactive-short-read-03}{IS}{3}        &                                                %
        \bf \queryRefCard{interactive-short-read-04}{IS}{4}        &                                                %
        \bf \queryRefCard{interactive-short-read-05}{IS}{5}        &                                                %
        \bf \queryRefCard{interactive-short-read-06}{IS}{6}        &                                                %
        \bf \queryRefCard{interactive-short-read-07}{IS}{7}                                                         \\ \hline
        \bf \queryRefCard{interactive-complex-read-01}{IC}{1}      & \yes & \yes & \yes &      &      &      &      \\ \hline
        \bf \queryRefCard{interactive-complex-read-02}{IC}{2}      & \yes & \yes & \yes & \yes & \yes & \yes & \yes \\ \hline
        \bf \queryRefCard{interactive-complex-read-03}{IC}{3}      & \yes & \yes & \yes &      &      &      &      \\ \hline
        \bf \queryRefCard{interactive-complex-read-07}{IC}{7}      & \yes & \yes & \yes & \yes & \yes & \yes & \yes \\ \hline
        \bf \queryRefCard{interactive-complex-read-08}{IC}{8}      & \yes & \yes & \yes & \yes & \yes & \yes & \yes \\ \hline
        \bf \queryRefCard{interactive-complex-read-09}{IC}{9}      & \yes & \yes & \yes & \yes & \yes & \yes & \yes \\ \hline
        \bf \queryRefCard{interactive-complex-read-10}{IC}{10}     & \yes & \yes & \yes &      &      &      &      \\ \hline
        \bf \queryRefCard{interactive-complex-read-11}{IC}{11}     & \yes & \yes & \yes &      &      &      &      \\ \hline
        \bf \queryRefCard{interactive-complex-read-12}{IC}{12}     & \yes & \yes & \yes &      &      &      &      \\ \hline
        \bf \queryRefCard{interactive-complex-read-14-v1}{IC}{14} & \yes & \yes & \yes &      &      &      &      \\ \hline
        \bf \queryRefCard{interactive-short-read-02}{IS}{2}         & \yes & \yes & \yes & \yes & \yes & \yes & \yes \\ \hline
        \bf \queryRefCard{interactive-short-read-03}{IS}{3}         & \yes & \yes & \yes &      &      &      &      \\ \hline
        \bf \queryRefCard{interactive-short-read-05}{IS}{5}         & \yes & \yes & \yes &      &      &      &      \\ \hline
        \bf \queryRefCard{interactive-short-read-06}{IS}{6}         & \yes & \yes & \yes &      &      &      &      \\ \hline
        \bf \queryRefCard{interactive-short-read-07}{IS}{7}         & \yes & \yes & \yes & \yes & \yes & \yes & \yes \\ \hline
    \end{tabular}
    \caption{Short read queries (columns) potentially triggered after given complex/short read queries (rows).}
    \label{tab:short-read-queries-triggered}
\end{table}

The specified frequencies, implicitly define the query ratios between queries
of different types, as well as a default target throughput. However, the Test
Sponsor may specify a different target throughput to test,  by ``squeezing''
together or ``stretching'' apart the queries of the workload. This is
achieved by means of the ``Time Compression Ratio'' that is multiplied by the
frequencies (see \autoref{table:freqs}).  Therefore, different
throughputs can be tested while maintaining the relative ratios between the
different query types.

\textbf{Warning.} Note that in the current implementation of SNB Interactive v1, short queries are only produced if updates are enabled. In the absence of updates, no short queries will be executed.

\chapter{Interactive v2 Workload}
\label{sec:interactive-v2}

\begin{quote}
    \textit{This chapter is based on the TPCTC~2023 paper ``The LDBC Social Network Benchmark Interactive Workload v2: A Transactional Graph Query Benchmark with Deep Delete Operations''~\cite{DBLP:journals/corr/abs-2307-04820}, co-authored by several members of the SNB task force.}
\end{quote}

\subsection*{Work-in-Progress}

The \interactivevtwo workload is currently work-in-progress.
As of January 2024, commissioning audits for this workload is not yet possible.

\subsection*{Related Software Components}

\begin{itemize}
    \item Datagen (Spark-based): \url{https://github.com/ldbc/ldbc_snb_datagen_spark}
    \item Driver: \url{https://github.com/ldbc/ldbc_snb_interactive_v2_driver}
    \item Reference implementations: \url{https://github.com/ldbc/ldbc_snb_interactive_v2_impls}
\end{itemize}


\section{Overview}

\begin{figure}[htb]
    \centering
    \includegraphics[scale=\yedscale]{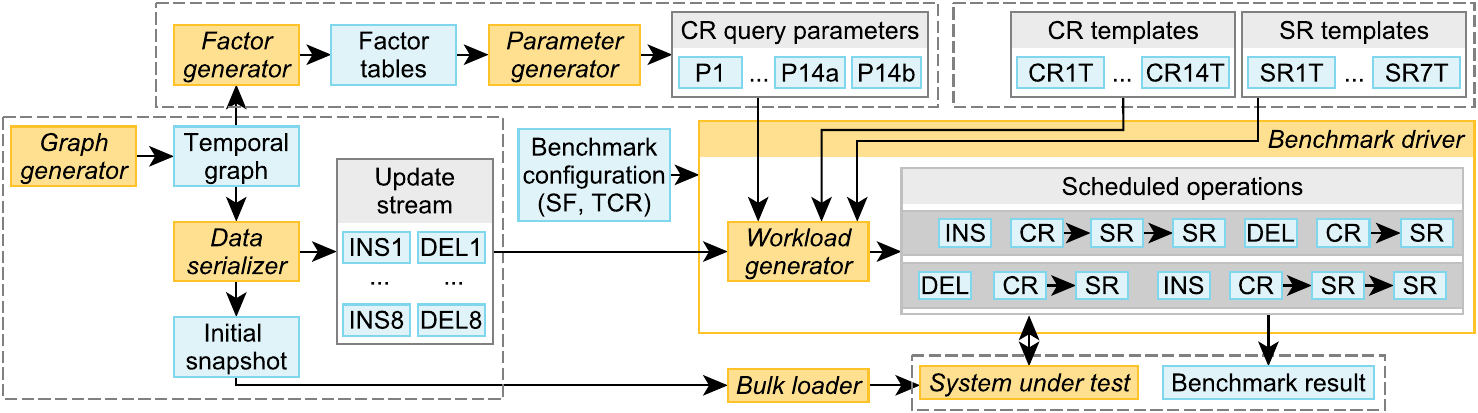}
    \caption{
        Components and workflow of the Interactive~v2 workload.
        The corresponding sections are shown in green circles \Circled{\textsf{\scriptsize \S}}.
        Legend:
        \fcolorbox{mydarkyellow}{mylightyellow}{{\scriptsize \textit{\textsf{Software component}}}}
        \fcolorbox{mydarkblue}{mylightblue}{{\scriptsize \textsf{Data artifact\vphantom{p}}}}
    }
    \label{fig:interactive-components}
\end{figure}

\section{Operations}

The LDBC \snbinteractivevtwo workload uses four types of operations.
There are 14~complex and 7~short read queries.
Update operations include
8~inserts and, newly introduced in the \interactivevtwo workload, 8~deletes.
The workload mix consists of approximately
8\% complex read,
72\% short read,
20\% insert, and
0.2\% delete operations.
The complex reads and the short reads are identical to the ones in \interactivevone, except for query 14, which was replaced to cover the \emph{Cheapest path-finding} choke point.%
\footnote{
    The term \emph{shortest paths} refers to the problem of finding \emph{unweighted shortest paths}, which can be computed with BFS.
    The term \emph{cheapest paths} refers to the \emph{weighted shortest paths} problem, which can be solved using \eg Dijkstra's algorithm.
}

\paragraph{Cheapest path-finding}
While we strived to keep the changes to the queries minimal, we replaced Q14 due to two reasons.
First, we found the original query in \interactivevone to be ill-suited to the workload as it required the enumeration of \emph{all shortest paths} between two \Persons, which can be prohibitively expensive on large scale factors.
%
%
%
%
%
Second, we introduced a new choke point,
\textsf{CP-7.6}
\emph{Cheapest path-finding,}
a key computational kernel and a language opportunity for GQL~\cite{DBLP:conf/sigmod/DeutschFGHLLLMM22}.
Therefore, we changed Q14 to use \emph{cheapest paths} instead of \emph{all shortest paths}.

\subsection{Complex Reads}
\label{sec:interactive-v2-complex-reads}

\iftoggle{StandaloneWorkloadSpecification}{

\renewcommand*{\arraystretch}{1.1}

\subsection*{Interactive / complex / 14v2}
\label{sec:interactive-complex-read-14-v2}

\let\oldemph\emph
\renewcommand{\emph}[1]{{\footnotesize \sf #1}}
\let\oldtextbf\textbf
\renewcommand{\textbf}[1]{{\it #1}}\renewcommand{\currentQueryCard}{interactive-complex-read-14-v2}
\marginpar{
	\raggedleft
	\vspace{0.22ex}

	\queryRefCard{interactive-complex-read-01}{IC}{1}\\
	\queryRefCard{interactive-complex-read-02}{IC}{2}\\
	\queryRefCard{interactive-complex-read-03}{IC}{3}\\
	\queryRefCard{interactive-complex-read-04}{IC}{4}\\
	\queryRefCard{interactive-complex-read-05}{IC}{5}\\
	\queryRefCard{interactive-complex-read-06}{IC}{6}\\
	\queryRefCard{interactive-complex-read-07}{IC}{7}\\
	\queryRefCard{interactive-complex-read-08}{IC}{8}\\
	\queryRefCard{interactive-complex-read-09}{IC}{9}\\
	\queryRefCard{interactive-complex-read-10}{IC}{10}\\
	\queryRefCard{interactive-complex-read-11}{IC}{11}\\
	\queryRefCard{interactive-complex-read-12}{IC}{12}\\
	\queryRefCard{interactive-complex-read-13}{IC}{13}\\
	\queryRefCard{interactive-complex-read-14-v1}{IC}{14v1}\\
	\queryRefCard{interactive-complex-read-14-v2}{IC}{14v2}\\
}

\noindent\begin{tabularx}{\queryCardWidth}{|>{\queryPropertyCell}p{\queryPropertyCellWidth}|X|}
	\hline
	query & Interactive / complex / 14v2 \\ \hline
	title & Trusted connection paths (v2) \\ \hline
	pattern & \centering \includegraphics[scale=\patternscale,margin=0cm .2cm]{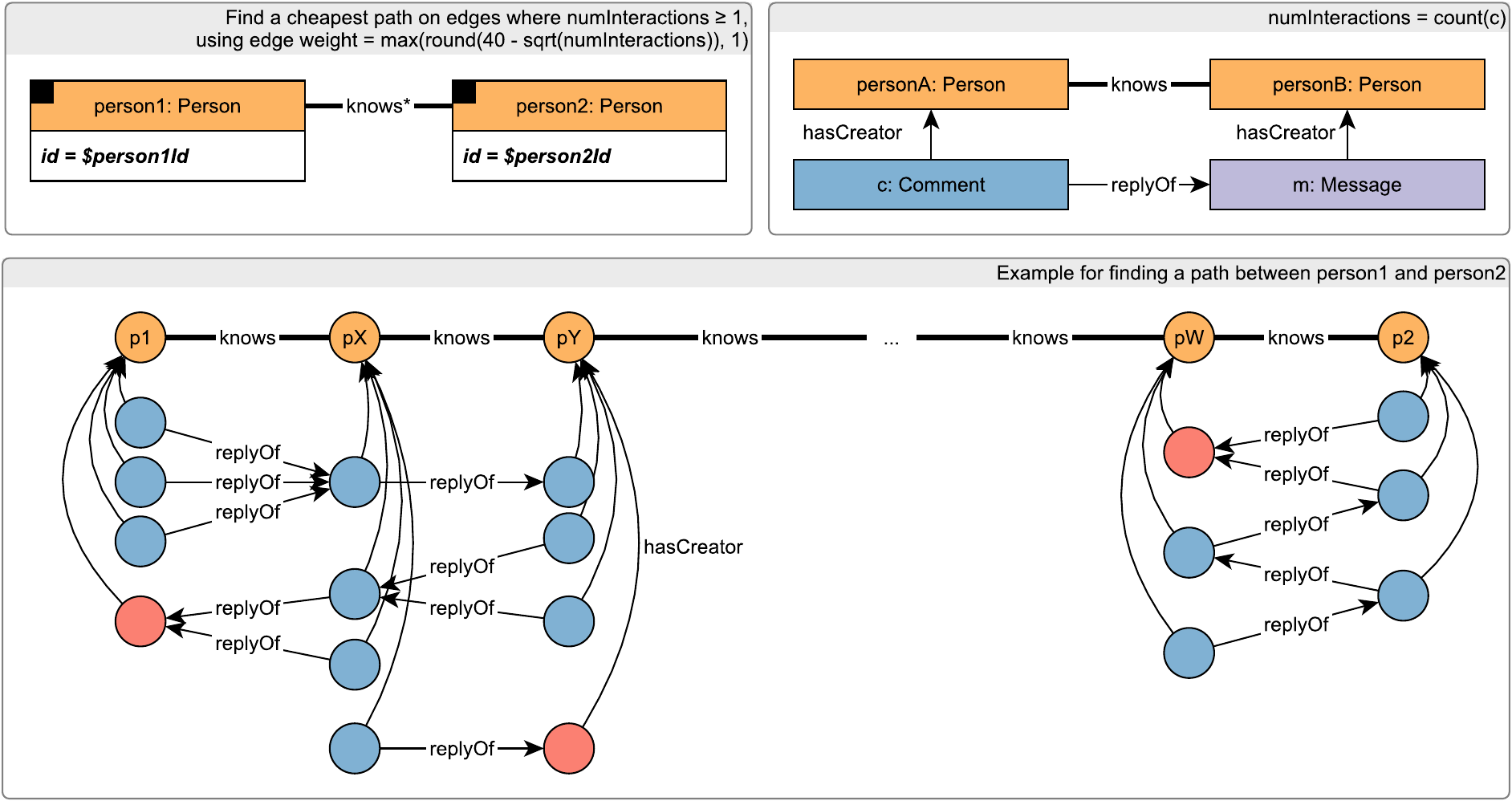} \tabularnewline \hline
	description & \textbf{This query is used in SNB Interactive v2.}

Find a cheapest path between two given \emph{Persons} with IDs
\texttt{\$person1Id} and \texttt{\$person2Id} in the interaction
subgraph. If there are multiple cheapest paths, any of them can be
returned. Do not return any rows if there is no path between the
\emph{Persons}. The interaction subgraph is based on a projection of the
\emph{Person}-\emph{knows}-\emph{Person} graph. In this projection, only
those \emph{knows} edges are kept whose endpoint \emph{Persons} have at
least one interaction between them. An interaction is defined as a
direct reply \emph{Comment} (by one of the \emph{Persons}) to a
\emph{Message} (by the other \emph{Person}). The weights are defined as:
\(\max(\mathrm{round}( 40 - \sqrt{\textit{numInteractions}} ), 1)\)

\textbf{Note:} Interactions are counted both ways, e.g.~if Alice
\emph{knows} Bob, Alice writes 2 reply \emph{Comments} to Bob's
\emph{Messages} and Bob writes 3 reply \emph{Comments} to Alice's
\emph{Messages}, their total number of interactions is 5 and the weight
of the knows edge is 38.

\textbf{Remark:} Determinism is ensured by using square root followed by
rounding. For all integers between 1 and \(\numprint{100000}\), the
square root's fractional part is more than 10e-5 from 0.5, where the
rounding could be non-deterministic based on floating point
inaccuracies. As 10e-5 is significantly larger than the machine epsilon
of IEEE 754 floats (both 32- and 64-bit), the floating point
inaccuracies have no chance to affect the derived integer edge weights.
 \\ \hline

		params &
		\innerCardVSpace{\begin{tabularx}{\attributeCardWidth}{|>{\paramNumberCell}C{\attributeNumberWidth}|>{\varNameCell}M|>{\typeCell}m{\typeWidth}|Y|} \hline
		$\mathsf{1}$ & \$person1Id
 & ID
 & \texttt{(b)} There are no paths between the two \emph{Persons}

\texttt{(b)} There is a 4-hop path between the two \emph{Persons}
 \\ \hline
		$\mathsf{2}$ & \$person2Id
 & ID
 &  \\ \hline
		\end{tabularx}}\innerCardVSpace \\ \hline

		result &
		\innerCardVSpace{\begin{tabularx}{\attributeCardWidth}{|>{\resultNumberCell}C{\attributeNumberWidth}|>{\varNameCell}M|>{\typeCell}m{\typeWidth}|>{\resultOriginCell}c|Y|} \hline
		$\mathsf{1}$ & personIdsInPath & {[}ID{]} & C &
				Identifiers representing an ordered sequence of the \emph{Persons} in
the path
 \\ \hline
		$\mathsf{2}$ & pathWeight & 64-bit Integer & C &
				 \\ \hline
		\end{tabularx}}\innerCardVSpace \\ \hline

	CPs &
	\multicolumn{1}{>{\raggedright}l|}{
		\chokePoint{3.3}, 
		\chokePoint{5.3}, 
		\chokePoint{7.6}, 
		\chokePoint{7.7}, 
		\chokePoint{7.8}, 
		\chokePoint{8.1}, 
		\chokePoint{8.2}, 
		\chokePoint{8.3}, 
		\chokePoint{8.6}
		} \\ \hline
	relevance &
		\footnotesize This query tests the performance of cheapest path (weighted shortest
path) computation.
 \\ \hline%
\end{tabularx}
\queryCardVSpace

\let\emph\oldemph
\let\textbf\oldtextbf

\renewcommand{\currentQueryCard}{0}

}{
    
}

\subsection{Short Reads}
\label{sec:interactive-v2-short-reads}
    
\iftoggle{StandaloneWorkloadSpecification}{
    
}{
    The short reads operations are identical to the ones in \interactivevone, see \autoref{sec:interactive-v1-short-reads}.
}

\subsection{Insert Operations}
\label{sec:interactive-v2-insert-operations}
\iftoggle{StandaloneWorkloadSpecification}{
    
}{
    See \autoref{sec:insert-operations}.
}

\subsection{Delete Operations}
\label{sec:interactive-v2-delete-operations}
\iftoggle{StandaloneWorkloadSpecification}{
    
}{
    See \autoref{sec:delete-operations}.
}

\section{Parameter Curation}
\label{sec:parameter-curation}


To prevent caching query results, the \snbinteractivevtwo driver instantiates the parameterized complex read (\CR) query templates with different \emph{substitution parameters} (\aka parameter bindings).
However, the na\"ive approach (using a uniform random sampling of parameters and ignoring updates)
leads to unstable runtimes,
which compromise both the benchmark's understandability and reproducibility.
To ensure stable runtimes, LDBC invented \emph{parameter curation} techniques, which select parameters that produce query runtimes with a unimodal (preferably Gaussian) distribution~\cite{DBLP:conf/tpctc/GubichevB14,DBLP:journals/pvldb/SzarnyasWSSBWZB22}.

\subsection{Building Blocks for Parameter Curation}

\paragraph{Temporal bucketing}
\label{sec:temporal-bucketing}
To ensure that operations are always executable, \ie they avoid targeting nodes that are yet to be inserted or ones that are already deleted, the parameter curation process in \interactivevtwo employs \emph{temporal bucketing}.
Namely, we create a parameter bucket for \emph{each day in the simulation time of the update streams},
\ie each day in the simulation time has its own distinct set of parameters.
This is a novel feature in \interactivevtwo{} -- previous SNB benchmarks lacked this feature and only selected parameters from the \emph{initial snapshot}.

\paragraph{Factor tables}
As shown in \autoref{fig:interactive-components}, the parameter generation is a two-step process.
The \emph{factor generator} produces \emph{factor tables}, which contain data cube-like summary statistics~\cite{DBLP:journals/datamine/GrayCBLRVPP97} of the temporal graph such as the number of \Messages for friends.
The factor generator is executed in a distributed setup using Spark as this computation includes expensive joins over large tables,
\eg $\knows(\snbperson, \snbfriend) \bowtie \hasCreator(\snbperson, \snbcomment)$.


\subsection{Parameter Curation for Relational Queries}

For relational queries (without path-finding), we based our parameter generation on two techniques.

\paragraph{(1)~Selecting windows}
To select the parameters that are expected to yield similar runtimes, we look for windows with the smallest variance for a given value using SQL window functions.
The parameters are first sorted and grouped together based on their difference in frequency.
Groups that are smaller than a given minimum threshold are discarded to select a group of
parameters large enough to generate a sufficient amount of parameters.
From the latter, we select
the group with the smallest standard deviation.

\paragraph{(2)~Selecting distributions}
For queries where we want to select parameters that are correlated or anti-correlated, we use factor tables encoding possible combinations (\eg \texttt{countryPairsNumFriends} for \CR[3]):
we select values near a high percentile for the correlated and a low percentile for the anti-correlated case.

\paragraph{Generating the parameters}
The parameter candidates discovered by the previous approaches are stored in temporary tables.
The parameter generation step uses these tables to select parameters for each day in the update stream.



\subsection{Parameter Curation for Path-Finding Queries}
\label{sec:path-curation}

\paragraph{The effect of deletes}
A key distinguishing feature of graph data management systems is their first-class support for path queries~\cite{DBLP:journals/csur/AnglesABHRV17}.
We demonstrate why ensuring stable query runtimes for path queries is particularly challenging through the example of \autoref{fig:paths}, where we query for the (unweighted) shortest path between \emph{Ada} and \emph{Bob} over a dynamic graph.
Initially, at $t = 1$, the length of the shortest path is 4~hops.
Then, the edge between \emph{Carl} and \emph{Dan} is deleted, making \emph{Ada} and \emph{Bob} unreachable from each other at $t = 2$.
Finally, a new edge is inserted between \emph{Carl} and \emph{Bob}, yielding a shortest path of length 3 at $t = 3$.
This illustrates how a given input parameter (a pair of \Persons) can oscillate between being reachable and being in disjoint connected components over a short period.
To ensure stable query runtimes for path queries in the presence of inserts and deletes, \interactivevtwo introduces a novel \emph{path curation} algorithm, which produces pairs of \Person nodes whose shortest path length from each other is guaranteed to be exactly $k$ hops at any point during a given day.

\paragraph{Graph construction}
The parameter curation algorithm builds two variants of the \Person--\knows--\Person subgraph for each day based on the \emph{temporal graph}:
graph $G_1$ has the inserts applied until the beginning of the day and the deletes applied until the end of the day,
while $G_2$ has the deletes applied until the beginning of the day and the inserts applied until the end of the day.
For a given pair of \Person nodes, their shortest path length in $G_1$ is an upper bound $k_\mathrm{upper}$ on their shortest path length at any point in the day -- when the inserts during the day are gradually applied, the shortest path length can only become shorter.
Conversely, $G_2$ gives a lower bound $k_\mathrm{lower}$ for the shortest path -- the deletes can only make the shortest path length become longer.

\paragraph{Parameter selection}
The bounds provided by $G_1$ and $G_2$ guarantee for the shortest path length $k$ that $k_\mathrm{lower} \leq k \leq k_\mathrm{upper}$ will hold at any point during the day.
We can ensure that $k$ will stay constant during the day by selecting \Person pairs where $k_\mathrm{lower} = k_\mathrm{upper}$ holds.
To this end, we select pairs who are exactly 4~hops apart in both $G_1$ and $G_2$, hence they will be always~4 hops apart during the given day.
Unreachable pairs of nodes can be generated by calculating the connected components of $G_2$ and selecting nodes from disjoint components.
The path curation for both the reachable and the unreachable cases is implemented using the NetworKit graph algorithm library~\cite{lit:networkit}.

\newsavebox{\largestimage}
\begin{figure}[htb]
    \savebox{\largestimage}{\includegraphics[width=.61\textwidth]{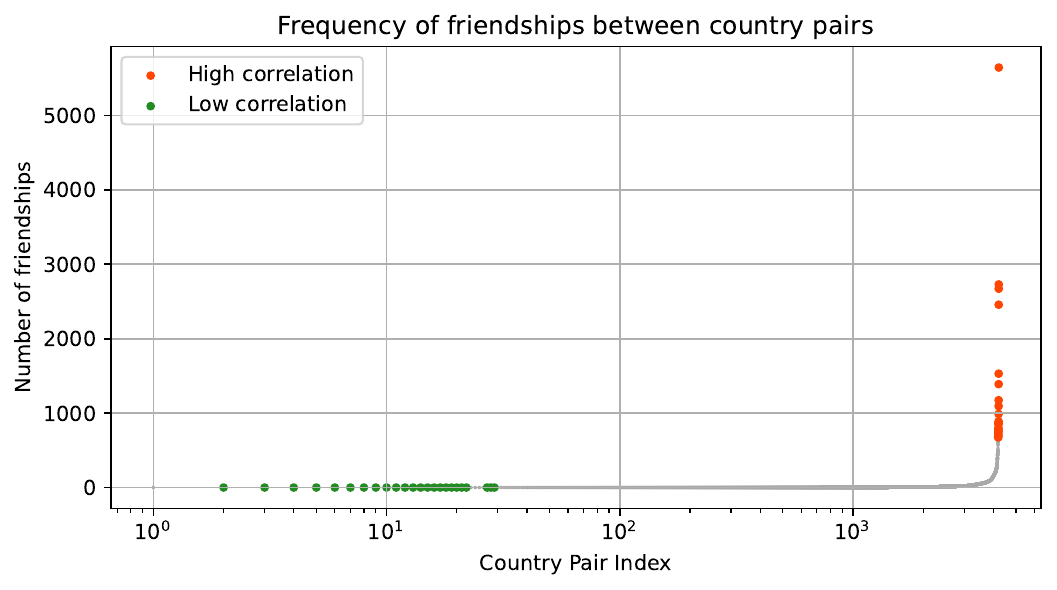}}%
    \begin{subfigure}{0.39\textwidth} 
        \centering
        \raisebox{\dimexpr.5\ht\largestimage-.5\height}{%
            \includegraphics[width=\textwidth]{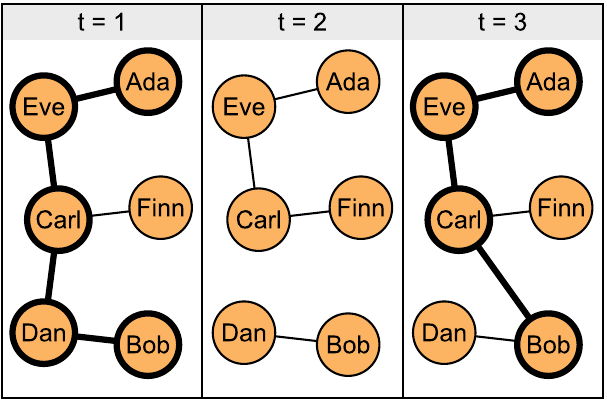}
        }
        \caption{
            Shortest path (denoted with thick lines) between \emph{Ada} and \emph{Bob} in the presence of updates.
        }
        \label{fig:paths}
    \end{subfigure}
    \hspace{2mm}
    \begin{subfigure}{0.58\textwidth}
        \centering
        \includegraphics[width=\linewidth]{figures/paramgen/paramgen-frequency-disc-countries}
        \vspace{-3ex}
        \caption{
            Pairs of \Countries in the \texttt{countryPairsNum\-Friends} factor table representing the number of friendships between both \Countries.
        }
        \label{fig:paramgen-frequency-disc-countries}
    \end{subfigure}
    \caption{Example graph and distribution for path curation.}
    \label{y}
    \vspace{-4.5ex}
\end{figure}

\subsection{Query Variants}
\label{sec:interactive-v2-query-variants}

The new workload introduces variants for three queries:
\queryRefCard{interactive-complex-read-03}{IC}{3},
\queryRefCard{interactive-complex-read-13}{IC}{13},
\queryRefCard{interactive-complex-read-14-v2}{IC}{14v2}.

\paragraph{Complex read 3: Correlated \vs anti-correlated Countries}
\label{sec:cr3-variants}
For \queryRefCard{interactive-complex-read-03}{IC}{3},
variant \CR[3(a)] starts from \Countries that have a high correlation in the friendship network,
while
variant \CR[3(b)] starts from \Countries that have a low correlation of friendships between.
To generate these inputs, we use the \texttt{countryPairsNumFriends} factor table visualized in \autoref{fig:paramgen-frequency-disc-countries} and select values at percentile~1.00 for variant~\texttt{(a)} and percentile~0.01 for variant~\texttt{(b)}.

\paragraph{Complex reads 13 and 14: Reachable \vs unreachable Persons}
Path queries are expected to have different runtimes if there is a path \vs when there is no path.
While the performance characteristics vary highly between systems, in principle, the ``no path'' case should be simpler in the SNB graph, where one of the nodes is always in a small connected component.
To distinguish between these cases, we have two variants for the two path queries \queryRefCard{interactive-complex-read-13}{IC}{13} and \queryRefCard{interactive-complex-read-14-v2}{IC}{14v2}.
For variants~\snbOperation{(a)} we select \Person pairs which \emph{do not have a path},
and for variants~\snbOperation{(b)} we select pairs which \emph{have} a path of length~4.

\subsection{Parameter Generator Implementation}
\label{sec:paramgen-implementation}

%
The parameter generator is implemented in Python using NetworKit~\cite{lit:networkit} and SQL queries executed by DuckDB~\cite{DBLP:conf/sigmod/RaasveldtM19}.
%
%
Based on our experiments in~\cite[Figure~4.3]{david-puroja-msc}, the new parameter generator is scalable.
Even with the significant extra work performed for temporal bucketing,
it outperforms the old parameter generator by more than $100\times$ on SF\numprint{1000},
and finishes in less than 1.5~hours on SF\numprint{10000}.

\section{Workload Scheduling and Benchmark Driver}
\label{sec:workload-and-driver}

In this section, we explain how operations are scheduled in the SNB Interactive workload, how the driver operates, and how the final \emph{throughput} metric is determined.
In all cases, we assume that the system-under-test has been populated with the \emph{initial snapshot} using a \emph{bulk loader} before the driver runs the operations.

\subsection{Scheduling Operations}
\label{sec:scheduling}

\paragraph{TCR (total compression ratio)}
The scheduling follows the \emph{simulation time} of the temporal social network graph.
The user-provided \emph{total compression ratio} (TCR) value controls the speed at which the simulation is replayed.
For example, a TCR value of $0.02$ means that the simulation is replayed $50\times$ faster, \ie for every 20~milliseconds in wall clock time, 1~second passes in the simulation time.

\paragraph{Update operations}
The driver replays the update operations starting from the cutoff date, Nov 29, 2012.
The operations are scheduled according to the distance of their start time from this date, adjusted by the TCR.
They are then used to set the cadence of the schedule for the complex reads and, in turn, the short read queries, as we will explain momentarily.

\paragraph{Complex read queries}
The \emph{complex read queries} differ significantly in their expected runtimes as they touch on different amounts of data.
As each query instance contributes equally to the output metric,%
\footnote{Unlike in \tpcH~\cite{tpch} and \snbbi~\cite{DBLP:journals/pvldb/SzarnyasWSSBWZB22}, which use \emph{geometric mean} in their metrics.}
we balance them such that each query type is expected to take the same amount of time to execute.
For example, \CR[14 (new)] is expected to be more difficult than \CR[13], therefore it is scheduled less frequently.
Frequencies vary based on the SF as the relative difficulties of queries change with the data size
(\eg three-hop neighbourhood queries grow faster on larger SFs than one-hop ones).

\paragraph{Short read queries}
Short read queries are triggered by complex read queries and other short read queries, and use their output as their input.
For example, both \CR[3] and \CR[14] trigger \SR[2], which also triggers itself.
This mimics the real-life scenario of a user retrieving more information about \Person profiles based on the result of the earlier queries.
To see which short read queries are potentially triggered after given short read and complex read queries, see \autoref{tab:short-read-queries-triggered}.

\subsection{Driver}
\label{sec:driver}

\begin{figure}[htb]
    \centering
    \begin{subfigure}{\linewidth}
        \centering
        \includegraphics[scale=\yedscale]{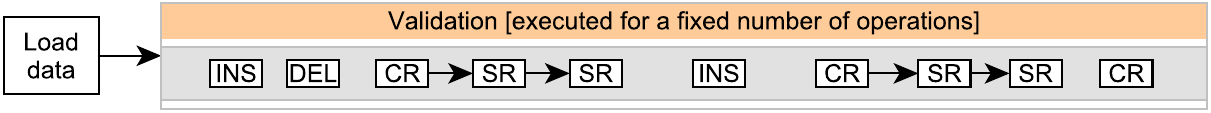}
        \caption{Validation workflow running on a single thread.}
        \label{fig:interactive-validation-workflow}
    \end{subfigure}
    \begin{subfigure}{\linewidth}
        \centering
        \includegraphics[scale=\yedscale]{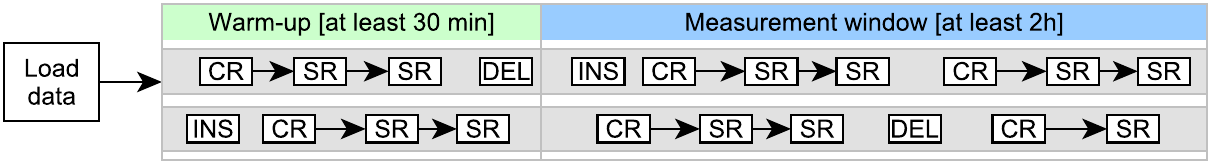}
        \caption{Benchmark workflow using multiple threads.}
        \label{fig:interactive-benchmark-workflow}
    \end{subfigure}
    \caption{
        Workflow of driver modes in SNB Interactive v2.
    }
    \label{fig:interactive-workflows}
    \vspace{-2.5ex}
\end{figure}

\paragraph{Driver modes}
The SNB driver has two key modes of operation.
In \emph{cross-validation mode} (\autoref{fig:interactive-validation-workflow})m
the driver tests an implementation against the output of another implementation.
To ensure deterministic results, operations in this mode are executed sequentially with no overlap between queries and updates.
In \emph{benchmark mode} (\autoref{fig:interactive-benchmark-workflow}),
the driver performs a benchmark run where queries and updates are issued concurrently from multiple threads.
The run starts with a 30-minute warm-up period, followed by a 2-hour \emph{measurement window}.
This mode does not perform validation as query results may differ (slightly) due to concurrent updates.

\paragraph{Dependency tracking}
To ensure that updates are executable, concurrent threads must be synchronized so that an operation is only executed when its dependencies exist in the network (\eg two \Persons can only become friends if both of them already exist).
This is achieved via maintaining a global clock in the driver and performing \emph{dependency tracking} for the updates~\cite{DBLP:conf/sigmod/ErlingALCGPPB15}: each update operation has a timestamp denoting the creation time of the last operation it depends on.
The data generator calculates these timestamp during generation and ensures that there is a minimum time separation,
$T_\textit{safe}$,
between dependent entities to reduce synchronization overhead in the driver when executing operations.
The driver then only needs to check every $T_\textit{safe}$ time whether a given update operation can be executed.
By default, $T_\textit{safe}$ is set to 10 seconds in the simulation time.

\paragraph{Latency requirements}
The workload simulates a highly transactional scenario where operations are subject to (soft) latency requirements.
To incorporate this in the workload, it prescribes the \emph{95\% on-time requirement}:
for a benchmark run to be successful, 95\% of the operations must start \emph{on-time}, \ie within 1~second of their scheduled start time.
Benchmark runs where the system-under-test falls behind too much from the schedule are considered invalid.

\paragraph{Throughput}
The throughput of a run is the total number of operations
(\CR, \SR, \INS, \DEL)
executed per second.
A lower TCR value implies a higher throughput.

\paragraph{Individual execution times}
To facilitate deeper analyis, the benchmark driver also collects all individual query execution times.
Based on these, the benchmark reports must include statics for each operation type (min, max, mean, $P_{50}$, $P_{90}$, $P_{95}$, and $P_{99}$ of the execution times).

\paragraph{Driver implementation in v2}
The \interactivevtwo is implemented in Java~17.
It consists of \numprint{26500} lines of code for the core project and an additional \numprint{18000} lines of test code.
The new version contains several patches including bug fixes, usability improvements, and performance optimizations.


\chapter{Business Intelligence Workload}
\label{sec:bi}

The Business Intelligence (BI) workload is the SNB's analytical (OLAP) workload.
As such, it defines complex read queries that touch a significant portion of the data (see \autoref{sec:bi-reads}).
Additionally, it defines daily batches of updates over a 33-day period
(see
\autoref{sec:bi-insert-operations} for inserts and
\autoref{sec:bi-delete-operations} for deletes).

\subsection*{Related Publications}

The BI workload was published in PVLDB 2022~\cite{DBLP:journals/pvldb/SzarnyasWSSBWZB22}.

\subsection*{Related Software Components}

\begin{itemize}
    \item Datagen (Spark-based): \url{https://github.com/ldbc/ldbc_snb_datagen_spark}
    \item Driver and reference implementations: \url{https://github.com/ldbc/ldbc_snb_bi}
\end{itemize}


\section{Overview}
\label{sec:bi-benchmark-overview}

\begin{figure}[htb]
    \centering
    \includegraphics[scale=\yedscale]{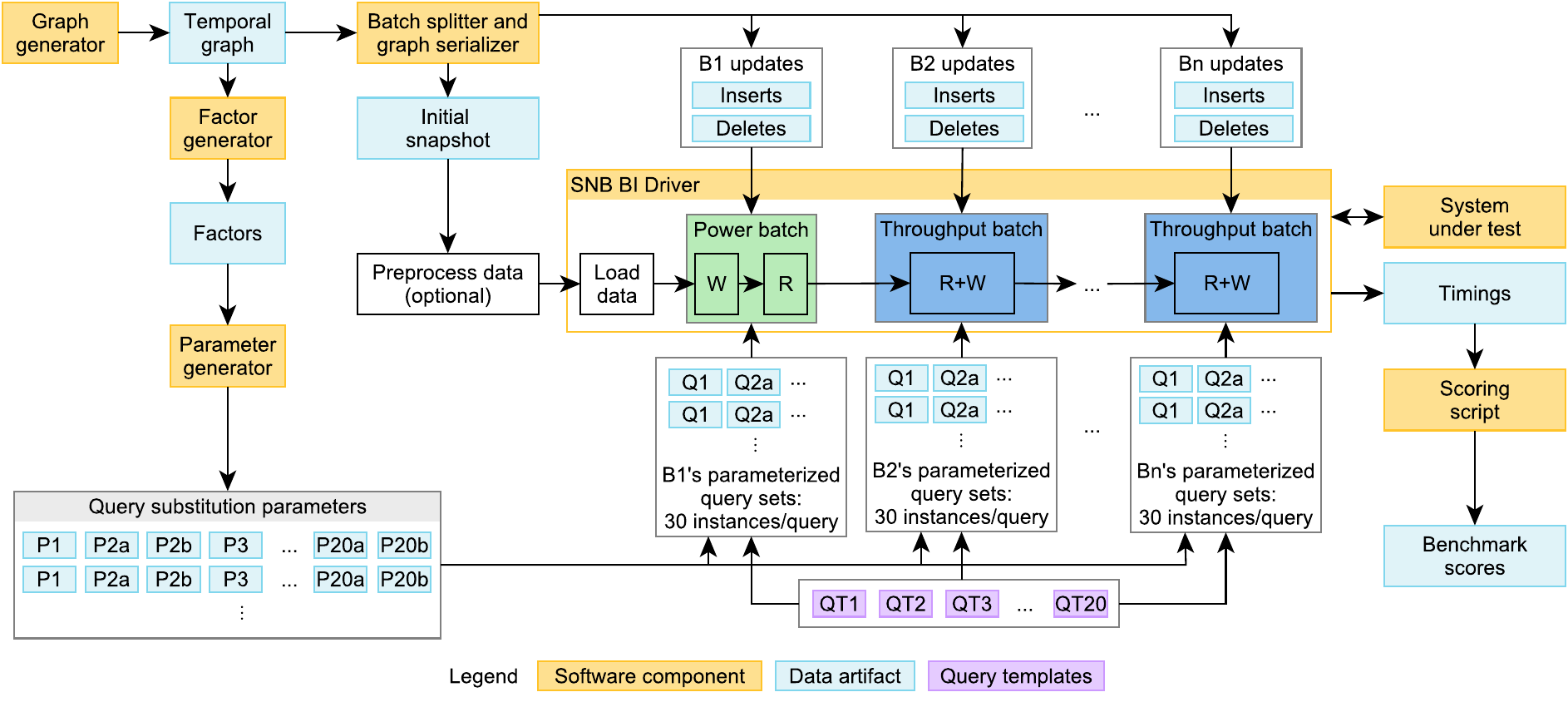}
    \caption{Main software components and data artifacts of the benchmark and their connection to the workflow executed by the BI benchmark driver.}
    \label{fig:bi-workflow}
\end{figure}

An overview of the BI workload is shown in \autoref{fig:bi-workflow}.
The rules for auditing workload implementations are given in \autoref{sec:auditing-bi-workload-audit}.

\section{Read Query Templates}
\label{sec:queries}


\snbbi consists of 20~parameterized \emph{read query templates}, referred to as \emph{queries}. These search for graph patterns (often implying join-heavy operations on many-to-many edges), traverse hierarchies, and compute cheapest paths (\aka weighted shortest paths).
Additionally, they include filtering, grouping, aggregation, and sorting operators.
While all queries explore a large portion of the graph, they only return the top-$k$ (typically 20 or 100) results,
keeping their result sizes compact to avoid emphasizing the client-server network protocol's role in the benchmark~\cite{DBLP:journals/pvldb/RaasveldtM17}.

\subsection{Choke Point-Based Design Methodology}

LDBC's query design process relies on the use of \emph{choke points} (\autoref{sec:choke-points}),
\ie challenging aspects of query processing.
\snbbi includes 38~choke points divided into 9~categories:
aggregation performance,
join performance,
data access locality,
expression calculation,
correlated subqueries,
parallelism and concurrency,
graph specifics,
language features,
and
update operations.
Their coverage is shown in \autoref{tab:query_choke_point}.
In the following, we discuss two challenges that are particularly prevalent in graph workloads.

\subsubsection{Explosive and redundant multi-joins}
In recent years it has become clear that graph pattern matching, or equivalent multi-join queries over many-to-many relationships, typically generate very large intermediate results when executed with traditional join algorithms. This is especially the case for cyclical join graphs (corresponding to cyclic graph queries). 
It was proven in theory~\cite{DBLP:journals/sigmod/NgoRR13} and shown in practice~\cite{DBLP:journals/corr/abs-1210-0481,DBLP:journals/pvldb/MhedhbiS19,DBLP:journals/pvldb/FreitagBSKN20} that ``worst-case optimal'' {\em multi-}join algorithms can avoid these large intermediates and outperform traditional joins. Following this, there has been increased attention on {\em redundancy} in join results (even when produced by worst-case optimal joins), which can be eliminated using {\em factorized} query processing techniques~\cite{DBLP:journals/pvldb/BakibayevOZ12,DBLP:journals/sigmod/OlteanuS16,DBLP:journals/pvldb/GuptaMS21}.
Graph pattern matching queries that contain large join patterns will trigger these phenomena.

\subsubsection{Expressive path finding}
\snbbi contains queries that require an efficient implementation of shortest path finding between many pairs.
Expressing such queries requires a query language which supports either path finding or recursion. The underlying system implementation must then handle this with an optimized execution strategy, as recursing to try all paths will not scale.
As some of this path finding includes on-the-fly computed edges (joins) between nodes, the queries can benefit from {\em path expressions}, as proposed in Oracle's PGQL language~\cite{DBLP:conf/grades/RestHKMC16} and as part of the GQL and SQL/PGQ languages~\cite{DBLP:conf/sigmod/DeutschFGHLLLMM22}.
The path finding required by \snbbi not only tests connectivity (as supported in SPARQL), but also requires returning the {\em cheapest cost} along weighted paths (necessitating SPARQL extensions~\cite{DBLP:conf/bigdataconf/MizellMR14}).

\subsection{Analysis of Selected Queries}
\label{sec:example-queries}

In order to defeat trivializing complex query performance by query caching, benchmarks can use both frequent updates (which require invalidating caches or maintaining cached intermediates) as well as parameterized query templates.
The BI workload features update batches, so parametrized \emph{read query templates} are necessary to guard against this between the batches.
In this section, we analyze four read query templates.

\emph{Notation:}
We denote the query parameters with the \param{}~symbol and discuss their generation in \autoref{sec:paramgen}.

\subsubsection{Q11: Friend triangles}

\queryRefCard{bi-read-11}{BI}{11} imposes two key difficulties.
First, systems should efficiently filter the \tKnows edges based on the location of their endpoint \tPersons (\tCountry \param{country}) and the date range.
Second, given a large number of \tKnows edges even after filtering,
efficient enumeration of \texttt{personA}--\texttt{personB}--\texttt{personC} triangles (a cyclic subgraph query)
requires worst-case optimal multi-joins.

\subsubsection{Q14: International dialog}

\queryRefCard{bi-read-14}{BI}{14} imposes different challenges depending on whether
\tCountries \param{country1} and \param{country2} are correlated or anti-correlated
(\autoref{sec:paramgen-correlations}).
For the ranking, \emph{top-k pushdown} can be exploited:
once a result for a \tCity in \param{country1} is obtained,
extra restrictions in a selection can be added based on the value of this element.
As the score of two \tPersons does not depend on any query parameters,
precomputing and maintaining it as an attribute on the \tKnows edge can be beneficial.

\subsubsection{Q18: Friend recommendation}

\queryRefCard{bi-read-18}{BI}{18} is inspired by Twitter's recommendation algorithm~\cite{DBLP:conf/www/GuptaGLSWZ13}.
Implementations of this query can exploit factorization:
systems can count the number of mutual friends without explicitly enumerating all
\textsf{<}\texttt{person1}, \texttt{personM}, \texttt{person2}\textsf{>} tuples.

\subsubsection{Q20: Recruitment}

\queryRefCard{bi-read-20}{BI}{20} performs \emph{graph projection}~\cite{DBLP:conf/sigmod/AnglesABBFGLPPS18}.
Instead of materializing this graph in the database,
systems may represent it using a compact in-memory structure such as CSR (Compressed Sparse Row)~\cite{DBLP:books/daglib/0009092}.
To perform the cheapest path computation, a single-source shortest path algorithm
(starting from \param{person2}), such as Dijkstra's algorithm, can be used.
As the projected graph 
is independent of query parameters, precomputing and maintaining it can be beneficial.

\section{Parameter Curation for BI Queries}
\label{sec:paramgen}

\subsection{The Need for Parameter Curation}

A disadvantage of executing the same read query template with different parameters is that the intermediate results and runtimes can be severely influenced by the parameter values.
This is particularly the case in \snbbi with its explosive joins, skewed out-degrees, skewed value distributions, correlated value distributions, and structural correlations.
Moreover, the updates (including cascading deletes) can significantly change the portion of the graph reached by the same query executed at different times. 
In order to keep query performance understandable we need to actively {\em curate} parameters, such that different parameters executed at different logical times 
still lead to stable and, therefore, understandable results.
We achieve this through \emph{parameter curation}~\cite{DBLP:conf/tpctc/GubichevB14,DBLP:conf/sigmod/ErlingALCGPPB15}, a data mining process of looking for parameter values with suitably similar characteristics.

\subsection{Parameter Generation Steps}
\label{sec:parameter-curation-method}

Our parameter curation process is a two-step process:
we first generate \emph{factors} followed by the \emph{parameters} (\autoref{fig:bi-workflow}).
These components are executed for each scale factor and are independent of the serialization format/layout of the data set.

\subsubsection{Factor Generator}
The factor generator produces 21~\emph{factor tables} containing summary statistics from the temporal graph,
\eg
the number of \tPersons per \tCity
or
the number of \tMessages per day for each \tTag.

\subsubsection{Parameter Generation}
\label{sec:parameter-generation-query}

To find suitable substitution parameters that (presumably) lead to the same amount of data access and thus similar runtimes,
we first identify the factor table containing the summary statistics of the query's parameters.
For example, Q14's template uses the parameters \tCountry \param{country1} and \tCountry \param{country2}.
Therefore, we use the \texttt{countryPairsNumFriends} factor table which contains \param{country1}, \param{country2} pairs and the number of friendships on \tPerson lives in \param{country1} and the other lives in \param{country2}.
Using this table, we select the $p$th percentile from the distribution as the \emph{anchor},
then rank the rest of the distribution based on their absolute difference from the anchor and take the top-$k$ values.
We shuffle the values using a hash function to avoid introducing artificial locality, where \eg subsequent queries start in nodes from the same ID range.
\autoref{lst:q14-parameter-generation} shows the SQL query implementing the parameter generation for Q14\variantA.

\subsection{Parameter Curation for Graph Queries}

We discuss two parameter curation cases that are particularly important in graph data management.

\subsubsection{Correlated \vs Anti-Correlated Parameters}
\label{sec:paramgen-correlations}

Our parameter curation provides a straightforward way of selecting start entities which
are affected by (structural or attribute-level) correlation \vs anti-correlation:
corresponding parameters can be found by selecting a high \vs low percentile as the anchor
in the parameter generation query.
For example, for Q14 (\autoref{sec:bi-read-14}),
we selected
variant~\variantA to $p=0.98$ (correlated) and
variant~\variantB to $p=0.03$ (anti-correlated).

\subsubsection{Path Queries}
\label{sec:path-queries}

\snbbi queries Q15, Q19, Q20 include cheapest path finding queries computed between given (sets of) \tPersons.
These queries are particularly challenging for parameter curation:
if there is no path between the two endpoints, query runtimes are significantly higher as the search has to traverse an entire connected component to ensure that no path exists.
Moreover, the presence of a path between two nodes \emph{at a given time} does not guarantee that it will always exist during the benchmark execution as deletions can render the endpoints of a path unreachable.

\subsection{Query Variants}
\label{sec:bi-query-variants}

12~queries have a single variant, while 8~queries have two variants, yielding a total of 28~query variants.
As a rule of thumb, variants \variantA are expected to produce a longer runtime
while variants \variantB are expected to be simpler.
Variants of Q2, Q8, Q16 are parametrized with a flashmob \vs a non-flashmob date.
Variants of Q14 and Q19 select correlated \vs non-correlated \tCountries/\tCities.
Q10's variants differ in degree (a start \tPerson with an average number of friends \vs only a few friends), while
Q15's variants have different path lengths and time intervals (4 hops and one week \vs 2 hops and one month).
Q20\variantA selects endpoints where it is guaranteed that \emph{no path exists}, while Q20\variantB selects ones where there is guaranteed that a path exists.

\subsection{Scalability and Reproducibility}
\label{sec:paramgen-scalability}


\subsubsection{Scalability}
The \emph{factor generator} is part of the SNB Datagen and runs after the \textit{temporal graph} has been created.
It is implemented in Spark for distributed execution.
While its computations use expensive, aggregration-heavy queries, the derived factor tables are \emph{compact}, \eg SF\numprint{10000} has only 20~GiB of factors in compressed Parquet format, the equivalent of approximately 100~GiB in CSV format, \ie 1\% of the total data set size.
The \emph{parameter generator} queries are executed in DuckDB~\cite{DBLP:conf/sigmod/RaasveldtM19},
which supports vertical scalability and is capable of running the parameter generation for SF\numprint{10000} using less than 512~GiB memory.

\lstset{language=sql,morekeywords={WITHIN}}
\begin{lstlisting}[label=lst:q14-parameter-generation,caption=Parameter generation SQL query for Q14\variantA.]
SELECT country1, country2
FROM (
  SELECT
    country1,
    country2,
    abs(frequency - (
      SELECT percentile_disc(0.98) WITHIN GROUP (ORDER BY frequency) AS anchor FROM countryPairsNumFriends
    )) AS diff
  FROM countryPairsNumFriends
  ORDER BY diff, country1, country2
)
ORDER BY md5(concat(country1, country2))
LIMIT 50
\end{lstlisting}

\subsubsection{Reproducibility}
It is important to guarantee that the parameter curation process is reproducible.
To this end, we leverage that the Datagen and, consequently, the factor generator are reproducible.
To ensure that the parameter generation queries yield deterministic results we define a total ordering in each query.
To provide deterministic shuffling we base the ordering on MD5 hashes (instead of the actual attribute values), see \autoref{lst:q14-parameter-generation}.


\section{Reads}
\label{sec:bi-reads}

\renewcommand*{\arraystretch}{1.1}

\subsection*{BI / read / 1}
\label{sec:bi-read-01}

\let\oldemph\emph
\renewcommand{\emph}[1]{{\footnotesize \sf #1}}
\let\oldtextbf\textbf
\renewcommand{\textbf}[1]{{\it #1}}

\renewcommand{\currentQueryCard}{bi-read-01}
\marginpar{
	\vspace{0.22ex}
	\raggedleft

	\queryRefCard{bi-read-01}{BI}{1}\\
	\queryRefCard{bi-read-02}{BI}{2}\\
	\queryRefCard{bi-read-03}{BI}{3}\\
	\queryRefCard{bi-read-04}{BI}{4}\\
	\queryRefCard{bi-read-05}{BI}{5}\\
	\queryRefCard{bi-read-06}{BI}{6}\\
	\queryRefCard{bi-read-07}{BI}{7}\\
	\queryRefCard{bi-read-08}{BI}{8}\\
	\queryRefCard{bi-read-09}{BI}{9}\\
	\queryRefCard{bi-read-10}{BI}{10}\\
	\queryRefCard{bi-read-11}{BI}{11}\\
	\queryRefCard{bi-read-12}{BI}{12}\\
	\queryRefCard{bi-read-13}{BI}{13}\\
	\queryRefCard{bi-read-14}{BI}{14}\\
	\queryRefCard{bi-read-15}{BI}{15}\\
	\queryRefCard{bi-read-16}{BI}{16}\\
	\queryRefCard{bi-read-17}{BI}{17}\\
	\queryRefCard{bi-read-18}{BI}{18}\\
	\queryRefCard{bi-read-19}{BI}{19}\\
	\queryRefCard{bi-read-20}{BI}{20}\\
}

\noindent\begin{tabularx}{\queryCardWidth}{|>{\queryPropertyCell}p{\queryPropertyCellWidth}|X|}
	\hline
	query & BI / read / 1 \\ \hline
	title & Posting summary \\ \hline
	pattern & \centering \includegraphics[scale=\patternscale,margin=0cm .2cm]{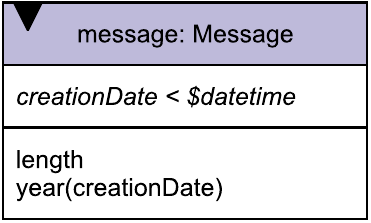} \tabularnewline \hline
	description & Given a \texttt{\$datetime}, find all \emph{Messages} created before
that moment. Group them by a 3-level grouping:

\begin{enumerate}
\def\labelenumi{\arabic{enumi}.}
\tightlist
\item
  by year of creation
\item
  for each year, group into \emph{Message} types: is \emph{Comment} or
  not
\item
  for each year-type group, split into four groups based on length of
  their content

  \begin{itemize}
  \tightlist
  \item
    \texttt{0}: \(0 \leq \text{length} < 40\) (short)
  \item
    \texttt{1}: \(40 \leq \text{length} < 80\) (one liner)
  \item
    \texttt{2}: \(80 \leq \text{length} < 160\) (tweet)
  \item
    \texttt{3}: \(160 \leq \text{length}\) (long)
  \end{itemize}
\end{enumerate}
 \\ \hline

		params &
		\innerCardVSpace{\begin{tabularx}{\attributeCardWidth}{|>{\paramNumberCell}C{\attributeNumberWidth}|>{\varNameCell}M|>{\typeCell}m{\typeWidth}|Y|} \hline
		$\mathsf{1}$ & \$datetime
 & DateTime
 &  \\ \hline
		\end{tabularx}}\innerCardVSpace \\ \hline

		result &
		\innerCardVSpace{\begin{tabularx}{\attributeCardWidth}{|>{\resultNumberCell}C{\attributeNumberWidth}|>{\varNameCell}M|>{\typeCell}m{\typeWidth}|>{\resultOriginCell}c|Y|} \hline
		$\mathsf{1}$ & year & 32-bit Integer & R &
				\texttt{year(message.creationDate)}
 \\ \hline
		$\mathsf{2}$ & isComment & Boolean & M &
				\texttt{True} for \emph{Comments}, \texttt{False} for \emph{Posts}
 \\ \hline
		$\mathsf{3}$ & lengthCategory & 32-bit Integer & C &
				\texttt{0} for short, \texttt{1} for one-liner, \texttt{2} for tweet,
\texttt{3} for long
 \\ \hline
		$\mathsf{4}$ & messageCount & 64-bit Integer & A &
				Total number of \emph{Messages} in that group
 \\ \hline
		$\mathsf{5}$ & averageMessageLength & 32-bit Float & A &
				Average length of the \emph{Message} content in that group
 \\ \hline
		$\mathsf{6}$ & sumMessageLength & 64-bit Integer & A &
				Sum of all \emph{Message} content lengths
 \\ \hline
		$\mathsf{7}$ & percentageOfMessages & 32-bit Float & A &
				Number of \emph{Messages} in group as a percentage of all messages
created before the given date
 \\ \hline
		\end{tabularx}}\innerCardVSpace \\ \hline

		sort		&
		\innerCardVSpace{\begin{tabularx}{\attributeCardWidth}{|>{\sortNumberCell}C{\attributeNumberWidth}|>{\varNameCell}M|>{\directionCell}c|Y|} \hline
		$\mathsf{1}$ & year
 & $\desc
$ &  \\ \hline
		$\mathsf{2}$ & isComment
 & $\asc
$ & \texttt{False} \textless{} \texttt{True}, i.e.~\emph{Posts} come first
and \emph{Comments} second
 \\ \hline
		$\mathsf{3}$ & lengthCategory
 & $\asc
$ &  \\ \hline
		\end{tabularx}}\innerCardVSpace \\ \hline
	limit & n/a \\ \hline
	CPs &
	\multicolumn{1}{>{\raggedright}l|}{
		\chokePoint{1.2}, 
		\chokePoint{3.2}, 
		\chokePoint{4.1}, 
		\chokePoint{4.2}, 
		\chokePoint{8.5}
		} \\ \hline
\end{tabularx}
\queryCardVSpace

\let\emph\oldemph
\let\textbf\oldtextbf

\renewcommand{\currentQueryCard}{0}
\renewcommand*{\arraystretch}{1.1}

\subsection*{BI / read / 2}
\label{sec:bi-read-02}

\let\oldemph\emph
\renewcommand{\emph}[1]{{\footnotesize \sf #1}}
\let\oldtextbf\textbf
\renewcommand{\textbf}[1]{{\it #1}}

\renewcommand{\currentQueryCard}{bi-read-02}
\marginpar{
	\vspace{0.22ex}
	\raggedleft

	\queryRefCard{bi-read-01}{BI}{1}\\
	\queryRefCard{bi-read-02}{BI}{2}\\
	\queryRefCard{bi-read-03}{BI}{3}\\
	\queryRefCard{bi-read-04}{BI}{4}\\
	\queryRefCard{bi-read-05}{BI}{5}\\
	\queryRefCard{bi-read-06}{BI}{6}\\
	\queryRefCard{bi-read-07}{BI}{7}\\
	\queryRefCard{bi-read-08}{BI}{8}\\
	\queryRefCard{bi-read-09}{BI}{9}\\
	\queryRefCard{bi-read-10}{BI}{10}\\
	\queryRefCard{bi-read-11}{BI}{11}\\
	\queryRefCard{bi-read-12}{BI}{12}\\
	\queryRefCard{bi-read-13}{BI}{13}\\
	\queryRefCard{bi-read-14}{BI}{14}\\
	\queryRefCard{bi-read-15}{BI}{15}\\
	\queryRefCard{bi-read-16}{BI}{16}\\
	\queryRefCard{bi-read-17}{BI}{17}\\
	\queryRefCard{bi-read-18}{BI}{18}\\
	\queryRefCard{bi-read-19}{BI}{19}\\
	\queryRefCard{bi-read-20}{BI}{20}\\
}

\noindent\begin{tabularx}{\queryCardWidth}{|>{\queryPropertyCell}p{\queryPropertyCellWidth}|X|}
	\hline
	query & BI / read / 2 \\ \hline
	title & Tag evolution \\ \hline
	pattern & \centering \includegraphics[scale=\patternscale,margin=0cm .2cm]{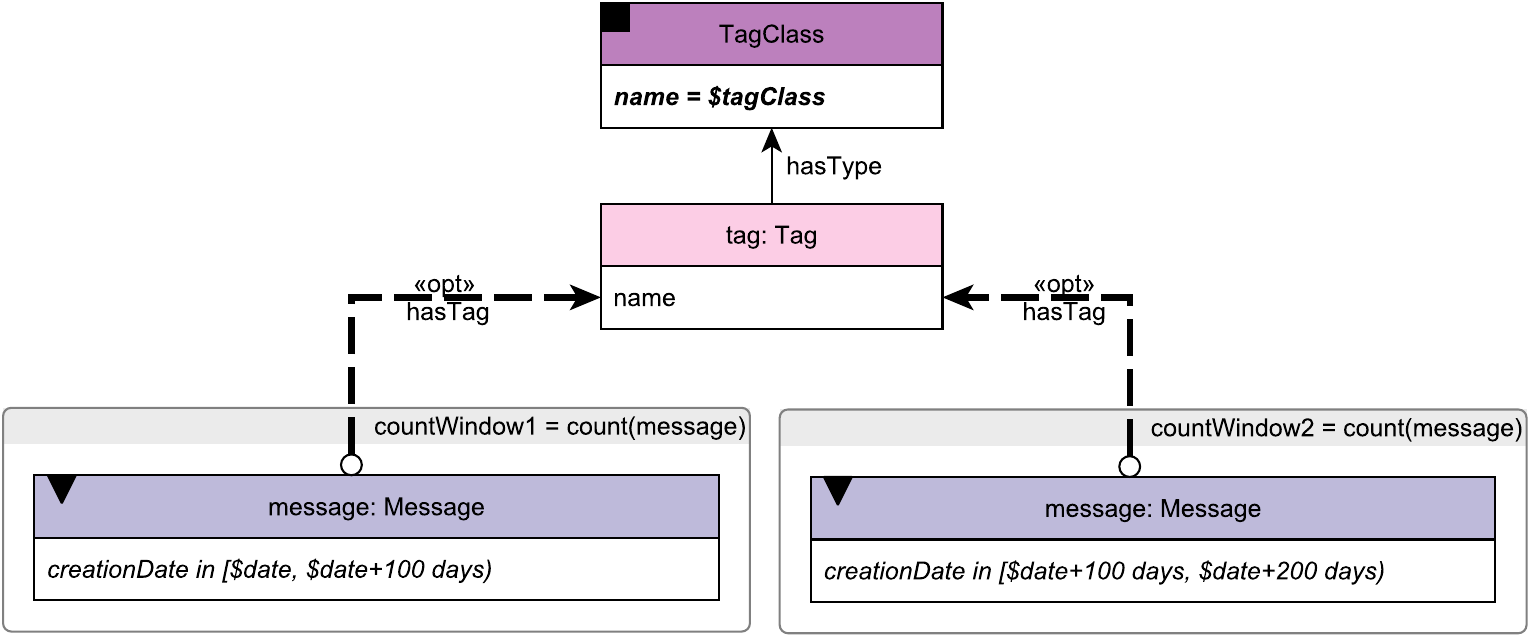} \tabularnewline \hline
	description & Find the \emph{Tags} under a given \texttt{\$tagClass} that were used in
\emph{Messages} during in the 100-day time window starting at
\texttt{\$date} and compare it with the 100-day time window that
follows. For the \emph{Tags} and for both time windows, compute the
count of \emph{Messages}.
 \\ \hline

		params &
		\innerCardVSpace{\begin{tabularx}{\attributeCardWidth}{|>{\paramNumberCell}C{\attributeNumberWidth}|>{\varNameCell}M|>{\typeCell}m{\typeWidth}|Y|} \hline
		$\mathsf{1}$ & \$date
 & Date
 & Based on the creation day -- \emph{TagClass} -- number of
\emph{Messages} factor table:

\texttt{(a)} A flashmob date

\texttt{(b)} A non-flashmob date
 \\ \hline
		$\mathsf{2}$ & \$tagClass
 & Long String
 & For both \texttt{(a)} and \texttt{(b)}, \emph{TagClasses} with a similar
amount of \emph{Messages} are selected
 \\ \hline
		\end{tabularx}}\innerCardVSpace \\ \hline

		result &
		\innerCardVSpace{\begin{tabularx}{\attributeCardWidth}{|>{\resultNumberCell}C{\attributeNumberWidth}|>{\varNameCell}M|>{\typeCell}m{\typeWidth}|>{\resultOriginCell}c|Y|} \hline
		$\mathsf{1}$ & tag.name & Long String & R &
				 \\ \hline
		$\mathsf{2}$ & countWindow1 & 32-bit Integer & A &
				Occurrences of the \texttt{tag} during the first time window
 \\ \hline
		$\mathsf{3}$ & countWindow2 & 32-bit Integer & A &
				Occurrences of the \texttt{tag} during the second time window
 \\ \hline
		$\mathsf{4}$ & diff & 32-bit Integer & A &
				Absolute difference of \texttt{countWindow1} and \texttt{countWindow2}
 \\ \hline
		\end{tabularx}}\innerCardVSpace \\ \hline

		sort		&
		\innerCardVSpace{\begin{tabularx}{\attributeCardWidth}{|>{\sortNumberCell}C{\attributeNumberWidth}|>{\varNameCell}M|>{\directionCell}c|Y|} \hline
		$\mathsf{1}$ & diff
 & $\desc
$ &  \\ \hline
		$\mathsf{2}$ & tag.name
 & $\asc
$ &  \\ \hline
		\end{tabularx}}\innerCardVSpace \\ \hline
	limit & 100 \\ \hline
	CPs &
	\multicolumn{1}{>{\raggedright}l|}{
		\chokePoint{2.4}, 
		\chokePoint{3.1}, 
		\chokePoint{3.2}, 
		\chokePoint{4.1}, 
		\chokePoint{4.2}, 
		\chokePoint{4.3}, 
		\chokePoint{5.3}, 
		\chokePoint{6.1}, 
		\chokePoint{8.2}, 
		\chokePoint{8.5}
		} \\ \hline
\end{tabularx}
\queryCardVSpace

\let\emph\oldemph
\let\textbf\oldtextbf

\renewcommand{\currentQueryCard}{0}
\renewcommand*{\arraystretch}{1.1}

\subsection*{BI / read / 3}
\label{sec:bi-read-03}

\let\oldemph\emph
\renewcommand{\emph}[1]{{\footnotesize \sf #1}}
\let\oldtextbf\textbf
\renewcommand{\textbf}[1]{{\it #1}}

\renewcommand{\currentQueryCard}{bi-read-03}
\marginpar{
	\vspace{0.22ex}
	\raggedleft

	\queryRefCard{bi-read-01}{BI}{1}\\
	\queryRefCard{bi-read-02}{BI}{2}\\
	\queryRefCard{bi-read-03}{BI}{3}\\
	\queryRefCard{bi-read-04}{BI}{4}\\
	\queryRefCard{bi-read-05}{BI}{5}\\
	\queryRefCard{bi-read-06}{BI}{6}\\
	\queryRefCard{bi-read-07}{BI}{7}\\
	\queryRefCard{bi-read-08}{BI}{8}\\
	\queryRefCard{bi-read-09}{BI}{9}\\
	\queryRefCard{bi-read-10}{BI}{10}\\
	\queryRefCard{bi-read-11}{BI}{11}\\
	\queryRefCard{bi-read-12}{BI}{12}\\
	\queryRefCard{bi-read-13}{BI}{13}\\
	\queryRefCard{bi-read-14}{BI}{14}\\
	\queryRefCard{bi-read-15}{BI}{15}\\
	\queryRefCard{bi-read-16}{BI}{16}\\
	\queryRefCard{bi-read-17}{BI}{17}\\
	\queryRefCard{bi-read-18}{BI}{18}\\
	\queryRefCard{bi-read-19}{BI}{19}\\
	\queryRefCard{bi-read-20}{BI}{20}\\
}

\noindent\begin{tabularx}{\queryCardWidth}{|>{\queryPropertyCell}p{\queryPropertyCellWidth}|X|}
	\hline
	query & BI / read / 3 \\ \hline
	title & Popular topics in a country \\ \hline
	pattern & \centering \includegraphics[scale=\patternscale,margin=0cm .2cm]{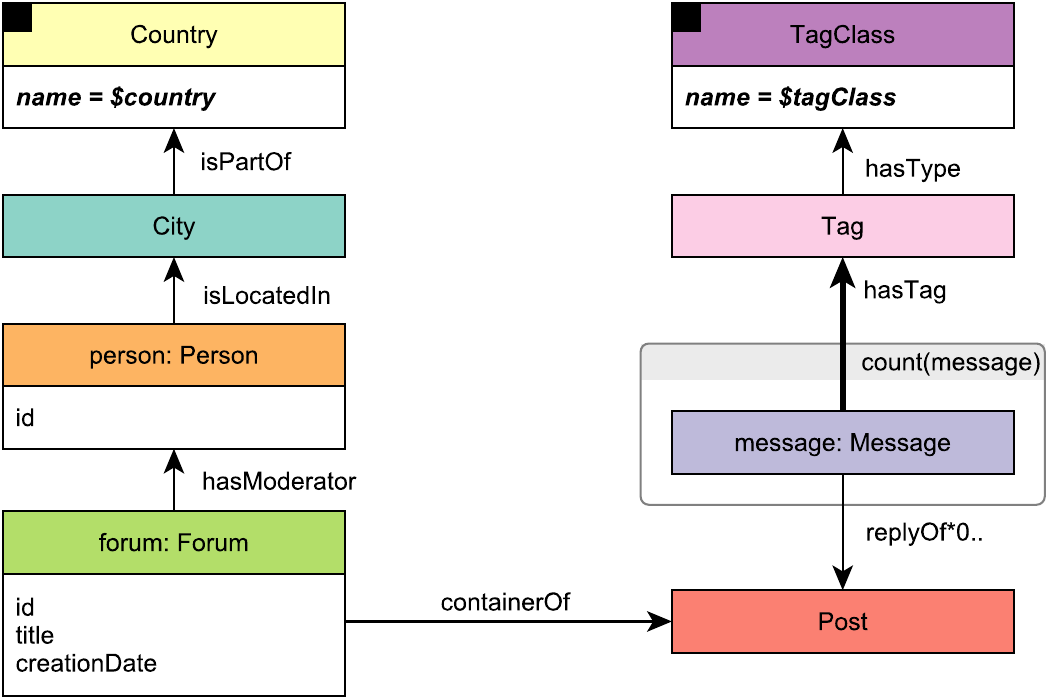} \tabularnewline \hline
	description & Given a \texttt{\$tagClass} and a \texttt{\$country}, find all the
\emph{Forums} created in the given \texttt{\$country}, containing at
least one \emph{Message} with \emph{Tags} belonging directly to the
given \texttt{\$tagClass}, and count the \emph{Messages} by the
\emph{Forum} which contains them.

The location of a \emph{Forum} is identified by the location of the
\emph{Forum}'s moderator.
 \\ \hline

		params &
		\innerCardVSpace{\begin{tabularx}{\attributeCardWidth}{|>{\paramNumberCell}C{\attributeNumberWidth}|>{\varNameCell}M|>{\typeCell}m{\typeWidth}|Y|} \hline
		$\mathsf{1}$ & \$tagClass
 & Long String
 & \emph{TagClasses} with a similar amount of \emph{Messages} are selected
 \\ \hline
		$\mathsf{2}$ & \$country
 & Long String
 & Big \emph{Countries} are selected
 \\ \hline
		\end{tabularx}}\innerCardVSpace \\ \hline

		result &
		\innerCardVSpace{\begin{tabularx}{\attributeCardWidth}{|>{\resultNumberCell}C{\attributeNumberWidth}|>{\varNameCell}M|>{\typeCell}m{\typeWidth}|>{\resultOriginCell}c|Y|} \hline
		$\mathsf{1}$ & forum.id & ID & R &
				 \\ \hline
		$\mathsf{2}$ & forum.title & Long String & R &
				 \\ \hline
		$\mathsf{3}$ & forum.creationDate & DateTime & R &
				 \\ \hline
		$\mathsf{4}$ & person.id & ID & R &
				 \\ \hline
		$\mathsf{5}$ & messageCount & 32-bit Integer & A &
				 \\ \hline
		\end{tabularx}}\innerCardVSpace \\ \hline

		sort		&
		\innerCardVSpace{\begin{tabularx}{\attributeCardWidth}{|>{\sortNumberCell}C{\attributeNumberWidth}|>{\varNameCell}M|>{\directionCell}c|Y|} \hline
		$\mathsf{1}$ & messageCount
 & $\desc
$ &  \\ \hline
		$\mathsf{2}$ & forum.id
 & $\asc
$ &  \\ \hline
		\end{tabularx}}\innerCardVSpace \\ \hline
	limit & 20 \\ \hline
	CPs &
	\multicolumn{1}{>{\raggedright}l|}{
		\chokePoint{1.1}, 
		\chokePoint{1.2}, 
		\chokePoint{1.3}, 
		\chokePoint{2.1}, 
		\chokePoint{2.2}, 
		\chokePoint{2.4}, 
		\chokePoint{3.3}, 
		\chokePoint{8.2}
		} \\ \hline
\end{tabularx}
\queryCardVSpace

\let\emph\oldemph
\let\textbf\oldtextbf

\renewcommand{\currentQueryCard}{0}
\renewcommand*{\arraystretch}{1.1}

\subsection*{BI / read / 4}
\label{sec:bi-read-04}

\let\oldemph\emph
\renewcommand{\emph}[1]{{\footnotesize \sf #1}}
\let\oldtextbf\textbf
\renewcommand{\textbf}[1]{{\it #1}}

\renewcommand{\currentQueryCard}{bi-read-04}
\marginpar{
	\vspace{0.22ex}
	\raggedleft

	\queryRefCard{bi-read-01}{BI}{1}\\
	\queryRefCard{bi-read-02}{BI}{2}\\
	\queryRefCard{bi-read-03}{BI}{3}\\
	\queryRefCard{bi-read-04}{BI}{4}\\
	\queryRefCard{bi-read-05}{BI}{5}\\
	\queryRefCard{bi-read-06}{BI}{6}\\
	\queryRefCard{bi-read-07}{BI}{7}\\
	\queryRefCard{bi-read-08}{BI}{8}\\
	\queryRefCard{bi-read-09}{BI}{9}\\
	\queryRefCard{bi-read-10}{BI}{10}\\
	\queryRefCard{bi-read-11}{BI}{11}\\
	\queryRefCard{bi-read-12}{BI}{12}\\
	\queryRefCard{bi-read-13}{BI}{13}\\
	\queryRefCard{bi-read-14}{BI}{14}\\
	\queryRefCard{bi-read-15}{BI}{15}\\
	\queryRefCard{bi-read-16}{BI}{16}\\
	\queryRefCard{bi-read-17}{BI}{17}\\
	\queryRefCard{bi-read-18}{BI}{18}\\
	\queryRefCard{bi-read-19}{BI}{19}\\
	\queryRefCard{bi-read-20}{BI}{20}\\
}

\noindent\begin{tabularx}{\queryCardWidth}{|>{\queryPropertyCell}p{\queryPropertyCellWidth}|X|}
	\hline
	query & BI / read / 4 \\ \hline
	title & Top message creators by country \\ \hline
	pattern & \centering \includegraphics[scale=\patternscale,margin=0cm .2cm]{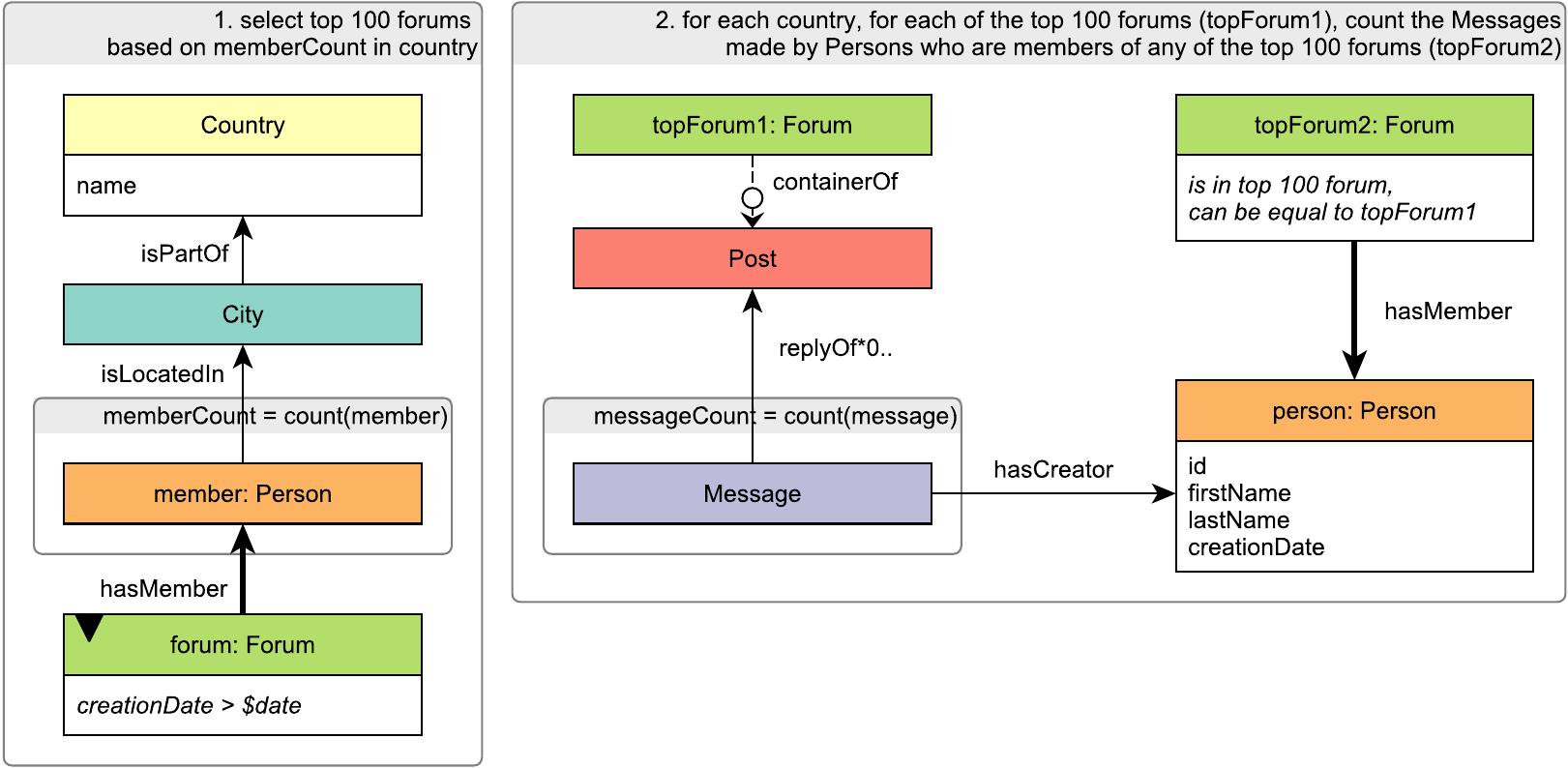} \tabularnewline \hline
	description & Find the most popular \emph{Forums} by \emph{Country}, where the
popularity of a \emph{Forum} is measured by the number of members that
\emph{Forum} has from a given \emph{Country} and the \emph{Forum} was
created after a given \texttt{\$date}.

Calculate the top 100 most popular \emph{Forums}. If a \emph{Forum} is
popular in multiple countries, it should only be calculated once with
its largest membership. In case of a tie, the \emph{Forum} with the
smaller id value should be selected.

For each member \emph{Person} of the 100 most popular \emph{Forums},
count the number of \emph{Messages} (\texttt{messageCount}) they made in
any of those (most popular) \emph{Forums}. Also include those member
\emph{Persons} who have not posted any \emph{Messages} (have a
\texttt{messageCount} of 0).
 \\ \hline

		params &
		\innerCardVSpace{\begin{tabularx}{\attributeCardWidth}{|>{\paramNumberCell}C{\attributeNumberWidth}|>{\varNameCell}M|>{\typeCell}m{\typeWidth}|Y|} \hline
		$\mathsf{1}$ & \$date
 & Date
 & Selected from the first 30 days of the network
 \\ \hline
		\end{tabularx}}\innerCardVSpace \\ \hline

		result &
		\innerCardVSpace{\begin{tabularx}{\attributeCardWidth}{|>{\resultNumberCell}C{\attributeNumberWidth}|>{\varNameCell}M|>{\typeCell}m{\typeWidth}|>{\resultOriginCell}c|Y|} \hline
		$\mathsf{1}$ & person.id & ID & R &
				 \\ \hline
		$\mathsf{2}$ & person.firstName & String & R &
				 \\ \hline
		$\mathsf{3}$ & person.lastName & String & R &
				 \\ \hline
		$\mathsf{4}$ & person.creationDate & DateTime & R &
				 \\ \hline
		$\mathsf{5}$ & messageCount & 32-bit Integer & A &
				 \\ \hline
		\end{tabularx}}\innerCardVSpace \\ \hline

		sort		&
		\innerCardVSpace{\begin{tabularx}{\attributeCardWidth}{|>{\sortNumberCell}C{\attributeNumberWidth}|>{\varNameCell}M|>{\directionCell}c|Y|} \hline
		$\mathsf{1}$ & messageCount
 & $\desc
$ &  \\ \hline
		$\mathsf{2}$ & person.id
 & $\asc
$ &  \\ \hline
		\end{tabularx}}\innerCardVSpace \\ \hline
	limit & 100 \\ \hline
	CPs &
	\multicolumn{1}{>{\raggedright}l|}{
		\chokePoint{1.2}, 
		\chokePoint{1.3}, 
		\chokePoint{2.1}, 
		\chokePoint{2.2}, 
		\chokePoint{2.3}, 
		\chokePoint{2.4}, 
		\chokePoint{3.3}, 
		\chokePoint{5.3}, 
		\chokePoint{6.1}, 
		\chokePoint{8.2}, 
		\chokePoint{8.4}
		} \\ \hline
\end{tabularx}
\queryCardVSpace

\let\emph\oldemph
\let\textbf\oldtextbf

\renewcommand{\currentQueryCard}{0}
\renewcommand*{\arraystretch}{1.1}

\subsection*{BI / read / 5}
\label{sec:bi-read-05}

\let\oldemph\emph
\renewcommand{\emph}[1]{{\footnotesize \sf #1}}
\let\oldtextbf\textbf
\renewcommand{\textbf}[1]{{\it #1}}

\renewcommand{\currentQueryCard}{bi-read-05}
\marginpar{
	\vspace{0.22ex}
	\raggedleft

	\queryRefCard{bi-read-01}{BI}{1}\\
	\queryRefCard{bi-read-02}{BI}{2}\\
	\queryRefCard{bi-read-03}{BI}{3}\\
	\queryRefCard{bi-read-04}{BI}{4}\\
	\queryRefCard{bi-read-05}{BI}{5}\\
	\queryRefCard{bi-read-06}{BI}{6}\\
	\queryRefCard{bi-read-07}{BI}{7}\\
	\queryRefCard{bi-read-08}{BI}{8}\\
	\queryRefCard{bi-read-09}{BI}{9}\\
	\queryRefCard{bi-read-10}{BI}{10}\\
	\queryRefCard{bi-read-11}{BI}{11}\\
	\queryRefCard{bi-read-12}{BI}{12}\\
	\queryRefCard{bi-read-13}{BI}{13}\\
	\queryRefCard{bi-read-14}{BI}{14}\\
	\queryRefCard{bi-read-15}{BI}{15}\\
	\queryRefCard{bi-read-16}{BI}{16}\\
	\queryRefCard{bi-read-17}{BI}{17}\\
	\queryRefCard{bi-read-18}{BI}{18}\\
	\queryRefCard{bi-read-19}{BI}{19}\\
	\queryRefCard{bi-read-20}{BI}{20}\\
}

\noindent\begin{tabularx}{\queryCardWidth}{|>{\queryPropertyCell}p{\queryPropertyCellWidth}|X|}
	\hline
	query & BI / read / 5 \\ \hline
	title & Most active posters of a given topic \\ \hline
	pattern & \centering \includegraphics[scale=\patternscale,margin=0cm .2cm]{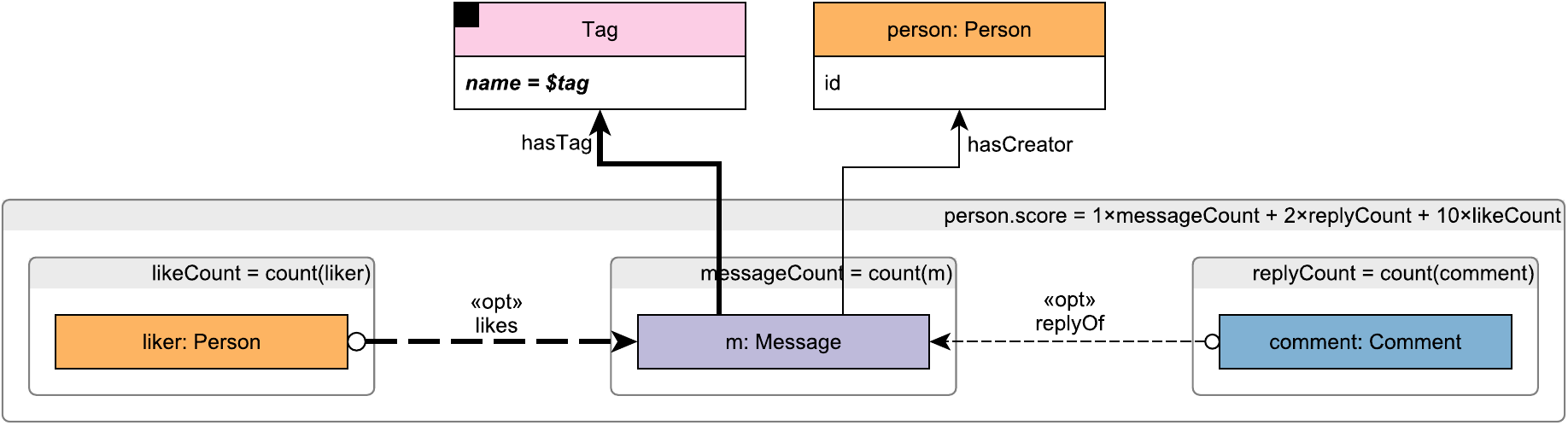} \tabularnewline \hline
	description & Get each \emph{Person} (\texttt{person}) who has created a
\emph{Message} (\texttt{message}) with a given \texttt{\$tag} (direct
relation, not transitive). Considering only these \emph{Messages}, for
each \emph{Person}~node:

\begin{itemize}
\tightlist
\item
  Count its \emph{Messages} (\texttt{messageCount}).
\item
  Count \emph{likes} (\texttt{likeCount}) to its \emph{Messages}.
\item
  Count \emph{Comments} (\texttt{replyCount}) in reply to its
  \emph{Messages}.
\end{itemize}

The \texttt{score} is calculated according to the following formula:
\(1 \times \texttt{messageCount} + 2 \times \texttt{replyCount} + 10 \times \texttt{likeCount}\).
 \\ \hline

		params &
		\innerCardVSpace{\begin{tabularx}{\attributeCardWidth}{|>{\paramNumberCell}C{\attributeNumberWidth}|>{\varNameCell}M|>{\typeCell}m{\typeWidth}|Y|} \hline
		$\mathsf{1}$ & \$tag
 & Long String
 & \emph{Tags} with a similar amount of \emph{Messages} are selected. To
avoid caching, different \emph{Tags} should be used than the ones in Q6
and Q7.
 \\ \hline
		\end{tabularx}}\innerCardVSpace \\ \hline

		result &
		\innerCardVSpace{\begin{tabularx}{\attributeCardWidth}{|>{\resultNumberCell}C{\attributeNumberWidth}|>{\varNameCell}M|>{\typeCell}m{\typeWidth}|>{\resultOriginCell}c|Y|} \hline
		$\mathsf{1}$ & person.id & ID & R &
				 \\ \hline
		$\mathsf{2}$ & replyCount & 32-bit Integer & A &
				 \\ \hline
		$\mathsf{3}$ & likeCount & 32-bit Integer & A &
				 \\ \hline
		$\mathsf{4}$ & messageCount & 32-bit Integer & A &
				 \\ \hline
		$\mathsf{5}$ & score & 32-bit Integer & A &
				 \\ \hline
		\end{tabularx}}\innerCardVSpace \\ \hline

		sort		&
		\innerCardVSpace{\begin{tabularx}{\attributeCardWidth}{|>{\sortNumberCell}C{\attributeNumberWidth}|>{\varNameCell}M|>{\directionCell}c|Y|} \hline
		$\mathsf{1}$ & score
 & $\desc
$ &  \\ \hline
		$\mathsf{2}$ & person.id
 & $\asc
$ &  \\ \hline
		\end{tabularx}}\innerCardVSpace \\ \hline
	limit & 100 \\ \hline
	CPs &
	\multicolumn{1}{>{\raggedright}l|}{
		\chokePoint{1.2}, 
		\chokePoint{2.3}, 
		\chokePoint{2.6}, 
		\chokePoint{8.2}
		} \\ \hline
\end{tabularx}
\queryCardVSpace

\let\emph\oldemph
\let\textbf\oldtextbf

\renewcommand{\currentQueryCard}{0}
\renewcommand*{\arraystretch}{1.1}

\subsection*{BI / read / 6}
\label{sec:bi-read-06}

\let\oldemph\emph
\renewcommand{\emph}[1]{{\footnotesize \sf #1}}
\let\oldtextbf\textbf
\renewcommand{\textbf}[1]{{\it #1}}

\renewcommand{\currentQueryCard}{bi-read-06}
\marginpar{
	\vspace{0.22ex}
	\raggedleft

	\queryRefCard{bi-read-01}{BI}{1}\\
	\queryRefCard{bi-read-02}{BI}{2}\\
	\queryRefCard{bi-read-03}{BI}{3}\\
	\queryRefCard{bi-read-04}{BI}{4}\\
	\queryRefCard{bi-read-05}{BI}{5}\\
	\queryRefCard{bi-read-06}{BI}{6}\\
	\queryRefCard{bi-read-07}{BI}{7}\\
	\queryRefCard{bi-read-08}{BI}{8}\\
	\queryRefCard{bi-read-09}{BI}{9}\\
	\queryRefCard{bi-read-10}{BI}{10}\\
	\queryRefCard{bi-read-11}{BI}{11}\\
	\queryRefCard{bi-read-12}{BI}{12}\\
	\queryRefCard{bi-read-13}{BI}{13}\\
	\queryRefCard{bi-read-14}{BI}{14}\\
	\queryRefCard{bi-read-15}{BI}{15}\\
	\queryRefCard{bi-read-16}{BI}{16}\\
	\queryRefCard{bi-read-17}{BI}{17}\\
	\queryRefCard{bi-read-18}{BI}{18}\\
	\queryRefCard{bi-read-19}{BI}{19}\\
	\queryRefCard{bi-read-20}{BI}{20}\\
}

\noindent\begin{tabularx}{\queryCardWidth}{|>{\queryPropertyCell}p{\queryPropertyCellWidth}|X|}
	\hline
	query & BI / read / 6 \\ \hline
	title & Most authoritative users on a given topic \\ \hline
	pattern & \centering \includegraphics[scale=\patternscale,margin=0cm .2cm]{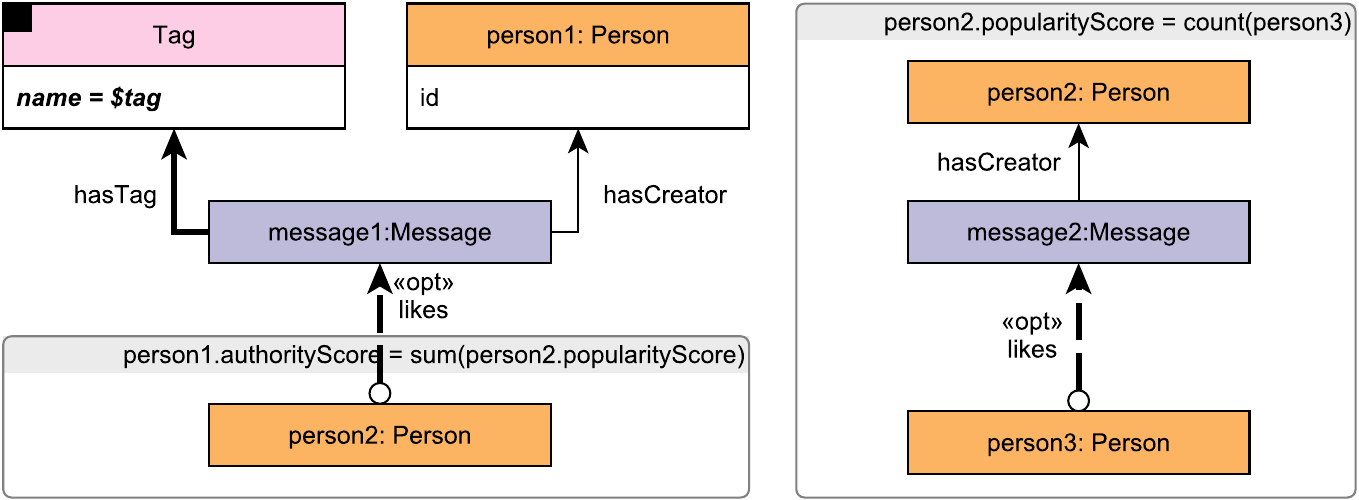} \tabularnewline \hline
	description & Given a \texttt{\$tag}, find all \emph{Persons} (\texttt{person1}) that
ever created a \emph{Message} with the \texttt{\$tag}. For each of these
\emph{Persons} (\texttt{person1}) compute their ``authority score'' as
follows:

\begin{itemize}
\tightlist
\item
  The ``authority score'' is the sum of ``popularity scores'' of the
  \emph{Persons} (\texttt{person2}) that liked any of that
  \emph{Person}'s \emph{Messages} with the given \texttt{\$tag} (same
  criterion as for \texttt{message1}).
\item
  A \emph{Person}'s (\texttt{person2}) ``popularity score'' is defined
  as the total number of likes (by any \emph{Person} \texttt{person3})
  on any of their \emph{Messages} (\texttt{message2}).
\end{itemize}
 \\ \hline

		params &
		\innerCardVSpace{\begin{tabularx}{\attributeCardWidth}{|>{\paramNumberCell}C{\attributeNumberWidth}|>{\varNameCell}M|>{\typeCell}m{\typeWidth}|Y|} \hline
		$\mathsf{1}$ & \$tag
 & Long String
 & \emph{Tags} with a similar amount of \emph{Messages} are selected. To
avoid caching, different \emph{Tags} should be used than the ones in Q5
and Q7.
 \\ \hline
		\end{tabularx}}\innerCardVSpace \\ \hline

		result &
		\innerCardVSpace{\begin{tabularx}{\attributeCardWidth}{|>{\resultNumberCell}C{\attributeNumberWidth}|>{\varNameCell}M|>{\typeCell}m{\typeWidth}|>{\resultOriginCell}c|Y|} \hline
		$\mathsf{1}$ & person1.id & ID & R &
				 \\ \hline
		$\mathsf{2}$ & authorityScore & 32-bit Integer & A &
				 \\ \hline
		\end{tabularx}}\innerCardVSpace \\ \hline

		sort		&
		\innerCardVSpace{\begin{tabularx}{\attributeCardWidth}{|>{\sortNumberCell}C{\attributeNumberWidth}|>{\varNameCell}M|>{\directionCell}c|Y|} \hline
		$\mathsf{1}$ & authorityScore
 & $\desc
$ &  \\ \hline
		$\mathsf{2}$ & person1.id
 & $\asc
$ &  \\ \hline
		\end{tabularx}}\innerCardVSpace \\ \hline
	limit & 100 \\ \hline
	CPs &
	\multicolumn{1}{>{\raggedright}l|}{
		\chokePoint{1.2}, 
		\chokePoint{2.3}, 
		\chokePoint{2.6}, 
		\chokePoint{3.3}, 
		\chokePoint{6.1}, 
		\chokePoint{8.2}
		} \\ \hline
	relevance &
		\footnotesize Computing the authority scores might involve computing the popularity
score for the same \emph{Person} multiple times. Implementations are
advised to avoid such redundant computations.
 \\ \hline%
\end{tabularx}
\queryCardVSpace

\let\emph\oldemph
\let\textbf\oldtextbf

\renewcommand{\currentQueryCard}{0}
\renewcommand*{\arraystretch}{1.1}

\subsection*{BI / read / 7}
\label{sec:bi-read-07}

\let\oldemph\emph
\renewcommand{\emph}[1]{{\footnotesize \sf #1}}
\let\oldtextbf\textbf
\renewcommand{\textbf}[1]{{\it #1}}

\renewcommand{\currentQueryCard}{bi-read-07}
\marginpar{
	\vspace{0.22ex}
	\raggedleft

	\queryRefCard{bi-read-01}{BI}{1}\\
	\queryRefCard{bi-read-02}{BI}{2}\\
	\queryRefCard{bi-read-03}{BI}{3}\\
	\queryRefCard{bi-read-04}{BI}{4}\\
	\queryRefCard{bi-read-05}{BI}{5}\\
	\queryRefCard{bi-read-06}{BI}{6}\\
	\queryRefCard{bi-read-07}{BI}{7}\\
	\queryRefCard{bi-read-08}{BI}{8}\\
	\queryRefCard{bi-read-09}{BI}{9}\\
	\queryRefCard{bi-read-10}{BI}{10}\\
	\queryRefCard{bi-read-11}{BI}{11}\\
	\queryRefCard{bi-read-12}{BI}{12}\\
	\queryRefCard{bi-read-13}{BI}{13}\\
	\queryRefCard{bi-read-14}{BI}{14}\\
	\queryRefCard{bi-read-15}{BI}{15}\\
	\queryRefCard{bi-read-16}{BI}{16}\\
	\queryRefCard{bi-read-17}{BI}{17}\\
	\queryRefCard{bi-read-18}{BI}{18}\\
	\queryRefCard{bi-read-19}{BI}{19}\\
	\queryRefCard{bi-read-20}{BI}{20}\\
}

\noindent\begin{tabularx}{\queryCardWidth}{|>{\queryPropertyCell}p{\queryPropertyCellWidth}|X|}
	\hline
	query & BI / read / 7 \\ \hline
	title & Related topics \\ \hline
	pattern & \centering \includegraphics[scale=\patternscale,margin=0cm .2cm]{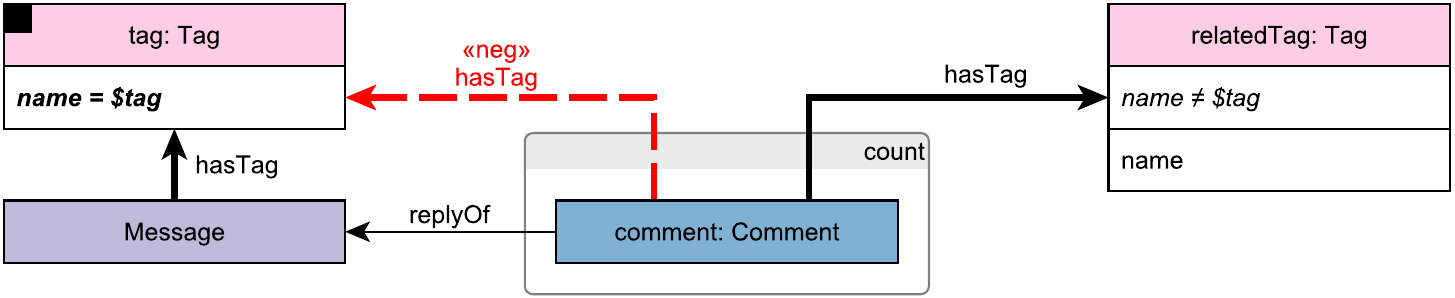} \tabularnewline \hline
	description & Find all \emph{Messages} that have a given \texttt{\$tag}. Find the
related \emph{Tags} attached to (direct) reply \emph{Comments} of these
\emph{Messages}, but only of those reply \emph{Comments} that do not
have the given \texttt{\$tag}.

Group the related \emph{Tags} by name, and get the count of replies in
each group.
 \\ \hline

		params &
		\innerCardVSpace{\begin{tabularx}{\attributeCardWidth}{|>{\paramNumberCell}C{\attributeNumberWidth}|>{\varNameCell}M|>{\typeCell}m{\typeWidth}|Y|} \hline
		$\mathsf{1}$ & \$tag
 & Long String
 & \emph{Tags} with a similar amount of \emph{Messages} are selected. To
avoid caching, different \emph{Tags} should be used than the ones in Q5
and Q6.
 \\ \hline
		\end{tabularx}}\innerCardVSpace \\ \hline

		result &
		\innerCardVSpace{\begin{tabularx}{\attributeCardWidth}{|>{\resultNumberCell}C{\attributeNumberWidth}|>{\varNameCell}M|>{\typeCell}m{\typeWidth}|>{\resultOriginCell}c|Y|} \hline
		$\mathsf{1}$ & relatedTag.name & Long String & R &
				 \\ \hline
		$\mathsf{2}$ & count & 32-bit Integer & A &
				 \\ \hline
		\end{tabularx}}\innerCardVSpace \\ \hline

		sort		&
		\innerCardVSpace{\begin{tabularx}{\attributeCardWidth}{|>{\sortNumberCell}C{\attributeNumberWidth}|>{\varNameCell}M|>{\directionCell}c|Y|} \hline
		$\mathsf{1}$ & count
 & $\desc
$ &  \\ \hline
		$\mathsf{2}$ & relatedTag.name
 & $\asc
$ &  \\ \hline
		\end{tabularx}}\innerCardVSpace \\ \hline
	limit & 100 \\ \hline
	CPs &
	\multicolumn{1}{>{\raggedright}l|}{
		\chokePoint{1.4}, 
		\chokePoint{3.3}, 
		\chokePoint{5.2}, 
		\chokePoint{8.1}
		} \\ \hline
\end{tabularx}
\queryCardVSpace

\let\emph\oldemph
\let\textbf\oldtextbf

\renewcommand{\currentQueryCard}{0}
\renewcommand*{\arraystretch}{1.1}

\subsection*{BI / read / 8}
\label{sec:bi-read-08}

\let\oldemph\emph
\renewcommand{\emph}[1]{{\footnotesize \sf #1}}
\let\oldtextbf\textbf
\renewcommand{\textbf}[1]{{\it #1}}

\renewcommand{\currentQueryCard}{bi-read-08}
\marginpar{
	\vspace{0.22ex}
	\raggedleft

	\queryRefCard{bi-read-01}{BI}{1}\\
	\queryRefCard{bi-read-02}{BI}{2}\\
	\queryRefCard{bi-read-03}{BI}{3}\\
	\queryRefCard{bi-read-04}{BI}{4}\\
	\queryRefCard{bi-read-05}{BI}{5}\\
	\queryRefCard{bi-read-06}{BI}{6}\\
	\queryRefCard{bi-read-07}{BI}{7}\\
	\queryRefCard{bi-read-08}{BI}{8}\\
	\queryRefCard{bi-read-09}{BI}{9}\\
	\queryRefCard{bi-read-10}{BI}{10}\\
	\queryRefCard{bi-read-11}{BI}{11}\\
	\queryRefCard{bi-read-12}{BI}{12}\\
	\queryRefCard{bi-read-13}{BI}{13}\\
	\queryRefCard{bi-read-14}{BI}{14}\\
	\queryRefCard{bi-read-15}{BI}{15}\\
	\queryRefCard{bi-read-16}{BI}{16}\\
	\queryRefCard{bi-read-17}{BI}{17}\\
	\queryRefCard{bi-read-18}{BI}{18}\\
	\queryRefCard{bi-read-19}{BI}{19}\\
	\queryRefCard{bi-read-20}{BI}{20}\\
}

\noindent\begin{tabularx}{\queryCardWidth}{|>{\queryPropertyCell}p{\queryPropertyCellWidth}|X|}
	\hline
	query & BI / read / 8 \\ \hline
	title & Central person for a tag \\ \hline
	pattern & \centering \includegraphics[scale=\patternscale,margin=0cm .2cm]{patterns/bi-read-08} \tabularnewline \hline
	description & Given a \texttt{\$tag}, find all \emph{Persons} that are interested in
the \texttt{\$tag} and/or have written a \emph{Message} (\emph{Post} or
\emph{Comment}) with a \texttt{creationDate} after a given
\texttt{\$startDate} and that has a given \texttt{\$tag}. For each
\emph{Person}, compute the \texttt{score} as the sum of the following
two aspects:

\begin{itemize}
\tightlist
\item
  100, if the \emph{Person} has this \texttt{\$tag} as their interest,
  or 0 otherwise
\item
  number of \emph{Messages} by this \emph{Person} with the given
  \texttt{\$tag}
\end{itemize}

Also, for each \emph{Person}, compute the sum of the score of the
\emph{Person}'s friends (\texttt{friendsScore}).
 \\ \hline

		params &
		\innerCardVSpace{\begin{tabularx}{\attributeCardWidth}{|>{\paramNumberCell}C{\attributeNumberWidth}|>{\varNameCell}M|>{\typeCell}m{\typeWidth}|Y|} \hline
		$\mathsf{1}$ & \$tag
 & Long String
 & \emph{Tags} with a similar amount of \emph{Messages} are selected
 \\ \hline
		$\mathsf{2}$ & \$startDate
 & Date
 & \texttt{(a)}: A range during which a flashmob event happened (it should
yield at least a 5× difference)

\texttt{(b)}: A regular range (does not include a flashmob event)
 \\ \hline
		$\mathsf{3}$ & \$endDate
 & Date
 &  \\ \hline
		\end{tabularx}}\innerCardVSpace \\ \hline

		result &
		\innerCardVSpace{\begin{tabularx}{\attributeCardWidth}{|>{\resultNumberCell}C{\attributeNumberWidth}|>{\varNameCell}M|>{\typeCell}m{\typeWidth}|>{\resultOriginCell}c|Y|} \hline
		$\mathsf{1}$ & person.id & ID & R &
				 \\ \hline
		$\mathsf{2}$ & score & 32-bit Integer & A &
				 \\ \hline
		$\mathsf{3}$ & friendsScore & 32-bit Integer & A &
				The sum of the score of the \texttt{person}'s friends
 \\ \hline
		\end{tabularx}}\innerCardVSpace \\ \hline

		sort		&
		\innerCardVSpace{\begin{tabularx}{\attributeCardWidth}{|>{\sortNumberCell}C{\attributeNumberWidth}|>{\varNameCell}M|>{\directionCell}c|Y|} \hline
		$\mathsf{1}$ & score + friendsScore
 & $\desc
$ &  \\ \hline
		$\mathsf{2}$ & person.id
 & $\asc
$ &  \\ \hline
		\end{tabularx}}\innerCardVSpace \\ \hline
	limit & 100 \\ \hline
	CPs &
	\multicolumn{1}{>{\raggedright}l|}{
		\chokePoint{1.2}, 
		\chokePoint{2.1}, 
		\chokePoint{2.3}, 
		\chokePoint{3.2}, 
		\chokePoint{5.3}, 
		\chokePoint{8.2}, 
		\chokePoint{8.4}, 
		\chokePoint{8.5}
		} \\ \hline
	relevance &
		\footnotesize Similarly to BI 16, there are two major ways to compute this query: (1)
creating an induced subgraph of the interested \emph{Persons} and their
friends and performing the scoring on this graph or (2) performing the
scoring without creating an induced subgraph and scoring the friends of
a \emph{Person} on-the-fly. The first approach is more efficient as it
avoids redundant computations, however, specifying it needs support for
composable graph queries.
 \\ \hline%
\end{tabularx}
\queryCardVSpace

\let\emph\oldemph
\let\textbf\oldtextbf

\renewcommand{\currentQueryCard}{0}
\renewcommand*{\arraystretch}{1.1}

\subsection*{BI / read / 9}
\label{sec:bi-read-09}

\let\oldemph\emph
\renewcommand{\emph}[1]{{\footnotesize \sf #1}}
\let\oldtextbf\textbf
\renewcommand{\textbf}[1]{{\it #1}}

\renewcommand{\currentQueryCard}{bi-read-09}
\marginpar{
	\vspace{0.22ex}
	\raggedleft

	\queryRefCard{bi-read-01}{BI}{1}\\
	\queryRefCard{bi-read-02}{BI}{2}\\
	\queryRefCard{bi-read-03}{BI}{3}\\
	\queryRefCard{bi-read-04}{BI}{4}\\
	\queryRefCard{bi-read-05}{BI}{5}\\
	\queryRefCard{bi-read-06}{BI}{6}\\
	\queryRefCard{bi-read-07}{BI}{7}\\
	\queryRefCard{bi-read-08}{BI}{8}\\
	\queryRefCard{bi-read-09}{BI}{9}\\
	\queryRefCard{bi-read-10}{BI}{10}\\
	\queryRefCard{bi-read-11}{BI}{11}\\
	\queryRefCard{bi-read-12}{BI}{12}\\
	\queryRefCard{bi-read-13}{BI}{13}\\
	\queryRefCard{bi-read-14}{BI}{14}\\
	\queryRefCard{bi-read-15}{BI}{15}\\
	\queryRefCard{bi-read-16}{BI}{16}\\
	\queryRefCard{bi-read-17}{BI}{17}\\
	\queryRefCard{bi-read-18}{BI}{18}\\
	\queryRefCard{bi-read-19}{BI}{19}\\
	\queryRefCard{bi-read-20}{BI}{20}\\
}

\noindent\begin{tabularx}{\queryCardWidth}{|>{\queryPropertyCell}p{\queryPropertyCellWidth}|X|}
	\hline
	query & BI / read / 9 \\ \hline
	title & Top thread initiators \\ \hline
	pattern & \centering \includegraphics[scale=\patternscale,margin=0cm .2cm]{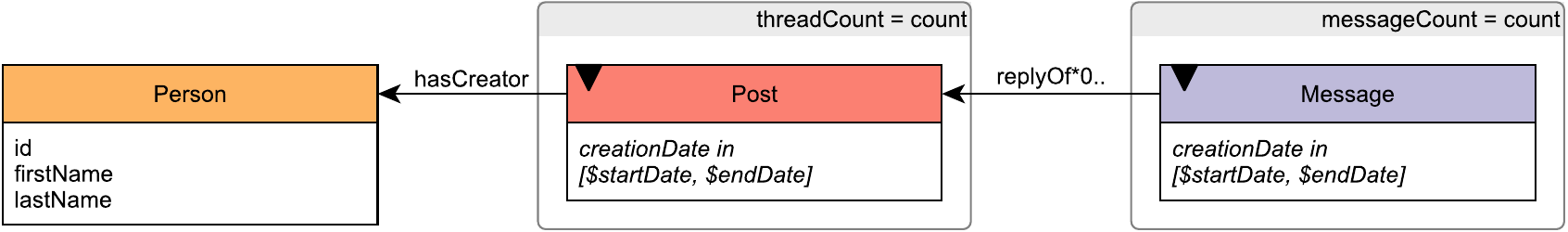} \tabularnewline \hline
	description & For each \emph{Person}, count the number of \emph{Posts} they created in
the time interval \texttt{{[}\$startDate,\ \$endDate{]}} (equivalent to
the number of threads they initiated) and the number of \emph{Messages}
in each of their (transitive) reply trees, including the root
\emph{Post} of each tree. When calculating \emph{Message} counts only
consider \emph{Messages} created within the given time interval.

Return each \emph{Person}, number of \emph{Posts} they created, and the
count of all \emph{Messages} that appeared in the reply trees (including
the \emph{Post} at the root of tree).
 \\ \hline

		params &
		\innerCardVSpace{\begin{tabularx}{\attributeCardWidth}{|>{\paramNumberCell}C{\attributeNumberWidth}|>{\varNameCell}M|>{\typeCell}m{\typeWidth}|Y|} \hline
		$\mathsf{1}$ & \$startDate
 & Date
 & Selected around the same date
 \\ \hline
		$\mathsf{2}$ & \$endDate
 & Date
 & 80-100 days after the \texttt{\$startDate}
 \\ \hline
		\end{tabularx}}\innerCardVSpace \\ \hline

		result &
		\innerCardVSpace{\begin{tabularx}{\attributeCardWidth}{|>{\resultNumberCell}C{\attributeNumberWidth}|>{\varNameCell}M|>{\typeCell}m{\typeWidth}|>{\resultOriginCell}c|Y|} \hline
		$\mathsf{1}$ & person.id & ID & R &
				 \\ \hline
		$\mathsf{2}$ & person.firstName & String & R &
				 \\ \hline
		$\mathsf{3}$ & person.lastName & String & R &
				 \\ \hline
		$\mathsf{4}$ & threadCount & 32-bit Integer & A &
				The number of \emph{Posts} created by that \emph{Person} (the number of
threads initiated)
 \\ \hline
		$\mathsf{5}$ & messageCount & 32-bit Integer & A &
				The number of \emph{Messages} created in all the threads this
\emph{Person} initiated
 \\ \hline
		\end{tabularx}}\innerCardVSpace \\ \hline

		sort		&
		\innerCardVSpace{\begin{tabularx}{\attributeCardWidth}{|>{\sortNumberCell}C{\attributeNumberWidth}|>{\varNameCell}M|>{\directionCell}c|Y|} \hline
		$\mathsf{1}$ & messageCount
 & $\desc
$ &  \\ \hline
		$\mathsf{2}$ & person.id
 & $\asc
$ &  \\ \hline
		\end{tabularx}}\innerCardVSpace \\ \hline
	limit & 100 \\ \hline
	CPs &
	\multicolumn{1}{>{\raggedright}l|}{
		\chokePoint{1.2}, 
		\chokePoint{2.2}, 
		\chokePoint{2.3}, 
		\chokePoint{2.6}, 
		\chokePoint{3.2}, 
		\chokePoint{7.2}, 
		\chokePoint{7.3}, 
		\chokePoint{7.4}, 
		\chokePoint{8.1}, 
		\chokePoint{8.5}
		} \\ \hline
\end{tabularx}
\queryCardVSpace

\let\emph\oldemph
\let\textbf\oldtextbf

\renewcommand{\currentQueryCard}{0}
\renewcommand*{\arraystretch}{1.1}

\subsection*{BI / read / 10}
\label{sec:bi-read-10}

\let\oldemph\emph
\renewcommand{\emph}[1]{{\footnotesize \sf #1}}
\let\oldtextbf\textbf
\renewcommand{\textbf}[1]{{\it #1}}

\renewcommand{\currentQueryCard}{bi-read-10}
\marginpar{
	\vspace{0.22ex}
	\raggedleft

	\queryRefCard{bi-read-01}{BI}{1}\\
	\queryRefCard{bi-read-02}{BI}{2}\\
	\queryRefCard{bi-read-03}{BI}{3}\\
	\queryRefCard{bi-read-04}{BI}{4}\\
	\queryRefCard{bi-read-05}{BI}{5}\\
	\queryRefCard{bi-read-06}{BI}{6}\\
	\queryRefCard{bi-read-07}{BI}{7}\\
	\queryRefCard{bi-read-08}{BI}{8}\\
	\queryRefCard{bi-read-09}{BI}{9}\\
	\queryRefCard{bi-read-10}{BI}{10}\\
	\queryRefCard{bi-read-11}{BI}{11}\\
	\queryRefCard{bi-read-12}{BI}{12}\\
	\queryRefCard{bi-read-13}{BI}{13}\\
	\queryRefCard{bi-read-14}{BI}{14}\\
	\queryRefCard{bi-read-15}{BI}{15}\\
	\queryRefCard{bi-read-16}{BI}{16}\\
	\queryRefCard{bi-read-17}{BI}{17}\\
	\queryRefCard{bi-read-18}{BI}{18}\\
	\queryRefCard{bi-read-19}{BI}{19}\\
	\queryRefCard{bi-read-20}{BI}{20}\\
}

\noindent\begin{tabularx}{\queryCardWidth}{|>{\queryPropertyCell}p{\queryPropertyCellWidth}|X|}
	\hline
	query & BI / read / 10 \\ \hline
	title & Experts in social circle \\ \hline
	pattern & \centering \includegraphics[scale=\patternscale,margin=0cm .2cm]{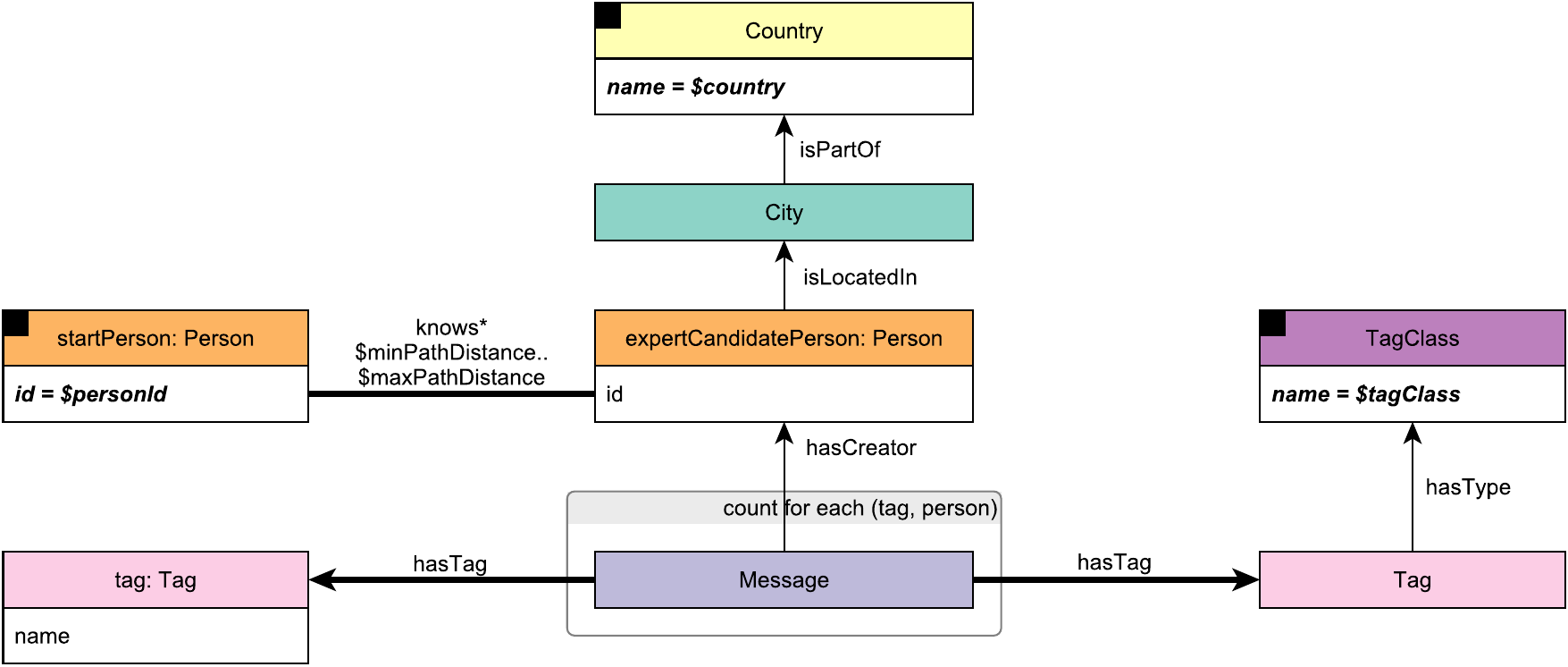} \tabularnewline \hline
	description & Given a \emph{Person} \texttt{startPerson} with ID \texttt{\$personID},
find all other \emph{Persons} (\texttt{expertCandidatePerson}) that live
in a given \texttt{\$country} and are connected to the
\texttt{startPerson} on a \textbf{shortest path} with length in range
\texttt{{[}\$minPathDistance,\ \$maxPathDistance{]}} through the
\emph{knows} relation.

For each of these \texttt{expertCandidatePerson} nodes, retrieve all of
their \emph{Messages} that contain at least one \emph{Tag} belonging to
a given \texttt{\$tagClass} (direct relation not transitive). For each
\emph{Message}, retrieve all of its \emph{Tags}.

Group the results by \emph{Persons} and \emph{Tags}, then count the
\emph{Messages} by a certain \emph{Person} having a certain \emph{Tag}.
 \\ \hline

		params &
		\innerCardVSpace{\begin{tabularx}{\attributeCardWidth}{|>{\paramNumberCell}C{\attributeNumberWidth}|>{\varNameCell}M|>{\typeCell}m{\typeWidth}|Y|} \hline
		$\mathsf{1}$ & \$personId
 & ID
 & \texttt{(a)} \emph{Persons} with an average degree of \emph{knows} edges
are selected

\texttt{(b)} \emph{Persons} who have only one friend and that
\emph{Person} has two friends in total (including the original
\emph{Person})
 \\ \hline
		$\mathsf{2}$ & \$country
 & String
 & Select mid-sized \emph{Countries}
 \\ \hline
		$\mathsf{3}$ & \$tagClass
 & Long String
 & \emph{TagClasses} with a similar degree of \emph{hasType} edges are
selected
 \\ \hline
		$\mathsf{4}$ & \$minPathDistance
 & 32-bit Integer
 & 3
 \\ \hline
		$\mathsf{5}$ & \$maxPathDistance
 & 32-bit Integer
 & 4
 \\ \hline
		\end{tabularx}}\innerCardVSpace \\ \hline

		result &
		\innerCardVSpace{\begin{tabularx}{\attributeCardWidth}{|>{\resultNumberCell}C{\attributeNumberWidth}|>{\varNameCell}M|>{\typeCell}m{\typeWidth}|>{\resultOriginCell}c|Y|} \hline
		$\mathsf{1}$ & expertCandidatePerson.id & ID & R &
				 \\ \hline
		$\mathsf{2}$ & tag.name & Long String & R &
				 \\ \hline
		$\mathsf{3}$ & messageCount & 32-bit Integer & A &
				Number of \emph{Messages} created by that \emph{Person} containing that
\emph{Tag}
 \\ \hline
		\end{tabularx}}\innerCardVSpace \\ \hline

		sort		&
		\innerCardVSpace{\begin{tabularx}{\attributeCardWidth}{|>{\sortNumberCell}C{\attributeNumberWidth}|>{\varNameCell}M|>{\directionCell}c|Y|} \hline
		$\mathsf{1}$ & messageCount
 & $\desc
$ &  \\ \hline
		$\mathsf{2}$ & tag.name
 & $\asc
$ &  \\ \hline
		$\mathsf{3}$ & expertCandidatePerson.id
 & $\asc
$ &  \\ \hline
		\end{tabularx}}\innerCardVSpace \\ \hline
	limit & 100 \\ \hline
	CPs &
	\multicolumn{1}{>{\raggedright}l|}{
		\chokePoint{1.2}, 
		\chokePoint{1.3}, 
		\chokePoint{2.3}, 
		\chokePoint{2.4}, 
		\chokePoint{2.6}, 
		\chokePoint{3.3}, 
		\chokePoint{5.3}, 
		\chokePoint{7.1}, 
		\chokePoint{7.2}, 
		\chokePoint{7.3}, 
		\chokePoint{8.1}, 
		\chokePoint{8.6}
		} \\ \hline
\end{tabularx}
\queryCardVSpace

\let\emph\oldemph
\let\textbf\oldtextbf

\renewcommand{\currentQueryCard}{0}
\renewcommand*{\arraystretch}{1.1}

\subsection*{BI / read / 11}
\label{sec:bi-read-11}

\let\oldemph\emph
\renewcommand{\emph}[1]{{\footnotesize \sf #1}}
\let\oldtextbf\textbf
\renewcommand{\textbf}[1]{{\it #1}}

\renewcommand{\currentQueryCard}{bi-read-11}
\marginpar{
	\vspace{0.22ex}
	\raggedleft

	\queryRefCard{bi-read-01}{BI}{1}\\
	\queryRefCard{bi-read-02}{BI}{2}\\
	\queryRefCard{bi-read-03}{BI}{3}\\
	\queryRefCard{bi-read-04}{BI}{4}\\
	\queryRefCard{bi-read-05}{BI}{5}\\
	\queryRefCard{bi-read-06}{BI}{6}\\
	\queryRefCard{bi-read-07}{BI}{7}\\
	\queryRefCard{bi-read-08}{BI}{8}\\
	\queryRefCard{bi-read-09}{BI}{9}\\
	\queryRefCard{bi-read-10}{BI}{10}\\
	\queryRefCard{bi-read-11}{BI}{11}\\
	\queryRefCard{bi-read-12}{BI}{12}\\
	\queryRefCard{bi-read-13}{BI}{13}\\
	\queryRefCard{bi-read-14}{BI}{14}\\
	\queryRefCard{bi-read-15}{BI}{15}\\
	\queryRefCard{bi-read-16}{BI}{16}\\
	\queryRefCard{bi-read-17}{BI}{17}\\
	\queryRefCard{bi-read-18}{BI}{18}\\
	\queryRefCard{bi-read-19}{BI}{19}\\
	\queryRefCard{bi-read-20}{BI}{20}\\
}

\noindent\begin{tabularx}{\queryCardWidth}{|>{\queryPropertyCell}p{\queryPropertyCellWidth}|X|}
	\hline
	query & BI / read / 11 \\ \hline
	title & Friend triangles \\ \hline
	pattern & \centering \includegraphics[scale=\patternscale,margin=0cm .2cm]{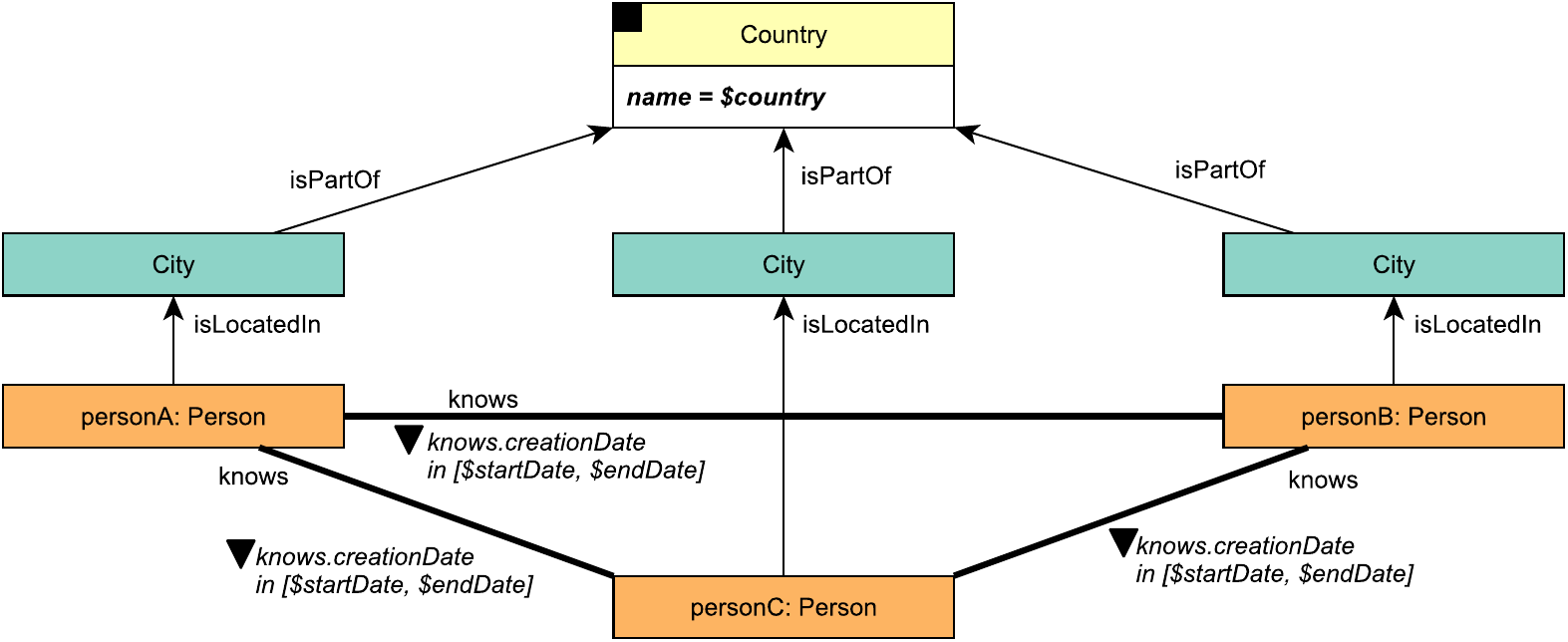} \tabularnewline \hline
	description & For a given \texttt{\$country}, count all the distinct triples of
\emph{Persons} such that:

\begin{itemize}
\tightlist
\item
  \texttt{personA} is friend of \texttt{personB},
\item
  \texttt{personB} is friend of \texttt{personC},
\item
  \texttt{personC} is friend of \texttt{personA},
\end{itemize}

and these friendships were created in the range
\texttt{{[}\$startDate,\ \$endDate{]}}.

Distinct means that given a triple \(t_1\) in the result set \(R\) of
all qualified triples, there is no triple \(t_2\) in \(R\) such that
\(t_1\) and \(t_2\) have the same set of elements.
 \\ \hline

		params &
		\innerCardVSpace{\begin{tabularx}{\attributeCardWidth}{|>{\paramNumberCell}C{\attributeNumberWidth}|>{\varNameCell}M|>{\typeCell}m{\typeWidth}|Y|} \hline
		$\mathsf{1}$ & \$country
 & Long String
 & Selected from the largest \emph{Countries} (India, China)
 \\ \hline
		$\mathsf{2}$ & \$startDate
 & Date
 & Selected from a 30-day interval towards the end of the simulation time
 \\ \hline
		$\mathsf{3}$ & \$endDate
 & Date
 & Selected to yield around a 100-day interval
 \\ \hline
		\end{tabularx}}\innerCardVSpace \\ \hline

		result &
		\innerCardVSpace{\begin{tabularx}{\attributeCardWidth}{|>{\resultNumberCell}C{\attributeNumberWidth}|>{\varNameCell}M|>{\typeCell}m{\typeWidth}|>{\resultOriginCell}c|Y|} \hline
		$\mathsf{1}$ & count & 64-bit Integer & A &
				 \\ \hline
		\end{tabularx}}\innerCardVSpace \\ \hline

	limit & n/a \\ \hline
	CPs &
	\multicolumn{1}{>{\raggedright}l|}{
		\chokePoint{2.3}, 
		\chokePoint{2.5}, 
		\chokePoint{3.2}
		} \\ \hline
\end{tabularx}
\queryCardVSpace

\let\emph\oldemph
\let\textbf\oldtextbf

\renewcommand{\currentQueryCard}{0}
\renewcommand*{\arraystretch}{1.1}

\subsection*{BI / read / 12}
\label{sec:bi-read-12}

\let\oldemph\emph
\renewcommand{\emph}[1]{{\footnotesize \sf #1}}
\let\oldtextbf\textbf
\renewcommand{\textbf}[1]{{\it #1}}

\renewcommand{\currentQueryCard}{bi-read-12}
\marginpar{
	\vspace{0.22ex}
	\raggedleft

	\queryRefCard{bi-read-01}{BI}{1}\\
	\queryRefCard{bi-read-02}{BI}{2}\\
	\queryRefCard{bi-read-03}{BI}{3}\\
	\queryRefCard{bi-read-04}{BI}{4}\\
	\queryRefCard{bi-read-05}{BI}{5}\\
	\queryRefCard{bi-read-06}{BI}{6}\\
	\queryRefCard{bi-read-07}{BI}{7}\\
	\queryRefCard{bi-read-08}{BI}{8}\\
	\queryRefCard{bi-read-09}{BI}{9}\\
	\queryRefCard{bi-read-10}{BI}{10}\\
	\queryRefCard{bi-read-11}{BI}{11}\\
	\queryRefCard{bi-read-12}{BI}{12}\\
	\queryRefCard{bi-read-13}{BI}{13}\\
	\queryRefCard{bi-read-14}{BI}{14}\\
	\queryRefCard{bi-read-15}{BI}{15}\\
	\queryRefCard{bi-read-16}{BI}{16}\\
	\queryRefCard{bi-read-17}{BI}{17}\\
	\queryRefCard{bi-read-18}{BI}{18}\\
	\queryRefCard{bi-read-19}{BI}{19}\\
	\queryRefCard{bi-read-20}{BI}{20}\\
}

\noindent\begin{tabularx}{\queryCardWidth}{|>{\queryPropertyCell}p{\queryPropertyCellWidth}|X|}
	\hline
	query & BI / read / 12 \\ \hline
	title & How many persons have a given number of messages \\ \hline
	pattern & \centering \includegraphics[scale=\patternscale,margin=0cm .2cm]{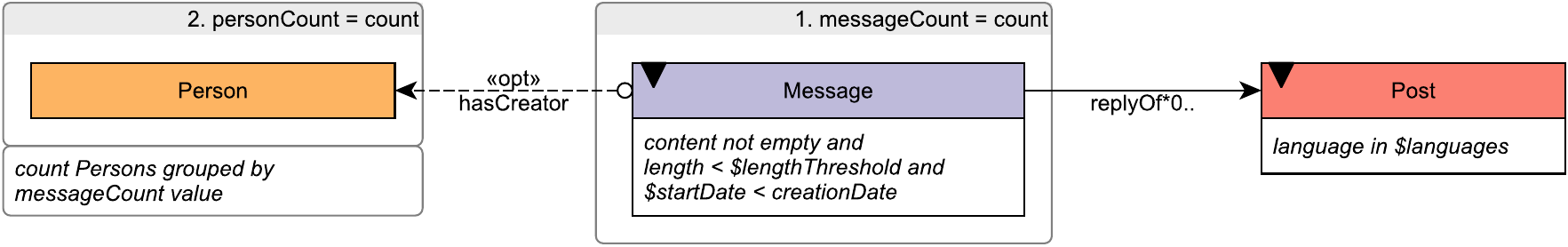} \tabularnewline \hline
	description & For each \emph{Person}, count the number of \emph{Messages} they made
(\texttt{messageCount}). Only count \emph{Messages} with the following
attributes:

\begin{itemize}
\item
  Its \texttt{content} is not empty (and consequently, the
  \texttt{imageFile} attribute is empty for \emph{Posts}).
\item
  Its \texttt{creationDate} is after \texttt{\$startDate} (exclusive,
  equality is not allowed).
\item
  Its \texttt{length} is below the \texttt{\$lengthThreshold}
  (exclusive, equality is not allowed).
\item
  It is written in any of the given \texttt{\$languages}.

  \begin{itemize}
  \tightlist
  \item
    The language of a \emph{Post} is defined by its \texttt{language}
    attribute.
  \item
    The language of a \emph{Comment} is that of the \emph{Post} that
    initiates the thread where the \emph{Comment} replies to.
  \end{itemize}

  The \emph{Post} and \emph{Comments} in the reply tree's path (from the
  \emph{Message} to the \emph{Post}) do not have to satisfy the
  constraints for \texttt{content}, \texttt{length}, and
  \texttt{creationDate}.
\end{itemize}

For each \texttt{messageCount} value, count the number of \emph{Persons}
with exactly \texttt{messageCount} \emph{Messages} (with the required
attributes).
 \\ \hline

		params &
		\innerCardVSpace{\begin{tabularx}{\attributeCardWidth}{|>{\paramNumberCell}C{\attributeNumberWidth}|>{\varNameCell}M|>{\typeCell}m{\typeWidth}|Y|} \hline
		$\mathsf{1}$ & \$startDate
 & Date
 & Selected randomly from a 60-day interval.
 \\ \hline
		$\mathsf{2}$ & \$lengthThreshold
 & 32-bit Integer
 & Balanced against \texttt{startDate} to filter around 30\% of the
\emph{Messages} within a language and keep the variance low.

The selection of this parameter uses a factor table of bucketed
\emph{Message} lengths and creation dates.
 \\ \hline
		$\mathsf{3}$ & \$languages
 & \{String\}
 & Only the most frequently used languages
 \\ \hline
		\end{tabularx}}\innerCardVSpace \\ \hline

		result &
		\innerCardVSpace{\begin{tabularx}{\attributeCardWidth}{|>{\resultNumberCell}C{\attributeNumberWidth}|>{\varNameCell}M|>{\typeCell}m{\typeWidth}|>{\resultOriginCell}c|Y|} \hline
		$\mathsf{1}$ & messageCount & 32-bit Integer & A &
				Number of \emph{Messages} created
 \\ \hline
		$\mathsf{2}$ & personCount & 32-bit Integer & A &
				Number of \emph{Persons} with \texttt{messageCount} \emph{Messages}
 \\ \hline
		\end{tabularx}}\innerCardVSpace \\ \hline

		sort		&
		\innerCardVSpace{\begin{tabularx}{\attributeCardWidth}{|>{\sortNumberCell}C{\attributeNumberWidth}|>{\varNameCell}M|>{\directionCell}c|Y|} \hline
		$\mathsf{1}$ & personCount
 & $\desc
$ &  \\ \hline
		$\mathsf{2}$ & messageCount
 & $\desc
$ &  \\ \hline
		\end{tabularx}}\innerCardVSpace \\ \hline
	limit & n/a \\ \hline
	CPs &
	\multicolumn{1}{>{\raggedright}l|}{
		\chokePoint{1.1}, 
		\chokePoint{1.2}, 
		\chokePoint{1.4}, 
		\chokePoint{2.6}, 
		\chokePoint{3.2}, 
		\chokePoint{4.2}, 
		\chokePoint{4.3}, 
		\chokePoint{8.1}, 
		\chokePoint{8.2}, 
		\chokePoint{8.3}, 
		\chokePoint{8.4}, 
		\chokePoint{8.5}
		} \\ \hline
\end{tabularx}
\queryCardVSpace

\let\emph\oldemph
\let\textbf\oldtextbf

\renewcommand{\currentQueryCard}{0}
\renewcommand*{\arraystretch}{1.1}

\subsection*{BI / read / 13}
\label{sec:bi-read-13}

\let\oldemph\emph
\renewcommand{\emph}[1]{{\footnotesize \sf #1}}
\let\oldtextbf\textbf
\renewcommand{\textbf}[1]{{\it #1}}

\renewcommand{\currentQueryCard}{bi-read-13}
\marginpar{
	\vspace{0.22ex}
	\raggedleft

	\queryRefCard{bi-read-01}{BI}{1}\\
	\queryRefCard{bi-read-02}{BI}{2}\\
	\queryRefCard{bi-read-03}{BI}{3}\\
	\queryRefCard{bi-read-04}{BI}{4}\\
	\queryRefCard{bi-read-05}{BI}{5}\\
	\queryRefCard{bi-read-06}{BI}{6}\\
	\queryRefCard{bi-read-07}{BI}{7}\\
	\queryRefCard{bi-read-08}{BI}{8}\\
	\queryRefCard{bi-read-09}{BI}{9}\\
	\queryRefCard{bi-read-10}{BI}{10}\\
	\queryRefCard{bi-read-11}{BI}{11}\\
	\queryRefCard{bi-read-12}{BI}{12}\\
	\queryRefCard{bi-read-13}{BI}{13}\\
	\queryRefCard{bi-read-14}{BI}{14}\\
	\queryRefCard{bi-read-15}{BI}{15}\\
	\queryRefCard{bi-read-16}{BI}{16}\\
	\queryRefCard{bi-read-17}{BI}{17}\\
	\queryRefCard{bi-read-18}{BI}{18}\\
	\queryRefCard{bi-read-19}{BI}{19}\\
	\queryRefCard{bi-read-20}{BI}{20}\\
}

\noindent\begin{tabularx}{\queryCardWidth}{|>{\queryPropertyCell}p{\queryPropertyCellWidth}|X|}
	\hline
	query & BI / read / 13 \\ \hline
	title & Zombies in a country \\ \hline
	pattern & \centering \includegraphics[scale=\patternscale,margin=0cm .2cm]{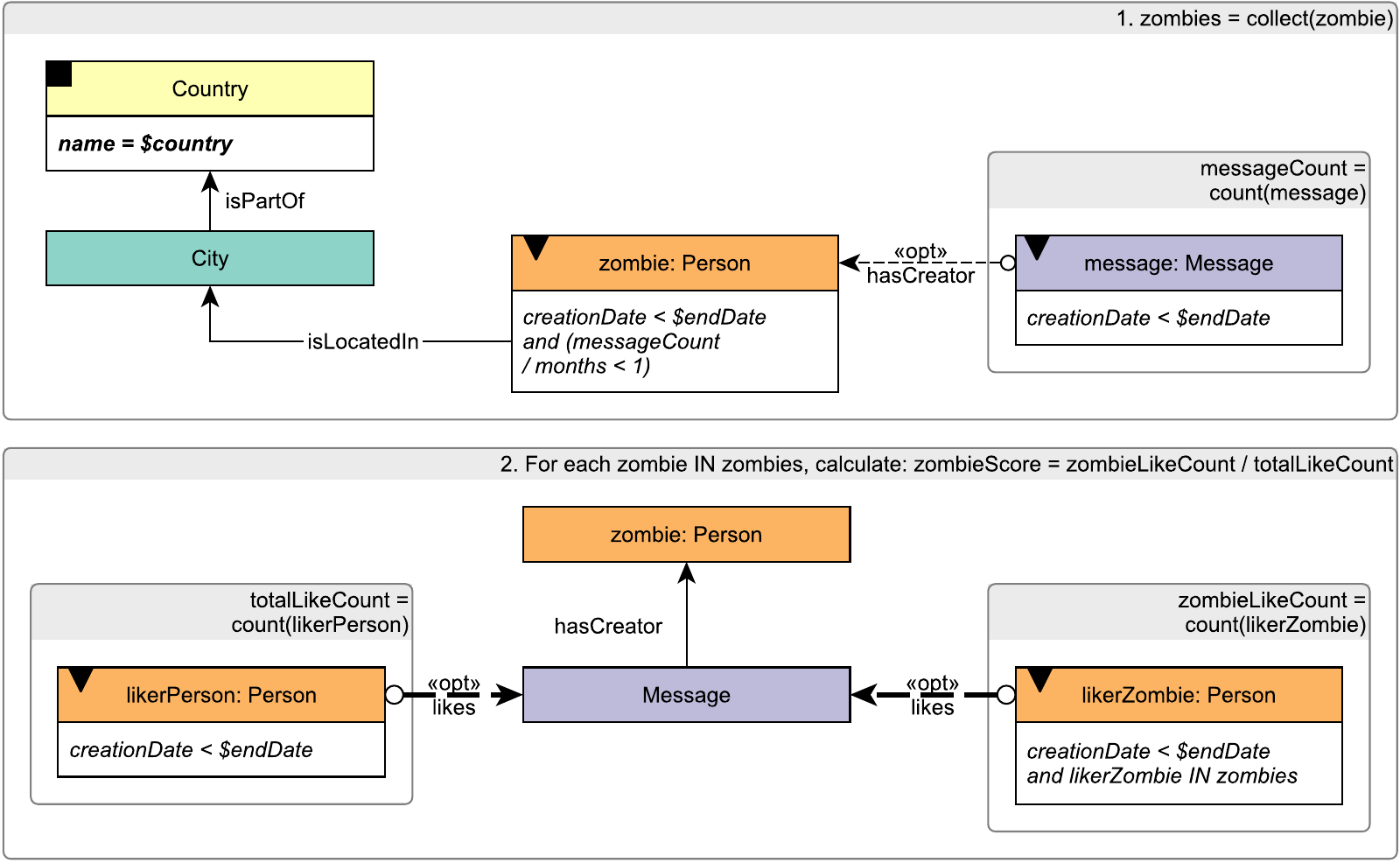} \tabularnewline \hline
	description & Find zombies within the given \texttt{\$country}, and return their
zombie scores. A \texttt{zombie} is a \emph{Person} created before the
given \texttt{\$endDate}, which has created an average of
\texttt{{[}0,\ 1)} \emph{Messages} per month, during the time range
between profile's \texttt{creationDate} and the given
\texttt{\$endDate}. The number of months spans the time range from the
\texttt{creationDate} of the profile to the \texttt{\$endDate} with
partial months on both end counting as one month (e.g.~a
\texttt{creationDate} of Jan 31 and an \texttt{\$endDate} of Mar 1
result in 3 months).

For each \texttt{zombie}, calculate the following:

\begin{itemize}
\tightlist
\item
  \texttt{zombieLikeCount}: the number of \emph{likes} received from
  other zombies.
\item
  \texttt{totalLikeCount}: the total number of \emph{likes} received.
\item
  \texttt{zombieScore}: \texttt{zombieLikeCount} /
  \texttt{totalLikeCount}. If the value of \texttt{totalLikeCount} is 0,
  the \texttt{zombieScore} of the \texttt{zombie} should be 0.0.
\end{itemize}

For both \texttt{zombieLikeCount} and \texttt{totalLikeCount}, only
consider \emph{likes} received from profiles that were created before
the given \texttt{\$endDate}.
 \\ \hline

		params &
		\innerCardVSpace{\begin{tabularx}{\attributeCardWidth}{|>{\paramNumberCell}C{\attributeNumberWidth}|>{\varNameCell}M|>{\typeCell}m{\typeWidth}|Y|} \hline
		$\mathsf{1}$ & \$country
 & Long String
 & Selected from the largest \emph{Countries} (India, China)
 \\ \hline
		$\mathsf{2}$ & \$endDate
 & Date
 & Selected from the last days of the initial data set
 \\ \hline
		\end{tabularx}}\innerCardVSpace \\ \hline

		result &
		\innerCardVSpace{\begin{tabularx}{\attributeCardWidth}{|>{\resultNumberCell}C{\attributeNumberWidth}|>{\varNameCell}M|>{\typeCell}m{\typeWidth}|>{\resultOriginCell}c|Y|} \hline
		$\mathsf{1}$ & zombie.id & ID & R &
				 \\ \hline
		$\mathsf{2}$ & zombieLikeCount & 32-bit Integer & A &
				 \\ \hline
		$\mathsf{3}$ & totalLikeCount & 32-bit Integer & A &
				 \\ \hline
		$\mathsf{4}$ & zombieScore & 32-bit Float & A &
				Determined as \texttt{zombieLikeCount} / \texttt{totalLikeCount}
 \\ \hline
		\end{tabularx}}\innerCardVSpace \\ \hline

		sort		&
		\innerCardVSpace{\begin{tabularx}{\attributeCardWidth}{|>{\sortNumberCell}C{\attributeNumberWidth}|>{\varNameCell}M|>{\directionCell}c|Y|} \hline
		$\mathsf{1}$ & zombieScore
 & $\desc
$ &  \\ \hline
		$\mathsf{2}$ & zombie.id
 & $\asc
$ &  \\ \hline
		\end{tabularx}}\innerCardVSpace \\ \hline
	limit & 100 \\ \hline
	CPs &
	\multicolumn{1}{>{\raggedright}l|}{
		\chokePoint{1.2}, 
		\chokePoint{2.1}, 
		\chokePoint{2.3}, 
		\chokePoint{2.4}, 
		\chokePoint{2.6}, 
		\chokePoint{3.2}, 
		\chokePoint{3.3}, 
		\chokePoint{4.2}, 
		\chokePoint{5.1}, 
		\chokePoint{5.3}, 
		\chokePoint{8.2}, 
		\chokePoint{8.4}, 
		\chokePoint{8.5}
		} \\ \hline
\end{tabularx}
\queryCardVSpace

\let\emph\oldemph
\let\textbf\oldtextbf

\renewcommand{\currentQueryCard}{0}
\renewcommand*{\arraystretch}{1.1}

\subsection*{BI / read / 14}
\label{sec:bi-read-14}

\let\oldemph\emph
\renewcommand{\emph}[1]{{\footnotesize \sf #1}}
\let\oldtextbf\textbf
\renewcommand{\textbf}[1]{{\it #1}}

\renewcommand{\currentQueryCard}{bi-read-14}
\marginpar{
	\vspace{0.22ex}
	\raggedleft

	\queryRefCard{bi-read-01}{BI}{1}\\
	\queryRefCard{bi-read-02}{BI}{2}\\
	\queryRefCard{bi-read-03}{BI}{3}\\
	\queryRefCard{bi-read-04}{BI}{4}\\
	\queryRefCard{bi-read-05}{BI}{5}\\
	\queryRefCard{bi-read-06}{BI}{6}\\
	\queryRefCard{bi-read-07}{BI}{7}\\
	\queryRefCard{bi-read-08}{BI}{8}\\
	\queryRefCard{bi-read-09}{BI}{9}\\
	\queryRefCard{bi-read-10}{BI}{10}\\
	\queryRefCard{bi-read-11}{BI}{11}\\
	\queryRefCard{bi-read-12}{BI}{12}\\
	\queryRefCard{bi-read-13}{BI}{13}\\
	\queryRefCard{bi-read-14}{BI}{14}\\
	\queryRefCard{bi-read-15}{BI}{15}\\
	\queryRefCard{bi-read-16}{BI}{16}\\
	\queryRefCard{bi-read-17}{BI}{17}\\
	\queryRefCard{bi-read-18}{BI}{18}\\
	\queryRefCard{bi-read-19}{BI}{19}\\
	\queryRefCard{bi-read-20}{BI}{20}\\
}

\noindent\begin{tabularx}{\queryCardWidth}{|>{\queryPropertyCell}p{\queryPropertyCellWidth}|X|}
	\hline
	query & BI / read / 14 \\ \hline
	title & International dialog \\ \hline
	pattern & \centering \includegraphics[scale=\patternscale,margin=0cm .2cm]{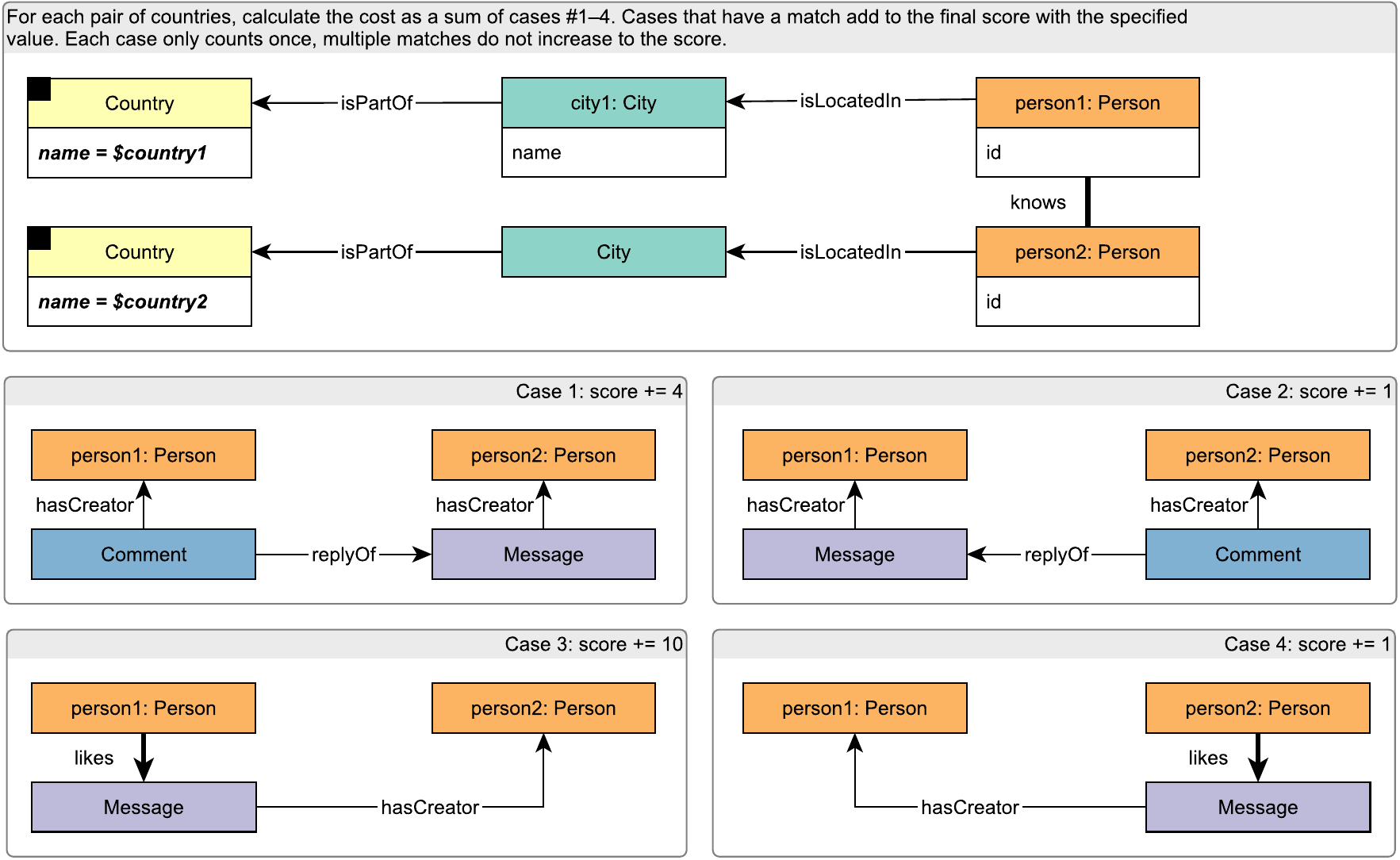} \tabularnewline \hline
	description & Consider all pairs of people \texttt{(person1,\ person2)} such that (1)
they know each other, (2) one is located in a \emph{City} of
\texttt{\$country1}, and (3) the other is located in a \emph{City} of
\texttt{\$country2}. For each \emph{City} of \texttt{\$country1}, return
the highest scoring pair. If there are multiple top-scoring pairs in a
city, return the pair with the lowest (\texttt{person1.id},
\texttt{person2.id}) using a lexicographical ordering.

The score of a pair is defined as the sum of the subscores awarded for
the following kinds of interaction. The initial value is
\texttt{score\ =\ 0}.

\begin{enumerate}
\def\labelenumi{\arabic{enumi}.}
\tightlist
\item
  \texttt{person1} has created a reply \emph{Comment} to at least one
  \emph{Message} by \texttt{person2}: \texttt{score\ +=\ 4}
\item
  \texttt{person1} has created at least one \emph{Message} that
  \texttt{person2} has created a reply to: \texttt{score\ +=\ 1}
\item
  \texttt{person1} liked at least one \emph{Message} by
  \texttt{person2}: \texttt{score\ +=\ 10}
\item
  \texttt{person1} has created at least one \emph{Message} that was
  liked by \texttt{person2}: \texttt{score\ +=\ 1}
\end{enumerate}

Consequently, the maximum score a pair can obtain is:
\texttt{4\ +\ 1\ +\ 10\ +\ 1\ =\ 16}.
 \\ \hline

		params &
		\innerCardVSpace{\begin{tabularx}{\attributeCardWidth}{|>{\paramNumberCell}C{\attributeNumberWidth}|>{\varNameCell}M|>{\typeCell}m{\typeWidth}|Y|} \hline
		$\mathsf{1}$ & \$country1
 & Long String
 & \texttt{(a)} Correlated with parameter \texttt{country2}, i.e.~the
\emph{Countries} are close and there are many \emph{Persons} knowing
each other

\texttt{(b)} Uncorrelated with parameter \texttt{country2}, i.e.~the
\emph{Countries} are afar and there are few \emph{Persons} knowing each
other
 \\ \hline
		$\mathsf{2}$ & \$country2
 & Long String
 &  \\ \hline
		\end{tabularx}}\innerCardVSpace \\ \hline

		result &
		\innerCardVSpace{\begin{tabularx}{\attributeCardWidth}{|>{\resultNumberCell}C{\attributeNumberWidth}|>{\varNameCell}M|>{\typeCell}m{\typeWidth}|>{\resultOriginCell}c|Y|} \hline
		$\mathsf{1}$ & person1.id & ID & R &
				 \\ \hline
		$\mathsf{2}$ & person2.id & ID & R &
				 \\ \hline
		$\mathsf{3}$ & city1.name & Long String & R &
				 \\ \hline
		$\mathsf{4}$ & score & 32-bit Integer & C &
				 \\ \hline
		\end{tabularx}}\innerCardVSpace \\ \hline

		sort		&
		\innerCardVSpace{\begin{tabularx}{\attributeCardWidth}{|>{\sortNumberCell}C{\attributeNumberWidth}|>{\varNameCell}M|>{\directionCell}c|Y|} \hline
		$\mathsf{1}$ & score
 & $\desc
$ &  \\ \hline
		$\mathsf{2}$ & person1.id
 & $\asc
$ &  \\ \hline
		$\mathsf{3}$ & person2.id
 & $\asc
$ &  \\ \hline
		\end{tabularx}}\innerCardVSpace \\ \hline
	limit & 100 \\ \hline
	CPs &
	\multicolumn{1}{>{\raggedright}l|}{
		\chokePoint{1.3}, 
		\chokePoint{1.4}, 
		\chokePoint{2.1}, 
		\chokePoint{3.1}, 
		\chokePoint{3.3}, 
		\chokePoint{5.1}, 
		\chokePoint{5.2}, 
		\chokePoint{5.3}, 
		\chokePoint{8.3}, 
		\chokePoint{8.4}
		} \\ \hline
\end{tabularx}
\queryCardVSpace

\let\emph\oldemph
\let\textbf\oldtextbf

\renewcommand{\currentQueryCard}{0}
\renewcommand*{\arraystretch}{1.1}

\subsection*{BI / read / 15}
\label{sec:bi-read-15}

\let\oldemph\emph
\renewcommand{\emph}[1]{{\footnotesize \sf #1}}
\let\oldtextbf\textbf
\renewcommand{\textbf}[1]{{\it #1}}

\renewcommand{\currentQueryCard}{bi-read-15}
\marginpar{
	\vspace{0.22ex}
	\raggedleft

	\queryRefCard{bi-read-01}{BI}{1}\\
	\queryRefCard{bi-read-02}{BI}{2}\\
	\queryRefCard{bi-read-03}{BI}{3}\\
	\queryRefCard{bi-read-04}{BI}{4}\\
	\queryRefCard{bi-read-05}{BI}{5}\\
	\queryRefCard{bi-read-06}{BI}{6}\\
	\queryRefCard{bi-read-07}{BI}{7}\\
	\queryRefCard{bi-read-08}{BI}{8}\\
	\queryRefCard{bi-read-09}{BI}{9}\\
	\queryRefCard{bi-read-10}{BI}{10}\\
	\queryRefCard{bi-read-11}{BI}{11}\\
	\queryRefCard{bi-read-12}{BI}{12}\\
	\queryRefCard{bi-read-13}{BI}{13}\\
	\queryRefCard{bi-read-14}{BI}{14}\\
	\queryRefCard{bi-read-15}{BI}{15}\\
	\queryRefCard{bi-read-16}{BI}{16}\\
	\queryRefCard{bi-read-17}{BI}{17}\\
	\queryRefCard{bi-read-18}{BI}{18}\\
	\queryRefCard{bi-read-19}{BI}{19}\\
	\queryRefCard{bi-read-20}{BI}{20}\\
}

\noindent\begin{tabularx}{\queryCardWidth}{|>{\queryPropertyCell}p{\queryPropertyCellWidth}|X|}
	\hline
	query & BI / read / 15 \\ \hline
	title & Trusted connection paths through forums created in a given timeframe \\ \hline
	pattern & \centering \includegraphics[scale=\patternscale,margin=0cm .2cm]{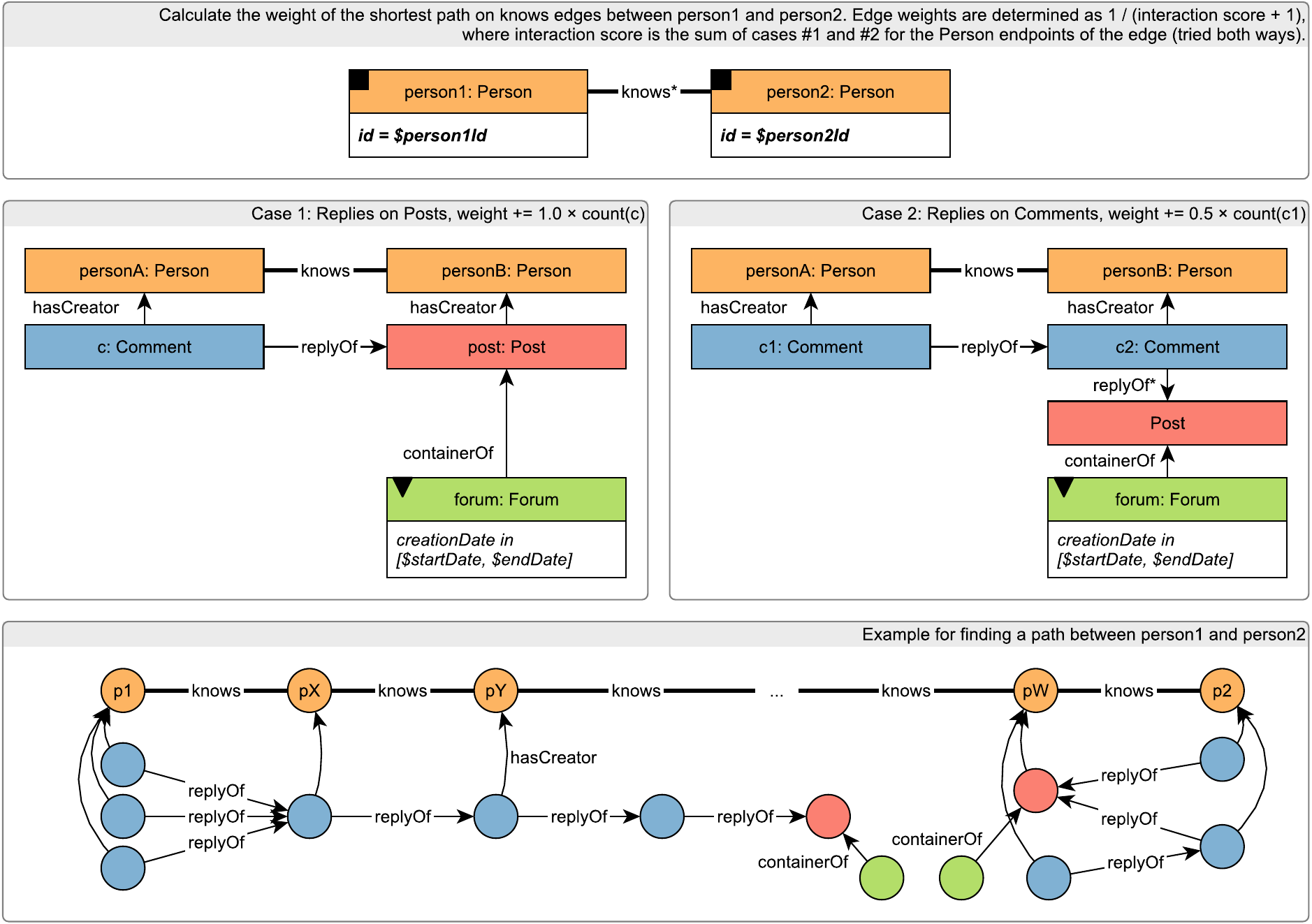} \tabularnewline \hline
	description & Given two \emph{Persons} with IDs \texttt{\$person1Id} and
\texttt{\$person2Id}, calculate the cost of the weighted shortest path
between these two \emph{Persons}, in the subgraph induced by the
\emph{knows} relationship. The interaction score of a \emph{knows} edge
is calculated based on the interactions of its \emph{Person} endpoints:

\begin{itemize}
\tightlist
\item
  Every direct reply (by one of the \emph{Persons}) to a \emph{Post} (by
  the other \emph{Person}) is 1.0 point.
\item
  Every direct reply (by one of the \emph{Persons}) to a \emph{Comment}
  (by the other \emph{Person}) is 0.5 points.
\end{itemize}

Only consider \emph{Messages} that were created in a \emph{Forum} that
was created within the timeframe (interval)
\texttt{{[}\$startDate,\ \$endDate{]}}. Note that for \emph{Comments},
the containing \emph{Forum} is that of the \emph{Post} that the comment
(transitively) replies to. Also note that interactions are counted both
ways.

The weight for the shortest path algorithm is determined as
\(\frac{1}{\textit{interaction score} + 1}\).

The result of the query is a single number, the cost of the weighted
shortest path. If no such path exists, the query should return \(-1.0\).
 \\ \hline

		params &
		\innerCardVSpace{\begin{tabularx}{\attributeCardWidth}{|>{\paramNumberCell}C{\attributeNumberWidth}|>{\varNameCell}M|>{\typeCell}m{\typeWidth}|Y|} \hline
		$\mathsf{1}$ & \$person1Id
 & ID
 & \texttt{(a)} \texttt{\$person1Id} -- \texttt{\$person2Id} pair with a
distance of 4 hops

\texttt{(b)} \texttt{\$person1Id} -- \texttt{\$person2Id} pair with a
distance of 2 hops
 \\ \hline
		$\mathsf{2}$ & \$person2Id
 & ID
 &  \\ \hline
		$\mathsf{3}$ & \$startDate
 & Date
 & \texttt{(a)} Small interval (approx. one week)

\texttt{(b)} Big interval (approx. one month)
 \\ \hline
		$\mathsf{4}$ & \$endDate
 & Date
 &  \\ \hline
		\end{tabularx}}\innerCardVSpace \\ \hline

		result &
		\innerCardVSpace{\begin{tabularx}{\attributeCardWidth}{|>{\resultNumberCell}C{\attributeNumberWidth}|>{\varNameCell}M|>{\typeCell}m{\typeWidth}|>{\resultOriginCell}c|Y|} \hline
		$\mathsf{1}$ & weight & 32-bit Float & C &
				 \\ \hline
		\end{tabularx}}\innerCardVSpace \\ \hline

	limit & n/a \\ \hline
	CPs &
	\multicolumn{1}{>{\raggedright}l|}{
		\chokePoint{1.2}, 
		\chokePoint{2.1}, 
		\chokePoint{2.2}, 
		\chokePoint{2.4}, 
		\chokePoint{3.3}, 
		\chokePoint{5.1}, 
		\chokePoint{5.3}, 
		\chokePoint{7.2}, 
		\chokePoint{7.3}, 
		\chokePoint{7.6}, 
		\chokePoint{7.7}, 
		\chokePoint{8.1}, 
		\chokePoint{8.2}, 
		\chokePoint{8.3}, 
		\chokePoint{8.4}, 
		\chokePoint{8.5}, 
		\chokePoint{8.6}
		} \\ \hline
\end{tabularx}
\queryCardVSpace

\let\emph\oldemph
\let\textbf\oldtextbf

\renewcommand{\currentQueryCard}{0}
\renewcommand*{\arraystretch}{1.1}

\subsection*{BI / read / 16}
\label{sec:bi-read-16}

\let\oldemph\emph
\renewcommand{\emph}[1]{{\footnotesize \sf #1}}
\let\oldtextbf\textbf
\renewcommand{\textbf}[1]{{\it #1}}

\renewcommand{\currentQueryCard}{bi-read-16}
\marginpar{
	\vspace{0.22ex}
	\raggedleft

	\queryRefCard{bi-read-01}{BI}{1}\\
	\queryRefCard{bi-read-02}{BI}{2}\\
	\queryRefCard{bi-read-03}{BI}{3}\\
	\queryRefCard{bi-read-04}{BI}{4}\\
	\queryRefCard{bi-read-05}{BI}{5}\\
	\queryRefCard{bi-read-06}{BI}{6}\\
	\queryRefCard{bi-read-07}{BI}{7}\\
	\queryRefCard{bi-read-08}{BI}{8}\\
	\queryRefCard{bi-read-09}{BI}{9}\\
	\queryRefCard{bi-read-10}{BI}{10}\\
	\queryRefCard{bi-read-11}{BI}{11}\\
	\queryRefCard{bi-read-12}{BI}{12}\\
	\queryRefCard{bi-read-13}{BI}{13}\\
	\queryRefCard{bi-read-14}{BI}{14}\\
	\queryRefCard{bi-read-15}{BI}{15}\\
	\queryRefCard{bi-read-16}{BI}{16}\\
	\queryRefCard{bi-read-17}{BI}{17}\\
	\queryRefCard{bi-read-18}{BI}{18}\\
	\queryRefCard{bi-read-19}{BI}{19}\\
	\queryRefCard{bi-read-20}{BI}{20}\\
}

\noindent\begin{tabularx}{\queryCardWidth}{|>{\queryPropertyCell}p{\queryPropertyCellWidth}|X|}
	\hline
	query & BI / read / 16 \\ \hline
	title & Fake news detection \\ \hline
	pattern & \centering \includegraphics[scale=\patternscale,margin=0cm .2cm]{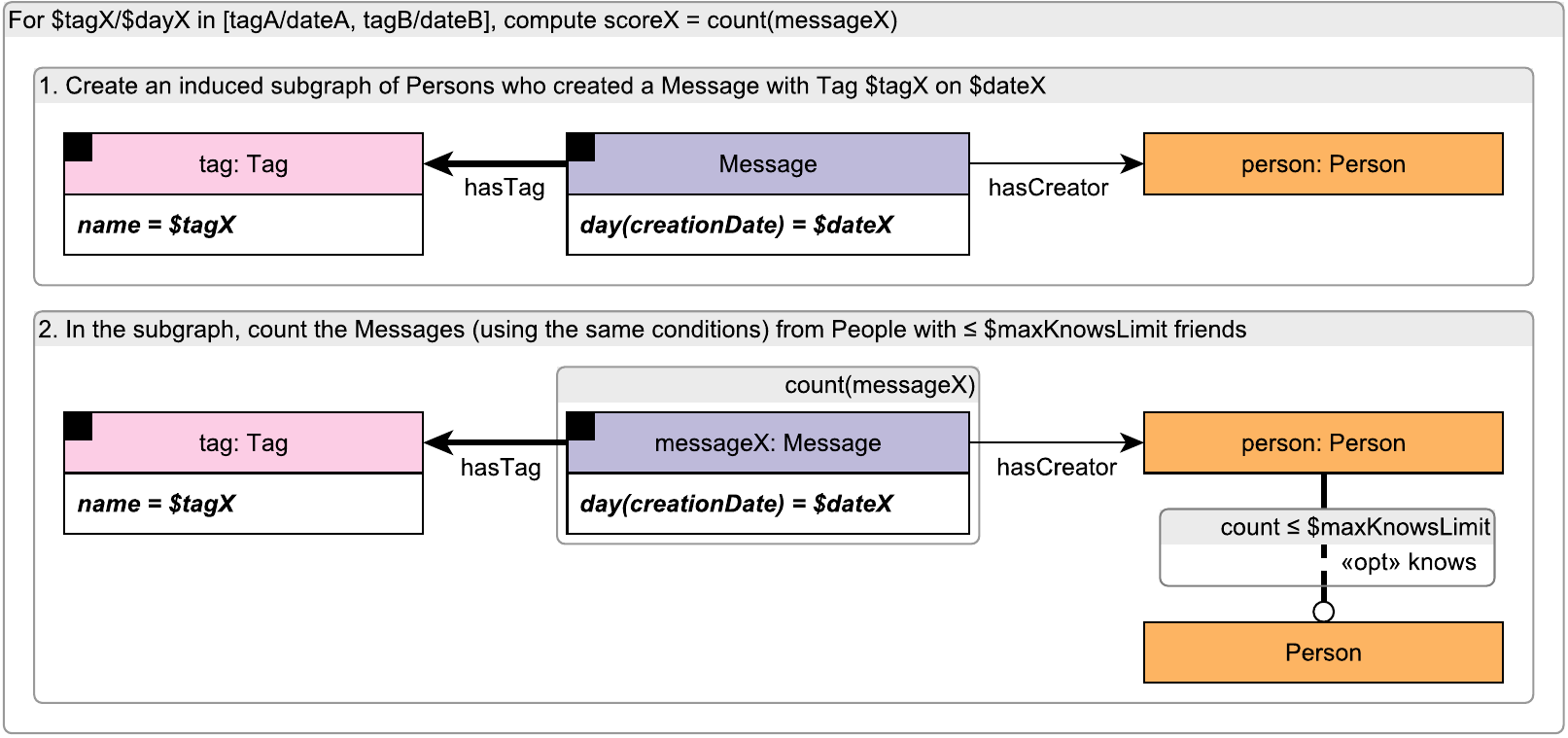} \tabularnewline \hline
	description & Given two \emph{Tag}/date pairs (\texttt{\$tagA}/\texttt{\$dateA} and
\texttt{\$tagB}/\texttt{\$dateB}), for each pair
\texttt{\$tagX}/\texttt{\$dateX}:

\begin{itemize}
\tightlist
\item
  Create an induced subgraph between \emph{Persons} where for each pair
  of \emph{Persons} \texttt{person1}/\texttt{person2}, both have created
  a \emph{Message} on the day of \texttt{\$dateX} with \emph{Tag}
  \texttt{\$tagX}.
\item
  In the induced subgraph, only keep pairs of \emph{Persons} who have at
  most \texttt{maxKnowsLimit} friends (in the induced subgraph).
\item
  For these \emph{Persons}, count the number of \emph{Messages} created
  on \texttt{\$dateX} with \emph{Tag} \texttt{\$tagX}.
\end{itemize}

Return \emph{Persons} who had at least one \emph{Messages} for both
\texttt{\$tagA}/\texttt{\$dateA} and \texttt{\$tagB}/\texttt{\$dateB}
ranked by their total number of \emph{Messages} (descending).
 \\ \hline

		params &
		\innerCardVSpace{\begin{tabularx}{\attributeCardWidth}{|>{\paramNumberCell}C{\attributeNumberWidth}|>{\varNameCell}M|>{\typeCell}m{\typeWidth}|Y|} \hline
		$\mathsf{1}$ & \$tagA
 & Long String
 & \texttt{(a)} \texttt{\$tagA}/\texttt{\$dateA},
\texttt{\$tagB}/\texttt{\$dateB} are both selected to be a flashmob
\emph{Tag}/date combination

\texttt{(b)} \texttt{\$tagA}/\texttt{\$dateA},
\texttt{\$tagB}/\texttt{\$dateB} are both selected to be a non-flashmob
\emph{Tag}/date combination
 \\ \hline
		$\mathsf{2}$ & \$dateA
 & Date
 &  \\ \hline
		$\mathsf{3}$ & \$tagB
 & Long String
 &  \\ \hline
		$\mathsf{4}$ & \$dateB
 & Date
 &  \\ \hline
		$\mathsf{5}$ & \$maxKnowsLimit
 & 32-bit Integer
 & Selected between 3 and 6
 \\ \hline
		\end{tabularx}}\innerCardVSpace \\ \hline

		result &
		\innerCardVSpace{\begin{tabularx}{\attributeCardWidth}{|>{\resultNumberCell}C{\attributeNumberWidth}|>{\varNameCell}M|>{\typeCell}m{\typeWidth}|>{\resultOriginCell}c|Y|} \hline
		$\mathsf{1}$ & person.id & ID & R &
				 \\ \hline
		$\mathsf{2}$ & messageCountA & 32-bit Integer & A &
				Message count for \texttt{\$tagA}/\texttt{\$dateA}
 \\ \hline
		$\mathsf{3}$ & messageCountB & 32-bit Integer & A &
				Message count for \texttt{\$tagB}/\texttt{\$dateB}
 \\ \hline
		\end{tabularx}}\innerCardVSpace \\ \hline

		sort		&
		\innerCardVSpace{\begin{tabularx}{\attributeCardWidth}{|>{\sortNumberCell}C{\attributeNumberWidth}|>{\varNameCell}M|>{\directionCell}c|Y|} \hline
		$\mathsf{1}$ & messageCountA + \mbox{messageCountB}
 & $\desc
$ &  \\ \hline
		$\mathsf{2}$ & person.id
 & $\asc
$ &  \\ \hline
		\end{tabularx}}\innerCardVSpace \\ \hline
	limit & 20 \\ \hline
	CPs &
	\multicolumn{1}{>{\raggedright}l|}{
		\chokePoint{5.3}, 
		\chokePoint{8.4}, 
		\chokePoint{8.5}
		} \\ \hline
	relevance &
		\footnotesize There are two major ways to compute this query: (1) create the induced
subgraph as suggested by the specification (either as a view or in
materialized form), or (2) skip creating the induced subgraph and
perform on-the-fly check for the number of friends (who also posted at
least one \emph{Message} with the given \emph{Tag} on the given date).
The latter approach is easier to express in systems which do not provide
graph views but might result in redundant computations (the query engine
might repeatedly check whether a \emph{Person} has at least one
\emph{Message} that satifies the conditions).
 \\ \hline%
\end{tabularx}
\queryCardVSpace

\let\emph\oldemph
\let\textbf\oldtextbf

\renewcommand{\currentQueryCard}{0}
\renewcommand*{\arraystretch}{1.1}

\subsection*{BI / read / 17}
\label{sec:bi-read-17}

\let\oldemph\emph
\renewcommand{\emph}[1]{{\footnotesize \sf #1}}
\let\oldtextbf\textbf
\renewcommand{\textbf}[1]{{\it #1}}

\renewcommand{\currentQueryCard}{bi-read-17}
\marginpar{
	\vspace{0.22ex}
	\raggedleft

	\queryRefCard{bi-read-01}{BI}{1}\\
	\queryRefCard{bi-read-02}{BI}{2}\\
	\queryRefCard{bi-read-03}{BI}{3}\\
	\queryRefCard{bi-read-04}{BI}{4}\\
	\queryRefCard{bi-read-05}{BI}{5}\\
	\queryRefCard{bi-read-06}{BI}{6}\\
	\queryRefCard{bi-read-07}{BI}{7}\\
	\queryRefCard{bi-read-08}{BI}{8}\\
	\queryRefCard{bi-read-09}{BI}{9}\\
	\queryRefCard{bi-read-10}{BI}{10}\\
	\queryRefCard{bi-read-11}{BI}{11}\\
	\queryRefCard{bi-read-12}{BI}{12}\\
	\queryRefCard{bi-read-13}{BI}{13}\\
	\queryRefCard{bi-read-14}{BI}{14}\\
	\queryRefCard{bi-read-15}{BI}{15}\\
	\queryRefCard{bi-read-16}{BI}{16}\\
	\queryRefCard{bi-read-17}{BI}{17}\\
	\queryRefCard{bi-read-18}{BI}{18}\\
	\queryRefCard{bi-read-19}{BI}{19}\\
	\queryRefCard{bi-read-20}{BI}{20}\\
}

\noindent\begin{tabularx}{\queryCardWidth}{|>{\queryPropertyCell}p{\queryPropertyCellWidth}|X|}
	\hline
	query & BI / read / 17 \\ \hline
	title & Information propagation analysis \\ \hline
	pattern & \centering \includegraphics[scale=\patternscale,margin=0cm .2cm]{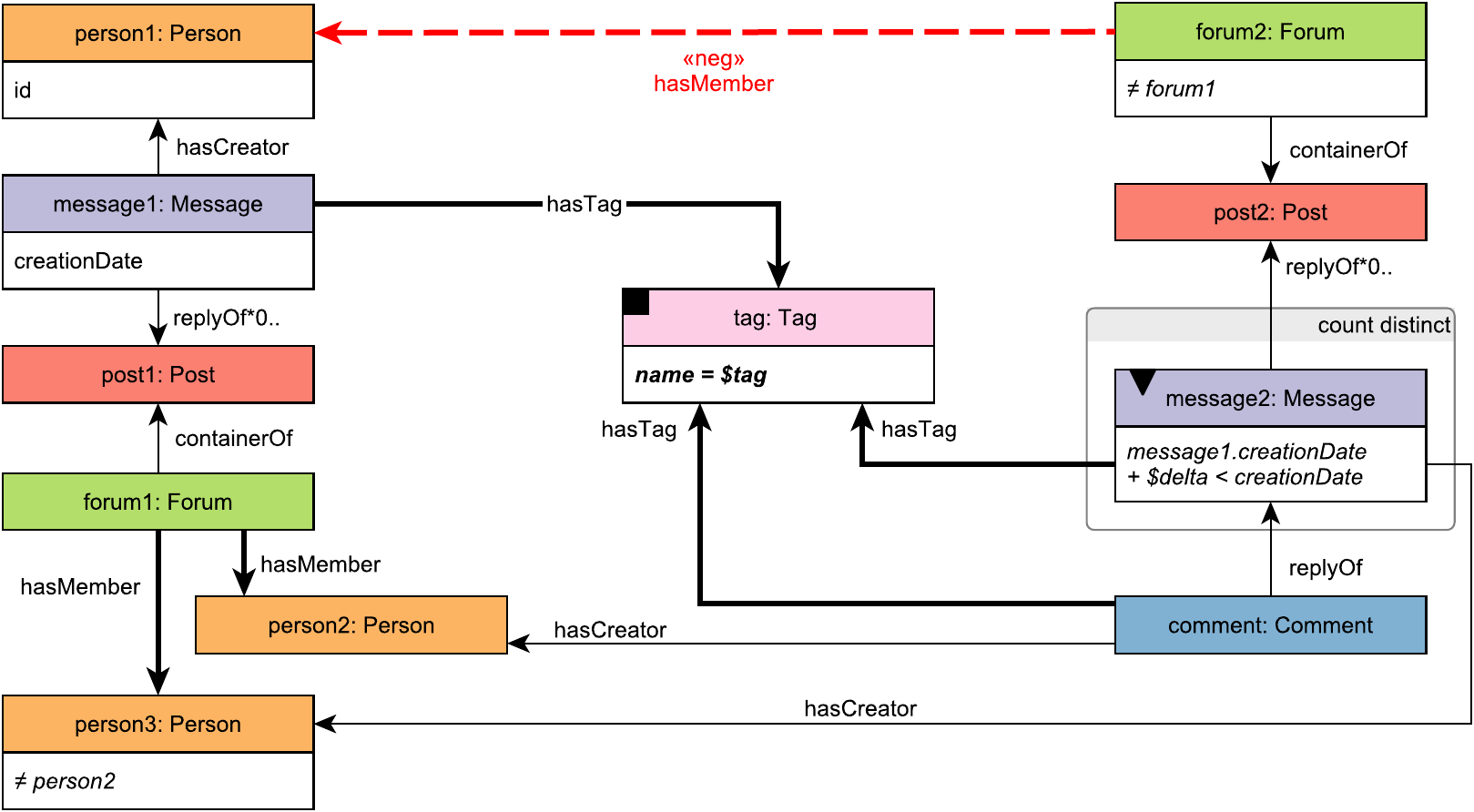} \tabularnewline \hline
	description & This query aims to identify instances of ``information propagation''
when a \emph{Person} (\texttt{person1}) submits a \emph{Message}
(\texttt{message1}) with a given \texttt{\$tag} to a \emph{Forum}
(\texttt{forum1}). This is read by other members of \texttt{forum1},
\emph{Persons} \texttt{person2} and \texttt{person3} (who must be
different \emph{Persons}). Some time later (specified by the
\texttt{\$delta} parameter), these persons have a discussion with the
same \texttt{\$tag} in a different \emph{Forum} (\texttt{forum2}) where
\texttt{person1} is not a member. The discussion consists of a
\emph{Message} (\texttt{message2}) by \texttt{person2} and a direct
reply \emph{Comment} (\texttt{comment}) by \texttt{person3}.

Return IDs of \texttt{person1} with the number of interactions their
\emph{Messages} (might have) caused.
 \\ \hline

		params &
		\innerCardVSpace{\begin{tabularx}{\attributeCardWidth}{|>{\paramNumberCell}C{\attributeNumberWidth}|>{\varNameCell}M|>{\typeCell}m{\typeWidth}|Y|} \hline
		$\mathsf{1}$ & \$tag
 & Long String
 & \emph{Tags} with a similar amount of \emph{Messages} are selected
 \\ \hline
		$\mathsf{2}$ & \$delta
 & 32-bit Integer
 & Measured in hours, selected to be between 8 and 16 hours.
 \\ \hline
		\end{tabularx}}\innerCardVSpace \\ \hline

		result &
		\innerCardVSpace{\begin{tabularx}{\attributeCardWidth}{|>{\resultNumberCell}C{\attributeNumberWidth}|>{\varNameCell}M|>{\typeCell}m{\typeWidth}|>{\resultOriginCell}c|Y|} \hline
		$\mathsf{1}$ & person1.id & ID & R &
				 \\ \hline
		$\mathsf{2}$ & messageCount & 32-bit Integer & A &
				 \\ \hline
		\end{tabularx}}\innerCardVSpace \\ \hline

		sort		&
		\innerCardVSpace{\begin{tabularx}{\attributeCardWidth}{|>{\sortNumberCell}C{\attributeNumberWidth}|>{\varNameCell}M|>{\directionCell}c|Y|} \hline
		$\mathsf{1}$ & messageCount
 & $\desc
$ &  \\ \hline
		$\mathsf{2}$ & person1.id
 & $\asc
$ &  \\ \hline
		\end{tabularx}}\innerCardVSpace \\ \hline
	limit & 10 \\ \hline
	CPs &
	\multicolumn{1}{>{\raggedright}l|}{
		\chokePoint{2.1}, 
		\chokePoint{2.3}, 
		\chokePoint{2.5}, 
		\chokePoint{2.6}, 
		\chokePoint{8.1}
		} \\ \hline
\end{tabularx}
\queryCardVSpace

\let\emph\oldemph
\let\textbf\oldtextbf

\renewcommand{\currentQueryCard}{0}
\renewcommand*{\arraystretch}{1.1}

\subsection*{BI / read / 18}
\label{sec:bi-read-18}

\let\oldemph\emph
\renewcommand{\emph}[1]{{\footnotesize \sf #1}}
\let\oldtextbf\textbf
\renewcommand{\textbf}[1]{{\it #1}}

\renewcommand{\currentQueryCard}{bi-read-18}
\marginpar{
	\vspace{0.22ex}
	\raggedleft

	\queryRefCard{bi-read-01}{BI}{1}\\
	\queryRefCard{bi-read-02}{BI}{2}\\
	\queryRefCard{bi-read-03}{BI}{3}\\
	\queryRefCard{bi-read-04}{BI}{4}\\
	\queryRefCard{bi-read-05}{BI}{5}\\
	\queryRefCard{bi-read-06}{BI}{6}\\
	\queryRefCard{bi-read-07}{BI}{7}\\
	\queryRefCard{bi-read-08}{BI}{8}\\
	\queryRefCard{bi-read-09}{BI}{9}\\
	\queryRefCard{bi-read-10}{BI}{10}\\
	\queryRefCard{bi-read-11}{BI}{11}\\
	\queryRefCard{bi-read-12}{BI}{12}\\
	\queryRefCard{bi-read-13}{BI}{13}\\
	\queryRefCard{bi-read-14}{BI}{14}\\
	\queryRefCard{bi-read-15}{BI}{15}\\
	\queryRefCard{bi-read-16}{BI}{16}\\
	\queryRefCard{bi-read-17}{BI}{17}\\
	\queryRefCard{bi-read-18}{BI}{18}\\
	\queryRefCard{bi-read-19}{BI}{19}\\
	\queryRefCard{bi-read-20}{BI}{20}\\
}

\noindent\begin{tabularx}{\queryCardWidth}{|>{\queryPropertyCell}p{\queryPropertyCellWidth}|X|}
	\hline
	query & BI / read / 18 \\ \hline
	title & Friend recommendation \\ \hline
	pattern & \centering \includegraphics[scale=\patternscale,margin=0cm .2cm]{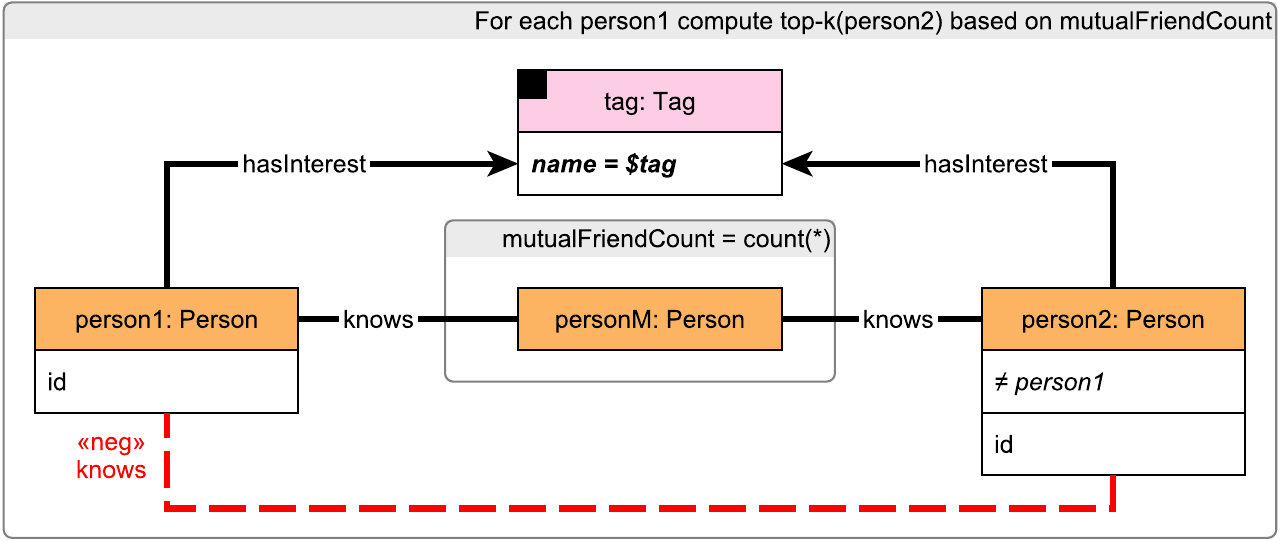} \tabularnewline \hline
	description & For a given \texttt{\$tag}, for each \texttt{person1} interested in
\texttt{\$tag}, recommend new friends (\texttt{person2}) who

\begin{itemize}
\tightlist
\item
  do not yet know \texttt{person1}
\item
  have at least one mutual friend with \texttt{person1}
\item
  are also interested in \texttt{\$tag}.
\end{itemize}

Rank \emph{Persons} \texttt{person2} based on the number of mutual
friends with \texttt{person1}.
 \\ \hline

		params &
		\innerCardVSpace{\begin{tabularx}{\attributeCardWidth}{|>{\paramNumberCell}C{\attributeNumberWidth}|>{\varNameCell}M|>{\typeCell}m{\typeWidth}|Y|} \hline
		$\mathsf{1}$ & \$tag
 & Long String
 & \emph{Tags} with a similar amount of \emph{Persons} are selected
 \\ \hline
		\end{tabularx}}\innerCardVSpace \\ \hline

		result &
		\innerCardVSpace{\begin{tabularx}{\attributeCardWidth}{|>{\resultNumberCell}C{\attributeNumberWidth}|>{\varNameCell}M|>{\typeCell}m{\typeWidth}|>{\resultOriginCell}c|Y|} \hline
		$\mathsf{1}$ & person1.id & ID & R &
				 \\ \hline
		$\mathsf{2}$ & person2.id & ID & R &
				 \\ \hline
		$\mathsf{3}$ & mutualFriendCount & 32-bit Integer & A &
				 \\ \hline
		\end{tabularx}}\innerCardVSpace \\ \hline

		sort		&
		\innerCardVSpace{\begin{tabularx}{\attributeCardWidth}{|>{\sortNumberCell}C{\attributeNumberWidth}|>{\varNameCell}M|>{\directionCell}c|Y|} \hline
		$\mathsf{1}$ & mutualFriendCount
 & $\desc
$ &  \\ \hline
		$\mathsf{2}$ & person1.id
 & $\asc
$ &  \\ \hline
		$\mathsf{3}$ & person2.id
 & $\asc
$ &  \\ \hline
		\end{tabularx}}\innerCardVSpace \\ \hline
	limit & 20 \\ \hline
	CPs &
	\multicolumn{1}{>{\raggedright}l|}{
		\chokePoint{2.5}, 
		\chokePoint{2.6}, 
		\chokePoint{8.1}
		} \\ \hline
\end{tabularx}
\queryCardVSpace

\let\emph\oldemph
\let\textbf\oldtextbf

\renewcommand{\currentQueryCard}{0}
\renewcommand*{\arraystretch}{1.1}

\subsection*{BI / read / 19}
\label{sec:bi-read-19}

\let\oldemph\emph
\renewcommand{\emph}[1]{{\footnotesize \sf #1}}
\let\oldtextbf\textbf
\renewcommand{\textbf}[1]{{\it #1}}

\renewcommand{\currentQueryCard}{bi-read-19}
\marginpar{
	\vspace{0.22ex}
	\raggedleft

	\queryRefCard{bi-read-01}{BI}{1}\\
	\queryRefCard{bi-read-02}{BI}{2}\\
	\queryRefCard{bi-read-03}{BI}{3}\\
	\queryRefCard{bi-read-04}{BI}{4}\\
	\queryRefCard{bi-read-05}{BI}{5}\\
	\queryRefCard{bi-read-06}{BI}{6}\\
	\queryRefCard{bi-read-07}{BI}{7}\\
	\queryRefCard{bi-read-08}{BI}{8}\\
	\queryRefCard{bi-read-09}{BI}{9}\\
	\queryRefCard{bi-read-10}{BI}{10}\\
	\queryRefCard{bi-read-11}{BI}{11}\\
	\queryRefCard{bi-read-12}{BI}{12}\\
	\queryRefCard{bi-read-13}{BI}{13}\\
	\queryRefCard{bi-read-14}{BI}{14}\\
	\queryRefCard{bi-read-15}{BI}{15}\\
	\queryRefCard{bi-read-16}{BI}{16}\\
	\queryRefCard{bi-read-17}{BI}{17}\\
	\queryRefCard{bi-read-18}{BI}{18}\\
	\queryRefCard{bi-read-19}{BI}{19}\\
	\queryRefCard{bi-read-20}{BI}{20}\\
}

\noindent\begin{tabularx}{\queryCardWidth}{|>{\queryPropertyCell}p{\queryPropertyCellWidth}|X|}
	\hline
	query & BI / read / 19 \\ \hline
	title & Interaction path between cities \\ \hline
	pattern & \centering \includegraphics[scale=\patternscale,margin=0cm .2cm]{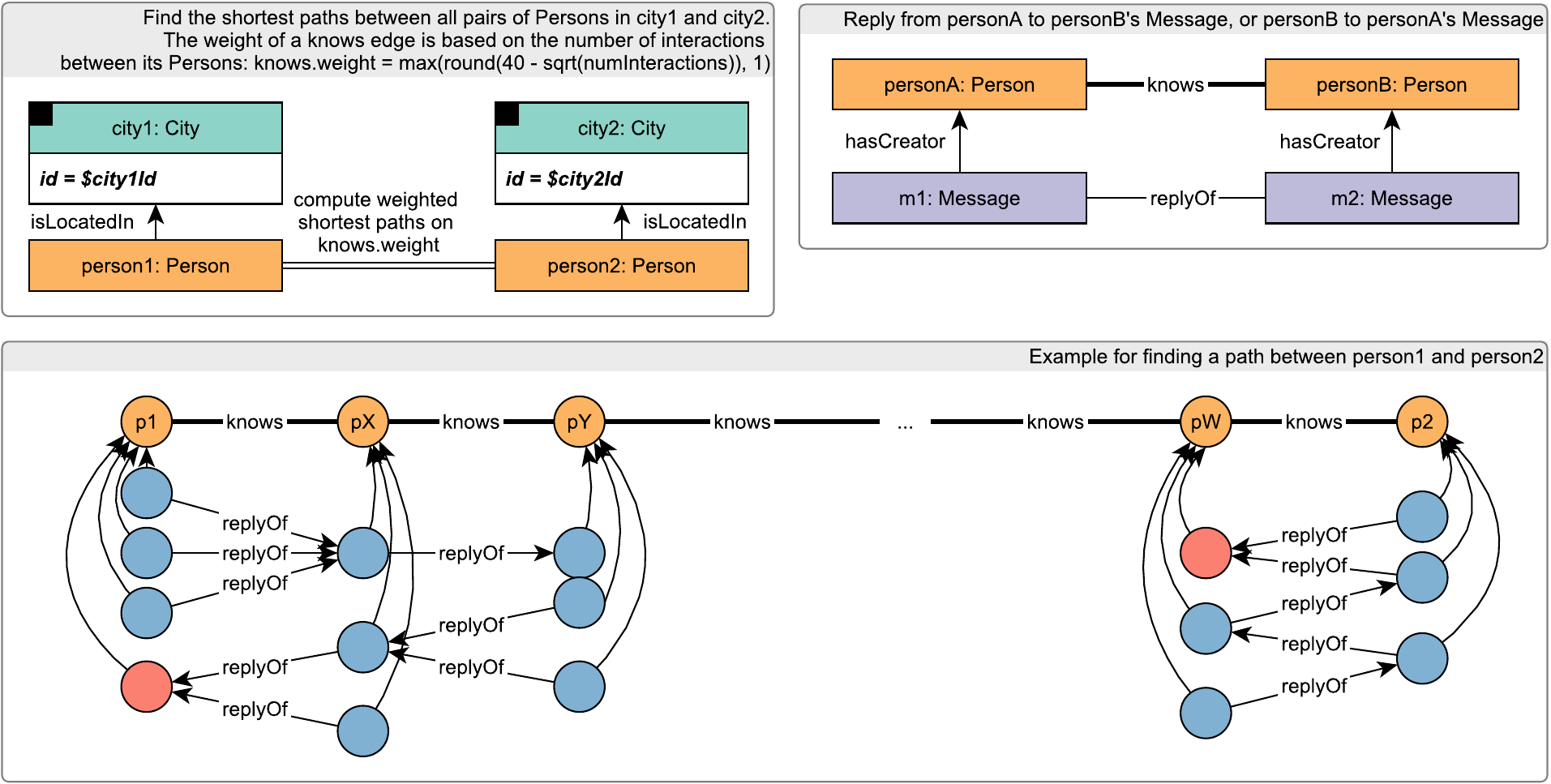} \tabularnewline \hline
	description & Given two \emph{Cities} with IDs \texttt{\$city1Id}, \texttt{\$city2Id},
find \emph{Persons} \texttt{person1}, \texttt{person2} living in these
\emph{Cities} (respectively) with the \textbf{cheapest} interaction path
between them.

The cheapest path is equivalent to the \textbf{weighted shortest} path.
It is computed on a subgraph of the \emph{Person-knows-Person} graph
with the edge weights based on the number of interactions. An
\textbf{interaction} is a direct reply \emph{Comments} from one
\emph{Person} to \emph{Messages} by the other \emph{Person}. Only
\emph{knows} edges with at least one interaction between their endpoint
\emph{Persons} are considered. For these, the weight of a \emph{knows}
edge is defined as:
\(\max(\mathrm{round}( 40 - \sqrt{\textit{numInteractions}} ), 1)\)

If there are multiple pairs of people with cheapest paths that have the
same total weight, return all of them.

\textbf{Note:} Interactions are counted both ways, e.g.~if Alice
\emph{knows} Bob, Alice writes 2 reply \emph{Comments} to Bob's
\emph{Messages} and Bob writes 3 reply \emph{Comments} to Alice's
\emph{Messages}, their total number of interactions is 5 and the weight
of the knows edge is 38.

\textbf{Remark:} Determinism is ensured by using square root followed by
rounding. For all integers between 1 and \(\numprint{100000}\), the
square root's fractional part is more than 10e-5 from 0.5, where the
rounding could be non-determinstic based on floating point inaccuracies.
As 10e-5 is significantly larger than the machine epsilon of IEEE 754
floats (both 32- and 64-bit), the floating point inaccuracies have no
chance to affect the derived integer edge weights.
 \\ \hline

		params &
		\innerCardVSpace{\begin{tabularx}{\attributeCardWidth}{|>{\paramNumberCell}C{\attributeNumberWidth}|>{\varNameCell}M|>{\typeCell}m{\typeWidth}|Y|} \hline
		$\mathsf{1}$ & \$city1Id
 & ID
 & \texttt{(a)} Small \emph{Cities} within the same \emph{Country}

\texttt{(b)} Larger \emph{Cities} from different \emph{Countries}
 \\ \hline
		$\mathsf{2}$ & \$city2Id
 & ID
 &  \\ \hline
		\end{tabularx}}\innerCardVSpace \\ \hline

		result &
		\innerCardVSpace{\begin{tabularx}{\attributeCardWidth}{|>{\resultNumberCell}C{\attributeNumberWidth}|>{\varNameCell}M|>{\typeCell}m{\typeWidth}|>{\resultOriginCell}c|Y|} \hline
		$\mathsf{1}$ & person1.id & ID & R &
				 \\ \hline
		$\mathsf{2}$ & person2.id & ID & R &
				 \\ \hline
		$\mathsf{3}$ & totalWeight & 32-bit Integer & C &
				 \\ \hline
		\end{tabularx}}\innerCardVSpace \\ \hline

		sort		&
		\innerCardVSpace{\begin{tabularx}{\attributeCardWidth}{|>{\sortNumberCell}C{\attributeNumberWidth}|>{\varNameCell}M|>{\directionCell}c|Y|} \hline
		$\mathsf{1}$ & person1.id
 & $\asc
$ &  \\ \hline
		$\mathsf{2}$ & person2.id
 & $\asc
$ &  \\ \hline
		\end{tabularx}}\innerCardVSpace \\ \hline
	limit & n/a \\ \hline
	CPs &
	\multicolumn{1}{>{\raggedright}l|}{
		\chokePoint{3.3}, 
		\chokePoint{7.6}, 
		\chokePoint{7.7}, 
		\chokePoint{8.4}, 
		\chokePoint{8.6}
		} \\ \hline
	relevance &
		\footnotesize To find the weighted shortest paths efficiently, the system can use
e.g.~a bidirectional Dijkstra algorithm. As the edge weights do not
depend on any parameter, systems can pre-compute them (if they do not
interleave reads and writes).
 \\ \hline%
\end{tabularx}
\queryCardVSpace

\let\emph\oldemph
\let\textbf\oldtextbf

\renewcommand{\currentQueryCard}{0}
\renewcommand*{\arraystretch}{1.1}

\subsection*{BI / read / 20}
\label{sec:bi-read-20}

\let\oldemph\emph
\renewcommand{\emph}[1]{{\footnotesize \sf #1}}
\let\oldtextbf\textbf
\renewcommand{\textbf}[1]{{\it #1}}

\renewcommand{\currentQueryCard}{bi-read-20}
\marginpar{
	\vspace{0.22ex}
	\raggedleft

	\queryRefCard{bi-read-01}{BI}{1}\\
	\queryRefCard{bi-read-02}{BI}{2}\\
	\queryRefCard{bi-read-03}{BI}{3}\\
	\queryRefCard{bi-read-04}{BI}{4}\\
	\queryRefCard{bi-read-05}{BI}{5}\\
	\queryRefCard{bi-read-06}{BI}{6}\\
	\queryRefCard{bi-read-07}{BI}{7}\\
	\queryRefCard{bi-read-08}{BI}{8}\\
	\queryRefCard{bi-read-09}{BI}{9}\\
	\queryRefCard{bi-read-10}{BI}{10}\\
	\queryRefCard{bi-read-11}{BI}{11}\\
	\queryRefCard{bi-read-12}{BI}{12}\\
	\queryRefCard{bi-read-13}{BI}{13}\\
	\queryRefCard{bi-read-14}{BI}{14}\\
	\queryRefCard{bi-read-15}{BI}{15}\\
	\queryRefCard{bi-read-16}{BI}{16}\\
	\queryRefCard{bi-read-17}{BI}{17}\\
	\queryRefCard{bi-read-18}{BI}{18}\\
	\queryRefCard{bi-read-19}{BI}{19}\\
	\queryRefCard{bi-read-20}{BI}{20}\\
}

\noindent\begin{tabularx}{\queryCardWidth}{|>{\queryPropertyCell}p{\queryPropertyCellWidth}|X|}
	\hline
	query & BI / read / 20 \\ \hline
	title & Recruitment \\ \hline
	pattern & \centering \includegraphics[scale=\patternscale,margin=0cm .2cm]{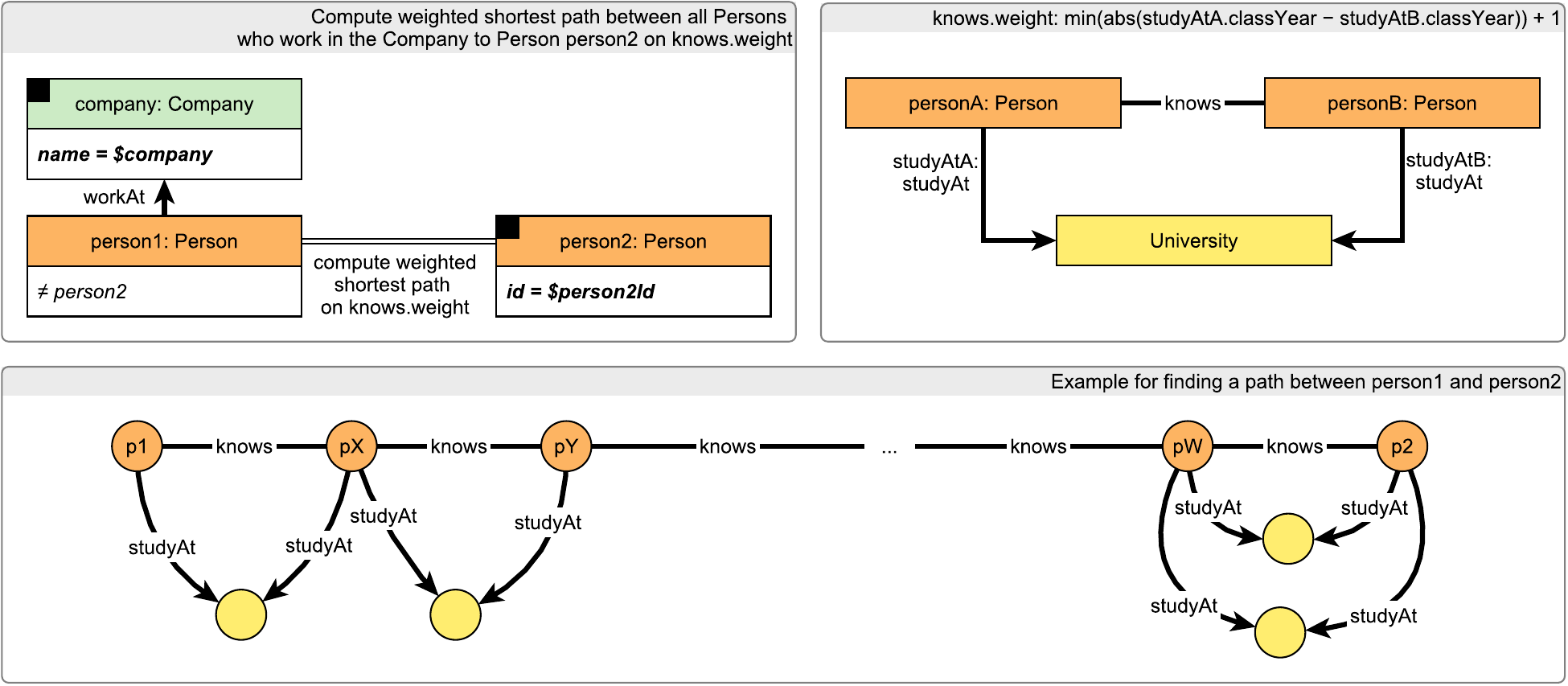} \tabularnewline \hline
	description & Consider \emph{knows} edges where the endpoint \emph{Persons} attended
the same \emph{University} and set the weight of the edge to the
absolute difference between the year of enrolment plus 1. If the
\emph{Persons} attended multiple universities, we select the smallest
(\texttt{min}) value. Formally: \[
w = \min_{\textsf{studyAt}_\textsf{A}, \textsf{studyAt}_\textsf{B}}
  \left|
    \textsf{studyAt}_\textsf{A}.\textsf{classYear} -
    \textsf{studyAt}_\textsf{B}.\textsf{classYear}
  \right|
  + 1
\]

Given a \texttt{\$company} and a \emph{Person} \texttt{person2} with ID
\texttt{\$person2Id} (who is not working and has not worked at
\texttt{\$company}), find a different \emph{Person} (\texttt{person1})
who works or at some point worked in \texttt{\$company} and is reachable
from \texttt{person2} through people who have studied together through
the shortest weighted path.

If there are multiple \emph{Person} \texttt{person1} nodes with the same
shortest path length, return all of them.
 \\ \hline

		params &
		\innerCardVSpace{\begin{tabularx}{\attributeCardWidth}{|>{\paramNumberCell}C{\attributeNumberWidth}|>{\varNameCell}M|>{\typeCell}m{\typeWidth}|Y|} \hline
		$\mathsf{1}$ & \$company
 & Long String
 & \emph{Companies} with a similar number of employees (former or current)
are selected
 \\ \hline
		$\mathsf{2}$ & \$person2Id
 & ID
 & \texttt{(a)} There is guaranteed to be no path between any
\texttt{person1} working at \texttt{company} and \texttt{person2}

\texttt{(b)} There is guaranteed to be a 2-hop path between at least one
\texttt{person1} working at \texttt{company} and \texttt{person}
 \\ \hline
		\end{tabularx}}\innerCardVSpace \\ \hline

		result &
		\innerCardVSpace{\begin{tabularx}{\attributeCardWidth}{|>{\resultNumberCell}C{\attributeNumberWidth}|>{\varNameCell}M|>{\typeCell}m{\typeWidth}|>{\resultOriginCell}c|Y|} \hline
		$\mathsf{1}$ & person1.id & ID & R &
				 \\ \hline
		$\mathsf{2}$ & totalWeight & 32-bit Integer & C &
				 \\ \hline
		\end{tabularx}}\innerCardVSpace \\ \hline

		sort		&
		\innerCardVSpace{\begin{tabularx}{\attributeCardWidth}{|>{\sortNumberCell}C{\attributeNumberWidth}|>{\varNameCell}M|>{\directionCell}c|Y|} \hline
		$\mathsf{1}$ & totalWeight
 & $\asc
$ &  \\ \hline
		$\mathsf{2}$ & person1.id
 & $\asc
$ &  \\ \hline
		\end{tabularx}}\innerCardVSpace \\ \hline
	limit & 20 \\ \hline
	CPs &
	\multicolumn{1}{>{\raggedright}l|}{
		\chokePoint{3.3}, 
		\chokePoint{7.6}, 
		\chokePoint{7.7}, 
		\chokePoint{7.8}, 
		\chokePoint{8.4}, 
		\chokePoint{8.6}
		} \\ \hline
	relevance &
		\footnotesize To find the weighted shortest paths efficiently, the system can use
e.g.~a bidirectional Dijkstra algorithm. As the edge weights do not
depend on any parameter, systems can pre-compute them (if they do not
interleave reads and writes).
 \\ \hline%
\end{tabularx}
\queryCardVSpace

\let\emph\oldemph
\let\textbf\oldtextbf

\renewcommand{\currentQueryCard}{0}


\section{Insert Operations}
\label{sec:bi-insert-operations}

Insert operations consist of individual inserts for each entity type.
Implementations typically use the same format as the one for loading the initial snapshot of the data set.


\section{Delete Operations}
\label{sec:bi-delete-operations}

\iftoggle{StandaloneWorkloadSpecification}{
    
}{
    See \autoref{sec:delete-operations}.
}

\chapter{Auditing Policies}
\label{sec:auditing}

\begin{quote}
    This chapter contains the auditing policies for the LDBC Social Network Benchmark. The initial draft of the auiting policies were published in the EU project deliverable D6.3.3 ``LDBC Benchmark Auditing Policies''.
\end{quote}


This chapter is divided in the following parts:
\begin{itemize}
    \item Motivation of benchmark result auditing
    \item General discussion of auditable aspects of benchmarks
    \item Specific checklists and running rules for the Social Network Benchmark's workloads (Interactive, Business Intelligence)
\end{itemize}

Many definitions and general considerations are shared between the benchmarks, hence it is justified to present the principles first and to refer to these in the context of the benchmark specific rules.

The auditing process, including the auditor certification exams, the possibility of challenging audited results, \etc, are defined in the LDBC Byelaws~\cite{ldbc_byelaws}. Please refer to the latest Byelaws document when conducting audits.


\section{Rationale and General Principles}


The purpose of benchmark auditing is to improve the \emph{credibility} and \emph{reproducibility} of benchmark claims by involving a set of detailed execution rules and third party verification of compliance with these.

Rules may exist separately of auditing but auditing is not meaningful unless the rules are adequately precise.
Aspects like auditor training and qualification cannot be addressed separately from a discussion of the matters the
auditor is supposed to verify. Thus the credibility of the entire process hinges on clear and shared understanding
of what a benchmark is expected to demonstrate and on the auditor being capable of understanding the process
and of verifying that the benchmark execution is fair and does not abuse the rules or pervert the objectives of
the benchmark.

Due to the open-ended nature of technology and the agenda of furthering innovation via measurement, it is
not feasible or desirable to over-specify the limits of benchmark implementation. Hence there will always remain
judgement calls for borderline cases. In this respect auditing and the LDBC are not separate. It is expected that
issues of compliance as well as of maintenance of rules will come before the LDBC as benchmark claims are
made.


\section{Auditing Rules Overview}

\subsection{Auditor Training, Certification, and Selection}
\subsubsection{Auditor Training}
Auditor training consists of familiarisation with the benchmark and existing implementations thereof. This involves the auditor candidate running the reference implementations of the benchmark in order to see what is normal behaviour and practice in the workload. The training and practice may involve communication with the benchmark task force for clarifying intent and details of the benchmark rules. This produces feedback for the task force for further specification of the rules.

\subsubsection{Auditor Certification}
The auditor certification and qualification is done in the form of an examination administered by the task force responsible for the benchmark being audited. The examination may be carried out by teleconference. The task force will subsequently vote on accepting each auditor, by simple majority. An auditor is certified for a particular benchmark by the task force maintaining the benchmark in question.

\subsubsection{Auditor Selection}
In the default auditor selection, the task force responsible for the benchmark being audited appoints a third party, impartial auditor. The task force may in special cases appoint itself as auditor of a particular result. This is not, however, the preferred course of action but may be done if no suitable third party auditor is available

\subsection{Auditing Process Stages}
\subsubsection{Getting Ready for a Benchmark Audit}
A benchmark result can be audited if it is a \emph{complete implementation} of an LDBC benchmark workload. This includes implementing all operations (reads and updates) correctly, using official data sets, using the official LDBC driver (if available), and complying with the auditing rules of the workload (\eg workloads may have different rules regarding query languages, the allowance of materialized views, \etc).
Workloads may specify further requirements such as ACID-compliance (checked using the LDBC ACID test suite).

\subsubsection{Performing a Benchmark Audit}
A benchmark result is to be audited by an LDBC appointed auditor or the LDBC task force managing the benchmark. An LDBC audit may be performed by remote login and does not require the auditor's physical presence on site. The test sponsor shall grant the auditor any access necessary for validating the benchmark run. This will typically include administrator access to the SUT hardware.

\subsubsection{Benchmark-Specific Checklist}
Each benchmark specifies a checklist to be verified by the auditor. The benchmark run shall be performed by the auditor. The auditor shall take copies of relevant configuration files and test results for future checking and insertion into the full disclosure report.

\subsubsection{Producing the FDR}
The FDR is produced by the auditor or auditors, with any required input from the test sponsor. Each non-default configuration parameter needs to be included in the FDR and justification needs to be provided why the given parameter was changed.
The auditor produces an attestation letter that verifies authenticity of the presented results. This letter is to be included into the FDR as an addendum. The attestation letter has no specific format requirements but shall state that the auditor has established compliance with a specified version of the benchmark specification.

\subsubsection{Publishing the FDR}
The FDR and any benchmark specific summaries thereof shall be published on the LDBC website, \url{https://ldbcouncil.org/}.

\subsection{Challenge Procedure}

A benchmark result may be \emph{challenged} for non-compliance with LDBC rules. The benchmark task force responsible for maintenance of the benchmark will rule on matters of compliance. A result found to be non-compliant will be withdrawn from the list of official LDBC benchmark results.


\section{Auditable Properties of Systems and Benchmark Implementations}


\subsection{Validation of Query Results}
\label{sec:validation}
A benchmark should be published with a deterministically reproducible validation data set. Validation queries applied to the validation data set will deterministically produce a set of correct answers. This is used in the first stage of benchmark run to test for the correctness of an SUT or benchmark implementation. This validation stage is not timed.

\paragraph{Inputs for validation}
The validation takes the form of a set of data generator parameters, a set of test queries that at least include one instance of each of the workload query templates and the expected results.

\paragraph{Approximate results and error margin}
In certain cases the results may be approximate. This may happen in cases of non-unique result ordering keys, imprecise numeric data types, random behaviours in certain graph analytics algorithms etc. Therefore, a validation set shall specify the degree of allowable error: For example, for counts, the value must be exact, for sums, averages and the like, at least 8 significant digits are needed, for statistical measures like graph centralities, the result must be within 1\% of the reference result. Each benchmark shall specify its expectation in an unambiguously verifiable manner.

\subsection{ACID Compliance}
\label{sec:acid-compliance}

As part of the auditing process for the Interactive workload and for certain systems in the BI workload, the auditors ascertain that the SUT satisfies the ACID properties,
\ie it provides atomic transactions, complies with its claimed isolation level, and ensures durability in case of failures.
This section outlines transactional behaviours of SUTs which are checked in the course of auditing an SUT in a given benchmark.

A benchmark specifies transactional semantics that may be required for different parts of the workload. The requirements will typically be different for initial bulk load of data and for the workload itself. Different sections of the workload may further be subject to different transactionality requirements.

No finite series of tests can prove that the ACID properties are fully supported. Passing the specified tests is a necessary, but not sufficient, condition for meeting the ACID requirements. However, for fairness of reporting, only the tests specified here are required and must appear in the FDR for a benchmark. (This is taken exactly from the \mbox{TPC-C} specification~\cite{tpcc}.)

The properties for ACID compliance are defined as follows:

\paragraph{Atomicity}
Either all of the effects of the transaction are in effect after the transaction or none of the effects
is in effect. This is by definition only verifiable after a transaction has finished.

\paragraph{Consistency}
ADS such as secondary indices will be consistent among themselves as well as with the table or other PDS, if any. Such a consistency (compliance to all constraints, if these are declared in the schema, \eg primary key constraint, foreign key constraints and cardinality constraints) may be verified
after the commit or rollback of a transaction. If a single thread of control runs within a transaction, then
subsequent operations are expected to see consistent state across all data indices pertaining to a table
or similar object. Multiple threads which may share a transaction context are not required to observe a
consistent state at all times during the execution of the transaction. Consistency will however always be
verifiable after the commit or rollback of any transaction, regardless of the number of threads that have
either implicitly or explicitly participated in the transaction. Any intra-transaction parallelism introduced
by the SUT will preserve transactional semantics statement-by-statement. If explicit, application created
sessions share a transaction context, then this definition of consistency does not hold: for example, if
two threads insert into the same table at the same time in the same transaction context, these may or may
not see a consistent image of (E)ADS for the parts affected by the other thread. All things will be
consistent after the commit or rollback, however, regardless of the number of threads, implicit or explicit
that have participated in the transaction.

\paragraph{Isolation}
Isolation is defined as the set of phenomena that may (or may not) be observed by operations running within a single transaction context. The levels of isolation are defined as follows:

\begin{description}
\item[Read uncommitted] No guarantees apply.
\item[Read committed] A transaction will never read a value that has at no point in time been part of a
    committed state.
\item[Repeatable read] If a transaction reads a value several times during its execution, then it will see
    the original state with its own modifications so far applied to it. If the transaction itself consists of
    multiple reading and updating threads then the ambiguities that may arise are beyond the scope of transaction isolation.
\item[Serializable] The transactions see values that correspond to a fully serial execution of
    all client transactions. This is like repeatable read except that if the transaction reads something, and
    repeats the read, it is guaranteed that no new values will appear for the same search condition on a
    subsequent read in the same transaction context. For example, a row that was seen not to exist when
    first checked will not be seen by a subsequent read. Likewise, counts of items will not be seen to
    change.
\end{description}

\paragraph{Durability}
Durability means that once the SUT has confirmed a successful commit, the committed state
will survive any instantaneous failure of the SUT (\eg a power failure, software crash, reboot or
the like). Durability is tied to atomicity in that if one part of the changes made by a transaction survives then
all parts must survive. 


\subsection{Data Schema}

A benchmark may specify restrictions on schema. For example, \mbox{TPC-H} and \mbox{TPC-DS} specify that only certain indices may be declared. In the LDBC context, the matter is more complex since the range of possible SUTs is much broader, including diverse combinations of schema first and schema-less systems and configurations.

\subsubsection{Schema Declaration}
By default, a system may declare no schema at all, as may be the case with RDF or graph DBMSs. If EADSs are declared, then these must be consistently applied to all data within the same workload for a given scale factor. The nature of prohibited EADSs, if any, depends on the benchmark and may be stated in the benchmark specification.

\subsubsection{Schema-Optional}

RDF and graph databases may sometimes be adopted due to their support for schema-last or schema-less operation. It is known that for many cases of RDF with a regular structure, a 1:1 mapping to a relational schema may exist. A benchmark may prohibit the use of such a mapping with the rationale that if the data were purely relational in structure then there would be no point in using RDF or graph DB in the first place. The example of such mapping is Sparqlify (or D2RQ), where SPARQL is directly translated to SQL and run against a relational database.

\paragraph{Use of EADS in a schema-less data model}
A benchmark may allow use of EADS with a schema-less data model such as RDF with the condition that whilst some data structures may become more efficient, no data structure is prohibited. The schema-less nature may persist but some common structures may benefit from more efficient physical representation.

\paragraph{Benchmarks enforcing schema-first semantics}
A benchmark may also state that it allows strict schema-first semantics, \eg SQL, and that the SUT need not make any specific provisions for schema change during the run. For an RDF system this would mean a priori imposing compliance with a data shape or ontology, not with OWL semantics but with semantics close to those of SQL DDL. In such a case, the ontology or data shape may as such be construed to be a valid hint for creation of application specific EADS.

\paragraph{Disclosure of data schema in the FDR}
In any case, a benchmark must state its policy concerning presence or absence of schema and enforcement thereof. If implementations declare a schema then any schema must be disclosed in full as part of the FDR.

\subsection{Data Format and Preprocessing}
\label{sec:auditing-data-format}

When producing the data sets, implementers are allowed to use custom formatting options (\eg use or omission of quotes, separator character, datetime format, \etc).
It is also allowed to convert the output of the Datagen into a format (\eg Parquet) that is loadable by the test-specific implementation of the data importer.
Additional preprocessing steps are also allowed, including adjustments to the CSV files (\eg with shell scripts), splitting and concatenating files, compressing and decompressing files, \etc
However, the preprocessing step shall not include a precomputation of (partial) query results.

\subsection{Data Access Transparency}

A benchmark may specify that an implementation is not allowed the use of explicit access paths. For example, explicitly specifying which EADS or IADS should be used for any given operation may be prohibited. Furthermore, in scale-out systems, explicit references to data location (other than via values of partitioning keys) may be prohibited. In general, references to internal data representation of an entity, \eg row in a table, should be prohibited. Reference should take place via column names in a schema or property URIs in RDF, not via physical offsets or the like.

\subsection{Query Languages}
\label{sec:query-languages}

In typical RDBMS benchmarks, online transaction processing (OLTP) benchmarks are allowed to be implementated via stored procedures, effectively amounting to explicit query plans.
Meanwhile, online analytical processing (OLAP) benchmarks prohibit the use of using general-purpose programming languages (\eg C, C\texttt{++}, Java) for query implementations and only allow domain-specific query languages.

In the graph processing space, there is currently (as of 2022) no standard query language and the systems are considerably more heterogeneous.
Therefore, the LDBC situation regarding declarativity is not as simple as that of for example the \mbox{TPC-H} (where queries should be specified in SQL with the additional constraint of omitting any hints for OLAP workloads) and individual SNB workloads specify their policy of either requiring a domain-specific query language or allowing the implementation of the queries in a general-purpose programming language.

In the case of domain-specific languages, systems are allowed to implement an SNB query as a sequence of multiple queries.
A typical example of this is the following sequence:
(1)~create projected graph,
(2)~run query,
(3)~drop projected graph.
However, it is not allowed to use subqueries in an unrealistic and contrived manner, \ie the goal of overcoming optimization issue, \eg hard-coding a certain join order in a declarative query language.
It is the responsibility of the auditor to determine whether a sequence of queries can be considered realistic w.r.t.\ how a user would formulate their queries in the language provided by the system.

\subsubsection{Rules for Imperative Implementations Using a General-Purpose Programming Language}
An implementation where the queries are written in a general-purpose programming language (including imperative and ``API-based'' implementations) may choose between semantically equivalent implementations of an operation based on the query parameters. This simulates the behaviour of a query optimizer in the presence of literal values in the query. If an implementation does this, all the code must be disclosed as part of the FDR and the decision must be based on values extracted from the database, not on hard-coded threshold values in the implementation.

The auditor must be able to reliably assess the compliance of an implementation to guidelines specifying these matters. The actual specification remains benchmark-dependent. Borderline cases may be brought to the task force responsible for arbitration.

\subsubsection{Disclosure of Query Implementations in the FDR}
Benchmarks allowing imperative expression of workload should require full disclosure of all query implementation code.

\subsection{Materialization}

The mix of read and update operations in a workload will determine to which degree precomputation of results is beneficial. The auditor must check that materialised results are kept consistent at the end of each transaction.

\subsection{Steady State}

An online workload must be able to indefinitely keep up the reported throughput. The benchmark definition may put specific restrictions on the duration of individual parts of the workload.

\subsubsection{Bringing the SUT into Steady State} One implication of this is that an SUT must be able to accommodate inserts at a specific rate for a realistic length of time. For example, if the workload is of an online nature then the SUT should be sized so as not to run out of space for new data for a reasonable duration of time. The \mbox{TPC-C} 180-day rule is an example of this. An analytical benchmark that primarily bulk loads data does not need to reserve as much space for new data. Each benchmark shall state its specific requirements in this respect.

\subsection{Query Mix}

A benchmark consists of multiple different operations that may vary in frequency and duration of individual
instances of each operation may vary in function of parameter selection. A benchmark must specify an operation
mix and a minimum count of operations that constitutes a compliant benchmark execution.

The auditor must ascertain from the records of a benchmark execution that a sufficient number of operations has indeed taken place for the report. For example, a 1000~GB \mbox{TPC-H} must have at least 7 streams in the throughput test and the workload is to be run twice following bulk load. For LDBC SNB, the run must be at least 2 hours of wall clock, measured time and the count of successful transactions of each type must be in a strictly set ratio with the count of other operations.

Benchmarks shall each specify a minimum count of operations and relative frequencies of operations for a qualifying
execution.

\subsubsection{Post-Processing of Query Results and Compression During Transmission}

All computing required for a given query needs to happen in the DBMS. The SUT's test driver shall not post-process query results in a way that changes their value. For example, it is not allowed to return floating-point values with a precision of 0.5 that are encoded as integers and divided by 2 on the client side.

Note that \emph{lossless compression} during the communication between the test driver and the DBMS is allowed. For instance, as long as the DBMS uses a data type that conforms with the schema requirements for a given attribute, one can apply compression to send it back to/from the driver and decompress it. For example, for complex query Q14 in the Interactive v1 workload, the implementation should ultimately produce a floating point score.

The same applies for query parameters. At both the client's and the server's endpoint, the correct fully qualified datatype must occur, but during transmission, it is allowed to apply compression.

\subsection{System Configuration and System Pricing}
\label{sec:system-config}


A benchmark execution shall produce a full disclosure report which specifies the hardware and software of the SUT, the benchmark implementation version and any specifics that are detailed in the benchmark specification. This clause gives a general minimum for disclosure for the SUT.

\subsubsection{Details of Machines Driving and Running the Workload}
An SUT may consist of one or more pieces of physical hardware. An SUT may include virtual or bare-metal machines in a cloud service.
For each distinct configuration, the FDR shall disclose the number of units of the type as well as the following:

\begin{enumerate}
    \item The used cloud provider (including the region where machines reside, if applicable).
    \item Common name of the item, \eg Dell PowerEdge xxxx or i3.2xlarge instance.
    \item Type and number of CPUs, cores \& threads per CPU, clock frequency, cache size.
    \item Amount of memory, type of memory and memory frequency, \eg 64GB DDR3 1333MHz.
    \item Disk controller or motherboard type if disk controller is on the motherboard.
    \item For each distinct type of secondary storage device, the number and specification of the device, \eg 4xSeagate Constellation 2TB SATA 6Gbit/s.
    \item Number and type of network controllers, \eg 1x Mellanox QDR InfiniBand HCA, PCIE 2.0, 2x1GbE on motherboard. If the benchmark execution is entirely contained on a single machine, it must be stated, and the description of network controllers can be omitted.
    \item Number and type of network switches. If multiple switches are used, the wiring between the switches should be disclosed.
    Only the network switches and interfaces that participate in the run need to be reported. If the benchmark execution is entirely contained on a single machine, it must be stated, and the description of network switches can be omitted.
    \item Date of availability of the system as a whole, \ie the latest date of availability of any part.
\end{enumerate}

\subsubsection{System Pricing}
The price of the hardware in question must be disclosed. For cloud setups, the price of a dedicated instance for 3 years must be disclosed. The price should reflect the single quantity list price that any buyer could expect when purchasing one system with the given specification. The price may be either an item by item price or a package price if the system is sold as a package.
Reported prices should adhere the TPC Pricing Specification 2.9.0~\cite{pricing,tpc-pricing}.
It is particularly important to ensure that the maintenance contract guarantees 24/7 support and 4~hour response time for problem recognition.
If the benchmark driver is running on a separate machine, the price of this machine should not be included in the total system price.

\subsubsection{Details of Software Components in the System}
The SUT software must be described at least as follows:
\begin{enumerate}
    \item The units of the SUT software are typically the DBMS and operating system.
    \item Name and version of each separately priced piece of the SUT software.
    \item If the price of the SUT software is tied to platform or count of concurrent users, these parameters must be disclosed.
    \item Price of the SUT software.
    \item Date of availability.
\end{enumerate}
Reported prices should adhere the TPC Pricing Specification 2.5.0~\cite{pricing,tpc-pricing}.

The configuration of the SUT must be reported so as to include the following:
\begin{enumerate}
    \item The used LDBC specification, driver and data generator version.
    \item Complete configuration files of the DBMS, including any general server configuration files, any configuration scripts run on the DBMS for setting up the benchmark run etc.
    \item Complete schema of the DBMS, including eventual specification of storage layout.
    \item Any OS configuration parameters if other than default, \eg \verb+vm.swappiness+, \verb+vm.max_map_count+ in Linux.
    \item Complete source code of any server-side logic, \eg stored procedures, triggers.
    \item Complete source code of driver-side benchmark implementation.
    \item Description of the benchmark environment, including software versions, OS kernel version, DBMS version as well as versions of other major software components used for running the benchmark (Docker, Java Virtual Machine, Python, etc.).
    \item The SUT's highest configurable isolation level and the isolation level used for running the benchmark.
\end{enumerate}

\subsubsection{Audit of System Configuration}
The auditor must ascertain that a reported run has indeed taken place on the SUT in the disclosed configuration.
The full disclosure shall contain any relevant parameters of the benchmark execution itself, including:
\begin{enumerate}
    \item Parameters, switches, configuration file for data generation.
    \item Complete text of any data loading script or program.
    \item Parameters, switches, configuration files for any test driver. If the test driver is not an LDBC supplied open source package or is a modification of such, then the complete text or diff against a specific LDBC package must be disclosed.
    \item Test driver output files shall be part of the disclosure. In general, these must at least detail the following:
    \begin{enumerate}[label=\roman*)]
        \item Time and duration of data load and the timed portion of the benchmark execution.
        \item Count of each workload item (\eg query, transaction) successfully executed within the measurement window.
        \item Min/average/max execution time of each workload item, the specific benchmark shall specify additional details.
    \end{enumerate}
\end{enumerate}

Given this information, the number of concurrent database sessions at each point in the execution must be clearly stated. In the case of a cluster database, the possible spreading of connections across multiple server processes must be disclosed.

All parameters included in this section must be reported in the full disclosure report to guarantee that the benchmark run can be reproduced exactly in the future. Similarly, the test sponsor will inform the auditor the scale factor to test. Finally, a clean test system with enough space to store the initial data set, the update streams, substitution parameters and anything that is part of the input and output as well as the benchmark run must be provided.

\subsection{Benchmark Specifics}

Similarly to TPC benchmarks, the LDBC benchmarks prohibit so-called benchmark specials (\ie extra software modules implemented in the core DBMS logic just to make a selected benchmark run faster are disallowed). Furthermore, upon request of the auditor, the test sponsor must provide all the source code relevant to the benchmark.


\section{Auditing Rules for the Interactive Workload}

\begin{figure}[h]
    \centering
    \includegraphics[scale=\yedscale]{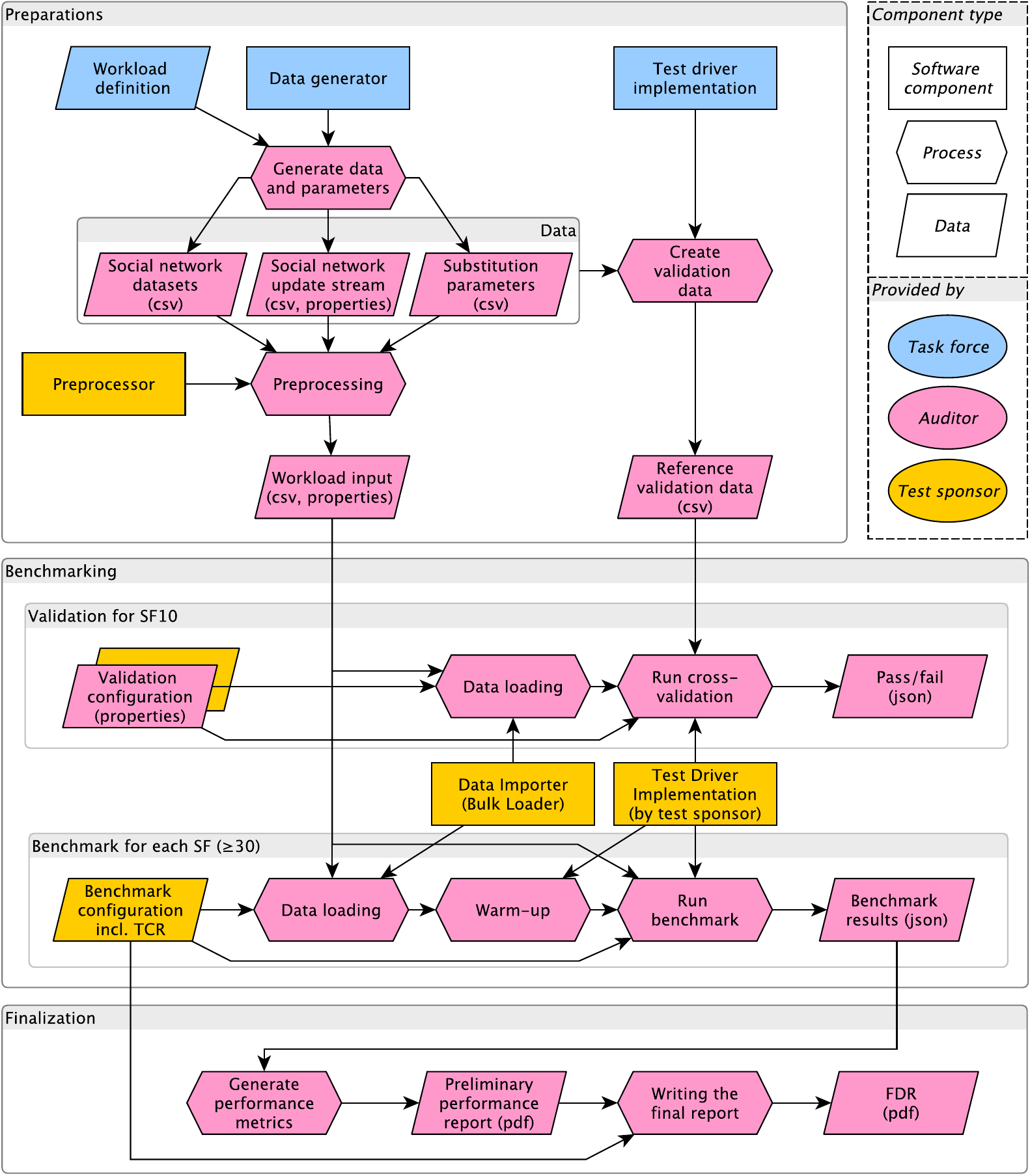}
    \caption{Benchmark execution and auditing workflow. For non-audited runs, the implementers perform the steps of the auditor.}
    \label{fig:audit-workflow}
\end{figure}

This section specifies a checklist (in the form of individual sections) that a benchmark audit shall cover in case of the SNB Interactive workload. An overview of the benchmark audit workflow is shown in \autoref{fig:audit-workflow}. The three major phases of the audit are preparing the input data and validation query results (captured by \emph{Preparations} in the figure), validating the correctness of query results returned by the SUT using the validation scale factor and running the benchmark with all the prescribed workloads (\emph{Benchmarking}), and creating the FDR (\emph{Finalization}). The colour codes capture the responsibilities of performing a step or providing some data in the workflow.

A key objective of the auditing guidelines for the Interactive workload is to \emph{allow a broad range of systems} to implement the benchmark.
Therefore, they do not impose constraints on the data model
(graph, relational, triple, \etc representations are allowed)
or on the query language
(both declarative and imperative languages are allowed).

\subsection{Scaling}
\label{sec:int-scaling}

\subsubsection{Scale Factors}

The scale factor of an SNB data set is the size of the data set in GiB of CSV (comma-separated values) files.
The size of a data set is characterized by scale factors: SF10, SF30, SF100 \etc (see \autoref{sec:scale-factors}).
All data sets contain data for three years of social network activity.

The \emph{validation run} shall be performed on the SF10 data set (see \autoref{sec:int-validation-data-set}) and use at least \numprint{100000} operations.
Note that the auditor may perform additional validation runs of the benchmark implementation using smaller data sets (\eg SF1) and issue queries.\footnote{%
An example test could be to issue complex reads with parameters such as \texttt{personId} and \texttt{messageId} selected from the \textsf{Person}/\textsf{Message} entities inserted from the update streams and cross-validate these against other systems. (The substitution parameters are taken from the initial snapshot of the graph so these nodes are not targeted by the regular workload executed by the driver.)%
}

Audited \emph{benchmark runs} of the Interactive workload shall use SF30 or larger data sets. The rationale behind this decision is to ensure that there is a sufficient number of update operations available to guarantee 2.5~hours of continuous execution (see \autoref{sec:int-measurement-window}).


\subsubsection{Social Network data sets}
\label{sec:int-data-sets}

\paragraph{Initial data set}
The data set is divided into a bulk loadable initial database population (90\%) and an update stream (10\%). These are generated by the SNB data generator. The data generator has options for splitting the data set into any number of files.

\paragraph{Dependencies between messages in the update stream}
The update stream contains the latest 10\% of the events in the simulated social network. These events form a single serializable sequence in time. Some events will depend on preceding events, for example a message must exist before a reply comment to the message is created. The data generator guarantees that these are separated by at least 10 seconds of simulation time.

\paragraph{Parallel updates}
The update stream may be broken into arbitrarily many sub-streams. The partition scheme is created by the \datagen. During benchmark execution, the driver preserves dependencies between update operations, such as ensuring not to refer to non-existent entities in updates (\eg a like is not added to a message which has not been inserted yet).

\subsection{Data Model and Data Loading}

\subsubsection{Supported Data Models}

SNB may be implemented with different data models (\eg relational, RDF, and different graph data models). The reference schema is provided in the specification using a UML-like notation.

\subsubsection{Generated Input Data}
\label{sec:generated-data}

\paragraph{Storage}
The data generator produces comma-separated values (CSV) for all data models.

\paragraph{Data format}
A single attribute has a single data type, as follows:
\begin{description}
    \item [Identifier] This is an integer value foreign key or a URI in RDF. If this is an integer column, the implementation data type should support at least $2^{50}$ distinct values.
    \item [Date] A date should support a date range from 0000 to 9999 in the year field.
    \item [DateTime] A datetime should support a date range from 0000 to 9999 in the year field, with at least millisecond precision.
    \item [Short string] The string column for names may have a variable length and may have a declared maximum length, \eg 40 characters.
    \item [Long string] For example a message content may be a long string that is often short in the data but may not declare a maximum length and must support data sizes of up to 1~MB.
\end{description}

The above is stated in further detail in the benchmark specification, and it shall take precedence over the
above in the case of conflict.

A single attribute in the reference schema may not be divided into multiple attributes in the target schema.

\paragraph{Database schema}
A schema on the DBMS is optional. An RDF implementation for example may work without one. An RDF implementation is allowed to load the RDF reference schema and to take advantage of the data type and cardinality statements therein. 

\paragraph{Configuration parameters}
\datagen configuration parameters, including SF, distributions, number of persons, serialiser (\eg CsvSingularMergedFK) should be reported.

\paragraph{Primary data structures}
An RDF, relational, or graph schema may specify system specific options affecting DBMS storage layout. These may for example specify vertical partitioning. Vertical partitioning means anything from a column store layout with per-column allocated storage space to use of explicit column groups. Any mix of row or column-wise storage structures is allowed as long as this is declaratively specified on a per data structure-basis.

\paragraph{Auxiliary data structures}
Covering indices and clustered indices are allowed. If these are defined, then all replications of data implied by these must be maintained statement by statement, \ie each auxiliary data structure must be consistent with any other data structures of the table after each data manipulation operation.

A covering index is an index which materialises a specific order of a specific subset or possibly all columns of a table. 
A clustered index is an index which materialises all columns of a table in a specific order, which order may or may not be that of the primary key of the table. A clustered or covering index may be the primary or only representation of a table.

Any subset of the columns on a covering or clustered index may be used for ordering the data. A hash based index or a combination of a hash based and tree based index are all allowed, in row or column-wise or hybrid forms.

\paragraph{Loading the data}

We expect the SUT to provide some means to bulk load the data set either in the form of a dedicated offline loader component or an online loader that allows bulk inserting into a database.
The total of the bulk load time and the time for subsequent operations (indexing, computing statistics, \etc) must be reported in the FDR (see \autoref{sec:int-benchmark-workflow}).
As loading can be an expensive operation, it is allowed to conduct the audit such that the loading is only performend once, and the validation/benchmarking phases use the resulting database instance.
In practice, this can look like as follows:
(1)~load the data,
(2)~compute statistics, uniqueness constraints, keys, indices, \etc,
(3)~shut down the SUT,
(4)~create a backup of the database (\eg by copying the directory of the database).
For all subsequent runs, the database shall be restored from the backup.

\subsection{Precomputation}

Precomputation of query results (both interim and end results) is allowed. However, systems must ensure that precomputed results (\eg materialized views) are kept consistent upon updates.

\subsection{Benchmark Software Components}
\label{sec:snb-software-components}
LDBC provides a test driver, data generator, and summary reporting scripts. Benchmark implementations shall use a stable version (\eg 0.3.6) of the test driver. The SUT's database software should be a stable version that is available publicly or can be purchased at the time of the release of the audit.

\subsubsection{Adaptation of the Test Driver to a DBMS}
\label{sec:test-driver}
A qualifying run must use a test driver that adapts the provided test driver to interface with the SUT. Such an implementation, if needed, must be provided by the test sponsor. The parameter generation, result recording, and workload scheduling parts of the test driver should not be changed. The auditor must be given access to the test driver source code used in the reported run.

The test driver produces the following artefacts for each execution as a by product of the run: Start and end timestamps in wall clock time, recorded with microsecond precision. The identifier of the operation and any substitution parameters.

\subsubsection{Summary of Benchmark Results}
\label{sec:performance-metrics}
A separate test summary tool provided with the test driver analyses the test driver log(s) after a measurement window is completed. 

The tool produces for each of the distinct queries and transactions the following summary:
\begin{itemize}
    \item Run time of query in wall clock time.
    \item Count of executions.
    \item Minimum/mean/percentiles/maximum execution time.
    \item Standard deviation from the average execution time.
\end{itemize}
The tool produces for the complete run the following summary:
\begin{itemize}
    \item Operations per second for a given SF (throughput). This is the primary metric of this workload.
    \item The total execution time in wall clock time.
    \item The total number of completed operations.
\end{itemize}

\subsection{Implementation Language and Data Access Transparency}

The queries and updates may be implemented in a domain-specific query language or as procedural code written in a general-purpose programming language (\eg using the API of the database).

\subsubsection{Implementations Using a Domain-Specific Query Language}
\label{sec:snb-domain-specific-query-language}

If a domain-specific query language is used, \eg GQL, SPARQL, SQL, SQL/PGQ, Cypher, or Gremlin, then explicit query plans are prohibited in all the read-only queries.%
\footnote{If the queries are not clearly declarative, the auditor must ensure that they do not specify explicit query plans by investigating their source code and experimenting with the query planner of the system (\eg using SQL's \texttt{EXPLAIN} command).}
The update transactions may still consist of multiple statements, effectively amounting to explicit plans.

Explicit query plans include but are not limited to:
\begin{itemize}
    \item Directives or hints specifying a join order or join type
    \item Directives or hints specifying an access path, \eg which index to use
    \item Directives or hints specifying an expected cardinality, selectivity, fanout or any other information that pertains to the expected number or results or cost of all or part of the query.
\end{itemize}

\begin{quote}
    \emph{Rationale behind the applied restrictions.} The updates are effectively OLTP and, therefore, the customary freedoms apply, including the use of stored procedures, however subject to access transparency. Declarative queries in a benchmark implementation should be such that they could plausibly be written by an application developer. Therefore, their formulation should not contain system specific aspects that an application developer would be unlikely to know. In other words, making a benchmark implementation should not require uncommon sophistication on behalf of the developer. This is regular practice in analytical benchmarks, \eg \mbox{TPC-H}.
\end{quote}

\subsubsection{Implementations Using a General-Purpose Programming Language}
\label{sec:snb-general-purpose-programming-language}

Implementations using a general-purpose programming language for specifying the queries (including procedural, imperative, and API-based implementations) are expected to respect the rules described in \autoref{sec:query-languages}.
For these implementations, the rules in \autoref{sec:snb-domain-specific-query-language} do not apply.

\subsection{Correctness of Benchmark Implementation}

\subsubsection{Validation data set}
\label{sec:int-validation-data-set}
The scale factor 10 shall be used as validation data set.

\subsubsection{ACID Compliance}
\label{sec:int-acid-compliance}

The Interactive workload requires full ACID support (\autoref{sec:acid-compliance}) from the SUT.
This is tested using the LDBC ACID test suite.
For the specification of this test suite, see \autoref{sec:acid-test-suite} and the related software repository at \url{https://github.com/ldbc/ldbc_acid}.

\paragraph{Expected level of isolation}
If a transaction reads the database with intent to update, the DBMS must guarantee that repeating the same read within the same transaction will return the same data. This also means that no more and no less data rows must be returned. In other words, this corresponds to snapshot or to serializable isolation. If the database is accessed without transaction context or without intent to update, then the DBMS should provide read committed semantics, \eg repeating the same read may produce different results but these results may never include effects of pending uncommitted transactions.

\paragraph{Durability and checkpoints}

A checkpoint is defined as the operation which causes data persisted in a transaction log to become durable outside of the transaction log. Specifically, this means that an SUT restart after instantaneous failure following the completion of the checkpoint may not have recourse to transaction log entries written before the end of the checkpoint.

A checkpoint typically involves a synchronisation barrier at which all data committed prior too the moment is required to be in durable storage that does not depend on the transaction log.
Not all DBMSs use a checkpointing mechanism for durability. For example a system may rely on redundant storage of data for durability guarantees against instantaneous failure of a single server.

The measurement window may contain a checkpoint. If the measurement window does not contain one, then the restart test will involve redoing all the updates in the window as part of the recovery test.

The timed window ends with an instantaneous failure of the SUT. Instantaneously killing all the SUT process(es) is adequate for simulating instantaneous failure. All these processes should be killed within one second of each other with an operating system action equivalent to the Unix \verb+kill -9+. If such is not available, then powering down each separate SUT component that has an independent power supply is also possible.

The restart test consists of restarting the SUT process(es) and finishes when the SUT is back online with all its functionality and the last successful update logged by the driver can be seen to be in effect in the database.

If the SUT hardware was powered down, the recovery period does not include the reboot and possible file system check time. The recovery time starts when the DBMS software is restarted.

\paragraph{Recovery} 
The SUT is to be restarted after the measurement window and the auditor will verify that the SUT contains the entirety of the last update recorded by the test driver(s) as successfully committed. The driver or the implementation have to make this information available. The auditor may also check the \emph{audit log} of the SUT (if available) to confirm that the operations issued by the driver were saved.

Once an official run has been validated, the recovery capabilities of the system must be tested. The system and the driver must be configured in the same way as in during the benchmark execution. After a warm-up period, an execution of the benchmark will be performed under the same terms as in the previous measured run.

\paragraph{Measuring recovery time}
At an arbitrary point close to 2 hours of wall clock time during the run, the machine will be shut down. Then, the auditor will restart the database system and will check that the last committed update (in the driver log file) is actually in the database. The auditor will measure the time taken by the system to recover from the failure. Also, all the information about how durability is ensured must be disclosed. If checkpoints are used, these must be performed with a period of 10 minutes at most.

\subsection{Benchmarking Workflow}
\label{sec:int-benchmark-workflow}

A benchmark execution is divided into the following processes (these processes are also shown in \autoref{fig:audit-workflow}):

\begin{description}
    \item[Generate data] This includes running the data generator, placing the generated files in a staging area, configuring storage, setting up the SUT configuration and preparing any data partitions in the SUT. This may include preallocating database space but may not include loading any data or defining any schema having to do with the benchmark. The \verb|ldbc.snb.interactive.update_interleave| driver parameter must come from the \verb|updateStream.properties| file, which is created by the data generator. That parameter should never be set manually. This parameter signifies the average distance of update operations in the workload.
    \item[Preprocessing] If needed, the output from the data generator is to preprocess the data set (\autoref{sec:auditing-data-format}).
    \item[Create validation data] Using one of the reference implementations of the benchmark, the reference validation data is obtained in .json format.
    \item[Data loading] The test sponsor must provide all the necessary documentation and scripts to load the data set into the database to test.
    This includes defining the database schema, if any, loading the initial database population, making this durably stored and gathering any optimiser statistics.
    The system under test must support the different data types needed by the benchmark for each of the attributes at their specified precision. No data can be filtered out, everything must be loaded. The test sponsor must provide a tool to perform arbitrary checks of the data or a shell to issue queries in a declarative language if the system supports it.
    \item[Run cross-validation] This step uses the data loader to populate the database, but the load is not timed. The validation data set is used to verify the correctness of the SUT. The auditor must load the provided data set and run the driver in validation mode, which will test that the queries provide the official results.  The benchmarking workflow will not go beyond this point unless results match the expected output.
    \item[Warm-up] Benchmark runs are preceded by a warm-up which must be performed using the LDBC driver.
    \item[Run benchmark] The bulk load time is reported and is equal to the amount of elapsed wall clock time between starting the schema definition and receiving the confirmation message of the end of statistics gathering. The workflow runs begin after the bulk load is completed. If the run does not directly follow the bulk load, it must start at a point in the update stream that has not previously been played into the database. In other words, a run may only include update events whose timestamp is later than the latest message creation date in the database prior to start of run. The run starts when the first of the test drivers send its first message to the SUT. If the SUT is running in the same process as the driver, the window starts
    when the driver starts. Also, make sure that the \verb|-rl/--results_log| is enabled. Make sure that all operations are enabled and the frequencies are those for the selected scale factor (see the exact specification of the frequencies in \autoref{sec:sf-statistics}).
\end{description}

\subsubsection{Query Timing During Benchmark Run}
\label{sec:ontime-requirements}
A valid benchmark run must last at least 2 hours of wall clock time and at most 2 hours and 15 minutes.
In order to be valid, a benchmark run needs to meet the ``95\% on-time requirement''.
The \texttt{results\_log.csv} file contains the $\mathsf{actual\_start\_time}$ and the $\mathsf{scheduled\_start\_time}$ of each of the issued queries. In order to have a valid run, 95\% of the queries must meet the following condition:
\begin{equation*}
\mathsf{actual\_start\_time} - \mathsf{scheduled\_start\_time} < 1\
\mathrm{second}
\end{equation*}

If the execution of the benchmark is valid, the auditor must retrieve all the files from directory specified by \verb|--results_dir| which includes configuration settings used, results log and results summary. All of which must be disclosed.

\subsubsection{Measurement Window}
\label{sec:int-measurement-window}

\begin{figure}[h]
    \centering
    \includegraphics[width=.7\linewidth]{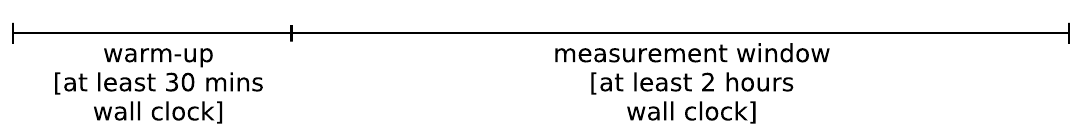}
    \caption{Warm-up and measurement window for benchmark run.}
    \label{fig:measurement-window-selection}
\end{figure}

Benchmark runs execute the workload on the SUT in two phases (\autoref{fig:measurement-window-selection}).
First, the SUT must undergo a warm-up period that takes at least 30 minutes and at most 35 minutes. The goal of this is to put the system in a steady state which reflects how it would behave in a normal operating environment. The performance of the operations during warm-up is not considered.
Next, the SUT is benchmarked during a two-hour measurement window. Operation times are recorded and checked to ensure the ``95\% on-time requirement'' is satisfied.

The SNB \datagen produces 3~years worth data of which 10\% is used for updates (\autoref{sec:int-data-sets}), \ie approximately $3 \times 365 \times 0.1 = 109.5~\text{days} = 2628~\text{hours}$.
To ensure that the 2.5~hour wall clock period has enough input data, the lower bound of TCR is defined as 0.001 (if $2628$ hours of updates are played back at more than $1000\times$ speed, the benchmark framework runs out of updates to execute). System that can achieve a better compression (\ie lower TCR value) on a given scale factor should use larger SFs for their benchmark runs -- otherwise their total runs will be less than 2.5~hours, making them unsuitable for auditing.


\subsection{Full Disclosure Report}
\label{sec:int-fdr}

Upon successful completion of the audit, an FDR is compiled. In addition to the general requirements, the full disclosure shall cover the following:

\begin{itemize}
    \item General terms: an executive summary and declaration of the credibility of the audit
    \item System description and pricing summary: see \autoref{sec:system-config}
    \item Data generation and data loading: see \autoref{sec:generated-data}
    \item Test driver details: see \autoref{sec:test-driver}
    \item Performance metrics: see \autoref{sec:performance-metrics}
    \item Validation results: see \autoref{sec:int-validation-data-set}
    \item ACID compliance: see \autoref{sec:acid-compliance}
    \item List of supplementary materials
\end{itemize}

To ensure reproducibility of the audited results, a supplementary package is attached to the full disclosure report. This package should contain:

\begin{itemize}
    \item A README file with instructions specifying how to set up the system and run the benchmark
    \item Configuration files of the database, including database-level configuration such as buffer size and schema descriptors (if necessary)
    \item Source code or binary of a generic driver that can be used to interact with the DBMS
    \item SUT-specific LDBC driver implementation (similarly to the projects in \url{https://github.com/ldbc/ldbc_snb_interactive_v1_impls}, \url{https://github.com/ldbc/ldbc_snb_interactive_v2_impls}, \url{https://github.com/ldbc/ldbc_snb_bi})
    \item Script or instructions to compile the LDBC Java driver implementation
    \item Instructions on how to the reach the server through CLI and/or web UI (if applicable), \eg the URL (including port number), user name and password
    \item LDBC configuration files (\texttt{.properties}), including the \texttt{time\_compression\_ratio} values used in the audited runs
    \item Scripts to preprocess the input files (if necessary) and to load the data sets into the database
    \item Scripts to create validation data sets and to run the benchmark
    \item The implementations of the queries and the update operations, including their complete source code (\eg declarative queries specifications, stored procedures, \etc)
    \item Implementation of the ACID test suite
    \item Binary package of the DBMS (\eg \texttt{.deb} or \texttt{.rpm})
\end{itemize}


\section{Auditing Rules for the Business Intelligence Workload}
\label{sec:auditing-bi-workload-audit}


The following section describes the auditing rules specific to the Business Intelligence (BI) workload.

\subsection{Overview}
\label{sec:auditing-bi-audit-overview}

Implementing the BI workload requires the following key capabilities:

\begin{itemize}
    \item Loading the initial snapshot of the social network graph
    \item Evaluating the BI read queries (\autoref{sec:bi-reads})
    \item Evaluating the BI write operations: inserts (\autoref{sec:bi-insert-operations}) and deletes (\autoref{sec:bi-delete-operations})
    \item Performing concurrent reads and writes (\autoref{sec:auditing-bi-audit-workflow}) (optional, only allowed if ACID compliance is guaranteed)
\end{itemize}

\subsection{Workflow}
\label{sec:auditing-bi-audit-workflow}

\begin{figure}[htbp]
    \centering
    \includegraphics[width=\textwidth]{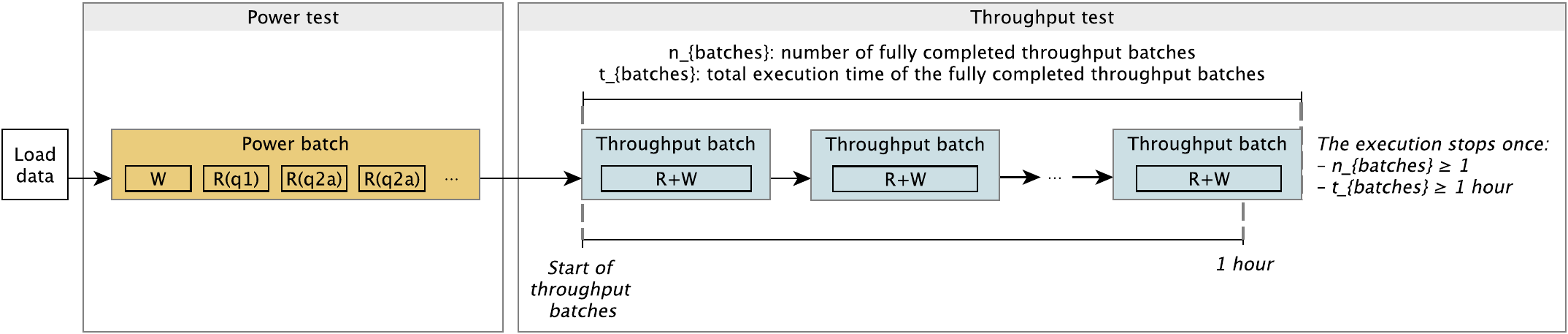}
    \caption{Tests and batches (power and throughput) executed in the BI workload's workflow.}
    \label{fig:bi-batches}
\end{figure}

The write operations and read queries are run in \emph{daily batches}.
In each batch, each query variant
(Q1, Q2\variantA, Q2\variantB, Q3, \ldots, Q20\variantA, Q20\variantB)
is executed using 30~different substitution parameters.
The BI workflow (\autoref{fig:bi-batches}) consists of two key parts:
the \emph{power test} (\autoref{sec:power-test}) and
the \emph{throughput test} (\autoref{sec:throughput-test}).

\subsubsection{Power Test}
\label{sec:power-test}

The \emph{power test} runs a single \emph{power batch}. This test first runs the write operations, followed by a sequential execution of individual read query variants.
The writes perform a day of inserts and deletes in the simulated social network,
while a total of $28 \times 30 = 840$ read queries are executed.

\subsubsection{Throughput Test}
\label{sec:throughput-test}

The \emph{throughput test} consists of multiple \emph{throughput batches}.
Each throughput batch runs the same type and the same amount of operations as the power batch.
However, they allow concurrent execution of the write operations and read queries in a given batch.

The execution of the throughput test during audits is the \emph{throughput measurement window}.
This window must span at least for 1~hour and
it must include at least one fully completed batch (see \autoref{fig:bi-batches}).

The workload defines two execution modes for the \emph{throughput batches}:

\begin{description}
    \item[Disjoint RW mode]
    In \emph{disjoint RW (read-write) mode} the system performs the reads and writes separately. It first executes the writes, then evaluates the read queries. Concurrency between the read operations is allowed.\\
    This mode is aimed at \emph{read-optimized data analytical systems} which do not support concurrent reads and writes. Implementations may also opt to use this mode for simplicity. For these implementations, passing the \emph{LDBC ACID compliance test} (\autoref{sec:acid-test-suite}) is not required.

    \item[Concurrent RW mode]
    In \emph{concurrent RW (read-write) mode} the system is allowed to run reads and writes concurrently. This requires the systems to be capable of handling \emph{transactions}. Implementations using this mode are required to pass the \emph{LDBC ACID compliance test} (\autoref{sec:acid-test-suite}).
\end{description}

\subsection{Runtimes}

The runtimes should be reported as follows:

\begin{itemize}
    \item The \emph{load time} ($t_\textit{load}$) denotes the time to load the data into the SUT and initialize auxiliary data structures (if applicable). For audited runs, we require that $t_\textit{load} < 24\textrm{ hours}$.
    \item The \emph{power test time} ($t_\textit{power test}$) denotes the time to perform the power test.
    \item The \emph{throughput measurement window time} ($t_\textit{throughput\ measurement}$) denotes the time to perform the throughput test, including the last (potentially unfinished) batch.
    \item The \emph{full throughput batches time} ($t_\textit{full\ throughput\ batches}$) denotes the time to evaluate the fully completed batches during the throughput measurement window.
\end{itemize}

Note that a warm-up period is not allowed (unlike the Interactive workload where such a period is required, see \autoref{sec:int-measurement-window}).

\subsection{Scoring Metrics}
\label{sec:auditing-bi-scoring-metrics}

SNBI BI provides four scoring metrics:
the \emph{power score}, the \emph{throughput score}, and their price-adjusted variants,
the \emph{per-\$ power score} and the \emph{per-\$ throughput score}.
All scores include the scale factor, denoted with ``@SF''.

\subsubsection{Price}
\label{sec:price-metrics}

We follow TPC's specification for reporting prices~\cite{tpc-pricing}.
The price is established as the \emph{total cost of ownership ($\textit{TCO}$)} for the SUT used in the benchmark,
reported as a breakdown of
machine cost,
software license costs,
and maintenance costs for 3 years.
In case of cloud deployments, the cost of running a 3-year reserved instance should be reported.
When establishing the price, the ``upfront payment'' option available at certain cloud providers should not be considered.

\subsubsection{Power Scores}
\label{sec:power-score}

The definition of \snbbi's power score follows \tpcH in using a geometric mean, ensuring that there is an incentive to improve all queries, no matter their running time.
Formally, the power score is based on the time to perform the writes
and
the time spent for executing each variant with 30~different substitution parameters, measured in seconds:
$$
\textit{power@SF} =
    \frac{\numprint{3600}}{\sqrt[29]{
        w
        \cdot q_{1}
        \cdot q_{2\variant{a}}
        \cdot q_{2\variant{b}}
        \cdot \ldots
        \cdot q_{18}
        \cdot q_{19\variant{a}}
        \cdot q_{19\variant{b}}
        \cdot q_{20\variant{a}}
        \cdot q_{20\variant{b}}
    }}
    \cdot
    \textit{SF}
$$


To determine the price-adjusted power score, we factor in the $\textit{TCO}$\emph{:}
$$ \textit{power@SF/\$} = \textit{power@SF} \cdot \frac{\numprint{1000}}{\textit{TCO}} $$

\subsubsection{Throughput Scores}
\label{sec:throughput-score}

The throughput score is based on $t_\textit{load}$, measured in hours,
and the cumulative execution time and number of the throughput batches executed:
$$
\textit{throughput@SF} =
    (24\text{ hours} - t_\textit{load})
    \cdot
    \frac{n_\textit{batches}}{t_\textit{batches}}
    \cdot
    \textit{SF}
$$

The subtraction of $t_\textit{load}$ ensures that the scoring rewards systems with efficient bulk loaders (unlike in \tpcH and \tpcDS which do not include load performance in their metrics).
The price-adjusted throughput score is determined analogously: 
$$ \textit{throughput@SF/\$} = \textit{throughput@SF} \cdot \frac{\numprint{1000}}{\textit{TCO}} $$

\subsection{Implementation Rules}
\label{sec:auditing-bi-implementation-rules}

\subsubsection{Correctness}
\label{sec:auditing-bi-correcntess}

The SUT shall evaluate all operators correctly.
The auditor shall ascertain correctness on the SF10 data set. However, they are allowed to also use data sets of different scale factors, as well as issue custom operations (both reads and writes) to test for the correctness of the implemenation.

The validation of correctness is performed on the output of the \emph{power test} step.
The rationale for using this only step is that during concurrent execution of R/W operations in the \emph{throughput test}, it is not possible to guarantee deterministic query results, making validation impossible. Moreover, this step already includes a write batch, therefore the query results indirectly test the correctness of the implementation of write operations.

\subsubsection{Auxiliary Data Structures}
\label{sec:auditing-bi-auxiliary-data-structures}

Using auxiliary data structures (\eg indices, materialized views) is allowed if they are kept in an up-to-date state after each write operation. The full disclosure report should enumerate the auxiliary data structures used by the SUT. 

\subsubsection{Query Declarativity}
\label{sec:auditing-bi-query-declarativity}

Systems should use a domain-specific query language (\eg Cypher, Gremlin, GQL, GSQL SQL/PGQ) for the implementation, including the read queries and the update operations.
General-purpose programming languages (\eg C, C\texttt{++}, Java, Julia) are not allowed.

Implementations shall not use \emph{query-specific stored procedures written in a general-purpose programming language} (\eg a given procedure which implements BI Q5).
Using the stored procedure libraries considered to be the ``standard libraries'' of the SUT is allowed.%
\footnote{These libraries often include features such as weighted shortest path algorithms.}
Implementations may use \emph{stored procedures written in a domain-specific language}.
In cases when the categorization of the approach used by the SUT's query implementations is uncertain, it is the auditor's responsibility to decide whether the SUT complies with this rule.

\subsubsection{Query Variants}
\label{sec:auditing-bi-query-variants}

Several queries (\eg \queryRefCard{bi-read-14}{BI}{14}) use \variantA and \variantB variants with different sets of input parameters.
The SUT should not receive any hints on which variant it is currently evaluating (\eg Q14\variantA or Q14\variantB).
Moreover, it is not allowed for the query implementations to contain code that aims to detect the query variant used.

\subsection{Scaling}
\label{sec:auditing-bi-scaling}

Audited \emph{benchmark runs} of the BI workload shall use SF30 or larger data sets.
The rationale behind this decision is to ensure that there should be a sufficient number of write operations available to guarantee the execution during the duration of the measurement window (see \autoref{fig:bi-batches}).

\subsection{Full Disclosure Report}
\label{sec:auditing-bi-fdr}

The \emph{full disclosure report} (FDR) and the \emph{supplementary package} shall contain
the same information as for SNB Interactive (\autoref{sec:int-fdr}),
including, if applicable (\autoref{sec:auditing-bi-implementation-rules}), the ACID compliance report (\autoref{sec:acid-compliance}).

\chapter{ACID Test Suite}
\label{sec:acid-test-suite}

\newcommand{\bl}[1]{\textcolor{blue}{#1}}
\newcommand{\rd}[1]{\textcolor{red}{#1}}
\newcommand{\gn}[1]{\textcolor{green}{#1}}
\newcommand{\gy}[1]{\textcolor{grey}{\textit{#1}}}

\newcommand{\level}[1]{\textsf{#1}}
\newcommand{\anomaly}[1]{\rd{#1}}
\newcommand{\anolong}[1]{\emph{\rd{#1}}}
\newcommand{\tx}[1]{#1}

\newcommand{\cmark}{\ding{51}}
\newcommand{\xmark}{\ding{55}}

\begin{quote}
  \textit{This chapter is based on the TPCTC~2020 paper ``Towards Testing ACID Compliance in the LDBC Social Network Benchmark''~\cite{DBLP:conf/tpctc/WaudbySKMBS20}, co-authored by several members of the SNB task force.}

  \textit{The framework and reference implementations of the ACID test suite are available at \url{https://github.com/ldbc/ldbc_acid}.}
\end{quote}

Verifying ACID compliance is an important step in the benchmarking process for enabling fair comparison between systems.
The performance benefits of operating with weaker safety guarantees are well established~\cite{DBLP:conf/ds/GrayLPT76} but this can come at the cost of application correctness.
To enable apples vs. apples performance comparisons between systems it is expected they uphold the ACID properties.
Currently, LDBC provides no mechanism for validating ACID compliance within the SNB Interactive workflow.
A simple solution would be to outsource the responsibility of demonstrating ACID compliance to benchmark implementors.
However, the safety properties claimed by a system often do not match observable behaviour~\cite{kingsbury}.
To mitigate this problem, benchmarks such as \mbox{TPC-C}~\cite{tpcc} include a number of ACID tests to be executed as part of the benchmarking auditing process.
However, we found these tests cannot readily be applied to our context, as they assume lock-based concurrency control and an interactive query API that provides clients with explicit control over a transaction's lifecyle.
Modern data systems often use optimistic concurrency control mechanisms~\cite{DBLP:journals/sigmod/PavloA16} and offer a restricted query API, such as only executing transactions as stored procedures~\cite{DBLP:conf/vldb/StonebrakerMAHHH07}.
Further, tests that trigger and test row-level locking phenomena, for instance, do not readily map on graph database systems.
Lastly, we found these tests are limited in the range of isolation anomalies they cover.

This chapter presents the design of an implementation-agnostic ACID-compliance test suite for the Interactive workload\footnote{We acknowledge verifying ACID-compliance with a finite set of tests is not possible. However, the goal is not an exhaustive quality assurance test of a system's safety properties but rather to demonstrate that ACID guarantees are supported.}.
Our guiding design principle was to be agnostic of system-level implementation details, relying solely on client observations to determine the occurrence of non-transactional behaviour.
Thus all systems can be subjected to the same tests and fair comparisons between SNB Interactive performance results can be drawn.
Tests are described in the context of a graph database employing the property graph data model~\cite{DBLP:journals/csur/AnglesABHRV17}.
Reference implementations are given in Cypher~\cite{DBLP:conf/sigmod/FrancisGGLLMPRS18}, the \emph{de facto} standard graph query language.

Particular emphasis is given to testing isolation, covering 10~known anomalies including recently discovered anomalies such as \anolong{Observed Transaction Vanishes}~\cite{DBLP:journals/pvldb/BailisDFGHS13} and \anolong{Fractured Reads}~\cite{DBLP:journals/tods/BailisFGHS16}.
The test suite has been implemented for 5~database systems.\footnote{Available at \url{https://github.com/ldbc/ldbc_acid}.}
A conscious decision was made to keep tests relatively lightweight, as to not add significant overhead to the benchmarking process.

\section{Background}

\begin{figure}
  \centering
  \includegraphics[scale=\yedscale]{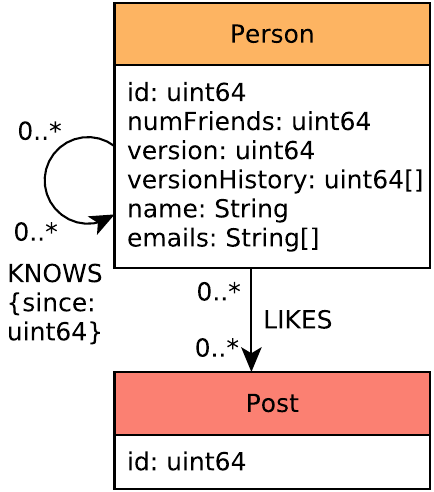}
  \caption{Graph schema for the ACID test queries.}
  \label{fig:core-schema}
\end{figure}

The tests presented in this chapter are defined on a small core of LDBC SNB schema (extended with properties for versioning) given in \autoref{fig:core-schema}.

\begin{figure}
  \centering
  \scriptsize
\begin{tikzpicture}[xscale=0.77,yscale=0.47,trim left=0.8cm]
\node[text width=4cm,align=center] (RU) at (3,1) {
\level{Read Uncommitted}  \\
\anomaly{G0}
};

\node[text width=4cm,align=center] (RC) at (3,3) {
\level{Read Committed}  \\
\anomaly{+ G1\{a-c\}}
};

\node[text width=4cm,align=center] (ICI) at (9,1) {
\level{Item Cut Isolation}  \\
\anomaly{IMP}
};

\node[text width=4cm,align=center] (PCI) at (9,4) {
\level{Predicate Cut Isolation}  \\
\anomaly{+ PMP}
};

\node[text width=1.5cm,align=center] (MAV) at (6,5) {
\level{Monotonic Atomic View}  \\
\anomaly{+ OTV}
};

\node[text width=4cm,align=center] (CS) at (3,6.5) {
\level{Cursor Stability}  \\
\anomaly{+ G-Cursor(x), LU}
};

\node[text width=2.5cm,align=center] (RA) at (9,7) {
\level{Read Atomic}  \\
\anomaly{+ FR}
};

\node[text width=3.3cm,align=center] (SI) at (9,9) {
\level{Snapshot Isolation}  \\
\anomaly{+ LU}
};

\node[text width=2.2cm,align=center] (RR) at (3,9) {
\level{Repeatable Read}  \\
\anomaly{+ WS (G2-Item)}
}; 

\node[text width=4cm,align=center] (S) at (6,11) {
  \level{Serializability}
};

\draw [->,>=stealth] (RU) -- (RC);
\draw [->,>=stealth] (ICI) -- (PCI);
\draw [->,>=stealth] (RC) -- (MAV);
\draw (ICI) -- (MAV);
\draw [->,>=stealth] (MAV) -- (RR);
\draw [->,>=stealth] (PCI) -- (RA);
\draw [->,>=stealth] (MAV) -- (RA);
\draw [->,>=stealth] (RA) -- (RR);
\draw [->,>=stealth] (RA) -- (SI);
\draw [->,>=stealth] (SI) -- (S);
\draw [->,>=stealth] (RR) -- (S);
\draw [->,>=stealth] (CS) -- (RR);
\draw [->,>=stealth] (RC) -- (CS);

\end{tikzpicture}
\caption{Hierarchy of isolation levels as described in~\cite{DBLP:journals/tods/BailisFGHS16}. All anomalies are covered except \anomaly{G-Cursor(x)}.}
\label{fig:isolation}
\end{figure}

\section{Atomicity}
\label{sec:atomicity}

\emph{Atomicity} ensures that either all of a transaction's actions are performed, or none are.
Two atomicity tests have been developed.
\textbf{Atomicity-C} checks for every successful commit message a client receives that any data items inserted or modified are subsequently visible.
\textbf{Atomicity-RB} checks for every aborted transaction that all its modifications are not visible.
Tests are executed as follows:
(i) load a graph of \texttt{Person} nodes (\autoref{fig:ainitial}) each with a unique \texttt{id} and a set of \texttt{emails};
(ii) a client executes a full graph scan counting the number of nodes, edges and emails (\autoref{fig:acheck}) using the result to initialize a counter \texttt{committed};
(iii) $N$ transaction instances (\autoref{fig:ac}, \autoref{fig:arb}) of the required test are then executed, \texttt{committed} is incremented for each successful commit;
(iii) repeat the full graph scan, storing the result in the variable \texttt{finalState};
(iv) perform the anomaly check: \texttt{committed=finalState}.

The \textbf{Atomicity-C} transaction (\autoref{fig:ac}) randomly selects a \texttt{Person}, creates a new \texttt{Person}, inserts a \texttt{KNOWS} edge and appends an \texttt{email}. The \textbf{Atomicity-RB} transaction (\autoref{fig:arb}) randomly selects a \texttt{Person}, appends an \texttt{email} and attempts to insert a \texttt{Person} only if it does not exist.
Note, for \textbf{Atomicity-RB} if the query API does not offer a \texttt{ROLLBACK} statement constraints such as node uniqueness can be utilized to trigger an abort.

\begin{figure}[htb]
\centering

\begin{lstlisting}[language=cypher,label=fig:ainitial,caption=Cypher query for creating initial data for the \tx{Atomicity} transactions.]
CREATE (:Person {id: 1, name: 'Alice', emails: ['alice@aol.com']}),
       (:Person {id: 2, name: 'Bob', emails: ['bob@hotmail.com', 'bobby@yahoo.com']})
\end{lstlisting}

\begin{minipage}{0.41\linewidth}
\begin{lstlisting}[language=cypher,label=fig:ac,caption=\tx{Atomicity-C Tx.}]
<<BEGIN>>
MATCH (p1:Person {id: $person1Id})
CREATE (p1)-[k:KNOWS]->(p2:Person)
SET
  p1.emails = p1.emails + [$newEmail],
  p2.id = $person2Id,
  k.creationDate = $creationDate
<<COMMIT>>
\end{lstlisting}
\end{minipage}
\quad
\begin{minipage}{0.52\linewidth}
\begin{lstlisting}[language=cypher,label=fig:arb,caption=\tx{Atomicity-RB Tx.}]
<<BEGIN>>
MATCH (p1:Person {id: $person1Id})
SET p1.emails = p1.emails + [$newEmail]
<<IF>> MATCH (p2:Person {id: $person2Id}) exists
<<THEN>> <<ABORT>> <<ELSE>>
CREATE (p2:Person {id: $person2Id, emails: []})
<<END>>
<<COMMIT>>
\end{lstlisting}
\end{minipage}

\begin{lstlisting}[language=cypher,label=fig:acheck,caption=\tx{Atomicity-C/Atomicity-RB:} counting entities in the graph.]
MATCH (p:Person)
RETURN count(p) AS numPersons, count(p.name) AS numNames, sum(size(p.emails)) AS numEmails
\end{lstlisting}
\end{figure}

\section{Isolation}
\label{sec:isolation}

The gold standard isolation level is \level{Serializability}, which offers protection against all possible \emph{anomalies} that can occur from the concurrent execution of transactions.
Anomalies are occurrences of non-serializable behaviour.
Providing \level{Serializability} can be detrimental to performance~\cite{DBLP:conf/ds/GrayLPT76}.
Thus systems offer numerous weak isolation levels such as \level{Read Committed} and \level{Snapshot Isolation} that allow a higher degree of concurrency at the cost of potential non-serializable behaviour.
As such, isolation levels are defined in terms of the anomalies they prevent~\cite{DBLP:conf/ds/GrayLPT76,DBLP:journals/pvldb/BailisDFGHS13}.
\autoref{fig:isolation} relates isolation levels to the anomalies they proscribe.

SNB Interactive does not require systems to provide \level{Serializability}.
However, to allow fair comparison systems must disclose the isolation level used during benchmark execution.
The purpose of these isolation tests is to verify that the claimed isolation level matches the expected behaviour.
To this end, tests have been developed for each anomaly presented in~\cite{DBLP:journals/tods/BailisFGHS16}.
Formal definitions for each anomaly are reproduced from~\cite{adya1999weak,DBLP:journals/tods/BailisFGHS16} using their system model which is described below.
General design considerations are discussed before each test is described.

\subsection{System Model}
\label{sec:system-model}

Transactions consist of an ordered sequence of read and write operations to an arbitrary set of data items, book-ended by a \texttt{BEGIN} operation and a \texttt{COMMIT} or an \texttt{ABORT} operation.
In a graph database data items are nodes, edges and properties.
The set of items a transaction reads from and writes to is termed its \emph{item read set} and \emph{item write set}.
Each write creates a \emph{version} of an item, which is assigned a unique timestamp taken from a totally ordered set (\eg natural numbers) version $i$ of item $x$ is denoted $x_i$.
All data items have an initial \emph{unborn} version $\bot$ produced by an initial transaction $\tx{T_{\bot}}$.
The unborn version is located at the start of each item's version order.
An execution of transactions on a database is represented by a \emph{history}, H, consisting of (i) each transaction's read and write operations, (ii) data item versions read and written and (iii) commit or abort operations.

There are three types of dependencies between transactions, which capture the ways in which transactions can \emph{directly} conflict.
\emph{Read dependencies} capture the scenario where a transaction reads another transaction's write.
\emph{Antidependencies} capture the scenario where a transaction overwrites the version another transaction reads.
\emph{Write dependencies} capture the scenario where a transaction overwrites the version another transaction writes. Their definitions are as follows:

\begin{description}
  \item[Read-Depends]
    Transaction $\tx{T_j}$ \emph{directly read-depends} (\textsf{wr}) on $\tx{T_i}$ if $\tx{T_i}$ writes some version $x_k$ and $\tx{T_j}$ reads $x_k$.
  \item[Anti-Depends]
    Transaction $\tx{T_j}$ \emph{directly anti-depends} (\textsf{rw}) on $\tx{T_i}$ if $\tx{T_i}$ reads some version $x_k$ and $\tx{T_j}$ writes $x$'s next version after $x_k$ in the version order.
  \item[Write-Depends]
    Transaction $\tx{T_j}$ \emph{directly write-depends} (\textsf{ww}) on $\tx{T_i}$ if $\tx{T_i}$ writes some version $x_k$ and $\tx{T_j}$ writes $x$'s next version after $x_k$ in the version order.
\end{description}

Using these definitions, from a history $H$ a \emph{direct serialization graph} $\textit{DSG}(H)$ is constructed.
Each node in the $\textit{DSG}$ corresponds to a committed transaction and edges correspond to the types of direct conflicts between transactions.
Anomalies can then be defined by stating properties about the $\textit{DSG}$.

The above \emph{item-based} model can be extended to handle \emph{predicate-based} operations~\cite{adya1999weak}.
Database operations are frequently performed on set of items provided a certain condition called the \emph{predicate}, $P$ holds.
When a transaction executes a read or write based on a predicate $P$, the database selects a version for each
item to which $P$ applies, this is called the version set of the predicate-based denoted as $\textit{Vset}(P)$.
A transaction $\tx{T_j}$ changes the matches of a predicate-based read $r_i(P_i)$ if $\tx{T_i}$ overwrites a version in $\textit{Vset}(P_i)$.

\subsection{General Design}
\label{sec:design-cons}

Isolation tests begin by loading a \emph{test graph} into the database.
Configurable numbers of \emph{write clients} and \emph{read clients} then execute a sequence of transactions on the database for some configurable time period.
After execution, results from read clients are collected and an \emph{anomaly check} is performed.
In some tests an additional full graph scan is performed after the execution period in order to collect information required for the anomaly check.

The guiding principle behind test design was the preservation of data item's version history -- the key ingredient needed in the system model formalization which is often not readily available to clients, if preserved at all.
Several anomalies are closely related, tests therefore had to be constructed such that other anomalies could not interfere with or mask the detection of the targeted anomaly.
Test descriptions provide
(i) informal and formal anomaly definitions,
(ii) the required test graph,
(iii) description of transaction profiles write and read clients execute, and
(iv) reasoning for why the test works.

\subsection{Dirty Write}
\label{sec:dirty-write}

Informally, a \anolong{Dirty Write} (Adya's \anomaly{G0}~\cite{adya1999weak})
occurs when updates by conflicting transactions are interleaved.
For example, say $\tx{T_i}$ and $\tx{T_j}$ both modify items $\{x,y\}$.
If version $x_i$ precedes version $x_j$ and $y_j$ precedes version $y_i$ a \anomaly{G0} anomaly has occurred.
Preventing \anomaly{G0} is especially important in a graph database in to order to maintain \emph{Reciprocal Consistency}~\cite{Waudby2020}.

\paragraph{Definition.}
A history $H$ exhibits phenomenon \anomaly{G0} if $\textit{DSG}(H)$ contains a directed cycle consisting entirely of write-dependency edges.

\paragraph{Test.}
Load a test graph containing pairs of \texttt{Person} nodes connected by a \texttt{KNOWS} edge.
Assign each \texttt{Person} a unique \texttt{id} and each \texttt{Person} and \texttt{KNOWS} edge a \texttt{versionHistory} property of type list (initially empty).
During the execution period, write clients execute a sequence of \tx{G0 $T_\mathrm{W}$} instances, \autoref{fig:dw1}.
This transaction appends its ID to the \texttt{versionHistory} property for each entity in the \texttt{Person} pair it matches.
Note, transaction IDs are assumed to be globally unique.
After execution, a read client issues a \tx{G0 $T_\mathrm{R}$} for each \texttt{Person} pair in the graph, \autoref{fig:dw2}.
Retrieving the \texttt{versionHistory} for each entity (2 \texttt{Persons} and 1 \texttt{KNOWS} edge) in a \texttt{Person} pair.

\paragraph{Anomaly check.}
For each \texttt{Person} pair in the test graph: (i) prune each \texttt{versionHistory} list to remove any version numbers that do not appear in all lists; needed to account for interference from \anolong{Lost Update} anomalies (\autoref{sec:lost-update}), (ii) perform an element-wise comparison between \texttt{versionHistory} lists for each entity, (iii) if lists do not agree a \anomaly{G0} anomaly has occurred.

\paragraph{Why it works.}
Each \tx{G0 $T_\mathrm{W}$} effectively creates a new version of a \texttt{Person} pair.
Appending the transaction ID preserves the version history of each entity in the \texttt{Person} pair.
In a system that prevents \anomaly{G0}, each entity of the \texttt{Person} pair should experience the \emph{same} updates, in the \emph{same} order.
Hence, each position in the \texttt{versionHistory} lists should be equivalent.
The additional pruning step is needed as \anolong{Lost Updates}
overwrite a version, effectively erasing it from the history of a data item.

\begin{figure}[htb]
  \centering
  \begin{minipage}{0.53\linewidth}
\begin{lstlisting}[language=cypher,label=fig:dw1,caption=\tx{Dirty Write (G0) $T_\mathrm{W}$}.]
MATCH
  (p1:Person {id: $person1Id})
  -[k:KNOWS]->(p2:Person {id: $person2Id})
SET p1.versionHistory = p1.versionHistory + [$tId]
SET p2.versionHistory = p2.versionHistory + [$tId]
SET k.versionHistory  = k.versionHistory  + [$tId]
\end{lstlisting}
\end{minipage}
\quad
\begin{minipage}{0.431\linewidth}
\begin{lstlisting}[language=cypher,label=fig:dw2,caption=\tx{Dirty Write (G0) $T_\mathrm{R}$}.]
MATCH (p1:Person {id: $person1Id})
-[k:KNOWS]->(p2:Person {id: $person2Id})
RETURN
  p1.versionHistory AS p1VersionHistory,
  k.versionHistory  AS kVersionHistory,
  p2.versionHistory AS p2VersionHistory
\end{lstlisting}
\end{minipage}
\end{figure}

\subsection{Dirty Reads}
\label{sec:dirty-reads-1}

\subsection*{Aborted Reads}

Informally, an \anolong{Aborted Read} (\anomaly{G1a}) anomaly occurs when a transaction reads the updates of a transaction that later aborts.

\paragraph{Definition.}
A history $H$ exhibits phenomenon \anomaly{G1a} if $H$ contains an aborted transaction $\tx{T_i}$ and a committed transaction $\tx{T_j}$ such that $\tx{T_j}$ reads a version written by $\tx{T_i}$.

\paragraph{Test.}
Load a test graph containing only \texttt{Person} nodes into the database.
Assign each \texttt{Person} a unique \texttt{id} and \texttt{version} initialized to 1; any odd number will suffice.
During execution, write clients execute a sequence of \tx{G1a $T_\mathrm{W}$} instances, \autoref{fig:ar1}.
Selecting a random \texttt{Person} \texttt{id} to populate each instance.
This transaction attempts to set \texttt{version=2} (any even number will suffice) but always aborts.
Concurrently, read clients execute a sequence of \tx{G1a $T_\mathrm{R}$} instances, \autoref{fig:ar2}.
This transaction retrieves the \texttt{version} property of a \texttt{Person}.
Read clients store results, which are pooled after execution has finished.

\paragraph{Anomaly check.}
Each read should return \texttt{version=1} (or any odd number).
Otherwise, a \anomaly{G1a} anomaly has occurred.

\paragraph{Why it works.}
Each transaction that attempts to set \texttt{version} to an even number \emph{always} aborts.
Therefore, if a transaction reads \texttt{version} to be an even number, it must have read the write of an aborted transaction.

\begin{figure}[htb]
\centering
\begin{minipage}{0.45\linewidth}
\begin{lstlisting}[language=cypher,label=fig:ar1,caption=\tx{Aborted Read (G1a) $T_\mathrm{W}$}.]
MATCH (p:Person {id: $personId})
SET p.version = 2
<<SLEEP($sleepTime)>>
<<ABORT>>
\end{lstlisting}

\begin{lstlisting}[language=cypher,label=fig:ar2,caption=\tx{Aborted Read (G1a) $T_\mathrm{R}$}.]
MATCH (p:Person {id: $personId})
RETURN p.version
\end{lstlisting}
\end{minipage}
\quad
\begin{minipage}{0.45\linewidth}
\begin{lstlisting}[language=cypher,label=fig:ir1,caption=\tx{Interm. Read (G1b) $T_\mathrm{W}$}.]
MATCH (p:Person {id: $personId})
SET p.version = $even
<<SLEEP($sleepTime)>>
SET p.version = $odd
\end{lstlisting}

\begin{lstlisting}[language=cypher,label=fig:ir2,caption=\tx{Interm. Read (G1b) $T_\mathrm{R}$}.]
MATCH (p:Person {id: $personId})
RETURN p.version
\end{lstlisting}
\end{minipage}
\end{figure}

\subsection*{Intermediate Reads}

Informally, an \anolong{Intermediate Read} (Adya's \anomaly{G1b}~\cite{adya1999weak}) anomaly occurs when a transaction reads the intermediate modifications of other transactions.

\paragraph{Definition.}
A history $H$ exhibits phenomenon \anomaly{G1b} if $H$ contains a committed transaction $\tx{T_i}$ that reads a version of an object $x_m$ written by transaction $\tx{T_j}$, and $\tx{T_j}$ also wrote a version $x_n$ such that $m < n$ in $x$'s version order.

\paragraph{Test.}
Load a test graph containing only \texttt{Person} nodes into the database. Assign each \texttt{Person} a unique \texttt{id} and \texttt{version} initialized to 1; any odd number will suffice.
During execution, write clients execute a sequence of \tx{G1b $T_\mathrm{W}$} instances, \autoref{fig:ir1}.
This transaction sets \texttt{version} to an even number, then an odd number before committing.
Concurrently read-clients execute a sequence of \tx{G1b $T_\mathrm{R}$} instances, \autoref{fig:ir2}.
Selecting a \texttt{Person} by \texttt{id} and retrieving its \texttt{version} property.
Read clients store results which are collected after execution has finished.

\paragraph{Anomaly check.}
Each read of \texttt{version} should be an odd number.
Otherwise, a \anomaly{G1b} anomaly has occurred.

\paragraph{Why it works.}
The final version installed by an \tx{G1b $T_\mathrm{W}$} instance is \emph{never} an even number.
Therefore, if a transaction reads \texttt{version} to be an even number it must have read an intermediate version.

\subsection*{Circular Information Flow}

Informally, a \anolong{Circular Information Flow} (Adya's \anomaly{G1c}~\cite{adya1999weak}) anomaly occurs when two transactions affect each other; \ie both transactions write information the other reads.
For example, transaction $\tx{T_i}$ reads a write by transaction $\tx{T_j}$ and transaction $\tx{T_j}$ reads a write by $\tx{T_i}$.

\paragraph{Definition.}
A history $H$ exhibits phenomenon \anomaly{G1c} if $\textit{DSG}(H)$ contains a directed cycle that consists entirely of read-dependency and write-dependency edges.

\paragraph{Test.}
Load a test graph containing only \texttt{Person} nodes into the database.
Assign each \texttt{Person} a unique \texttt{id} and \texttt{version} initialized to 0.
Read-write clients are required for this test, executing a sequence of \tx{G1c $T_\mathrm{RW}$}, \autoref{fig:cif1}.
This transaction selects two different \texttt{Person} nodes, setting the \texttt{version} of one \texttt{Person} to the transaction ID and retrieving the \texttt{version} from the other.
Note, transaction IDs are assumed to be globally unique.
Transaction results are stored in format \texttt{(txn.id, versionRead)} and collected after execution.

\begin{figure}
\begin{lstlisting}[language=cypher,label=fig:cif1,caption=\tx{G1c $T_\mathrm{RW}$}.]
MATCH (p1:Person {id: $person1Id}) SET p1.version = $transactionId
MATCH (p2:Person {id: $person2Id}) RETURN p2.version
\end{lstlisting}
\end{figure}

\paragraph{Anomaly check.}
For each result, check the result of the transaction the \texttt{versionRead} corresponds to, did not read the transaction of that result.
If so a \anomaly{G1c} anomaly has occurred.

\paragraph{Why it works.}
Consider the result set:
\texttt{\{($T_\mathrm{1}$, $T_\mathrm{2}$), ($T_\mathrm{2}$, $T_\mathrm{3}$), ($T_\mathrm{3}$, $T_\mathrm{2}$)\}}.
$T_\mathrm{1}$ reads the version written by $T_\mathrm{2}$ and $T_\mathrm{2}$  reads the version written by $T_\mathrm{3}$.
Here information flow is unidirectional from $T_\mathrm{1}$ to $T_\mathrm{2}$.
However, $T_\mathrm{2}$ reads the version written by $T_\mathrm{3}$ and $T_\mathrm{3}$  reads the version written by $T_\mathrm{2}$.
Here information flow is circular from $T_\mathrm{2}$ to $T_\mathrm{3}$ and $T_\mathrm{3}$ to $T_\mathrm{2}$.
Thus a \anomaly{G1c} anomaly has been detected.

\subsection{Cut Anomalies}

\subsection*{Item-Many-Preceders}
\label{sec:cut-anomalies}

Informally, an \anolong{Item-Many-Preceders} (\anomaly{IMP}) anomaly~\cite{DBLP:journals/pvldb/BailisDFGHS13} occurs if a transaction observes multiple versions of the same item (\eg transaction $\tx{T_i}$ reads versions $x_1$ and $x_2$).
In a graph database this can be multiple reads of a node, edge, property or label.
Local transactions (involving a single data item) occur frequently in graph databases, \eg in \emph{``Retrieve content of a message''} \queryRefCard{interactive-short-read-04}{IS}{4}.

\paragraph{Definition.}
A history $H$ exhibits \anomaly{IMP} if $\textit{DSG}(H)$ contains a transaction $\tx{T_i}$ such that $\tx{T_i}$ directly \emph{item-read-depends} on $x$ by more than one other transaction.

\paragraph{Test.}
Load a test graph containing \texttt{Person} nodes.
Assign each \texttt{Person} a unique \texttt{id} and \texttt{version} initialized to 1.
During execution write clients execute a sequence of \tx{IMP $T_\mathrm{W}$} instances, \autoref{fig:ic1}.
Selecting a random \texttt{id} and installing a new version of the \texttt{Person}.
Concurrently read clients execute a sequence of \tx{IMP $T_\mathrm{R}$} instances, \autoref{fig:ic2}.
Performing multiple reads of the same \texttt{Person}; successive reads can be separated by some artificially injected wait time to make conditions more favourable for detecting an anomaly.
Both reads within an \tx{IMP $T_\mathrm{R}$} transaction are returned, stored and collected after execution.

\paragraph{Anomaly check.}
Each \tx{IMP $T_\mathrm{R}$} result set \texttt{(firstRead, secondRead)} should contain the \emph{same} \texttt{Person} version.
Otherwise, an \anomaly{IMP} anomaly has occurred.

\paragraph{Why it works.}
By performing successive reads within the same transaction this test checks that a system ensures consistent reads of the same data item.
If the version changes then a concurrent transaction has modified the data item and the reading transaction is not protected from this change.

\begin{figure}[htb]
\centering
\begin{minipage}{0.35\linewidth}
\begin{lstlisting}[language=cypher,label=fig:ic1,caption=\tx{IMP $T_\mathrm{W}$}.]
MATCH (p:Person {id: $personId})
SET p.version = p.version + 1
\end{lstlisting}
\begin{lstlisting}[language=cypher,label=fig:ic2,caption=\tx{IMP $T_\mathrm{R}$}.]
MATCH (p1:Person {id: $personId})
WITH p1.version AS firstRead
<<SLEEP($sleepTime)>>
MATCH (p2:Person {id: $personId})
RETURN firstRead,
  p2.version AS secondRead
\end{lstlisting}
\end{minipage}
\quad
\begin{minipage}{0.61\linewidth}
\begin{lstlisting}[language=cypher,label=fig:pc1,caption=\tx{PMP $T_\mathrm{W}$}.]
MATCH (pe:Person {id: $personId}), (po:Post {id: $postId)
CREATE (pe)-[:LIKES]->(po)
\end{lstlisting}
\begin{lstlisting}[language=cypher,label=fig:pc2,caption=\tx{PMP $T_\mathrm{R}$}.]
MATCH (po1:Post {id: $postId})<-[:LIKES]-(pe1:Person)
WITH count(pe1) AS firstRead
<<SLEEP($sleepTime)>>
MATCH (po2:Post {id: $postId})<-[:LIKES]-(pe2:Person)
RETURN firstRead,
  count(pe2) AS secondRead
\end{lstlisting}
\end{minipage}
\end{figure}

\subsection*{Predicate-Many-Preceders}

Informally, a \anolong{Predicate-Many-Preceders} (\anomaly{PMP}) anomaly~\cite{DBLP:journals/pvldb/BailisDFGHS13} occurs if a transaction observes different versions resulting from the same predicate read
(\eg $\tx{T_i}$ reads
$\textit{Vset}(P_i) =  \{x_1\}$ and
$\textit{Vset}(P_i) = \{x_1,y_2\}$).
Pattern matching is a common predicate read operation in a graph database, \eg query \emph{``Find friends and friends of friends that have been to given countries''} \queryRefCard{interactive-complex-read-03}{IC}{3}.

\paragraph{Definition.}
A history $H$ exhibits the phenomenon \anomaly{PMP} if, for all predicate-based reads $r_i(P_i : \textit{Vset}(P_i))$ and $r_j(P_j : \textit{Vset}(P_j))$ in $\tx{T_k}$ such that the logical ranges of $P_i$ and $P_j$ overlap (call it $P_o$), the set of transactions that change the matches of $P_o$ for $r_i$ and $r_j$ differ.

\paragraph{Test.}
Load a test graph containing \texttt{Person} and \texttt{Post} nodes.
Within each node type assign unique IDs.
During execution write clients execute a sequence of \tx{PMP $T_\mathrm{W}$} instances, inserting a \texttt{LIKES} edge between a randomly selected \texttt{Person} and \texttt{Post}, shown in \autoref{fig:pc1}.
Concurrently read clients execute a sequence of \tx{PMP $T_\mathrm{R}$} instances, \autoref{fig:pc2}.
Performing multiple reads of the pattern \texttt{(po:Post)<-[:LIKES]-(p:Person)} and counting the number of \texttt{LIKES} edges; successive reads can be separated by some artificially injected wait time to make conditions more favourable for detecting an anomaly.
Both predicate reads within a \tx{PMP $T_\mathrm{R}$} transaction are returned, stored and collected after test execution.

\paragraph{Anomaly check.}
For each \tx{PMP $T_\mathrm{R}$} transaction result set \texttt{(firstRead, secondRead)}, the \texttt{firstRead} should be equal to \texttt{secondRead}.
Otherwise, a \anomaly{PMP} anomaly has occurred.

\paragraph{Why it works.}
By performing successive predicate reads and counting the number of \texttt{LIKES} edges within the same transaction this test checks that a system ensures consistent reads of the same predicate.
If the number of \texttt{LIKES} edges changes then a concurrent transaction has inserted a new \texttt{LIKES} edge and the reading transaction is not protected from this change.

\subsection{Observed Transaction Vanishes}
\label{sec:observ-trans-vanish}
Informally, an \anolong{Observed Transaction Vanishes} (\anomaly{OTV}) anomaly~\cite{DBLP:journals/pvldb/BailisDFGHS13} occurs when a transaction observes part of another transaction's updates but not all of them (\eg $\tx{T_1}$ writes $x_1$ and $y_1$ and $\tx{T_2}$ reads $x_1$ and $y_\bot$).
Before formally defining \anomaly{OTV} the \emph{Unfolded Serialization Graph (USG)} must be introduced~\cite{adya1999weak}.
The $\textit{USG}$ is specified for an individual transaction, $\tx{T_i}$ and a history, $H$ and is denoted by $\textit{USG}(H,\tx{T_i})$.
In a \emph{USG} the $\tx{T_i}$ node is split into multiple nodes, one for each action read $r_i(\cdot)$ or  write $w_i(\cdot)$  within the transaction.
The dependency edges are now incident on the relevant event of $\tx{T_i}$.
Additionally, actions within $\tx{T_i}$ are connected by  an \emph{order edge} \eg if $T_i$ reads object $y_j$ then immediately writes on object $x$ an order edge exists from $w_i(x_i)$ to $r_i(y_j)$.

\paragraph{Definition.}
A history $H$ exhibits phenomenon \anomaly{OTV} if $\textit{USG}(H,T_i)$ contains a directed cycle consisting of
(i)~exactly one read dependency edge induced by data item $x$ from $\tx{T_j}$ to $\tx{T_i}$ and
(ii)~a set of edges induced by data item $y$ containing at least one anti dependency edge from $\tx{T_i}$ to $\tx{T_j}$.
Additionally, $\tx{T_i}$'s read from $y$ precedes its read from $x$.

\paragraph{Test.}
Load a test graph containing a set of cycles of length 4 of \texttt{Persons} with same \texttt{name} connected by \texttt{Knows} edges.
Assign each \texttt{Person} an \texttt{id}, \texttt{name} and \texttt{version} property (initialized to 1).
Note, \texttt{id} must be unique across nodes and \texttt{name} must be unique across cycles.
During execution write clients select a \texttt{name}, \texttt{id} and executes a sequence of \tx{OTV  $T_\mathrm{W}$} instances, \autoref{fig:otvfr1}.
This transaction effectively creates a new version of a given cycle.
Concurrently read-clients execute a sequence of \tx{OTV $T_\mathrm{R}$} instances, \autoref{fig:otvfr2}.
Matching a given cycle and performing multiple reads.
Both reads within an \tx{OTV $T_\mathrm{R}$} are returned, stored and collected after execution.

\paragraph{Anomaly check.}
For each \tx{OTV $T_\mathrm{R}$} result set \texttt{(firstRead,secondRead)}, the maximum \texttt{version} in the \texttt{firstRead} should be less than or equal to the minimum \texttt{version} in the \texttt{secondRead}.
Otherwise, an \anomaly{OTV} anomaly has occurred.

\paragraph{Why it works.}
\tx{OTV $T_\mathrm{W}$} installs a new version of a cycle by updating the \texttt{version} property of each \texttt{Person}.
Therefore when matching a cycle once a transaction has observed some \texttt{version} it should \emph{at least} observe this version for every remaining entity in the cycle.
Unfortunately, this cannot be deduced from a single read of the cycle as results from matching cycles often does not preserve the order in which graph entities were read.
This is solved by making multiple reads of the cycle.
The maximum \texttt{version} of the \texttt{firstRead} determines the minimum \texttt{version} of \texttt{secondRead}.
If this condition is violated then a transaction has observed the effects of a transaction in the \texttt{firstRead} then subsequently failed to observe it in the \texttt{secondRead} -- the observed transaction has vanished!


\begin{figure}[htb]
\centering
\begin{minipage}{0.33\linewidth}
\begin{lstlisting}[language=cypher,label=fig:otvfr1,caption=\tx{OTV/FR $T_\mathrm{W}$}.]
MATCH path =
  (n:Person {id: $personId})
  -[:KNOWS*..4]->(n)
UNWIND nodes(path)[0..4] AS p
SET p.version = p.version + 1
\end{lstlisting}
\end{minipage}
\quad
\begin{minipage}{0.60\linewidth}
\begin{lstlisting}[language=cypher,label=fig:otvfr2,caption=\tx{OTV/FR $T_\mathrm{R}$}.]
MATCH p1=(n1:Person {id: $personId})-[:KNOWS*..4]->(n1)
RETURN extract(p IN nodes(p1) | p.version) AS firstRead
<<SLEEP($sleepTime)>>
MATCH p2=(n2:Person {id: $personId})-[:KNOWS*..4]->(n2)
RETURN extract(p IN nodes(p2) | p.version) AS secondRead
\end{lstlisting}
\end{minipage}
\end{figure}

\subsection{Fractured Read}
\label{sec:fractured-reads}

Informally, a \anolong{Fractured Read} (\anomaly{FR}) anomaly~\cite{DBLP:journals/tods/BailisFGHS16} occurs when a transaction reads \emph{across} transaction boundaries.
For example, if $\tx{T_1}$ writes $x_1$ and $y_1$ and $\tx{T_3}$ writes $x_3$.
If $\tx{T_2}$ reads $x_1$ and $y_1$, then repeats its read of $x$ and reads $x_3$ a fractured read has occurred.

\paragraph{Definition.}
A transaction $\tx{T_j}$ exhibits phenomenon \anomaly{FR} if transaction $\tx{T_i}$ writes versions $x_a$ and $y_b$ (in any order, where $x$ and $y$ may or may not be distinct items), $\tx{T_j}$ reads version $x_a$ and version $y_c$, and $c < b$.

\paragraph{Test.}
Same as the \anomaly{OTV} test.

\paragraph{Anomaly check.}
For each \tx{FR $T_\mathrm{R}$}  (\autoref{fig:otvfr2}) result set \texttt{(firstRead, secondRead)}, all
\texttt{versions} across both version sets should be equal.
Otherwise, an \anomaly{FR} anomaly has occurred.

\paragraph{Why it works.}
\tx{FR $T_\mathrm{W}$} installs a new version of a cycle by updating the \texttt{version} properties on each \texttt{Person}.
When \tx{FR $T_\mathrm{R}$} observes a \texttt{version} every subsequent read in that cycle should read the \emph{same} \texttt{version} as \tx{FR $T_\mathrm{W}$} (\autoref{fig:otvfr1}) installs the same \texttt{version} for all \texttt{Person} nodes in the cycle.
Thus, if it observes a different \texttt{version} it has observed the effect of a different transaction and has read across transaction boundaries.

\subsection{Lost Update}
\label{sec:lost-update}

Informally, a \anolong{Lost Update} (\anomaly{LU}) anomaly~\cite{DBLP:journals/tods/BailisFGHS16} occurs when two transactions concurrently attempt to make conditional modifications to the same data item(s).

\paragraph{Definition.}
A history $H$ exhibits phenomenon \anomaly{LU} if $\textit{DSG}(H)$ contains a directed cycle having one or more antidependency edges and all edges are induced by the same data item $x$.

\paragraph{Test.}
Load a test graph containing \texttt{Person} nodes.
Assign each \texttt{Person} a unique \texttt{id} and a property \texttt{numFriends} (initialized to 0).
During execution write clients execute a sequence of \tx{LU $T_\mathrm{W}$} instances, \autoref{fig:lu1}.
Choosing a random \texttt{Person} and incrementing its \texttt{numFriends} property.
Clients store local counters (\texttt{expNumFriends}) for each \texttt{Person}, which is incremented each time a \texttt{Person} is selected \emph{and} the \tx{LU $T_\mathrm{W}$} instance successfully commits.
After the execution period the \texttt{numFriends} is retrieved for each \texttt{Person} using \tx{LU $T_\mathrm{R}$} in \autoref{fig:lu2} and \texttt{expNumFriends} are pooled from write clients for each \texttt{Person}.

\paragraph{Anomaly check.}
For each \texttt{Person} its \texttt{numFriends} property should be equal to the (global) \texttt{expNumFriends} for that \texttt{Person}.

\paragraph{Why it works.}
Clients know how many successful \tx{LU $T_\mathrm{W}$} instances were issued for a given \texttt{Person}.
The observable \texttt{numFriends} should reflect this ground truth, otherwise, an \anomaly{LU} anomaly must have occurred.

\begin{figure}[htb]
\centering
\begin{minipage}{0.41\linewidth}
\begin{lstlisting}[language=cypher,label=fig:lu1,caption=\tx{Lost Update $T_\mathrm{W}$}.]
MATCH (p:Person {id: $personId})
SET p.numFriends = p.numFriends + 1
\end{lstlisting}
\end{minipage}
\quad
\begin{minipage}{0.52\linewidth}
\begin{lstlisting}[language=cypher,label=fig:lu2,caption=\tx{Lost Update $T_\mathrm{R}$}.]
MATCH (p:Person {id: $personId})
RETURN p.numFriends AS numFriends
\end{lstlisting}
\end{minipage}
\end{figure}

\subsection{Write Skew}
\label{sec:write-skew}

Informally, \anolong{Write Skew} (\anomaly{WS}) occurs when two transactions simultaneously attempted to make \emph{disjoint} conditional modifications to the same data item(s).
It is referred to as \anomaly{G2-Item} in~\cite{adya1999weak,DBLP:journals/tods/FeketeLOOS05}.

\paragraph{Definition.}
A history $H$ exhibits \anomaly{WS} if $\textit{DSG}(H)$ contains a directed cycle having one or more antidependency edges.

\paragraph{Test.}
Load a test graph containing $n$ pairs of \texttt{Person} nodes \texttt{(p1, p2)} for $k = 0, \ldots, n-1$, where the $k$th pair gets IDs \texttt{p1.id = 2*k+1} and \texttt{p2.id = 2*k+2}, and values \texttt{p1.value = 70} and \texttt{p2.value = 80}.
There is a constraint: \texttt{p1.value + p2.value > 0}.
During execution write clients execute a sequence of \tx{WS $T_\mathrm{W}$} instances, \autoref{fig:ws1}.
Selecting a random \texttt{Person} pair and decrementing the \texttt{value} property of one \texttt{Person} provided doing so would not violate the constraint.
After execution the database is scanned using \tx{WS $T_\mathrm{R}$}, \autoref{fig:ws2}.

\paragraph{Anomaly check.}
For each \texttt{Person} pair the constraint should hold true, otherwise, a \anomaly{WS} anomaly has occurred.

\paragraph{Why it works.}
Under no \level{Serializable} execution of WS $T_\mathrm{W}$ instances would the constraint \texttt{p1.value + p2.value > 0} be violated.
Therefore, if \tx{WS $T_\mathrm{R}$} returns a violation of this constraint it is clear a \anomaly{WS} anomaly has occurred.

\begin{figure}[htb]
\centering
\begin{minipage}{0.55\linewidth}
\begin{lstlisting}[language=cypher,label=fig:ws1,caption=\tx{WS $T_\mathrm{W}$}.]
MATCH (p1:Person {id: $person1Id}),
      (p2:Person {id: $person2Id})
<<IF (p1.value+p2.value < 100)>> <<THEN>> <<ABORT>> <<END>>
<<SLEEP($sleepTime)>>
pId = <<pick randomly between personId1, personId2>>
MATCH (p:Person {id: $pId})
SET p.value = p.value - 100
<<COMMIT>>
\end{lstlisting}
\end{minipage}
\quad
\begin{minipage}{0.33\linewidth}
\begin{lstlisting}[language=cypher,label=fig:ws2,caption=\tx{WS $T_\mathrm{R}$}.]
MATCH (p1:Person),
      (p2:Person {id: p1.id+1})
WHERE p1.value + p2.value <= 0
RETURN
  p1.id AS p1id,
  p1.value AS p1value,
  p2.id AS p2id,
  p2.value AS p2value
\end{lstlisting}
\end{minipage}
\end{figure}






\section{Consistency and Durability Tests}
\label{sec:cd}

While this chapter mainly focused on \emph{atomicity} and \emph{isolation} from the ACID properties, we provide a short overview of the other two aspects.

{\bf Durability} is a hard requirement for SNB Interactive and checking it is part of the 
auditing process. 
The durability test requires the execution of the SNB Interactive workload and uses the LDBC workload driver.
Note, the database and the driver must be configured in the same way as would be used in the performance run.
First, the database is subject to a warm-up period.
Then after 2 hours of simulation execution, the database processes will be terminated, possibly by disconnecting the entire machine or by a hard process kill.
Note that turning the machine off may not be possible in cloud tests.
The database system is then restarted and each client issues a read for the value of the last entity (node or edge) it received a successful commit message for, that should produce a correct response.

{\bf Consistency} is defined in terms of constraints: the database remains 
consistent under updates; i.e. no constraint is violated.
Relational database systems usually support primary- and foreign-key 
constraints, as well as domain constraints on column values and 
sometimes also support simple within-row constraints.
Graph database systems have a diversity of interfaces and generally do not
support constraints, beyond sometimes domain and primary key constraints 
(in case indices are supported).
As such, we leave them out of scope for LDBC SNB. 
However, we do note that we anticipate that graph database 
systems will evolve to support constraints in the future. 
Beyond equivalents of the relational ones, property graph systems 
might introduce graph-specific constraints, such as (partial) compliance to
a schema formulated on top of property graphs, rules that guide the 
presence of labels or structural graph constraints such as
connectedness of the graph, absence of cycles, 
or arbitrary well-formedness constraints~\cite{DBLP:journals/sosym/SemerathBHSV17}.

\chapter{Related Work}
\label{sec:related-work}

\begin{quote}
    \textit{A detailed list of LDBC publications is curated at~\url{https://ldbcouncil.org/publications}.}
\end{quote}


\section{ACID Tests in Other Benchmarks}

The challenge of verifying ACID-compliance has been addressed before by transactional benchmarks.
For example, TPC-C~\cite{tpcc} provides a suite of ACID tests.
However, the isolation tests are reliant on lock-based concurrency control, hence are not generalizable across systems.
Also, the transactional anomaly test coverage is limited to only four anomalies.
The authors of~\cite{DBLP:conf/icde/DeyFNR14} augment the popular YCSB framework for benchmarking transactional NewSQL systems, including a \emph{validation phase} that detects and quantifies consistency anomalies.
They permit the definition of arbitrary integrity constraints, checking they hold before and after a benchmark run.
Such an approach is not possible within SNB Interactive due to the restrictive nature of transactional updates and the distinct lack of application-level constraints.

The Hermitage project~\cite{Hermitage} with the goal of improving understanding of weak isolation, developed a range of hand-crafted isolation tests.
This test suite has much higher anomaly coverage but suffers from a problem similar to TPC-C.
Test execution is performed by hand, opening multiple terminals to step through the tests.\footnote{We initially experimented with Hermitage but found it difficult to induce anomalies that relied on fast timings due to some graph databases offering limited client-side control over transactions, with all statements submitted in one batch.}
The Jepsen project~\cite{kingsbury} is not a benchmark rather it addresses correctness testing, traditionally focusing on distributed systems under various failure modes.
Most of Jepsen's transactional tests adopt a similar approach to us, executing a suite of transactions with hand-proven invariants.
However recently, the project has spawned Elle~\cite{DBLP:journals/corr/abs-2003-10554} a black-box transactional anomaly checker.
Elle does not rely on hand-crafted tests and can detect every anomaly in \cite{adya1999weak} (except for predicate-based anomalies) from an arbitrary transaction history.


\section{Graph Processing Benchmarks}

Recent graph benchmarking initiatives focus on three key areas:

\begin{enumerate}
\item transactional workloads consisting of interactive read and update queries (OLTP) aiming at graph databases that explore small portions of the graph in each query~\cite{DBLP:conf/cidr/BarahmandG13,DBLP:conf/sigmod/ArmstrongPBC13,DBLP:journals/ase/DayarathnaS14,DBLP:conf/sigmod/ErlingALCGPPB15,DBLP:journals/pvldb/LissandriniBV18},
\item graph analysis algorithms (\eg PageRank) computed in bulk, typically expressed in cluster frameworks with graph APIs, rather than high-level queries~\cite{DBLP:conf/hipc/BaderM05,DBLP:conf/bigdataconf/ElserM13,DBLP:conf/sc/NaiXTKL15,DBLP:journals/pvldb/IosupHNHPMCCSAT16},
\item pattern matching and inferencing on semantic data~\cite{DBLP:journals/ws/GuoPH05,DBLP:books/sp/virgilio09/SchmidtHMPL09,DBLP:conf/semweb/MorseyLAN11,DBLP:conf/semweb/AlucHOD14,DBLP:journals/sosym/SzarnyasIRV18}.
\end{enumerate}

The SIGMOD 2014 Programming Contest defined queries on the Social Network Benchmark schema with a mix of subgraph projection and graph analytics~\cite{DBLP:journals/corr/abs-2010-12243}.

The challenges of using benchmarks correctly are described in~\cite{DBLP:conf/sigmod/RaasveldtHGM18}.

The Interactive queries were used in paper~\cite{DBLP:conf/grades/PacaciZLO17} to compare the performance of Gremlin, Cypher, SQL and SPARQL query engines.

The Labelled Subgraph Query Benchmark (LSQB)~\cite{DBLP:conf/sigmod/MhedhbiLKWS21} uses graphs produced by the LDBC SNB Datagen but simplifies them by omitting all attributes. It defines join-heavy subgraph queries to perform graph pattern matching.


\section{Scalable Graph Generators}


A recent survey~\cite{DBLP:journals/csur/BonifatiHPS20} studied 38 graph generators, finding that only 4 of them supported generating updates and, intriguingly, even these generators only yield insertions and simple deletions at best.
\emph{LinkBench}~\cite{DBLP:conf/sigmod/ArmstrongPBC13} defines primitive delete operations targeting a single node or a single edge.
\emph{XGDBench}~\cite{DBLP:journals/ase/DayarathnaS14} defines an operation that deletes a single node.
The \emph{Social Network Intelligence BenchMark} (SIB)~\cite{SIB} (a precursor to LDBC SNB) requires the deletion of individual nodes (posts/photos).

\printbibliography[heading=bibintoc]

\appendix

\chapter{Choke Points}
\label{sec:choke-points}

\newcommand{\tpcCard}[1]{\colorbox{lightgray}{\tt TPC-H #1}}
\newcommand{\cpSection}[4][]{%
	\subsection*{%
		CP-#2: [#3] #4%
		\ifthenelse{\equal{#1}{}}{}{\hfill \tpcCard{#1}}%
	}%
	\label{choke_point_#2}}


\section*{Introduction}

Choke points capture particularly challenging aspects of queries.
The correlations between choke points and read queries are displayed in \autoref{tab:query_choke_point}.

{
\setlength{\tabcolsep}{.05em}
\begin{table}[htbp]
\scriptsize
\centering
\begin{tabular}{|l||c|c|c|c||c|c|c|c|c|c||c|c|c||c|c|c||c|c|c||c||c|c|c|c|c|c|c|c||c|c|c|c|c|c||c|c|c|c|c|} \hline
& \chokePoint{1.1}& \chokePoint{1.2}& \chokePoint{1.3}& \chokePoint{1.4}& \chokePoint{2.1}& \chokePoint{2.2}& \chokePoint{2.3}& \chokePoint{2.4}& \chokePoint{2.5}& \chokePoint{2.6}& \chokePoint{3.1}& \chokePoint{3.2}& \chokePoint{3.3}& \chokePoint{4.1}& \chokePoint{4.2}& \chokePoint{4.3}& \chokePoint{5.1}& \chokePoint{5.2}& \chokePoint{5.3}& \chokePoint{6.1}& \chokePoint{7.1}& \chokePoint{7.2}& \chokePoint{7.3}& \chokePoint{7.4}& \chokePoint{7.5}& \chokePoint{7.6}& \chokePoint{7.7}& \chokePoint{7.8}& \chokePoint{8.1}& \chokePoint{8.2}& \chokePoint{8.3}& \chokePoint{8.4}& \chokePoint{8.5}& \chokePoint{8.6} \\ \hline\hline
        \queryRefCard{bi-read-01}{BI}{1}
        &  &  \yes&  &  &  &  &  &  &  &  &  &  \yes&  &  \yes&  \yes&  &  &  &  &  &  &  &  &  &  &  &  &  &  &  &  &  &  \yes&   \\ \hline
        \queryRefCard{bi-read-02}{BI}{2}
        &  &  &  &  &  &  &  &  \yes&  &  &  \yes&  \yes&  &  \yes&  \yes&  \yes&  &  &  \yes&  \yes&  &  &  &  &  &  &  &  &  &  \yes&  &  &  \yes&   \\ \hline
        \queryRefCard{bi-read-03}{BI}{3}
        &  \yes&  \yes&  \yes&  &  \yes&  \yes&  &  \yes&  &  &  &  &  \yes&  &  &  &  &  &  &  &  &  &  &  &  &  &  &  &  &  \yes&  &  &  &   \\ \hline
        \queryRefCard{bi-read-04}{BI}{4}
        &  &  \yes&  \yes&  &  \yes&  \yes&  \yes&  \yes&  &  &  &  &  \yes&  &  &  &  &  &  \yes&  \yes&  &  &  &  &  &  &  &  &  &  \yes&  &  \yes&  &   \\ \hline
        \queryRefCard{bi-read-05}{BI}{5}
        &  &  \yes&  &  &  &  &  \yes&  &  &  \yes&  &  &  &  &  &  &  &  &  &  &  &  &  &  &  &  &  &  &  &  \yes&  &  &  &   \\ \hline
        \queryRefCard{bi-read-06}{BI}{6}
        &  &  \yes&  &  &  &  &  \yes&  &  &  \yes&  &  &  \yes&  &  &  &  &  &  &  \yes&  &  &  &  &  &  &  &  &  &  \yes&  &  &  &   \\ \hline
        \queryRefCard{bi-read-07}{BI}{7}
        &  &  &  &  \yes&  &  &  &  &  &  &  &  &  \yes&  &  &  &  &  \yes&  &  &  &  &  &  &  &  &  &  &  \yes&  &  &  &  &   \\ \hline
        \queryRefCard{bi-read-08}{BI}{8}
        &  &  \yes&  &  &  \yes&  &  \yes&  &  &  &  &  \yes&  &  &  &  &  &  &  \yes&  &  &  &  &  &  &  &  &  &  &  \yes&  &  \yes&  \yes&   \\ \hline
        \queryRefCard{bi-read-09}{BI}{9}
        &  &  \yes&  &  &  &  \yes&  \yes&  &  &  \yes&  &  \yes&  &  &  &  &  &  &  &  &  &  \yes&  \yes&  \yes&  &  &  &  &  \yes&  &  &  &  \yes&   \\ \hline
        \queryRefCard{bi-read-10}{BI}{10}
        &  &  \yes&  \yes&  &  &  &  \yes&  \yes&  &  \yes&  &  &  \yes&  &  &  &  &  &  \yes&  &  \yes&  \yes&  \yes&  &  &  &  &  &  \yes&  &  &  &  &  \yes \\ \hline
        \queryRefCard{bi-read-11}{BI}{11}
        &  &  &  &  &  &  &  \yes&  &  \yes&  &  &  \yes&  &  &  &  &  &  &  &  &  &  &  &  &  &  &  &  &  &  &  &  &  &   \\ \hline
        \queryRefCard{bi-read-12}{BI}{12}
        &  \yes&  \yes&  &  \yes&  &  &  &  &  &  \yes&  &  \yes&  &  &  \yes&  \yes&  &  &  &  &  &  &  &  &  &  &  &  &  \yes&  \yes&  \yes&  \yes&  \yes&   \\ \hline
        \queryRefCard{bi-read-13}{BI}{13}
        &  &  \yes&  &  &  \yes&  &  \yes&  \yes&  &  \yes&  &  \yes&  \yes&  &  \yes&  &  \yes&  &  \yes&  &  &  &  &  &  &  &  &  &  &  \yes&  &  \yes&  \yes&   \\ \hline
        \queryRefCard{bi-read-14}{BI}{14}
        &  &  &  \yes&  \yes&  \yes&  &  &  &  &  &  \yes&  &  \yes&  &  &  &  \yes&  \yes&  \yes&  &  &  &  &  &  &  &  &  &  &  &  \yes&  \yes&  &   \\ \hline
        \queryRefCard{bi-read-15}{BI}{15}
        &  &  \yes&  &  &  \yes&  \yes&  &  \yes&  &  &  &  &  \yes&  &  &  &  \yes&  &  \yes&  &  &  \yes&  \yes&  &  &  \yes&  \yes&  &  \yes&  \yes&  \yes&  \yes&  \yes&  \yes \\ \hline
        \queryRefCard{bi-read-16}{BI}{16}
        &  &  &  &  &  &  &  &  &  &  &  &  &  &  &  &  &  &  &  \yes&  &  &  &  &  &  &  &  &  &  &  &  &  \yes&  \yes&   \\ \hline
        \queryRefCard{bi-read-17}{BI}{17}
        &  &  &  &  &  \yes&  &  \yes&  &  \yes&  \yes&  &  &  &  &  &  &  &  &  &  &  &  &  &  &  &  &  &  &  \yes&  &  &  &  &   \\ \hline
        \queryRefCard{bi-read-18}{BI}{18}
        &  &  &  &  &  &  &  &  &  \yes&  \yes&  &  &  &  &  &  &  &  &  &  &  &  &  &  &  &  &  &  &  \yes&  &  &  &  &   \\ \hline
        \queryRefCard{bi-read-19}{BI}{19}
        &  &  &  &  &  &  &  &  &  &  &  &  &  \yes&  &  &  &  &  &  &  &  &  &  &  &  &  \yes&  \yes&  &  &  &  &  \yes&  &  \yes \\ \hline
        \queryRefCard{bi-read-20}{BI}{20}
        &  &  &  &  &  &  &  &  &  &  &  &  &  \yes&  &  &  &  &  &  &  &  &  &  &  &  &  \yes&  \yes&  \yes&  &  &  &  \yes&  &  \yes \\ \hline
        \queryRefCard{interactive-complex-read-01}{IC}{1}
        &  &  &  &  &  \yes&  &  &  &  &  &  &  &  &  &  &  &  &  &  \yes&  &  &  &  &  &  &  &  &  &  &  \yes&  &  &  &   \\ \hline
        \queryRefCard{interactive-complex-read-02}{IC}{2}
        &  \yes&  &  &  &  &  \yes&  \yes&  &  &  &  &  \yes&  &  &  &  &  &  &  &  &  &  &  &  &  &  &  &  &  &  &  &  &  \yes&   \\ \hline
        \queryRefCard{interactive-complex-read-03}{IC}{3}
        &  &  &  &  &  \yes&  &  &  &  &  &  \yes&  &  &  &  &  &  \yes&  &  &  &  &  &  &  &  &  &  &  &  &  \yes&  &  &  \yes&   \\ \hline
        \queryRefCard{interactive-complex-read-04}{IC}{4}
        &  &  &  &  &  &  &  \yes&  &  &  &  &  &  &  &  &  &  &  &  &  &  &  &  &  &  &  &  &  &  &  \yes&  &  &  \yes&   \\ \hline
        \queryRefCard{interactive-complex-read-05}{IC}{5}
        &  &  &  &  &  &  &  \yes&  &  &  &  &  &  \yes&  &  &  &  &  &  &  &  &  &  &  &  &  &  &  &  &  \yes&  &  &  \yes&   \\ \hline
        \queryRefCard{interactive-complex-read-06}{IC}{6}
        &  &  &  &  &  &  &  &  &  &  &  &  &  &  &  &  &  \yes&  &  &  &  &  &  &  &  &  &  &  &  &  \yes&  &  &  &   \\ \hline
        \queryRefCard{interactive-complex-read-07}{IC}{7}
        &  &  &  &  &  &  \yes&  \yes&  &  &  &  &  &  \yes&  &  &  &  \yes&  &  &  &  &  &  &  &  &  &  &  &  \yes&  &  \yes&  &  &   \\ \hline
        \queryRefCard{interactive-complex-read-08}{IC}{8}
        &  &  &  &  &  &  &  &  \yes&  &  &  &  &  \yes&  &  &  &  &  &  \yes&  &  &  &  &  &  &  &  &  &  &  &  &  &  &   \\ \hline
        \queryRefCard{interactive-complex-read-09}{IC}{9}
        &  \yes&  \yes&  &  &  &  \yes&  \yes&  &  &  &  &  \yes&  \yes&  &  &  &  &  &  &  &  &  &  &  &  &  &  &  &  &  &  &  &  \yes&   \\ \hline
        \queryRefCard{interactive-complex-read-10}{IC}{10}
        &  &  &  &  &  &  &  \yes&  &  &  &  &  &  \yes&  \yes&  \yes&  &  \yes&  \yes&  &  \yes&  \yes&  &  &  &  &  &  &  &  &  &  &  &  &  \yes \\ \hline
        \queryRefCard{interactive-complex-read-11}{IC}{11}
        &  &  &  \yes&  &  &  &  \yes&  \yes&  &  &  &  &  \yes&  &  \yes&  &  &  &  &  &  &  &  &  &  &  &  &  &  &  &  &  &  &   \\ \hline
        \queryRefCard{interactive-complex-read-12}{IC}{12}
        &  &  &  &  &  &  &  &  &  &  &  &  &  \yes&  &  &  &  &  &  &  &  &  \yes&  \yes&  &  &  &  &  &  &  \yes&  &  &  &   \\ \hline
        \queryRefCard{interactive-complex-read-13}{IC}{13}
        &  &  &  &  &  &  &  &  &  &  &  &  &  \yes&  &  &  &  &  &  &  &  &  \yes&  \yes&  &  \yes&  &  &  \yes&  \yes&  &  &  &  &  \yes \\ \hline
        \queryRefCard{interactive-complex-read-14-v1}{IC}{14v1}
        &  &  &  &  &  &  &  &  &  &  &  &  &  \yes&  &  &  &  &  &  \yes&  &  &  \yes&  \yes&  &  \yes&  &  \yes&  &  \yes&  \yes&  \yes&  &  &  \yes \\ \hline
        \queryRefCard{interactive-complex-read-14-v2}{IC}{14v2}
        &  &  &  &  &  &  &  &  &  &  &  &  &  \yes&  &  &  &  &  &  \yes&  &  &  &  &  &  &  \yes&  \yes&  \yes&  \yes&  \yes&  \yes&  &  &  \yes \\ \hline

\end{tabular}
\caption{Coverage of choke points by queries.}
\label{tab:query_choke_point}
\end{table}
}


\section{Aggregation Performance}


\cpSection[1.2]{1.1}{QOPT}{Interesting orders}

This choke point tests the ability of the query optimizer to exploit the interesting orders induced by some operators. Apart from clustered indices providing key order, other operators also preserve or even induce tuple orderings.
Sort-based operators create new orderings, typically the probe-side of a hash join conserves its order, etc.

\paragraph{Queries}
{\raggedright
\queryRefCard{bi-read-03}{BI}{3}
\queryRefCard{bi-read-12}{BI}{12}
\queryRefCard{interactive-complex-read-02}{IC}{2}
\queryRefCard{interactive-complex-read-09}{IC}{9}

}


\cpSection[1.1]{1.2}{QEXE}{High cardinality group-by performance}

This choke point tests the ability of the execution engine to parallelize group-by operations with a large number of groups. Some queries require performing large group-by operations.
In such a case, if an aggregation produces a significant number of groups, intra-query parallelization can be exploited as each thread may make its own partial aggregation.
Then, to produce the result, these have to be re-aggregated. In order to avoid this, the tuples entering the aggregation operator may be partitioned by a hash of the grouping key and be sent to the appropriate partition.
Each partition would have its own thread so that only that thread would write the aggregation, hence avoiding costly critical sections as well. A high cardinality distinct modifier in a query is a special case of this choke point.
It is amenable to the same solution with intra-query parallelization and partitioning as the group-by.
We further note that scale-out systems have an extra incentive for partitioning since this will distribute the CPU and memory pressure over multiple machines, yielding better platform utilization and scalability.

\paragraph{Queries}
{\raggedright
\queryRefCard{bi-read-01}{BI}{1}
\queryRefCard{bi-read-03}{BI}{3}
\queryRefCard{bi-read-04}{BI}{4}
\queryRefCard{bi-read-05}{BI}{5}
\queryRefCard{bi-read-06}{BI}{6}
\queryRefCard{bi-read-08}{BI}{8}
\queryRefCard{bi-read-09}{BI}{9}
\queryRefCard{bi-read-10}{BI}{10}
\queryRefCard{bi-read-12}{BI}{12}
\queryRefCard{bi-read-13}{BI}{13}
\queryRefCard{bi-read-15}{BI}{15}
\queryRefCard{interactive-complex-read-09}{IC}{9}

}


\cpSection{1.3}{QOPT}{Top-k pushdown}

This choke point tests the ability of the query optimizer to perform optimizations based on top-$k$ selections. Many times queries demand for returning the top-$k$ elements based on some property.
Engines can exploit that once $k$ results are obtained, extra restrictions in a selection can be added based on the properties of the $k$th element currently in the top-$k$, being more restrictive as the query advances, instead of sorting all elements and picking the highest $k$.

\paragraph{Queries}
{\raggedright
\queryRefCard{bi-read-03}{BI}{3}
\queryRefCard{bi-read-04}{BI}{4}
\queryRefCard{bi-read-10}{BI}{10}
\queryRefCard{bi-read-14}{BI}{14}
\queryRefCard{interactive-complex-read-11}{IC}{11}

}


\cpSection[1.3]{1.4}{QEXE}{Low cardinality group-by performance}

This choke point tests the ability to efficiently perform group-by evaluation
when only a very limited set of groups is available.  This can require special
strategies for parallelization, \eg pre-aggregation when possible. This case also allows using special strategies for grouping like using array lookup if the domain of keys is small.

\paragraph{Queries}
{\raggedright
\queryRefCard{bi-read-07}{BI}{7}
\queryRefCard{bi-read-12}{BI}{12}
\queryRefCard{bi-read-14}{BI}{14}

}


\section{Join Performance}


\cpSection[2.3]{2.1}{QOPT}{Rich join order optimization}

This choke point tests the ability of the query optimizer to find optimal join orders. A graph can be traversed in different ways. In the relational model, this is equivalent to different join orders.
The execution time of these orders may differ by orders of magnitude. Therefore, finding an efficient join (traversal) order is important, which in general, requires enumeration of all the possibilities.
The enumeration is complicated by operators that are not freely re-orderable like semi-, \mbox{anti-,} and outer-joins. Because of this difficulty most join enumeration algorithms do not enumerate all possible plans, and therefore can miss the optimal join order. Therefore, this choke point tests the ability of the query optimizer to find optimal join (traversal) orders.

\paragraph{Queries}
{\raggedright
\queryRefCard{bi-read-03}{BI}{3}
\queryRefCard{bi-read-04}{BI}{4}
\queryRefCard{bi-read-08}{BI}{8}
\queryRefCard{bi-read-13}{BI}{13}
\queryRefCard{bi-read-14}{BI}{14}
\queryRefCard{bi-read-15}{BI}{15}
\queryRefCard{bi-read-17}{BI}{17}
\queryRefCard{interactive-complex-read-01}{IC}{1}
\queryRefCard{interactive-complex-read-03}{IC}{3}

}


\cpSection[2.4]{2.2}{QOPT}{Late projection}

This choke point tests the ability of the query optimizer to delay the projection of unneeded attributes until late in the execution. Queries where certain columns are only needed late in the query.
In such a situation, it is better to omit them from initial table scans, as fetching them later by row-id with a separate scan operator, which is joined to the intermediate query result, can save temporal space, and therefore I/O.
Late projection does have a trade-off involving locality, since late in the plan the tuples may be in a different order, and scattered I/O in terms of tuples/second is much more expensive than sequential I/O.
Late projection specifically makes sense in queries where the late use of these columns happens at a moment where the amount of tuples involved has been considerably reduced;
for example after an aggregation with only few unique group-by keys or a top-$k$ operator.

\paragraph{Queries}
{\raggedright
\queryRefCard{bi-read-03}{BI}{3}
\queryRefCard{bi-read-04}{BI}{4}
\queryRefCard{bi-read-09}{BI}{9}
\queryRefCard{bi-read-15}{BI}{15}
\queryRefCard{interactive-complex-read-02}{IC}{2}
\queryRefCard{interactive-complex-read-07}{IC}{7}
\queryRefCard{interactive-complex-read-09}{IC}{9}

}


\cpSection{2.3}{QOPT}{Join type selection}

This choke point tests the ability of the query optimizer to select the proper join operator type, which implies accurate estimates of cardinalities.
Depending on the cardinalities of both sides of a join, a hash or an index-based join operator is more appropriate.
This is especially important with column stores, where one usually has an index on everything. Deciding to use a hash join requires a good estimation of cardinalities on both the probe and build sides.
In TPC-H, the use of hash join is almost a foregone conclusion in many cases, since an implementation will usually not even define an index on foreign key columns.
There is a break even point between index and hash based plans, depending on the cardinality on the probe and build sides.

\paragraph{Queries}
{\raggedright
\queryRefCard{bi-read-04}{BI}{4}
\queryRefCard{bi-read-05}{BI}{5}
\queryRefCard{bi-read-06}{BI}{6}
\queryRefCard{bi-read-08}{BI}{8}
\queryRefCard{bi-read-09}{BI}{9}
\queryRefCard{bi-read-10}{BI}{10}
\queryRefCard{bi-read-11}{BI}{11}
\queryRefCard{bi-read-13}{BI}{13}
\queryRefCard{bi-read-17}{BI}{17}
\queryRefCard{interactive-complex-read-02}{IC}{2}
\queryRefCard{interactive-complex-read-04}{IC}{4}
\queryRefCard{interactive-complex-read-05}{IC}{5}
\queryRefCard{interactive-complex-read-07}{IC}{7}
\queryRefCard{interactive-complex-read-09}{IC}{9}
\queryRefCard{interactive-complex-read-10}{IC}{10}
\queryRefCard{interactive-complex-read-11}{IC}{11}

}


\cpSection[2.2]{2.4}{QOPT}{Sparse foreign key joins}

This choke point tests the performance of join operators when the join is sparse. Sometimes joins involve relations where only a small percentage of rows in one of the tables is required to satisfy a join. When tables are larger, typical join methods can be sub-optimal. Partitioning the sparse table, using Hash Clustered indices or implementing Bloom-filter tests inside the join are techniques to improve the performance in such situations~\cite{DBLP:journals/csur/Graefe93}.

\paragraph{Queries}
{\raggedright
\queryRefCard{bi-read-02}{BI}{2}
\queryRefCard{bi-read-03}{BI}{3}
\queryRefCard{bi-read-04}{BI}{4}
\queryRefCard{bi-read-10}{BI}{10}
\queryRefCard{bi-read-13}{BI}{13}
\queryRefCard{bi-read-15}{BI}{15}
\queryRefCard{interactive-complex-read-08}{IC}{8}
\queryRefCard{interactive-complex-read-11}{IC}{11}

}


\cpSection{2.5}{QEXE}{Worst-case optimal joins}

This choke point tests the query engine's ability to use multi-way, worst-case optimal joins to evaluate cyclic queries which are required to efficiently compute some dense subgraphs such as the triangle, the 4-cycle, and the diamond (4-cycle with a cross-edge).
The absence of multi-way joins (\eg in systems which only support binary joins), implies that join performance will be provably suboptimal for cyclic queries.

\paragraph{Queries}
{\raggedright
\queryRefCard{bi-read-11}{BI}{11}
\queryRefCard{bi-read-17}{BI}{17}
\queryRefCard{bi-read-18}{BI}{18}

}

\cpSection{2.6}{QEXE}{Factorized query execution}

Query results produced by many-to-many joins often have redundancies when represented as tuples.
Factorization~\cite{DBLP:journals/sigmod/OlteanuS16} provides a more compact (relational) representation by eliminating repetitions,
while still allowing some operations (\eg aggregation) to be performed without flattening the relation.

\paragraph{Queries}
{\raggedright
\queryRefCard{bi-read-05}{BI}{5}
\queryRefCard{bi-read-06}{BI}{6}
\queryRefCard{bi-read-09}{BI}{9}
\queryRefCard{bi-read-10}{BI}{10}
\queryRefCard{bi-read-12}{BI}{12}
\queryRefCard{bi-read-13}{BI}{13}
\queryRefCard{bi-read-17}{BI}{17}
\queryRefCard{bi-read-18}{BI}{18}

}


\section{Data Access Locality}


\cpSection[3.3]{3.1}{QOPT}{Detecting correlation}

This choke point tests the ability of the query optimizer to detect data correlations and exploiting them. If a schema rewards creating clustered indices, the question then is which of the date or data columns to use as key.
In fact it should not matter which column is used, as range-propagation between correlated attributes of the same table is relatively easy. One way is through the creation of multi-attribute histograms after detection of attribute correlation.
With MinMax indices, range-predicates on any column can be translated into qualifying tuple position ranges. If an attribute value is correlated with tuple position, this reduces the area to scan roughly equally to predicate selectivity.

\paragraph{Queries}
{\raggedright
\queryRefCard{bi-read-02}{BI}{2}
\queryRefCard{bi-read-14}{BI}{14}
\queryRefCard{interactive-complex-read-03}{IC}{3}

}


\cpSection{3.2}{STORAGE}{Dimensional clustering}

This choke point tests suitability of the identifiers assigned to entities by the storage system to better exploit data locality. A data model where each entity has a unique synthetic identifier,
\eg RDF or graph models, has some choice in assigning a value to this identifier.
The properties of the entity being identified may affect this, \eg type (label), other dependent properties,
\eg geographic location, date, position in a hierarchy, etc., depending on the application. Such identifier choice may create locality which in turn improves efficiency of compression or index access.

\paragraph{Queries}
{\raggedright
\queryRefCard{bi-read-01}{BI}{1}
\queryRefCard{bi-read-02}{BI}{2}
\queryRefCard{bi-read-08}{BI}{8}
\queryRefCard{bi-read-09}{BI}{9}
\queryRefCard{bi-read-11}{BI}{11}
\queryRefCard{bi-read-12}{BI}{12}
\queryRefCard{bi-read-13}{BI}{13}
\queryRefCard{interactive-complex-read-02}{IC}{2}
\queryRefCard{interactive-complex-read-09}{IC}{9}

}


\cpSection{3.3}{QEXE}{Scattered index access patterns}

This choke point tests the performance of indices when scattered accesses are performed. The efficiency of index lookup is very different depending on the locality of keys coming to the indexed access.
Techniques like vectoring non-local index accesses by simply missing the cache in parallel on multiple lookups vectored on the same thread may have high impact.
Also detecting absence of locality should turn off any locality dependent optimizations if these are costly when there is no locality. A graph neighbourhood traversal is an example of an operation with random access without predictable locality.

\paragraph{Queries}
{\raggedright
\queryRefCard{bi-read-03}{BI}{3}
\queryRefCard{bi-read-04}{BI}{4}
\queryRefCard{bi-read-06}{BI}{6}
\queryRefCard{bi-read-07}{BI}{7}
\queryRefCard{bi-read-10}{BI}{10}
\queryRefCard{bi-read-13}{BI}{13}
\queryRefCard{bi-read-14}{BI}{14}
\queryRefCard{bi-read-15}{BI}{15}
\queryRefCard{bi-read-19}{BI}{19}
\queryRefCard{bi-read-20}{BI}{20}
\queryRefCard{interactive-complex-read-05}{IC}{5}
\queryRefCard{interactive-complex-read-07}{IC}{7}
\queryRefCard{interactive-complex-read-08}{IC}{8}
\queryRefCard{interactive-complex-read-09}{IC}{9}
\queryRefCard{interactive-complex-read-10}{IC}{10}
\queryRefCard{interactive-complex-read-11}{IC}{11}
\queryRefCard{interactive-complex-read-12}{IC}{12}
\queryRefCard{interactive-complex-read-13}{IC}{13}
\queryRefCard{interactive-complex-read-14-v1}{IC}{14v1}
\queryRefCard{interactive-complex-read-14-v2}{IC}{14v2}

}


\section{Expression Calculation}


\cpSection[4.2a]{4.1}{QOPT}{Common subexpression elimination}

This choke point tests the ability of the query optimizer to detect common sub-expressions and reuse their results. A basic technique helpful in multiple queries is common subexpression elimination (CSE).
CSE should recognize also that \lstinline{avg} aggregates can be derived afterwards by dividing a \lstinline{sum} by the \lstinline{count} when those have been computed.

\paragraph{Queries}
{\raggedright
\queryRefCard{bi-read-01}{BI}{1}
\queryRefCard{bi-read-02}{BI}{2}
\queryRefCard{interactive-complex-read-10}{IC}{10}

}


\cpSection[4.2d]{4.2}{QOPT}{Complex boolean expression joins and selections}

This choke point tests the ability of the query optimizer to reorder the execution of boolean expressions to improve the performance. Some boolean expressions are complex, with possibilities for alternative optimal evaluation orders.
For instance, the optimizer may reorder conjunctions to test first those conditions with larger selectivity~\cite{DBLP:conf/vldb/Moerkotte98}.

\paragraph{Queries}
{\raggedright
\queryRefCard{bi-read-01}{BI}{1}
\queryRefCard{bi-read-02}{BI}{2}
\queryRefCard{bi-read-12}{BI}{12}
\queryRefCard{bi-read-13}{BI}{13}
\queryRefCard{interactive-complex-read-10}{IC}{10}
\queryRefCard{interactive-complex-read-11}{IC}{11}

}


\cpSection{4.3}{QEXE}{Low overhead expressions interpretation}

This choke point tests the ability of efficiently evaluating simple expressions on a large number of values. A typical example could be simple arithmetic expressions, mathematical functions like floor and absolute or date functions like extracting a year.

\paragraph{Queries}
{\raggedright
\queryRefCard{bi-read-02}{BI}{2}
\queryRefCard{bi-read-12}{BI}{12}

}

%
%


\section{Correlated Sub-Queries}


\cpSection[5.1]{5.1}{QOPT}{Flattening sub-queries}

This choke point tests the ability of the query optimizer to flatten execution plans when there are correlated sub-queries. Many queries have correlated sub-queries and their query plans can be flattened,
such that the correlated sub-query is handled using an equi-join, outer-join or anti-join.
In TPC-H Q21, for instance, there is an \lstinline{EXISTS} clause (for orders with more than one supplier) and a \lstinline{NOT EXISTS} clause (looking for an item that was received too late).
To execute this query well, systems need to flatten both sub-queries, the first into an equi-join plan, the second into an anti-join plan.
Therefore, the execution layer of the database system will benefit from implementing these extended join variants.

The ill effects of repetitive tuple-at-a-time sub-query execution can also be mitigated if execution systems by using vectorized, or blockwise query execution, allowing to run sub-queries with thousands of input parameters instead of one.
The ability to look up many keys in an index in one API call creates the opportunity to benefit from physical locality, if lookup keys exhibit some clustering.

\paragraph{Queries}
{\raggedright
\queryRefCard{bi-read-13}{BI}{13}
\queryRefCard{bi-read-14}{BI}{14}
\queryRefCard{bi-read-15}{BI}{15}
\queryRefCard{interactive-complex-read-03}{IC}{3}
\queryRefCard{interactive-complex-read-06}{IC}{6}
\queryRefCard{interactive-complex-read-07}{IC}{7}
\queryRefCard{interactive-complex-read-10}{IC}{10}

}


\cpSection[5.3]{5.2}{QEXE}{Overlap between outer and sub-query}

This choke point tests the ability of the execution engine to reuse results when there is an overlap between the outer query and the sub-query. In some queries, the correlated sub-query and the outer query have the same joins and selections.
In this case, a non-tree, rather DAG-shaped~\cite{DBLP:conf/btw/NeumannM09} query plan would allow to execute the common parts just once, providing the intermediate result stream to both the outer query and correlated sub-query,
which higher up in the query plan are joined together (using normal query decorrelation rewrites).
As such, the benchmark rewards systems where the optimizer can detect this and the execution engine supports an operator that can buffer intermediate results and provide them to multiple parent operators.

\paragraph{Queries}
{\raggedright
\queryRefCard{bi-read-07}{BI}{7}
\queryRefCard{bi-read-14}{BI}{14}
\queryRefCard{interactive-complex-read-10}{IC}{10}

}


\cpSection[5.2]{5.3}{QEXE}{Intra-query result reuse}

This choke point tests the ability of the execution engine to reuse sub-query results when two sub-queries are mostly identical.
Some queries have almost identical sub-queries, where some of their internal results can be reused in both sides of the execution plan, thus avoiding to repeat computations.

\paragraph{Queries}
{\raggedright
\queryRefCard{bi-read-02}{BI}{2}
\queryRefCard{bi-read-04}{BI}{4}
\queryRefCard{bi-read-08}{BI}{8}
\queryRefCard{bi-read-10}{BI}{10}
\queryRefCard{bi-read-13}{BI}{13}
\queryRefCard{bi-read-14}{BI}{14}
\queryRefCard{bi-read-15}{BI}{15}
\queryRefCard{bi-read-16}{BI}{16}
\queryRefCard{interactive-complex-read-01}{IC}{1}
\queryRefCard{interactive-complex-read-08}{IC}{8}
\queryRefCard{interactive-complex-read-14-v1}{IC}{14v1}
\queryRefCard{interactive-complex-read-14-v2}{IC}{14v2}

}


\section{Parallelism and Concurrency}


\cpSection[6.3]{6.1}{QEXE}{Inter-query result reuse}

This choke point tests the ability of the query execution engine to reuse results from different queries. Sometimes with a high number of streams a significant amount of identical queries emerge in the resulting workload.
The reason is that certain parameters, as generated by the workload generator, have only a limited amount of parameters bindings.
This weakness opens up the possibility of using a query result cache, to eliminate the repetitive part of the workload.
A further opportunity that detects even more overlap is the work on recycling, which does not only cache final query results, but also intermediate query results of a ``high worth''.
Here, worth is a combination of partial-query result size, partial-query evaluation cost, and observed (or estimated) frequency of the partial-query in the workload.

\paragraph{Queries}
{\raggedright
\queryRefCard{bi-read-02}{BI}{2}
\queryRefCard{bi-read-04}{BI}{4}
\queryRefCard{bi-read-06}{BI}{6}
\queryRefCard{interactive-complex-read-10}{IC}{10}

}


\section{Graph Specifics}


\cpSection{7.1}{QEXE}{Incremental path computation}

This choke point tests the ability of the execution engine to reuse work across
graph traversals. For example, when computing paths within a range of distances,
it is often possible to incrementally compute longer paths by reusing paths of
shorter distances that were already computed.

\paragraph{Queries}
{\raggedright
\queryRefCard{bi-read-10}{BI}{10}
\queryRefCard{interactive-complex-read-10}{IC}{10}

}


\cpSection{7.2}{QOPT}{Cardinality estimation of transitive paths}

This choke point tests the ability of the query optimizer to properly estimate the cardinality of intermediate results when executing transitive paths. A transitive path may occur in a ``fact table'' or a ``dimension table'' position.
A transitive path may cover a tree or a graph, \eg descendants in a geographical hierarchy \vs graph neighbourhood or transitive closure in a many-to-many connected social network.
In order to decide proper join order and type, the cardinality of the expansion of the transitive path needs to be correctly estimated.
This could for example take the form of executing on a sample of the data in the
cost model or of gathering special statistics, \eg the depth and fan-out of a tree. In the case of hierarchical dimensions,
\eg geographic locations or other hierarchical classifications, detecting the cardinality of the transitive path will allow one to go to a star schema plan with scan of a fact table with a selective hash join.
Such a plan will be on the other hand very bad for example if the hash table is much larger than the ``fact table'' being scanned.

\paragraph{Queries}
{\raggedright
\queryRefCard{bi-read-09}{BI}{9}
\queryRefCard{bi-read-10}{BI}{10}
\queryRefCard{bi-read-15}{BI}{15}
\queryRefCard{interactive-complex-read-12}{IC}{12}
\queryRefCard{interactive-complex-read-13}{IC}{13}
\queryRefCard{interactive-complex-read-14-v1}{IC}{14v1}

}


\cpSection{7.3}{QEXE}{Execution of a transitive step}

This choke point tests the ability of the query execution engine to efficiently execute transitive steps. Graph workloads may have transitive operations, for example finding a shortest path between nodes.
This involves repeated execution of a short lookup, often on many values at the
same time, while usually having an end condition, \eg the target node being reached or having reached the border of a search going in the opposite direction.
For the best efficiency, these operations can be merged or tightly coupled to
the index operations themselves. Also parallelization may be possible but may
need to deal with a global state, \eg set of visited nodes.
There are many possible tradeoffs between generality and performance.

\paragraph{Queries}
{\raggedright
\queryRefCard{bi-read-09}{BI}{9}
\queryRefCard{bi-read-10}{BI}{10}
\queryRefCard{bi-read-15}{BI}{15}
\queryRefCard{interactive-complex-read-12}{IC}{12}
\queryRefCard{interactive-complex-read-13}{IC}{13}
\queryRefCard{interactive-complex-read-14-v1}{IC}{14v1}

}


\cpSection{7.4}{QEXE}{Efficient evaluation of termination criteria for transitive queries}

This tests the ability of a system to express termination criteria for transitive queries so that not the whole transitive relation has to be evaluated as well as efficient testing for termination.

\paragraph{Queries}
{\raggedright
\queryRefCard{bi-read-09}{BI}{9}

}


\cpSection{7.5}{QEXE}{Unweighted shortest paths}

A common problem in graph queries is determining the distance between a node and a set of nodes. To compute the distance values, systems may employ BFS or a single-source shortest path algorithm with uniform weights.
To compute the distance between two given node, systems can use bidirectional search algorithms.

\paragraph{Queries}
{\raggedright
\queryRefCard{interactive-complex-read-13}{IC}{13}
\queryRefCard{interactive-complex-read-14-v1}{IC}{14v1}

}


\cpSection{7.6}{QEXE}{Cheapest paths (weighted shortest paths)}

Computing \emph{cheapest paths} (weighted shortest paths) is a fundamental problem in graph queries.
While there are well-known algorithms to compute it, \eg Dijkstra's algorithm, Bellman--Ford, and delta-stepping~\cite{DBLP:journals/jal/MeyerS03}, system often use na\"ive approaches such as enumerating all paths which makes these queries unnecessarily complex.

\paragraph{Queries}
{\raggedright
\queryRefCard{bi-read-15}{BI}{15}
\queryRefCard{bi-read-19}{BI}{19}
\queryRefCard{bi-read-20}{BI}{20}
\queryRefCard{interactive-complex-read-14-v2}{IC}{14v2}

}


\cpSection{7.7}{QEXE}{Composition of graph queries}

In many cases, it is desirable to specify multiple graph queries, where the first one defines an induced subgraph or an overlay graph on the original graph, which is then passed two the next query, and so on.
Expressing such computations as a sequence of composable graph queries would be desirable from both usability, optimization, and execution aspects. However, currently many graph dabases lack support for composable graph queries.

The \mbox{G-CORE}~\cite{DBLP:conf/sigmod/AnglesABBFGLPPS18} design language tackled problem this by introducing the \emph{path property graph} data model (consisting of nodes, edges, and paths) and defining queries such that they return a graph (while also providing means to return a tabular output).

\paragraph{Queries}
{\raggedright
\queryRefCard{bi-read-15}{BI}{15}
\queryRefCard{bi-read-19}{BI}{19}
\queryRefCard{bi-read-20}{BI}{20}
\queryRefCard{interactive-complex-read-14-v1}{IC}{14v1}
\queryRefCard{interactive-complex-read-14-v2}{IC}{14v2}

}

\cpSection{7.8}{QEXE}{Reachability between disconnected components}

For path finding queries, the result is often that the specified path does not exist in the graph.
For example, for a single-source single-destination search, when one of the endpoints is in a small component (\eg the endpoint is an isolated node), systems using a bidirectional search algorithm can quickly determine that there is no path to be found.

\paragraph{Queries}
{\raggedright
\queryRefCard{bi-read-20}{BI}{20}
\queryRefCard{interactive-complex-read-13}{IC}{13}
\queryRefCard{interactive-complex-read-14-v2}{IC}{14v2}

}


\section{Language Features}


\cpSection{8.1}{LANG}{Complex patterns}

\paragraph{Description.}

A natural requirement for graph query systems is to be able to express complex
graph patterns.

\paragraph{Transitive edges.} Transitive closure-style computations are common in graph query systems, both with fixed bounds
(\eg get nodes that can be reached through at least 3 and at most 5 \textsf{knows} edges),
and without fixed bounds
(\eg get all \textsf{Messages} that a \textsf{Comment} replies to).

\paragraph{Negative edge conditions.} Some queries define \emph{negative pattern conditions}. For example, the condition that a certain \textsf{Message} does not have a certain \textsf{Tag} is represented in the graph as the absence of a \textsf{hasTag} edge between the two nodes. Thus, queries looking for cases where this condition is satisfied check for negative patterns, also known as negative application conditions (NACs) in graph transformation literature~\cite{DBLP:journals/fuin/HabelHT96}.

\paragraph{Queries}
{\raggedright
\queryRefCard{bi-read-07}{BI}{7}
\queryRefCard{bi-read-09}{BI}{9}
\queryRefCard{bi-read-10}{BI}{10}
\queryRefCard{bi-read-12}{BI}{12}
\queryRefCard{bi-read-15}{BI}{15}
\queryRefCard{bi-read-17}{BI}{17}
\queryRefCard{bi-read-18}{BI}{18}
\queryRefCard{interactive-complex-read-07}{IC}{7}
\queryRefCard{interactive-complex-read-13}{IC}{13}
\queryRefCard{interactive-complex-read-14-v1}{IC}{14v1}
\queryRefCard{interactive-complex-read-14-v2}{IC}{14v2}

}


\cpSection{8.2}{LANG}{Complex aggregations}

\paragraph{Description.}

BI workloads are heavy on aggregation, including queries with \emph{subsequent
aggregations}, where the results of an aggregation serves as the input of
another aggregation. Expressing such operations requires some sort of query
composition or chaining (see also CP-8.4). It is also common to \emph{filter on
aggregation results} (similarly to the \lstinline[language=sql]{HAVING} keyword of SQL).

\paragraph{Queries}
{\raggedright
\queryRefCard{bi-read-02}{BI}{2}
\queryRefCard{bi-read-03}{BI}{3}
\queryRefCard{bi-read-04}{BI}{4}
\queryRefCard{bi-read-05}{BI}{5}
\queryRefCard{bi-read-06}{BI}{6}
\queryRefCard{bi-read-08}{BI}{8}
\queryRefCard{bi-read-12}{BI}{12}
\queryRefCard{bi-read-13}{BI}{13}
\queryRefCard{bi-read-15}{BI}{15}
\queryRefCard{interactive-complex-read-01}{IC}{1}
\queryRefCard{interactive-complex-read-03}{IC}{3}
\queryRefCard{interactive-complex-read-04}{IC}{4}
\queryRefCard{interactive-complex-read-05}{IC}{5}
\queryRefCard{interactive-complex-read-06}{IC}{6}
\queryRefCard{interactive-complex-read-12}{IC}{12}
\queryRefCard{interactive-complex-read-14-v1}{IC}{14v1}
\queryRefCard{interactive-complex-read-14-v2}{IC}{14v2}

}


\cpSection{8.3}{LANG}{Ranking-style queries}

\paragraph{Description.}

Along with aggregations, BI workloads often use \emph{window functions},
which perform aggregations without grouping input tuples to a single output
tuple.  A common use case for windowing is \emph{ranking}, \ie selecting the top
element with additional values in the tuple (nodes, edges or
attributes).\footnote{PostgreSQL defines the \lstinline[language=sql]{OVER}
keyword to use aggregation functions as window functions, and the
\lstinline[language=sql]{rank()} function to produce numerical ranks, see
\url{https://www.postgresql.org/docs/9.1/static/tutorial-window.html} for
details.}

\paragraph{Queries}
{\raggedright
\queryRefCard{bi-read-12}{BI}{12}
\queryRefCard{bi-read-14}{BI}{14}
\queryRefCard{bi-read-15}{BI}{15}
\queryRefCard{interactive-complex-read-07}{IC}{7}
\queryRefCard{interactive-complex-read-14-v1}{IC}{14v1}
\queryRefCard{interactive-complex-read-14-v2}{IC}{14v2}

}


\cpSection{8.4}{LANG}{Query composition}

\paragraph{Description.}

Numerous use cases require \emph{composition} of queries, including the reuse of
query results (\eg nodes, edges) or using scalar subqueries (\eg selecting a
threshold value with a subquery and using it for subsequent filtering
operations).

\paragraph{Queries}
{\raggedright
\queryRefCard{bi-read-04}{BI}{4}
\queryRefCard{bi-read-08}{BI}{8}
\queryRefCard{bi-read-12}{BI}{12}
\queryRefCard{bi-read-13}{BI}{13}
\queryRefCard{bi-read-14}{BI}{14}
\queryRefCard{bi-read-15}{BI}{15}
\queryRefCard{bi-read-16}{BI}{16}
\queryRefCard{bi-read-19}{BI}{19}
\queryRefCard{bi-read-20}{BI}{20}

}


\cpSection{8.5}{LANG}{Dates and times}

\paragraph{Description.}

Handling dates and times is a fundamental requirement for production-ready
database systems. It is particularly important in the context of BI queries as
these often calculate aggregations on certain periods of time (\eg on entities created during the course of a month).

\paragraph{Queries}
{\raggedright
\queryRefCard{bi-read-01}{BI}{1}
\queryRefCard{bi-read-02}{BI}{2}
\queryRefCard{bi-read-08}{BI}{8}
\queryRefCard{bi-read-09}{BI}{9}
\queryRefCard{bi-read-12}{BI}{12}
\queryRefCard{bi-read-13}{BI}{13}
\queryRefCard{bi-read-15}{BI}{15}
\queryRefCard{bi-read-16}{BI}{16}
\queryRefCard{interactive-complex-read-02}{IC}{2}
\queryRefCard{interactive-complex-read-03}{IC}{3}
\queryRefCard{interactive-complex-read-04}{IC}{4}
\queryRefCard{interactive-complex-read-05}{IC}{5}
\queryRefCard{interactive-complex-read-09}{IC}{9}

}


\cpSection{8.6}{LANG}{Handling paths}

\paragraph{Description.}

Handling paths as first-class citizens is one of the key distinguishing features of graph database 
systems~\cite{DBLP:conf/sigmod/AnglesABBFGLPPS18}. Hence, additionally to
reachability-style checks, a language should be able to express
queries that operate on elements of a path, \eg calculate a score on each edge
of the path.
Also, some use cases specify uniqueness constraints on paths~\cite{DBLP:journals/csur/AnglesABHRV17}:
\emph{arbitrary path},
\emph{shortest path},
\emph{no-repeated-node semantics} (also known as \emph{simple paths}), and
\emph{no-repeated-edge semantics} (also known as \emph{trails}).
Other variants are also used in rare cases, such as \emph{maximal} (non-expandable) or \emph{minimal} (non-contractable) paths.

\paragraph{Note on terminology.}

The \emph{Glossary of graph theory terms} page of Wikipedia\footnote{\url{https://en.wikipedia.org/wiki/Glossary_of_graph_theory_terms}} defines \emph{paths} as follows: ``A path may either be a walk (a sequence of nodes and edges, with both endpoints of an edge appearing adjacent to it in the sequence) or a simple path (a walk with no repetitions of nodes or edges), depending on the source.''
In this work, we use the first definition, which is more common in modern graph database systems and is also followed in a recent survey on graph query languages~\cite{DBLP:journals/csur/AnglesABHRV17}.

\paragraph{Queries}
{\raggedright
\queryRefCard{bi-read-10}{BI}{10}
\queryRefCard{bi-read-15}{BI}{15}
\queryRefCard{bi-read-19}{BI}{19}
\queryRefCard{bi-read-20}{BI}{20}
\queryRefCard{interactive-complex-read-10}{IC}{10}
\queryRefCard{interactive-complex-read-13}{IC}{13}
\queryRefCard{interactive-complex-read-14-v1}{IC}{14v1}
\queryRefCard{interactive-complex-read-14-v2}{IC}{14v2}

}


\section{Update Operations}


\cpSection{9.1}{UPD}{Insert node}

This choke point tests the ability of the database to insert a node.

\paragraph{Queries}
{\raggedright
\queryRefCard{insert-01}{INS}{1}
\queryRefCard{insert-04}{INS}{4}
\queryRefCard{insert-05}{INS}{5}
\queryRefCard{insert-06}{INS}{6}
\queryRefCard{insert-07}{INS}{7}

}


\cpSection{9.2}{UPD}{Insert edge}

This choke point tests the ability of the database to insert an edge.

\paragraph{Queries}
{\raggedright
\queryRefCard{insert-01}{INS}{1}
\queryRefCard{insert-02}{INS}{2}
\queryRefCard{insert-03}{INS}{3}
\queryRefCard{insert-04}{INS}{4}
\queryRefCard{insert-05}{INS}{5}
\queryRefCard{insert-06}{INS}{6}
\queryRefCard{insert-07}{INS}{7}
\queryRefCard{insert-08}{INS}{8}

}


\cpSection{9.3}{UPD}{Delete node}

This choke point tests the ability of the database to delete a node.

\paragraph{Queries}
{\raggedright
\queryRefCard{delete-01}{DEL}{1}
\queryRefCard{delete-04}{DEL}{4}
\queryRefCard{delete-06}{DEL}{6}
\queryRefCard{delete-07}{DEL}{7}

}


\cpSection{9.4}{UPD}{Delete edge}

This choke point tests the ability of the database to delete an edge.

\paragraph{Queries}
{\raggedright
\queryRefCard{delete-01}{DEL}{1}
\queryRefCard{delete-02}{DEL}{2}
\queryRefCard{delete-03}{DEL}{3}
\queryRefCard{delete-04}{DEL}{4}
\queryRefCard{delete-05}{DEL}{5}
\queryRefCard{delete-06}{DEL}{6}
\queryRefCard{delete-07}{DEL}{7}
\queryRefCard{delete-08}{DEL}{8}

}


\cpSection{9.5}{UPD}{Delete recursively}

This choke point tests the ability of the database to recursively perform a delete operation, \eg delete an entire message thread.

\paragraph{Queries}
{\raggedright
\queryRefCard{delete-01}{DEL}{1}
\queryRefCard{delete-04}{DEL}{4}
\queryRefCard{delete-06}{DEL}{6}
\queryRefCard{delete-07}{DEL}{7}

}

\chapter{Scale Factor Statistics}
\label{sec:sf-statistics}

\section{Number of Entities for SNB Interactive v1.0}

\begin{table}[htb]
    \setlength{\tabcolsep}{.3em}
    \centering
    {
        \tiny
        \begin{tabular}{|>{\sffamily}c|>{\tt}l|r|r|r|r|r|r|r|r|r|r|r|r|r|}
            \hline
            \tableHeaderFirst{C}                  & \tableHeader{File}               & \tableHeader{SF0.1} & \tableHeader{SF0.3} & \tableHeader{SF1}   & \tableHeader{SF3}   & \tableHeader{SF10}   & \tableHeader{SF30}   & \tableHeader{SF100}   & \tableHeader{SF300}   & \tableHeader{SF\numprint{1000}} \\ \hline
            \hline
            N                                     & organisation                     & \numprint{7955}     & \numprint{7955}     & \numprint{7955}     & \numprint{7955}     & \numprint{7955}      & \numprint{7955}      & \numprint{7955}       & \numprint{7955}       & \numprint{7955}                 \\
            E                                     & organisation\_isLocatedIn\_place & \numprint{7955}     & \numprint{7955}     & \numprint{7955}     & \numprint{7955}     & \numprint{7955}      & \numprint{7955}      & \numprint{7955}       & \numprint{7955}       & \numprint{7955}                 \\ \hline
            N                                     & place                            & \numprint{1460}     & \numprint{1460}     & \numprint{1460}     & \numprint{1460}     & \numprint{1460}      & \numprint{1460}      & \numprint{1460}       & \numprint{1460}       & \numprint{1460}                 \\
            E                                     & place\_isPartOf\_place           & \numprint{1454}     & \numprint{1454}     & \numprint{1454}     & \numprint{1454}     & \numprint{1454}      & \numprint{1454}      & \numprint{1454}       & \numprint{1454}       & \numprint{1454}                 \\ \hline
            N                                     & tag                              & \numprint{16080}    & \numprint{16080}    & \numprint{16080}    & \numprint{16080}    & \numprint{16080}     & \numprint{16080}     & \numprint{16080}      & \numprint{16080}      & \numprint{16080}                \\
            E                                     & tag\_hasType\_tagclass           & \numprint{16080}    & \numprint{16080}    & \numprint{16080}    & \numprint{16080}    & \numprint{16080}     & \numprint{16080}     & \numprint{16080}      & \numprint{16080}      & \numprint{16080}                \\ \hline
            N                                     & tagclass                         & \numprint{71}       & \numprint{71}       & \numprint{71}       & \numprint{71}       & \numprint{71}        & \numprint{71}        & \numprint{71}         & \numprint{71}         & \numprint{71}                   \\
            E                                     & tagclass\_isSubclassOf\_tagclass & \numprint{70}       & \numprint{70}       & \numprint{70}       & \numprint{70}       & \numprint{70}        & \numprint{70}        & \numprint{70}         & \numprint{70}         & \numprint{70}                   \\ \hline\hline
            N                                     & comment                          & \numprint{203354}   & \numprint{682061}   & \numprint{2581736}  & \numprint{7882971}  & \numprint{26540464}  & \numprint{80390821}  & \numprint{261475982}  & \numprint{767719169}  & \numprint{2550634137}           \\
            E                                     & comment\_hasCreator\_person      & \numprint{203354}   & \numprint{682061}   & \numprint{2581736}  & \numprint{7882971}  & \numprint{26540464}  & \numprint{80390821}  & \numprint{261475982}  & \numprint{767719169}  & \numprint{2550634137}           \\
            E                                     & comment\_hasTag\_tag             & \numprint{232524}   & \numprint{807266}   & \numprint{3145443}  & \numprint{9688491}  & \numprint{32922873}  & \numprint{100818244} & \numprint{330756583}  & \numprint{975122821}  & \numprint{3253337649}           \\
            E                                     & comment\_isLocatedIn\_place      & \numprint{203354}   & \numprint{682061}   & \numprint{2581736}  & \numprint{7882971}  & \numprint{26540464}  & \numprint{80390821}  & \numprint{261475982}  & \numprint{767719169}  & \numprint{2550634137}           \\
            E                                     & comment\_replyOf\_comment        & \numprint{103552}   & \numprint{346553}   & \numprint{1310385}  & \numprint{3997838}  & \numprint{13465094}  & \numprint{40789548}  & \numprint{132671059}  & \numprint{389555963}  & \numprint{1294311108}           \\
            E                                     & comment\_replyOf\_post           & \numprint{99802}    & \numprint{335508}   & \numprint{1271351}  & \numprint{3885133}  & \numprint{13075370}  & \numprint{39601273}  & \numprint{128804923}  & \numprint{378163206}  & \numprint{1256323029}           \\ \hline
            N                                     & forum                            & \numprint{16818}    & \numprint{38050}    & \numprint{110347}   & \numprint{271226}   & \numprint{727502}    & \numprint{1835458}   & \numprint{4982966}    & \numprint{12560110}   & \numprint{36086326}             \\
            E                                     & forum\_containerOf\_post         & \numprint{168873}   & \numprint{404531}   & \numprint{1237554}  & \numprint{3200561}  & \numprint{9119229}   & \numprint{24346116}  & \numprint{70420477}   & \numprint{188400071}  & \numprint{575768804}            \\
            E                                     & forum\_hasMember\_person         & \numprint{266965}   & \numprint{861079}   & \numprint{3345548}  & \numprint{10352102} & \numprint{35510056}  & \numprint{110335311} & \numprint{362933964}  & \numprint{1070304327} & \numprint{3570974603}           \\
            E                                     & forum\_hasModerator\_person      & \numprint{16818}    & \numprint{38050}    & \numprint{110347}   & \numprint{271226}   & \numprint{727502}    & \numprint{1835458}   & \numprint{4982966}    & \numprint{12560110}   & \numprint{36086326}             \\
            E                                     & forum\_hasTag\_tag               & \numprint{54288}    & \numprint{124186}   & \numprint{354943}   & \numprint{878307}   & \numprint{2364249}   & \numprint{5941428}   & \numprint{16147466}   & \numprint{40642813}   & \numprint{116757400}            \\ \hline
            N                                     & person                           & \numprint{1700}     & \numprint{3900}     & \numprint{11000}    & \numprint{27000}    & \numprint{73000}     & \numprint{184000}    & \numprint{499000}     & \numprint{1254000}    & \numprint{3600000}              \\
            A                                     & person\_email\_emailaddress      & \numprint{3690}     & \numprint{8393}     & \numprint{23372}    & \numprint{57419}    & \numprint{155585}    & \numprint{392497}    & \numprint{1064135}    & \numprint{2675881}    & \numprint{7681772}              \\
            E                                     & person\_hasInterest\_tag         & \numprint{39170}    & \numprint{90036}    & \numprint{255596}   & \numprint{634081}   & \numprint{1709747}   & \numprint{4289970}   & \numprint{11663500}   & \numprint{29336703}   & \numprint{84271074}             \\
            E                                     & person\_isLocatedIn\_place       & \numprint{1700}     & \numprint{3900}     & \numprint{11000}    & \numprint{27000}    & \numprint{73000}     & \numprint{184000}    & \numprint{499000}     & \numprint{1254000}    & \numprint{3600000}              \\
            E                                     & person\_knows\_person            & \numprint{18074}    & \numprint{57179}    & \numprint{226515}   & \numprint{704246}   & \numprint{2431407}   & \numprint{7514541}   & \numprint{24842767}   & \numprint{73448777}   & \numprint{245296255}            \\
            E                                     & person\_likes\_comment           & \numprint{96865}    & \numprint{412010}   & \numprint{1946260}  & \numprint{6868912}  & \numprint{25596818}  & \numprint{84821954}  & \numprint{301042048}  & \numprint{947303146}  & \numprint{3357196350}           \\
            E                                     & person\_likes\_post              & \numprint{97638}    & \numprint{328473}   & \numprint{1303778}  & \numprint{4120299}  & \numprint{14228924}  & \numprint{44582924}  & \numprint{149809880}  & \numprint{451827331}  & \numprint{1540438666}           \\
            A                                     & person\_speaks\_language         & \numprint{3771}     & \numprint{8595}     & \numprint{24246}    & \numprint{59609}    & \numprint{160992}    & \numprint{405234}    & \numprint{1099519}    & \numprint{2763100}    & \numprint{7933284}              \\
            E                                     & person\_studyAt\_organisation    & \numprint{1337}     & \numprint{3089}     & \numprint{8808}     & \numprint{21586}    & \numprint{58439}     & \numprint{147527}    & \numprint{399487}     & \numprint{1003543}    & \numprint{2880284}              \\
            E                                     & person\_workAt\_organisation     & \numprint{3732}     & \numprint{8561}     & \numprint{24079}    & \numprint{58912}    & \numprint{159511}    & \numprint{401230}    & \numprint{1086041}    & \numprint{2730945}    & \numprint{7836570}              \\ \hline
            N                                     & post                             & \numprint{168873}   & \numprint{404531}   & \numprint{1237554}  & \numprint{3200561}  & \numprint{9119229}   & \numprint{24346116}  & \numprint{70420477}   & \numprint{188400071}  & \numprint{575768804}            \\
            E                                     & post\_hasCreator\_person         & \numprint{168873}   & \numprint{404531}   & \numprint{1237554}  & \numprint{3200561}  & \numprint{9119229}   & \numprint{24346116}  & \numprint{70420477}   & \numprint{188400071}  & \numprint{575768804}            \\
            E                                     & post\_hasTag\_tag                & \numprint{59862}    & \numprint{207814}   & \numprint{816048}   & \numprint{2521635}  & \numprint{8584195}   & \numprint{26346801}  & \numprint{86600144}   & \numprint{255541805}  & \numprint{852679225}            \\
            E                                     & post\_isLocatedIn\_place         & \numprint{168873}   & \numprint{404531}   & \numprint{1237554}  & \numprint{3200561}  & \numprint{9119229}   & \numprint{24346116}  & \numprint{70420477}   & \numprint{188400071}  & \numprint{575768804}            \\ \hline\hline
            \multicolumn{2}{|l|}{\bf Total nodes}                                    & \numprint{416311}   & \numprint{1154108}  & \numprint{3966203}  & \numprint{11407324} & \numprint{36485761}  & \numprint{106781961} & \numprint{337403991}  & \numprint{969958916}  & \numprint{3166114833}           \\
            \multicolumn{2}{|l|}{\bf Total edges}                                    & \numprint{2031213}  & \numprint{6226978}  & \numprint{23031794} & \numprint{69422952} & \numprint{231371359} & \numprint{701455758} & \numprint{2286478782} & \numprint{6729459600} & \numprint{22450588784}          \\ \hline
        \end{tabular}
    }
    \caption{The number of entities per SF and per file in the Interactive workload (produced by the Hadoop-based generator and measured based on the output of the CsvBasic serializer).
        To derive these numbers, 100\% of the network was generated as an initial bulk data set with no update streams.
        Notation -- \textsf{C}: entity category, \textsf{N}: node, \textsf{E}: edge.}
    \label{tab:number-of-entities-interactive-v1}
\end{table}

\section{Number of Entities for SNB BI v1.0}

\begin{table}[htb]
    \setlength{\tabcolsep}{.3em}
    \centering
    \tiny
    \begin{tabular} {|>{\sffamily}c|>{\tt}l|r|r|r|r|r|r|r|r|r|r|}
        \hline
        \tableHeaderFirst{C}                  & \tableHeader{File}               & \tableHeader{SF1}   & \tableHeader{SF3}    & \tableHeader{SF10}   & \tableHeader{SF30}    & \tableHeader{SF100}   & \tableHeader{SF300}    & \tableHeader{SF\numprint{1000}} & \tableHeader{SF\numprint{3000}} & \tableHeader{SF\numprint{10000}} & \tableHeader{SF\numprint{30000}} \\ \hline
        N                                     & Organisation                     & \numprint{7955}     & \numprint{7955}      & \numprint{7955}      & \numprint{7955}       & \numprint{7955}       & \numprint{7955}        & \numprint{7955}                 & \numprint{7955}                 & \numprint{7955}                  & \numprint{7955}                  \\
        E                                     & Organisation\_isLocatedIn\_Place & \numprint{7955}     & \numprint{7955}      & \numprint{7955}      & \numprint{7955}       & \numprint{7955}       & \numprint{7955}        & \numprint{7955}                 & \numprint{7955}                 & \numprint{7955}                  & \numprint{7955}                  \\ \hline
        N                                     & Place                            & \numprint{1460}     & \numprint{1460}      & \numprint{1460}      & \numprint{1460}       & \numprint{1460}       & \numprint{1460}        & \numprint{1460}                 & \numprint{1460}                 & \numprint{1460}                  & \numprint{1460}                  \\
        E                                     & Place\_isPartOf\_Place           & \numprint{1454}     & \numprint{1454}      & \numprint{1454}      & \numprint{1454}       & \numprint{1454}       & \numprint{1454}        & \numprint{1454}                 & \numprint{1454}                 & \numprint{1454}                  & \numprint{1454}                  \\ \hline
        N                                     & Tag                              & \numprint{16080}    & \numprint{16080}     & \numprint{16080}     & \numprint{16080}      & \numprint{16080}      & \numprint{16080}       & \numprint{16080}                & \numprint{16080}                & \numprint{16080}                 & \numprint{16080}                 \\
        E                                     & Tag\_hasType\_TagClass           & \numprint{16080}    & \numprint{16080}     & \numprint{16080}     & \numprint{16080}      & \numprint{16080}      & \numprint{16080}       & \numprint{16080}                & \numprint{16080}                & \numprint{16080}                 & \numprint{16080}                 \\ \hline
        N                                     & TagClass                         & \numprint{71}       & \numprint{71}        & \numprint{71}        & \numprint{71}         & \numprint{71}         & \numprint{71}          & \numprint{71}                   & \numprint{71}                   & \numprint{71}                    & \numprint{71}                    \\
        E                                     & TagClass\_isSubclassOf\_TagClass & \numprint{70}       & \numprint{70}        & \numprint{70}        & \numprint{70}         & \numprint{70}         & \numprint{70}          & \numprint{70}                   & \numprint{70}                   & \numprint{70}                    & \numprint{70}                    \\ \hline \hline
        N                                     & Comment                          & \numprint{1739438}  & \numprint{5343582}   & \numprint{18196074}  & \numprint{54737515}   & \numprint{185495476}  & \numprint{554017340}   & \numprint{1876785283}           & \numprint{5656073047}           & \numprint{18880439325}           & \numprint{58666958815}           \\
        E                                     & Comment\_hasCreator\_Person      & \numprint{1739438}  & \numprint{5343582}   & \numprint{18196074}  & \numprint{54737515}   & \numprint{185495476}  & \numprint{554017340}   & \numprint{1876785283}           & \numprint{5656073047}           & \numprint{18880439325}           & \numprint{58666958815}           \\
        E                                     & Comment\_hasTag\_Tag             & \numprint{2176131}  & \numprint{6754220}   & \numprint{23113520}  & \numprint{70035650}   & \numprint{238074593}  & \numprint{714772017}   & \numprint{2426657766}           & \numprint{7330444735}           & \numprint{24505161117}           & \numprint{76236094545}           \\
        E                                     & Comment\_isLocatedIn\_Country    & \numprint{1739438}  & \numprint{5343582}   & \numprint{18196074}  & \numprint{54737515}   & \numprint{185495476}  & \numprint{554017340}   & \numprint{1876785283}           & \numprint{5656073047}           & \numprint{18880439325}           & \numprint{58666958815}           \\
        E                                     & Comment\_replyOf\_Comment        & \numprint{789020}   & \numprint{2425043}   & \numprint{8274158}   & \numprint{25130258}   & \numprint{85829276}   & \numprint{258292038}   & \numprint{883936628}            & \numprint{2688432865}           & \numprint{9045050101}            & \numprint{28244723682}           \\
        E                                     & Comment\_replyOf\_Post           & \numprint{950418}   & \numprint{2918539}   & \numprint{9921916}   & \numprint{29607257}   & \numprint{99666200}   & \numprint{294572950}   & \numprint{992848655}            & \numprint{2967640182}           & \numprint{9835389224}            & \numprint{30422235133}           \\ \hline
        N                                     & Forum                            & \numprint{100827}   & \numprint{245524}    & \numprint{667545}    & \numprint{1659632}    & \numprint{4611436}    & \numprint{11642786}    & \numprint{33168124}             & \numprint{87364322}             & \numprint{257338738}             & \numprint{728629666}             \\
        E                                     & Forum\_containerOf\_Post         & \numprint{1121226}  & \numprint{2873419}   & \numprint{8273491}   & \numprint{21651342}   & \numprint{64029217}   & \numprint{171283445}   & \numprint{519738978}            & \numprint{1440235348}           & \numprint{4461342990}            & \numprint{13148296221}           \\
        E                                     & Forum\_hasMember\_Person         & \numprint{2909768}  & \numprint{8780738}   & \numprint{30201123}  & \numprint{90198118}   & \numprint{303838931}  & \numprint{898932504}   & \numprint{3004740356}           & \numprint{8909683066}           & \numprint{29398116490}           & \numprint{90652090014}           \\
        E                                     & Forum\_hasModerator\_Person      & \numprint{100827}   & \numprint{245524}    & \numprint{667545}    & \numprint{1659632}    & \numprint{4611436}    & \numprint{11642786}    & \numprint{33168124}             & \numprint{87364322}             & \numprint{257338738}             & \numprint{728629666}             \\
        E                                     & Forum\_hasTag\_Tag               & \numprint{328584}   & \numprint{809991}    & \numprint{2207525}   & \numprint{5467942}    & \numprint{15195472}   & \numprint{38372330}    & \numprint{109341702}            & \numprint{288057168}            & \numprint{848359157}             & \numprint{2401607343}            \\ \hline
        N                                     & Person                           & \numprint{10295}    & \numprint{25066}     & \numprint{68673}     & \numprint{170654}     & \numprint{473001}     & \numprint{1193579}     & \numprint{3399580}              & \numprint{8955552}              & \numprint{26384952}              & \numprint{74689437}              \\
        E                                     & Person\_hasInterest\_Tag         & \numprint{238052}   & \numprint{589533}    & \numprint{1608653}   & \numprint{3978964}    & \numprint{11057039}   & \numprint{27923123}    & \numprint{79573188}             & \numprint{209648434}            & \numprint{617405426}             & \numprint{1747667501}            \\
        E                                     & Person\_isLocatedIn\_City        & \numprint{10295}    & \numprint{25066}     & \numprint{68673}     & \numprint{170654}     & \numprint{473001}     & \numprint{1193579}     & \numprint{3399580}              & \numprint{8955552}              & \numprint{26384952}              & \numprint{74689437}              \\
        E                                     & Person\_knows\_Person            & \numprint{173014}   & \numprint{528896}    & \numprint{1839354}   & \numprint{5524302}    & \numprint{18655515}   & \numprint{55656915}    & \numprint{187247788}            & \numprint{559360185}            & \numprint{1854528925}            & \numprint{5734470022}            \\
        E                                     & Person\_likes\_Comment           & \numprint{1109813}  & \numprint{3826649}   & \numprint{14586377}  & \numprint{48651549}   & \numprint{184325690}  & \numprint{605620715}   & \numprint{2249224980}           & \numprint{7279159053}           & \numprint{25779776654}           & \numprint{83352563279}           \\
        E                                     & Person\_likes\_Post              & \numprint{760455}   & \numprint{2417873}   & \numprint{8546995}   & \numprint{26908834}   & \numprint{98423296}   & \numprint{314778935}   & \numprint{1140808487}           & \numprint{3619661715}           & \numprint{12593759314}           & \numprint{40072928363}           \\
        E                                     & Person\_studyAt\_University      & \numprint{8309}     & \numprint{20113}     & \numprint{55066}     & \numprint{136614}     & \numprint{378582}     & \numprint{955425}      & \numprint{2719877}              & \numprint{7165145}              & \numprint{21108848}              & \numprint{59758459}              \\
        E                                     & Person\_workAt\_Company          & \numprint{22044}    & \numprint{54135}     & \numprint{149581}    & \numprint{371634}     & \numprint{1029492}    & \numprint{2598384}     & \numprint{7398286}              & \numprint{19491928}             & \numprint{57416114}              & \numprint{162518922}             \\ \hline
        N                                     & Post                             & \numprint{1121226}  & \numprint{2873419}   & \numprint{8273491}   & \numprint{21651342}   & \numprint{64029217}   & \numprint{171283445}   & \numprint{519738978}            & \numprint{1440235348}           & \numprint{4461342990}            & \numprint{13148296221}           \\
        E                                     & Post\_hasCreator\_Person         & \numprint{1121226}  & \numprint{2873419}   & \numprint{8273491}   & \numprint{21651342}   & \numprint{64029217}   & \numprint{171283445}   & \numprint{519738978}            & \numprint{1440235348}           & \numprint{4461342990}            & \numprint{13148296221}           \\
        E                                     & Post\_hasTag\_Tag                & \numprint{751933}   & \numprint{2305927}   & \numprint{7865279}   & \numprint{23426338}   & \numprint{78380259}   & \numprint{231621916}   & \numprint{769380657}            & \numprint{2273989086}           & \numprint{7454473533}            & \numprint{22896875734}           \\
        E                                     & Post\_isLocatedIn\_Country       & \numprint{1121226}  & \numprint{2873419}   & \numprint{8273491}   & \numprint{21651342}   & \numprint{64029217}   & \numprint{171283445}   & \numprint{519738978}            & \numprint{1440235348}           & \numprint{4461342990}            & \numprint{13148296221}           \\ \hline\hline
        \multicolumn{2}{|l|}{\bf Total nodes}                                    & \numprint{2997352}  & \numprint{8513157}   & \numprint{27231349}  & \numprint{78244709}   & \numprint{254634696}  & \numprint{738162716}   & \numprint{2433117531}           & \numprint{7192653835}           & \numprint{23625531571}           & \numprint{72618599705}           \\
        \multicolumn{2}{|l|}{\bf Total edges}                                    & \numprint{17196776} & \numprint{51035227}  & \numprint{170343945} & \numprint{505722361}  & \numprint{1703042944} & \numprint{5078844191}  & \numprint{17203259133}          & \numprint{51881931133}          & \numprint{173439201772}          & \numprint{539565683952}          \\ \hline
    \end{tabular}
    \caption{The number of entities per SF and per file in the \emph{initial data set} used in the BI workload.
        Notation -- \textsf{C}: entity category, \textsf{N}: node, \textsf{E}: edge.}
    \label{tab:number-of-entities-bi-initial}
\end{table}

\begin{table}[htb]
    \setlength{\tabcolsep}{.3em}
    \centering
    \tiny
    \begin{tabular} {|>{\sffamily}c|>{\sffamily}c|>{\tt}l|r|r|r|r|r|r|r|r|r|r|}
        \hline
        \tableHeaderFirst{T}                                   & \tableHeader{C}    & \tableHeader{File}            & \tableHeader{SF1}   & \tableHeader{SF3}    & \tableHeader{SF10}   & \tableHeader{SF30}    & \tableHeader{SF100}   & \tableHeader{SF300}   & \tableHeader{SF\numprint{1000}} & \tableHeader{SF\numprint{3000}} & \tableHeader{SF\numprint{10000}} & \tableHeader{SF\numprint{30000}} \\
        \hline
        I                                                      & N                  & Comment                       & \numprint{652269}   & \numprint{1932347}   & \numprint{6122166}   & \numprint{17233922}   & \numprint{53364420}   & \numprint{144700167}  & \numprint{428355986}            & \numprint{1132241525}           & \numprint{3323091103}            & \numprint{9411625366}            \\
        I                                                      & E                  & Comment\_hasCreator\_Person   & \numprint{652269}   & \numprint{1932347}   & \numprint{6122166}   & \numprint{17233922}   & \numprint{53364420}   & \numprint{144700167}  & \numprint{428355986}            & \numprint{1132241525}           & \numprint{3323091103}            & \numprint{9411625366}            \\
        I                                                      & E                  & Comment\_hasTag\_Tag          & \numprint{727839}   & \numprint{2203748}   & \numprint{7079778}   & \numprint{20150855}   & \numprint{62861828}   & \numprint{171071832}  & \numprint{508165623}            & \numprint{1339365204}           & \numprint{3909017913}            & \numprint{11014456527}           \\
        I                                                      & E                  & Comment\_isLocatedIn\_Country & \numprint{652269}   & \numprint{1932347}   & \numprint{6122166}   & \numprint{17233922}   & \numprint{53364420}   & \numprint{144700167}  & \numprint{428355986}            & \numprint{1132241525}           & \numprint{3323091103}            & \numprint{9411625366}            \\
        I                                                      & E                  & Comment\_replyOf\_Comment     & \numprint{408491}   & \numprint{1216510}   & \numprint{3872589}   & \numprint{10961292}   & \numprint{33981364}   & \numprint{92048927}   & \numprint{272358361}            & \numprint{718336661}            & \numprint{2102670076}            & \numprint{5946782323}            \\
        I                                                      & E                  & Comment\_replyOf\_Post        & \numprint{243778}   & \numprint{715837}    & \numprint{2249577}   & \numprint{6272630}    & \numprint{19383056}   & \numprint{52651240}   & \numprint{155997625}            & \numprint{413904864}            & \numprint{1220421027}            & \numprint{3464843043}            \\
        I                                                      & N                  & Forum                         & \numprint{5767}     & \numprint{14105}     & \numprint{38084}     & \numprint{94700}      & \numprint{265314}     & \numprint{671285}     & \numprint{1915909}              & \numprint{5047113}              & \numprint{14895929}              & \numprint{42218181}              \\
        I                                                      & E                  & Forum\_containerOf\_Post      & \numprint{71716}    & \numprint{182738}    & \numprint{507826}    & \numprint{1297451}    & \numprint{3735615}    & \numprint{9741528}    & \numprint{28453210}             & \numprint{76669773}             & \numprint{231949432}             & \numprint{671846867}             \\
        I                                                      & E                  & Forum\_hasMember\_Person      & \numprint{350924}   & \numprint{1050322}   & \numprint{3436445}   & \numprint{9978585}    & \numprint{32960000}   & \numprint{93286265}   & \numprint{295103572}            & \numprint{825253679}            & \numprint{2554550825}            & \numprint{7479070111}            \\
        I                                                      & E                  & Forum\_hasModerator\_Person   & \numprint{5767}     & \numprint{14105}     & \numprint{38084}     & \numprint{94700}      & \numprint{265314}     & \numprint{671285}     & \numprint{1915909}              & \numprint{5047113}              & \numprint{14895929}              & \numprint{42218181}              \\
        I                                                      & E                  & Forum\_hasTag\_Tag            & \numprint{13456}    & \numprint{31162}     & \numprint{86525}     & \numprint{214373}     & \numprint{592043}     & \numprint{1495805}    & \numprint{4280777}              & \numprint{11235864}             & \numprint{33142429}              & \numprint{94020588}              \\
        I                                                      & N                  & Person                        & \numprint{325}      & \numprint{804}       & \numprint{2127}      & \numprint{5296}       & \numprint{14699}      & \numprint{36921}      & \numprint{105420}               & \numprint{276448}               & \numprint{815048}                & \numprint{2310563}               \\
        I                                                      & E                  & Person\_hasInterest\_Tag      & \numprint{8014}     & \numprint{17861}     & \numprint{50568}     & \numprint{124969}     & \numprint{341426}     & \numprint{861441}     & \numprint{2470258}              & \numprint{6465213}              & \numprint{19061544}              & \numprint{54112770}              \\
        I                                                      & E                  & Person\_isLocatedIn\_City     & \numprint{325}      & \numprint{804}       & \numprint{2127}      & \numprint{5296}       & \numprint{14699}      & \numprint{36921}      & \numprint{105420}               & \numprint{276448}               & \numprint{815048}                & \numprint{2310563}               \\
        I                                                      & E                  & Person\_knows\_Person         & \numprint{46436}    & \numprint{139535}    & \numprint{465597}    & \numprint{1356282}    & \numprint{4461290}    & \numprint{12657067}   & \numprint{39877751}             & \numprint{111602193}            & \numprint{347323797}             & \numprint{1028845782}            \\
        I                                                      & E                  & Person\_likes\_Comment        & \numprint{507078}   & \numprint{1642981}   & \numprint{5814742}   & \numprint{17739535}   & \numprint{59010156}   & \numprint{170613836}  & \numprint{547019411}            & \numprint{1522602131}           & \numprint{4738606525}            & \numprint{14044004355}           \\
        I                                                      & E                  & Person\_likes\_Post           & \numprint{84089}    & \numprint{242012}    & \numprint{781367}    & \numprint{2228761}    & \numprint{7227562}    & \numprint{21174383}   & \numprint{69394102}             & \numprint{203079530}            & \numprint{664408922}             & \numprint{2040369359}            \\
        I                                                      & E                  & Person\_studyAt\_University   & \numprint{253}      & \numprint{642}       & \numprint{1711}      & \numprint{4215}       & \numprint{11684}      & \numprint{29520}      & \numprint{84408}                & \numprint{221160}               & \numprint{651833}                & \numprint{1848819}               \\
        I                                                      & E                  & Person\_workAt\_Company       & \numprint{722}      & \numprint{1691}      & \numprint{4541}      & \numprint{11473}      & \numprint{32135}      & \numprint{79806}      & \numprint{228835}               & \numprint{601641}               & \numprint{1772442}               & \numprint{5025385}               \\
        I                                                      & N                  & Post                          & \numprint{71716}    & \numprint{182738}    & \numprint{507826}    & \numprint{1297451}    & \numprint{3735615}    & \numprint{9741528}    & \numprint{28453210}             & \numprint{76669773}             & \numprint{231949432}             & \numprint{671846867}             \\
        I                                                      & E                  & Post\_hasCreator\_Person      & \numprint{71716}    & \numprint{182738}    & \numprint{507826}    & \numprint{1297451}    & \numprint{3735615}    & \numprint{9741528}    & \numprint{28453210}             & \numprint{76669773}             & \numprint{231949432}             & \numprint{671846867}             \\
        I                                                      & E                  & Post\_hasTag\_Tag             & \numprint{26578}    & \numprint{78669}     & \numprint{247471}    & \numprint{690212}     & \numprint{2192065}    & \numprint{6197708}    & \numprint{19682903}             & \numprint{56322268}             & \numprint{180509835}             & \numprint{545993292}             \\
        I                                                      & E                  & Post\_isLocatedIn\_Country    & \numprint{71716}    & \numprint{182738}    & \numprint{507826}    & \numprint{1297451}    & \numprint{3735615}    & \numprint{9741528}    & \numprint{28453210}             & \numprint{76669773}             & \numprint{231949432}             & \numprint{671846867}             \\
        \hline
        \multicolumn{3}{|l|}{\bf Total insert node operations}                                                      & \numprint{730077}   & \numprint{2129994}   & \numprint{6670203}   & \numprint{18631369}   & \numprint{57380048}   & \numprint{155149901}  & \numprint{458830525}            & \numprint{1214234859}           & \numprint{3570751512}            &  \numprint{10128000977}          \\
        \multicolumn{3}{|l|}{\bf Total insert edge operations}                                                      & \numprint{3943436}  & \numprint{11768787}  & \numprint{37898932}  & \numprint{108193375}  & \numprint{341270307}  & \numprint{941500954}  & \numprint{2858756557}           & \numprint{7708806338}           & \numprint{23129878647}           &  \numprint{66602692431}          \\
        \multicolumn{3}{|l|}{\bf Total insert operations}                                                           & \numprint{4673513}  & \numprint{13898781}  & \numprint{44569135}  & \numprint{126824744}  & \numprint{398650355}  & \numprint{1096650855} & \numprint{3317587082}           & \numprint{8923041197}           & \numprint{26700630159}           &  \numprint{76730693408}          \\
        \hline\hline
        D                                                      & N                  & Comment                       & \numprint{11966}    & \numprint{35147}     & \numprint{110712}    & \numprint{309712}     & \numprint{959810}     & \numprint{2597282}    & \numprint{7704534}              & \numprint{20373985}             & \numprint{59821497}              &  \numprint{169401271}            \\
        D                                                      & N                  & Forum                         & \numprint{212}      & \numprint{459}       & \numprint{1252}      & \numprint{3220}       & \numprint{8975}       & \numprint{22699}      & \numprint{64932}                & \numprint{172181}               & \numprint{506906}                &  \numprint{1440207}              \\
        D                                                      & E                  & Forum\_hasMember\_Person      & \numprint{2004}     & \numprint{5002}      & \numprint{12857}     & \numprint{31647}      & \numprint{86820}      & \numprint{221834}     & \numprint{609738}               & \numprint{1565418}              & \numprint{4544009}               &  \numprint{12773538}             \\
        D                                                      & N                  & Person                        & \numprint{54}       & \numprint{122}       & \numprint{264}       & \numprint{510}        & \numprint{1265}       & \numprint{2827}       & \numprint{7285}                 & \numprint{18234}                & \numprint{48251}                 &  \numprint{123926}               \\
        D                                                      & E                  & Person\_knows\_Person         & \numprint{5548}     & \numprint{16704}     & \numprint{57638}     & \numprint{168900}     & \numprint{560741}     & \numprint{1604444}    & \numprint{5140980}              & \numprint{14599090}             & \numprint{46275390}              &  \numprint{139258201}            \\
        D                                                      & E                  & Person\_likes\_Comment        & \numprint{12220}    & \numprint{39660}     & \numprint{138268}    & \numprint{420001}     & \numprint{1394595}    & \numprint{4040199}    & \numprint{12955551}             & \numprint{36066934}             & \numprint{112313459}             &  \numprint{332839378}            \\
        D                                                      & E                  & Person\_likes\_Post           & \numprint{1992}     & \numprint{5869}      & \numprint{18835}     & \numprint{52070}      & \numprint{169649}     & \numprint{498070}     & \numprint{1634887}              & \numprint{4788019}              & \numprint{15655650}              &  \numprint{48054670}             \\
        D                                                      & N                  & Post                          & \numprint{1908}     & \numprint{5004}      & \numprint{13566}     & \numprint{34948}      & \numprint{100375}     & \numprint{263354}     & \numprint{767998}               & \numprint{2067056}              & \numprint{6267076}               &  \numprint{18141667}             \\
        \hline
        \multicolumn{3}{|l|}{\bf Total delete node operations}                                                      & \numprint{14140}   & \numprint{40732}      & \numprint{125794}    & \numprint{348390}     & \numprint{1070425}    & \numprint{2886162}    & \numprint{8544749}              & \numprint{22631456}             & \numprint{66643730}              &  \numprint{189107071}            \\
        \multicolumn{3}{|l|}{\bf Total delete edge operations}                                                      & \numprint{21764}   & \numprint{67235}      & \numprint{227598}    & \numprint{672618}     & \numprint{2211805}    & \numprint{6364547}    & \numprint{20341156}             & \numprint{57019461}             & \numprint{178788508}             &  \numprint{532925787}            \\
        \multicolumn{3}{|l|}{\bf Total delete operations}                                                           & \numprint{35904}   & \numprint{107967}     & \numprint{353392}    & \numprint{1021008}    & \numprint{3282230}    & \numprint{9250709}    & \numprint{28885905}             & \numprint{79650917}             & \numprint{245432238}             &  \numprint{722032858}            \\
        \hline
    \end{tabular}
    \caption{The number of entities per SF and per file in the \emph{update data sets} used in the BI workload.
        Notation -- \textsf{T}: update type, \textsf{I}: insert, \textsf{D}: delete; \textsf{C}: entity category, \textsf{N}: node, \textsf{E}: edge.}
    \label{tab:number-of-entities-bi-updates}
\end{table}

\section{Factor Tables}

\begin{table}[htb]
    \setlength{\tabcolsep}{.3em}
    \centering
    \begin{tabular}{|r|r|}
        \hline
        \tableHeaderFirst{Scale Factor} & \tableHeader{Size} \\\hline
        \numprint{1}                    &  8.6M              \\
        \numprint{3}                    &   18M              \\
        \numprint{10}                   &   41M              \\
        \numprint{30}                   &  100M              \\
        \numprint{100}                  &  259M              \\
        \numprint{300}                  &  656M              \\
        \numprint{1000}                 &  1.9G              \\
        \numprint{3000}                 &  5.1G              \\
        \numprint{10000}                &   16G              \\
        \numprint{30000}                &   47G              \\\hline
    \end{tabular}
    \caption{The total size of the factor tables.}
    \label{tab:factor-table-sizes}
\end{table}

\chapter{Benchmark Checklist}
\label{sec:benchmark-checklist}

We expect LDBC benchmarks to be used in many scenarios.
For most research papers, fully audited results are unrealistic and even unaudited results can provide insight into the performance of the systems under test (SUT). However, we ask authors to include the following information in their papers:

\begin{itemize}
\item Were the results cross-validated for at least one scale factor?
\item Were the results cross-validated for all scale factors used in the benchmark?
\item Does the SUT have a persistent storage?
\item Does the SUT provide ACID transactions?
\item Does the SUT provide any level of fault-tolerance?
\item How many warm-up rounds were performed?
\item How many execution rounds were performed?
\item How were the execution times summarized?\footnote{Paper~\cite{DBLP:conf/sc/HoeflerB15} provides an excellent overview on how to summarize benchmark results.}
\item Is the loading phase included in the query execution times?\footnote{This might be relevant for systems without persistent storage, or systems providing lazy/incremental computation.}
\item If the SUT is not your own system, did you contact its developers or experts to help optimizing the queries?\footnote{For a research prototype tool, the tuning knobs are usually not well documented. Hence, it is worth contacting the tool's authors, who are generally keen to help. For more mature systems (\eg most established RDBMSs), there is a large body of knowledge available, in the form of books and online forums, which should help your optimization efforts. It is also possible to contact experienced DBAs who can assist with fine tuning the system.}
\end{itemize}

These results will help the reader to put the results in context. For example, a non-ACID compliant, non-fault-tolerant system working on read-only graphs and offering no persistent storage is expected to have significantly better results than a fully-fledged disk-based DBMS.

We also suggest the reader to take a look at the checklist presented in~\cite{DBLP:conf/sigmod/RaasveldtHGM18}.

\chapter{Legacy Data Sets for the Interactive workload}
\label{sec:legacy-data-sets}

The Interactive workload uses the legacy version of the data sets. These can be generated using the Hadoop-based \datagen hosted at \url{https://github.com/ldbc/ldbc_snb_datagen_hadoop/}. This chapter documents these data sets.

The SNB data sets are available in the SURF/CWI LDBC SNB data repository~\cite{cwi:snb} at \url{https://repository.surfsara.nl/datasets/cwi/ldbc-snb-interactive-v1-datagen-v100}.

\begin{itemize}
  \item Serializers:
    \texttt{csv\_basic},
    \texttt{csv\_basic-longdateformatter},
    \texttt{csv\_composite},
    \texttt{csv\_composite-longdateformatter},
    \texttt{csv\_composite\_merge\_foreign},
    \texttt{csv\_composite\_merge\_foreign-longdateformatter},
    \texttt{csv\_merge\_foreign},
    \texttt{csv\_merge\_foreign-longdateformatter},
    \texttt{ttl}
  \item Partition numbers: $2^k$ (1, 2, 4, 8, 16, 32, 64, 128, 256, 512, 1024) and $6 \times 2^k$ (24, 48, 96, 192, 384, 768).
\end{itemize}

The \textbf{key differences from the latest (BI and Interactive v2) data sets} are the following:

\begin{itemize}
    \item DateTime values follow the format \texttt{yyyy-mm-ddTHH:MM:ss.sss+0000}, i.e. their offset string is \texttt{0000} instead of \texttt{00:00}. This implies that they are not compatible with the recommendations of RFC-3339\footnote{\url{https://tools.ietf.org/html/rfc3339}}.
    \item The Forum-hasModerator-Person edge type has an \emph{exactly one} cardinality on the Person's end.
\end{itemize}

\section{Output Data}

For each scale factor, \datagen produces three different artefacts:
\begin{itemize}
  \item \textbf{Dataset:} The dataset to be bulk loaded by the SUT. In the Interactive workload, it corresponds to roughly the 90\% of the total generated network.
  \item \textbf{Update Streams:} A set of update streams containing update
    queries, which are used by the driver to generate the update queries of the
    workloads. This update
    streams correspond to the remaining 10\% of the generated dataset.
  \item \textbf{Substitution Parameters:} A set of files containing the
    different parameter bindings that will be used by the driver to generate the
    read queries of the workloads.
\end{itemize}

\subsection{Scale Factors}
\label{sec:legacy-scale-factors}

\ldbcsnb defines a set of scale factors (SFs), targeting systems of different sizes and budgets.
SFs are computed based on the ASCII size in Gibibytes of the generated output files using the CsvMergeForeign serializer (see \autoref{sec:legacy-serializers}) and default settings, \ie both the 90\% initial data and the 10\% update streams count towards the total size.
For example, SF1 takes roughly 1~GiB in CSV format, SF3 weighs roughly 3~GiB and so on and so forth.
It is important to note that for a given scale factor, data sets generated using different serializers contain exaclty the same data, the only difference is in how they are represented.%
\footnote{Naturally, there are slight differences in the disk usage of the data sets created with different serializers. For example, for a given scale factor, the disk usage of the data set serialized with the CsvBasic serializer is expected to be higher, while with the CsvMergeForeignComposite, it is expected to be lower.}
The provided SFs are the following: 1, 3, 10, 30, 100, 300, 1000.
Additionally, two small data sets, 0.1, and 0.3 are provided to help initial validation efforts.

The Test Sponsor may select the SF that better fits their needs, by properly configuring the \datagen, as described in \autoref{sec:data_generation}.
The size of the resulting dataset is mainly affected by the following configuration parameters: the number of persons and the number of years simulated.
By default, all SFs are defined over a period of three years, starting from 2010, and SFs are computed by scaling the number of Persons in the network.
\autoref{tab:snsize-interactive} shows some metrics of SFs 0.1, \ldots, 1000 data sets.

\begin{table}[htb]
    \small
    \setlength{\tabcolsep}{.5em}
    \centering
    \begin{tabular}{|l||r|r|r|r|r|r|r|r|r|r|r|}
        \hline
        \bf Scale Factor & \bf SF0.1 & \bf SF0.3 & \bf SF1 & \bf SF3 & \bf SF10 & \bf SF30 & \bf SF100 & \bf SF300 & \bf SF\numprint{1000} \\ \hline\hline
        \# Persons       & 1.5K    & 3.5K    & 11K   & 27K   & 73K    & 182K   & 499K    & 1.25M   & 3.6M                \\ \hline\hline
        \# nodes         & 327.6K  & 908K    & 3.2M  & 9.3M  & 30M    & 88.8M  & 282.6M  & 817.3M  & 2.7B                \\ \hline
        \# edges         & 1.5M    & 4.6M    & 17.3M & 52.7M & 176.6M & 540.9M & 1.8B    & 5.3B    & 17B                 \\ \hline
    \end{tabular}
    \centering
    \caption{Properties of data sets for each scale factor in the Interactive workload produced by the Hadoop-based generator.
        For detailed statistics, see \autoref{tab:number-of-entities-interactive-v1}}
    \label{tab:snsize-interactive}
\end{table}

\begin{table}[htb]
    \setlength{\tabcolsep}{.3em}
    \centering
    \scriptsize
    \begin{tabular}{|c|l|r|r|r|r|r|r|r|r|r|r|r|r|r|r|}
        \hline
        \tableHeaderFirst{INS} & \tableHeader{Operation} & \tableHeader{SF0.1} & \tableHeader{SF0.3} & \tableHeader{SF1}  & \tableHeader{SF3}  & \tableHeader{SF10}  & \tableHeader{SF30}   & \tableHeader{SF100}  & \tableHeader{SF300}  & \tableHeader{SF\numprint{1000}} \\
        \hline
        1                      & Add person              & \numprint{172}      & \numprint{386}      & \numprint{1108}    & \numprint{2672}    & \numprint{7355}     & \numprint{18570}     & \numprint{50374}     & \numprint{125931}    & \numprint{360960}               \\
        2                      & Add like to post        & \numprint{44313}    & \numprint{132041}   & \numprint{494410}  & \numprint{1460471} & \numprint{4875874}  & \numprint{14378128}  & \numprint{45633086}  & \numprint{129721727} & \numprint{410899721}            \\
        3                      & Add like to comment     & \numprint{30395}    & \numprint{105061}   & \numprint{460487}  & \numprint{1450891} & \numprint{5210730}  & \numprint{16114277}  & \numprint{54990638}  & \numprint{163624084} & \numprint{539128029}            \\
        4                      & Add forum               & \numprint{3059}     & \numprint{6913}     & \numprint{19757}   & \numprint{49223}   & \numprint{131439}   & \numprint{330288}    & \numprint{898185}    & \numprint{2257347}   & \numprint{6479509}              \\
        5                      & Add forum membership    & \numprint{126615}   & \numprint{405441}   & \numprint{1566914} & \numprint{4874316} & \numprint{16647977} & \numprint{51095793}  & \numprint{165881862} & \numprint{478826826} & \numprint{1543247540}           \\
        6                      & Add post                & \numprint{32610}    & \numprint{78164}    & \numprint{229614}  & \numprint{592875}  & \numprint{1655168}  & \numprint{4304447}   & \numprint{12236177}  & \numprint{32109577}  & \numprint{96023955}             \\
        7                      & Add comment             & \numprint{46969}    & \numprint{144917}   & \numprint{490328}  & \numprint{1372420} & \numprint{4414427}  & \numprint{12588582}  & \numprint{39547415}  & \numprint{112862922} & \numprint{362292612}            \\
        8                      & Add friendship          & \numprint{3197}     & \numprint{10337}    & \numprint{40124}   & \numprint{122714}  & \numprint{431916}   & \numprint{1304053}   & \numprint{4252839}   & \numprint{12047072}  & \numprint{36762818}             \\
        \hline
        \multicolumn{2}{|l|}{\bf Total insert operations}
                                                         & \numprint{287330}   & \numprint{883260}   & \numprint{3302742} & \numprint{9925582} & \numprint{33374886} & \numprint{100134138} & \numprint{323490576} & \numprint{931575486} & \numprint{2995195144}           \\
        \hline
    \end{tabular}
    \caption{Update stream statistics for SNB Interactive v1.0}
    \label{table:interactive-v1-update-stream-statistics}
\end{table}

\begin{table}[htb]
    \centering
    \begin{tabular}{|l|c|cc|cc|}
        \hline
        \multicolumn{1}{|c|}{\multirow{2}{*}{\bf Serializer}} & \multirow{2}{*}{\bf Nodes} & \multicolumn{2}{c|}{\bf Attributes} & \multicolumn{2}{c|}{\bf Edges}                                      \\
                                                              &                            & \bf single-valued                   & \bf multi-valued               & \bf one-to-many & \bf many-to-many \\ \hline
        CsvBasic                                              & \yes                       & \no                                 & \yes                           & \yes            & \yes             \\
        CsvComposite                                          & \yes                       & \no                                 & \no                            & \yes            & \yes             \\
        CsvMergeForeign                                       & \yes                       & \no                                 & \yes                           & \no             & \yes             \\
        CsvCompositeMergeForeign                              & \yes                       & \no                                 & \no                            & \no             & \yes             \\ \hline
    \end{tabular}
    \centering
    \caption{Attributes and edges serialized to separate files the different CSV serializers.}
    \label{tab:legacy-csv-serializers}
\end{table}

\autoref{tab:legacy-csv-serializers} show how each CSV serializer handles attributes/edges of different cardinalities. The data shows why CsvBasic has the most files and CsvCompositeMergeForeign has the least number of files.

\begin{table}[htb]
    \scriptsize
    \centering
    \begin{tabularx}{\linewidth}{|>{\sffamily}c|>{\tt}l|>{\tt}X|}
        \hline
        \tableHeaderFirst{C} & \tableHeader{File}                      & \tableHeader{Content}                                                                             \\
        \hline\hline
        N                    & organisation\_*.csv                     & id | type | name | url                                                                            \\
        E                    & organisation\_isLocatedIn\_place\_*.csv & Organisation.id | Place.id                                                                        \\
        \hline
        N                    & place\_*.csv                            & id | name | url | type                                                                            \\
        E                    & place\_isPartOf\_place\_*.csv           & Place.id | Place.id                                                                               \\
        \hline
        N                    & tag\_*.csv                              & id | name | url                                                                                   \\
        E                    & tag\_hasType\_tagclass\_*.csv           & Tag.id | TagClass.id                                                                              \\
        \hline
        N                    & tagclass\_*.csv                         & id | name | url                                                                                   \\
        E                    & tagclass\_isSubclassOf\_tagclass\_*.csv & TagClass.id | TagClass.id                                                                         \\
        \hline\hline
        N                    & comment\_*.csv                          & creationDate | id | locationIP | browserUsed | content | length                                   \\
        E                    & comment\_hasCreator\_person\_*.csv      & creationDate | Comment.id | Person.id                                                             \\
        E                    & comment\_hasTag\_tag\_*.csv             & creationDate | Comment.id | Tag.id                                                                \\
        E                    & comment\_isLocatedIn\_place\_*.csv      & creationDate | Comment.id | Place.id                                                              \\
        E                    & comment\_replyOf\_comment\_*.csv        & creationDate | Comment.id | ParentComment.id                                                      \\
        E                    & comment\_replyOf\_post\_*.csv           & creationDate | Comment.id | ParentPost.id                                                         \\
        \hline
        N                    & forum\_*.csv                            & creationDate | id | title | type                                                                  \\
        E                    & forum\_containerOf\_post\_*.csv         & creationDate | Forum.id | Post.id                                                                 \\
        E                    & forum\_hasMember\_person\_*.csv         & creationDate | Forum.id | Person.id | type                                                        \\
        E                    & forum\_hasModerator\_person\_*.csv      & creationDate | Forum.id | Person.id                                                               \\
        E                    & forum\_hasTag\_tag\_*.csv               & creationDate | Forum.id | Tag.id                                                                  \\
        \hline
        N                    & person\_*.csv                           & creationDate | id | firstName | lastName | gender | birthday | locationIP | browserUsed           \\
        A                    & person\_email\_emailaddress\_*.csv      & creationDate | Person.id | email                                                                  \\
        E                    & person\_hasInterest\_tag\_*.csv         & creationDate | Person.id | Tag.id                                                                 \\
        E                    & person\_isLocatedIn\_place\_*.csv       & creationDate | Person.id | Place.id                                                               \\
        E                    & person\_knows\_person\_*.csv            & creationDate | Person1.id | Person2.id                                                            \\
        E                    & person\_likes\_comment\_*.csv           & creationDate | Person.id | Comment.id                                                             \\
        E                    & person\_likes\_post\_*.csv              & creationDate | Person.id | Post.id                                                                \\
        A                    & person\_speaks\_language\_*.csv         & creationDate | Person.id | language                                                               \\
        E                    & person\_studyAt\_organisation\_*.csv    & creationDate | Person.id | Organisation.id | classYear                                            \\
        E                    & person\_workAt\_organisation\_*.csv     & creationDate | Person.id | Organisation.id | workFrom                                             \\
        \hline
        N                    & post\_*.csv                             & creationDate | id | imageFile | locationIP | browserUsed | language | content | length | Forum.id \\
        E                    & post\_hasCreator\_person\_*.csv         & creationDate | Post.id | Person.id                                                                \\
        E                    & post\_hasTag\_tag\_*.csv                & creationDate | Post.id | Tag.id                                                                   \\
        E                    & post\_isLocatedIn\_place\_*.csv         & creationDate | Post.id | Place.id                                                                 \\
        \hline
    \end{tabularx}
    \caption{Files output by the CsvBasic serializer (33 in total). The first part of the table contains the static entites, the second part contains the dynamic ones.
        Notation -- \textsf{C}: entity category, \textsf{N}: node, \textsf{E}: edge.}
    \label{table:csv_basic}
\end{table}

\begin{table}[htb]
    \scriptsize
    \centering
    \begin{tabularx}{\linewidth}{|>{\sffamily}c|>{\tt}l|>{\tt}X|}
        \hline
        \tableHeaderFirst{C} & \tableHeader{File}                   & \tableHeader{Content}                                                                                               \\
        \hline\hline
        N                    & organisation\_*.csv                  & id | type | name | url | place                                                                                      \\        \hline
        N                    & place\_*.csv                         & id | name | url | type | isPartOf                                                                                   \\\hline
        N                    & tag\_*.csv                           & id | name | url | hasType                                                                                           \\\hline
        N                    & tagclass\_*.csv                      & id | name | url | isSubclassOf                                                                                      \\\hline\hline
        N                    & comment\_*.csv                       & id | creationDate | locationIP | browserUsed | content | length | creator | place | replyOfPost | replyOfComment    \\
        E                    & comment\_hasTag\_tag\_*.csv          & Comment.id | Tag.id                                                                                                 \\\hline
        N                    & forum\_*.csv                         & id | title | creationDate | moderator                                                                               \\
        E                    & forum\_hasMember\_person\_*.csv      & Forum.id | Person.id | joinDate                                                                        \\
        E                    & forum\_hasTag\_tag\_*.csv            & Forum.id | Tag.id                                                                                                   \\\hline
        N                    & person\_*.csv                        & id | firstName | lastName | gender | birthday | creationDate | locationIP | browserUsed | place                     \\
        A                    & person\_email\_emailaddress\_*.csv   & Person.id | email                                                                                                   \\
        E                    & person\_hasInterest\_tag\_*.csv      & Person.id | Tag.id                                                                                                  \\
        E                    & person\_knows\_person\_*.csv         & Person.id | Person.id | creationDate                                                                                \\
        E                    & person\_likes\_comment\_*.csv        & Person.id | Post.id | creationDate                                                                                  \\
        E                    & person\_likes\_post\_*.csv           & Person.id | Post.id | creationDate                                                                                  \\
        A                    & person\_speaks\_language\_*.csv      & Person.id | language                                                                                                \\
        E                    & person\_studyAt\_organisation\_*.csv & Person.id | Organisation.id | classYear                                                                             \\
        E                    & person\_workAt\_organisation\_*.csv  & Person.id | Organisation.id | workFrom                                                                              \\\hline
        N                    & post\_*.csv                          & id | imageFile | creationDate | locationIP | browserUsed | language | content | length | creator | Forum.id | place \\
        E                    & post\_hasTag\_tag\_*.csv             & Post.id | Tag.id                                                                                                    \\\hline
    \end{tabularx}
    \caption{Files output by the CsvMergeForeign serializer (20 in total). The first part of the table contains the static entites, the second part contains the dynamic ones.
        Notation -- \textsf{C}: entity category, \textsf{N}: node, \textsf{E}: edge.}
    \label{table:csv_merge_foreign}
\end{table}

\begin{table}[htb]
    \scriptsize
    \centering
    \begin{tabularx}{\linewidth}{|>{\sffamily}c|>{\tt}l|>{\tt}X|}
        \hline
        \tableHeaderFirst{C} & \tableHeader{File}                      & \tableHeader{Content}                                                                                      \\
        \hline\hline
        N                    & organisation\_*.csv                     & id | type | name | url                                                                                     \\
        E                    & organisation\_isLocatedIn\_place\_*.csv & Organisation.id | Place.id                                                                                 \\
        \hline
        N                    & place\_*.csv                            & id | name | url | type                                                                                     \\
        E                    & place\_isPartOf\_place\_*.csv           & Place.id | Place.id                                                                                        \\
        \hline
        N                    & tag\_*.csv                              & id | name | url                                                                                            \\
        E                    & tag\_hasType\_tagclass\_*.csv           & Tag.id | TagClass.id                                                                                       \\
        \hline
        N                    & tagclass\_*.csv                         & id | name | url                                                                                            \\
        E                    & tagclass\_isSubclassOf\_tagclass\_*.csv & TagClass.id | TagClass.id                                                                                  \\
        \hline\hline
        N                    & comment\_*.csv                          & id | creationDate | locationIP | browserUsed | content | length                                            \\
        E                    & comment\_hasCreator\_person\_*.csv      & Comment.id | Person.id                                                                                     \\
        E                    & comment\_hasTag\_tag\_*.csv             & Comment.id | Tag.id                                                                                        \\
        E                    & comment\_isLocatedIn\_place\_*.csv      & Comment.id | Place.id                                                                                      \\
        E                    & comment\_replyOf\_comment\_*.csv        & Comment.id | Comment.id                                                                                    \\
        E                    & comment\_replyOf\_post\_*.csv           & Comment.id | Post.id                                                                                       \\
        \hline
        N                    & forum\_*.csv                            & id | title | creationDate                                                                                  \\
        E                    & forum\_containerOf\_post\_*.csv         & Forum.id | Post.id                                                                                         \\
        E                    & forum\_hasMember\_person\_*.csv         & Forum.id | Person.id | joinDate                                                               \\
        E                    & forum\_hasModerator\_person\_*.csv      & Forum.id | Person.id                                                                                       \\
        E                    & forum\_hasTag\_tag\_*.csv               & Forum.id | Tag.id                                                                                          \\
        \hline
        N                    & person\_*.csv                           & id | firstName | lastName | gender | birthday | creationDate | locationIP | browserUsed | language | email \\
        E                    & person\_hasInterest\_tag\_*.csv         & Person.id | Tag.id                                                                                         \\
        E                    & person\_isLocatedIn\_place\_*.csv       & Person.id | Place.id                                                                                       \\
        E                    & person\_knows\_person\_*.csv            & Person.id | Person.id | creationDate                                                                       \\
        E                    & person\_likes\_comment\_*.csv           & Person.id | Post.id | creationDate                                                                         \\
        E                    & person\_likes\_post\_*.csv              & Person.id | Post.id | creationDate                                                                         \\
        E                    & person\_studyAt\_organisation\_*.csv    & Person.id | Organisation.id | classYear                                                                    \\
        E                    & person\_workAt\_organisation\_*.csv     & Person.id | Organisation.id | workFrom                                                                     \\
        \hline
        N                    & post\_*.csv                             & id | imageFile | creationDate | locationIP | browserUsed | language | content | length                     \\
        E                    & post\_hasCreator\_person\_*.csv         & Post.id | Person.id                                                                                        \\
        E                    & post\_hasTag\_tag\_*.csv                & Post.id | Tag.id                                                                                           \\
        E                    & post\_isLocatedIn\_place.csv            & Post.id | Place.id                                                                                         \\
        \hline
    \end{tabularx}
    \caption{Files output by the CsvComposite serializer (31 in total). The first part of the table contains the static entites, the second part contains the dynamic ones.
        Notation -- \textsf{C}: entity category, \textsf{N}: node, \textsf{E}: edge.}
    \label{table:csv_composite}
\end{table}

\begin{table}[htb]
    \scriptsize
    \centering
    \begin{tabularx}{\linewidth}{|>{\sffamily}c|>{\tt}l|>{\tt}X|}
        \hline
        \tableHeaderFirst{C} & \tableHeader{File}                   & \tableHeader{Content}                                                                                               \\
        \hline\hline
        N                    & organisation\_*.csv                  & id | type | name | url | place                                                                                      \\
        \hline
        N                    & place\_*.csv                         & id | name | url | type | isPartOf                                                                                    \\
        \hline
        N                    & tag\_*.csv                           & id | name | url | hasType                                                                                           \\
        \hline
        N                    & tagclass\_*.csv                      & id | name | url | isSubclassOf                                                                                      \\
        \hline\hline
        N                    & comment\_*.csv                       & id | creationDate | locationIP | browserUsed | content | length | creator | place | replyOfPost | replyOfComment    \\
        E                    & comment\_hasTag\_tag\_*.csv          & Comment.id | Tag.id                                                                                                 \\
        \hline
        N                    & forum\_*.csv                         & id | title | creationDate | moderator                                                                               \\
        E                    & forum\_hasMember\_person\_*.csv      & Forum.id | Person.id | joinDate                                                                        \\
        E                    & forum\_hasTag\_tag\_*.csv            & Forum.id | Tag.id                                                                                                   \\
        \hline
        N                    & person\_*.csv                        & id | firstName | lastName | gender | birthday | creationDate | locationIP | browserUsed | place | language | email  \\
        E                    & person\_hasInterest\_tag\_*.csv      & Person.id | Tag.id                                                                                                  \\
        E                    & person\_knows\_person\_*.csv         & Person.id | Person.id | creationDate                                                                                \\
        E                    & person\_likes\_comment\_*.csv        & Person.id | Post.id | creationDate                                                                                  \\
        E                    & person\_likes\_post\_*.csv           & Person.id | Post.id | creationDate                                                                                  \\
        E                    & person\_studyAt\_organisation\_*.csv & Person.id | Organisation.id | classYear                                                                             \\
        E                    & person\_workAt\_organisation\_*.csv  & Person.id | Organisation.id | workFrom                                                                              \\
        \hline
        N                    & post\_*.csv                          & id | imageFile | creationDate | locationIP | browserUsed | language | content | length | creator | Forum.id | place \\
        E                    & post\_hasTag\_tag\_*.csv             & Post.id | Tag.id                                                                                                    \\
        \hline
    \end{tabularx}
    \caption{Files output by the CsvCompositeMergeForeign serializer (18 in total). The first part of the table contains the static entites, the second part contains the dynamic ones.
        Notation -- \textsf{C}: entity category, \textsf{N}: node, \textsf{E}: edge.}
    \label{table:csv_composite_merge_foreign}
\end{table}

\subsection{Serializers}
\label{sec:legacy-serializers}

The datasets are generated in the \texttt{social\_network/} directory, split into static and dynamic parts (\autoref{fig:schema}).
The filenames (without the extension) end in \texttt{\_i\_j} where \texttt{i} is the block id and \texttt{j} is the partition id (set by \texttt{numThreads}).
The SUT has to take care only of the generated Dataset to be bulk loaded. Using \texttt{NULL} values for storing optional values is allowed.

Datagen's CSV (Comma Separated Values) serializers produce text files which use the pipe character ``\texttt{|}'' as the primary field separator and the semicolon character ``\texttt{;}'' as a separator for multi-valued attributes (for the composite serializers).
The CSV files are stored in two subdirectories: \texttt{static/} and \texttt{dynamic/}.
Depending on the number of threads used for generating the dataset, the number of files varies, since there is a file generated per thread. We indicate this with ``\texttt{*}'' in the specification.

The following CSV variants are supported:
    \begin{itemize}
      \item \textbf{CsvBasic:}
      Each entity, relation and attribute with a cardinality larger than one (including attributes \texttt{Person.email} and \texttt{Person.speaks}), are output in a separate file.
      Generated files are summarized in \autoref{table:csv_basic}.

      \item \textbf{CsvMergeForeign:}
      This serializer is similar to CsvBasic, but relations that have a cardinality of 1-to-N are merged in the entity files as a foreign keys.
      There are 13~such relations in total:
      \begin{itemize}
          \item comment\_hasCreator\_person, comment\_isLocatedIn\_place, comment\_replyOf\_comment, comment\_replyOf\_post (merged to Comment);
          \item forum\_hasModerator\_person (merged to Forum);
          \item organisation\_isLocatedIn\_place (merged to Organisation);
          \item person\_isLocatedIn\_place (merged to Person);
          \item place\_isPartOf\_place (merged to Place);
          \item post\_hasCreator\_person, post\_isLocatedIn\_place, forum\_containerOf\_post (merged to Post);
          \item tag\_hasType\_tagclass (merged to Tag);
          \item tagclass\_isSubclassOf\_tagclass (merged to TagClass)
      \end{itemize}
      Generated files are summarized in \autoref{table:csv_merge_foreign}.

      \item \textbf{CsvComposite:}
      Similar to the CsvBasic format but each entity, and relations with a cardinality larger than one, are output in a separate file.
      Multi-valued attributes (\texttt{Person.email} and \texttt{Person.speaks}) are stored as composite values.
      Generated files are summarized in \autoref{table:csv_composite}.

      \item \textbf{CsvCompositeMergeForeign:}
      Has the traits of both the CsvComposite and the CsvMergeForeign formats.
      Multi-valued attributes (\texttt{Person.email} and \texttt{Person.speaks}) are stored as composite values.
      Generated files are summarized in \autoref{table:csv_composite_merge_foreign}.
    \end{itemize}

Additionally, the Hadoop \datagen can generate files in Turtle format (\texttt{.ttl}).

\paragraph{Inheritance}

The inheritance hierarchies in the schema have two important characteristics
(1)~all subclasses use the same id space, \eg there cannot be a Comment and a Post with id 1 at the same time,
(2)~they are serialized to CSVs using either the \emph{map hierarchy to single table} or \emph{map each concrete class to its own table} strategies\footnote{\url{http://www.agiledata.org/essays/mappingObjects.html}}:

\begin{description}
    \item[Message = Comment | Post]
    \emph{Map each concrete class to its own table} is used \ie separate CSV files are used for the Comment and the Post classes.

    \item[Place = City | Country | Continent]
    \emph{Map hierarchy to single table} is used, \ie all Place node are serialized in a single file. A discriminator attribute ``type'' is used with the value denoting the concrete class, \eg ``Country''.

    \item[Organisation = Company | University]
    \emph{Map hierarchy to single table} is used, \ie all Organisation nodes are serialized in a single fiel. A discriminator attribute ``type'' is used with the value denoting the concrete class, \eg ``Company''.
\end{description}

\subsection{Update Streams}

\begin{table}[htb]
    \scriptsize
    \centerline{\begin{tabular}{|c|c|c|c|}
        \hline
        start time ($t_\text{s}$) & dependant time ($t_\textrm{d}$) & operation id & \ldots \\
        \hline
    \end{tabular}}
    \caption{
        Generic schema of update (insert) stream files.
        The start time ($t_\text{s}$) is identical to the creationDate attribute (repeated later in the row).}
    \label{table:update_stream_generic_schema}
\end{table}

\begin{table}[htb]
	\scriptsize
	\centerline{\begin{tabular}{|l|}
		\hline
		$t_\text{s}$ | $t_\textrm{d}$ | 1 | personId | personFirstName | personLastName | gender | birthday | creationDate | locationIP | browserUsed | cityId | languages | emails | tagIds | studyAt | workAt \\\hline
		$t_\text{s}$ | $t_\textrm{d}$ | 2 | personId | postId | creationDate \\\hline
		$t_\text{s}$ | $t_\textrm{d}$ | 3 | personId | commentId | creationDate \\\hline
		$t_\text{s}$ | $t_\textrm{d}$ | 4 | forumId | forumTitle | creationDate | moderatorPersonId | tagIds \\\hline
		$t_\text{s}$ | $t_\textrm{d}$ | 5 | forumId | personId | creationDate \\\hline
		$t_\text{s}$ | $t_\textrm{d}$ | 6 | postId | imageFile | creationDate | locationIP | browserUsed | language | content | length | authorPersonId | forumId | countryId | tagIds \\\hline
		$t_\text{s}$ | $t_\textrm{d}$ | 7 | commentId | creationDate | locationIP | browserUsed | content | length | authorPersonId | countryId | replyToPostId | replyToCommentId | tagIds \\\hline
		$t_\text{s}$ | $t_\textrm{d}$ | 8 | person1Id | person2Id | creationDate \\
		\hline
	\end{tabular}}
	\caption{Schemas of the lines in the update stream (insert stream) files.}
	\label{table:update_stream_schemas}
\end{table}

The generic schema is given in \autoref{table:update_stream_generic_schema}, while the concrete schema of each insert operation is given in \autoref{table:update_stream_schemas}.
The update stream files are generated in the \texttt{social\_network/} directory and are grouped as follows:

\begin{itemize}
    \item \texttt{updateStream\_*\_person.csv} files contain update operation 1: \queryRefCard{insert-01}{INS}{1}
    \item \texttt{updateStream\_*\_forum.csv} files contain update operations 2--8: %
    \queryRefCard{insert-02}{INS}{2}
    \queryRefCard{insert-03}{INS}{3}
    \queryRefCard{insert-04}{INS}{4}
    \queryRefCard{insert-05}{INS}{5}
    \queryRefCard{insert-06}{INS}{6}
    \queryRefCard{insert-07}{INS}{7}
    \queryRefCard{insert-08}{INS}{8}
\end{itemize}

Remark: update streams in \interactivevone only contain inserts. Delete operations are being designed and will be released later.

\subsection{Substitution Parameters}

The substitution parameters are generated in the \texttt{substitution\_parameters/} directory.
Each parameter file is named \texttt{\{interactive|bi\}\_<id>\_param.txt}, corresponding to an operation of
Interactive complex reads (\queryRefCard{interactive-complex-read-01}{IC}{1}--\queryRefCard{interactive-complex-read-14-v1}{IC}{14v1}) and
BI reads (\queryRefCard{bi-read-01}{BI}{1}--\queryRefCard{bi-read-20}{BI}{20}).
The schemas of these files are defined by the operator, \eg the schema of \queryRefCard{interactive-complex-read-01}{IC}{1} is ``\texttt{personId|firstName}''.

\textbf{Warning.} Note that the substitution parameter files use UNIX epoch timestamps (which should be converted to a datetime value in GMT+0).

\chapter{Example graph}
\label{sec:example-graph}

\begin{figure}[ht]
    \centering
    \includegraphics[width=\textwidth]{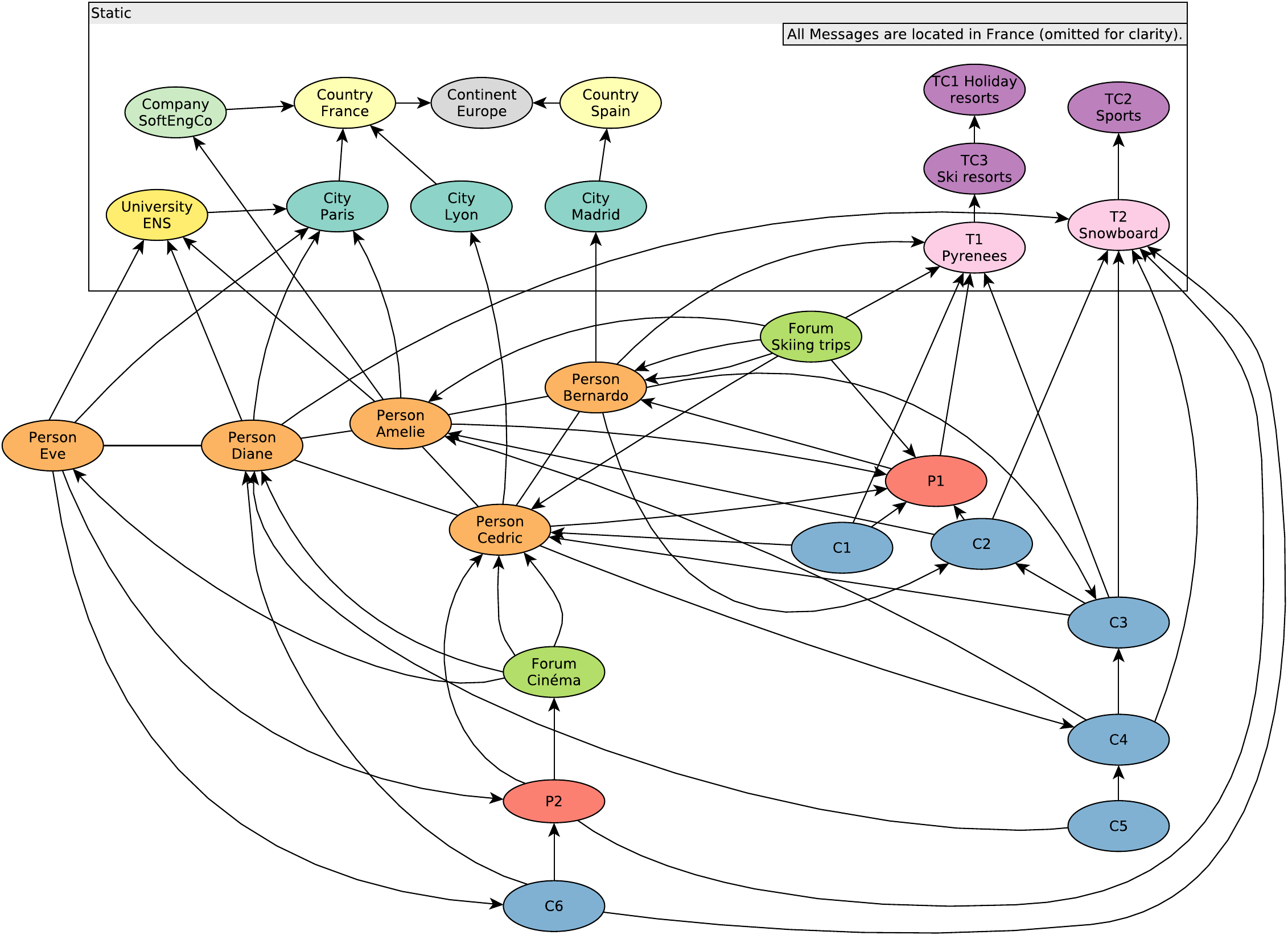}
    \caption{Example graph snapshot (without update operations).}
    \label{fig:example-graph-without-updates}
\end{figure}

\begin{figure}[ht]
    \centering
    \includegraphics[width=\textwidth]{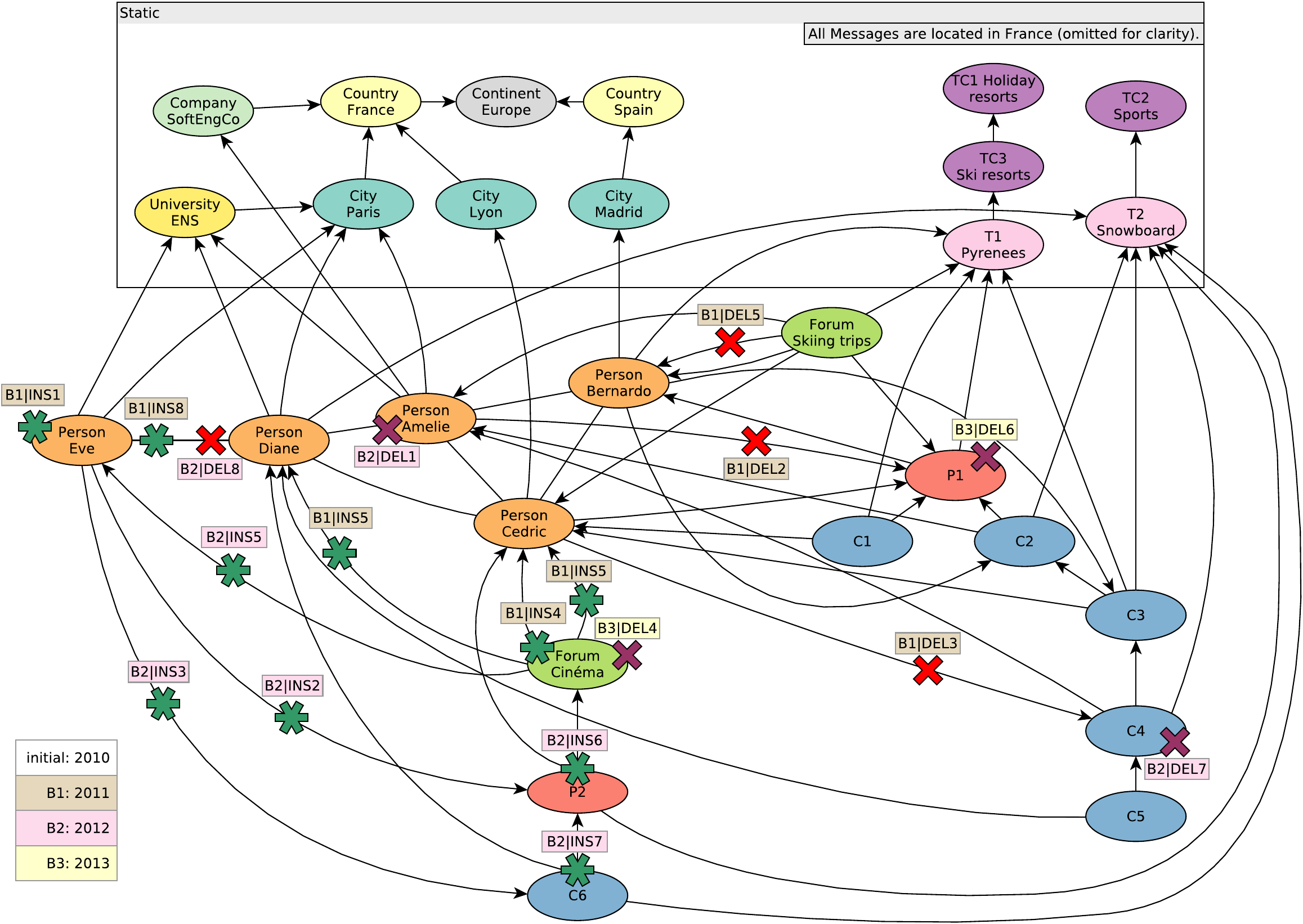}
    \caption{Example graph with update operations.}
    \label{fig:example-graph-with-updates}
\end{figure}

\autoref{fig:example-graph-without-updates} shows a static snapshot of an example graph, while \autoref{fig:example-graph-with-updates} shows an example graph with update operations.
Insertions are denoted with a green asterisk~\includegraphics[scale=0.25]{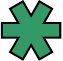}.
Deletions of a single element are denoted with a red cross~\includegraphics[scale=0.25]{patterns/delete-single},
while recursive deletions are denoted with a purple cross~\includegraphics[scale=0.25]{patterns/delete-recursively}.

\end{document}